\pdfoutput=1
\documentclass[twoside,letterpaper,titlepage,nofootinbib,prd,12pt,superscriptaddress,floatfix]{revtex4-1}

\usepackage[english]{babel}
\usepackage{color}
\usepackage{multirow}
\usepackage{subfigure}
\usepackage{graphicx}
\usepackage{epsfig}
\usepackage{amsmath}
\usepackage{amssymb}
\usepackage{amsfonts}
\usepackage{wasysym}
\usepackage{slashed}
\usepackage{bbm}
\usepackage{ctable,longtable}
\usepackage{cancel}
\usepackage{txfonts}
\usepackage[tight]{units} 
\usepackage{bigstrut}
\usepackage{multirow}

\def\IC{{\scriptscriptstyle{\rm IC}}}

\def\SU2U1{{\rm SU}(2)\times{\rm U}(1)}

\def\exp{{\rm exp}}

\mathchardef\qsm=63
\mathchardef\pls=43
\mathchardef\mns=512
\mathchardef\plm=518
\mathchardef\eql=61
\mathchardef\smallleft=300
\mathchardef\smallright=301
\mathchardef\perslsh=47
\mathchardef\les=316
\mathchardef\gre=318
\mathchardef\leq=532
\mathchardef\grq=533
\chardef\usc=95
\chardef\til=126

\def\sqr#1#2#3{{\vcenter{\hrule height.#3ex\hbox{\vrule width.#2ex height#1ex
    \kern#1ex\vrule width.#3ex}\hrule height.#2ex}}}

\def\angleto{\vrule width.035em height2.1ex depth-.56ex\unskip\kern-.6ex\to}
\def\perchc#1{{\raise.4ex\hbox{$\mkern4mu#1{\it\perslsh}_
             {\mkern-5mu\scriptscriptstyle{{\rm o}\!{\rm o}}}^
             {\mkern-12.8mu\scriptscriptstyle{\rm o}}$}}}

\catcode`\@=11 
\def\parenbar{\mathpalette\p@renb@r}
\def\p@renb@r#1#2{\vbox{%
  \ifx#1\scriptscriptstyle \dimen@.7em\dimen@ii.2em\else
  \ifx#1\scriptstyle \dimen@.8em\dimen@ii.25em\else
  \dimen@1em\dimen@ii.4em\fi\fi \offinterlineskip
  \ialign{\hfill##\hfill\cr
    \vbox{\hrule width\dimen@ii}\cr
    \noalign{\vskip-.3ex}%
    \hbox to\dimen@{$\mathchar300\hfil\mathchar301$}\cr
    \noalign{\vskip-.3ex}%
    $#1#2$\cr}}}
\catcode`\@=12 

\newbox\struttbox
\setbox\struttbox=\hbox{\vrule height1.65ex depth.485ex width0pt}
\def\strutt{\relax\ifmmode\copy\struttbox\else\unhcopy\struttbox\fi}
\def\stru#1#2{\relax\ifmmode\hbox{\vrule height#1 depth#2 width0pt}
\else\vrule height#1 depth#2 width0pt\fi}

\def\ronum#1{\uppercase\expandafter{\romannumeral#1}}
\def\ronuml#1{\expandafter{\romannumeral#1}}

\DeclareMathAlphabet{\mathbf}{OT1}{cmr}{bx}{sl}

\newcommand{\lsim}      {\mbox{\raisebox{-0.4ex}{$\;\stackrel{<}{\scriptstyle \sim}\;$}}}

 {\end{list}}
 {\end{list}}
 {\end{list}}

\catcode`\@=11 
\newlength{\@fninsert}
\newlength{\@fnwidth}
\renewcommand{\@makefntext}[1]%
  {\noindent\makebox[\@fninsert][r]{\@makefnmark}\hfil%
  \parbox[t]{\@fnwidth}{#1}}
\catcode`\@=12 
\catcode`\@=11 
\renewcommand\section{\@startsection{section}{1}{\z@}%
                                   {-3.5ex \@plus -1ex \@minus -.2ex}%
                                   {2.3ex \@plus.2ex}%
                                   {\normalfont\Large\bfseries}}
\renewcommand\subsection{\@startsection{subsection}{2}{\z@}%
                                   {-3.25ex\@plus -1ex \@minus -.2ex}%
                                   {1.5ex \@plus .2ex}%
                                   {\normalfont\large\bfseries}}
\renewcommand\subsubsection{\@startsection{subsubsection}{3}{\z@}%
                                   {-3.25ex\@plus -1ex \@minus -.2ex}%
                                   {1.5ex \@plus .2ex}%
                                   {\normalfont\normalsize\bfseries}}
\renewcommand\paragraph{\@startsection{paragraph}{4}{\z@}%
                                   {3.25ex \@plus1ex \@minus.2ex}%
                                   {1.2ex \@plus .2ex}%
                                   {\normalfont\normalsize\bfseries}}
\catcode`\@=12 

\newcommand{\eps}{\varepsilon}

\newcommand{\eVq}  {\text{eV}^2}
\newcommand{\Dmq}  {\Delta m^2}

\newcommand{\bi}{\begin{itemize}}
\newcommand{\ei}{\end{itemize}}
\newcommand{\be}{\begin{equation}}
\newcommand{\ee}{\end{equation}}
\newcommand{\bea}{\begin{eqnarray}}
\newcommand{\eea}{\end{eqnarray}}

\newcommand{\beq}{\begin{eqnarray}}
\newcommand{\eeq}{\end{eqnarray}}

\newcommand{\bmp}{\noindent\begin{minipage}{16cm}}
\newcommand{\emp}{\end{minipage}\vskip 7mm} 

\def\drawbox#1#2{\hrule height#2pt
        \hbox{\vrule width#2pt height#1pt \kern#1pt
              \vrule width#2pt}
              \hrule height#2pt}

\def\Asym#1#2{\vcenter{\vbox{\drawbox{#1}{#2}
              \kern-#2pt 
              \drawbox{#1}{#2}}}}

\def\simge{\mathrel{%
   \rlap{\raise 0.511ex \hbox{$>$}}{\lower 0.511ex \hbox{$\sim$}}}}

\def\simle{\mathrel{
   \rlap{\raise 0.511ex \hbox{$<$}}{\lower 0.511ex \hbox{$\sim$}}}}

\def\s#1{\setbox0=\hbox{$#1$}%
\rlap{\ifdim\wd0>.7em\kern.22\wd0\else\kern.1\wd0\fi /}#1}

\def\nuebar{\bar\nu_e}

\def\gtwid{\mathrel{\raise.3ex\hbox{$>$\kern-.75em\lower1ex\hbox{$\sim$}}}}
\def\ltwid{\mathrel{\raise.3ex\hbox{$<$\kern-.75em\lower1ex\hbox{$\sim$}}}}

\def\nubar{\bar{\nu}}
\def\inbar{\,\vrule height1.5ex width.4pt depth0pt}
\def\IR{\relax{\rm I\kern-.18em R}}
\def\IC{\relax\hbox{$\inbar\kern-.3em{\rm C}$}}

\def\weakangle{\sin^2{\theta}_W}

\newcommand{\isotope}[2]{$^{#2}{\rm #1}$}

\begin{document}
\selectlanguage{english}
\makeatletter

\pagestyle{empty}
\thispagestyle{empty}
\title{Light Sterile Neutrinos: A White Paper}
\date{\today}
\clearpage
\author{K.~N.~Abazajian\footnotemark[1]\footnotetext[1]{Section editor}}
\affiliation{University of California, Irvine}
\author{M.~A.~Acero}
\affiliation{Instituto de Ciencias Nucleares, Universidad Nacional Aut\'onoma de M\'exico}
\author{S.~K.~Agarwalla}
\affiliation{Instituto de Fisica Corpuscular, CSIC and Universidad de Valencia}
\author{A.~A.~Aguilar-Arevalo}
\affiliation{Instituto de Ciencias Nucleares, Universidad Nacional Aut\'onoma de M\'exico}
\author{C.~H.~Albright}
\affiliation{Northern Illinois University }
\affiliation{Fermi National Accelerator Laboratory}
\author{S.~Antusch}
\affiliation{University of Basel}
\author{C.~A.~Arg\"uelles}
\affiliation{Pontificia Universidad Cat\'{o}lica del Per\'{u}}
\author{A.~B.~Balantekin}
\affiliation{University of Wisconsin, Madison}
\author{G.~Barenboim\footnotemark[1]}
\affiliation{Instituto de Fisica Corpuscular, CSIC and Universidad de Valencia}
\author{V.~Barger}
\affiliation{University of Wisconsin, Madison}
% ...

\maketitle

\clearpage

\thispagestyle{empty}
~~\\

\vspace{8.5cm} 

~~\\

\begin{center}
{\it In memoriam}\\
Ramaswami ``Raju'' S. Raghavan\\
\oldstylenums{1937} -- \oldstylenums{2011}
\end{center}
\clearpage

\parindent 10pt
\pagenumbering{roman}                   
\setcounter{page}{1}
\thispagestyle{plain}
\pagestyle{plain}
\tableofcontents
\clearpage

\pagenumbering{arabic}                   
\setcounter{page}{1}

\section*{Executive summary}

This white paper addresses the hypothesis of sterile neutrinos based 
on recent anomalies observed in neutrino experiments.  It is by no 
means certain that sterile neutrinos are responsible for the set of
anomalies which have triggered the current effort, but the 
extraordinary consequence of such a possibility justifies a detailed 
assessment of status of the field. Decades of experimentation have 
produced a vast number of results in neutrino physics and astrophysics, 
some of which are in perfect agreement with only three 
active\footnote{Active neutrinos are those which couple to $Z$ and $W$ 
bosons.} neutrinos, while a small subset calls for physics beyond 
the standard model\footnote{Here, the standard model is to be 
understood to include massive neutrinos.}. The first, and individually 
still most significant, piece pointing towards new physics is the LSND 
result, where electron antineutrinos were observed in a pure muon 
antineutrino beam. The most straightforward interpretation of the LSND 
result is antineutrino oscillation with a mass squared difference, 
$\Delta m^2$, of about $1\,\mathrm{eV}^2$\@. Given that solar neutrino 
oscillations correspond to $\Delta m^2_\mathrm{sol}\simeq 7 
\times10^{-5}\,\mathrm{eV}^2$ and atmospheric neutrino oscillations 
correspond to $\Delta m^2_\mathrm{atm}\simeq 2.3\times10^{-3}\,
\mathrm{eV}^2$, the LSND $\Delta m^2$ requires a fourth neutrino. 
However, the results from the Large Electron Positron collider (LEP) at 
CERN on the invisible decay width of the $Z$ boson show that there are 
only three neutrinos with a mass below one half of the mass of the $Z$ 
boson, which couple to the $Z$ boson, and therefore the fourth neutrino, 
if it indeed exists, can not couple to the $Z$ boson and hence is a 
sterile neutrino, {\it i.e.}~a Standard Model gauge singlet.

A new anomaly supporting the sterile neutrino hypothesis emerges from 
the recent re-evaluations of reactor antineutrino fluxes, which find a 
3\% increased flux of antineutrinos relative to the previous 
calculations.  At the same time, the experimental value for the neutron 
lifetime became significantly smaller, which in turn implies a larger 
inverse $\beta$-decay cross section. In combination with the
previously-neglected effects from long-lived isotopes which do not
reach equilibrium in a nuclear reactor, the overall expectation value
for antineutrino events from nuclear reactors increased by roughly
6\%.  As a result, more than 30 years of data from reactor neutrino
experiments, which formerly agreed well with the flux prediction, have 
become the observation of an apparent 6\% deficit of electron 
antineutrinos. This is known as the reactor antineutrino anomaly and is 
compatible with sterile neutrinos having a $\Delta m^2_\mathrm{sterile}
>1\,\mathrm{eV}^2$. 

Another hint consistent with sterile neutrinos comes from the source
calibrations performed for radio-chemical solar neutrino experiments
based on gallium. In these calibrations very intense sources of $^{51}$Cr 
and $^{37}$Ar, which both decay via electron capture and emit 
mono-energetic electron neutrinos, were placed in proximity to the
detector and the resulting event rate were measured. Both the source
strength and reaction cross section are known with some precision and
a 5-20\% deficit of the measured to expected count rate was observed.
Again, this result would find a natural explanation by a sterile
neutrino with $\Delta m^2_\mathrm{sterile}>1\,\mathrm{eV}^2$, which would 
allow some of the electron neutrinos from the source to ``disappear'' 
before they can interact.  This anomaly persists even if one used the 
minimum cross section compatible with the precisely know $^{71}$Ga 
lifetime.

The aforementioned results suggesting a sterile neutrino with a mass
around $1\,\mathrm{eV}$ have to be contrasted with a number of results
which clearly disfavor this interpretation. The strongest constraints 
derive from the non-observation of muon neutrino disappearance by 
accelerator experiments like CDHSW or MINOS. Bounds on the disappearance 
of electron neutrinos are obtained from KARMEN and LSND, as well.  The 
MiniBooNE neutrino result, a non-observation of electron neutrino 
appearance in a muon neutrino beam, is incompatible with the LSND 
appearance result, if CP is conserved. On the other hand, the 
antineutrino result from the same experiment is fully compatible with 
the LSND result. A further difficulty in interpreting experimental 
evidence in support of a light sterile neutrino is that the effects are 
purely in count rates.  The energy and distance-dependent characteristic 
of the oscillation phenomena associated with sterile neutrinos remains to 
be observed.

The existing data from oscillation experiments, both for and against the 
sterile neutrino hypothesis, are summarized in  Section~\ref{sec:oscillation} 
of this report, while Section~\ref{sec:global} discusses the compatibility 
of and tensions between the various oscillation data sets. 

Cosmological data, mainly from observations of the cosmic microwave
background and large scale structure favor the existence of a fourth light 
degree-of-freedom which could be a sterile neutrino.  At the same time 
the standard cosmological evolution model prefers this neutrino to be lighter 
than $1\,\mathrm{eV}$\@.  The relevant cosmological and astrophysical evidence is 
discussed in Section~\ref{sec:astro}.

From a theoretical point of view (see Section~\ref{sec:theory}) the
existence of sterile neutrinos is a rather natural consequence of
neutrinos having a non-zero mass.  Sterile neutrinos are gauge
singlets and as a result no {\it a priori} scale for their mass is
set.  Once neutrino mass generation via the seesaw mechanism is put
into the wider context of grand unification and leptogenesis, light
sterile neutrinos are slightly less natural.  However, in this
context, expectations for the mass of sterile neutrinos are ultimately
based on an attempt to predict the eigenvalues of a 6$\times$6 matrix 
in which we do not know any entries and therefore, having one
or several sterile neutrinos in the 1~eV mass range is certainly far
from being a surprise. The same is true for keV sterile neutrinos,
which are warm dark matter candidates.

In summary, there are a number of experimental results that appear anomalous 
in the context of the standard 3 neutrino framework, and can be explained by 
a sterile neutrino with mass around 1\,eV\@.  At the same time, there are a
number of results which are in conflict with this interpretation. The data 
collected to date present an incomplete, perhaps even contradictory picture, 
where 2-3$\,\sigma$ agreement {\em in favor of} and {\em in contradiction to} 
the existence of sterile neutrinos is present.  The need thus arises to 
provide a more ardent and complete test of the sterile neutrino hypothesis, 
which will unambiguously confirm ir refute the interpretation of past 
experimental results.  This white paper documents the currently available 
evidence for and against sterile neutrinos, it highlights the theoretical 
implications of their existence and provides an overview of possible future 
experiments (see the appendix) to test the LSND result and the sterile neutrino 
hypothesis. The overriding goal is to provide the motivation for a new round of
measurements so that the questions laid out here can be definitively answered.

\clearpage

\section{Theory and Motivation}
\label{sec:theory}

In this Chapter we provide an overview of the theoretical
background of sterile neutrinos. We will set the stage by defining the
term ``sterile neutrino'', and will present model building aspects of these
entities. Sterile neutrinos are naturally present in
many theories beyond the standard model, in particular in several manifestations 
of the seesaw mechanism. We will be often interested in directly observable sterile 
neutrinos,  and will categorize different 
possibilities for naturally accommodating them. More concretely, the so-called 
low energy seesaw will be analyzed in some detail. 
Particle physics aspects of keV scale neutrinos as warm dark matter
candidates will be focussed on as well. Finally, for the sake of
completeness, other non-standard neutrino physics fields, often
connected to sterile neutrinos or used as an alternative explanation
to some of the current experimental hints, are summarized. 

\subsection{Introduction: What is a Sterile Neutrino?}
A sterile neutrino is a neutral lepton with no ordinary weak
interactions except those induced by mixing. They are present in most
extensions of the standard model, and in principle can have any
mass. Very heavy sterile neutrinos are utilized in the minimal type I
seesaw model~\cite{Minkowski:1977sc,gmrs79,Mohapatra:1979ia,Yanagida:1979as,Schechter:1980gr}
and play a pivotal role in leptogenesis~\cite{Fukugita:1986hr,Davidson:2008bu}. However,
here we are especially concerned with relatively light sterile
neutrinos that mix significantly  with ordinary neutrinos and those
that are relevant to neutrino oscillation experiments, astrophysics,
dark matter, etc. We first introduce the necessary terminology and
formalism to discuss sterile neutrinos and their possible mixing with
ordinary neutrinos, following the notation
of~\cite{Langacker:1226768}.

A massless fermion can be described by a Weyl spinor, which 
necessarily has two components of opposite chirality related
(up to Dirac indices) by Hermitian conjugation. Chirality coincides
with helicity for a massless particle and is associated with the
chiral projections  $P_{L,R}= \frac{1}{2} (1\mp \gamma^5)$. Thus, a
left-chiral Weyl spinor $\psi_L=P_L \psi_L$, which annihilates a
left-chiral particle,  is related to 
its right-chiral CP conjugate  $\psi^c_R= P_R \psi^c_R$, which
annihilates a right-chiral antiparticle, by   
\begin{equation} \psi^c_R=\mathcal{C} \gamma^{0T}
\psi_L^\ast. \label{weyl} \end{equation} 
 In (\ref{weyl}) $\mathcal{C}$ is the charge conjugation
matrix\footnote{$\mathcal{C}=i \gamma^2 \gamma^0$ in the Pauli-Dirac
representation.} and $\psi^\ast_{L\alpha}\equiv\left(
\psi_{L\alpha}\right)^\dagger$. $ \psi^c_R$  exists 
 even if CP is not conserved. One can just as well define a
right-chiral Weyl spinor $\psi'_R$ and its associated 
 left-chiral anti-spinor $\psi'^c_L$. It is a matter of convention
which is called the particle and which the antiparticle.

A Weyl neutrino is said to be active (or ordinary or doublet) if it
participates in the  conventional charged and neutral current  weak interactions. This
means that it transforms as a member of a doublet together with a charged lepton partner
under the weak $SU(2)$ gauge group of the standard model. There are
three light active neutrinos in nature, $\nu_{eL}$, $\nu_{\mu L}$, and
$\nu_{\tau L}$, which are associated respectively with $e^-_L$,
$\mu^-_L$, and $\tau^-_L$ in charged current transitions. Their CP
conjugates, $\nu^c_{\alpha R}, \, \alpha= e, \mu, \tau$, are the weak partners of the
$\alpha^+_R$. Including small neutrino masses, the $\nu_{\alpha L}$ (and
$\nu^c_{\alpha R}$) are linear combinations of the mass eigenstates. The
observed $Z$-boson decay width implies that any additional active neutrinos are quite heavy, $m_\nu \gtrsim M_Z/2$.

A  {sterile} (or singlet or right-handed)  neutrino is an $SU(2)$
singlet which does not take part in weak interactions 
 except those induced by mixing with active neutrinos. It may participate
in Yukawa interactions involving  the Higgs boson or in interactions involving new
physics. Many extensions of the standard model introduce a
right-chiral sterile $\nu_R$, which has a left-chiral CP conjugate
$\nu^c_L$.

Chirality can be a conserved quantum number for massless
fermions, depending on the nature of their interactions. However, fermion mass terms violate chirality and also break
its equivalence to helicity (which is not  a Lorentz
invariant quantity). For example,  a weak transition involving, {\it e.g.}, a massive
but relativistic $\nu_L$, may produce the ``wrong helicity'' state, with
amplitude $\propto m_\nu/E_\nu$. The chirality violation is due  to
the fact that a fermion mass term describes a transition between right
and left chiral Weyl spinors:
\begin{equation} 
-\mathcal{L}= m  \left(  \bar{\psi}_L \psi_R+
\bar{\psi}_R\psi_L \right),\label{fmass} 
\end{equation}
where we have taken $m$ to be real and non-negative by choosing an
appropriate  relative $\psi_{L,R}$ phase. There are two important
types of fermion mass terms, Dirac and Majorana, depending on whether
or not $\psi_R$ is distinct from $\psi^c_R$, the CP conjugate  of
$\psi_L$.

\begin{figure}
\centerline{
\includegraphics[width=0.9\textwidth]{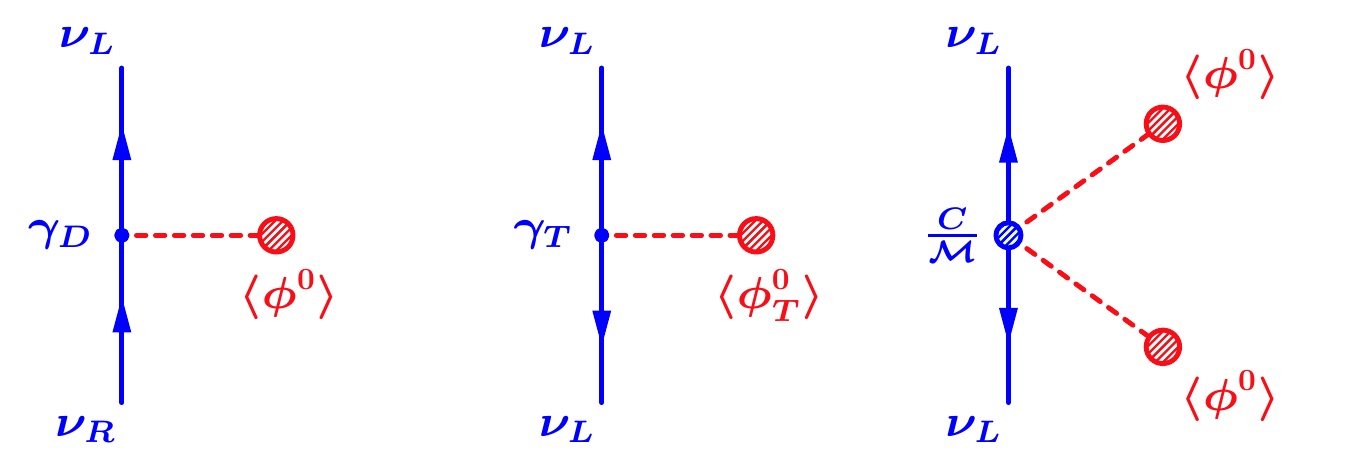}}
\caption{Left: Dirac mass term generated by a Higgs doublet.
Center: Majorana mass term generated by a Higgs triplet. 
Right: Majorana mass term generated by a higher-dimensional operator.}
\label{pgl_fig1}
\end{figure}

A {Dirac neutrino mass} term relates two distinct Weyl spinors:
\begin{equation} -\mathcal{L}_D= m_D \left( \bar{\nu}_L \nu_R
+\bar{\nu}_{R}\nu_{L}\right)= m_D \bar \nu_D \nu_D, \label{diracmass}
\end{equation}
where, in the second form, $\nu_D \equiv \nu_L+\nu_R$ is the {Dirac field}. $\nu_D$ (and its
conjugate $\nu^c_D=\nu^c_R+\nu^c_L$) have four components, $\nu_L,
\nu_R, \nu^c_R$, and $\nu^c_L$.  One can define a global  lepton number
($L=1$ for  $\nu_D$ and $L=-1$ for   $\nu^c_D$), which is conserved by
$\mathcal{L}_D$. In most theories $\nu_L$ and $\nu_R$ in a Dirac mass
term are respectively active and sterile. In this case,
$\mathcal{L}_D$  violates the third component of weak isospin by
$\Delta t^3_L=\pm \frac{1}{2}$. $m_D$ can be generated by the
vacuum expectation value (VEV) of the neutral component $\phi^0$ of a
Higgs  doublet, as illustrated in  Figure \ref{pgl_fig1},  yielding 
\begin{equation}   
m_D =y  \langle{\phi^0}\rangle=y v /\sqrt{2},
\label{Higgsdirac}
\end{equation}
where $y$ is the Yukawa coupling and $v \simeq 246$ GeV is the weak
scale. This is analogous to the generation of quark and charged lepton
masses in the standard model. In order to accommodate the observed neutrino masses,
 $y$ must be extremely small, {\it e.g.}~$y \sim 10^{-12}$
for $m_D \sim 0.1$ eV, as compared to $y_e\sim 3\times  10^{-6}$
for the electron. These small Dirac neutrino mass are often considered unnatural, or as evidence that some more
subtle underlying mechanism is at play. For example, one may interpret the tiny neutrino Yukawa couplings as 
evidence that the elementary Yukawa  couplings
$y$ are required to vanish by some new discrete, global, or
gauge symmetry. A small effective Yukawa coupling can then be
generated by  higher-dimension
operators~\cite{Langacker:1998ut,ArkaniHamed:2000bq} or by
non-perturbative effects such as  string
instantons~\cite{Cvetic:2008hi,Blumenhagen:2009qh}. Another
possibility is that a small Dirac coupling is due to wave function
overlaps between $\nu_L$ and $\nu_R$ in theories involving large
and/or warped extra
dimensions~\cite{Dienes:1998sb,ArkaniHamed:1998vp}.  

A {Majorana mass} term requires only one Weyl spinor. Majorana masses are forbidden by color and electromagnetic gauge invariance for quarks and charged leptons, but are possible for both active (after electroweak symmetry breaking) and sterile neutrinos if there is no conserved lepton number. For an active neutrino $\nu_L$,
\begin{equation} 
-\mathcal{L}_T  =  \frac{m_{L}}{2}\left(\bar{\nu}_{L}\nu_{R}^{c}+\bar{\nu}_{R}^{c}\nu_{L}\right)
= \frac{m_{L}}{2}\left(\bar{\nu}_{L}\mathcal{C}\bar{\nu}_{L}^{T}+\nu_L^T\mathcal{C} \nu_L\right)= 
\frac{m_{L}}{2}\bar{\nu}_M\nu_M, \label{majoranaactive}
\end{equation}
where $\nu_M\equiv\nu_{L}+\nu_{R}^{c}=\nu^c_M$ is a (self-conjugate)
two-component Majorana field. $\mathcal{L}_T $ violates lepton number
by two units, $\Delta L=\pm 2$, and weak isospin by one unit, $\Delta
t^3_L = \pm 1$ (hence one sometimes writes $m_L$ as $m_T$, with $T$ for triplet). $m_L$  can be
generated by a Higgs triplet, with a small Yukawa coupling $\gamma_T$
and/or a small VEV $\langle{\phi^0_T}\rangle$, as illustrated in
Figure \ref{pgl_fig1}. It could instead be associated with  a
higher-dimensional operator (the Weinberg
operator~\cite{Weinberg:1980bf}) involving two Higgs doublets, with a
coefficient $C/\mathcal{M}$. The scale $\mathcal{M}$ represents some
heavy new physics which has been integrated out, such as the exchange
of a very heavy Majorana sterile neutrino (the type I 
seesaw)~\cite{Minkowski:1977sc,gmrs79,Mohapatra:1979ia,Yanagida:1979as,Schechter:1980gr}, 
a heavy scalar triplet (the type II seesaw)~\cite{Hambye:2000ui}, 
a fermion triplet (the type III seesaw)~\cite{Foot:1988aq}, or new
degrees of freedom in a string theory~\cite{Langacker:2011bi}. The second form in
(\ref{majoranaactive}) emphasizes that $\mathcal{L}_T$ also describes
the creation or annihilation of two neutrinos. 
A sterile neutrinos $\nu_R$ can also have a Majorana mass term,
\begin{equation} 
   -\mathcal{L}_S =  \frac{M_{R}}{2}\left(\bar{\nu}_{L}^{c} \nu_{R}+\bar{\nu}_{R} \nu_{L}^{c}\right)
   =
\frac{M_{R}}{2}\left(\bar{\nu}^c_{L}\mathcal{C}\bar{\nu}_{L}^{cT}+\nu_L^{cT}\mathcal{C}
\nu^c_L\right)=\frac{M_{R}}{2}\bar{\nu}_{M_S}\nu_{M_S},\label{Majoranasterile}
\end{equation} 
where $\nu_{M_S} \equiv \nu^c_L+\nu_R= \nu^c_{M_S}$ is a Majorana
field. $\mathcal{L}_S$ also violates lepton number by $\Delta L=\pm
2$, but does not violate weak isospin ($S$ denotes singlet, one often
writes $M_R$ as $m_S$) and can in principle occur as a bare mass term. However, in some models a bare
mass for the right-handed neutrinos is forbidden by new physics, 
and $M_R$ is instead generated by the VEV of, say, a SM Higgs singlet field $S$, or by  a
higher-dimension operator.

When Dirac and Majorana mass terms are both present, one must
diagonalize the resulting mass matrix in order to identify the mass-eigenstates, which will, in general, be linear combinations of
$\nu_L$ and $\nu_R^c$. For one active neutrino
$\nu^0_L \xrightarrow[CP]{}\nu_{R}^{0c}$ and one sterile neutrino
$\nu^0_R  \xrightarrow[CP]{}\nu_{L}^{0c}$ (the superscripts imply weak
eigenstates) 
\begin{equation} 
-\mathcal{L}=\frac{1}{2}
\left(\begin{array}{cc}
\bar{\nu}_{L}^{0} & \bar{\nu}_{L}^{0c}\end{array}\right)
\left(\begin{array}{cc}
m_{L} & m_{D}\\
m_{D} & M_{R}
\end{array}\right)
\left(\begin{array}{c}\nu_{R}^{0c} \\ \nu_{R}^{0}\end{array}\right) + h.c.
\label{simultaneous}  
\end{equation}
The  mass matrix can be diagonalized by a unitary matrix ${\mathcal U}$,
\begin{equation}  
{\mathcal U}^\dagger 
{ \left(\begin{array}{cc}
m_{L} & m_{D}\\
m_{D} & M_{R}
\end{array}\right)}
{\mathcal U}^\ast=  \left(\begin{array}{cc}
m_{1} & 0\\
0 & m_{2}
\end{array}\right), \label{masseigenvalues}
\end{equation}
because the matrix is symmetric. The 
mass eigenvalues $m_{1,2}$ can be taken to be real and positive by
appropriate phase choices. The corresponding eigenvectors represent
two Majorana mass eigenstates, $
\nu_{iM}=\nu_{iL}+\nu_{iR}^c=\nu_{iM}^{c}, \ i=1,2$, where 
\begin{equation} 
\left(\begin{array}{c}\nu_{1L} \\ \nu_{2L} \end{array}\right)=
{\mathcal U}^\dagger \left(\begin{array}{c}\nu_{L}^{0} \\ \nu_{L}^{0c}\end{array}\right), \qquad
\left(\begin{array}{c}\nu_{1R}^c \\ \nu_{2R}^c \end{array}\right)=
{\mathcal U}^{T} \left(\begin{array}{c}\nu_{R}^{0c} \\ \nu_{R}^{0}\end{array}\right).
 \end{equation}

There are a number of important special cases and limits of
(\ref{simultaneous}):
\begin{itemize} 
\item[(a)] The pure Majorana case, $m_D=0$. There is no
mixing between the active and sterile states, and the sterile neutrino
decouples (unless there are new interactions); 
\item[(b)] The pure Dirac case, $m_L=M_R=0$. This leads to two
degenerate Majorana neutrinos which can be combined to form a Dirac
neutrino with a conserved lepton number; 
\item[(c)] The seesaw limit, $M_R \gg m_{D,T}$.  There is one, mainly
sterile, state with $m_2 \simeq M_R$, which decouples at low energy, and
one light, mainly active, state, with mass $m_1 \simeq m_L -
m_D^2/M_R$. For $m_L=0$, this yields an elegant explanation for why
$|m_1| \ll m_D$. The sterile state may be integrated out, leading to
an effective higher-dimensional operator for the active Majorana mass
with $\mathcal{ M}/C= \langle{\phi^0}\rangle^2 M_R/m_D^2$; 
\item[(d)] The pseudo-Dirac limit, $m_D \gg m_{L},M_R$, which leads to a small shift in
the mass eigenvalues $ |m_{1,2}|=m_D\pm (m_L + M_R)/2$ (we are taking
the masses here to be real and positive for simplicity).  Another
possibility for the pseudo-Dirac limit occurs when $m_L \simeq M_R$;
\item[(e)] The active-sterile mixed case, in which $m_D$ is comparable to $M_R$
and/or $m_L$. The mass eigenstates contain significant admixtures
of active and sterile. 
\end{itemize}
The mass matrix in (\ref{simultaneous}) can  be generalized to 3
active  and $n_R$ sterile neutrinos, where $n_R$ need not be 3. In case
(a), and in cases (b) and (c) with $n=3$,  the observable  spectrum
consists of three active or essentially active neutrinos. The
left-chiral components $\nu_{iL}$  are related to the weak eigenstate
fields $\nu_{\alpha L}, \  \alpha= e, \mu, \tau,$ by  $\nu_{ \alpha L}= \sum_{i=1}^3
U_{ \alpha i} \,  \nu_{iL}$, where $U$ is the unitary PMNS
matrix~\cite{Pontecorvo:1967fh,Maki:1962mu}, and similarly for their
CP conjugates $\nu^c_{iR}$. 

In the case of $n_R$ singlet neutrinos, the full neutral fermion mass
matrix takes the form 
\begin{eqnarray} \label{eq:mnu}
{\mathcal M}_\nu = \left(\begin{array}{ll} m_L &m_D\\
m_D^T & M_R \end{array}\right),
\end{eqnarray}
where $M_R$ can, without loss of generality, be made diagonal $M_R={\mathrm
Diag}(M_1,M_2,\ldots)$, while $m_D$ is a generic $3\times n_R$ complex
matrix. This matrix is diagonalized by a unitary $(3+n_R) \times
(3+n_R)$ matrix ${\mathcal U}$. From now on we will neglect $m_L$, unless otherwise noted. 
This scenario is the result of the most general renormalizable Lagrangian consistent with the standard model
augmented by gauge-singlet (right-handed neutrino) fermion fields $N_i$, written as 
\begin{equation}
{\mathcal L}_{\nu} \supset - \frac{{M_R}_{ij}}{2} N_i N_j - y^{\alpha i} L_{\alpha} N_i H  + h.c.
\label{eq:lnu}
\end{equation}

For sterile neutrinos to be relevant to neutrino oscillations, as
suggested by LSND and MiniBooNE, or for most astrophysical and
cosmological implications, there has to be non-negligible mixing between active and
sterile states of the same chirality. 
That does not occur in the pure
Majorana, pure Dirac, or very high energy seesaw limits, but only for the pseudo-Dirac
and active-sterile cases (of course, hybrid possibilities can occur
for three families and $n_R>1$). In fact, the pseudo-Dirac case is
essentially excluded unless $m_{L}, M_R \lesssim 10^{-9}$ eV. Otherwise,
there would be significant oscillations of solar neutrinos into
sterile states, contrary to observations~\cite{deGouvea:2009fp}
(mass-squared differences as small as $10^{-18}$ eV$^2$ could be
probed with neutrino telescopes \cite{Beacom:2003eu}). An elegant
method to parameterize ${\mathcal U}$ for pseudo-Dirac neutrinos can be
found in \cite{Kobayashi:2000md}. 
Therefore, significant active-sterile mixing requires that at least
some Dirac masses ($m_D$) and some Majorana masses ($M_R$ and/or
$m_L$) are simultaneously very small but non-zero\footnote{More
accurately, two distinct types of neutrino masses must be
simultaneously small but non-zero.  For example, there could be
simultaneous Dirac mass terms between active and sterile neutrinos, as
well as  between sterile left and right-chiral neutrinos, as occurs in
some extra-dimensional theories. See,
{\it e.g.}~\cite{Davoudiasl:2002fq}.}, presenting an interesting challenge
to theory.  One possibility is that some new symmetry  forbids all of
the mass terms at the perturbative level, but allows, {\it e.g.}~$M_R \sim 1$
eV and $m_D \sim 0.1$ eV due to higher-dimensional operators (as in the
mini-seesaw). A list of models and some of the general aspects
associated with building them will be discussed below.

In the case $M_R\gg m_D$, the so-called seesaw limit, which is an
excellent approximation for $M_R \in [1~{\mathrm eV},10^{15}~{\mathrm GeV}]$, it
is instructive to express $\mathcal{U}$ in terms of 4 submatrices, 
\begin{equation}
{\mathcal U} =\left(\begin{array}{cc} U & \Theta \\ \Theta'^T & V_s \end{array}
\right),
\end{equation}
where $U$ is a $3\times 3$ matrix, $V_s$ is an $n_R\times
n_R$ matrix, $\Theta,\Theta'$ are $3\times n_R$ matrices, and we will
refer to $\Theta$ as the ``active--sterile'' mixing matrix. In the
case of interest, the elements of $U$ are very similar to
those of the standard neutrino mixing matrix, which are measured with
different degrees of precision \cite{Nakamura:2010zzi}. The elements
of $\Theta$ are parametrically smaller than those of $U$. In the seesaw limit, $U$ is approximately unitary:
$U U^{\dagger}\sim \mathbbm{1}_{3\times 3}$. In the absence
of interactions beyond those in Eq.~(\ref{eq:lnu}), $V_s$ is not
physical, while the elements of $\Theta'$ are proportional to those of
$\Theta$. For more details on how to parameterize ${\mathcal U}$ in the case at
hand see, for example,
\cite{deGouvea:2008nm,deGouvea:2010iv,Asaka:2011pb,Blennow:2011vn}.

Eq.~(\ref{eq:mnu}) is expressed in the so-called flavor basis: $\nu_e,
\nu_{\mu}, \nu_{\tau}, \nu_{s_1}, \ldots,\nu_{s_n}$. This is related
to the neutrino mass basis -- $\nu_1,\nu_2,\ldots,\nu_{3+n}$ -- via
the $(3+n)\times (3+n)$ neutrino unitary mixing matrix $U$:  
\begin{equation}
\nu_{\alpha}={\mathcal U}_{\alpha i}\nu_i,~~~~{\mathcal U}{\mathcal U}^{\dagger}={\mathcal U}^{\dagger}{\mathcal U}=\mathbbm{1},
\end{equation}
$\alpha=e,\mu,\tau,s_1,\ldots,s_n$, $i=1,2,\ldots,n_R+3$ such that
\begin{equation}
{\mathcal M}_\nu=\left(\begin{array}{cc} 0 & m_D \\ m_D^T & M_R \end{array}
\right)={\mathcal U}^*\left(\begin{array}{cccc} m_1 &  & &  \\  & m_2 & &  \\  &  & \ddots & \\ & & & m_{3+n_R}\end{array}\right){\mathcal U}^{\dagger},
\label{eq:flavor_mass}
\end{equation}
where $m_1,m_2,\ldots,m_{3+n_R}$ are the neutrino masses.

\begin{figure}
\includegraphics[width=0.85\textwidth]{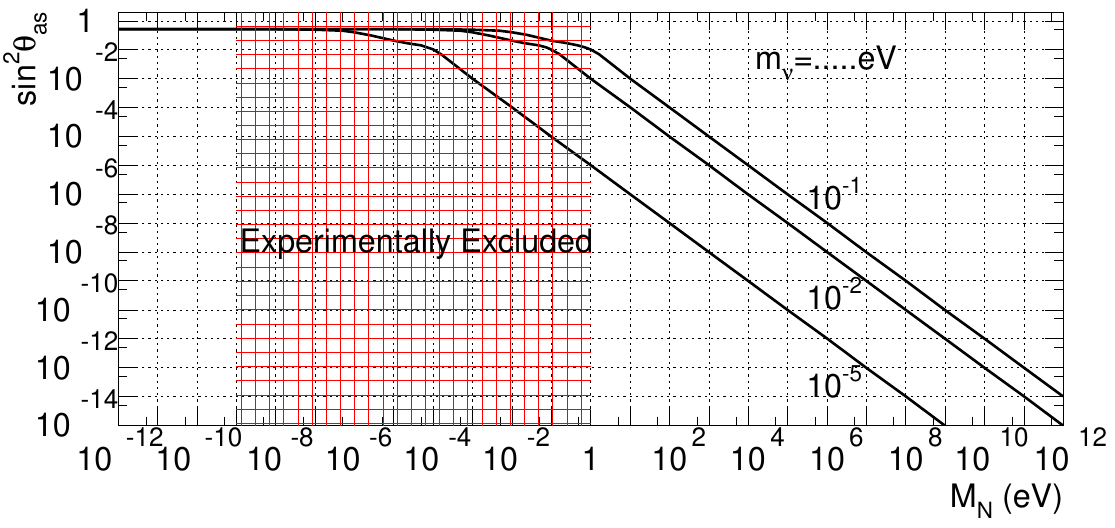}
\caption{Estimate of the magnitude of the mixing between active and sterile neutrinos 
$\sin^2\theta_{as}$ as a function of the right-handed neutrino mass $M_R$, for different 
values of the mostly active neutrino masses, $m_{\nu}=10^{-1},~10^{-2},$ and  
$10^{-5}$~eV. The hatched region qualitatively indicates the values of $M_R$ that are 
currently excluded by the world's particle physics data. From \cite{deGouvea:2009fp}.}
\label{fig:summary}
\end{figure}

In the seesaw limit, there are 3 light, mostly active neutrinos with
mass matrix $m_\nu = -m_D \, M_R^{-1} \, m_D^T$ and $n_R$ heavy, mostly
sterile, neutrinos with mass matrix $M_R$. The distribution of masses
is parametrically bimodal: $m_{1,2,3}\ll m_{4,5,\ldots}\sim M_R$. 
It is easy to show (see, for example, \cite{deGouvea:2005er}) that 
\begin{equation}
\Theta^2 \sim \frac{m_{1,2,3}}{m_{4,5,\ldots}} \sim \frac{m_D^2}{M_R^2}.
\label{eq:theta}
\end{equation} 
Our ability to observe the mostly sterile states is proportional to
$\Theta^2$. Given $m_{1,2,3}\lesssim 10^{-1}$~eV, the mostly sterile
states are very weakly coupled unless $m_{4,5,\ldots}\lesssim 10$~eV
and $\Theta^2\gtrsim 10^{-2}$. This generic behavior of the
active--sterile mixing angles is also depicted in
Fig.~\ref{fig:summary}. 
In general, in a theory with sterile neutrinos, the leptonic neutral
   current is non-diagonal in mass eigenstates
\cite{Schechter:1980gr,Lee:1977tib}. This is a special case of 
   a theorem \cite{Lee:1977tib} that the necessary and sufficient condition that the leptonic
   neutral current is diagonal in mass eigenstates is that all leptons of the
   same chirality and charge must have the same weak total and third
component isospin. This condition is violated in a theory with sterile neutrinos. 

In the context of the sterile neutrino parameters we are dealing with
in this paper, there are two important statements one can make from
Eq.~(\ref{eq:theta}): if there is a sterile neutrino of mass around eV
and mixing around 0.1, then its contribution to the active neutrino
masses is of order $m_D^2 /M_R \sim \Theta^2 M_R \sim 0.1$ eV, and can
lead to interesting effects, see {\it e.g.}~\cite{Smirnov:2006bu} for a detailed
analysis. In case a sterile neutrino with  mass around keV 
and mixing around $10^{-4}$ exists, then its contribution to the active neutrino
masses is of order $m_D^2 /M_R \sim \Theta^2 M_R \sim 10^{-5}$ eV, and
completely negligible. This also implies that there is a massless active neutrinos
in this case.

An example of a theory with sterile neutrinos and the seesaw mechanism
is the $SO(10)$ grand unified theory (GUT), in which a sterile
neutrino is the $SU(5)$-singlet 
member of the 16-dimensional spinor representation for each SM generation,
which has $SU(5)$ decomposition $16_L = 10_L + \bar 5_L + 1_L$.  In the
original $SO(10)$ GUT, the number of sterile neutrinos, $n_R$, is thus equal to
the number $N_{\mathrm gen}=3$ of SM fermion generations, although in other models,
$n_R$ may differ from $N_{\mathrm gen}$ (there have been, and continue to be
searches for quarks and leptons in a possible fourth SM generation, but since
these are negative so far, we take $N_{\mathrm gen}=3$). In an analogous manner,
sterile neutrinos appear naturally as the spinor component of the corresponding
chiral superfields in supersymmetric $SO(10)$ GUTs.  At the simplest level, in
the $SO(10)$ theory, SM fermion mass terms involve the bilinear $16 \times 16$,
with the Clebsch-Gordon decomposition $16 \times 16 = 10_s + 120_a + 126_s$ and
arise from Yukawa couplings to corresponding Higgs fields transforming as 10,
120, and 126 dimensional representations.  Dirac masses arise from terms
transforming according to each of these representations, while Majorana mass
terms for the sterile neutrinos occur in the terms transforming according to
the 126-dimensional representation. Obviously, another aspect of the
seesaw mechanism which makes it even more appealing is the remarkable
fact that the decays of the right-handed neutrinos in the early
Universe can provide an explanation of the baryon asymmetry of the
Universe, in the leptogenesis mechanism~\cite{Fukugita:1986hr}.

\subsection{Theoretical Motivations and Symmetries Behind the Existence of Light Sterile Neutrinos}

In this Section the theoretical aspects entering models of sterile
neutrinos will be discussed. 
It is important to stress that none of these imply that light sterile neutrinos must exist, 
and none specify the sterile neutrino mass scale. Most models allow for the possibility of light,
mostly sterile states, and can accommodate a subset of the
short-baseline anomalies and/or provide a (warm) dark matter candidate. 
This does not imply, of course, that sterile neutrinos do not exist. Recall that large mixing in
the lepton sector, now a well established fact, was not predicted by
theorists.

\subsubsection{Sterile neutrinos in ``standard approaches''}

We have seen that sterile neutrinos are a natural ingredient of the
most popular and appealing mechanism to generate neutrino masses, the
conventional or type I seesaw mechanism. No equally appealing
alternative has emerged in the thirty years since it was discovered.
In what follows, we will attempt to give a flavor of the required
model building ingredients that can be introduced in order to generate or
accommodate (light) sterile neutrinos in the mass range of eV or
keV. Note that these scales are not far away from each other, and the
same techniques apply for both cases.

\begin{itemize}
\item{\bf The split seesaw mechanism}

The split seesaw model~\cite{Kusenko:2010ik} consists of the Standard
Model with three right-handed neutrinos and a spontaneously broken
$U(1)_{B-L}$ gauge symmetry.  It is also assumed that space-time
is a five-dimensional space compactified to four dimensions at some
scale $M_c=1/l\sim 10^{16}$~GeV.  If one of the right-handed neutrinos
is localized on a brane separated from the standard-model brane by a
distance $l$, then the wave function overlap between this right-handed
neutrino and the other fields is very small.  This leads to a
suppression of both the Yukawa coupling and the sterile neutrino mass
in the low-energy effective four-dimensional theory, but the
successful seesaw formula is preserved.  This opens a possibility for
one of the right-handed, or sterile, neutrinos to serve as the
dark-matter particle, and the observed abundance can be achieved for
some values of parameters. 

More specifically, in this model, one can express the effective mass
$M_s$ and the Yukawa coupling $y$ of the sterile neutrino in terms of the
fundamental (5-dimensional) Planck mass $M$, the distance $l$ between
the branes, and the right-handed neutrino bulk mass $\tilde m$:  
\bea
M_s =  \kappa_i v_{B-L} \, \frac{2\tilde{m}}{M(e^{2 \tilde{m} \ell}-1)}~,~~~
\label{5d_mass}
y =\tilde \lambda   \sqrt{\frac{2 \tilde{m}}{M(e^{2 \tilde{m}
\ell}-1)}} , 
\label{5d_yukawa}
\eea
where $\tilde \lambda $ and $\kappa $ are couplings in the underlying
five-dimensional theory.  Clearly, the exponential suppression in
$M_s$ is the square of the term which appears in $y$,  
so  the ratio $y^2/M_s$ is unchanged, while both
the Yukawa coupling and the mass are much smaller than $M$ or
$v_{B-L}$. The idea has been extended with an $A_4$ flavor symmetry in
\cite{Adulpravitchai:2011rq}.

The split seesaw makes vastly different scales equally likely, because a
small difference in the bulk masses or a small change in the distance
$l$ between the two branes can lead to exponentially different results
for the sterile neutrino mass.\footnote{It is also possible to  
have multiple light sterile neutrinos because small changes in the
bulk masses can result in big changes in $y$ and $M$.  In particular, 
it is possible to reproduce the mass spectrum of the model
$\nu$MSM~\cite{Asaka:2005an,Boyarsky:2009ix}, in which the lightest
dark-matter  
sterile neutrino is accompanied by two GeV sterile neutrinos which are
nearly degenerate in mass.}  This provides a strong motivation for
considering the sterile neutrino as a dark matter candidate. Indeed, the
split seesaw model provides at least two ways of generating the
correct dark-matter abundance~\cite{Kusenko:2010ik}.  Furthermore,
each of these production scenarios generates dark matter that is
substantially colder than the warm dark matter produced in neutrino
oscillations~\cite{Dodelson:1993je,Shi:1998km,Laine:2008pg}.  The
reason this dark matter is cold enough to agree with the structure
formation constraints is simple: sterile neutrinos produced out of
equilibrium at temperatures above the electroweak scale suffer
redshifting when the standard model degrees of freedom go out of
equilibrium and entropy is produced~\cite{Kusenko:2006rh}.

The emergence of a sterile neutrino as a dark-matter particle gives a
new meaning to the fact that the standard model has three generations
of fermions. Three generations of fermions allow for CP violation in
the Cabibbo-Kobayashi-Maskawa matrix, but, quantitatively, this CP
violation is too small to play any role in generating the
matter-antimatter asymmetry of the Universe.  In the context of
leptogenesis, CP violation in the neutrino mass matrix would have been
possible even with two generations, thanks to the Majorana phase.
Hence, the existence of three families of leptons, apparently, has no
purpose in the standard model, except for anomaly cancellation. However, if the lightest sterile
neutrino is the dark-matter particle (responsible for the formation of
galaxies from primordial fluctuations), while the two heavier sterile
neutrinos are responsible for generating the matter-antimatter
asymmetry via leptogenesis, the existence of three lepton families
acquires a new, significant meaning.

\item{\bf Symmetries leading to a vanishing mass}

An appealing way to have a strong hierarchy in the neutrino masses is
to introduce a symmetry that implies one vanishing sterile neutrino
mass. A slight breaking of this symmetry will lead to a naturally small
mass compared to those allowed by the symmetry. A very
popular Ansatz is the flavor symmetry $L_e-L_\mu-L_\tau$ \cite{Petcov:1982ya}.

The first suggestion to apply the $L_e-L_\mu-L_\tau$ symmetry in the
context of keV sterile neutrinos was made in~\cite{Shaposhnikov:2006nn},
and a concrete model was presented later on in Ref.~\cite{Lindner:2010wr}. The
basic idea is the following: the $L_e-L_\mu-L_\tau$
symmetry leads to a very characteristic mass pattern for active
neutrinos, namely to one neutrino being exactly massless, while the
other two are exactly degenerate, denoted by $(0,m,m)$. Applying the
symmetry to three right-handed neutrinos, results in an analogous pattern
$(0,M,M)$ for the heavy neutrino masses. 

The flavor symmetry must be broken, which can be conveniently
parameterized by \emph{soft breaking terms} of the order of a new mass
scale $S$. Any explicit or spontaneous symmetry breaking will result
in terms of that form, and their general effect is to lift the
degeneracies and to make the massless state massive, $(0,M,M) \to
\left( \mathcal{O}(S), M-\mathcal{O}(S), M+\mathcal{O}(S) \right)$,
cf.\ left panel of Fig.~\ref{fig:schemes}. A similar effect is observed
for the light neutrinos. Since the symmetry breaking scale $S$ must be
smaller than the symmetry preserving scale $M$ (we would not speak of
a symmetry otherwise), this mechanism gives a motivation for $S\sim
{\mathrm keV}$ or eV, while $M\gtrsim \mathcal{O}({\mathrm GeV})$ or heavier. Note,
however, that $L_e-L_\mu-L_\tau$ symmetry tends to predict bimaximal
mixing, which is incompatible with data~\cite{Schwetz:2011zk}. This
problem can be circumvented by taking into account the mixing from the
charged lepton sector~\cite{Lindner:2010wr,Petcov:2004rk}. In this
case, an additional bonus is the prediction of a non-zero value of the
small mixing angle $\theta_{13}$, in accordance with recent
experimental indications~\cite{Ahn:2012nd,Abe:2011fz,An:2012eh}. 

\begin{figure}[t]
\centering
\begin{tabular}{lr}
\includegraphics[width=5.5cm]{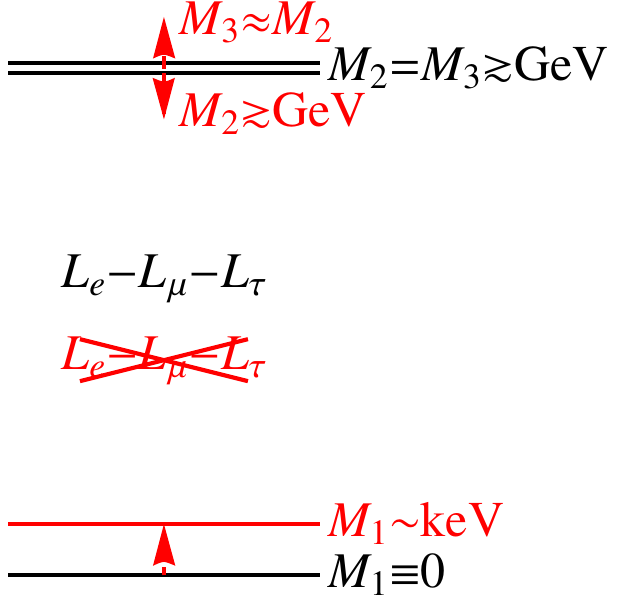}\hspace{0.5cm}
& \includegraphics[width=7.2cm]{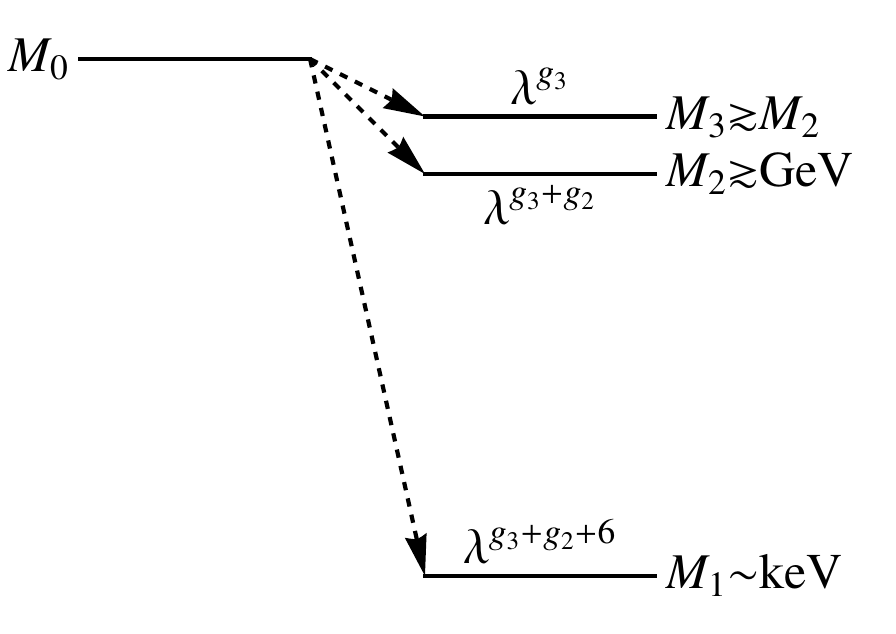}
\end{tabular}
\caption{\label{fig:schemes} The mass shifting schemes for the
$L_e-L_\mu-L_\tau$ and the Froggatt-Nielsen models (figures taken
from Ref.~\cite{Merle:2011yv}). }
\end{figure}

The potential of $L_e-L_\mu-L_\tau$ to accommodate eV-scale neutrinos
within the type I seesaw was noted earlier in~\cite{Mohapatra:2001ns}. 
Suppose we have a symmetry in the theory that leads to det$M_R=0$ 
with only one of the eigenvalues zero. If the same symmetry also
guarantees that det$M_D~=~0$, then one can ``take out'' the zero 
mass eigenstates from both $M_R$ and $M_D$ and then use the
seesaw formula for the $2\times 2$ sub-system.\footnote{The case of
vanishing determinant of $M_R$ is called singular seesaw
\cite{Glashow:1990dg}, and leads to two light, two intermediate and
two heavy masses
\cite{Chun:1998qw,Chikira:1998qf,Liu:1998qp,McKellar:2001hi,McDonald:2004pa,McDonald:2004qx}.}
The spectrum will consist of two light Majorana neutrinos
(predominantly left-handed), two heavy right-handed Majorana
neutrinos; one massless right-handed neutrino, which will play
the role of the sterile neutrino of the model, and a massless 
left-handed neutrino. Breaking the symmetry  very weakly either by 
loop effects or by higher dimensional operators can lead to a nonzero mass for the light sterile
neutrino as well as its mixing with the $\nu_{e,\mu}$, so that
one can have oscillations between sterile and active states. 
The simplest realization makes use of an $L_e-L_\mu-L_\tau$ times a $Z_2$
symmetry for the right-handed sector, and yields~\cite{Mohapatra:2001ns} 
\begin{eqnarray}
M_R= \begin{pmatrix}0 & M & M\cr M & 0 & 0 \cr M & 0 & 0
\end{pmatrix}  ~,~~
m_D = \begin{pmatrix} m_{11} & 0 & 0\cr 0 & m_{22} & m_{22} \cr 0 & m_{32} & 
m_{32} \end{pmatrix}
\end{eqnarray}
It is easy to see that in this case the two massless neutrino states
are: $(\nu_\mu-\nu_\tau)/\sqrt{2}$ and $(N_\mu - N_\tau)/\sqrt{2}$, where we 
are using the symbol $N$ for the SM singlet RH neutrino states. 
The second of the above states is a sterile state and can be identified 
as $\nu_s$. Excluding these two modes, the mass matrix reduces 
to a $2\times 2$ seesaw form and one gets an inverted mass hierarchy for 
the active neutrino with $(\nu_\mu-\nu_\tau)/\sqrt{2}$ being the lightest
active state. Note that a similar method for generating a suitable mass pattern is to use discrete instead of continuous
symmetries, which do not automatically suffer from problems with
Goldstone bosons. One example from the literature is \cite{Araki:2011zg}, where the
authors have used a $Q_6$ symmetry to forbid a leading order mass for
one of the sterile neutrinos.

\item{\bf The Froggatt-Nielsen mechanism}

Another idea to associate small right-handed neutrino masses to ultra-high 
new physics scales is to use the Froggatt-Nielsen
(FN) mechanism~\cite{Froggatt:1978nt}. The FN mechanism introduces an
implicit high-energy sector of scalars and fermions, which are all
suitably charged under a new global $U(1)_{\mathrm FN}$ symmetry. Although
this has the drawback of lacking an explicit
UV-completion~\cite{Mavromatos:2011ur}, it has the big advantage of
leading to a very strong (exponential) generation-dependent suppression
of mass eigenvalues, and thus to strong hierarchies, cf.\ right panel
of Fig.~\ref{fig:schemes}. This is the reason why this mechanism is so
popular when it comes to explaining the pattern of quark masses and mixing, and it is equally well
applicable to the problem of generating a keV or eV scale for right-handed
neutrinos.

Recent applications of the FN mechanism to sterile states were presented
in \cite{Merle:2011yv,Barry:2011wb,Barry:2011fp}, the latter work
combining it with an explicit flavor symmetry in order to generate
(close-to) tri-bimaximal mixing for the neutrinos. Effects of the
leading and next-to-leading order breaking terms have been carefully
analyzed and it was shown that, {\it e.g.}~the strong constraints on the
active-sterile mixing as well as a non-zero mixing angle $\theta_{13}$
are consistent with the model. Corrections to the leading order seesaw
formula \cite{Grimus:2000vj,Hettmansperger:2011bt} have also been
addressed.  In addition, many constraints arising
from different requirements, such as anomaly cancellation, embedding
into Grand Unified Theories, or bounds on lepton flavor violation have
been taken into account~\cite{Merle:2011yv}. 
The combination of all of these requirements may render
the FN models predictive: For example, they would be incompatible with
the left-right symmetry used for keV neutrino DM production in~\cite{Bezrukov:2009th}. Finally, in~\cite{Barry:2011fp} the FN
charges of all 3 right-handed neutrinos were varied such that several
different results could be achieved: two super heavy neutrinos for
leptogenesis and one for keV warm dark matter; one eV-scale neutrino for the reactor
anomaly/LSND and one for warm dark matter, etc. 

Similar to the split seesaw case, the FN models presented accommodate only 
slightly modified versions of the high scale seesaw mechanism, even though keV or
eV scale right-handed neutrinos are involved. The reason for this is that
the charges of the right-handed neutrinos under any global $U(1)$ -- be
lepton number or the $U(1)_{\mathrm FN}$ -- drop out of the seesaw
formula: if a right-handed neutrino has charge $F_1$ under the
$U(1)_{\mathrm FN}$, then its mass is suppressed by $\lambda^{2 F_1}$,
where $\lambda = \langle \Theta \rangle/\Lambda$ is the ratio of the
FN field VEV and the cutoff scale of the theory. The respective
column of $m_D$ is suppressed by $\lambda^{F_1}$, hence the seesaw
formula is constant.

\item{\bf Extended seesaw mechanisms}

Other approaches to the problem extend the seesaw mechanism. One
possibility is similar to the ``take out zero mass eigenstates'' idea
discussed above, and was put forward in~\cite{Mohapatra:2005wk}. 
Considering the gauge group $SU(2)_L\times U(1)_{I_{3R}}\times
U(1)_{B-L}$ and various new weak singlets leads to an extended (inverse) seesaw
mass matrix \cite{Mohapatra:1986aw,Mohapatra:1986bd}
\begin{eqnarray} 
{\mathcal M}_{\nu NS} = \begin{pmatrix} 0 & m_D & 0\cr m^T_D & 0 & M\cr 0 &
M^T & \mu \end{pmatrix} , 
\end{eqnarray}
where 
\begin{eqnarray}   
\label{eq4xxx} 
m_D = \begin{pmatrix} m_{11} & m_{12} & m_{12}\cr m_{21} & m_{22} &
m_{22} \cr m_{21}& m_{22}& m_{22} \end{pmatrix}~,~~
M = \begin{pmatrix}M_{11} & M_{12} & M_{13}\cr M_{21} & M_{22} & M_{23}
\cr M_{21}& M_{22}& M_{23} \end{pmatrix} \end{eqnarray} 
The matrix $\mu$ has an arbitrary (symmetric) form. Decoupling first the
fields associated with $\mu$ gives 
\begin{eqnarray} 
{\mathcal M}_{\nu NS} = \begin{pmatrix} 0 & m_D \cr m^T_D & M\mu^{-1}M^T \end{pmatrix}
\end{eqnarray}
The determinant of the $(N_e,N_\mu,N_\tau)$ mass matrix  $M_R = M
\mu^{-1}M^T$ vanishes and the zero mass eigenstate is given by
$\frac{1}{\sqrt{2}}(N_\mu-N_\tau)$. We ``take out'' this state and the
remaining right-handed neutrino mass matrix is $2\times 2$ involving only the states 
$(N_e,\frac{1}{\sqrt{2}}(N_\mu+N_\tau))$. Similar to the discussion
above, the seesaw matrix is now a two generation matrix
yielding two light Majorana neutrino states. The other active
neutrino state is massless, together with $\nu_s\equiv
\frac{1}{\sqrt{2}}(N_\mu-N_\tau)$. 

Another model is given in Ref.~\cite{Fong:2011xh}, where 
an attempt was made to understand the tiny Majorana mass required in
the inverse seesaw mechanism from warped extra-dimensional
models. Parity anomaly cancellation requires the total number of bulk
fermions that couple to gauge and gravity fields to be even, thus a
fourth singlet fermion was needed. Three of the singlet fields pair up
with the three right-handed Majorana neutrino fields to make the three
pseudo-Dirac fermions of inverse seesaw, and the remaining singlet
in the end becomes the sterile neutrino with mass in the 
eV range. 

A related approach, first discussed in \cite{Chun:1995bb} and studied
in detail in \cite{Zhang:2011vh}, introduces in addition to the three
generations of left-handed active neutrinos $\nu_{Li}$ and three
generations of right-handed sterile neutrinos $N_{Rj}$, another
singlet right-handed field $S_R$. This field carries a non-trivial
charge under a new discrete $A_4 \times Z_4$ auxiliary symmetry, 
different from the charges of the $N_{Rj}$. The effect of these
chirality and charge assignments is to forbid any direct Dirac or
Majorana masses for the field $S$, which only obtains a mass by coupling to
one generation of ordinary right-handed neutrinos, resulting in a
mixed mass term of size $M_S$. The full mass matrix is 
\begin{eqnarray}
M_\nu^{7 \times 7}  = \left(\begin{matrix}  0 & m_D & 0 \cr m^T_D &
M_R & M^T_S \cr 0 & M_S & 0
\end{matrix} \right)  .
\end{eqnarray}
With $M_R \gg M_S$ and $M_S \gg m_D$, one block-diagonalizes this
matrix and finds active neutrinos roughly of order $m_D^2/M_R$ and
a sterile neutrino of order $M_S^2/M_R$. In such a
framework, both active and sterile neutrino masses are
suppressed via the seesaw mechanism, and thus an eV~scale
sterile neutrino together with sizable active-sterile
mixing is accommodated without the need of 
inserting small mass scales or Yukawa couplings.

\end{itemize}

Very often the $4\times4$ mass matrix for the active plus sterile
neutrinos obeys an approximate $\mu$--$\tau$ symmetry, which naturally
generates small $\theta_{13}$ and close-to maximal $\theta_{23}$. For
instance~\cite{Mohapatra:2005wk},
\begin{eqnarray}  
M_{\nu} = \left(\begin{matrix} m_{11}&m_{12}&m_{12}&f_1\cr
m_{12}&m_{22}&m_{22}&f_2\cr m_{12}&m_{22}&m_{22}&f_2\cr 
f_1&f_2&f_2& f_4 \end{matrix}\right).
\end{eqnarray}
Due to small mixing with the active states the $f_i$ are typically
smaller than the $m_{ij}$, and one can diagonalize the mass matrix
with 
\begin{eqnarray} 
U \simeq \left(\begin{matrix}c&s&0&\delta_1 \cr \frac{-s}{\sqrt 
2}&\frac{c}{\sqrt 2}&\frac{1}{\sqrt 2}&\delta_2 \cr 
\frac{-s}{\sqrt 2}&\frac{c}{\sqrt 2}&\frac{-1}{\sqrt 
2}&\delta_2\cr -\delta_1c+\sqrt2\delta_2 s &-\delta_1  
s-\sqrt2\delta_2 c&0&1 \end{matrix}\right) 
\end{eqnarray}
where $\delta_1\simeq \frac{f_1}{f_4}$ and
$\delta_2\simeq\frac{f_2}{f_4}$.

There are other related approaches capable of ``explaining'' or at least 
accommodating light sterile sterile neutrinos. For example, a general
effective approach is possible: in many flavor symmetry models to
generate the unusual lepton mixing scheme (see Refs.~\cite{Altarelli:2010gt,
Ishimori:2010au} for recent reviews) the mass terms are effective. Simply 
adding a sterile state to the particle content, and making it a singlet 
under the flavor symmetry group as well as the SM group, does the 
job~\cite{Barry:2011wb}. Other models which accommodate light sterile 
neutrinos include~\cite{Berezhiani:1995yi,Ma:1995xk,Langacker:1998ut,
Bando:1998ww,ArkaniHamed:1998pf,Shafi:1999rm,Babu:2004mj,Sayre:2005yh,
Chen:2006hn,Barger:2010iv,Ghosh:2010hy,McDonald:2010jm,Araki:2011zg,
Chen:2011ai,Geng:2012jm}.

\subsubsection{Sterile neutrinos in ``non-standard approaches''}

A variety of other ways to explain light sterile neutrinos exists, for
instance {\bf mirror models}. 
Starting from the famous Lee and Yang parity violation paper
\cite{Lee:1956qn} attempts were made to generalize
the concepts of mirror symmetry and parity by assuming the existence
of mirror images of our particles. 
 According to~\cite{Lee:1956qn}, these mirror
particles were supposed to have strong and electromagnetic
interactions with our particles. 
For a few years it seemed plausible that the role of these mirror
particles is played by anti-particles of ordinary matter  and that the
true mirror symmetry is the CP symmetry. But in 1964 violation of
CP symmetry  was discovered. Then Kobzarev, Okun and  Pomeranchuk \cite{okun}
returned to the idea of mirror particles, which are different from the
ordinary matter, and came to the conclusion that mirror particles
cannot have strong and electromagnetic interaction with ordinary
matter and hence could appear as dark matter. This ``dark matter 
mirror model'' was discussed by many authors, for instance~\cite{Okun:1967zz,
Pavsic:1974rq,Blinnikov:1982eh,Foot:1991bp,Foot:1991py,Silagadze:1995tr}.
For a review of some two hundred papers see~\cite{Okun:2006eb}.

Mirror model were invoked to understand neutrino puzzles 
in~\cite{Foot:1995pa,Berezhiani:1995yi,Berezinsky:2002fa} after the LSND 
results were announced.  In this picture, the mirror sector of the model 
has three new neutrinos which do not couple to the $Z$-boson and would 
therefore not have been seen at LEP, even if these are light. We will 
refer to the $\nu'_i$ as sterile neutrinos of which we now have three.  
The lightness of $\nu'_i$ is dictated by the mirror $B'-L'$ symmetry in a 
manner parallel to what happens in the standard model. The masses of the 
mirror (sterile) neutrinos could arise for example from a mirror analog of 
the seesaw mechanism.  

The two ``Universes'' communicate only via gravity or other forces that 
are very weakly coupled or associated to very heavy intermediate
states. This leads to a mixing between the neutrinos of the two 
Universes and can cause neutrino oscillation between, say, the $\nu_e$ of our
Universe to  the $\nu'_e$ of the parallel one in order to explain the LSND 
results without disturbing the three neutrino oscillation picture that 
explains the solar and the atmospheric data.

Such a picture appears quite natural in superstring theories which lead 
to $E_8\times E_8^\prime$ gauge theories below the Planck scale, where 
both $E_8$'s living on two separate branes are connected by gravity.

There are two classes of mirror models: (i) after symmetry breaking, the 
breaking scales in the visible sector are different from those in the mirror 
sector (this is called asymmetric mirror model); or (ii) all the scales are 
the same (symmetric mirror model). If all scales in both sectors are identical 
and the neutrinos mix, this scenario is a priori in contradiction with solar 
and atmospheric data since it leads to maximal mixing between active and 
sterile neutrinos.  In the asymmetric mirror model, one can either have 
different weak scales or different seesaw scales or both. Either way, these 
would yield different neutrino spectra in the two sectors, along with different 
mixing between the two sectors. 

As far as neutrino masses and mixings go, they can arise from a seesaw 
mechanism, with the mixings between sterile and active neutrinos given by
higher dimensional operators of the form $ (LH)(L'H')/M$ after both 
electroweak symmetries are broken.  The mixing between the two sectors 
could arise from a mixing between RH neutrinos in the two sectors {\it i.e.}, 
operators of the form $N^T\sigma_2 N'+ h.c.$  In this case the active-sterile 
neutrino mixing angles are of order $v/v'$ implying that if this is the 
dominant mechanism for neutrino masses, the mirror weak scale should be 
within a factor 10-30 of the known weak scale. In this case, all charged 
fermion masses would be scaled up by a common factor of $v'/v$.

Regarding warm dark matter, in the mirror model there is no need to use oscillation to 
generate dark matter keV steriles. Here the inflaton reheating produces
primordial sterile neutrinos. However, due to asymmetric inflation, 
the density of the keV steriles is down by a factor $(T'_R/T_R)^3$,
where $T_R$ ($T'_R$) is the reheating temperature of our (the mirror) sector. One 
has 
\begin{eqnarray}
\Omega_m h^2\sim~108
\left(\frac{T'_R}{T_R}\right)^3\frac{m_{\nu_s}}{10~\mathrm keV} \, .
\end{eqnarray}
To get $\Omega_m h^2\sim 0.12$, with $(T'_R/T_R)^3\sim 10^{-3}$, 
one needs $m_{\nu_s}\sim 10$ keV. 
Note that this is regardless of whether the active and sterile neutrinos 
mix at all. Since the X-ray constraints depend on the 
active-sterile mixing angles, the mirror model for warm dark matter need not be seriously constrained by 
X-ray data. In this picture, one of the two other $\nu'$s could be in the eV range to 
explain LSND if needed.

The bottom line here is that in this case, there are three sterile 
neutrinos with arbitrary masses and mixings. For example, one could 
accommodate a 3+2 solution to the LSND puzzle as well as a keV warm dark 
matter. \\

A completely different sterile ``neutrino'' candidate is the {\bf axino}. It has been
argued that this particle, arising from the supersymmetric version of the axion solution
to the strong CP problem, is a natural candidate for a sterile neutrino
in the framework of gauge mediated supersymmetry breaking.

A long-standing puzzle in the standard model is the smallness of the 
QCD vacuum angle $\bar\theta$ which appears in the Lagrangian
 \begin{equation} \label{GG}
  {\mathcal L}_{QCD} = {\bar\theta \over 32\pi^2} G^a_{\mu\nu}
 \tilde{G}^a_{\mu\nu} \,,
 \end{equation}
where $G^a_{\mu\nu}$  is the QCD gauge field strength. The current
upper limits of the neutron electric dipole moment severely constrain 
$|\bar\theta| \lesssim 10^{-11}$, which is a priori a free parameter.  
The apparently arbitrary ``smallness'' of $\bar\theta$ is referred to 
as the strong CP problem \cite{Kim:2008hd}.   An elegant solution is 
to promote the parameter $\bar\theta$ to a dynamical degree of freedom, 
the axion $a$, which is a pseudo-Goldstone boson of the Peccei-Quinn 
(PQ) symmetry.  The QCD potential of the axion sets its vacuum 
expectation value to zero; $ \bar\theta = \langle a \rangle/f_{PQ} =0$, 
and thus the strong CP problem is solved dynamically.  Here $f_{PQ}$ is 
the scale of the PQ symmetry breaking which is constrained to be 
$10^{10} \, \mbox{GeV} \lesssim f_{PQ} \lesssim 10^{12}\,  
\mbox{GeV}$~\cite{Kim:2008hd}. 

In supersymmetric theories, the axion is accompanied by its
fermionic partner, the axino $\tilde a$, which can remain very
light due to its quasi Goldstone fermion character~\cite{Chun:1995bb}.
In the following we argue that the axino can naturally be a
sterile neutrino in gauge mediated supersymmetry breaking 
models~\cite{Chun:1999kd,Choi:2001cm} implementing the PQ symmetry 
through the Kim-Nilles mechanism~\cite{Kim:1983dt,Chun:1991xm}.

We first discuss how the axino can be as light as ${\mathcal
O}(\mbox{eV})$.  As the superpartner of a Goldstone boson, the axino
is massless as long as supersymmetry is unbroken.  When the
supersymmetry breaking scale is higher than the PQ symmetry
breaking scale as in gravity-mediated supersymmetry breaking,  the
axino is expected to have generically a supersymmetry breaking
mass, that is the gravitino mass $m_{3/2}$ \cite{Chun:1992zk,Chun:1995hc}:
 \begin{equation}
 m_{\tilde a} \sim m_{3/2} \,.
 \end{equation}
 However,
if the PQ symmetry breaking occurs before supersymmetry
breaking as in gauge mediated supersymmetry breaking, the axino
mass is suppressed by the high PQ scale and thus can remain very
light even after supersymmetry breaking.  In the effective
Lagrangian below the PQ scale, the axion (axino) appears in
combination of $a/f_{PQ}$ ($\tilde a/f_{PQ}$) due to its Goldstone
nature.  This implies that the axino would get a mass of order
\begin{equation}
 m_{\tilde a} \sim {m_{SUSY}^3 \over f_{PQ}^2}\,,
\end{equation}
where $m_{SUSY}$ is a supersymmetry breaking scale \cite{Chun:1999kd}.
In the framework of gauge mediated supersymmetry breaking
\cite{Giudice:1998bp}, a hidden sector field breaking supersymmetry can be
charged under the PQ symmetry.  In this case, $m_{SUSY}$ is the
hidden sector supersymmetry breaking scale. Taking roughly
$m_{SUSY} \gtrsim 10^5$ GeV, one gets  $m_{\tilde a} \gtrsim 1$ eV
for $f_{PQ} \lesssim 10^{12}$ GeV.  Otherwise, the axino mass
would get a generic contribution from the supersymmetry breaking
scale $m_{SUSY} \sim 10^3$ GeV of the Supersymmetric Standard
Model (SSM) sector: $m_{\tilde a} \lesssim 10^{-2}$ eV for $f_{PQ}
\gtrsim 10^{10}$ GeV.  There exists also a supergravity
contribution
 \begin{equation}
 m_{\tilde a} \sim m_{3/2} \sim  1 \mbox{ eV}
 \end{equation}
where $m_{3/2} \sim m_{SUSY}^2/\overline M_{\mathrm Pl}$ with $m_{SUSY}\sim 10^5$ GeV
and $\overline M_{\mathrm Pl}=2.4\times 10^{18}$ GeV. This will be the main
contribution to the axino mass in the scenario \cite{Choi:2001cm}. 
If the axino is light enough, appropriate mixing with active neutrinos and flavor structure can be
constructed \cite{Choi:1998wc} and hence the supersymmetric axion solution to the strong CP problem
can provide a candidate of a sterile neutrino.

\subsection{The Low-Energy Seesaw and Minimal Models}

\subsubsection{General Aspects}

Let us focus in more detail on the low energy seesaw. Recall the most general renormalizable Lagrangian 
\begin{equation} 
\label{eq:Lhier}
{\mathcal L}_{\nu} \supset {\mathcal L}_{\mathrm old} - \frac{{M_R}_{ij}}{2} N_i N_j - y^{\alpha i} L_{\alpha} N_i H  + h.c.,
\end{equation}
where ${\mathcal L}_{\mathrm old}$ is the standard model Lagrangian in the
absence of gauge singlet fermions, $y^{\alpha i}$ are the neutrino Yukawa
couplings, and $M$ are the right-handed neutrino Majorana mass
parameters. Eq.~(\ref{eq:Lhier}) is expressed in the weak  basis where
the Majorana mass matrix for the right-handed neutrinos is diagonal.

The seesaw formula allows the mass of singlet neutrinos
to be a free parameter: Multiplying $m_D$ by any number $x$ and $M_R$ by $x^2$
does not change the right-hand side of the formula. Therefore, \emph{the
  choice of $M_R$ is a matter of theoretical prejudice} that cannot be fixed
by active-neutrino experiments alone.  A
possible approach is to choose these parameters so that they explain
certain phenomena and aspects beyond-the-standard model, for example, provide a dark
matter candidate or a mechanism of baryogenesis. 
The most often considered standard approach takes Yukawa couplings $y_{\alpha
  I}\sim 1$ and the Majorana masses in the range $M_N \sim
10^{10}-10^{15}$~GeV. Models with this choice of parameters 
give rise to baryogenesis through leptogenesis~\cite{Fukugita:1986hr}. For a 
review of the GUT-scale seesaw and the 
thermal leptogenesis scenario associated with it see
{\it e.g.}~\cite{Davidson:2008bu}.  Here we would like to focus on variants
at lower energy scales. 

\begin{figure}[t]
  \begin{tabular}{cc}
    {\includegraphics[width=.4\textwidth]{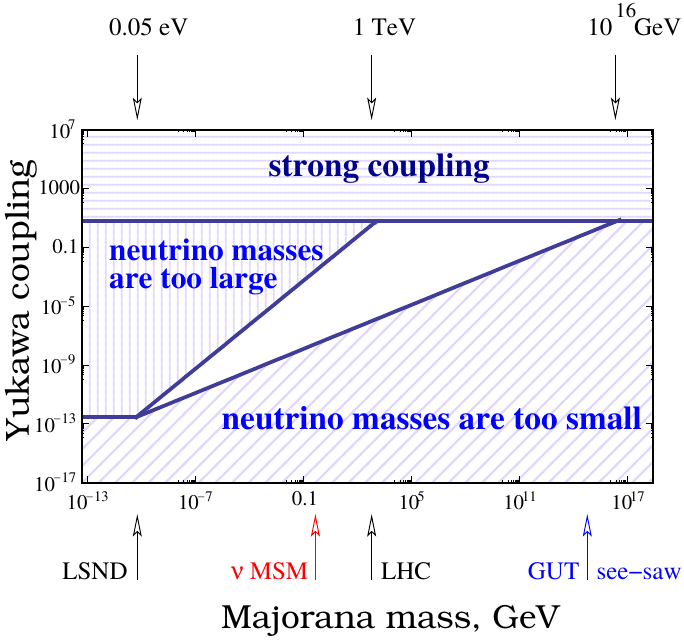}}&
    \raisebox{1.8cm}{\includegraphics[width=.5\textwidth]{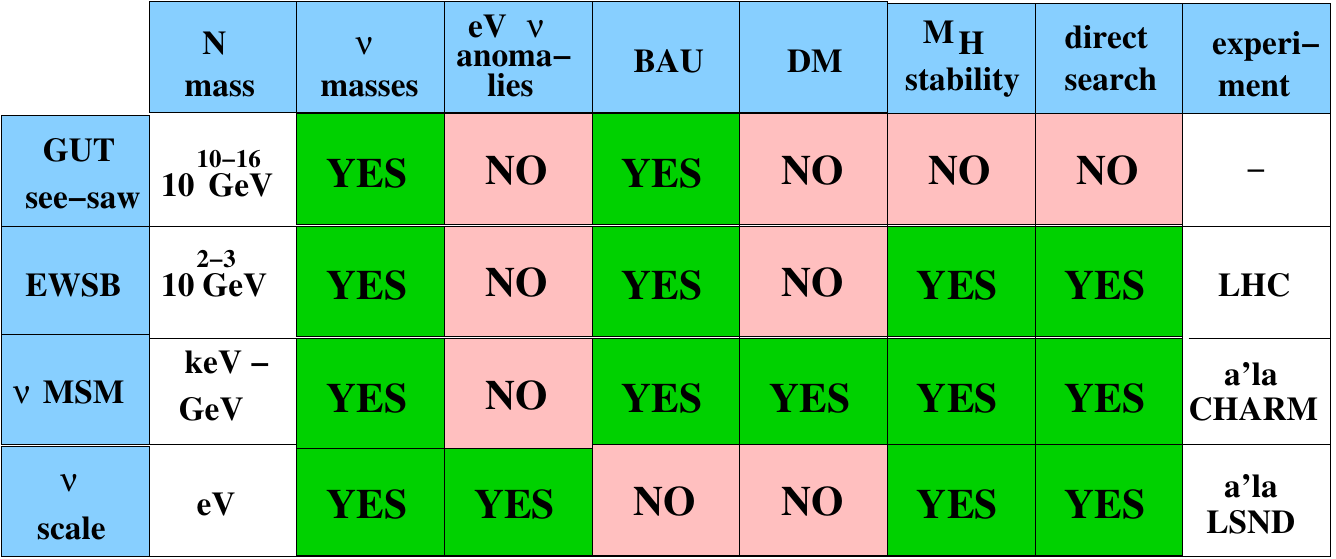}}
\end{tabular}
\caption{Left: possible values of the Yukawa couplings and 
  Majorana masses of the sterile neutrinos in seesaw models.  Right:
  the table shows whether the corresponding choice of the mass for Majorana
  fermions may explain neutrino masses and oscillations, accommodate eV
  neutrino anomalies, lead to baryogenesis, provide the dark matter candidate,
  ensure the stability of the Higgs mass against radiative corrections, and be
  directly searched at some experiments.}
\label{yukawa}
\end{figure}

Figure~\ref{yukawa} summarizes various choices of combination of mass/Yukawa
couplings of sterile neutrinos in seesaw models. The right panel
summarizes properties of resulting seesaw models, their ability to solve
various beyond-the-SM problems and anomalies, and their testability.

The main generic prediction of Eq.~(\ref{eq:Lhier}) is the existence of
$3+n_R$ Majorana neutrinos, most of them massive.  All of these
``contain'' the three active neutrino flavors and hence can, in
principle, be observed experimentally. One exception is the case
$M_R=0$. In this case, the massive Majorana neutrinos ``pair up'' into
at most three massive Dirac fermions.\footnote{In the case $n_R=2$,
there are two massive Dirac neutrinos and one massless neutrino. In
the case $n_R>3$, there are three massive Dirac neutrinos and $n_R-3$
massless gauge singlet, {\it bona fide} sterile, neutrinos that do not
mix with any of the active states and are completely unobservable.}
The neutrino data can determine all physically observable
values of  $yv=m_D$ -- the neutrino masses and the elements of the
neutrino mixing matrix, three angles and one CP-odd Dirac
phase. Qualitatively, the neutrino data require $m_D\sim 10^{-3}$~eV
to $\sim 10^{-1}$~eV. 
The case $M\ll m_D$, as far as observations are concerned, is similar
to the Dirac neutrino case. Here the neutrinos are Majorana fermions
but they still ``pair up'' into pseudo-Dirac fermions. These can be
distinguished from Dirac fermions if one is sensitive to the tiny mass
splitting between the components of the pseudo-Dirac pair, which are
proportional to $M$. Solar neutrino data require $M\lesssim
10^{-9}$~eV \cite{deGouvea:2009fp}, see also \cite{Donini:2011jh}.

One can also deduce an upper bound for $M$. Theoretical considerations
allow one to rule out $M\gtrsim 10^{15}$~GeV \cite{Maltoni:2000iq},
while a simple interpretation of the gauge hierarchy problem leads one
to favor $M\lesssim 10^{7}$~GeV, assuming there are no other new
states at or above the electroweak symmetry breaking scale
\cite{Casas:2004gh}. Naturalness is not a good guide when it comes to
picking a value for $M$ -- all values for $M$ are technically natural
in the sense that in the limit where all $M$ vanish the non-anomalous
global symmetries of Eq.~(\ref{eq:lnu}) are augmented by $U(1)_{B-L}$,
global baryon number minus lepton number (see, for example,
\cite{deGouvea:2005er}). Finally, direct experimental probes of
Eq.~(\ref{eq:lnu}) are possible for $M$ values below the TeV scale --
the reach of collider experiments. We argue, however, that
Eq.~(\ref{eq:Lhier}) can only be unambiguously tested for $M$ values
less than, very roughly, 10~eV, mostly via searches for the effects of
light sterile neutrinos in neutrino oscillations.

A few important points are worthy of note. Eq.~(\ref{eq:theta}),
describing the mixing between active and sterile states, can be
severely violated and much larger values of $\Theta$ are possible
(see, for example, \cite{deGouvea:2007uz}). These are, however, not
generic and require special choices for the entries in the neutrino
mass matrix. This implies that, under extraordinary circumstances, one
may be able to observe not-so-light seesaw sterile neutrinos for, say,
$M\sim 1$~MeV or $1$~GeV, but it is not possible to falsify
Eq.~(\ref{eq:Lhier}) if all $M$ are much larger than 10~eV. On the flip
side, the structure of the active--sterile mixing matrix is not
generic. This means that Eq.~(\ref{eq:Lhier}) can be ruled out (or ruled
``in'') if enough information concerning hypothetical sterile
neutrinos is experimentally collected. For all the details see
\cite{deGouvea:2011zz}.

\subsubsection{eV-scale Seesaw}

Here we will focus on the case that the seesaw scale corresponds to
the scale of light sterile neutrinos responsible for various short
baseline anomalies.

These light sterile neutrinos should be observable in short-baseline
neutrino oscillation experiments sensitive to disappearance at the few
percent level or appearance at the $10^{-4}$ level
\cite{deGouvea:2011zz}. It is interesting that these can also
accommodate $3+1$ and $3+2$ solutions (see, for example,
\cite{Kopp:2011qd,Giunti:2011gz}) to the so-called short baseline
neutrino anomalies from LSND \cite{Aguilar:2001ty}, MiniBooNE
\cite{AguilarArevalo:2007it,AguilarArevalo:2008rc,AguilarArevalo:2010wv}
and reactor data \cite{Mention:2011rk}. If that is indeed the case,
Eq.~(\ref{eq:lnu}) further predicts that $\nu_{\mu}\to\nu_{\tau}$
appearance at short baselines is just beyond the current experimental
upper bounds \cite{deGouvea:2011zz}.

Non-oscillation experiments sensitive to the low energy seesaw include
searches for double-beta decay \cite{deGouvea:2005er,deGouvea:2006gz},
precision measurements of the $\beta$-ray spectrum in nuclear
$\beta$-decay \cite{deGouvea:2005er,deGouvea:2006gz,Barrett:2011jg},
and cosmological bounds on the number of relativistic particle species
in the early Universe and bounds on the fraction of hot dark matter
\cite{deGouvea:2005er,deGouvea:2006gz,Hamann:2010bk,Hamann:2011ge}. An
interesting consequence of the low-energy seesaw (all masses well below 
100~MeV) is that while the
neutrinos are Majorana fermions, the rate for neutrinoless double-beta 
decay and all other potentially observable searches for
lepton-number violation, vanishes
\cite{deGouvea:2005er,deGouvea:2006gz}. It is also possible that there
is partial cancellation:  this means that the
contribution to neutrinoless double-beta decay of a light sterile
neutrino, which generates an active neutrino via seesaw, cancels the
contribution to double-beta decay from this active state
\cite{Barry:2011fp}.  
The observation of a finite
lifetime for neutrinoless double-beta decay would rule out
Eq.~(\ref{eq:Lhier}) unless at least one of the right-handed neutrino
mass parameters $M$ is above a few tens of MeV. A recent re-analysis of sterile
neutrino effects in double beta decay can be found in
\cite{Mitra:2011qr}. This is assuming that
no other mechanisms \cite{Rodejohann:2011mu} that can lead to double
beta decay is realized in nature.

If a global symmetry ({\it e.g.}~lepton number) is imposed on the Lagrangian
(\ref{eq:Lhier}), the general matrices $y^{\alpha i}$, as well as the in
general non-diagonal $M_R$ have a more constrained structure which depends 
on how the different fields transform under the global symmetry. 
It is clear that the complexity of these models increases very fast
with the number of extra Weyl fermions, $n_R$. This number is often
taken to be the same as the number of families, but there is no
fundamental reason why this should be the case. In
Table~\ref{tab:param} we summarize the neutrino mass spectrum and the
number of leptonic mixing parameters as a function of $n_R$, with and
without an exact global lepton number symmetry. Note that  various
lepton charge assignments are possible. 

 \begin{table}
\begin{center}
\begin{tabular}{|l|l|l|l|l|l|}
\hline
$n_R$ & $L_i$ & $\#$ zero modes & $\#$ masses & $\#$ angles & $\#$ CP phases \\
\hline
 1 & - & 2 & 2 & 2 & 0 \\
  & +1 & 2 & 1 & 2 & 0 \\
 \hline
 2 & - & 1 & 4 & 4 & 3 \\
  & (+1,+1) & 1 & 2 & 3 & 1 \\
  & (+1,$-1$) & 3 & 1 & 3  & 1  \\
  \hline
  3 & - & 0 & 6 & 6 & 6 \\
  & (+1,+1,+1) & 0 & 3 & 3 & 1 \\
  & (+1,$-1$,+1) & 2 & 2 & 6  & 4
    \\
  & (+1,$-1$,$-1$) & 4 & 1 & 4   & 1  \\
  \hline
\end{tabular}
\caption{\label{tab:param} Spectrum and number of independent angles and phases for the
models with $n_R=1, 2$ without and with  global lepton number
symmetries. The second column shows the lepton number charge
assignments of the extra singlets, $L_i$. }
\end{center}
\end{table}

A first step towards a systematic exploration of the phenomenology of
such models, in increasing order of complexity, in order to quantify
the constraints imposed by data, was presented in~\cite{Donini:2011jh}, 
see also~\cite{Blennow:2011vn,Fan:2012ca}.  For previous work on the 
subject, see also~\cite{deGouvea:2005er,deGouvea:2006gz,deGouvea:2009fp}.  

The requirement that two distinct mass splittings exist implies 
that the second, fifth and last entries on the table are excluded by
neutrino oscillation data.  The fourth and seventh entries correspond
to two and  three massive Dirac neutrinos, {\it i.e.}, to the standard three 
neutrino scenario. Here we consider the remaining cases for $n_R=1,2$. 
The addition of  just one additional singlet Weyl fermion, $n_R=1$, 
has in principle  enough free parameters (two mixing angles and two
mass differences) to fit the solar and atmospheric oscillation
data. The next-to-minimal choice requires two Weyl fermions,
$n_R=2$. Such a possibility is, of course, well known to give a good  
fit to the data, in the limits $M_R \rightarrow 0$ and $M_R
\rightarrow \infty$, where the standard three neutrino scenario is
recovered. In both cases, the physics spectrum contains one massless
neutrino and two massive ones. The main goal will now be the
exploration of the parameter space in between these two limits, in
search for other viable solutions that could accommodate at least the
solar and atmospheric oscillation, and maybe explain some of the
outliers ({\it e.g.}~LSND).

In the literature, several authors have devoted a lot of effort to 
study the implications of neutrino oscillation data on models with 
extra sterile neutrinos (some recent analyses are~\cite{Sorel:2003hf,
Akhmedov:2010vy,Kopp:2011qd,Giunti:2010jt}), usually referred to as 
$3+1, 3+2, ...3+N_s$. Most of these studies have been done with the 
motivation of trying to accommodate LSND~\cite{Aguilar:2001ty}, and 
MiniBooNE data \cite{AguilarArevalo:2008rc,AguilarArevalo:2010wv}.  It is
important to stress that these phenomenological models usually
correspond to a generic model with $3+N_s$ mass eigenstates, but {\it
do not} correspond in general to a model with $N_s$ extra Weyl
fermions. The number of  free parameters for  $N_s=n_R$ is typically
much larger  than what is shown in Table~\ref{tab:param}, either
because the number of Weyl fermions involved is larger ({\it e.g.}~$3+1$
Dirac fermions correspond in our context to $n_R=5$ and not to
$n_R=1$) or because couplings that are forbidden by gauge invariance
in our model, such as Majorana mass entries for the active neutrinos,
are included, {\it effectively}, in the phenomenological
models. Models for any $n_R$ can be parameterized as  the
phenomenological models with $N_s=n_R$, but in that case there are
generically correlations between parameters ({\it e.g.}~between mixings and
masses). The analyses performed in the context of the phenomenological
models do not take such correlations into account, and usually
restrict the number of parameters by assuming instead some hierarchies
between neutrino masses that could accommodate LSND. In order to
distinguish the $3+N_s$ phenomenological models from those in
Eq.~(\ref{eq:Lhier}), we refer from now on to the latter as $3+n_R$ {\it
minimal} models.  

There are many possible parametrizations of the mass matrix. A good
choice will usually be one that satisfies two properties: 1) it
contains all independent parameters and no more, 2) it is convenient
for imposing existing constraints. Without loss of generality, we can  
choose a basis where the mass matrix $M_R$ takes a diagonal form,
while $m_D$ is a generic $3\times n_R$ complex matrix.  
In the case when $m_D \ll M_N$, a convenient parametrization is the
one first introduced by Casas-Ibarra \cite{Casas:2001sr}, which
exploits the approximate decoupling of the light and heavy sectors,
using as parameters the light masses and mixings that have already
been measured. When this condition is not satisfied, we will use  the
following parametrization for the $n_R=1, 2$ cases: 
\begin{eqnarray}
n_R=1\;\;\; &  m_D &= U^* (\theta_{13},\theta_{23}) \left(\begin{array}{l}0 \\
0\\
m_D
\end{array}\right), \nonumber\\
n_R=2\;\;\; &  m_D &= U^*(\theta_{12},\theta_{13},\theta_{23},\delta) \left(\begin{array}{ll}0 & 0 \\
m_{D^-} & 0\\
0 & m_{D^+} \\
\end{array}\right) V^\dagger(\theta_{45},\alpha_1,\alpha_2),
\end{eqnarray}
where $U$ is a $3\times 3$ unitary matrix, with the same structure as
the PMNS matrix, while $V$ is a $2\times 2$ unitary matrix, depending
on two phases and one angle. In the degenerate limit for $n_R=2$, 
{\it i.e.}, when $M_1=M_2=M$, the matrix $V$ becomes unphysical and $U$
coincides with the PMNS matrix in the two limiting cases $M\rightarrow
0, \infty$.

\begin{figure}[t]
\begin{center}
\includegraphics[width=0.6\textwidth]{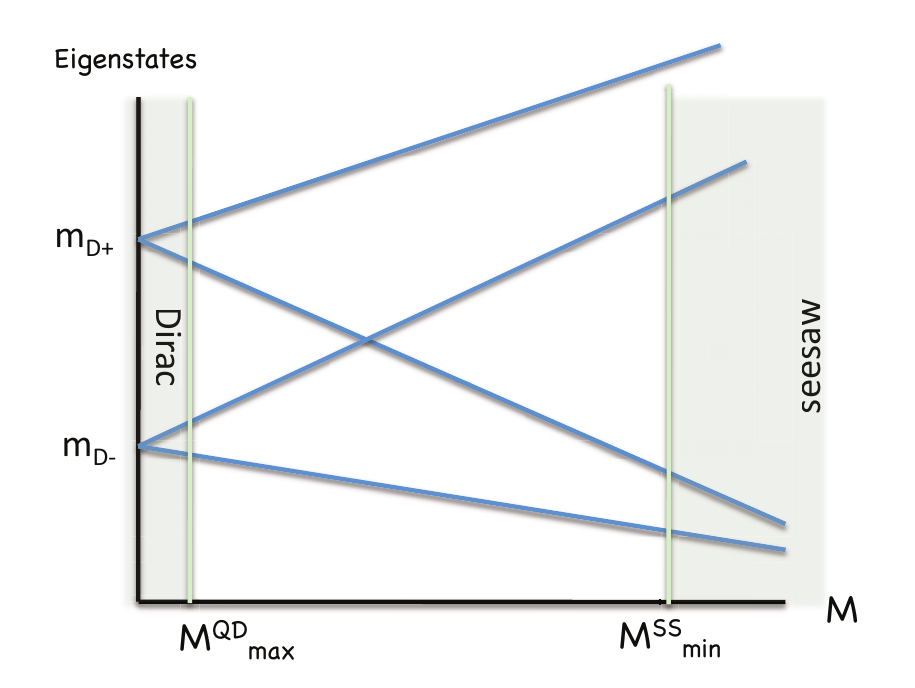}
\end{center}
\caption{Spectrum in the $3+2$ minimal model in the degenerate limit as a function of $M$. }
\label{fig:graph}
\end{figure}

It turns out that the $3+1$ is excluded \cite{Donini:2011jh}, even though in principle it
has sufficient parameters to fit two mass splittings and two mixing
angles. Thus, the most minimal working model is the 3+2 model, which
contains one massless neutrino, four massive states, four angles and 2
CP phases. As a first simplification,  the
degenerate case in the CP conserving limit was explored, where the two
eigenvalues of $M_R$ were assumed to be the degenerate, $M_1=M_2=M$ and no
phases. The number of extra parameters with respect to the standard
three-neutrino scenario is just one extra mass, the common Majorana
mass, $M$. As mentioned above, this model in fact reduces to the
standard three neutrino scenario in the two limits: $M\rightarrow 0$,
when the four eigenstates degenerate into two Dirac pairs, and the
opposite $M\rightarrow \infty$, when the two heavier states
decouple. It is clear therefore that the fit to neutrino data will be 
good for $M$ below the quasi-Dirac limit, $M \leq M_{QD}$ ($M_{QD}$
being the scale where the two massive neutrinos become Pseudo-Dirac particles) and also for
$M$ above the seesaw limit $M \geq M_{SS}$ ($M_{SS}$ 
being the scale where the mass of the active neutrinos is given by the
seesaw formula), see
Fig.~\ref{fig:graph}. As it turns out, only the
expected region $M\leq M_{QD}$ and $M\geq M_{SS}$ survive the fits to
oscillation data. The value of $M_{QD}$ is essentially fixed by solar
data to be very small. The seesaw limit is mostly determined by
long-baseline data.

Above and near the seesaw limit $M \sim M_{SS}$, the spectrum has two
almost degenerate states  with mass ${\mathcal O}$(eV). 
The heavy-light mixings in the minimal model with degenerate Majorana 
masses are completely fixed in terms of the parameters measured in 
oscillations and $M$ ($m_{D^\pm}$ are fixed in terms of $M$ and the 
atmospheric/solar mass splittings): 
\bea
\label{matrixelements}\nonumber
(U_{\mathrm mix})_{e4}= -s_{13} {m_{D^+} \over M} ~,~~
(U_{\mathrm mix})_{e5}= c_{13}s_{12} {m_{D^-} \over M}~,~~ \\ \nonumber
(U_{\mathrm mix})_{\mu 4} =  -c_{13}s_{23} {m_{D^+} \over M} ~,~~
(U_{\mathrm mix})_{\mu 5} = (c_{12}c_{23}- s_{12}s_{13}s_{23})  {m_{D^-}
\over M} .
\eea
Interestingly, for the inverse hierarchy, these mixings are in the 
right ballpark as indicated by the phenomenological $3+1$ and $3+2$ 
fits of all oscillation data. However, due to a strong cancellation of 
the probabilities related to the existence of two almost degenerate 
heavy states, a more 
detailed analysis shows that the degenerate case does not provide a
better fit to the anomalies than the standard three neutrino scenario.
A different situation is found away from the degenerate limit. 
In this case the heavy-light mixings depend on the
light masses and mixings, but also on the two heavy masses, and on a
new complex angle.

\begin{figure}
\includegraphics[width=0.4\textwidth]{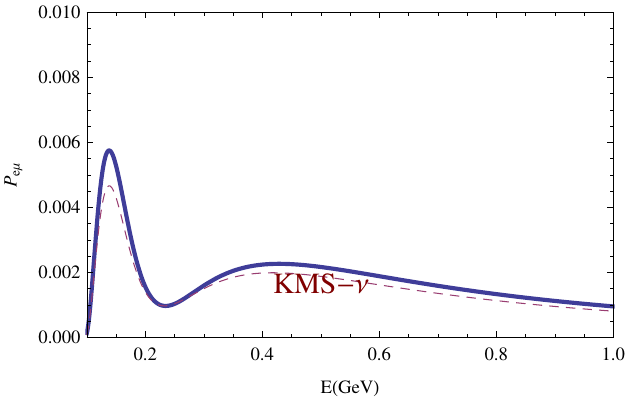} \includegraphics[width=0.4\textwidth]{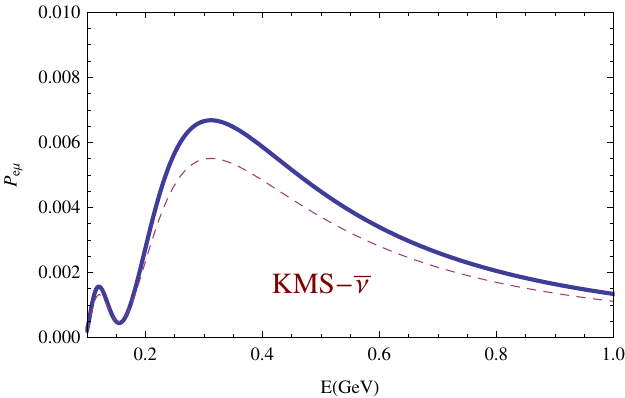}
\caption{\label{fig:usvskms} Oscillation probabilities relevant for MiniBooNE setup for the $3+2$ best fit of \cite{Kopp:2011qd} and 
the minimal $3+2$ model for some choice of the parameters $\theta_{13}, \delta$ and the free complex angle.}
\end{figure}

Varying the free parameters ({\it i.e.}, $\theta_{13}$, $\delta$, $M_1$,
$M_2$ and the extra complex angle) to match the best-fit  heavy-light
mixings and  splittings found in the $3+2$ phenomenological  fit of
\cite{Kopp:2011qd,Giunti:2010jt} we find that there are regions of
parameter space where the two minimal and phenomenological models are
rather close. The oscillation probabilities relevant for MiniBooNE are
compared in Fig.~\ref{fig:usvskms}. 

Although it looks like the $3+2$ minimal model could be as efficient
as the $3+2$ phenomenological model in fitting all oscillation data,
while being much more predictive, a definite conclusion regarding this
point requires a detailed fit of this more constrained model. In
particular, if this model were to explain the LSND/MiniBooNE anomaly,
it would do so only for relatively large value of $\theta_{13}$ and
particular values of the CP phase $\delta$. \\

\subsubsection{The $\nu$MSM}

One can choose masses of sterile neutrinos of the order of masses of other
leptons in the standard model, in keV--GeV range.  It turns out that with this choice of
parameters it is possible to explain simultaneously the data on neutrino
oscillations, the dark matter in the Universe and generate the correct
matter-antimatter asymmetry of the Universe without introducing any new
physics above the electro-weak scale~\cite{Asaka:2005an,Asaka:2005pn,Boyarsky:2009ix}. 
Of course, there is still no explanation for the various parameters (masses,
mixing parameters, couplings) of the extended SM. This model is
called \emph{the $\nu$MSM} -- Neutrino Minimal Standard
Model~\cite{Asaka:2005an,Asaka:2005pn}.\footnote{This theory can accommodate
  inflation, if one considers non-minimal coupling between the Higgs boson and
  gravity~\cite{Bezrukov:2007ep}. Finally, a scale-invariant extension of the
  model~\cite{Shaposhnikov:2008xb,Shaposhnikov:2008xi} that includes
  unimodular gravity may solve the problem of stability of the Higgs mass
  against radiative corrections, even if the Planck scale is included, and
  could lead to an explanation of dark energy and of the absence of the
  cosmological constant.}  The $\nu$MSM provides a new approach to the
``physics beyond the standard model'', concentrating on its observational
problems (see discussion in~\cite{Shaposhnikov:2007nj,Boyarsky:2009ix}).

\begin{figure}
  \centering %
  \includegraphics[height=4.8cm]{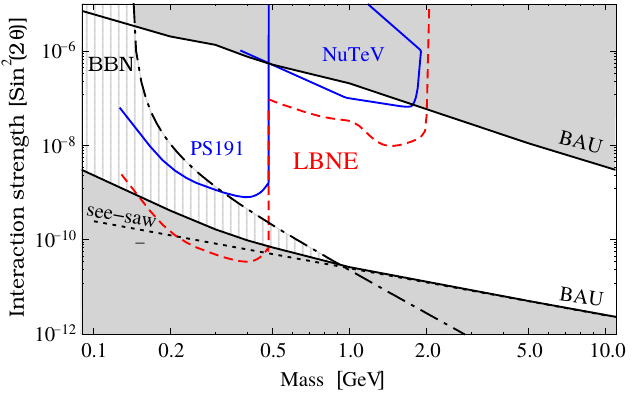}%
  \caption{The allowed region of parameters of two sterile neutrinos in the
    $\nu$MSM,  responsible for neutrino oscillations (region above dotted line
    ``seesaw'') and for baryo/leptogenesis (region between two black solid
    lines).  The hatched region ``BBN'' is excluded from primordial
    nucleosynthesis~\cite{Dolgov:2000pj,Dolgov:2000jw}. Accelerator
    experiments, searching for heavy neutral leptons exclude regions above
    blue solid lines (see~\cite{Ruchayskiy:2011aa} for up-to-date
    details). Prospects for sterile neutrino searches with the LBNE experiment
    are showed in red dashed line~\cite{Akiri:2011dv}. }
  \label{fig:heavy-neutrinos}
\end{figure}

\emph{The $\nu$MSM is testable already with existing experimental means}, as
the masses of all new particles are within the laboratory-accessible range,
including bounds on neutrinoless double beta-decay~\cite{Bezrukov:2005mx},
and possibilities to see the heavier non-DM neutrinos in rare decay
experiments \cite{Gorbunov:2007ak}.  The most direct experimental verification
of the $\nu$MSM would be a discovery of two heavier sterile neutrinos,
responsible for baryogenesis and generation of lepton asymmetry for resonant
production of sterile neutrino dark matter~\cite{Gorbunov:2007ak}. The current
experimental bounds on parameters of these particles are shown in
Fig.~\ref{fig:heavy-neutrinos}.  Future
experiments (for example, NA62 in
CERN\footnote{\url{http://na62.web.cern.ch/NA62}}, LBNE experiment in
FNAL\footnote{\url{http://lbne.fnal.gov}} or the LHCb experiment) are capable
of entering into cosmologically interesting regions of parameters of the $\nu$MSM
which might lead to the discovery of light neutral leptons (see
{\it e.g.}~\cite{Akiri:2011dv}).  If the LHC experiment will not find any particles
other than Higgs boson, these searches may provide the only way to learn about
the origin of neutrino masses and of mechanism of baryogenesis.

The final test of the $\nu$MSM would be to find in astrophysical X-ray
observation clear signatures of DM decay line (potentially distinguishable
from any signal of astrophysical
origin~\cite{Boyarsky:2006fg,Abazajian:2009hx,Herder:2009im}), with parameters
consistent with those of two heavier neutrinos.

\subsection{Sterile Neutrino Dark Matter}
\label{sec:d.-sterile-neutrino}

The nature of Dark Matter ({DM}) is among the most intriguing questions
of modern physics. Its resolution will have a profound impact on the
development of particle physics.  Indeed, massive neutrinos of the Standard
Model were among the first dark matter candidates. However, it was understood
that in a Universe filled with such \emph{hot} dark matter particles, large
scale structure would be incompatible with
observations~\cite{Davis:1985rj}. Standard model neutrinos turned out to be
too light to be viable dark matter candidate.  As a result, the DM particle
hypothesis necessarily implies an extension of the standard model. Candidate
DM particles in such hypothetical standard model extensions differ drastically
in their properties (such as mass, interaction strength, clustering
properties) and therefore in their observational signatures.

Sterile neutrinos (that have very weak interaction with the ordinary matter,
exactly as DM does) are attractive dark matter candidates \emph{if} the model
provides
\begin{description}
\item[Sufficient stability.] This originates from the requirement that the 
  dark matter candidate lives at least as long as the age of the Universe and, what is perhaps more important, from
  astrophysical X-ray bounds.  This requirement poses strict
  bound on the sterile neutrino Yukawa couplings and on interactions
  beyond those of the SM (if any are present).
\item[Production mechanism.] This may be the most complicated problem
  with the sterile neutrino DM.  Depending on the model, this may give
  further requirements on the DM neutrino Yukawa couplings, and
  constrain properties of other particles (for example
  other sterile neutrinos).
\item[Specific mass scale.] The mass scale of the DM neutrino is
  bounded from several considerations.  A lower bound is provided by
  the structure formation in the Universe, in a way that is dependent on the
  neutrino momentum distribution, which itself depends on the
  production mechanism.  In specific models, different
  additional requirements may appear for the masses of the non-DM
  sterile neutrinos.
\end{description}
The generation of the mass scale has been discussed before, below we review 
the other points. The masses of the sterile neutrinos can vary in a wide range. 
On the one hand, sterile neutrinos, as any fermionic dark matter, should 
satisfy the Tremaine-Gunn bound~\cite{Tremaine:1979we}, meaning that their 
masses should be above roughly $400$~eV~\cite{Boyarsky:2008ju,Gorbunov:2008ka}. 
On the other hand, because their decay width is proportional to 
$M_N^5$~\cite{Pal:1981rm} (at least in the models without extra light particles, 
{\it e.g.}~Majorons), high mass neutrinos require very small Yukawa couplings 
to warrant their stability.  The considerations of stability over the lifetime 
of the Universe favor relatively light (keV scale) neutrinos. More stringent 
constraints are imposed on the combination of the masses and mixing angles of 
the sterile neutrino dark matter by the non-observation of a dark matter decay 
line in X-ray spectra of DM-dominated objects~\cite{Abazajian:2001vt,Dolgov:2000ew}. 
The search for such a \emph{decaying dark matter} is very
promising~\cite{Herder:2009im,Abazajian:2009hx}. First of all, a positive
result would be conclusive, as the DM origin of any ``suspicious'' line can be
unambiguously checked.  Indeed, the decay signal is proportional to the
\emph{column density} --- the integral of DM distribution along the line of
sight ($ \int\rho_{DM}(r)dr$) and not to the $\int\rho^2_{DM}(r)dr$ (as it is
the case for annihilating DM).  As a result a vast variety of astrophysical
objects of different nature would produce a comparable decay signal
(cf.~\cite{Boyarsky:2006hr,Boyarsky:2006fg,Boyarsky:2009rb}).  Therefore
{(i)} one has the freedom of choosing the observational targets, avoiding
complicated astrophysical backgrounds; and {(ii)} if a candidate line is
found, its surface brightness profile may be measured (as it does not decay
quickly away from the centers of the objects), distinguished from
astrophysical lines (which usually decay in outskirts) and compared among
several objects with the same expected signal.  This allows one to distinguish the
decaying DM line from any possible astrophysical background rendering 
astrophysical searches for the decaying DM \emph{another type of a direct 
detection experiment}~\cite{Abazajian:2001vt,Dolgov:2000ew}.

\begin{figure}
  \centering
  \includegraphics[width=0.445\textwidth]{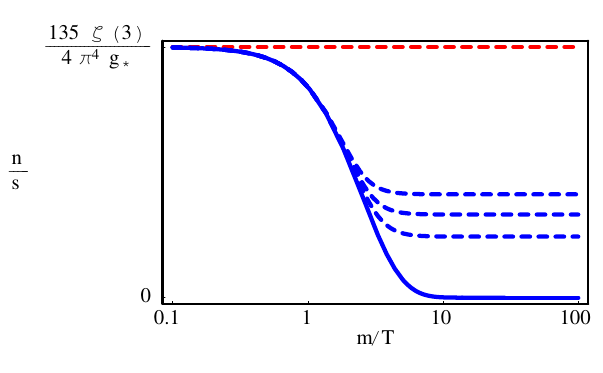}
  \hspace{0.03\textwidth}
  \includegraphics[width=0.503\textwidth]{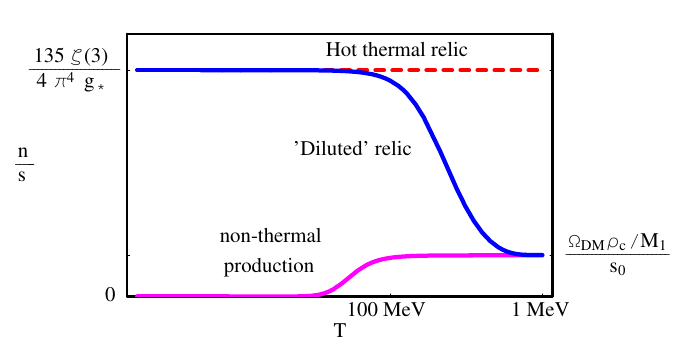}
  \caption{Schematic evolution of the relic abundance in the Universe.  Left
    plot depicts the standard non-relativistic freeze-out mechanism, where the
    abundance is controlled by the decoupling moment (several dashed curves).
    Right plot illustrates various non-thermal mechanisms.  The dashed line is
    a thermal relic decoupled while being relativistic (hot thermal relic),
    leading to the over closure of the Universe.  The blue decreasing line is
    the same hot thermal relic, but with the abundance diluted by rapid
    expansion of the Universe (entropy production), leading to correct DM
    abundance.  The lowest (magenta) line depicts the evolution of the
    non-thermally produced particle with zero primordial abundance.}
  \label{fig:DMgeneration}
\end{figure}

If the sterile neutrino dark matter interacts with other particles so that it
enters into thermal equilibrium in the early Universe and then freezes out
while relativistic (so called ``\emph{thermal relic DM}''~\cite{Bode:2000gq}),
then the contribution $\Omega_N$ of a thermal relic of mass $M_N$ to the
energy balance of the Universe is
\begin{equation}
  \frac{\Omega_N}{\Omega_\mathrm{DM}}
  \simeq
  \frac{1}{S}\left(\frac{10.75}{g_{*\mathrm{f}}}\right)
  \left(\frac{M_N}{1~\mathrm{keV}}\right)\times100
  ,
\end{equation}
where $\Omega_\mathrm{DM}=0.105h^{-2}$ is the observed DM density, $g_{*f}$ is
the effective number of degrees of freedom at freeze-out, and $S$ is the
additional entropy release after the sterile neutrino DM freezes out.  One can
see that in order to obtain the proper amount of DM one either needs to have significant
entropy release, or prevent the DM neutrino from entering thermal equilibrium
at all.\footnote{Reducing the mass of the thermal relic leads to Hot DM, in
  contradiction with data on large-scale structure formation.}  Three main possibilities of sterile
neutrino dark matter production will be reviewed below:
\begin{enumerate}
\item Production from mixing with SM neutrinos,
\item Thermal production with subsequent dilution,
\item Non-thermal production from additional beyond-SM physics.
\end{enumerate}
Some details on general aspects of relic abundance formation are displayed in
Fig.~\ref{fig:DMgeneration}. 
The first mechanism (weak reactions like $\ell\ell \to N\nu$
$qq'\to N\nu$ with interaction strength proportional to $G_F \times
\sin^2(2\theta)$) is always effective if the sterile neutrino mixes with active
ones~\cite{Dodelson:1993je,Shi:1998km,Abazajian:2002yz,Abazajian:2005gj,Asaka:2006rw,Asaka:2006nq,Shaposhnikov:2008pf,Laine:2008pg}.
Some amount of sterile neutrinos is \emph{always} produced via this mixing in
the early Universe (the so-called \emph{non-resonant production of sterile
  neutrinos},
{NRP}~\cite{Dodelson:1993je,Abazajian:2001nj,Abazajian:2002yz,%
  Abazajian:2005gj,Asaka:2006rw,Asaka:2006nq}), and allows one to set a rather robust upper bound on the
mixing angle (from the requirement that sterile neutrinos do not overclose the Universe). This
production is peaked at temperatures $T\gtrsim
150~\text{MeV}(M_N/\text{keV})^{1/3}$ and the resulting momentum distribution
function of DM particles is a rescaled
Fermi-Dirac distribution~\cite{Dolgov:2000ew,Asaka:2006nq}.  As a result, keV-scale sterile
neutrinos are \emph{created relativistic} and become non-relativistic deeply
in the radiation dominated epoch.

Such dark matter particles may have a significant free streaming horizon
(see~\cite{Boyarsky:2008xj} for discussion), which influences structure
formation.  These DM models (often called \emph{Warm Dark Matter}, WDM as opposed to
non-relativistic ``cold'' dark matter) are as compatible with the data on CMB
and large scale structure as the $\Lambda$CDM ``concordance''
model~\cite{Boyarsky:2008xj}.  The difference between cold and warm DM
particles would show up at approximately galactic (sub-Mpc) scales and 
only very recently such small scale effects are starting to be resolved both
theoretically and experimentally. There are several reasons for that. First of
all, dark matter-only simulations of structure formation require, in the case
of WDM, significantly larger number of particles to achieve a resolution
comparable with the CDM case~(see
{\it e.g.}~\cite{Wang:2007he,Polisensky:2010rw,Lovell:2011rd}). Additionally, at
scales of interest, the baryonic physics can hide (or mimic) the WDM
suppression of power~(see {\it e.g.}~\cite{Semboloni:2011fe}).

Experimentally, the effects of suppression of the matter power spectrum are
probed with the Lyman-$\alpha$ forest method.  This method was successfully
applied to ``warm'' thermal relics and to non-resonantly produced sterile neutrinos
({\it e.g.}~\cite{Hansen:2001zv,Viel:2005ha,Abazajian:2005xn,Seljak:2006qw,%
  Viel:2006kd,Viel:2007mv,Boyarsky:2008xj}).  The resulting Lyman-$\alpha$
restrictions are strong enough that free-streaming would affect only structures at
scales considerably less than galactic~(see
{\it e.g.}~\cite{Strigari:2006ue,deNaray:2009xj,Maccio':2009rx,Schneider:2011yu}).

The Lyman-$\alpha$ method was also extended in~\cite{Boyarsky:2008xj} to a
more realistic, particle physics motivated class of ``cold+warm DM models''
({CWDM}).  In this case, the Lyman-$\alpha$ constraints allow a
significant fraction of very warm DM particles~\cite{Boyarsky:2008xj}. Sterile
neutrino dark matter, produced via \emph{resonant oscillations} in the
presence of a significant lepton asymmetry in the
plasma~\cite{Shi:1998km,Shaposhnikov:2008pf,Laine:2008pg} (resonant
production, {RP}) resembles CWDM models at scales probed by the
Lyman-$\alpha$ method~\cite{Boyarsky:2008mt}.  RP DM particles can easily
evade Lyman-$\alpha$ constraints and light (1--2~keV) sterile neutrinos remain
viable dark matter candidates~\cite{Boyarsky:2008mt}, consistent with all
astrophysical and cosmological data down to galactic scales.

Recent results~\cite{Lovell:2011rd} demonstrate that these DM models 
modify the amount of substructure in Galaxy-sized halo along with their properties.
The discrepancy between the number of observed substructures with small masses
and those predicted by $\Lambda$CDM models~\cite{Klypin:1999uc,Moore:1999nt}
can simply mean that these substructures did not confine gas and are therefore
completely dark (see {\it e.g.}~\cite{Bullock:2000wn,Benson:2001at}).
This is not true for larger objects.  In particular, CDM numerical simulations
invariably predict several satellites ``too big'' to be masked by galaxy
formation processes, in contradiction with
observations~\cite{Klypin:1999uc,Moore:1999nt,Strigari:2010un,BoylanKolchin:2011de}. Resonantly
produced sterile neutrino DM turns out to be ``warm enough'' to amend these
issues~\cite{Lovell:2011rd} and ``cold enough'' to be in agreement with
Lyman-$\alpha$ bounds~\cite{Boyarsky:2008mt}.

Along with the Lyman-$\alpha$ forest method, weak gravitational lensing
can be used to probe primordial velocities of dark matter particles (see
{\it e.g.}~\cite{Markovic:2010te,Smith:2011ev,Semboloni:2011fe}), as the next
generation of weak lensing surveys (such as {\it e.g.}~KiDS, LSST, WFIRST, Euclid)
will be able to measure the matter power spectrum at scales down to
$1-10$~h/Mpc with a few percent accuracy~(see {\it e.g.}~\cite{vanDaalen:2011xb}). As in
the case of the Lyman-$\alpha$ forest method, the main challenge is to properly
take into account the baryonic effects on the matter power
spectrum~\cite{Semboloni:2011fe}.

\subsubsection{Dark matter in the $\nu$MSM}
\label{sec:numsm-dm}

\begin{figure}
  \centering %
  \subfigure[][]{\label{fig:sf-window}
    \includegraphics[width=0.31\textwidth]{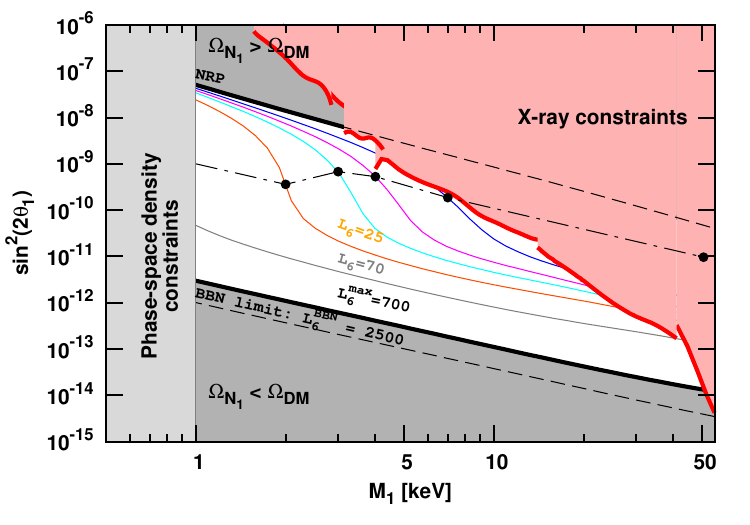}}
  \subfigure[][]{
    \includegraphics[width=0.31\textwidth]{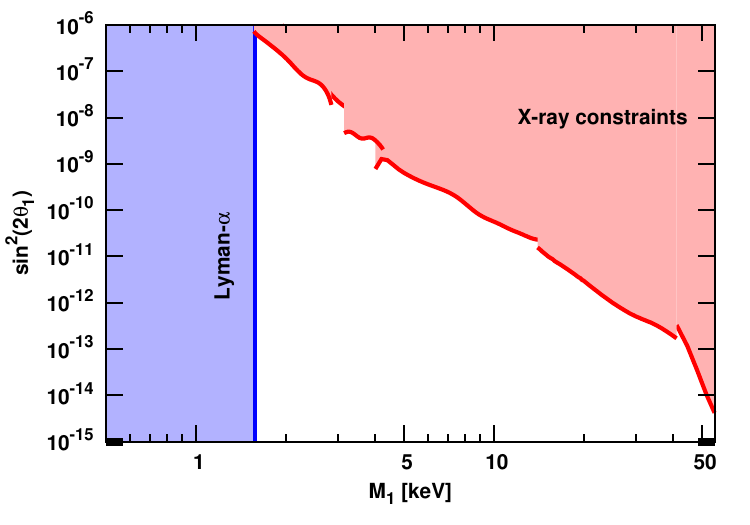}}
  \subfigure[][]{
    \includegraphics[width=0.31\textwidth]{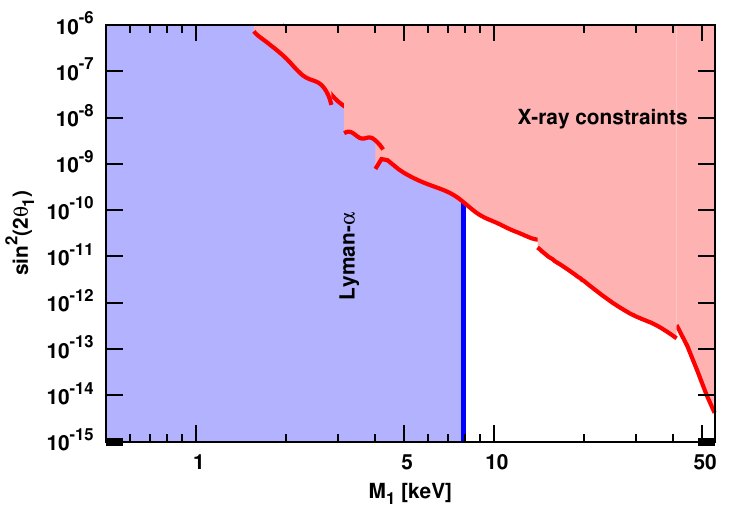}}
  \caption{Bounds on the mass $M_1$ and the mixing angle $\theta_1$ of the
    sterile neutrino dark matter for the models, discussed in
    Section~\ref{sec:d.-sterile-neutrino}: DM in the $\nu$MSM (Panel a, see text
    for details); DM produced in the model with entropy dilution (Panel b);
    and DM produced in the light singlet Higgs decays (Panel c). }
  \label{fig:DMconstraints}
\end{figure}

The lightest sterile neutrino in the $\nu$MSM can be coupled to the rest of
the matter weakly enough to provide a viable DM candidate.  In the $\nu$MSM
sterile neutrino dark matter particles can be produced \emph{only} via
oscillations with active
neutrinos~\cite{Asaka:2006rw,Asaka:2006nq,Laine:2008pg,Shaposhnikov:2008pf}.
Their abundance must correctly reproduce the measured density of DM.  The
allowed region lies between the two black thick lines in
Fig.~\ref{fig:sf-window}. The shaded (red) upper right corner corresponds to
a region excluded by searches for the DM decay line
from X-ray observations~\cite{Abazajian:2001vt,Dolgov:2000ew,Boyarsky:2005us,%
  Boyarsky:2006fg,Boyarsky:2006ag,Watson:2006qb,Abazajian:2006yn,%
  Boyarsky:2007ay,Abazajian:2006jc,RiemerSorensen:2006fh,Boyarsky:2007ge,%
  Loewenstein:2008yi,Loewenstein:2009cm}
neutrinos~\cite{Boyarsky:2006jm}.  
Finally, a lower limit on the mass of DM sterile neutrino $M_{DM} \gtrsim 1$~keV comes
from the analysis of the phase-space density of the Milky way's dwarf
spheroidal galaxies~\cite{Boyarsky:2008ju}.  For non-resonantly produced
sterile neutrinos the Lyman-$\alpha$ method provides a \emph{lower bound} on
the mass $M_{DM} \gtrsim 8$~keV (assuming that baryonic feedback on matter
distribution, discussed {\it e.g.}~in~\cite{Semboloni:2011fe} do not affect this
result significantly).  This result is in tension with the combination of the
X-ray limits, that put an \emph{upper bound} $M_{DM} \lesssim 3-4$~keV~(see
{\it e.g.}~\cite{Seljak:2006qw,Boyarsky:2007ay,Boyarsky:2008xj}).

However, in the $\nu$MSM there is an allowed region with the masses as low as
$1-2$~keV in which \emph{sterile neutrinos, produced resonantly, remain a
  viable DM candidate}~\cite{Boyarsky:2008mt}. The required large lepton
asymmetry is generated by two other sterile neutrinos of the $\nu$MSM, also
responsible for baryogenesis and neutrino oscillations.  We stress that the
resonant production of sterile neutrino DM requires the presence of large,
$\Delta L/L>2\times 10^{-3} $ lepton asymmetry at temperature $T \sim 100$ MeV
and it can only be produced in the $\nu$MSM \cite{Shaposhnikov:2008pf} (the
GUT, electro-weak, or eV scale seesaw cannot provide it).\footnote{See
  however~\cite{Gu:2010dg} where a possibility of cancellation between lepton
  asymmetries stored in different flavours was discussed.}  The thin colored
curves between thick black lines in Fig.~\ref{fig:sf-window} represent
production curves for different values of lepton asymmetry, $L_6\equiv 10^6
(n_{\nu_e}-n_{\bar\nu_e})/s$ (where $s$ -- entropy density). The resulting
parameter space of the $\nu$MSM is bounded, with the highest possible DM mass
$M_{DM} \lesssim 50$~keV~\cite{Boyarsky:2007ge}.  Therefore, in the
  $\nu$MSM the properties of dark matter particles are tightly related to the
  mechanism of baryogenesis and properties of two other sterile neutrinos.

\subsubsection{Neutrinos in gauge multiplets -- thermal production of DM neutrinos}
\label{sec:neutr-gauge-mult}

In this model sterile neutrinos are charged under some beyond the SM gauge
group \cite{Bezrukov:2009th}.  A natural candidate are here left-right 
symmetric theories, in which the sterile neutrinos are sterile only under 
the SM $SU(2)_L$ gauge group, but are active with respect to an additional 
$SU(2)_R$, under which the left-handed SM particles are sterile. The 
steriles couple in particular to a new gauge boson $W_R$, which belongs to
$SU(2)_R$.  One of the sterile neutrinos $N_1$ is light and plays the role 
of dark matter, entering in thermal equilibrium before freeze-out.  Other 
sterile neutrinos $N_{2,3}$ should dilute its abundance up to the correct 
amount via out-of-equilibrium decays.  This entropy production happens if 
there are heavy particles with long lifetimes, which first decouple while 
still relativistic and then decay when already 
non-relativistic~\cite{Scherrer:1984fd}.  The proper DM abundance is controlled 
by the properties of this long-lived particle through the entropy dilution factor
$S\simeq 0.76\frac{\bar{g}_{*}^{1/4}M_2}{g_{*f} \sqrt{\Gamma M_{\mathrm Pl}}}$,
where $g_*$ is an averaged number of d.o.f.\ during entropy generation, and
$M_2$ is the mass of the sterile neutrino, responsible for the dilution. 
The X-ray constraint here bounds the mixing angle $\theta_1$ of the DM
neutrino in the same way as for the $\nu$MSM.  The mixing between new and SM 
gauge bosons is also severely constrained.  The structure formation from the 
Lyman-$\alpha$ analysis constraints the DM neutrino mass:, $M_1>1.6$~keV, 
because its velocity distribution is that of the cooled thermal relic
\cite{Bezrukov:2009th,Boyarsky:2008xj}.  At the same time, this implies
that the DM in this model is cold (CDM).

All other constraints in this scenario apply to the heavier sterile neutrinos 
and to the new gauge sector. The correct abundance of the CDM sterile neutrino 
requires entropy dilution.  To properly provide the entropy dilution, $N_2$ 
should decouple while relativistic and has a decay width
\begin{equation}
  \Gamma \simeq
  0.50\times10^{-6}
  \frac{g_N^2}{4} \frac{g_{*\mathrm{f}}^2}{g_{*}^2}
  \bar{g}_{*}^{1/2}\frac{M_2^2}{M_{\mathrm Pl}}
  \left(\frac{1~\mathrm{keV}}{M_1}\right)^2
  .
\end{equation}
At the same time, the heavy neutrino $N_2$ should decay before BBN,
which bounds its lifetime to be shorter than approximately
$0.1\div2$~s.  Then, the proper entropy can be generated only if its
mass is larger than
\begin{equation}
  M_2 > \left(\frac{M_1}{1~\mathrm{keV}}\right)
  (1.7\div10)~\mathrm{GeV}
  .
\end{equation}
The entropy is effectively generated by out-of-equilibrium decays if
the particle decoupled while still relativistic.  The bound on the
decoupling temperature leads to a bound on the new gauge
boson mass
\begin{equation}
  M >
  \frac{1}{g_{*\mathrm{f}}^{1/8}}\left(
    \frac{M_2}{1~\mathrm{GeV}}\right)^{3/4}(10\div16)~\mathrm{TeV}
  .
\end{equation}

An alternative way of fighting the overproduction in a model with
sterile neutrinos charged under some gauge symmetry is analyzed in
\cite{Kusenko:2010ik}, where the mass of the corresponding gauge
boson is heavy, of the order of $10^{15}$~GeV.  In this case, if the
reheating temperature is also lower than the GUT scale, the
additional gauge interactions do not fully thermalize the SM sterile
neutrinos but still allow the generation of the required DM abundance. 

Note here that in most models with additional physics
responsible for the properties of sterile neutrinos, similar analyses
should be performed, since new sterile neutrino interactions may
lead to an increase in its decay width or bring it into thermal
equilibrium.

\subsubsection{Primordial generation of DM neutrinos}
\label{sec:prim-gener-dm}

Finally, the DM neutrinos may be produced by some other mechanism related to
interactions beyond the SM which do not lead to thermal equilibration of the 
DM sterile neutrino.  An example is provided by models with an additional 
scalar $\phi$, which provides Majorana masses to the sterile neutrinos when 
it acquires a vacuum expectation value
\[
\mathcal{L} = \sum_i {f_i \phi} {\bar N}^c_i N_i.
\]
Here, the neutrinos will get masses $m_i=f_i\langle\phi\rangle$.
At the same time, this interaction leads to the decay of the scalar
$\phi$ into sterile neutrinos.  This decay can be made responsible for
the generation of the DM.

A scenario where the scalar $\phi$ is also responsible for inflation in 
the early Universe was analyzed in~\cite{Shaposhnikov:2006xi,
Anisimov:2008qs,Bezrukov:2009yw}.  In this setup the inflaton decays 
during reheating mostly into SM particles, but some fraction of decays 
are directly into DM neutrino.  In this case, the production is not 
connected with $\theta_1$, and the only bounds present for the DM neutrino 
properties are the X-ray bound and the Lyman-$\alpha$ bound.  The DM 
neutrino in this model is produced with nearly thermal 
spectrum~\cite{Shaposhnikov:2006xi} (``thermal relic'' case 
of~\cite{Boyarsky:2008xj}), so its mass should be $M_1>8$~keV. In this 
scenario, the mass of the DM sterile neutrino can be related to the mass 
of the inflaton, which is quite light and can be searched for in B-meson 
decays~\cite{Bezrukov:2009yw}.

Note that it is also possible that a keV mass 
sterile DM neutrino (if produced from non-thermal decay) can belong to
a thermalized hidden sector which is not in thermal equilibrium with
SM. One particular example~\cite{Das:2010ts} is a lower limit of 1.5 keV (from
free-streaming and phase space density) set on hidden sector warm
dark matter which might be at a different temperature than the SM.

Another setup, where the additional scalar is produced itself from its
interaction with the Higgs boson is described in~\cite{Kusenko:2006rh,
Petraki:2007gq}.  The laboratory consequences for it in sterile neutrino 
decays are discussed in~\cite{Shoemaker:2010fg}.

Alternatively, one can also produce DM neutrinos from higher dimensional
operators appearing in the model.  This was analyzed in the context of the
$\nu$MSM and Higgs-inflationary models~\cite{Bezrukov:2011sz,Bezrukov:2008ut}.
In this mechanism it is possible to produce only neutrinos with higher masses,
of the order $10^4$~keV.

The generation of DM sterile neutrinos is also possible in inflationary
model with $R^2$ modification of gravity~\cite{Gorbunov:2010bn}.  In
such a model  an additional degree of freedom, the scalaron, effectively
appears from the higher order operator $R^2$, and serves as the
inflaton.  At a subsequent reheating, the scalaron decays into all
non-conformal fields, with the main decay products being the SM
particles (via the Higgs decay).  The right-handed neutrinos are also
produced, because their Majorana mass terms are not conformal, and can
provide for DM.  The required mass in this case is of the order
$10^7$~GeV, making it CDM.

\subsubsection{Other models}

Several other papers attempt to provide an explanation for the specific 
mass patterns of the sterile neutrinos with a keV DM neutrino and heavier 
sterile neutrinos (see {\it e.g.}~\cite{He:2009mv,Lindner:2010wr,Merle:2011yv,
Barry:2011fp,Chen:2011ai}).  However,  further research is needed in order to 
explicitly verify whether the sterile neutrino dark matter in these models is 
produced in the correct amount and does not contradict existing astrophysical 
and cosmological bounds.

\subsection{Light Sterile Neutrinos as Messengers of New Physics}
\label{sec:messengers}

Before discussing alternative scenarios to non-standard neutrino
physics, which can be invoked to explain hints usually interpreted in
terms of sterile neutrinos, let us set the stage with an overview.  

It is often said that, because neutrinos have only weak interactions, neutrino experiments can be
sensitive to small new physics effects that would not otherwise be
manifest. Indeed, the interferometric nature of neutrino oscillations
and their tiny masses make them excellent candidates to search for new
physics. As will be seen in the next subsection, however, any new 
physics affecting neutrinos must also affect at some level the
properties of charged leptons, since neutrinos and charged leptons are
related by $SU(2)$ gauge symmetry. For instance, neutrino non-standard
interactions (NSI) from $d= 6$ effective operators, which in principle
could affect neutrino production, detection, and matter effects in
propagation, are already strongly constrained by charged lepton
data. Instead of higher dimension effective operators, one can
introduce new light force carriers that couple to neutrinos; the
resulting flavor-dependent Long-Range Interactions (LRI) then affect
neutrino flavor oscillations, in a way that will appear as a
non-standard matter effect if sufficiently short-range, or as a
non-standard vacuum oscillation if sufficiently long-range. However
here again such flavor-dependent interactions are already highly
constrained by charged lepton data (and, for sufficiently long-range
forces, by tests of the equivalence
principle)~\cite{Lee:2011uh,Heeck:2010pg}. To put it mildly, it is a
challenge to theorists to formulate UV-complete scenarios of new
physics that satisfy existing bounds from the charged lepton and quark
sectors while making dramatic predictions for the neutrino sector.

The prospects for observing exotic physics in the neutrino sector brighten considerably
in scenarios with light steriles. Now one has the possibility of combining new interactions
with active-sterile mixing, such that the combined effects are less constrained by charged
lepton data while still detectable in present or future neutrino oscillation experiments.
Let us mention a few examples illustrating this synergy\footnote{An
aspect of sterile neutrinos which is similar to a non-standard 
neutrino physics aspect is a ``zero-distance'' effect \cite{Langacker:1988up}, which denotes
the property that in case of a non-unitary mixing matrix the
oscillation probabilities have the property $P(\nu_\alpha \to
\nu_\beta) \neq \delta_{\alpha\beta}$ in the limit $L \to 0$. In case
sterile neutrinos are present, the averaged out oscillation
corresponding to its large mass-squared difference mimics this effect.}.

In the analysis of Ref.~\cite{Akhmedov:2010vy,Karagiorgi:2012kw}, a single light sterile neutrino
is combined with NSI in an attempt to reconcile LSND with MiniBooNE short-baseline
antineutrino data. The NSI couplings are taken to be of the order of $10^{-2}$, and
to affect both neutrino production and detection. In the presence of NSI, the neutrino
transition probabilities exhibit CP violation even in the one-mass-scale-dominance limit
relevant to the short-baseline experiments; this allows to reconcile the antineutrino
data in a ($3$$+$$1$) oscillation scheme.

The analyses in Refs.~\cite{Nelson:2007yq,Engelhardt:2010dx} posit a light sterile neutrino
in combination with a new light gauge boson that couples to $B-L$.  In the absence of the
sterile neutrino the new force would have no effect on neutrino flavor oscillations, since all three 
active neutrino flavors have the same $B-L$ charge. In the presence of active-sterile mixing,
the light gauge boson produces a non-standard matter effect in
neutrino propagation that effectively violates CPT and is energy
dependent. For a coupling $< 10^{-5}$ and gauge boson mass on the
order of tens of keV, this new physics can reconcile LSND and
MiniBooNE data. A lighter gauge boson could instead produce an
apparent CPT asymmetry in MINOS long baseline muon-neutrino oscillations.

In the event that genuine CPT-violating mass differences are discovered in the neutrino sector, it would
be difficult not only to produce a plausible model for such an effect, but also to reconcile
CPT violation in the neutrino sector with the very strong upper bounds on CPT-violating
physics for quarks and charged leptons~\cite{Barenboim:2009ts}. These problems were
addressed in Ref.~\cite{Barenboim:2002tz}, which exhibited an explicit model for neutrinos
with a CPT-asymmetric mass spectrum. In this framework the active neutrinos are 
conventional Dirac fermions, while there is one or more species of exotic sterile neutrinos.
The steriles are nonlocal and respect CPT, but the implementation of CPT differs from
that for Dirac fermions. When the actives and steriles mix, the mismatch between CPT
operators induces a CPT asymmetry in the mass eigenstates. 

Light sterile fermions, through mixing with the active neutrino flavors, could be the
messengers of one or more varieties of exotic physics. As we see from these examples,
in such scenarios the leakage of the exotic
physics into phenomena with charged leptons is reduced, both because the steriles
are standard model singlets, and because they can differ in other respects from
the active neutrinos.

\subsection{Non-Standard Neutrino Interactions (NSI)}
\label{sec:NSI}

Non-standard neutrino interactions (NSI) are a very widespread and 
convenient way of parameterizing the effects of new physics in neutrino
oscillations~\cite{Wolfenstein:1977ue,Mikheev:1986gs}. NSI can lead to
non-diagonal neutral currents.  Heavy SM-singlet ({\it i.e.} sterile) 
neutrinos, for instance, induce certain types of NSI, as will be discussed 
in subsection~\ref{sec:modelsSA}. Recall that in the presence of sterile 
neutrinos  the leptonic neutral current is in general non-diagonal in mass
eigenstates, and therefore also in flavor eigenstates~\cite{Schechter:1980gr,
Lee:1977tib}.  The formalism, however, is model-independent.  Even though 
present data constrain NSI to be a subleading effect in neutrino oscillation 
experiments, the possibility of their eventual detection and interference 
with standard neutrino oscillation measurements has triggered a considerable 
interest in the community. Here we review the present bounds on NSI and 
discuss the possibility of saturating them in particular standard model (SM) 
extensions.

\subsubsection{CC-like NSI}
\label{sec:cc}

In this section we discuss NSI for source and detector charged-current processes, 
we will refer to these as charged-current-like NSI. Leptonic NSI are given by the
effective Lagrangian density \footnote{Note that for simplicity only vector 
couplings are assumed, neglecting the possibility of scalar, pseudoscalar, axial 
vector or tensor terms.}.
\begin{equation}
 \label{eq:leptonCCNSI}
 \mathcal L_{\mathrm NSI}^\ell =
 -2\sqrt{2} G_F \eps^{\alpha\beta P}_{\gamma\delta}
 [\bar\ell_\alpha \gamma^\mu P \ell_\beta] 
 [\bar\nu_\gamma \gamma_\mu P_L \nu_\delta],
\end{equation}
where $G_F$ is the Fermi constant and $P$ is either $P_L$ or $P_R$ and, due to Hermiticity,
$\eps^{\alpha\beta P}_{\gamma\delta} = \eps^{\beta\alpha
  P*}_{\delta\gamma}$. For NSI at neutrino production via muon decay 
$\alpha =\mu$ and $\beta = e$. Note that $\alpha = \beta = e$ would 
instead correspond to NSI with electrons in matter. 

In a similar fashion, the charged-current-like NSI with quarks are
given by 
\begin{equation}
 \mathcal L^q_{\mathrm NSI} = -2\sqrt{2} G_F \eps^{qq' P}_{\alpha\beta} V_{qq'} 
 [\bar q \gamma^\mu P q'][\bar\ell_\alpha \gamma_\mu P_L \nu_\beta] + h.c.,
\label{eq:quarkCCNSI}
\end{equation}
where $q$ is an up-type and $q'$ is a down-type quark. Only
$q = u$ and $q' = d$ are of practical interest for neutrino
oscillations, due to their contributions to charged-current
interactions with pions and nuclei. In Ref.~\cite{Biggio:2009nt} bounds on these production
and detection NSI were derived from constraints on the comparison of different measurements of $G_F$: through
$\mu$ decay (affected by leptonic NSI of Eq.~(\ref{eq:leptonCCNSI})), $\beta$ decays (affected by quark NSI of Eq.~(\ref{eq:quarkCCNSI})) and
the kinematic measurements of the masses of the gauge bosons $M_Z$ and $M_W$ (not affected by neutrino NSI). Universality tests of $G_F$ from $\pi$ and $\tau$
decays as well as short baseline neutrino oscillation experiments such as KARMEN or NOMAD, were also considered. Note that in order to derive those bounds only one NSI parameter was switched on at a time in order to avoid the relaxation of the bounds through cancellations among different parameters. Here we present a summary of the most stringent bounds derived:
\begin{eqnarray}
\nonumber
 |\eps^{\mu e}_{\alpha\beta}| 
 \;< \;
 \left( \begin{array}{ccc} 0.025 & 0.030 & 0.030 \\ 
                           0.025 & 0.030 & 0.030 \\ 
			   0.025 & 0.030 & 0.030
\end{array} \right), \quad
\label{matrixud}
 |\eps^{ud}_{\alpha\beta}|
 &<&
 \left( \begin{array}{ccc} 0.041 & 0.025 & 0.041 \\ 
\begin{array}{c}
1.8 \cdot 10^{-6} \\[-.1cm]
0.026
\end{array}
 & 0.078 & 0.013 \\ 
\begin{array}{c}
0.087 \\[-.1cm]
0.12
\end{array} &
\begin{array}c
0.013 \\[-.1cm]
0.018
\end{array} &
0.13 
\end{array} \right).
\label{eq:CCbounds}
\end{eqnarray}
Whenever two values are quoted, the upper value refers to left-handed NSI and 
the lower to right-handed NSI. These bounds stem from the direct effect of the 
corresponding effective operator and are rather weak. However, 
Eqs.~(\ref{eq:leptonCCNSI}) and~(\ref{eq:quarkCCNSI}) are not gauge invariant 
and can be related to flavour changing processes involving charged leptons when 
promoting the neutrino fields to full lepton doublets which would lead to much 
tighter constraints. In Sect.~\ref{sec:modelsSA} we will discuss possible ways 
to evade these stronger constraints.

\subsubsection{NC-like NSI}
\label{sec:nc}

We next review the current status of the
bounds on NSI matter effects, or neutral-current-like NSI defined as
\begin{equation}
 \mathcal L^{M}_{\mathrm NSI} = -2\sqrt{2} G_F \eps^{fP}_{\alpha\beta} 
[\bar f \gamma^\mu P f][\bar\nu_\alpha \gamma_\mu P_L \nu_\beta],
\label{matterNSI}
\end{equation}
where $f = e, u, d$. This type of NSI is the most extensively studied in the 
literature, since the constraints on the charged-current-like NSI are 
generally stronger. Indeed,the bounds from Refs~\cite{Davidson:2003ha,
Barranco:2005ps,Barranco:2007ej,Bolanos:2008km}, but discarding the loop 
constraints on $\eps^{fP}_{e\mu}$ given the discussion in~\cite{Biggio:2009kv}, 
result in the following $\mathcal O(10^{-1})$ constraints:
\begin{eqnarray}
\nonumber
 |\eps^e_{\alpha\beta}|
 \;<\;
  \left( \begin{array}{ccc}
\begin{array}c
0.06 \\[-.1cm]
0.14
\end{array} &
0.10 & 
\begin{array}c
0.4 \\[-.1cm]
0.27
\end{array} \\
0.10 & 0.03 & 0.10 \\
\begin{array}c
0.4 \\[-.1cm]
0.27
\end{array} &
0.10 &
\begin{array}c
0.16 \\[-.1cm]
0.4
\end{array} 
\end{array} \right),
\nonumber
 |\eps^u_{\alpha\beta}|
 <
  \left( \begin{array}{ccc}
\begin{array}c
1.0 \\[-.1cm]
0.7
\end{array} &
0.05 & 0.5 \\
0.05 & 
\begin{array}c
0.003 \\[-.1cm]
0.008
\end{array} & 0.05 \\
0.5   & 0.05 &
\begin{array}c
1.4 \\[-.1cm]
3
\end{array}
 \end{array} \right),
 |\eps^d_{\alpha\beta}|
 <
  \left( \begin{array}{ccc}
 \begin{array}c
0.3 \\[-.1cm]
0.6
\end{array}  & 0.05 &  0.5 \\
 0.05  & 
\begin{array}c
0.003 \\[-.1cm]
0.015
\end{array} & 0.05 \\
 0.5  & 0.05 & 
\begin{array}c
1.1 \\[-.1cm]
6
\end{array}
 \end{array} \right).
\label{eq:NCbounds}
\end{eqnarray}
Again, the operator in Eq.~(\ref{matterNSI}) is not gauge invariant. 

\subsubsection{Models to evade charged lepton NSI}
\label{sec:modelsSA}

In Refs.~\cite{Antusch:2008tz,Gavela:2008ra} extensions of the SM 
leading to neutrino NSI but avoiding their charged current counterparts were studied both 
through $d=6$ and $d=8$ operators.  Only two examples of SM extensions giving rise to 
matter NSI but avoiding their charged lepton counterpart exist at $d=6$. The most direct 
one involves an antisymmetric 4-lepton operator, generated from the exchange of virtual 
singly charged scalar fields. In the second possibility NSI are induced through the 
dimension 6 operator modifying the neutrino kinetic terms, generated by the exchange of 
virtual fermionic singlets. The latter operator generates the NSI in an indirect way, 
{\it i.e.}, after canonical normalization of the neutrino kinetic terms.  

In the first case, NSI with electrons in matter appear through the operator 
\cite{Bilenky:1993bt,Bergmann:1999pk,Cuypers:1996ia}
\begin{eqnarray}\label{Eq:AntisymmDim6}
{\mathcal L}^{d=6,as}_{NSI} =
c^{d=6,as}_{\alpha\beta\gamma\delta} (\overline{L^c}_\alpha i \sigma_2 L_\beta) (\bar L_\gamma i \sigma_2 L^c_\delta) 
\;,
\end{eqnarray} 
which is generated after integrating out a charged scalar $SU(2)$ singlet coupling to the SM lepton doublets:
\begin{equation}\label{Eq:IntS}
 \mathcal{L}^{ S }_{int} = - \lambda^i_{\alpha\beta} \overline{L}_{\alpha}^c i \sigma_2 L_\beta  S_i   = \lambda^i_{\alpha\beta}  S_i 
(\overline{\ell}^c_\alpha P_L \nu_\beta - \overline{\ell}^c_\beta P_L \nu_\alpha) 
\end{equation}
Integrating out the heavy scalars $S_i$ generates the dimension 6 operator of Eq.~(\ref{Eq:AntisymmDim6}) at tree level. 
We find that, given the antisymmetric flavour coupling of Eq.~(\ref{Eq:IntS}), the only NSI induced are those between $\nu_\mu$ and $\nu_\tau$ with electrons in matter. Bounds on these NSI can be derived from $\mu$ and $\tau$ decays:
\begin{eqnarray}\label{Eq:EpsBoundsDim6As}
|\varepsilon^{m,e_\mathrm{L}}_{\mu \mu}| < 8.2 \cdot 10^{-4}\;, 
|\varepsilon^{m,e_\mathrm{L}}_{\tau \tau}| < 8.4 \cdot 10^{-3}\;, 
|\varepsilon^{m,e_\mathrm{L}}_{\mu \tau}| < 1.9 \cdot 10^{-3}\;.
\end{eqnarray}
NSI in production and detection are also induced with similar strengths.

The second realization of NSI at $d=6$ is via the dimension 6 operator
\begin{eqnarray}\label{Eq:Dim6Kin}
{\mathcal L}^{d=6}_{kin} =
- c^{d=6,kin}_{\alpha\beta} (\bar L_\alpha \cdot H^\dagger) \,i\cancel{\partial}\, (H \cdot L_\beta) ,
\end{eqnarray} 
which induces non-canonical neutrino kinetic terms \cite{DeGouvea:2001mz,Broncano:2002rw,Antusch:2006vwa} though the vev of the Higgs field.  
After diagonalizing and normalizing the neutrino kinetic terms, 
a non-unitary lepton mixing matrix is produced from this operator and hence non-standard 
matter interactions as well as related non-standard interactions at the source and detector are induced.
The tree level generation of this operator, avoiding a similar contribution to charged
leptons that would lead to flavour changing neutral currents, requires the introduction of
SM-singlet fermions which couple to the Higgs and lepton doublets via the Yukawa couplings
(see {\it e.g.}~\cite{Abada:2007ux}). 
Electroweak decays set the following bounds on the operator of Eq.~(\ref{Eq:Dim6Kin}) due to the effects a deviation from unitarity of the mixing matrix would imply \cite{Tommasini:1995ii,Antusch:2006vwa,Antusch:2008tz}:
\begin{eqnarray}\label{Eq:D6MaboveEW}
 \frac{v^2}{2}|c^{d=6,kin}_{\alpha\beta}| < 
 \left( \begin{array}{ccc}
4.0 \cdot 10^{-3}  & 1.2 \cdot 10^{-4}  &  3.2 \cdot 10^{-3}\\
1.2 \cdot 10^{-4}  & 1.6 \cdot 10^{-3}  &  2.1 \cdot 10^{-3}\\
3.2 \cdot 10^{-3}  & 2.1 \cdot 10^{-3}  &  5.3 \cdot 10^{-3} 
\end{array} \right)\, .
\label{nndag}
\end{eqnarray}
These bounds can be translated to constraints in the matter NSI that would be induced 
by a non-unitary mixing:
\begin{eqnarray}
|\varepsilon_{\alpha\beta}| <\frac{v^2}{4} |\left(\frac{n_n}{n_e}-\delta_{\alpha e} -
\delta_{e \beta }\right)\,c^{d=6,kin}_{\alpha \beta}| .
\end{eqnarray}
Since the ratio of the neutron to electron density $\frac{n_n}{n_e}$ is in general 
close to $1$, this implies that the bounds on $|\varepsilon_{e \mu}|$ and 
$|\varepsilon_{e \tau}|$ are significantly stronger than the bounds on the 
individual $\varepsilon_{\alpha\beta}$. For the main constituents of the Earth's 
crust and mantle the factor $\frac{n_n}{n_e}-1$ means an additional suppression of 
two orders of magnitude of the NSI coefficient \cite{Antusch:2008tz}. Thus, even 
if two $d=6$ possibilities exist to induce neutrino NSI while avoiding the 
corresponding charged lepton NSI, they cannot saturate the mild direct bounds from 
Eqs.~(\ref{eq:CCbounds}) and (\ref{eq:NCbounds}) and stronger $\mathcal O(10^{-3})$ 
bounds apply.
 
In Refs.~\cite{Antusch:2008tz,Gavela:2008ra} the generation of matter NSI avoiding 
similar operators involving charged leptons was also studied through $d=8$ operators. 
In~\cite{Antusch:2008tz} the analysis was restricted to new physics realizations that 
did not involve cancellations between diagrams involving different messenger particles 
and, under those conditions, the bounds derived for the $d=6$ realizations can be 
translated also to the $d=8$ operators and, even if charged fermion NSI are avoided, no 
large neutrino NSI are obtained. In~\cite{Gavela:2008ra} it was shown that, allowing 
cancellation between diagrams involving different messengers, large neutrino NSI could 
be obtained by tuning away all other dangerous contributions to the charged lepton 
sector. However, this cancellation is only exact at zero momentum transfer, therefore 
constraints can be derived when probing larger energies at colliders, as studied in 
Ref.~\cite{Davidson:2011kr}, and strong constraints $\eps < 10^{-2} - 10^{-3}$ are 
again recovered through LEP2 data.

Finally, the analyses of Refs.~\cite{Antusch:2008tz,Gavela:2008ra} were limited to tree 
level effective operators. Therefore, they do not cover the possibility of loop-induced 
NSI or lighter mediators that cannot be integrated out of the theory. These can thus be 
ways out of the stringent constraints derived and potential sources of large NSI. 
Unfortunately, neither the loop generation nor lighter mediators seem to provide ways in 
which to evade gauge invariance and avoid the more stringent bounds leaking from the 
charged lepton sector. The generation of NSI via loop processes was studied in detail in 
Ref.~\cite{Bellazzini:2010gn} and applied in particular to the MSSM contributions. It was 
found that charged lepton flavor-violating processes constrain the induced NSI to the 
$\sim 10^{-3}$ level, as with the other alternatives explored. Concerning lighter 
mediators, this is a way out of the effective theory treatment presented 
in~\cite{Antusch:2008tz,Gavela:2008ra}, but gauge invariance still poses strong 
constraints and, indeed, flavor-conserving interactions, much more weakly constrained 
in the charged lepton sector than the flavor-violating counterparts, are usually 
considered instead.

\subsubsection{Summary of the present status on NSI}

The present model independent bounds on NSI were reviewed, which state
that the production and detection NSI are bounded to be $< \mathcal
O(10^{-2})$. Conversely the bounds on matter NSI are around one order
of magnitude weaker. Saturating these mild direct bounds could lead to
large observable signals at neutrino experiments.  Matter NSI are
however related to production and detection NSI and to flavour
changing operators for charged leptons through gauge invariance. 
Exploring gauge invariant realizations one finds that gauge invariant 
NSI are constraint to be $\mathcal O(10^{-3})$ making them very challenging, 
but not impossible, to probe at present~\cite{Alonso:2010wu,Antusch:2010fe} 
facilities with dedicated new detectors or future neutrino oscillation 
facilities such as the Neutrino Factory~\cite{FernandezMartinez:2007ms,
Goswami:2008mi,Kopp:2008ds,Antusch:2009pm,Gago:2009ij,Coloma:2011rq}.
Furthermore, the stringent constraints from gauge invariance mainly
apply to flavor-violating processes, thus, fully $SU(2)$ invariant
operators can be considered with $\mathcal O(10^{-2})$ entries in some
of the diagonal elements, see, {\it e.g.}~Ref.~\cite{Carpentier:2010ue}.

\subsection{Extra Forces}

The term ``sterile neutrino''   may be  misleading, and covers an extremely broad 
range of theoretical possibilities. The states which could be discovered in a sterile 
neutrino search may have little to do with the standard neutrinos from a fundamental 
theory point of view, and need not be ``sterile" when non-standard interactions are 
concerned. Consider that 96\% of the energy density  of our Universe today  resides in
mysterious  dark matter and dark energy, neither of which are comprised of standard 
model particles.  From a fundamental theory point of view, there is no reason to 
exclude additional ``sectors", {\it i.e.}, a field or set of fields which have not 
yet been discovered due to the extreme feebleness of their interactions with the 
fields of the standard model. Such sectors are ubiquitous in string-based constructions 
and theories with extra dimensions, and it is possible that the answer to the dark 
matter and/or dark energy puzzles resides in a hidden sector, which contains one or 
more  light fermions that manifests as a sterile neutrino or neutrinos that can
mix with the neutrinos of the standard model.  If such a hidden sector also contains 
at least one light boson which couples to the sterile neutrinos, sterile neutrino 
cosmology can be altered as well, in some cases alleviating the cosmological 
tension~\cite{Dolgov:2002wy,Hamann:2010bk} between more than one additional ultralight 
states or reconciling a  large  neutrino mass with cosmology \cite{Weiner:2005ac}, 
or even contributing to dark matter~\cite{Das:2006ht} or dark 
energy~\cite{Fardon:2003eh,Fardon:2005wc}.

The wealth of theoretical possibilities may at first seem overwhelming when one is 
trying to design experiments to search for   hidden sectors, but effective field theory 
and dimensional analysis provides a useful guideline for focussing the search. At low 
energy, one may write a generic interaction between the standard model and a hidden 
sector as 
\begin{equation}
\frac{\mathcal C}{ \Lambda^n} {\mathcal O}_H {\mathcal O}_{SM} , 
\end{equation}
where $\mathcal C$ is a dimensionless number, $\Lambda$ is a scale (typically large) 
associated with particles that can readily interact with both sectors, ${\mathcal O}_H$ 
is a hidden sector operator, ${\mathcal O}_{SM}$ is a gauge invariant operator in the 
standard model, and $n={\mathrm dim}{\mathcal O}_{SM}+{\mathrm dim}{\mathcal O}_{H}-4$. 
Dimensional analysis tells us to focus on operators of low dimension in both sectors, as 
these are likely to be the most important.

Neutrinos offer a uniquely sensitive window into hidden sectors, because neutrinos do 
not carry any electric or color charge, because their masses are so small, and because 
the operator $L H$ is the lowest dimension  gauge invariant  operator involving 
standard model fermions (here $L$ refers to any of the lepton doublets and $H$ to the 
Higgs doublet). For instance, if the hidden sector   contains  a composite fermion $N$ 
which can be created by a dimension $5/2$ operator ${\mathcal O}_H$, the operator 
\begin{equation}
\frac{\mathcal C}{ \Lambda } {\mathcal O}_H  LH  
\end{equation}
gives an effective Dirac-type mass term connecting the standard and hidden sectors. If  
the compositeness scale in the hidden sector is $f$, the size of the resulting mass term  
$m_D$ is of order $ {\mathcal C}\frac{f}{\Lambda} v$, where $v$ is the Higgs vev. For 
instance,  ${\mathcal C}\sim 1$, $f\sim$   10 eV, and $\Lambda\sim 10^4$ GeV would give 
$m_D\sim 10^{-1}$ eV. If there were no other mass terms in the theory, such a term could 
be the origin of the neutrino mass, which would be of the Dirac type. In the presence of 
a Majorana mass term for the hidden fermion, this term could  give rise to neutrino 
masses via a seesaw type mechanism, and provide mixing between active and sterile 
neutrinos~\cite{Langacker:1998ut}. Depending on the size of the Majorana and Dirac-type 
mass terms, the seesaw could be of the ``mini'' type, with sufficiently light sterile 
neutrinos to show up in neutrino oscillation experiments \cite{Fan:2012ca}.  Another 
possibility is that the hidden fermion has a Dirac-type mass of its own, of size $M$, 
in which case the mixing between neutrinos and hidden fermions would have a mixing angle 
of order $m_D/M$, unconnected with the size of the neutrino mass. The visible neutrino 
masses would then have to arise via an additional term, which could be via the 
conventional  seesaw   or could be a small, Majorana-type mass in the hidden sector 
(this latter type possibility is known as the ``inverse seesaw"). Such mixing with a 
heavy neutral fermion could show up as a non-unitary mixing matrix in the neutrino 
sector, potentially allowing for interesting observable effects such as CP violation 
in neutrino oscillations even with effective 2 flavor 
mixing~\cite{FernandezMartinez:2007ms, Nelson:2010hz}.

Mixing between neutrinos and hidden fermions provides a unique mechanism for the 
neutrinos to interact with any new dark forces that may be present.  Such forces could 
be responsible for an anomalous dependence of neutrino oscillation parameters on their 
environment, which could reconcile otherwise conflicting 
discoveries~\cite{Kaplan:2004dq,Zurek:2004vd,Weiner:2005ac}.

\subsection{Lorentz Violation}
\label{sec:Lorentz violation}

Lorentz invariance is a fundamental ingredient in the standard model
(SM) and General Relativity (GR). The discovery of the breakdown of
such a fundamental symmetry that underlies our successful descriptions
of nature would definitely impact our notions of space-time. The fact
that we do not observe Lorentz violation in our current experiments
indicate that if real, this violation should be small. Nonetheless,
effects of Lorentz violation could be sizable and remain undetected
because of couplings with suppressed effects, such as weak
gravitational fields~\cite{Kostelecky:2008in}.

The various anomalous experimental results described elsewhere in this
White Paper put challenges to the model of three active 
neutrinos. This has led to explore Lorentz violation as a possible
explanation, despite the fact that sterile neutrinos are a more 
popular solution. One point in which Lorentz violation as a possible
solution of neutrino anomalies can overtake the sterile neutrino
interpretation is the preservation of the number of neutrino states to
three active flavors. Additionally, the global models to be described
attempt to be alternatives to the massive model
instead of extensions of it. Moreover, these global models take
advantage of the unconventional energy dependence in the effective
Hamiltonian to minimize the number of necessary parameters.  

\subsubsection{Violation of Lorentz invariance in the standard model}

One of the most common reasons to accept Lorentz invariance as an
exact symmetry of nature is the large number of tests that relativity
has passed. Nevertheless, these dozens of tests have explored just a
tiny fraction of the possible observable signals that could appear in
the presence of Lorentz violation. Another reason is the misleading
idea that Lorentz violation implies observer dependence. Physical laws
are independent of any coordinate system; therefore, an observer
(characterized by a set of coordinates) cannot be privileged to
determine the physics of a system respect to another. This statement
is sometimes confused with Lorentz invariance, despite the fact that
Lorentz invariance is the symmetry of a theory under Lorentz
transformations acting on the fields of the theory (this
transformation is called {\it particle Lorentz transformation}) rather
than on the coordinates used to describe the system (this
transformation is called {\it observer Lorentz transformation})~\cite{Colladay:1996iz}. 

In order to guarantee coordinate independence, the construction of a
general theory would require SM operators coupled to general tensor
fields. The space-time structure of these tensor fields would make
them functions in a given frame, acting as background fields filling
the Universe. When properties such as spin and momentum of the fields
in the theory are boosted or rotated (particle Lorentz transformation)
the background fields remain unaffected, making the coupling between
them and the particle properties change under this type of
transformation. In other words, the physics of a theory of this type
is not invariant under {\it particle Lorentz transformations}; we say
that Lorentz invariance is broken. When a Lorentz transformation is
applied on the observer rather than particle properties, this is a
boost or rotation of the coordinate system, both background fields and
properties of the particles in the theory transform inversely, leaving
the coupling invariant under coordinate changes. We say that the
theory is invariant under {\it observer Lorentz transformations};
therefore, coordinate independent. 

In 1989, Kosteleck\'y and Samuel discovered that background fields
like the ones described above could arise as vacuum expectation values
of dynamic tensor fields string theory, in which Lorentz symmetry is
spontaneously broken~\cite{Kostelecky:1988zi,Kostelecky:1991ak}. Various 
developments led
to a theoretical robust framework describing  Lorentz violation for
all the particles in the SM known as the Standard-Model Extension
(SME)~\cite{Colladay:1996iz,Colladay:1998fq}. 
 The background fields, called coefficients for Lorentz violation, appear
coupled to conventional SM operators of the form 
$\mathcal{L}_\text{LV} \supset a_\alpha \overline{\psi} \gamma^\alpha
\psi$, where this term from the fermion sector exhibits coordinate
invariance via the properly contracted form of the space-time indices of the
coefficient for Lorentz violation $a_\alpha$ and the SM
operator. Since SM operators with odd number of free space-time
indices reverse sign under CPT, the SME contains all possible terms
that break CPT, which are a subset of the terms that break Lorentz
symmetry~\cite{Greenberg:2002uu}. For this reason, the SME is a
general framework to study Lorentz and CPT violation in field theory. 

SME preserves the $SU(3)\times SU(2)\times U(1)$ gauge structure of the SM. 
The spontaneous $SU(2)\times U(1)$ symmetry breaking is also maintained. Since 
space-time translations lie outside the Lorentz group, the violation of particle 
Lorentz invariance leaves unaffected the conservation of energy and 
momentum~\cite{Colladay:1998fq}.  Moreover, since the SME was created from a
Lagrangian its physics is not limited to phenomenologically-driven constructions 
or modified dispersion relations. In fact, any possible description that breaks 
Lorentz and CPT symmetry in a realistic field theory is a subset of the SME. 
For instance, in 1998 Coleman and Glashow~\cite{Coleman:1998ti} presented a 
popular framework for possible tests of Lorentz invariance at high energies that 
corresponds to an isotropic limit of the SME.  There are also alternative 
approaches that violate Lorentz symmetry lying outside local quantum field theory, 
see, {\it e.g.}~Refs.~\cite{Barenboim:2001ac,Mavromatos:2009ww}.

The development of the SME triggered theoretical developments and also boosted 
the experimental front because for a given experiment the SME serves a guide for 
the key observable signatures that would appear if Lorentz invariance is broken. 
The results are summarized in the annually-updated {\it Data tables for Lorentz 
and CPT violation}~\cite{Kostelecky:2008ts}. Review articles on experimental and
theoretical approaches to Lorentz and CPT violation can be found in
Ref.~\cite{Bluhm:2005uj,Diaz:2011tx}. 

\subsubsection{Lorentz-violating neutrinos}

The renormalizable neutrino sector of the SME is the starting point to study 
oscillations in the presence of Lorentz violation. The propagation of three 
left-handed neutrinos  is described by a $3\times 3$ effective Hamiltonian of the 
form~\cite{Kostelecky:2003cr}
\begin{equation}\label{h_effLV}
(h_\text{eff})_{ab} = E \delta_{ab} + \frac{m^2_{ab}}{2E} 
+ (a_L)^\alpha_{ab} \hat p_\alpha-(c_L)^{\alpha\beta}_{ab} \hat p_\alpha \hat p_\beta E,
\end{equation}
where the indices take the values $a,b=e,\mu,\tau$. The first two terms correspond 
to the conventional massive description of neutrino propagation, where the 
constant mass matrix appears with the standard $1/E$ energy dependence. The third 
term breaks both Lorentz and CPT symmetry and exhibits no energy dependence. 
Finally, the last term grows linearly with energy and breaks only Lorentz invariance. 
The fact that space-time indices are properly contracted with the neutrino direction 
of propagation $\hat p_\alpha=(1;-\hat p)$ manifests coordinate invariance. The 
explicit dependence of the effective Hamiltonian on neutrino direction of propagation 
reveals the loss of invariance under particle rotations. The corresponding Hamiltonian 
for antineutrinos is obtained by reversing the sign of the real part of 
$(a_L)^\alpha_{ab}$ and the imaginary part of $(c_L)^{\alpha\beta}_{ab}$. Notice that 
these coefficients also have flavor indices, meaning that these coefficients are 
$3\times3$ matrices in flavor space. The effective Hamiltonian (\ref{h_effLV}) has 
been used to develop formalisms for generic searches of observable signals of Lorentz 
violation~\cite{Kostelecky:2004hg,Diaz:2009qk}. Using these formalisms, 
LSND~\cite{Auerbach:2005tq}, MINOS~\cite{Adamson:2008ij,Adamson:2010rn,Adamson:2012hp}, 
IceCube~\cite{Abbasi:2010kx}, and MiniBooNE~\cite{AguilarArevalo:2011yi} collaborations 
have performed searches of this type~\cite{Kostelecky:2008ts}. 

\subsubsection{Global models using Lorentz violation}

The variety of observable effects introduced by the Lorentz-violating
terms in the effective Hamiltonian (\ref{h_effLV}) is very
rich. Ref.~\cite{Kostelecky:2003cr} presents a general classification
of the key signals of Lorentz violation in neutrinos. Characterized by
unconventional energy dependence, spectral anomalies are among the
generic signals that arise due to the presence of non-negative powers
of the energy in the effective Hamiltonian (\ref{h_effLV}). Even
though the most popular solution for neutrino anomalies is the
incorporation of $N$ sterile neutrino states that allow $2+N$
mass-square differences in a $(3+N)\times(3+N)$ effective Hamiltonian
for neutrinos, we can alternatively keep the three active neutrino
states by using (\ref{h_effLV}) and its unconventional energy
dependence. Since the oscillation length is proportional to the
difference of the Hamiltonian eigenvalues, positive powers of the
energy can reduce the oscillation length as the energy increases and
possibly produce oscillation signals in short-baseline experiments. 

One of the interesting features of the effective Hamiltonian
(\ref{h_effLV}) is that neutrinos can oscillate even in the absence of
masses. This idea was explored~\cite{Kostelecky:2003xn} in the
construction of a simple two-parameter description known as the {\it
bicycle} model, in which neutrinos are massless. One of the challenges
of a model with non-negative powers of the energy is the description
of atmospheric neutrinos observed in  Super-Kamiokande (SK)~\cite{Ashie:2004mr}, 
which shows an excellent agreement with an
oscillation length that grows linearly with energy. In the bicycle
model, the elements of the effective Hamiltonian have a mixed energy
dependence, specifically the non-vanishing coefficients  
${\mathaccent'27 a}$ and ${\mathaccent'27 c}E$ in the effective
Hamiltonian combine at high energy to form the dimensionless
oscillation phase ${\mathaccent'27 a}^2L/{\mathaccent'27 c}E$. This
Lorentz-violating seesaw mechanism~\cite{Kostelecky:2003cr,Kostelecky:2003xn} 
makes the ratio
${\mathaccent'27 a}^2/{\mathaccent'27 c}$ mimic the effect of a
neutrino mass-squared difference, leading to the conventional $L/E$
oscillatory signature observed in SK. In 2007 it was 
proved that the combination of recent long-baseline
accelerator, atmospheric, and reactor data excludes this
direction-dependent model~\cite{Barger:2007dc}. Even though the use of
new experimental data allows its exclusion, the bicycle model served
to show that simple descriptions of neutrino oscillations are possible
in the SME and that neutrino masses might be irrelevant at high
energies.  Recently Ref.~\cite{Barger:2011qj}studied general
isotropic textures based on the bicycle model. They conclude that
massless models can easily describe atmospheric data but a tension
appears between new solar and KamLAND data. 

In 2006, the structure of the bicycle model motivated~\cite{Katori:2006mz} 
to construct the so-called {\it tandem} model, a
hybrid three-parameter model that involves neutrino masses at low
energies and isotropic Lorentz violation that leads to a seesaw at
high energies. This model is consistent with long-baseline
accelerator, atmospheric, reactor, and solar data. Additionally, the
tandem model is consistent with the puzzling LSND signal and the null
result in KARMEN. As an extra feature, the tandem model predicted an
oscillation signal at low energies in MiniBooNE one year before its
observation.  

Until 2011, all studies of Lorentz-violating neutrinos only considered
the minimal SME, which means only renormalizable terms. Following the
general ideas of the tandem model, the so-called {\it puma} model was
proposed~\cite{Diaz:2010ft,Diaz:2011ia}. This isotropic
three-parameter model incorporates nonrenormalizable terms that appear
in the SME~\cite{Kostelecky:2011gq,Kostelecky:2009zp,Kostelecky:2008be}. The
breaking of Lorentz invariance is manifest by the different powers of
the energy that control neutrino oscillations at high energies. A
Lorentz-violating seesaw mechanism produces not only the appropriate
$L/E$ oscillatory signature but also leads to maximal
$\nu_\mu\to\nu_\tau$ mixing consistent with SK and accelerator
experiments. A mass term controls oscillations at low energies, in
perfect agreement with solar data and producing the $L/E$ signature
observed by KamLAND~\cite{Araki:2004mb,Abe:2008ee}. The description
that the puma model provides of neutrino data at low and high energies
is very similar to the massive model; nonetheless, due to the
unconventional energy dependence of the effective Hamiltonian, the
puma differs drastically from the massive model in the mid-energy
range. In particular, for the oscillation channel $\nu_\mu\to\nu_e$
the oscillation phase growths very rapidly with energy producing an
oscillation signal in MiniBooNE. While the oscillation phase gets
larger with neutrino energy, the seesaw mechanism suppresses the
oscillation phase at energies above 500 MeV, resulting in an
oscillation signal in MiniBooNE only at low energies, in perfect
agreement with the data. For the same reason, the model is consistent
with null results in short-baseline high-energy experiments. An
enhanced version of the puma model also produces a signal in LSND that
preserves all the features mentioned above. The different features of
this model make it a simple and effective alternative to the massive
model and consistent with all established data and some anomalous
results using less parameters. A list of many other models that make
use of the effective Hamiltonian (\ref{h_effLV}) is presented in
Refs.~\cite{Ribeiro:2007jq,Bernardini:2008ef,Altschul:2009zza,Ando:2009ts, 
Bustamante:2010nq,Yang:2009ge,Engelhardt:2010dx,Arias:2009fk,Bhattacharya:2010xj, 
Liu:2011nwa}.

\subsubsection{Some types of Lorentz violation: CPT violation,
non-standard interactions, and lepton-number-violating oscillations} 

One of the important features of the study of Lorentz violation is its
connection with CPT violation, that corresponds to a subset of the
terms in the SME that break Lorentz invariance. Note, however, that
there are no terms that break CPT preserving Lorentz symmetry~\cite{Greenberg:2002uu}. 
Additionally, note that as mentioned earlier,
CPT-odd operators have an odd number of space-time indices (properly
contracted with a coefficient with the same number of space-time
indices to preserve coordinate invariance); therefore, in the SME
particles and anti-particles have exactly the same masses, even in the
presence of CPT violation.  

Another interesting property of the study of Lorentz violation in
neutrinos is the connection to another popular approach to describe
neutrino oscillations, non-standard neutrino interactions (NSI)~\cite{Grossman:1995wx}, 
discussed in Section \ref{sec:NSI}. Independent of its origin, the effect of NSI
in neutrino oscillations is a constant contribution to the effective
Hamiltonian, producing energy-dependent mixing similar to neutrino
oscillations in matter. This constant contribution is usually
parameterized by a matrix $\varepsilon_{ab}$ in flavor space. We
emphasize that the resulting effective Hamiltonian is
indistinguishable from (\ref{h_effLV}) with nonzero
$(a_L)^T_{ab}$. This means that analyses using NSI are studying the
time component of the CPT-odd coefficient for Lorentz violation
$(a_L)^\alpha_{ab}$. 

Another important remark is that Lorentz-violating terms can also
produce neutrino-antineutrino oscillations~\cite{Kostelecky:2003cr,Diaz:2009qk}. 
The Hamiltonian (\ref{h_effLV})
corresponds to the $3\times3$ diagonal block of a $6\times6$
Hamiltonian. A similar $3\times3$ block parameterizes antineutrinos
that is related to (\ref{h_effLV}) by a CP transformation. When
Lorentz symmetry is broken there is, however, a $3\times3$
off-diagonal block that mixes neutrinos and antineutrinos given 
by~\cite{Kostelecky:2003cr} 
\begin{equation}\label{h_effLV(g,H)}
h_{a\bar b} =
i\sqrt2 (\epsilon_+)_\alpha \tilde{H}^\alpha_{a\bar b} - i\sqrt2 (\epsilon_+)_\alpha\hat p_\beta \,\tilde g^{\alpha\beta}_{a\bar b} E,
\end{equation}
where $(\epsilon_+)_\alpha$ is a complex 4-vector representing the
helicity state, and the flavor indices span $a=e,\mu,\tau$; $\bar
b=\bar e,\bar\mu,\bar\tau$. Notice that the two terms in
(\ref{h_effLV(g,H)}) are direction dependent, which exhibits
explicitly that Lorentz symmetry is broken through the loss of
invariance under rotations. The size of this breaking is parameterized
by the coefficients for Lorentz violation $\tilde{H}^\alpha_{a\bar b}$
and $\tilde g^{\alpha\beta}_{a\bar b}$. The second term in
(\ref{h_effLV(g,H)}) also breaks CPT. 
The comments on unconventional energy dependence described for
(\ref{h_effLV}) also apply for (\ref{h_effLV(g,H)}). These
coefficients cause lepton-number violating oscillations
($\nu\leftrightarrow\bar\nu$). These terms also produce lepton-number
preserving oscillations as a second-order effect. Details about the
effects of these coefficients and their application to different
experiments can be found in Ref.~\cite{Diaz:2009qk}. 
These lepton-number-violating oscillations have been studied as an
alternative to explain the LSND and MiniBooNE anomalies~\cite{Hollenberg:2009tr}. 

\subsection{CPT Violation in Neutrino Oscillations and the Early
Universe as an Alternative to Sterile Neutrinos}

One of  the most important questions of fundamental physics, that is
still unanswered today, is the reason for our existence, namely why
the Universe is made up mostly of matter. To put it in more
microscopic terms, the important unanswered question relates to a
theoretical understanding of the magnitude of the observed Baryon
Asymmetry in the Universe (BAU). According to the Big Bang theory,
matter and antimatter have been created at equal amounts in the early
Universe. The observed charge-parity (CP) violation in particle
physics~\cite{Christenson:1964fg}, prompted 
Sakharov~\cite{Sakharov:1967dj} to conjecture that non-equilibrium
physics in the early Universe produce Baryon number (B), charge (C)
and charge-parity (CP) violating, but CPT \emph{conserving},
interactions/decays of anti-particles in the early Universe, resulting
in the observed baryon--anti-baryon ($n_B - n_{\overline{B}}$)
asymmetry. In fact there are two types of non-equilibrium processes in
the early Universe that could produce this asymmetry: the first type
concerns processes generating asymmetries between leptons and
anti-leptons (\emph{Leptogenesis}), while the second produce asymmetries between
baryons and anti-baryons (\emph{Baryogenesis}).  

The almost 100\% observed asymmetry today, is estimated in the
Big Bang theory~\cite{Gamow:1946eb} to be of order:  
\begin{equation}\label{basym}
\Delta n (T \sim 1~{\mathrm GeV}) = \frac{n_{B} - n_{\overline{B}}}{n_{B} + n_{\overline{B}}} \sim \frac{n_{B} - n_{\overline{B}}}{s} = (8.4 - 8.9) \times 10^{-11} 
\end{equation}
at the early stages of the expansion, {\it e.g.}~for times $t < 10^{-6}$~s and 
temperatures $T > 1 $~GeV. In the above formula $n_B$
($n_{\overline{B}}$) denotes the (anti-)baryon density in the
Universe, and $s$ is the entropy density.  Unfortunately, the observed
CP violation within the standard model cannot reproduce
(\ref{basym})~\cite{Kuzmin:1985mm}.

A basic assumption in all current models employing sterile neutrinos is that 
the \emph{CPT symmetry} holds in the early Universe, and this produces
matter and antimatter in equal amounts initially. It is then, the CP
violating decays of the sterile neutrinos that produce the observed
BAU, as mentioned above. Such CPT invariance is a cornerstone of all
known \emph{local} effective \emph{relativistic} field theories
without gravity, which current particle physics phenomenology is based upon. 
An interesting idea, though, is that during the Big Bang, one or more
of the assumptions for the CPT theorem (Lorentz Invariance, unitarity
and/or locality of interactions) break down ({\it e.g.}~due to
quantum gravity influences, that may be strong at such early
times). This may result in \emph{CPT Violation} (CPTV) and a naturally
induced matter-antimatter asymmetry, without the need for extra
sources of CP violation, such as sterile neutrinos~\cite{Bertolami:1996cq}.  

What we examine below is whether such ideas are viable, in the sense
that they can be realized in concrete models, in a way consistent with
the known particle physics and cosmology today. The simplest 
possibility~\cite{Dolgov:2009yk}  for inducing such CPT violations in
the early Universe is through particle--anti-particle mass differences
$m \ne \overline{m}$.  These would affect the (anti) particle
phase-space distribution functions  
\begin{equation}
f(E, \mu) = [{\mathrm exp}(E-\mu)/T) \pm 1]^{-1}~, \quad E^2 = p^2 + m^2
\label{cptvf}
\end{equation}
and similarly for anti-matter, upon the replacements $m \rightarrow
\overline{m}, \, \mu \rightarrow \overline{\mu}$ (from now on,
overlined quantities refers to anti-particles). 
Mass differences between particles and anti-particles, $\overline{m}
\ne m$,  generate a matter anti-matter asymmetry in the relevant
densities  
\begin{equation}\label{nnbarcptv}
n-{\overline n} = g_{d.o.f.} \int \frac{d^3 p}{(2\pi)^3}[f(E,\mu) - f(\overline{E},\mu)]~,
\end{equation}
where $g_{d.o.f.}$ denotes the number of degrees of freedom of the
particle species under study.  

Upon making the quite reasonable assumption~\cite{Dolgov:2009yk}  that
the dominant contributions to baryon asymmetry come from
quark--anti-quark mass differences, and that their masses are
increasing, say, linearly with the temperature $m \sim g T$ (with $g$
the QCD coupling constant), one can provide estimates for the induced
baryon asymmetry by the fact that the maximum quark--anti-quark  
mass difference is bounded by the current experimental bound on
proton--anti-proton mass difference, $\delta m_p = | m_p -
\overline{m}_p|$, which is known to be less than $\delta m_p \, < \, 2
\cdot 10^{-9}$ GeV.  
This leads to the following estimate for the
baryon($n_B$)-to-photon($n_\gamma$)-density ratio in the Universe, at
temperature $T$~\cite{Dolgov:2009yk}: 
\begin{equation}
\beta_T = \frac{n_B}{n_\gamma} = 8.4 \times 10^{-3} \,\frac{m_u \,\delta m_u + 15 m_d \,\delta m_d }{T^2}~, \quad \delta m_q = |m_q - {\overline m}_q|~.
\end{equation}
Unfortunately, for $n_\gamma \sim 0.24\, T^3 $, which is the photon
equilibrium density at temperature $T$, this produces  too small BAU
compared to the observed one.  To reproduce the observed $\beta_{T=0}
\sim 6 \cdot 10^{-10}$ one would need  $\delta m_q (T=100~{\mathrm GeV})
\sim 10^{-5} - 10^{-6}~{\mathrm GeV}  \gg \delta m_p (T=0)$, which is rather
unnatural, although admittedly such a scenario  cannot be excluded on
theoretical grounds. 

However, active (\emph{light}) neutrino--antineutrino mass
differences alone may reproduce BAU; some phenomenological models in
this direction have been discussed in~\cite{Barenboim:2001ac},
considering, for instance, particle--anti-particle mass differences for
active neutrinos  compatible with current oscillation data. This leads
to the result
\begin{equation}
n_B = n_\nu - n_{{\overline \nu}} \simeq \frac{\mu_\nu \, T^2}{6}
\end{equation}
yielding $ n_B/s \sim \frac{\mu_\nu}{T} \sim 10^{-11}$ at $T \sim
100$~GeV, in agreement with the observed BAU. Above, $s$ denotes the
entropy density, and $n_\nu, \, \mu_\nu$ are the neutrino densities
and chemical potential respectively. 

\begin{figure}[t]
  \centering
    \includegraphics[width=0.6\textwidth]{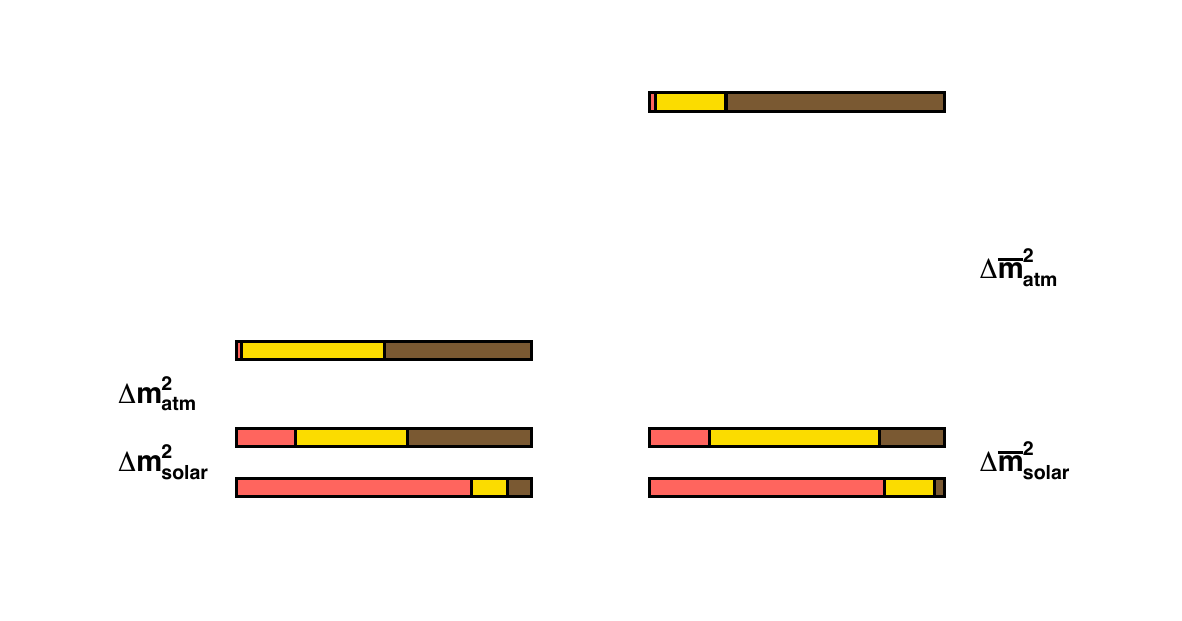}%
  \caption{A generic CPT-violating neutrino spectrum that can explain MINOS results.}
  \label{fig:neutus.pdf}
\end{figure}

It should also be noticed that CPT violating spectra can not only
accommodate the established oscillation experiments but also provide
distinguishing features to be observed in future ones. 
A CPT violation of the kind represented in Fig.~\ref{fig:neutus.pdf},
can accommodate the intriguing results of the MINOS data, indicating
 possible differences between the neutrino and antineutrino oscillation parameters.
The CPT-violating features in this case are encapsulated basically in the 
atmospheric channel :
\be
\Delta\bar{m}_{\mathrm atm}^2 \approx 0.0030{\mathrm\ eV}^2\;, \quad
{\mathrm sin}^2\,\bar{2\theta}_{23} \approx 0.8 \; ,
\ee
and
\be
\Delta\bar{m}_{\mathrm atm}^2 \approx 0.0025{\mathrm\ eV}^2\;, \quad
{\mathrm sin}^2\,2\theta_{23} \approx 1 \; .
\ee
Notice however that no CPT-violating spectra can easily accommodate LSND 
without conflicting with other existing experiments.

But particle--anti--particle mass differences may not be the only way
by which CPT is violated. As discussed in~\cite{Barenboim:2004ev,Barenboim:2004wu}, 
quantum gravity fluctuations
in the structure of space-time, that may be strong in the early
Universe, may act as an \emph{environment} inducing decoherence for
the (anti-)neutrinos, but with couplings between the particles and the
environment that are different between the neutrino and antineutrino
sectors. Once there is decoherence for a low-energy  (compared to the
Planck scale $M_{\mathrm Pl} \simeq 10^{19}$\, GeV) observer, the effective CPT
symmetry generator may be \emph{ill-defined} as a quantum mechanical
operator, according to a theorem by Wald~\cite{Wald:1980nm},
leading to an intrinsic violation of CPT symmetry. This type of
violation may characterize models of quantum gravity with stochastic
space-time fluctuations due, for instance, to gravitational space-time
defects, as is the case of certain brane
models~\cite{Ellis:2005ib,Ellis:2004ay,Bernabeu:2006av}. In such a
case, there may be a slight mismatch in the strength of the stochastic
space-time fluctuations between particle and anti-particle sectors,
leading to different decoherence parameters to describe the
interaction of the gravitational environment with matter.   

In~\cite{Barenboim:2004ev,Barenboim:2004wu}, simple models of Lindblad 
decoherence~\cite{Lindblad:1975ef}, conjectured to characterize also 
quantum-gravity-induced CPT violating decoherent situations~\cite{Ellis:1983jz, 
Ellis:1995xd}, have been considered for neutrinos~\cite{Benatti:2001fa}, 
assuming, though, non-trivial decoherence parameters \emph{only} in the 
anti-particle sector, consistently with the lack of any experimental evidence 
to date~\cite{Fogli:2007tx,Lisi:2000zt,Barenboim:2006xt} for vacuum decoherence 
in the particle sector.  The antineutrino decoherence parameters (with dimension 
of energy) had a mixed energy dependence. Some of the eight coefficients (necessary 
to parameterize, in a minimal way, a diagonal Lindblad decoherence matrix for 
three-generation neutrinos, assumed for simplicity in~\cite{Barenboim:2004ev, 
Barenboim:2004wu}) were proportional to the antineutrino energy  
\begin{equation}
\overline{\gamma}_i = \frac{T}{M_P} \, E~, \quad i = 1,2,4,5
\end{equation}
while the remaining ones (sub-dominant) were inversely proportional to it  
\begin{equation}
\overline{\gamma}_j = 10^{-24} \frac{1}{E}~,~ j = 3,6,7,8~,
\end{equation} 
in the notation of~\cite{Barenboim:2004ev,Barenboim:2004wu}.  The
model is phenomenological and its choice was originally motivated by
fitting the LSND ``anomalous data'' in the antineutrino
sector~\cite{Aguilar:2001ty} with the rest of the neutrino data (in
fact, for such a purpose, the  
coefficient $T/M_P$ in $\overline{\gamma}_i $, $i=1, \dots 5$ was
assumed to get a value $T/M_P \sim 10^{-18}$, {\it i.e.}, ``frozen''
at the Temperature range of Electroweak Symmetry Breaking (EWSB) in
the Universe). However, the model may be disentangled from this issue
and presented as a candidate model entailing CPT violation that can
lead to phenomenologically acceptable estimates for the BAU today.  

In this context, one can
derive~\cite{Barenboim:2004ev,Barenboim:2004wu} an active (light) $\nu
- \overline{\nu}$  asymmetry of order 
\begin{equation} \label{nnubar}
\mathcal{A} = \frac{n_\nu - n_{\overline{\nu}}}{n_\nu + n_{\overline{\nu}}} = 
\frac{\overline{\gamma}_1}{ \sqrt{\Delta m^2}} = \frac{T}{M_P} \cdot 
\frac{E}{\sqrt{\Delta m^2}}~,
\end{equation}
where $\Delta m^2 $ denotes the (atmospheric) neutrino mass squared
difference, which plays the role of a characteristic low mass scale
in the problem (analogous to the difference in widths $\Delta \Gamma$
between nearly mass generate Kaon states, entering the final
expression for the dimensionless decoherence parameters  $(\alpha,
\beta, \gamma) /\Delta \Gamma$  in the notation of the CPT violating
decoherence models of \cite{Ellis:1983jz,Ellis:1995xd}, which physical
processes depend upon).  
This lepton number violation is communicated to the baryon 
sector by means of baryon plus lepton ($B + L$) number conserving
sphaleron processes. The so-induced Baryon number in the Universe 
today may be estimated as~\cite{Barenboim:2004ev} 
\begin{equation}\label{baryon}
B = \frac{n_\nu - n_{\overline \nu}}{s} \sim \mathcal{A}\, 
\frac{n_\nu}{g^\star \, n_\gamma} ,
\end{equation}
with $n_\gamma$ the photon number density and $g^\star$ the effective number 
of degrees of freedom (at the temperature where the asymmetry is developed, in 
the model of \cite{Barenboim:2004ev} this is the EWSB temperature), which 
depends on the matter content of the model (with a typical range  $g^\star 
\in [10^2 - 10^3 ]$).  Taking into account that in the model considered 
in~\cite{Barenboim:2004ev} $\mathcal{A} \sim 10^{-6}$ at the EWSB
Temperature range, one can therefore see that the observed BAU may be
reproduced in this case without the need for extra sources of CP
violation and thus sterile neutrinos.  

Unfortunately, at present such models lack microscopic understanding,
although progress in this direction is currently being made. One issue
that arises with such models concerns the preferential role of
neutrinos on being involved in this early Universe
quantum-gravity-decoherence-induced CPT violation in contrast to other
particles in the standard model (SM). Within some microscopic models
of space-time foam, involving populations of point-like brane defects
(D-particles) puncturing three(or higher)-spatial dimension brane
worlds~\cite{Ellis:2005ib,Ellis:2004ay,Bernabeu:2006av}, such a
preferred role may be justified by electric charge conservation: the
representation of SM particles as open strings, with their ends
attached to the brane worlds, prevents capture and splitting of open
strings carrying electric fluxes by the D-particles. This is due to
the fact that the latter are electrically neutral and thus electric
charge would not have been conserved if such processes had taken
place. Hence, the D-particle foam is transparent to  
charged excitations of the SM, leaving neutral particles, in
particular neutrinos, to be vulnerable to the foam effects. Of course
to construct realistic cosmologies from such models is still an open
issue.

Before closing, we would like to discuss one more scenario~\cite{Debnath:2005wk, 
Mukhopadhyay:2005gb,Mukhopadhyay:2007vca,Sinha:2007uh} of 
neutrino--antineutrino CPT violation, induced by local curvature
effects in geometries of the early Universe, and its connection to
leptogenesis/baryogenesis. The scenario is based on the well-known
fact that fermions in curved space times exhibit a coupling of their
spin to the curvature of the background space time. In fact, the Dirac
Lagrangian density of a fermion can be re-written as (the analysis can
also be adapted to incorporate Majorana neutrinos, with similar
results):
\begin{equation}
\mathcal{L} = \sqrt{-g}\,  \overline{\psi}  \left( i \gamma^a \partial_a - m + 
\gamma^a \gamma^5 B_a \right) \psi~, 
\quad B^d = \epsilon^{abcd} e_{b\lambda} \left( \partial_a e^\lambda_{\,\, c} + 
\Gamma^\lambda_{\nu \mu}\, e^\nu_{\,\, c} \,  e^\mu_{\,\, a} \right)~,
\end{equation}
in a standard notation, where $e^\mu_{\,\, a}$ are the vielbeins,
$\Gamma^\mu_{\,\alpha\beta}$ is the Christoffel connection and Latin
(Greek) letters denote tangent space (curved space-time) indices. The
space-time curvature background has, therefore, the effect of inducing
an ``axial'' background field $B_a$ which can be non trivial in
certain axisymmetric (in general anisotropic) space time geometries,
such as certain Bianchi-type Cosmologies,  that may characterize
epochs of the early Universe, according to the assumptions of~\cite{Debnath:2005wk, 
Mukhopadhyay:2005gb,Mukhopadhyay:2007vca,Sinha:2007uh}.
If the background axial field $B_a$ is constant in some local frame,
this breaks CPT symmetry, and the dispersion relation of neutrinos in
such backgrounds differs from that of antineutrinos: 
\begin{equation}\label{nunubardr}
E = \sqrt{ (\vec{p} - \vec{B})^2 + m^2} + B_0 ~, \quad \overline{E} = \sqrt{ (\vec{p} + 
\vec{B})^2 + m^2} - B_0  .
\end{equation}
The relevant neutrino asymmetry is then generated through the same
steps (cf.~(\ref{cptvf}), (\ref{nnbarcptv})) as in the case
where CPT is violated through particle--anti-particle mass
differences, leading to the following neutrino-antineutrino density
differences in, say, Bianchi II
Cosmologies~\cite{Debnath:2005wk,Mukhopadhyay:2005gb,Mukhopadhyay:2007vca,Sinha:2007uh}:
\begin{equation}\label{bianchi}
\Delta n_\nu \equiv n_{\nu} - n_{\overline{\nu}} \sim g^\star \, T^3 
\left(\frac{B_0}{T}\right) ,
\end{equation}
with $g^\star$ the number of degrees of freedom for the (relativistic)
neutrino. At temperatures $T < T_d$, with $T_d$ the decoupling
temperature of the lepton-number violating processes, the ratio of the
net lepton number $\Delta L $ (neutrino asymmetry) to entropy density
(which scales as $T^3$) remains constant,  
\begin{equation}
\Delta L (T < T_d) = \frac{\Delta n_\nu }{s} \sim \frac{B_0}{T_d}
\end{equation}
which, for $T_d \sim 10^{15}$~GeV and $B_0 \sim 10^{5}$~GeV, implies a
lepton asymmetry (leptogenesis) of order $\Delta L \sim 10^{-10}$, in
agreement with observations.  The latter can then  be communicated to
the baryon sector to produce the observed BAU (baryogenesis), either
by a $B-L$ conserving symmetry in the context of Grand Unified Theories
(GUT),  or via $(B+L)$-conserving sphaleron processes, as in the
decoherence-induced CPT violating case of~\cite{Barenboim:2004ev,Barenboim:2004wu}, 
mentioned previously.  

From the above analysis, we therefore conclude that it is plausible to
have CPT violation in the early Universe, due to a variety of
reasons. These range from (exotic) quantum-gravity stochastic
space-time fluctuations, inducing decoherence on low-energy matter of
different strength between particle and anti-particle sectors, to
standard general relativity background effects due to anisotropic
space-time geometries that may characterize such early epochs of the
Universe. Such CPT violating scenarios may generate differences
between neutrino and antineutrino populations that can in turn
reproduce, via subsequent sphaleron processes or $B-L$ conserving GUT
symmetries, the observable BAU without the presence of sterile
neutrinos, unlike the standard CPT violating scenario involving only
particle--anti-particle mass differences. The cosmology and
particle-physics phenomenology of such scenarios are of course at
their infancy, but we believe they are worth pursuing.  

\clearpage
\section{Astrophysical Evidence}
\label{sec:astro}

\subsection{Cosmology}

In the early universe, neutrinos of left-handed chirality are held in
a state of thermal equilibrium with other standard model particles by
weak interactions.  As the universe expands and the thermal plasma
cools, interactions become increasingly infrequent.  When the plasma
cools to a temperature of order 1~MeV, the interaction rate per
neutrino drops below the universal expansion rate; the neutrinos
therefore ``decouples'' from the thermal plasma.  Because neutrino
decoupling happens while the neutrinos are ultra-relativistic, they
retain their relativistic Fermi-Dirac phase space distribution
parametrized by a temperature, and possibly a nonzero chemical
potential.

Shortly after neutrino decoupling is the event of electron-positron
annihilation at $T \sim 0.2$~MeV.  The energy released in this process
``reheats'' the photon population, but this reheating is not felt by
the neutrinos because they have already decoupled.  The net effect is
that the neutrinos emerge from the event a little colder than the
photons.  Using entropy conservation arguments and assuming
instantaneous neutrino decoupling, one can show that
\begin{equation}
\label{eq:nutemp}
T_\nu = \left(\frac{4}{11} \right)^{1/3} T_\gamma,
\end{equation}
where $T_\nu$ and $T_\gamma$ are the neutrino and the photon
temperatures, respectively, after $e^+e^-$ annihilation.

In current usage, the relation~(\ref{eq:nutemp}) is generally taken to
{\it define} the standard neutrino temperature after $e^+e^-$
annihilation, in spite of finite temperature QED effects and
non-instantaneous decoupling which render the actual neutrino
energy density and number density a little higher than is implied by this relation.
The extra energy density is
commonly absorbed into the definition of the {\it effective number of
  neutrino families} $N_{\mathrm eff}$ via
\begin{equation}
\label{eq:neff}
\rho_\nu =  N_{\mathrm eff} \frac{7 \pi^2}{120} T_\nu^4.
\end{equation}
Here, $\rho_\nu$ is the total neutrino energy density deep in the
radiation domination era (but after $e^+e^-$ annihilation) when the
neutrinos are ultra-relativistic, the factor $(7 \pi^2/120) T_\nu^4$
denotes the energy density in one family of relativistic neutrinos
with temperature $T_\nu$ as defined in equation~(\ref{eq:nutemp}), and
$N_{\mathrm eff}=3.046$ in the standard model with three generations of
fermions~\cite{Mangano:2005cc}. Note that because the parameter $N_{\mathrm eff}$ is defined in
relation to the neutrino energy density in an epoch when the neutrinos
are ultra-relativistic, there is no ambiguity in the
definition~(\ref{eq:neff}) even if the neutrinos should become
nonrelativistic at a later time because of their nonzero masses
(provided they do not exceed the eV-mass scale).

Cosmology is sensitive to the effective number of neutrino families
$N_{\mathrm eff}$ primarily because energy density in relativistic
particles affects directly the universe's expansion rate during the
radiation domination era.  In the epoch immediately after $e^+e^-$
annihilation, the expansion rate $H(t)$ is given by
\begin{equation}
\label{eq:hubble}
H^2(t) \simeq \frac{8 \pi G}{3} (\rho_\gamma + \rho_\nu).
\end{equation}
With the photon energy density $\rho_\gamma$ being extremely well
determined from measurements of the cosmic microwave background (CMB) temperature, constraints on
$H(t)$ in the early universe can be interpreted as bounds on
$\rho_\nu$ and hence $N_{\mathrm eff}$.  From the particle physics
perspective, a thermal population of light sterile neutrinos is one
possible cause of a non-standard $N_{\mathrm eff}$.  However, it is
important to emphasize that, as far as $H(t)$ is concerned, any
thermal background of light particles such as axions and axion-like
particles, hidden sector photons, majorons, or even gravitons will
contribute to the relativistic energy density.  Likewise, any process
that alters the thermal abundance of neutrinos ({\it e.g.}, a low reheating
temperature) or affects directly the expansion rate itself ({\it e.g.}, a
time-dependent Newton's constant $G$) can mimic a non-standard $N_{\mathrm
  eff}$ value.

In the following, we describe first the production of a thermal light
sterile neutrino population in the early universe.  We then discuss
the various signatures of light sterile neutrinos in cosmological
observables such as the light elemental abundances from big bang nucleosynthesis (BBN),
 the CMB
anisotropies and the large-scale structure (LSS) distribution, as well
as the strength of evidence for sterile neutrinos from the current
generation of observations.

\subsubsection*{Sterile neutrino thermalization}
A sterile neutrino that mixes with an active neutrino can augment the
relativistic energy density in the early universe through
oscillation-based thermal production in the early universe, leading to
an $N_{\mathrm eff}>3.046$.  The primary thermalization process is
collisional production from active neutrinos that oscillate into
sterile neutrinos, with collisions ``forcing'' the evolved neutrino
back into a flavor state.  The oscillation production is enhanced with
a larger mixing amplitude $\sin^2 2\theta$ and more rapid (shorter
wavelength) oscillations due to a larger $\Delta m^2$.  This effect and the
region of parameter space leading to thermalization of the sterile neutrino was
found by the early work of~\cite{Langacker:1989sv}.  If sterile
neutrinos do become thermalized, then we expect them to have the same
temperature as the active neutrinos.

Most analyses so far have focused exclusively on two-neutrino mixing
solutions of sterile neutrino thermalization, although multiple flavor
effects were studied in references~\cite{DiBari:2001ua,Abazajian:2002bj},
and in \cite{Cirelli:2004cz,Melchiorri:2008gq} using the momentum-averaged approximation.
The conclusions of these works were that the LSND neutrino mixing parameters would
inevitably lead to thermalization of the two or more sterile neutrinos associated with 
the signal.  We emphasize however that a full thermalization calculation involving 
momentum-dependent density matrices for 3 active and $N_s$ sterile flavors has 
never been performed, though studies of the range of physical effects
have been done~\cite{Abazajian:2004aj}.   

Thermalization of the sterile neutrino can be suppressed by the
presence of a large lepton asymmetry ($L \sim 10^{-5}$).  Such an asymmetry alters the matter potential of
the active-sterile system, thereby suppressing the effective
mixing angle and reducing or stopping the thermal production of
the sterile neutrinos~\cite{Foot:1995bm}.  The required lepton asymmetry may
be pre-existing, or, interestingly, dynamically generated by the
multi-flavored active-sterile system~\cite{Foot:1995qk}.  The
multi-flavored effects of lepton-number generation, the conversion of
this lepton number into other flavors, and thermalization was studied
in detail in reference~\cite{Abazajian:2004aj}.  However, as discussed
above, a full multi-flavor density matrix calculation has not been
performed.  The main conclusions of the work performed so far is that
the sterile neutrino or neutrinos associated with the LSND oscillation
interpretation would be thermalized except for limited cases of
inherent lepton number generation or in the presence of a pre-existing
lepton number.

\subsubsection*{Big Bang Nucleosynthesis}

\paragraph{Sterile neutrino and primordial elemental abundances} 
The formation of light elements from protons and neutrons commences at
a temperature of order 0.1 MeV with the synthesis of deuterium, which subsequently fuels
a number of nuclear reactions, thereby
enabling  the production of heavier nuclei.
Because of its large binding energy, almost all of the initially
available neutrons  end up bound in $^4$He nuclei, resulting in a helium-4
mass fraction
\begin{equation}
Y_p  \equiv \frac{4 n_{\mathrm He4}}{n_n+n_p}
\end{equation}
of order 0.25,  where $n_{\mathrm He4}$, $n_n$, and $n_p$ are the 
number densities of $^4$He, neutron and proton respectively.
  After freeze-out of the
nuclear processes and decay of unstable isotopes, small amounts of
deuterium and $^3$He (D/H $\sim$ $^3$He/H $\sim {\mathcal O}(10^{-5})$),
as well as traces of $^6$Li and $^7$Li remain in addition.  In standard BBN, 
barring experimental uncertainties in the nuclear reaction rates
and the free neutron lifetime,
the baryon-to-photon ratio $\eta$ alone determines the elemental abundances.

Neutrinos are important in this picture for
two reasons.  Firstly, electron neutrinos participate directly in the charged
current weak interactions that determine the neutron-to-proton ratio,
\begin{eqnarray}
\label{eq:weak}
\nu_e + n & \leftrightarrow & p + e^-, \nonumber \\
\bar{\nu}_e + p &\leftrightarrow& n + e^+.
\end{eqnarray}
Secondly, neutrinos of all flavors influence the expansion rate of the universe prior to and during 
BBN via equation~(\ref{eq:hubble}).
Changing this expansion rate alters the neutron-to-proton
ratio at the onset of BBN and hence the light element
abundances, most notably the abundance of $^4$He.

The presence of sterile neutrinos therefore has the potential to
change BBN in two ways: distorting the $\nu_e$ phase space distribution via flavor oscillations, and 
enhancing the Hubble expansion rate  by contributing to $N_{\mathrm eff}$. 
For sterile neutrinos with mixing parameters currently favored by terrestrial experiments,
however, only the second effect is at play.  This is because the large mass-square
difference  $\Delta m^2\sim 1 \ {\mathrm eV}^2$ 
and mixing $\sin^2 2 \theta \sim 10^{-3}$ between the sterile and the active (standard model) neutrino states essentially
guarantee full thermalization of the sterile species prior to the decoupling of the 
active neutrinos.  Therefore, the phase space distribution of the sterile states follows closely 
 those of the active neutrinos, namely, a relativistic Fermi-Dirac distribution, 
 thereby contributing $\Delta N_{\mathrm eff}=1$ per sterile species.  This also means 
that active-sterile flavor oscillations after decoupling
 will not cause the $\nu_e$ phase space distribution to deviate from 
  a relativistic Fermi-Dirac distribution.%
  \footnote{In the general case with arbitrary $\Delta m^2$ and $\sin^2 2 \theta$, it is necessary to solve 
the quantum kinetic equations through BBN in order to establish
both the total neutrino energy density {\it and} the exact form of the $\nu_e$ phase space 
distribution.}
Thus, under this approximation, the standard BBN scenario need only be extended with one
 extra parameter, namely, $N_{\mathrm eff}$, to incorporate the effects of sterile neutrinos.

\paragraph{Current observational constraints}  
  
The limiting factor in using BBN to constrain cosmology is the presence of significant
systematic uncertainties in the astrophysical measurements of all
primordial abundances.  These uncertainties essentially preclude the use of $^3$He and Li abundance measurements
for precision cosmology, leaving us with the helium-4 mass fraction $Y_{\mathrm p}$ and 
the deuterium abundance D/H as the only reasonable probes
(see, {\it e.g.},
\cite{Aver:2011bw} and \cite{Pettini:2008mq} for a discussion of
progress in the determination of these two abundances).

Generally, $Y_{\mathrm p}$ is regarded as the most sensitive probe of
$N_{\mathrm eff}$.  This can be understood in the following way:
increasing the radiation energy density leads to a higher expansion
rate, which means that the reactions~(\ref{eq:weak}) freeze-out at an earlier time, thus enhancing the 
neutron-to-proton ratio at the onset of BBN.  Because almost all neutrons eventually wind up 
bound in $^4$He nuclei, the resulting $Y_{\mathrm p}$ will be larger.  The deuterium abundance D/H, on the other hand, is mainly sensitive to
$\eta$.   Combination of the two therefore constrains the two free parameters of the standard BBN+sterile neutrino scenario.

Optionally, under the assumption that $\eta$ remains constant between the epochs of
BBN and of photon decoupling, the primordial abundance data
can be complemented by the very precise CMB constraint on the baryon-to-photon ratio
$\eta_{\mathrm CMB}$ (which is obtained by combining the
measurement of the photon energy density $\omega_\gamma$ via the CMB
energy spectrum, and the baryon energy density $\omega_{\mathrm b}$
inferred from the CMB anisotropy angular power spectra).

Current estimates suggest that $N_{\mathrm eff} \sim 3$--$4$ is compatible
with observations, with perhaps a slight preference for
higher-than-standard
values~\cite{Izotov:2010ca,Mangano:2011ar,Hamann:2011ge,Nollett:2011aa}
(see table~\ref{tab:neffbbn}).  Within the minimal BBN scenario, models
with more than one fully thermalized extra species seem at odds with
current BBN data.  These results hold even after folding the uncertainties in the nuclear
reaction rates and the free neutron lifetime into the analysis~\cite{Hamann:2011ge,Nollett:2011aa}.
If a higher value turns out to be required by late time 
observables such as the CMB anisotropies and the large-scale structure distribution, its explanation will
presumably require additional non-standard
physics such as the decay of massive particles after BBN.

Finally, we stress that the picture in which the sterile sector can be
described by just one parameter $N_{\mathrm eff}$ may not hold in more complicated
scenarios, such as models with a non-zero lepton asymmetry (see 
{\it e.g.}~\cite{Dolgov:2002ab,Abazajian:2004aj,Pastor:2008ti}).

\begin{table}[t]
\caption{Current constraints on $N_{\mathrm eff}$ from BBN, with 68\% (95\%) uncertainties.  
 Different results for the same nominal data set are due to different measurements of
 $Y_{\mathrm p}$.
  \label{tab:neffbbn}}
\begin{center}
{\small
\begin{tabular}{lllc}
\hline
\hline
Model & Data & $N_{\mathrm eff}$ & Ref. \\
\hline
$\eta$+$N_{\mathrm eff}$ & $\eta_{\mathrm CMB}$+$Y_{\mathrm p}$+D/H & $3.8^{(+0.8)}_{(-0.7)}$ &  \cite{Izotov:2010ca} \\
& $\eta_{\mathrm CMB}$+$Y_{\mathrm p}$+D/H & $< (4.05)$  & \cite{Mangano:2011ar} \\
& \multirow{3}{*}{$Y_{\mathrm p}$+D/H $\left\{ \begin{array}{l} \vphantom{0} \\ \vphantom{0} \\  \vphantom{0} \end{array} \right.$} & $3.85 \pm 0.26$  & \cite{Nollett:2011aa} \\
&  & $3.82 \pm 0.35$  & \cite{Nollett:2011aa} \\
&  & $3.13 \pm 0.21$  & \cite{Nollett:2011aa} \\
\hline
 $\eta$+$N_{\mathrm eff}$, $(\Delta N_{\mathrm eff} \equiv N_{\mathrm eff} - 3.046 \ge 0)$ &
  $\eta_{\mathrm CMB}$+D/H    & $3.8 \pm 0.6$  &  \cite{Hamann:2011ge}\\
&      $\eta_{\mathrm CMB}$+$Y_{\mathrm p}$   & $3.90^{+0.21}_{-0.58}$  &  \cite{Hamann:2011ge}\\
&      $Y_{\mathrm p}$+D/H   & $3.91^{+0.22}_{-0.55}$  &  \cite{Hamann:2011ge}\\
\hline
\hline
\end{tabular}
}
\end{center}
\end{table}

\subsubsection*{Cosmic microwave background and large-scale structure}

Unlike BBN, probes of the universe's late-time inhomogeneities
such as the CMB anisotropies and the 
large-scale structure distribution are not sensitive to the flavor
content of the neutrino sector, only to its contribution to the
stress-energy tensor.   If neutrinos are massless, then the $N_{\mathrm eff}$ parameter 
as defined in equation~(\ref{eq:neff})
alone characterizes their effects on the universe's evolution.
If neutrinos are massive, then in principle it is necessary to know the exact
form of the neutrino phase space distribution in order to solve the evolution equations for
the inhomogeneities exactly.   However, the current generation of late-time
cosmological probes are not sensitive to deviations of the neutrino phase space 
distribution from the relativistic Fermi-Dirac distribution~\cite{Cuoco:2005qr,Shiraishi:2009fu}.
Therefore, it suffices to discuss the neutrino sector only in terms of 
the neutrino masses $m_{\nu_i}$ and  the $N_{{\mathrm eff}}$ parameter.

\paragraph{Signatures of light sterile neutrinos  in the CMB}

Additional relativistic energy density due to a thermal population of
light particles affects the CMB anisotropies mainly through its effect
on the epoch around matter-radiation equality.  If the particle has a
rest mass substantially below the temperature of the thermal
population around matter-radiation equality (a rough estimate for
neutrinos might be $m_{\nu, i} \ll 0.1$~eV), then it alters the equality
redshift $z_{\mathrm eq}$ according to
\begin{equation}
\label{eq:zeq}
1+z_{\mathrm eq}=\frac{\omega_m}{\omega_{\gamma}}\frac{1}{1+0.227N_{\mathrm eff}},
\end{equation}
where $\omega_m$ and $\omega_\gamma$ are the present-day matter and 
photon energy densities respectively.  Because observations of the CMB
temperature anisotropies strongly constrain $z_{\mathrm eq}$ through the
ratio of the first and the third acoustic peaks~\cite{Komatsu:2010fb},
measuring $z_{\mathrm eq}$ essentially fixes the ratio of the energy
density in matter to the energy density in radiation.  However,
equation~(\ref{eq:zeq}) also implies that $N_{\mathrm eff}$ and the
physical matter density $\omega_m$ are strongly degenerate
parameters.

If the particle's rest mass is such that the thermal population
transits from an ultra-relativistic to a non-relativistic distribution
around matter-radiation equality (a crude estimate for neutrinos is $m_{\nu,i}
\sim 0.1$--1~eV), then $z_{\mathrm eq}$ depends also somewhat on the
particle's mass to temperature ratio.  In other words, we expect some
degeneracy of $N_{\mathrm eff}$ with the particle's rest mass.  Heavier
and/or colder particle populations ({\it e.g.}, $m_{\nu,i} \gg 10$~eV) that have
already become fully non-relativistic before matter-radiation equality
is a non-relativistic matter component in the context of CMB and LSS.
For this reason, the $N_{\mathrm eff}$ parameter does not encompass
keV-mass sterile neutrinos in CMB and LSS studies.

A non-standard $N_{\mathrm eff}$ also affects the 
sound horizon $r_s$ at the time of CMB decoupling $\tau_*$,
\begin{eqnarray}
\label{eq:rs}
r_s & =&  \int_0^{\tau_*} d \tau' c_s(\tau') \nonumber \\
& \simeq & \frac{1}{H_0} \sqrt{\frac{4}{3} \frac{a_{\mathrm eq}}{\Omega_m R(z_{\mathrm
eq})}} \ln \left[ \frac{\sqrt{1+R (z_*)} + \sqrt{R(z_*)+R(z_{\mathrm
eq})}}{1+\sqrt{R(z_{\mathrm eq})}} \right],
\end{eqnarray}
where $R (z) \equiv (3/4)(\rho_b/\rho_\gamma)=(3/4) (\omega_b/\omega_\gamma) a$, 
and $H_0 = 100\  h \ {\mathrm km} \ {\mathrm s}^{-1}  \ {\mathrm Mpc}^{-1}$ is the present-day Hubble parameter.
 The sound horizon in turn affects the angular position of the acoustic peaks via
\begin{equation}
\label{eq:thetas}
\theta_s = \frac{r_s}{D_A},
\end{equation}
where $D_A$ is the angular diameter distance to the last scattering
surface.

Since both $z_{\mathrm eq}$ and $\omega_b$---and hence $R(z_{\mathrm eq})$ and $R(z_*)$---are well-constrained 
parameters
\footnote{The decoupling redshift $z_*$ also depends  logarithmically on the Hubble parameter $h$.  We ignore this dependence here for our crude estimates.}
 ($\omega_b$ is constrained via the ratio of the odd to the even acoustic peaks), 
equations~(\ref{eq:rs}) and~(\ref{eq:thetas}) imply that
$\omega_m = \Omega_m h^2$---and hence $N_{\mathrm eff}$ through 
$z_{\mathrm eq}$---is exactly degenerate with those parameters governing the angular diameter distance $D_A$
in its effects on $\theta_s$.
 In the simplest $\Lambda$CDM model%
\footnote{The vanilla $\Lambda$CDM model assumes (i) a flat spatial
  geometry, (ii) an energy content of cold dark matter, baryons, dark
  energy due to a cosmological constant, and massless neutrinos, and
  (iii) adiabatic, scalar-only primordial perturbations described by a
  nearly scale-invariant power spectrum.}  extended with a freely
varying $N_{\mathrm eff}$ parameter, holding $\omega_b$ and $z_{\mathrm eq}$ fixed
we find from equations~(\ref{eq:rs}),
and~(\ref{eq:thetas}) that
\begin{equation}
\theta_s \propto \frac{\Omega_m^{-1/2}}{\int^1_{a_*} \frac{d a}{a^2 \sqrt{\Omega_m a^{-3} + (1 - \Omega_m)}}}.
\end{equation}
This relation implies that while $\theta_s$ constrains the parameter combination $\Omega_m = \omega_m/h^2$, it does not constrain $\omega_m$ and $h$ 
individually.  Hence, a strong correlation can be expected between $N_{\mathrm eff}$ and the present-day Hubble parameter $H_0$.  In more complex models,
degeneracies of $N_{\mathrm eff}$ with the spatial curvature $\Omega_k$
and the dark energy equation of state are also possible.

Because of the strong $N_{\mathrm eff}$--$\omega_m$ and $N_{\mathrm
  eff}$--$H_0$ degeneracies, measurements of the CMB acoustic peaks
alone generally do not completely constrain $N_{\mathrm eff}$.  However,
non-interacting, free-streaming relativistic particles such as
neutrinos leave another important imprint on the CMB anisotropies
through their anisotropic
stress~\cite{Hu:1998tk}.  This anisotropic
stress suppresses the amplitude of higher harmonics in the temperature
anisotropy spectrum ($\ell > 200$)~\cite{Bashinsky:2003tk} and is phenomenologically
somewhat degenerate with the effects of changing the primordial fluctuation amplitude.
However, because anisotropic stress also shifts slightly the higher
acoustic peak positions~\cite{Bashinsky:2003tk}, it has been possible since the measurement of the third
acoustic peak by WMAP after 5 years of observations to constrain $N_{\mathrm eff}$ from below using WMAP data alone,
with $N_{\mathrm eff}=0$ excluded at better
than 99.5\% confidence in a $\Lambda$CDM+$N_{\mathrm eff}$
fit~\cite{Dunkley:2008ie,Komatsu:2010fb}.  An upper limit on $N_{\mathrm eff}$ is however not yet possible with WMAP data alone.

More recently, measurements of the CMB damping tail 
  ($\ell \gtrsim 1000$) by ACT and SPT has provided us with an additional
handle on $N_{\mathrm eff}$ through the effect of $z_{\mathrm eq}$ on diffusion damping (or
Silk damping).   The diffusion damping scale $r_d$ can be approximated as~\cite{Hou:2011ec}
\begin{eqnarray}
r^2_d &\simeq& (2 \pi)^2 \int_0^{a_*} \frac{da}{a^3 \sigma_T n_e H} \left[ \frac{R^2 + (16/15) (1+R)}{6(1+R)^2} \right] \nonumber \\
& \simeq & \frac{(2 \pi)^2}{H_0 \sqrt{\Omega_m}} \int_0^{a_*} \frac{da}{a \sqrt{a + a_{\mathrm eq}} \sigma_T n_e} \left[ \frac{R^2 + (16/15) (1+R)}{6 (1+R)^2} \right]
\end{eqnarray}
where $\sigma_T$ is the Thomson scattering cross-section, and $n_e$ is the free electron number density.  
As with $\theta_s$, it is the angular diffusion scale $\theta_d = r_d/D_A$ that we measure.  Thus, 
assuming fixed values of $z_{\mathrm eq}$, $\omega_b$,  $a_*$,
and $n_e$, and using the definition of $\theta_s$ in equation~(\ref{eq:thetas}), we can show that
\begin{equation}
\label{eq:thetad}
\theta_d \propto  (\Omega_m H_0^2)^{1/4} \ \theta_s.
\end{equation}
In other words, once $\theta_s$ has been established by large angular-scale CMB experiments such as WMAP, measurement of $\theta_d$ in the CMB damping tail 
allows us to pin down the physical matter density $\omega_m$ and hence
$N_{\mathrm eff}$ through the $N_{\mathrm eff}$--$\omega_m$ degeneracy independently
of the details of $D_A$ ({\it e.g.}, spatial curvature, dark energy equation of state, etc.).  The
measurement of $\theta_d$ by ACT and SPT has allowed us to place constraints on $N_{\mathrm
  eff}$ also from above using CMB data
alone~\cite{Dunkley:2010ge,Hou:2011ec,Keisler:2011aw}.

In deriving equation~(\ref{eq:thetad}) we have had to assume fixed values of $a_*$ and
$n_e$.  These quantities can in principle be changed, while keeping the baryon density
$\omega_b$ constant, by altering the helium-4 mass fraction $Y_p$.  Therefore, we expect a
degeneracy between $N_{\mathrm eff}$ and $Y_p$.  At present, this degeneracy can be overcome
by either assuming a fixed value of $Y_p=0.24$, or by implementing a BBN consistency
relation linking each pair of $\{\omega_b,N_{\mathrm eff}\}$ to a specific value of $Y_p$~\cite{Hamann:2007sb}.

\paragraph{Signatures of light sterile neutrinos in the LSS}

Two quantities determine the general shape of the present-day LSS
matter power spectrum $P(k)$: the comoving wavenumber
\begin{eqnarray}
k_{\mathrm eq}&\equiv&a_{\mathrm eq} H(a_{\mathrm eq}) \nonumber \\
&\simeq& 4.7 \times 10^{-4} \sqrt{\Omega_m (1+z_{\mathrm eq})} \ h {\mathrm Mpc}^{-1}
\end{eqnarray}
at matter-radiation equality, and the baryon-to-matter fraction
\begin{equation}
f_b\equiv \frac{\omega_b}{\omega_m}.
\end{equation}
The former determines the position of the ``turning point'' of $P(k)$,
while the latter controls the  power suppression due to
baryon acoustic oscillations at $k > k_{\mathrm eq}$.  If some of the
neutrinos are massive and non-relativistic today, then a third
parameter
\begin{equation}
f_\nu\equiv\frac{\omega_\nu }{\omega_m} = \frac{\sum_i m_{\nu,i}/94 \ {\mathrm eV}}{\omega_m}
\end{equation}
controls power suppression at $k > k_{\mathrm eq}$ due to the neutrinos'
thermal velocity dispersion.

If there are extra neutrino species and all neutrino species are still
relativistic today, then consistency with the CMB-determined $z_{\mathrm
  eq}$ and $\Omega_m$ implies that $k_{\mathrm eq}$ is unaffected (recall that $k$ is 
  measured in units of $h{\mathrm Mpc}^{-1}$).
However, because the physical matter density $\omega_m$ is now
higher while $\omega_b$ remains unchanged, the baryon fraction
$f_b$ decreases.  This leads to a higher matter fluctuation amplitude
today at $k> k_b$, where $k_b$ corresponds roughly to the comoving Hubble rate at the 
baryon drag epoch ({\it i.e.}, when baryons decouple from photons).
Thus, measurements of the LSS matter power
spectrum can be used to break the strong $N_{\mathrm eff}$--$\omega_m$
degeneracy in the CMB data.

On the other hand, if some of the neutrinos---sterile or standard
model---are massive enough to be non-relativistic today, then they
will contribute to the expansion rate of the universe but not cluster
on small scales.  This leads to a $f_\nu$-dependent suppression of the 
matter fluctuation amplitude at $k > k_\nu$, where $k_\nu$ is the neutrino 
free-streaming scale at the time neutrinos become nonrelativistic, which
in turn obviates some of the enhancement due to the extra
relativistic energy density near CMB decoupling.  For this reason,
some degeneracy persists between $N_{\mathrm eff}$--$f_\nu$ in combined
analyses of CMB and LSS data.

\paragraph{Other cosmological observations}

Low redshift measurements of the expansion rate and of the distance
versus relation are not directly sensitive to the presence of light
sterile neutrinos: if these sterile neutrinos are still relativistic
today, then their energy density is completely negligible; if they are
non-relativistic, then they simply act like a matter component
indistinguishable from cold dark matter as far as the low-redshift
expansion history is concerned.  However, because of the strong
correlation between $N_{\mathrm eff}$ and $H_0$ from WMAP data alone,
direct measurements of $H_0$ and $H(z)$ at low redshifts can be a very
effective means to break this degeneracy and to constrain $N_{\mathrm
  eff}$~\cite{Reid:2009nq}.  In more complex models where the spatial
curvature and/or the dark energy equation of state are also allowed to
vary, geometric constraints from baryon acoustic oscillations (BAO)
and type Ia supernovae (SN) can be used to further break degeneracies.

\paragraph{Present constraints on light sterile neutrinos}

Table~\ref{tab:neff} shows a selection of recent constraints on the
relativistic energy density $N_{\mathrm eff}$. The salient features can be
summarized as follows:

\begin{table}[t]
\caption{A selection of recent constraints on $N_{\mathrm eff}$, with 68\% (95\%) uncertainties.  W-5 and W-7 stand for WMAP 5-year and 7-year data
respectively, $H_0$ refers to the constraint  $H_0 = 74.2 \pm 3.6 \  {\mathrm km} \  {\mathrm s}^{-1}$  from~\cite{Riess:2009pu}, 
LRG  the halo power spectrum determined from the luminous red galaxy sample of the SDSS data release 7~\cite{Reid:2009xm}, while
CMB denotes  a combination of small-scale CMB experiments such as ACBAR, BICEP and QUaD.  \label{tab:neff}}
\begin{center}
{\small
\begin{tabular}{lllc}
\hline
\hline
Model & Data & $N_{\mathrm eff}$ & Ref. \\
\hline
$N_{\mathrm eff}$ & W-5+BAO+SN+$H_0$ & $4.13^{+0.87 (+1.76)}_{-0.85 (-1.63)} $& \cite{Reid:2009nq} \\
& W-5+LRG+$H_0$ & $4.16^{+0.76 (+1.60)}_{-0.77 (-1.43)}$ & \cite{Reid:2009nq}\\
& W-5+CMB+BAO+XLF+$f_{\mathrm gas}$+$H_0$ & $3.4^{+0.6}_{-0.5}$ & \cite{Mantz:2009rj} \\
& W-5+LRG+maxBCG+$H_0$ & $3.77^{+0.67 (+1.37)}_{-0.67 (-1.24)}$ & \cite{Reid:2009nq} \\
& W-7+BAO+$H_0$ & $4.34^{+0.86}_{-0.88}$ & \cite{Komatsu:2010fb} \\
& W-7+LRG+$H_0$ & $4.25^{+0.76}_{-0.80}$ & \cite{Komatsu:2010fb} \\
& W-7+ACT & $5.3 \pm 1.3$ & \cite{Dunkley:2010ge} \\
& W-7+ACT+BAO+$H_0$ & $4.56 \pm 0.75$ & \cite{Dunkley:2010ge} \\
& W-7+SPT & $3.85 \pm 0.62$ & \cite{Keisler:2011aw} \\
& W-7+SPT+BAO+$H_0$ & $3.85 \pm 0.42$ & \cite{Keisler:2011aw} \\
& W-7+ACT+SPT+LRG+$H_0$  & $4.08^{(+0.71)}_{(-0.68)}$  & \cite{Archidiacono:2011gq} \\
& W-7+ACT+SPT+BAO+$H_0$  & $3.89 \pm 0.41$  & \cite{Smith:2011ab} \\
\hline
$N_{\mathrm eff}$+$f_\nu$ & W-7+CMB+BAO+$H_0$ & $4.47^{(+1.82)}_{(-1.74)}$ & \cite{Hamann:2010pw} \\
 & W-7+CMB+LRG+$H_0$ & $4.87^{(+1.86)}_{(-1.75)}$ & \cite{Hamann:2010pw} \\
\hline
$N_{\mathrm eff}$+$\Omega_k$ & W-7+BAO+$H_0$ & $4.61 \pm 0.96$ & \cite{Smith:2011ab} \\
& W-7+ACT+SPT+BAO+$H_0$ & $4.03 \pm 0.45$ & \cite{Hamann:2010pw} \\
\hline
$N_{\mathrm eff}$+$\Omega_k$+$f_\nu$ & W-7+ACT+SPT+BAO+$H_0$ & $4.00 \pm 0.43$ & \cite{Smith:2011ab} \\
 \hline
$N_{\mathrm eff}$+$f_\nu$+$w$ & W-7+CMB+BAO+$H_0$ & $3.68^{(+1.90)}_{(-1.84)}$ & \cite{Hamann:2010pw} \\
 & W-7+CMB+LRG+$H_0$ & $4.87^{(+2.02)}_{(-2.02)}$ & \cite{Hamann:2010pw} \\
 \hline
$N_{\mathrm eff}$+$\Omega_k$+$f_\nu$+$w$ & W-7+CMB+BAO+SN+$H_0$ & $4.2^{+1.10 (+2.00)}_{-0.61(-1.14)}$ & \cite{GonzalezGarcia:2010un} \\
 & W-7+CMB+LRG+SN+$H_0$ & $4.3^{+1.40 (+2.30)}_{-0.54 (-1.09)}$ & \cite{GonzalezGarcia:2010un}\\
\hline
\hline
\end{tabular}
}
\end{center}
\end{table}

\begin{enumerate}

\item Precision cosmological data since the WMAP 5-year data release
  have consistently shown a mild preference for an excess of
  relativistic energy density.  With one exception, all combinations
  of data find a $> 68$\% preference for $N_{\mathrm eff}>3$, although
  whether $N_{\mathrm eff}=3.046$ falls inside or outside the 95\%
  credible region depend on the exact data sets used.

\item The only set of observations that tend to prefer a more ``standard'' $N_{\mathrm
    eff}$ is the cluster catalog from the ROSAT All-Sky Survey/Chandra X-ray Observatory combined 
with WMAP-5+CMB+BAO+$H_0$.  Here, the preferred value is $N_{\mathrm eff}=3.4^{+0.6}_{-0.5}$~\cite{Mantz:2009rj}.

\item Measurement of the CMB damping tail by ACT has disfavored for
  the first time the standard value of $N_{\mathrm eff}=3.046$ at more
  than 95\% confidence within the $\Lambda$CDM+$N_{\mathrm eff}$ class of
  models, when ACT data are analyzed together with
  WMAP-7+BAO+$H_0$~\cite{Dunkley:2010ge,Hou:2011ec}. A similar
  measurement by SPT also finds a preference for a large $N_{\mathrm
    eff}=3.85 \pm 0.42$ from WMAP-7+SPT+BAO+$H_0$, albeit with a
  smaller statistical significance~\cite{Keisler:2011aw}.

\item Some earlier analyses have also reported a $>95\%$ significant
  deviation from $N_{\mathrm eff}=3.046$ without using either ACT or SPT
  data~\cite{Hamann:2010pw,GonzalezGarcia:2010un}.  This result
  can be traced to the fact that the massive neutrino fraction $f_\nu$
  was also allowed to vary in these analysis.  Because of the $N_{\mathrm
    eff}$--$f_\nu$ degeneracy discussed earlier, a non-zero $f_\nu$
  tends to drag up the preferred values of $N_{\mathrm eff}$ after
  marginalization.

\item The evidence for $N_{\mathrm eff}>3$ does not appear to be strongly
  dependent on the complexity of the cosmological model.  As discussed
  in the previous point, non-zero neutrino masses in fact tend to push
  up the preferred $N_{\mathrm eff}$.  Including other parameters
  degenerate with $N_{\mathrm eff}$ such as spatial curvature $\Omega_k$
  or a dark energy equation of state $w \neq 1$ also does not remove
  the data's preference for a large $N_{\mathrm eff}$.  Fitting
  WMAP-7+ACT+SPT+BAO+$H_0$, for example, the analysis
  of~\cite{Smith:2011ab} finds that $N_{\mathrm eff}=3.046$ remains
  disfavored at $>95$\% confidence in the $\Lambda$CDM+$N_{\mathrm
    eff}$+$\Omega_k$ model.

\end{enumerate}

Because most constraints listed in table~\ref{tab:neff} have been
obtained using Bayesian statistics, concern has been raised over their
sensitivity to the choice of priors on the model
parameters~\cite{GonzalezMorales:2011ty,Hamann:2011hu}.  However,
fitting a $\Lambda$CDM+$N_{\mathrm eff}$ model to WMAP-7+ACT, the analysis
of~\cite{Hamann:2011hu} finds no significant difference ($< 0.1 \sigma$)
in the $N_{\mathrm eff}$ constraints between using a uniform prior on
$H_0$, $\Omega_\Lambda$, or on $\theta_s$.  The difference becomes
even smaller when $H_0$ data are also included in the analysis.
Similarly, while prior-independent ``frequentist'' constraints
obtained from profile likelihood ratios tend to prefer smaller values
of $N_{\mathrm eff}$~\cite{GonzalezMorales:2011ty,Hamann:2011hu}, the
shift is only marginal: in the case of WMAP-7+ACT, roughly
$0.4\sigma$--$0.6\sigma$, while for WMAP-7+ACT+$H_0$, $< 0.3
\sigma$~\cite{Hamann:2011hu}.%
\footnote{The mapping between the profile likelihood ratios ${\mathcal
    L}/{\mathcal L}_{\mathrm max} \leq \exp(-1/2)$ and ${\mathcal L}/{\mathcal L}_{\mathrm
    max} \leq \exp(-4/2)$ and the frequentist $1\sigma$ and $2\sigma$
  confidence intervals is valid only in the limit Wilks' theorem is
  satisfied, {\it i.e.}, the profile likelihood approaches a Gaussian
  distribution.  If a mapping cannot be established, then the profile
  likelihood ratios have no statistical interpretation.}  As future
data become even more constraining, we expect the Bayesian and the
frequentist credible/confidence intervals to converge
completely~\cite{Hamann:2007pi,Reid:2009nq}.  Thus, we conclude that while
the current data do not yet imply a definitive detection of $N_{\mathrm
  eff}>3$, the preference for $N_{\mathrm eff}>3$ is not an artifact
driven by the choice of priors.

\paragraph{Connection to terrestrial neutrino oscillation experiments}

An important issue now is, if the excess relativistic energy density
is interpreted as a thermal population of light sterile neutrinos, do
these sterile neutrino have anything to do with the sterile
neutrino(s) required to explain the anomalies observed in the LSND,
MiniBooNE, and reactor neutrino oscillation experiments?

As we saw earlier, active-sterile neutrino oscillations with mixing
parameters required to explain the terrestrial anomalies inevitably
lead to sterile neutrino thermalization and hence a non-standard
$N_{\mathrm eff}$, unless some new physics is introduced to suppress their
production.  However, the fact that the anomalies point to a
mass-square splitting of $\Delta m^2 \sim 1~{\mathrm eV}^2$ implies that
the sterile neutrino must have a minimum mass of $\sim 1$~eV, which
could potentially come into conflict with cosmological neutrino mass
bounds.  Indeed, if we extend the $\Lambda$CDM model with one species
of thermalized sterile neutrino of some arbitrary mass $m_s$, then
WMAP-7+LRG+$H_0$ impose a 95\% constraint on $m_s$
of~\cite{Hamann:2010bk,Giusarma:2011zq}
\begin{equation}
\label{eq:mbound}
m_s < 0.45~{\mathrm eV}.
\end{equation}
A similar constraint on $m_s$ exists also in the case of two sterile
neutrino species each with mass
$m_s$~~\cite{Hamann:2010bk,Giusarma:2011zq}.  Thus, the sterile
neutrino oscillation interpretation of the neutrino oscillation
anomalies appear at first glance to be incompatible with precision
cosmology.

However, if we are willing to make further modifications to the
$\Lambda$CDM model, it is possible to circumvent these constraints to
some extent by exploiting parameter degeneracies.  One well-known
solution is to endow the dark energy component with an equation of
state parameter $w$ that is more negative than
$-1$~\cite{Hamann:2011ge,Kristiansen:2011mp}.  Another is to allow for
a non-flat
geometry~\cite{Smith:2011ab,GonzalezGarcia:2010un,Kristiansen:2011mp}
Yet another is to exploit the $N_{\mathrm eff}$--$f_\nu$ degeneracy and
allow for even more massless degrees of freedom~\cite{Hamann:2011ge}.
In all three cases, one species of sterile neutrinos with $m_s = 1$~eV
can be reasonably accommodated by precision cosmological data.
However, two species with $m_s=1$~eV each, or one species with
$m_s=2$~eV is still strongly disfavored.

\paragraph{Future cosmological probes}

As discussed above, because of the multitude of parameter degeneracies, 
no single cosmological observation at present is able to constrain 
$N_{\mathrm eff}$ on its own.  The ongoing Planck mission, however, will be able to do so,
with a projected 68\% sensitivity to $N_{\mathrm eff}$ at the 0.2--0.25 level~\cite{Bashinsky:2003tk,Hannestad:2006as}. 
With this level of sensitivity, Planck will be able to conclusively confirm or exclude the presence of a fully thermalized 
light sterile neutrino.  Combining Planck data with ongoing ground-based small-scale polarization
experiments and future, high precision CMB measurements, a sensitivity to $N_{\mathrm eff}$
of 0.05 may be possible~\cite{Galli:2010it}.   A comparable
precision may also be achieved by combining Planck data with future weak
lensing (cosmic shear) measurements by facilities such as LSST or EUCLID~\cite{Hannestad:2006as}.

Because the effect of a non-zero neutrino mass is important for the CMB anisotropies
only if the mass is comparable to the temperature of CMB decoupling ($T \sim 0.3$~MeV), Planck on is own is not 
expected to improve on the upper bound on the 
sterile (or active) neutrino mass~(\ref{eq:mbound}).
However, since even very small neutrino masses are
important for the subsequent evolution of the large-scale structure distribution, a non-zero
neutrino mass could well be detected through weak lensing of the CMB
signal.  Naturally, cosmic shear measurements at low redshifts have
an even greater potential for probing the mass of light sterile
neutrinos.   Combining Planck data with cosmic shear data from LSST or EUCLID, a 68\%
sensitivity to $m_s$ of 0.05~eV may be possible~\cite{Hannestad:2006as,Giusarma:2011zq}.
Future galaxy cluster surveys combined with Planck data might also reach a similar level of sensitivity to $m_s$~\cite{Wang:2005vr}.

\subsection{Core Collapse Supernovae}

The progenitors of core collapse supernovae are massive stars which
evolve in a few millions of years to the endpoint of their
evolution. Progenitor stars with masses in excess of about $12\,{\mathrm 
M}_\odot$ will evolve through a series of nuclear burning stages,
slowly and modestly neutronizing, and continually losing entropy
through weak interactions and neutrino emission, ultimately forming a
core composed of iron peak material and supported by electron
degeneracy pressure. The relativistic electron degeneracy means that
the cores of these stars are trembling on the verge of instability,
and one or all of general relativity, electron capture, or nuclear
statistical equilibrium shifts will lead to collapse. Lower progenitor
star masses, roughly in the range of $9\,{\mathrm M}_\odot$ to
$12\,{\mathrm M}_\odot$, will give rise to cores that suffer this
instability before they have burned all the way to iron, in fact while
still composed of O-Ne-Mg. In either case, since the pressure support
in these cores is coming mostly from relativistically degenerate
electrons, they will have roughly the Chandrasekhar mass, $\sim
1.4\,{\mathrm M}_\odot$. These cores will collapse in $\sim 1\,{\mathrm s}$ to
nuclear density where the collapse will be violently halted by nucleon
pressure, and a shock wave and proto-neutron star will be formed. The
shock wave will move out and, though initially very energetic, will
have its energy sapped by nuclear photo-dissociation, eventually
($\sim 100\,{\mathrm ms}$) forming a standing accretion shock $\sim
100\,{\mathrm km}$ from the neutron star center.

The shock wave is probably re-energized by neutrinos emitted from the
neutron star surface (neutrino-sphere) capturing on nucleons nearer
the shock front. In turn, this process is augmented by hydrodynamic
motion, {\it e.g.}, in the Standing Accretion Shock Instability 
(SASI)~\cite{Blondin:2006yw}.  Sufficiently re-energized, the shock moves 
out and blows up the star, removing the bulk of the envelope. Subsequently, 
the still-hot, neutrino emitting proto-neutron star can drive a 
neutrino-heated wind which is a candidate site for the production of 
heavy $r$-process elements~\cite{Arnould:2007gh}. Neutrinos play a
dominant role in every one of these aspects of the supernova.

\paragraph{Sterile neutrino oscillations in supernovae}

When the core collapses to a neutron star configuration,
initially with a radius $\sim 40\,{\mathrm km}$, but within a few seconds
shrinking down to $\sim 10\,{\mathrm km}$, ultimately some
${10}^{53}\,{\mathrm ergs}$ of gravitational binding energy ($10\%$ of the
core's rest mass) will be radiated as active neutrinos and
antineutrinos of all flavors. During the in-fall phase of the
collapse, neutrino mean free paths become smaller than the size of the
collapsing core once the central density reaches about one percent of
nuclear matter density. The proto-neutron star harbors trapped seas of
neutrinos of all types. These random-walk, with a diffusion timescale
of seconds, to the neutron star surface where they decouple and more
or less freely stream away.

As in other astrophysical systems ({\it e.g.}, the Sun), this free-streaming stage 
of the neutrinos' evolution is influenced by matter effects due to coherent 
forward scattering of the neutrinos on the ambient matter (electrons and 
nucleons).  In a supernova, however, one must consider in addition forward 
scattering on the neutrinos themselves.  Such neutrino-neutrino interactions 
turn out to play acrucial role in the flavor evolution of both neutrinos and 
antineutrinos, inducing large flavor conversions depending on the ratio among 
the neutrino fluxes of different flavors and the mass hierarchy 
(see~\cite{Duan:2010bg} for a recent review and references therein).  
Self-induced neutrino oscillations dominate the neutrino propagation at a 
radius much deeper than the conventional matter-induced MSW effect among active 
flavors and impact on $r$-process nucleosynthesis~\cite{Balantekin:2004ug,
Duan:2010af}.   For sterile neutrinos, however, the location of the 
active-sterile MSW resonance varies according to the sterile neutrino mass, 
possibly coupling active-sterile MSW conversion with collective oscillations 
among the active states.

Neutrino oscillations in a supernova are treated in terms of the usual matrices of
neutrino densities $\rho_E$ for each neutrino mode with energy $E$,where the 
diagonal elements are neutrino densities, and the off-diagonal elements encode 
phase information caused by flavor oscillations.  The radial flavor variation of 
the quasi-stationary neutrino flux is given by the ``Schr\"odinger equation''
\begin{equation}\label{eq:eom1}
i\partial_r\rho_E=[{\mathsf H}_{E},\rho_{E}]
\quad\hbox{and}\quad
i\partial_r\bar\rho_E=[\bar{\mathsf H}_{E},\bar\rho_{E}]\,,
\end{equation}
where an overbar refers to antineutrinos, and sans-serif letters
denote matrices in flavor space.  The Hamiltonian matrix
contains vacuum, matter, and neutrino--neutrino terms
\begin{equation}
{\mathsf H}_{E}= {\mathsf H}^{\mathrm vac}_{E}+{\mathsf H}^{\mathrm m}_{E}+{\mathsf H}^{\nu\nu}_{E}\ ,
\label{eq:ham}
\end{equation}
and  $\bar{\mathsf H}^{\mathrm vac}_{E} ={\mathsf H}^{\mathrm vac}_{E} $,  
$\bar{\mathsf H}^{\mathrm m}_{E}=-{\mathsf H}^{\mathrm m}_{E}$, and 
$\bar{\mathsf H}^{\nu\nu}_{E}=-{\mathsf H}^{\nu\nu}_{E}$.
In the flavor basis, the vacuum term is 
\begin{equation}
{\mathsf H}^{\mathrm vac}_{E} = {\mathsf U} \frac{{\mathsf M}^2}{2 E} {\mathsf U}^{\dagger}\ ,
\end{equation}
where  ${\mathsf M}^2$ is a diagonal matrix containing the the squared mass differences 
between the mass eigenstates, and 
${\mathsf U}$ is a unitary matrix containing the mixing angles that encode 
the transformation between the mass and the weak interaction bases.
In a supernova environment $\nu_\mu$ and $\nu_\tau$
are indistinguishable. Therefore, we can define $\nu_x$ as a linear combination
of $\nu_\mu$ and $\nu_\tau$, so that the matter term ${\mathsf H}^{\mathrm m}_{E}$ in
the flavor basis spanned by $(\nu_e,\nu_x,\nu_s)$  is 
\begin{eqnarray}
\label{lambda}
{\mathsf H}^{\mathrm m} &=& \sqrt{2}G_{\mathrm F}\;
{\mathrm diag}(N_{e}-\frac{N_{n}}{2},-\frac{N_{n}}{2},0)\;,
\end{eqnarray}
where  $N_{e}$ is the net electron number density (electrons minus
positrons), and $N_{n}$ the neutron number density. Note that ${\mathsf H}^{\mathrm m}_{E}$ here 
contains both charged-current (CC) and neutral-current (NC) contributions.

In all neutral media, $Y_e=Y_p$ and $Y_n=1-Y_e$, where $Y_j$ is the
number density of particle species $j$ relative to the baryon number density $N_b$. Therefore, we can rewrite the
local electron fraction as
\begin{equation}
\label{yedef}
Y_e(r) =\frac{N_e(r)}{N_e(r)+N_n(r)}\ ,
\end{equation}
so that the matter Hamiltonian becomes
\begin{eqnarray}
\label{Yelambda}
{\mathsf H}^{\mathrm m} &=& \sqrt{2}G_{\mathrm F} N_b \;
{\mathrm diag}\left(\frac{3}{2} Y_e -\frac{1}{2},\frac{1}{2} Y_e -\frac{1}{2},0\right)\;,
\end{eqnarray}
with $N_b= N_e+N_n$. Note that, in presence of sterile neutrinos, the matter potential can be
positive or negative depending on the size of $Y_e$. For $Y_e>1/3$, it is $\nu_e$ that can undergo
an active-sterile MSW resonance,
whereas for $Y_e < 1/3$ it is $\bar{\nu}_e$.

The last term in the Hamiltonian~(\ref{eq:ham}), ${\mathsf H}^{\nu\nu}_{E}$,
corresponding to  neutrino-neutrino
interactions  vanishes for all elements involving the sterile
neutrino flavor eigenstate~\cite{Sigl:1992fn}, {\it i.e.}, ${\mathsf H}^{\nu\nu}_{es}={\mathsf
H}^{\nu\nu}_{xs}={\mathsf H}^{\nu\nu}_{ss}=0$; only the
block involving the active flavors is non-zero. In
particular, the only non-vanishing off-diagonal elements are
the ones involving only active species.

\paragraph{Electron abundance evolution}
The material in a fluid element moving away from the supernova core will
experience three stages of nuclear evolution. Near the surface of
the neutron star,  the material is typically very hot, and essentially
all of the baryons are in the form of free nucleons. As the material
flows away from the neutron star, it cools.  When the temperature drops below
$\sim 1$~MeV, $\alpha$-particles begin to assemble. As the fluid flows
farther out and cools even more, heavier nuclei begin to form. Around
half of the nuclei with mass numbers $A>100$ are supposed to be created via the $r$-process.

Activation of the $r$-process requires a neutron-rich environment, {\it i.e.}, an electron
fraction per baryon $Y_e< 0.5$, sufficiently large entropy to favor
a high ratio of free neutrons, and sufficiently fast
timescales to lower the efficacy of $\alpha$-particle conversion to
heavier nuclei.   
The electron abundance $Y_e$ in the neutrino-heated material
flowing away from the neutron star is set by a competition between
the rates of, predominantly, the reactions
\begin{eqnarray}
\label{nue1}
\nu_e + n &\rightarrow& p + e^-\ ,\\
\label{anue1}
\bar{\nu}_e + p  &\rightarrow& n + e^+\ ,
\end{eqnarray}
and their associated reverse processes. Because of the slow time variations
of the outflow conditions during the cooling phase, a near
steady-state situation applies, and the
$Y_e$ rate-of-change within an
outflowing mass element may be written as~\cite{McLaughlin:1997qi}
\begin{equation}
\label{Yeeq}
\frac{dY_e}{dt} = v(r) \frac{dY_e}{dr} \simeq (\lambda_{\nu_e} + \lambda_{e^+}) Y_n^{\mathrm
f} - (\lambda_{\bar{\nu}_e} + \lambda_{e^-}) Y_p^{\mathrm f}\ ,
\end{equation}
where $v(r)$ is the velocity of the outflowing mass element, 
$Y_n^{\mathrm f}$ and $Y_p^{\mathrm f}$ are the abundances of free nucleons,
and $\lambda$ are the rates of the neutrino capture processes~(\ref{nue1}) and~(\ref{anue1}) and their reverse processes.  The forward rates (neutrino capture on free nucleons) 
are given by
\begin{eqnarray}
\label{lambdanue}
\lambda_{\nu_e} &\propto& \frac{L_{\nu_e}}{\langle E_{\nu_e}\rangle}\, \langle \sigma_{\nu_e n}(r)\rangle \ ,
\\
\label{lambdaantinue} \lambda_{\bar{\nu}_e} &\propto&
\frac{L_{\bar{\nu}_e}}{4 \pi r^2 \langle E_{\bar{\nu}_e} \rangle}\,
\langle \sigma_{\bar{\nu}_e p}(r)\rangle \ ,
\end{eqnarray}
with $L_{\nu_e\ (\bar{\nu}_e)}$ the $\nu_e$ ($\bar{\nu}_e$) luminosity,
and $\langle E_{\nu_e}\rangle$ ($\langle E_{\bar{\nu}_e} \rangle$) 
the $\nu_e$ ($\bar{\nu}_e$) mean energy.
The rates for the reverse processes (electron and positron capture
on free nucleons) are functions of  the relativistic electron chemical potential 
$\mu_e$ and the electron temperature $T_e$.
The details are generally quite complex, but $Y_e$ at small radii is
determined mainly by the $e^-$ and $e^+$ capture rates, whereas at larger
radii the neutrino-capture reactions dominate.  

The values of $Y_e$ found in standard ({\it i.e.}, no flavor oscillations) supernova 
simulations are generally too high to enable a successful $r$-process.  However, 
as it is clear from equation~(\ref{Yeeq}) that a variation of neutrino fluxes due 
to flavor oscillations can affect the evolution of $Y_e$, one possible solution to 
the high $Y_e$ problem is to remove the $\nu_e$ flux by way of active-sterile 
oscillations~\cite{Beun:2006ka,Keranen:2007ga,Fetter:2002xx,Fetter:2000kf,
McLaughlin:1999pd,Hidaka:2007se,Nunokawa:1997ct}.  On the other hand, a variation 
of $Y_e$ changes the matter potential felt  by the neutrinos according to 
equation~(\ref{Yelambda}), and can significantly alter the flavor evolution 
especially if $Y_e$ is driven to below $1/3$ (see discussion immediately after 
equation~(\ref{Yelambda})).  This can have important consequences for key aspects 
of energy transport in the proto-neutron star core.  We discuss these possibilities 
below.

\paragraph{Light sterile neutrinos in supernovae}

If the sterile neutrino rest mass in is 
the range $\sim 0.1\,{\mathrm eV}$ to few ${\mathrm eV}$, {\it i.e.}, overlapping with the 
preferred mass range of the MiniBooNE/LSND and the reactor-anomalies, two MSW
resonances are expected to occur close to the neutrino-sphere, one for neutrinos 
and one for antineutrinos.  However, because the matter potential is very steep 
as a result of a rapidly changing $Y_e$ at small radii, these resonances are not 
adiabatic.  At larger radii, a second and more adiabatic resonance occurs in the 
neutrino sector only. The net effect, therefore, is that a large fraction of 
$\nu_e$ is converted to $\nu_s$ while the $\bar{\nu}_e$  flux is only partially 
converted to $\bar{\nu}_s$.   The equilibrium of the neutrino capture 
processes~(\ref{nue1}) and~(\ref{anue1}) and their reverse processes then drives 
down the electron abundance $Y_e$ as a consequence, thereby creating a neutron-rich 
environment that enables the formation of elements via the $r$-process.  However, 
recent studies have shown that collective oscillations in the active sector due to 
neutrino-neutrino interactions become dominant especially in the late cooling phase. 
They act to repopulate the $\nu_e$ flux so that the resulting $Y_e$ is always lower 
than in the case with no active-sterile oscillations, but not always as low as in 
the ``naive'' case without considerations of the neutrino-neutrino matter 
effects~\cite{Tamborra:2011is}.  Therefore, the presence of light sterile neutrinos 
can alter the conditions for element formation, although active-sterile oscillations 
alone probably cannot activate the $r$-process.

Observations of ultra-metal poor halo stars with $r$-process excesses
reveal the curious feature that the nuclear mass $130$ and $195$
abundance peaks have similar overall abundances, suggesting that
fission cycling in the $r$-process may be operating, essentially tying
the abundances in these two peaks together in steady state
equilibrium. Such an equilibrium would require very large neutron
excesses and, so far, the only models that can stimulate such an excess
in neutrino-heated ejecta are those that invoke active-sterile
neutrino flavor conversion in the wind as outlined above. 
 Although the actual site of $r$-process nucleosynthesis
remains an open question, however, many neutron-rich environments,
like those associated with neutron star mergers and accretion disks,
also may have large neutrino fluxes and so be subject to some form of
the alpha effect. In any case, the connection between BBN and
$r$-process nucleosynthesis on the one hand, and active-sterile
neutrino conversion physics on the other, is tantalizing.

\paragraph{Heavy sterile neutrinos in supernovae}

Sterile neutrinos with rest masses in the keV region and
mixing angles with active species that make them dark matter
candidates  can also affect key aspects of energy transport in the
proto-neutron star core. In fact, the high matter density (essentially
nuclear matter density) and large electron lepton number in the core
and typical active neutrino energies of order $10\,{\mathrm MeV}$ mean
that keV-mass sterile neutrinos can be resonant~\cite{Abazajian:2001nj}.  
If flavor evolution through the resonance is sufficiently
adiabatic, then the $\nu_e$ will be turned into a sterile neutrino and
this neutrino likely will stream out of the core at near light speed---unless 
it encounters another MSW resonance further out that
re-converts it to an active neutrino. In fact, the dependence of the
potential on $Y_e$ can engineer just such an effect: at first as the
core collapses $\nu_e$'s are converted, and unbalanced electron capture
lowers $Y_e$, eventually driving it below $\frac{1}{3}$ so that the matter 
potential becomes negative as per equation~(\ref{Yelambda}), thereby causing 
the conversion of $\bar\nu_e$'s to sterile species. This will bring the 
potential back up if the collapse/expansion rate of the material in the core 
is slow enough, engineering a double resonance scenario where $\nu_e$'s are
regenerated from the sterile neutrinos but closer to the 
neutrino-sphere~\cite{Hidaka:2006sg,Hidaka:2007se}. The result of such a
scenario is that the effective transport time for $\nu_e$'s can be
dramatically reduced, bringing up the $\nu_e$ luminosity at the
neutrino-sphere, and thereby affecting nearly every aspect of
downstream supernova evolution, for example increasing the material
heating rate behind the shock. 

Moreover, active-to-sterile conversion
has been considered as a way to produce large space motions of
pulsar/neutron star remnants following core collapse~\cite{Kusenko:1998bk,
Fuller:2003gy,Fryer:2005sz}. Even heavier sterile
neutrinos, with rest masses in the $\sim 100\,{\mathrm MeV}$ range, could
aid supernova shock re-heating~\cite{Fuller:2009zz}, although these
may run afoul of cosmological constraints on nucleosynthesis or
measurements of $N_{\mathrm eff}$ derived from the cosmic microwave
background anisotropies~\cite{Dolgov:2000jw,Kusenko:2004qc,Kusenko:2009up,Fuller:2011qy}.

Invoking $\sim {\mathrm keV}$ rest mass sterile neutrinos and
active-to-sterile neutrino flavor conversion to engineer shock
re-heating is a dangerous gambit, however, because energy can be lost
from the core. Ultimately, this could prove to be in conflict with
what we know about neutrinos from the SN 1987A neutrino burst
observations~\cite{Raffelt:1992bs}. But there is plenty of cushion
here. The total kinetic energy of bulk motion plus optical energy in a
core collapse supernova explosion is only ${10}^{51}\,{\mathrm ergs}$,
some $1\%$ of the energy resident in the active neutrino energy in the
core.  Therefore, much energy could be \lq\lq thrown away\rq\rq\ in sterile
states escaping from the core and still there will be plenty of active
neutrino energy to power an explosion, especially if the transport of
this energy to the shock is augmented by efficient hydrodynamic
motion, as in SASI-like scenarios.

\clearpage
\section{Evidence from Oscillation Experiments}
\label{sec:oscillation}

While the standard three-flavor framework of neutrino oscillations is by now
well established, there are a number of oscillation experiments whose results
cannot be explained in this framework, but, should they stand up to further
scrutiny, might require the introduction of extra, sterile, neutrino species.
The first piece of evidence in favor of oscillations beyond the three-flavor
framework came from the LSND experiment, which performed a sensitive search for
neutrino oscillations and obtained $> 3\sigma$ evidence for $\bar \nu_\mu
\rightarrow \bar \nu_e$ oscillations with $\Delta m^2 > 0.2$ eV$^2$
\cite{Aguilar:2001ty}.  The MiniBooNE experiment, which was designed to test LSND, has
reported oscillation results in both neutrino mode and antineutrino mode.
Whereas the results obtained in neutrino mode disfavor most of the parameter
space preferred by LSND, the MiniBooNE antineutrino data, though not yet
conclusive, are consistent with the LSND signal and consistent with
oscillations at a $\Delta m^2 \sim 1$ eV$^2$.  Moreover, MiniBooNE reports an
excess of events at low energy, outside the energy range where LSND-like
oscillations are expected.  Further hints and evidence for the existence of
sterile neutrinos come from measurements of the neutrino flux from intense
radioactive sources, and from short-baseline reactor experiments. In
particular, a recent re-evaluation of the expected antineutrino flux from
nuclear reactor indicates that the measured flux is about 3\% below the
prediction, with a significance above $3\sigma$.

On the other hand, there are a number of experiments that do not support this
body of evidence for sterile neutrinos. The KARMEN experiment, which is very
similar to LSND, observed no such evidence. (But a joint analysis of the two
experiments \cite{Church:2002tc} shows that their data sets are compatible with
oscillations occuring either in a band from 0.2 to 1~eV$^2$ or in a region
around 7~eV$^2$.) Also, a number of $\nu_\mu$ disappearance searches in the
relevant mass range, as well as a MINOS search for disappearance of active 
neutrinos in neutral current events, produced negative results.

In this section, we will discuss the evidence for sterile neutrinos from LSND,
MiniBooNE, the radiocative source experiments and the reactor experiments, and
also the null results from KARMEN and MINOS. A combined global analysis of
these datasets (and others) will be presented in sec.~\ref{sec:global}.

\subsection{The LSND Signal}
\label{sec:lsnd}

\subsubsection*{Description of the Experiment}

\begin{figure}
  \centerline{\includegraphics[height=6.in,angle=0]{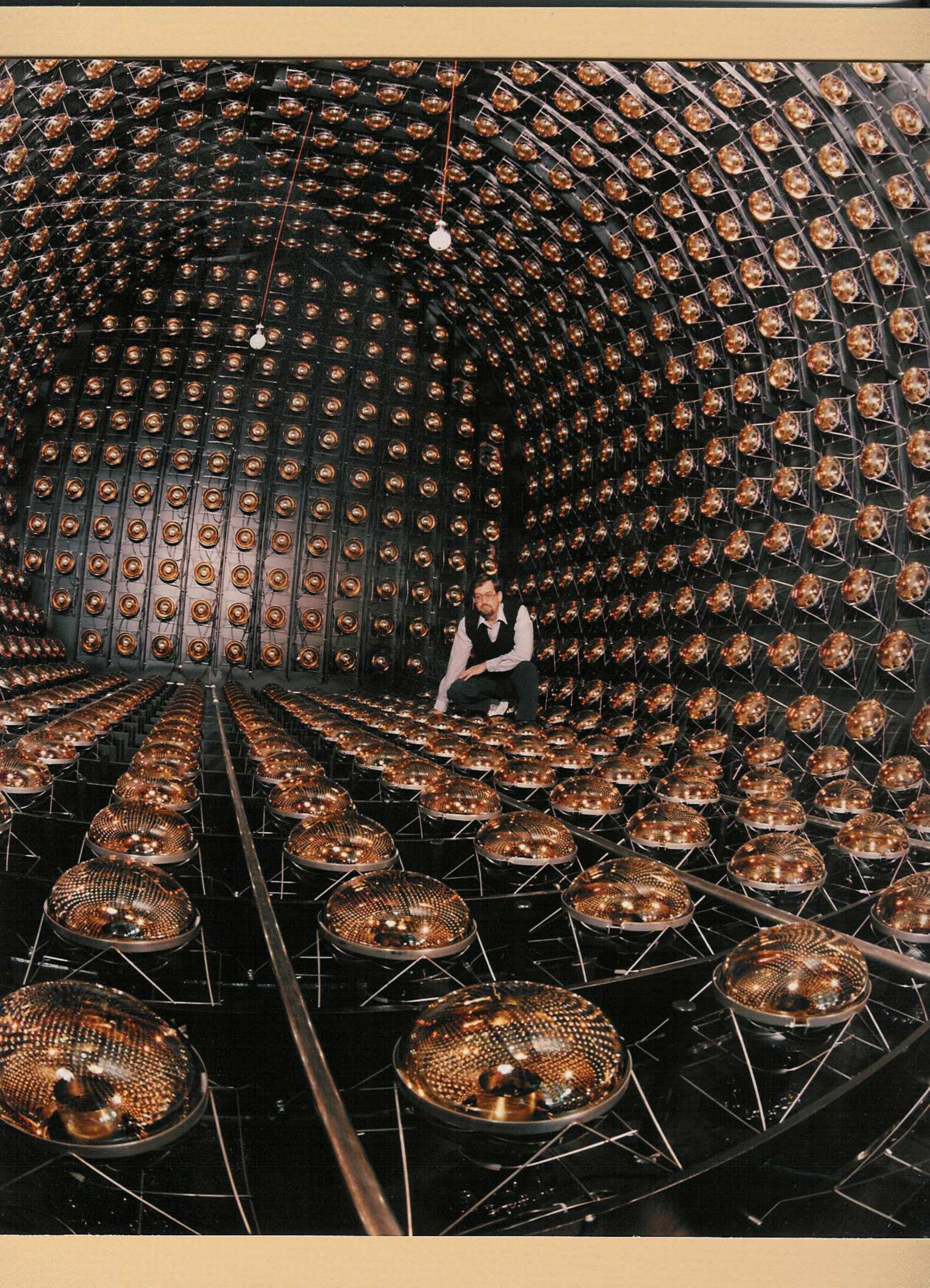}}
  \caption{A photograph of the inside of the LSND detector tank.}
  \label{fig:LSND_photo}
\end{figure}

The LSND experiment \cite{Athanassopoulos:1996ds} was designed to search for 
$\bar \nu_\mu \rightarrow \bar \nu_e$ oscillations with high sensitivity and to 
measure $\nu C$ cross sections.  A photograph of the inside of the detector tank 
is shown in Fig.~\ref{fig:LSND_photo}.  The main characteristics of the LSND 
experiment are given in Table~\ref{tab:compare}. LSND had the advantage of a 
very high proton intensity, a large detector mass, and good particle 
identification.  Also shown in Table~\ref{tab:compare} are the main 
characteristics of the KARMEN experiment, which had a lower duty factor than 
LSND and excellent energy resolution.  An important difference between the 
experiments is that in LSND the distance travelled by the neutrinos was 30~m, 
compared to 17.7~m for KARMEN, so that a combined analysis provides a more 
sensitive search for neutrino oscillations over a wider range of $\Delta m^2$ 
values. Both experiments made use of a high-intensity, 800~MeV proton beam 
that interacted in an absorber to produce a large number of pions. Most of the 
pions produced are $\pi^+$, which almost all decay. The $\pi^-$ are mainly 
absorbed and only a small fraction decay to $\mu^-$, which in turn are largely 
captured. Therefore, almost all of the neutrinos produced arise from $\pi^+ 
\rightarrow \mu^+ \nu_\mu$ and $\mu^+ \rightarrow e^+ \bar \nu_\mu \nu_e$ 
decays, where most of the decays ($>95\%$) are at rest and only a small 
fraction ($<5\%$) are in flight.

\begin{table}
  \begin{ruledtabular}
  \begin{tabular}{lcc}
    Property&LSND&KARMEN \\
    \colrule
    Proton Energy&798 MeV&800 MeV \\
    Proton Intensity&1000 $\mu$A&200 $\mu$A \\
    Protons on Target&28,896 C&9425 C \\
    Duty Factor&$6 \times 10^{-2}$&$1 \times 10^{-5}$ \\
    Total Mass&167 t&56 t \\
    Neutrino Distance&30 m&17.7 m \\
    Particle Identification&YES&NO \\
    Energy Resolution at 50 MeV&6.6\%&1.6\% \\
    Events for 100\% $\bar \nu_\mu \rightarrow \bar \nu_e$ Transmutation&33,300&
    14,000 \\
  \end{tabular}
  \end{ruledtabular}
  \caption{The main characteristics of the LSND and KARMEN experiments.}
  \label{tab:compare}
\end{table}

\begin{figure}
  \includegraphics[height=5.in]{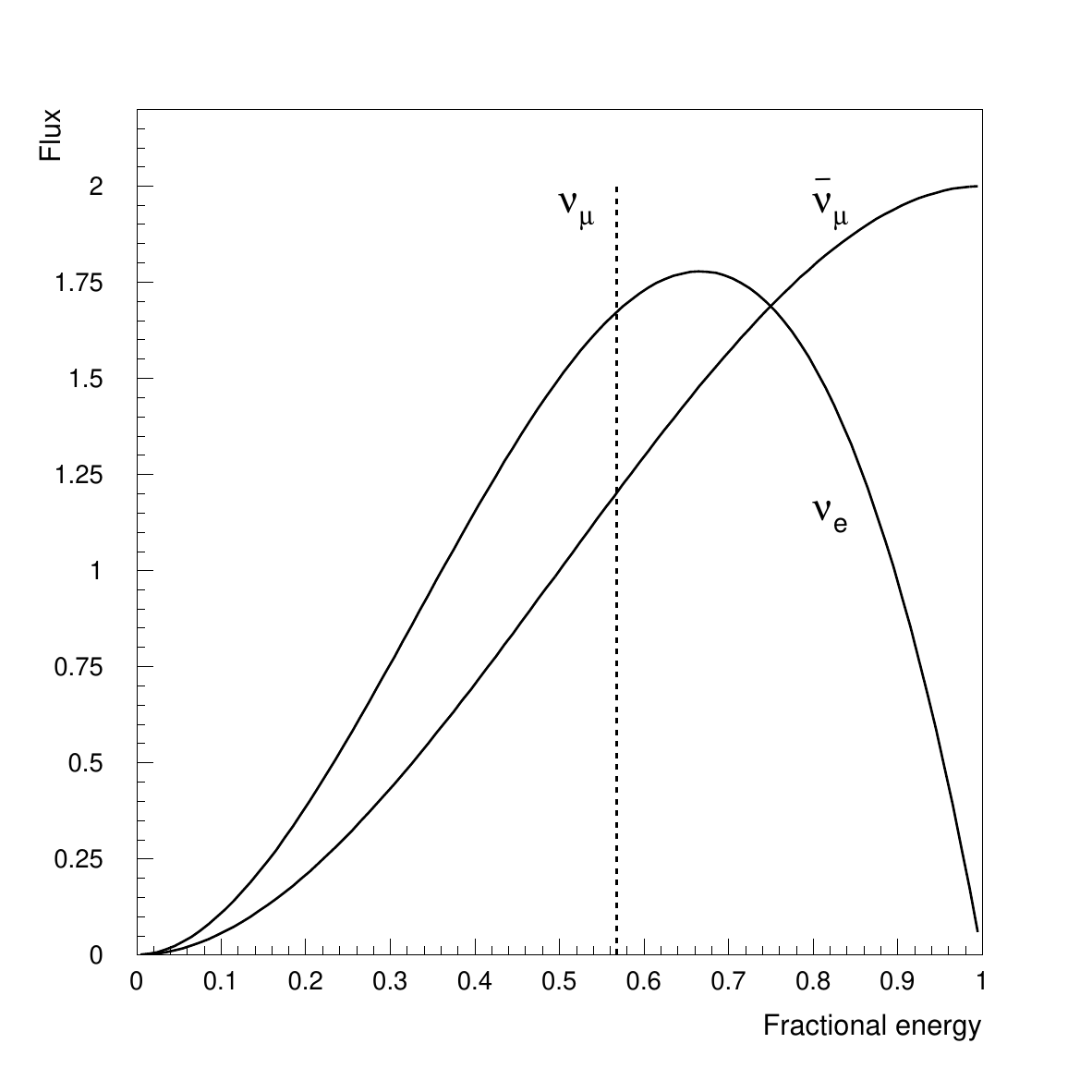}
  \vspace{-0.5cm}
  \caption{The neutrino energy spectra from $\pi^+$ and $\mu^+$ DAR.}
  \label{fig:dar_fluxes}
\end{figure}

LSND made use of the LAMPF accelerator, which was an intense source of low
energy neutrinos produced with a proton current of 1~mA at 798~MeV kinetic
energy.  For the 1993--1995 running period the production target consisted of a
30~cm long water target (20~cm in 1993) followed by a water-cooled Cu beam
dump, while for the 1996--1998 running period the production target was
reconfigured with the water target replaced by a close-packed, high-Z target.
The resulting DAR neutrino fluxes are well understood because almost all
detectable neutrinos arise from $\pi^+$ or $\mu^+$ decay; $\pi^-$ and $\mu^-$
that stop are readily captured in the Fe of the shielding and Cu of the beam
stop~\cite{Burman:1996gt}.  The production of kaons or heavier mesons is
negligible at these proton energies.  The $\bar \nu_e$ flux is calculated to be
only $\sim 8 \times 10^{-4}$ times as large as the $\bar \nu_{\mu}$ flux in the
$20 < E_{\nu} < 52.8$~MeV energy range, so that the observation of a $\bar
\nu_e$ event rate significantly above the calculated background would be
evidence for $\bar \nu_{\mu} \rightarrow \bar \nu_e$ oscillations.
Fig.~\ref{fig:dar_fluxes} shows the neutrino energy spectra from $\pi^+$ and
$\mu^+$ DAR.

\begin{figure}
  \includegraphics[height=3.in]{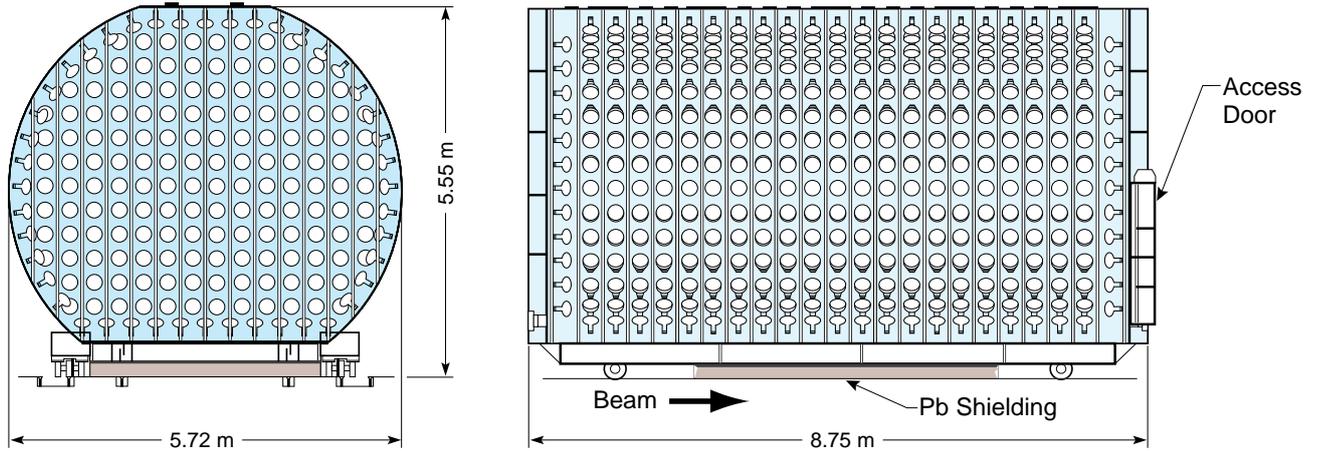}
  \vspace{-1.2cm}
  \caption{A schematic drawing of the LSND detector.}
  \label{fig:detector}
\end{figure}

The LSND detector \cite{Athanassopoulos:1996ds} consisted of an approximately cylindrical
tank 8.3~m long by 5.7~m in diameter. A schematic drawing of the detector is
shown in Fig.~\ref{fig:detector}.  The center of the detector was located 30~m
from the neutrino source.  On the inside surface of the tank, 1220 8-inch
Hamamatsu PMTs covered 25\% of the area with photocathode.  The tank was filled
with 167~t of liquid scintillator consisting of mineral oil and 0.031~g/l of
b-PBD.  This low scintillator concentration allows the detection of both
Cherenkov light and scintillation light and yields an attenuation length of
more than 20~ m for wavelengths greater than 400~nm~\cite{Reeder:1993ff}.  A typical
45~MeV electron created in the detector produced a total of $\sim 1500$
photoelectrons, of which $\sim 280$ photoelectrons were in the Cherenkov cone.
PMT time and pulse-height signals were used to reconstruct the track with an
average RMS position resolution of $\sim 14$~cm, an angular resolution of $\sim
12^\circ$ , and an energy resolution of $\sim 7\%$ at the Michel endpoint of
52.8~MeV.  The Cherenkov cone for relativistic particles and the time
distribution of the light, which is broader for non-relativistic particles
\cite{Athanassopoulos:1996ds}, gave excellent separation between electrons and particles
below Cherenkov threshold.  Identification of neutrons was accomplished through
the detection of the 2.2~MeV photon from neutron capture on a free proton.  The
veto shield enclosed the detector on all sides except the bottom.  Additional
counters were placed below the veto shield after the 1993 run to reduce
cosmic-ray background entering through the bottom support structure.  The main
veto shield~\cite{Napolitano:1989aj} consisted of a 15~cm layer of liquid scintillator 
in an external tank and 15~cm of lead shot in an internal tank.  This combination of
active and passive shielding tagged cosmic-ray muons that stopped in the lead
shot.  A veto inefficiency $<10^{-5}$ was achieved for incident charged
particles.

\subsubsection*{Event Selection}

The goal of the event selection is to reduce the cosmic-ray background to as
low a level as possible, while retaining a high efficiency for neutrino-induced
electron events.  The selection criteria and corresponding efficiencies are
shown in Table~\ref{tab:electron_selection}.  The energy range $20<E<200$~MeV
is chosen so as to accept both decay-at rest (DAR) $\bar\nu_\mu \rightarrow
\bar\nu_e$ and decay-in-flight (DIF) $\nu_\mu \rightarrow \nu_e$ oscillation
candidates.  The energy region $20<E_e<60$~MeV is required for the $\bar
\nu_\mu \rightarrow \bar \nu_e$ oscillation search and $60<E_e<200$~MeV for the
$\nu_\mu \rightarrow \nu_e$ oscillation search.  Below 20~MeV there are large
backgrounds from the $\beta$ decay of ${}^{12}B$ created by the capture of
stopped cosmic-ray $\mu^-$ on ${}^{12}C$.  Above 200~MeV the beam-related
backgrounds from $\pi^+ \rightarrow e^+\nu_e$ are large compared to any likely
oscillation signal.  Events with a previous activity within 12 $\mu$s, a future
activity within 8~$\mu$s, or a bottom veto counter hit are rejected in order to
eliminate cosmic-ray muon events.  To further minimize cosmic-ray background, a
tight electron particle identification is applied, $-1.5<\chi_{tot}^\prime <
0.5$, where the allowed range is chosen by maximizing the selection efficiency
divided by the square root of the beam-off background with a correlated
neutron.  The $\chi_{tot}^\prime$ parameter depends on the product of the
$\chi$ parameters defined in reference~\cite{Athanassopoulos:1996ds}.  Briefly, $\chi_r$ and
$\chi_a$ are the quantities minimized for the determination of the event
position and direction, and $\chi_t$ is the fraction of PMT hits that occur
more than 12~ns after the fitted event time. The dependence of the $\chi$
parameters on energy and position for Michel electrons was studied, and a
correction was developed that made $\chi_{tot}^\prime$ independent of energy or
position.  Additionally, no veto hit is allowed within 30~ns of the trigger
time and the reconstructed electron vertex is required to be inside a volume
35~cm from the faces of the photomultiplier tubes. Finally, the number of
associated $\gamma$s with $R_\gamma >10$ ($R_\gamma$ is discussed below) is
required to be $<2$ ($<1$) for events $<60$~($>60$)~MeV in order to reject
neutron-induced events, which tend to have many associated $\gamma$s.  In
addition to the electron reduction and selection efficiencies,
Table~\ref{tab:electron_selection} also shows the efficiencies due to the data
acquisition (DAQ) and veto deadtime. The total efficiency for electrons in the
fiducial volume with energies in the range $20 <E_e<60$ MeV is $0.42 \pm 0.03$.

\begin{table}
  \begin{ruledtabular}
  \begin{tabular}{lr}
    Criteria & Efficiency \\
    \colrule
    \multicolumn{2}{c}{Electron Reduction}\\
    \colrule
    Veto Hits $< 4$&$0.98 \pm 0.01$ \\
    Loose Electron PID&$0.96 \pm 0.01$ \\
    Cosmic Muon Cut &$0.92 \pm 0.01$ \\
    \colrule
    \multicolumn{2}{c}{Electron Selection}\\
    \colrule
    $\Delta t_{past}>12\mu$s        &  $0.96 \pm 0.01$  \\
    $\Delta t_{future}>8\mu$s       &  $0.99 \pm 0.01$  \\
    $-1.5<\chi_{tot}^\prime<0.5$            &  $0.84 \pm 0.01$ \\
    $0.3<\chi^{old}_{tot}<0.65$  (1993 only)      &  $0.98 \pm 0.01$    \\
    $\Delta t^{best}_{veto}>30\mathrm{ns}$    &  $0.97 \pm 0.01$   \\
    D $>35$ cm & $0.88 \pm 0.02$      \\
    $N_\gamma <1$, $E>60$ & $1.00$    \\
    $N_\gamma <2$, $E<60$ & $1.00$ \\
    \colrule
    \multicolumn{2}{c}{Deadtime}\\
    \colrule
    DAQ \& Tape Deadtime & $0.96  \pm 0.02$ \\
    Veto Deadtime & $0.76 \pm 0.02$ \\
    \colrule
    Total&$0.42 \pm 0.03$   \\
  \end{tabular}
  \end{ruledtabular}
  \caption{The LSND average efficiencies for electrons in the fiducial volume
    with energies in the range $20 <E_e<60$ MeV.}
  \label{tab:electron_selection}
\end{table}

\begin{figure}
  \includegraphics[height=5.in]{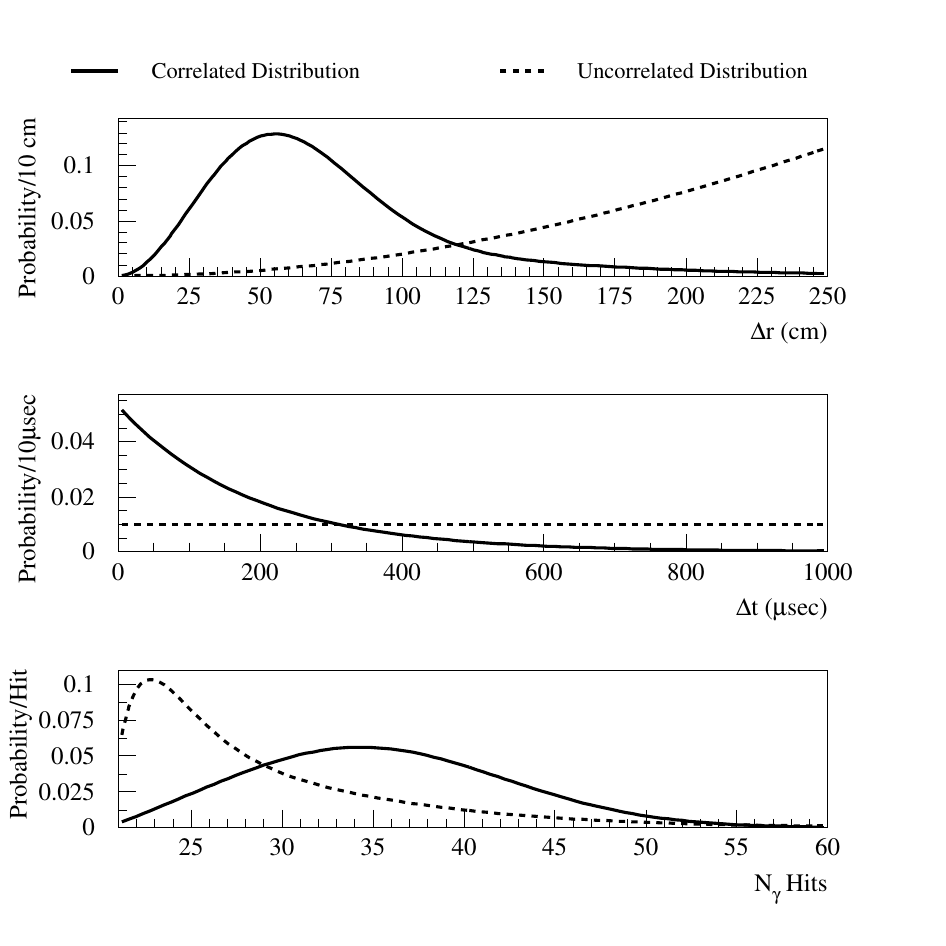}
  \vspace{-0.5cm}
  \caption{Distributions of correlated 2.2 MeV $\gamma$ (solid curves)
    and accidental $\gamma$ (dashed curves) from LSND. The top plot shows the
    distance between the reconstructed $\gamma$ position and positron position,
    $\Delta r$, the middle plot shows the time interval between the $\gamma$
    and positron, $\Delta t$, and the bottom plot shows the number of hit PMTs
    associated with the $\gamma$, $N_{hits}$.}
  \label{gamdist}
\end{figure}

Correlated 2.2~MeV $\gamma$ from neutron capture are distinguished from
accidental $\gamma$ from radioactivity by use of the likelihood ratio,
$R_\gamma$, which is defined to be the likelihood that the $\gamma$ is
correlated divided by the likelihood that the $\gamma$ is accidental.
$R_\gamma$ depends on three quantities: the number of hit PMTs associated with
the $\gamma$ (the multiplicity is proportional to the $\gamma$ energy), the
distance between the reconstructed $\gamma$ position and positron position, and
the time interval between the $\gamma$ and positron (neutrons have a capture
time in mineral oil of 186~$\mu$s, while the accidental $\gamma$ are uniform in
time). Fig.~\ref{gamdist} shows these distributions, which are obtained from
fits to the data, for both correlated 2.2 MeV~$\gamma$ (solid curves) and
accidental $\gamma$ (dashed curves). To determine $R_\gamma$, the product of
probabilities for the correlated distributions is formed and divided by the
product of probabilities for the uncorrelated distributions. The accidental
$\gamma$ efficiencies are measured from the laser-induced calibration events,
while the correlated $\gamma$ efficiencies are determined from the MC
simulation of the experiment. Similar results for the correlated $\gamma$
efficiencies are obtained from the cosmic-ray neutron events, whose high energy
gives them a slightly broader position distribution.  The systematic uncertainty
of these efficiencies is estimated to be $\pm 7\%$ of their values.  For
$R_\gamma >10$, the correlated and accidental efficiencies are 0.39 and 0.003,
respectively.

\subsubsection*{Neutrino Oscillation Signal and Background Reactions}

The primary oscillation search in LSND is for $\bar \nu_\mu \rightarrow \bar
\nu_e$ oscillations, where the $\bar \nu_\mu$ arise from $\mu^+$ DAR in the
beam stop and the $\bar \nu_e$ are identified through the reaction $\bar \nu_e
p \rightarrow e^+ n$.  This reaction allows a two-fold signature of a positron
with a 52.8~MeV endpoint and a correlated 2.2~MeV $\gamma$ from neutron capture
on a free proton. There are only two significant neutrino backgrounds with a
positron/electron and a correlated neutron. The first background is from
$\mu^-$ DAR in the beam stop followed by $\bar \nu_e p \rightarrow e^+ n$
scattering in the detector. As mentioned earlier, this background is highly
suppressed due to the requirements that a $\pi^-$ be produced, the $\pi^-$ DIF,
and the $\mu^-$ DAR prior to capture.  The second background is from $\pi^-$
DIF in the beam stop followed by $\bar \nu_\mu p \rightarrow \mu^+ n$
scattering in the detector.  (Additional contributions are from $\bar \nu_\mu C
\rightarrow \mu^+ n X$ and $\nu_\mu C \rightarrow \mu^- n X$ scattering.) This
background will mimic the oscillation reaction if the $\mu^+$ is sufficiently
low in energy that it is below the threshold of 18 hit PMTs, corresponding to
$E_\mu < 4$~MeV.  Table~\ref{sigback} shows the estimated number of events in
the $20<E_e<60$~MeV energy range satisfying the electron selection criteria for
100\% $\bar \nu_\mu \rightarrow \bar \nu_e$ transmutation and for the two
beam-related backgrounds with neutrons.

\begin{table}
  \begin{ruledtabular}
  \begin{tabular}{ccc}
    Neutrino Source&Reaction&Number of Events \\
    \colrule
    $\mu^+$ DAR&100\% $\bar \nu_\mu \rightarrow \bar \nu_e$&$33300 \pm 3300$ \\
    $\mu^-$ DAR&$\bar \nu_e p \rightarrow e^+ n$&$19.5 \pm 3.9$ \\
    $\pi^-$ DIF&$\bar \nu_\mu p \rightarrow \mu^+ n$&$10.5 \pm 4.6$ \\
  \end{tabular}
  \end{ruledtabular}
  \caption{The LSND estimated number of events in the $20<E_e<60$ MeV energy
    range due to 100\% $\bar \nu_\mu \rightarrow \bar \nu_e$ transmutation and to
    the two beam-related backgrounds with neutrons, $\mu^-$ DAR in the beam stop
    followed by $\bar \nu_e p \rightarrow e^+ n$ scattering in the detector and
    $\pi^-$ DIF in the beam stop followed by $\bar \nu_\mu p \rightarrow \mu^+ n$
    scattering.  The events must satisfy the electron selection criteria, but no
    correlated $\gamma$ requirement is imposed.}
\label{sigback}
\end{table}

\subsubsection*{LSND Oscillation Results}

\begin{figure}
  \centerline{\includegraphics[height=4.in]{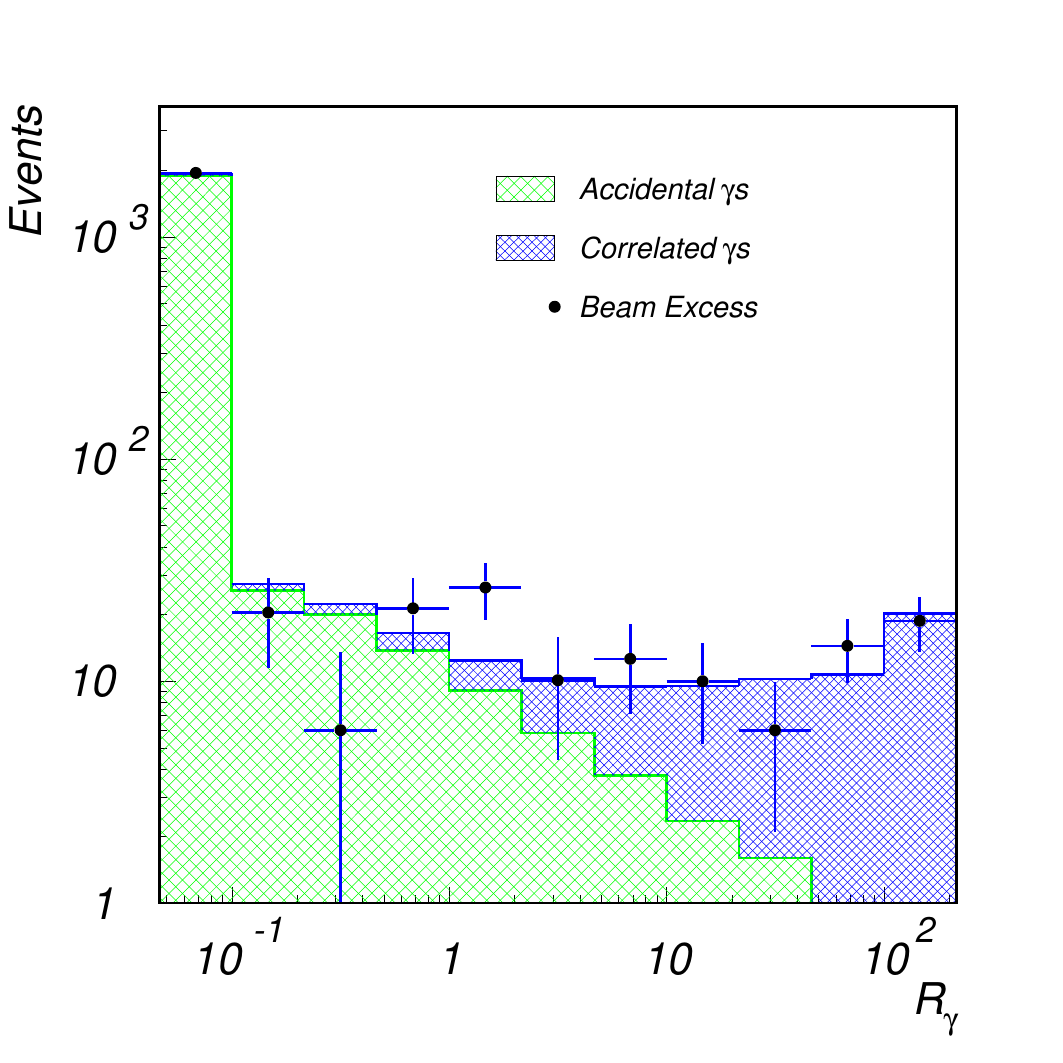}}
  \vspace{-0.5cm}
  \caption{The LSND $R_\gamma$ distribution for events that satisfy the
    selection criteria for the primary $\bar \nu_\mu \rightarrow \bar \nu_e$
    oscillation search.}
  \label{r2_dar}
\end{figure}

Table~\ref{tab:stats} shows the LSND statistics for events that satisfy the
selection criteria for the primary $\bar \nu_\mu \rightarrow \bar \nu_e$
oscillation search. An excess of events is observed over that expected from
beam-off and neutrino background that is consistent with neutrino oscillations.
A $\chi^2$ fit to the $R_\gamma$ distribution, as shown in Fig.~\ref{r2_dar},
gives $f_c = 0.0567 \pm 0.0108$ ($\chi^2 = 10.7/9$~DOF), which leads to a beam
on--off excess of $117.9 \pm 22.4$ events with a correlated neutron.
Subtracting the neutrino background from $\mu^-$ DAR followed by $\bar \nu_e p
\rightarrow e^+ n$ scattering ($19.5 \pm 3.9$ events) and $\pi^-$ DIF followed
by $\bar \nu_\mu p \rightarrow \mu^+ n$ scattering ($10.5 \pm 4.6$ events)
\footnote{This background also includes contributions from$\bar \nu_\mu C 
\rightarrow \mu^+ n X$ and $\nu_\mu C \rightarrow \mu^- n X$.} leads to a total 
excess of $87.9 \pm 22.4 \pm 6.0$ events.  This excess corresponds to an 
oscillation probability of $(0.264 \pm 0.067 \pm 0.045)\%$, where the first 
error is statistical and the second error is the systematic error arising from 
uncertainties in the backgrounds, neutrino flux (7\%), $e^+$ efficiency (7\%), 
and $\gamma$ efficiency (7\%).

\begin{table}
  \begin{ruledtabular}
  \begin{tabular}{ccccc}
    Selection &Beam-On Events&Beam-Off Background&$\nu$ Background&
    Event Excess \\
    \colrule
    $R_\gamma>1$ &205&$106.8\pm2.5$&$39.2 \pm 3.1$&$59.0\pm14.5 \pm 3.1$ \\
    $R_\gamma>10$ &86&$36.9\pm1.5$&$16.9 \pm 2.3$&$32.2\pm9.4 \pm 2.3$ \\
    $R_\gamma>100$ &27&$8.3\pm0.7$&$5.4 \pm 1.0$&$13.3\pm5.2 \pm 1.0$ \\
  \end{tabular}
  \end{ruledtabular}
  \caption{Numbers of LSND beam-on events that satisfy the selection criteria
    for the primary $\bar \nu_\mu \rightarrow \bar \nu_e$ oscillation search with
    $R_\gamma>1$, $R_\gamma>10$, and $R_\gamma>100$. Also shown are the beam-off
    background, the estimated neutrino background, and the excess of events that
    is consistent with neutrino oscillations.}
  \label{tab:stats}
\end{table}

\begin{figure}
  \centerline{\includegraphics[height=4.in]{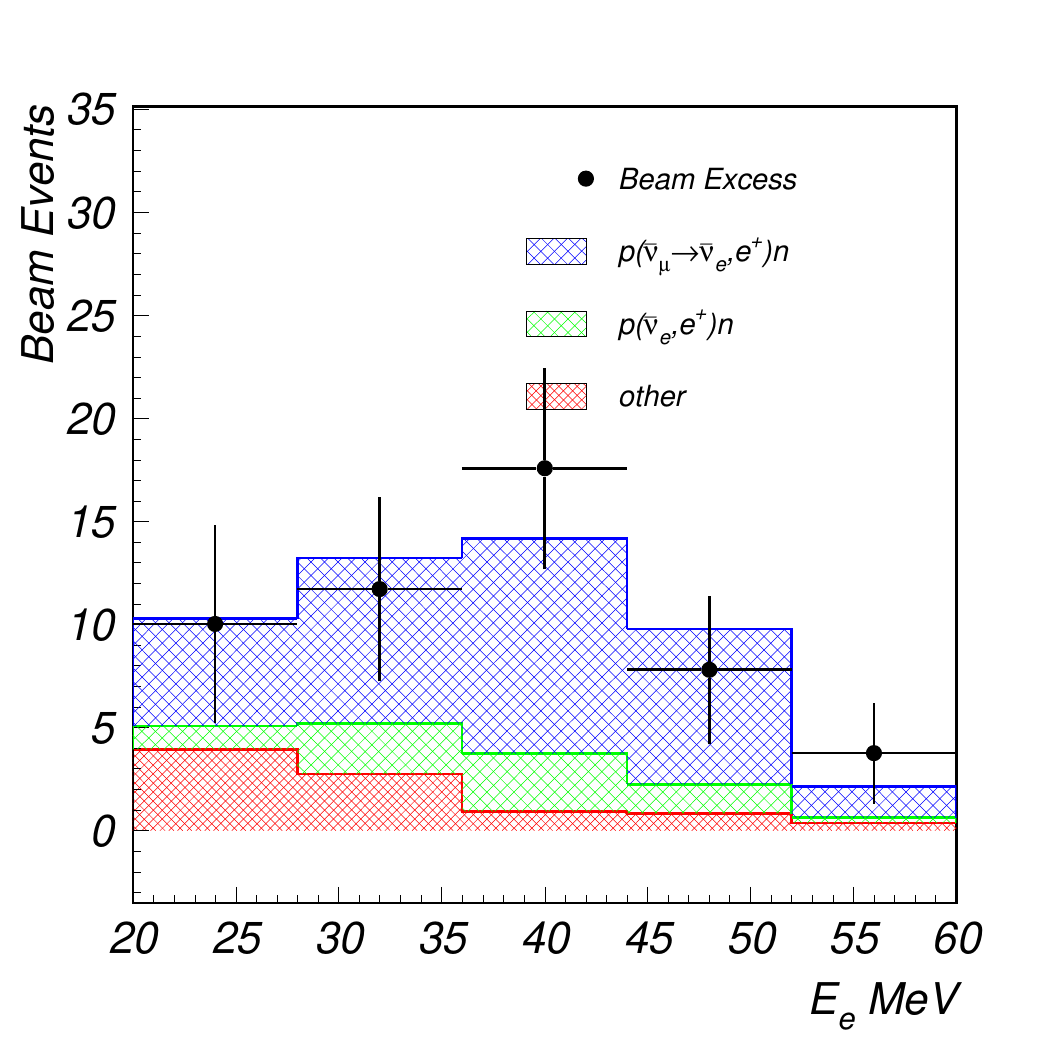}}
  \vspace{-0.5cm}
  \caption{The LSND energy distribution for events with $R_\gamma >10$.  The
    shaded region shows the expected distribution from a combination of neutrino
    background plus neutrino oscillations at low $\Delta m^2$.}
  \label{fig:elec_en_rcut}
\end{figure}

\begin{figure}
  \centerline{\includegraphics[height=4.in]{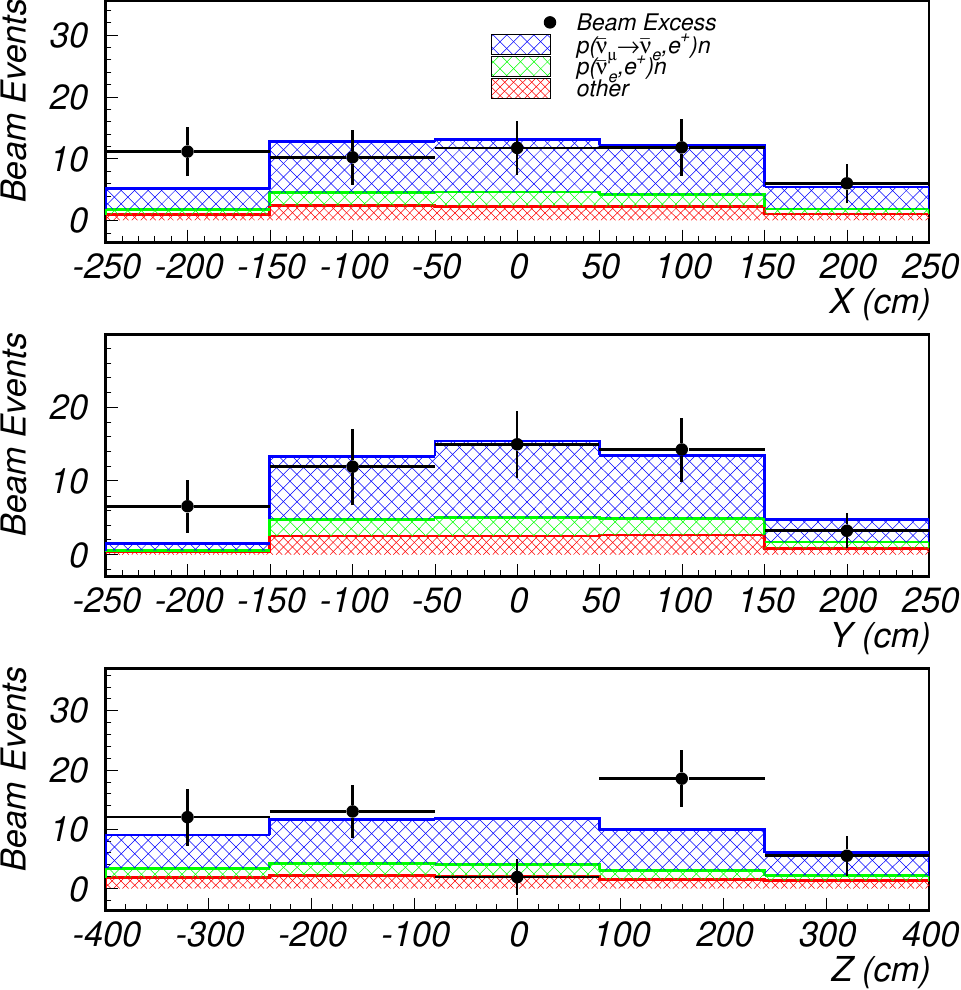}}
  \caption{The LSND spatial distributions for events with $R_\gamma >10$ and
    $20<E_e<60$~MeV.  The shaded region shows the expected distribution from a
    combination of neutrino background plus neutrino oscillations at low $\Delta
    m^2$.}
  \label{spatial}
\end{figure}

\begin{figure}
  \centerline{\includegraphics[height=4.in]{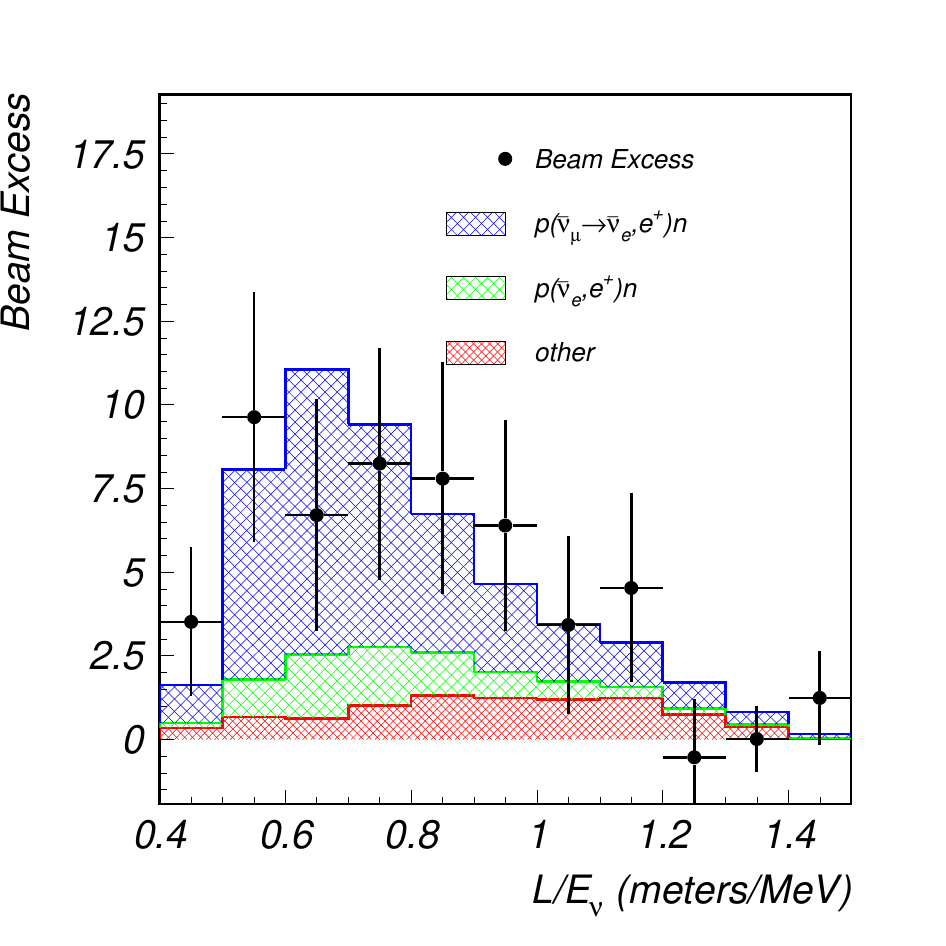}}
  \vspace{-0.5cm}
  \caption{The LSND $L_\nu/E_\nu$ distribution for events with $R_\gamma >10$
    and $20<E_e<60$~MeV, where $L_\nu$ is the distance travelled by the neutrino
    in meters and $E_\nu$ is the neutrino energy in MeV. The data agree well with
    the expectation from neutrino background and neutrino oscillations at low
    $\Delta m^2$.}
  \label{fig:elec_loe_rcut}
\end{figure}

A clean sample of oscillation candidate events can be obtained by requiring
$R_\gamma >10$, where as shown in Table~\ref{tab:stats}, the beam on-off excess
is $49.1 \pm 9.4$ events while the estimated neutrino background is only $16.9
\pm 2.3$ events.  Fig.~\ref{fig:elec_en_rcut} displays the energy distribution
of events with $R_\gamma >10$. The shaded regions show the combination of
neutrino background plus neutrino oscillations at low $\Delta m^2$. The data
agree well with the oscillation hypothesis.  Fig.~\ref{spatial} shows the
spatial distribution for events with $R_\gamma >10$ and $20<E_e<60$~MeV, where
z is along the axis of the tank (and approximately along the beam direction), y
is vertical, and x is transverse.  The shaded region shows the expected
distribution from a combination of neutrino background plus neutrino
oscillations.  Finally, Fig.~\ref{fig:elec_loe_rcut} shows the $L_\nu/E_\nu$
distribution for events with $R_\gamma >10$ and $20<E_e<60$~MeV, where $L_\nu$
is the distance travelled by the neutrino in meters and $E_\nu$ is the neutrino
energy in MeV determined from the measured positron energy and angle with
respect to the neutrino beam. The data agree well with the expectation from
neutrino background plus neutrino oscillations at low $\Delta m^2$
($\chi^2/\text{dof} = 4.9/8$) or high $\Delta m^2$ ($\chi^2/\text{dof} = 
5.8/8$).

The $(\sin^22\theta,\Delta m^2)$ likelihood ($\mathcal{L}$) fitter is applied
to beam-on events in the final oscillation sample and calculates a likelihood
in the $(\sin^22\theta,\Delta m^2)$ plane in order to extract the favored
oscillation parameters.  The $\mathcal{L}$ product in the
$(\sin^22\theta,\Delta m^2)$ plane is formed over the individual beam-on events
that pass the oscillation cuts.  This three-dimensional contour is sliced to
arrive finally at the LSND allowed oscillation region.  The beam-related
backgrounds are determined from Monte Carlo (MC) event samples for each
individual background contribution. The MC contains the trigger simulation and
generally very well reproduces the tank response to all particles of interest.
Agreement between the data and MC is excellent.

\begin{figure}
  \centerline{\includegraphics[height=4.in]{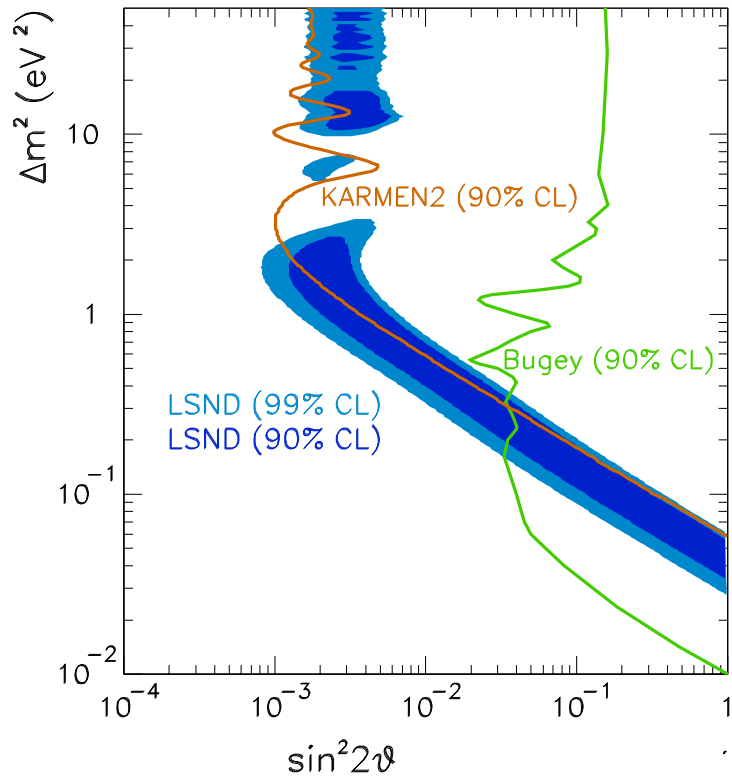}}
  \caption{The $(\sin^22\theta,\Delta m^2)$ oscillation parameter fit for the
    entire LSND data sample, $20<E_e<200$~MeV.  The inner and outer regions
    correspond to 90\% and 99\%~CL allowed regions, while the curves are 90\%~CL
    limits from the Bugey reactor experiment and the KARMEN experiment at ISIS.}
  \label{fit}
\end{figure}

The $(\sin^22\theta,\Delta m^2)$ oscillation parameter fit for the entire data
sample, $20<E_e<200$~MeV, is shown in Fig.~\ref{fit}. The fit includes both
$\bar \nu_\mu \rightarrow \bar \nu_e$ and $\nu_\mu \rightarrow \nu_e$
oscillations, as well as all known neutrino backgrounds. The inner and outer
regions correspond to 90\% and 99\%~CL allowed regions, while the curves are
90\%~CL limits from the Bugey reactor experiment \cite{Declais:1994su} and the KARMEN
experiment at ISIS \cite{Armbruster:2002mp} (see below).  The most favored allowed region is the
band from $0.2 - 2.0$~eV$^2$, although a region around $7$~eV$^2$ is also
possible.

\subsection{The KARMEN Constraint}

\subsubsection*{Description of the Experiment}

The KARMEN experiment \cite{Gemmeke:1990ix} made use of the ISIS rapid-cycling
synchrotron, which accelerates protons up to 800~MeV at an intensity of 200~$\mu$A.
The protons are extracted from the synchrotron at a frequency of 50~Hz
as a double pulse consisting of two 100~ns pulses separated by 325~ns. The two
bursts, therefore, occur within 600~ns and lead to an overall duty factor of
about $10^{-5}$. After extraction, the protons interact in a water-cooled
Ta-D$_2$O target, producing about $(0.0448 \pm 0.0030)$ $\pi^+$ per incident
proton~\cite{Burman:1996gt}. Due to the small duty factor, $\nu_\mu$ from $\pi^+$
decay can be clearly separated from the $\bar \nu_\mu$ and $\nu_e$ from $\mu^+$
decay. The $\bar \nu_e / \bar \nu_\mu$ background is estimated to be $6.4
\times 10^{-4}$ \cite{Burman:1996gt}, slightly smaller than for LSND.

\begin{figure}
  \centerline{\includegraphics[width=\textwidth]{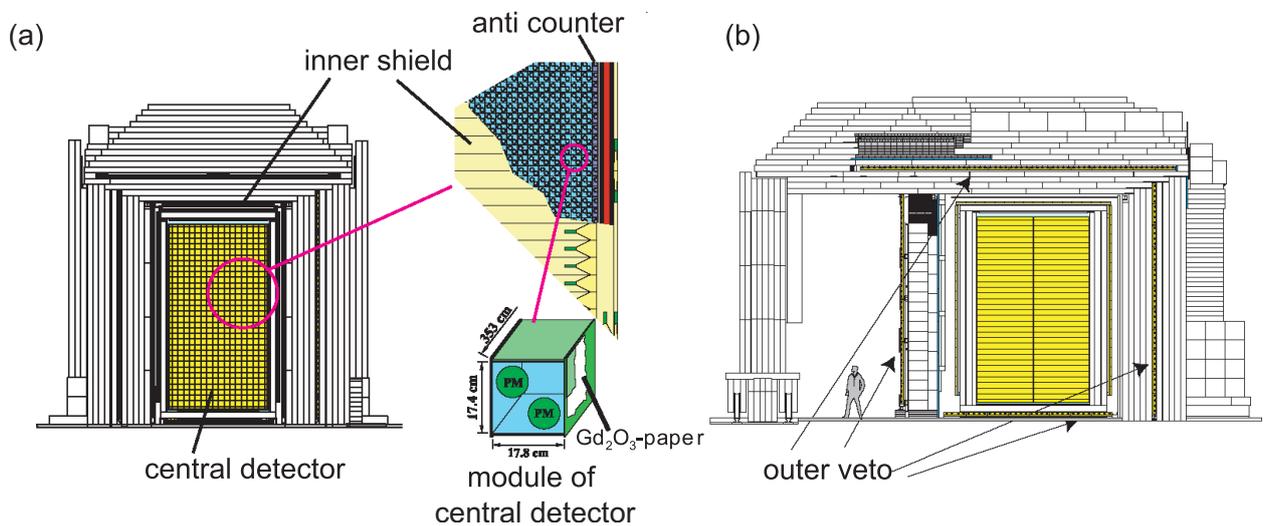}}
  \caption{(a) Front view of the KARMEN detector showing details of the central detector region.
  (b) Side view of the detector.}
  \label{fig:karmen_detector}
\end{figure}

The KARMEN detector, as shown in Fig.~\ref{fig:karmen_detector}, is a segmented
liquid scintillator calorimeter with 608 modules and a total mass of 56~t. The
liquid scintillator is made of paraffin oil (75\%~vol.), pseudocumene
(25\%~vol.), and PMP (2~g/l). The modules are read-out by pairs of 3-inch PMTs
and are enclosed by a tank with outside dimensions 3.53~m$\ \times\ $3.20~m$\
\times\ $5.96~m.  Excellent energy resolution is obtained for electrons
produced inside the detector and can be parametrized by $\sigma_E =
11.5\%/\sqrt{E / \text{MeV}}$.  Gadolinium-coated paper was inserted between
the modules for the detection of thermal neutrons.  The detector is enclosed by
a multilayer active veto system and 7000~t of steel shielding and is located
17.7~m from the neutrino source at an angle of $100^\circ$ to the incident
proton beam direction.

\subsubsection*{Event Selection}

KARMEN searches for $\bar \nu_\mu \rightarrow \bar \nu_e$ oscillations in the
same way as in LSND by looking for the reaction $\bar \nu_e p \rightarrow e^+
n$, which gives a two-fold signature of a $e^+$ followed by one or more
$\gamma$ from neutron capture.  The neutrino oscillation event selection and
corresponding efficiencies are summarized in Table~\ref{karmen_selection}. For
the $e^+$ candidate it is required that there be no previous activity in the
detector and veto, that the $e^+$ occur from 0.6~$\mu$s to 10.6~$\mu$s after
the beam spill, and that the $e^+$ energy be in the range from 16~MeV to 50~MeV.
For the $\gamma$ candidate it is required that the $\gamma$ occur from 5~$\mu$s
to 300~$\mu$s after the $e^+$, that the $\gamma$ have an energy less than
8.0~MeV, and that the $\gamma$ be reconstructed within a coincidence volume of
1.3~m$^3$. The total efficiency for the two-fold signature is estimated to be
$0.192 \pm 0.0145$.

\begin{table}
  \begin{ruledtabular}
  \begin{tabular}{ccc}
    Event& Selection&Efficiency \\
    \colrule
    $e^+$&no previous activity&0.709 \\
    $e^+$&$0.6 < t_{pr} < 300 \mu$s&0.840 \\
    $e^+$&$16 < E_{pr} < 50$ MeV&0.775 \\
    \colrule
    $(n,\gamma)$&$5 < \Delta t < 300 \mu$s& \\
    $(n,\gamma)$&$E_{del} < 8.0$ MeV&0.416 \\
    $(n,\gamma)$&$V_c = 1.3 m^3$& \\
  \end{tabular}
\end{ruledtabular}

  \caption{The KARMEN event selection and corresponding efficiencies for the
    $\bar \nu_\mu \rightarrow \bar \nu_e$ oscillation search.}
  \label{karmen_selection}
\end{table}

\subsubsection*{Neutrino Oscillation Signal and Background Reactions}

Table~\ref{karmen_sigback} shows the estimated number of events in the
$16 < E_e< 50$~MeV energy range for 100\% $\bar \nu_\mu \rightarrow \bar \nu_e$
transmutation. Also shown are the number of events for the four backgrounds
with apparent neutrons. The first background is the cosmic-induced background,
which is well measured from data collected with the beam off. The second
background is due to $\nu_e C \rightarrow e^- N_{gs}$, where the $N_{gs}$
$\beta$ decay mimics the $\gamma$ from neutron capture. The third background is
due to normal $\nu_e C \rightarrow e^- N^*$ inclusive interactions with an
accidental $\gamma$ coincidence. The final background is due to the intrinsic
$\bar \nu_e$ contamination in the beam from $\mu^-$ DAR. The total background
is estimated to be $15.8 \pm 0.5$ events.

\begin{table}
  \begin{ruledtabular}
  \begin{tabular}{cc}
    Process&Number of Events \\
    \colrule
    100\% $\bar \nu_\mu \rightarrow \bar \nu_e$&$5826 \pm 534$ \\
    Cosmic-induced Background&$3.9 \pm 0.2$ \\
    Charged-current Coincidences&$5.1 \pm 0.2$ \\
    $\nu_e$-induced Random Coincidences&$4.8 \pm 0.3$ \\
    Intrinsic $\bar \nu_e$ Background&$2.0 \pm 0.2$ \\
  \end{tabular}
  \end{ruledtabular}
  \caption{The KARMEN estimated number of events in the $16<E_e<50$~MeV energy
    range due to 100\% $\bar \nu_\mu \rightarrow \bar \nu_e$ transmutation and to
    the four backgrounds with apparent neutrons.}
  \label{karmen_sigback}
\end{table}

\subsubsection*{KARMEN Neutrino Oscillation Results}

\begin{figure}
  \centerline{\includegraphics[height=5.in]{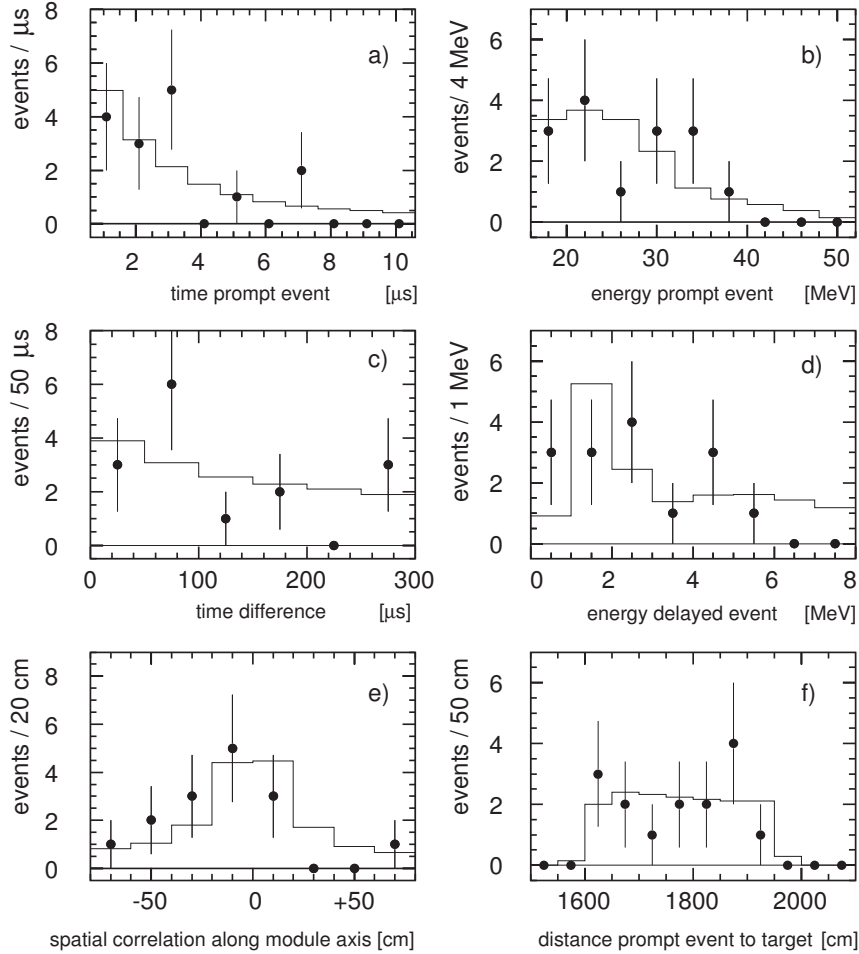}}
  \caption{The energy, time, and spatial distributions for the events observed
    by the KARMEN experiment: (a) time of prompt events, (b) energy of prompt
    events, (c) time difference between prompt and delayed events, (d) energy of
    delayed events, (e) spatial correlation, and (f) distance to target of prompt
    events. The 15 oscillation candidate events are in good agreement with the
    background expectation of 15.8 events (solid line).}
  \label{fig:karmen_distrib}
\end{figure}

KARMEN observes 15 events that pass the selection criteria discussed above,
which is consistent with the estimated background of $15.8 \pm 0.5$ events. The
energy, time, and spatial distributions for the 15 events are shown in
Fig.~\ref{fig:karmen_distrib}. Also shown are the shapes of the expected
backgrounds, which are in good agreement with the data. A maximum likelihood
fit to the data is performed~\cite{Armbruster:2002mp} to obtain the 90\% C.L. limits,
as shown in Fig.~\ref{fit}. The LSND oscillation region with $\Delta m^2 >
10$~eV$^2$ is ruled-out by the KARMEN data; however, in the regions $<
2$~eV$^2$ and around 7~eV$^2$ the KARMEN result is compatible with the LSND
oscillation evidence.

\subsection{Joint Analysis of LSND and KARMEN Data}

\begin{figure}
  \centerline{\includegraphics[height=2.5in]{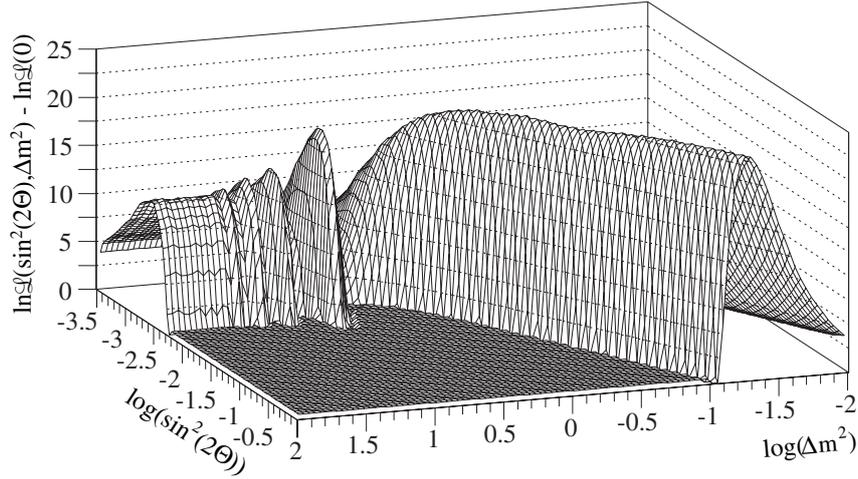}}
  \caption{The combined LSND and KARMEN log-likelihood function in terms
    of $\sin^22\theta$ and $\Delta m^2$.}
  \label{lik_joint}
\end{figure}

\begin{figure}
  \centerline{\includegraphics[height=5.in]{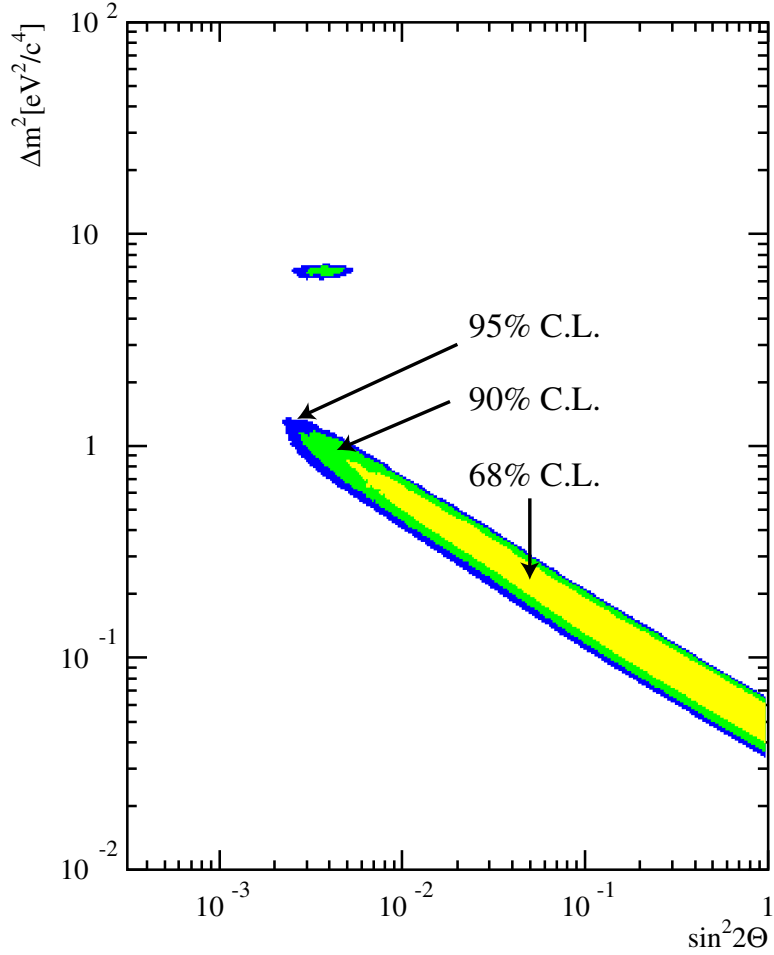}}
  \caption{The confidence regions of the oscillation parameters for the
    combined likelihood analysis, assuming statistical compatibility
    of LSND and KARMEN.}
  \label{joint_plot}
\end{figure}

A joint analysis of the LSND and KARMEN experiments has been performed
\cite{Church:2002tc} that is based on a frequentist approach following the
suggestions of reference~\cite{Feldman:1997qc}. For both experiments, the data are
analyzed with a maximum likelihood analysis followed by the extraction of
confidence levels in a unified approach. The two experiments are found to be
incompatible at a level of combined confidence of 36\%. For the cases of
statistical compatibility, Fig.~\ref{lik_joint} shows the combined LSND and
KARMEN log-likelihood function in terms of $\sin^22\theta$ and $\Delta m^2$.
The maximum log-likelihood function occurs at $\sin^22\theta = 1$ and $\Delta
m^2 = 0.05$ eV$^2$, which is 21.5 units of log-likelihood above the no
oscillation hypothesis. Fig.~\ref{joint_plot} shows the confidence regions of
the oscillation parameters for the combined likelihood analysis, assuming
statistical compatibility of LSND and KARMEN. By combining the two experiments,
the solutions with $\Delta m^2 > 10$~eV$^2$ are excluded, and there remain
essentially two solutions: one with $\Delta m^2 < 1$~eV$^2$ and the other with
$\Delta m^2 \sim 7$~eV$^2$.

\subsection{Sterile Neutrino Analysis of Super-K}

This section describes the sterile neutrino analysis done using the Super-K I 
and II atmospheric data. For detailed description of the Super-K experiment and 
the atmospheric neutrino data, one can refer to Ref.~\cite{Nakayama:2004dp}.  First 
there is a comparison of the results of the two oscillation models $\nu_{\mu}\to\nu_{\tau}$ 
and $\nu_{\mu} \to \nu_s$. Then there is a study of how much sterile admixture can 
be allowed in the atmospheric neutrino oscillation based on a 2+2 model.

\subsubsection{Comparison of $\nu_{\mu}\to\nu_{\tau}$ and $\nu_{\mu}\to\nu_{s}$ Oscillations}

\paragraph{Signatures of $\nu_{\mu}\to\nu_{s}$ oscillation}
For the two competing models, $\nu_{\mu}\to\nu_{\tau}$ and $\nu_{\mu}\to\nu_{s}$,
though the second neutrino is not identified, there are some signatures which 
can be used to tell two oscillations apart.  First, sterile neutrinos
do not interact with matter at all so they do not even make any neutral
current signals in the detector; and second, because the potentials experienced
by $\nu_{\mu}$ and $\nu_{s}$ in matter are different, the matter effect
modifies the survival probabilities of muon neutrinos.

To utilize the first signature, one must identify the neutral current
events. For a water Cherenkov detector, neutral current events can be enriched 
by identifying the $\pi^{0}$ particles which are produced in neutral current 
single-pion neutrino reactions~\cite{Nakayama:2004dp}.  Unlike in 
Ref.~\cite{Nakayama:2004dp} tight cuts are not used in this analysis, in order 
to keep as many events as possible.  As will be see later, even without  
tight cuts, the neutral current enhanced selection defined below can be used to 
differentiate two models quite well.  For sub-GeV events, the cuts use to 
enhance neutral current events are:
\begin{itemize}
\item Multi-ring events: $\pi^{0}$ particles produced in neutral current
reactions produce two $e$-like rings.
\item The most energetic ring is $e$-like: this cut is chosen to get rid
of CC $\nu_{\mu}$ events.
\item $400\,{\mathrm MeV}<E_{vis}<1330\,{\mathrm MeV}$: this cut is chosen to 
preserve the directional information of the parent neutrinos.
\end{itemize}
For the multi-GeV samples, the events rejected by the enhanced the charged 
current likelihood selection developed for the standard Super-K 
analysis are used.  Ref.~\cite{Ashie:2005ik} provides detailed description
on the charged and neutral current event enhancement method.

\begin{figure}
\begin{centering}
\includegraphics[width=0.4\textwidth]{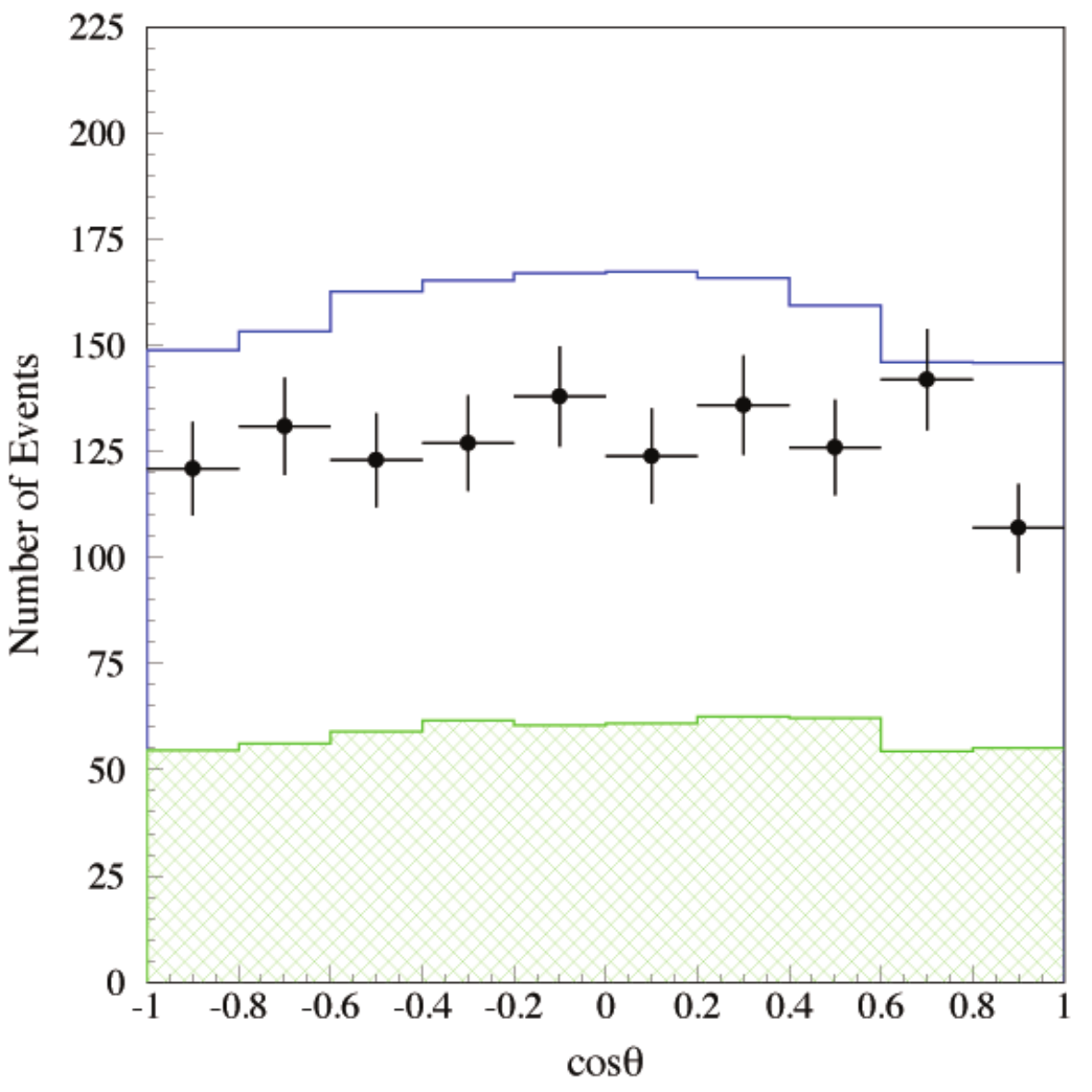}
\hspace{0.1\textwidth}
\includegraphics[width=0.4\textwidth]{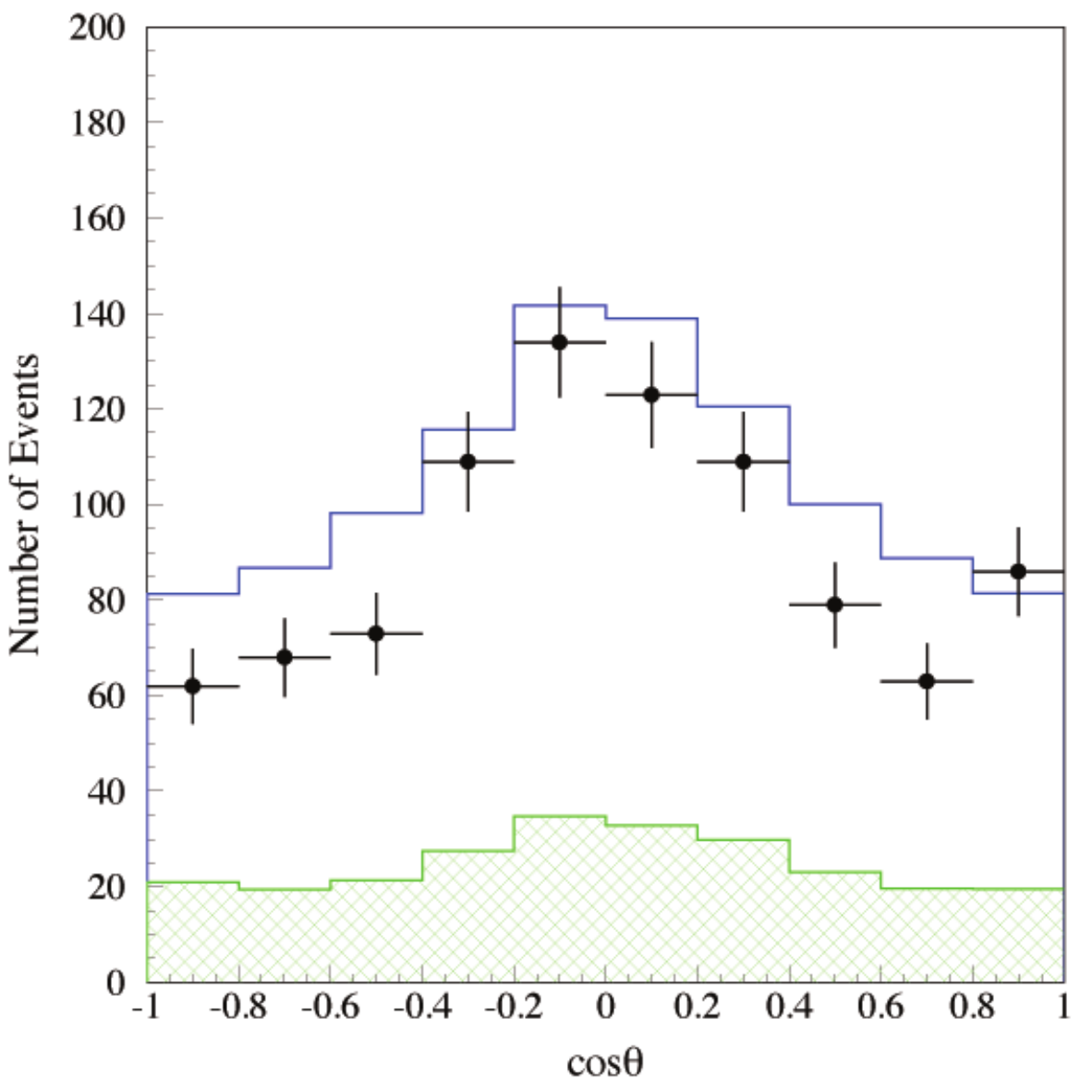}
\par\end{centering}

\caption{\label{fig:Neutral-current-enhanced}Zenith angle distributions of
neutral current enhanced samples. Left: sub-GeV; right: multi-GeV}
\end{figure}

The zenith distributions of neutral current enhanced samples selected by these 
cuts are shown in Fig.~\ref{fig:Neutral-current-enhanced}.
The hatched areas are neutral current events based on the Monte Carlo
simulation. The percentages of neutral current events and charge current
$\nu_{\mu}$ contaminations are shown in Table~\ref{tab:Portions-of-NC}.

\begin{table}
\caption{\label{tab:Portions-of-NC}Portions of NC and CC events in the NC
enhanced data samples}
\centering{}%
\begin{tabular}{|c|c|c|c|}
\hline 
 & NC events & CC $\nu_{\mu}$ & CC $\nu_{e}$\tabularnewline
\hline 
\hline 
Sub-GeV sample & 37\% & 22\% & 41\%\tabularnewline
\hline 
Multi-GeV sample & 24\% & 35\% & 41\%\tabularnewline
\hline 
\end{tabular}
\end{table}

The second signature of $\nu_{\mu}\to\nu_{s}$ oscillation is the matter
effect. It is known that in vacuum oscillation, the two-flavor survival
probability is:
$P_{survival}=1-\sin^22\theta_{atm}\sin^21.27\Delta m^2_{atm}({\mathrm eV}^2)L({\mathrm
km})/E({\mathrm GeV}).$
Where $\sin^22\theta_{atm}$ is the mixing angle term and $\Delta
m^2_{atm}$ is the mass-squared difference between the two mass
eigenstates. In matter with constant density, due to different forward
scattering potentials between $\nu_\mu/\bar{\nu_\mu}$ and $\nu_s$, the
mixing angle and the mass-squared difference are replaced by the
effective values in matter:
\begin{eqnarray}
\left\{ 
\begin{array}{l}
\sin^{2}2\theta_{M}=\frac{\sin^{2}2\theta}{(\cos2\theta-2\zeta/\Delta m^{2})^{2}+\sin^{2}2\theta}\\
\Delta m_{M}^{2}=\Delta m^{2}\sqrt{(\cos2\theta-2\zeta/\Delta
  m^{2})^{2}+\sin^{2}2\theta}
\end{array}\right..
\label{matter}
\end{eqnarray}
Where $\zeta=\mp\sqrt{2}G_{F}EN_{n}$ is the potential difference
between $\nu_{\mu}/\bar{\nu}_{\mu}$ and $\nu_{s}$. $G_{F}$ is the
Fermi constant, $E$ is the neutrino (or antineutrino) energy, and
$N_{n}$ is the number density of neutrons in matter.

\begin{figure}
\begin{centering}
\includegraphics[width=0.95\textwidth]{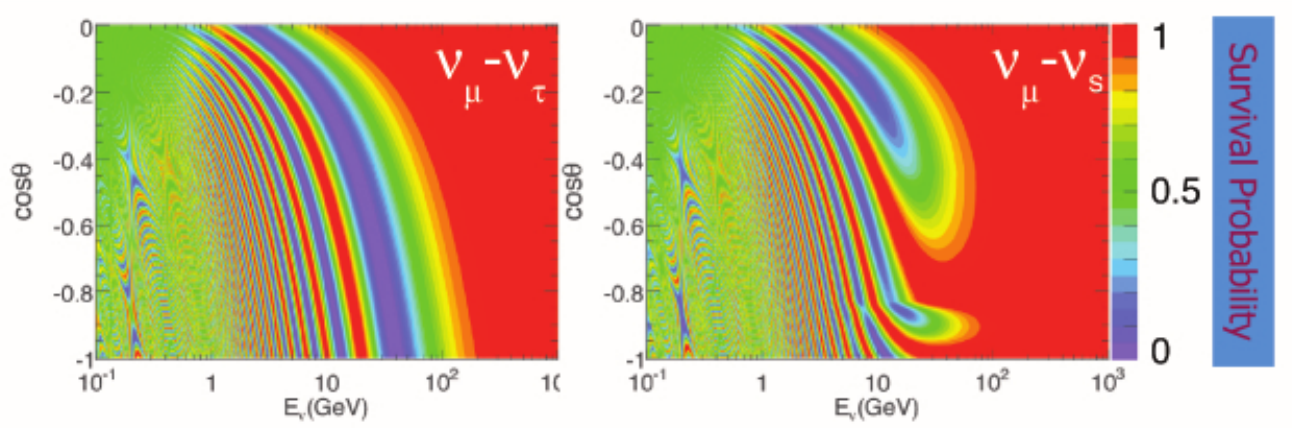}
\par\end{centering}

\caption{\label{fig:Survival-probabilities}Survival probabilities
of muon neutrinos crossing the Earth}
\end{figure}

The density of the Earth is not a constant~\cite{PREM:1981}, so for 
atmospheric neutrinos crossing the Earth, the two-flavor survival 
probability with modifications due to the matter effect as shown in 
Eq.~\ref{matter} is not valid any more.  However, one can divide the 
Earth into different sections along the trajectories of neutrinos 
and approximate each section with a constant density.
The final survival probability is calculated by diagonalizing the
Hamiltonian of each small section, the method is described in
Ref.~\cite{Barger:1980tf}.  Figure~\ref{fig:Survival-probabilities}
shows the survival probabilities of neutrinos crossing the Earth with
and without the matter effect using the density profile provided in
Ref.~\cite{PREM:1981}. As can be seen in
Fig.~\ref{fig:Survival-probabilities}, for atmospheric
neutrinos, the oscillation probability is suppressed by the
matter effect around and above 10~GeV.

\paragraph{Best-fits of two models}
These charged current and neutral current enhanced data samples can now
be used to do the oscillation analysis. Considering that the uncertainty
on the total number of multi-ring $e$-like events is relatively large,
the neutral current enhanced sample is given an independent flux normalization,
as was done for the charged current enhanced $e$-like events.
The sub-GeV neutral current events are treated as one energy bin and the
multi-GeV sample is divided into 4 energy bins based on $\log E_{vis}$
values since the parent neutrinos have wider energy spread. Both samples
are divided into 10 zenith angle bins based on the reconstructed event
direction.

\begin{figure}
\begin{centering}
\includegraphics[width=0.49\textwidth]{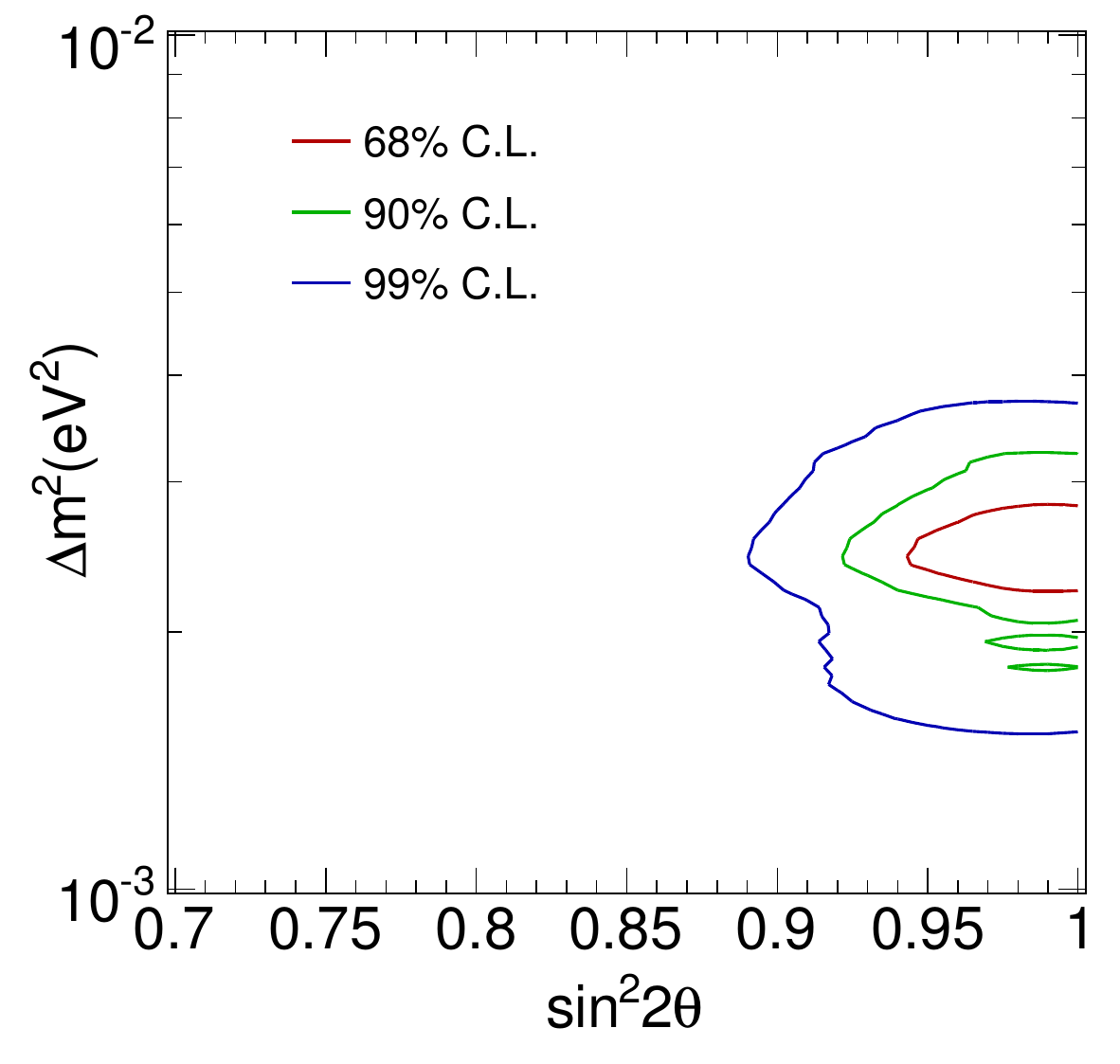}
\includegraphics[width=0.49\textwidth]{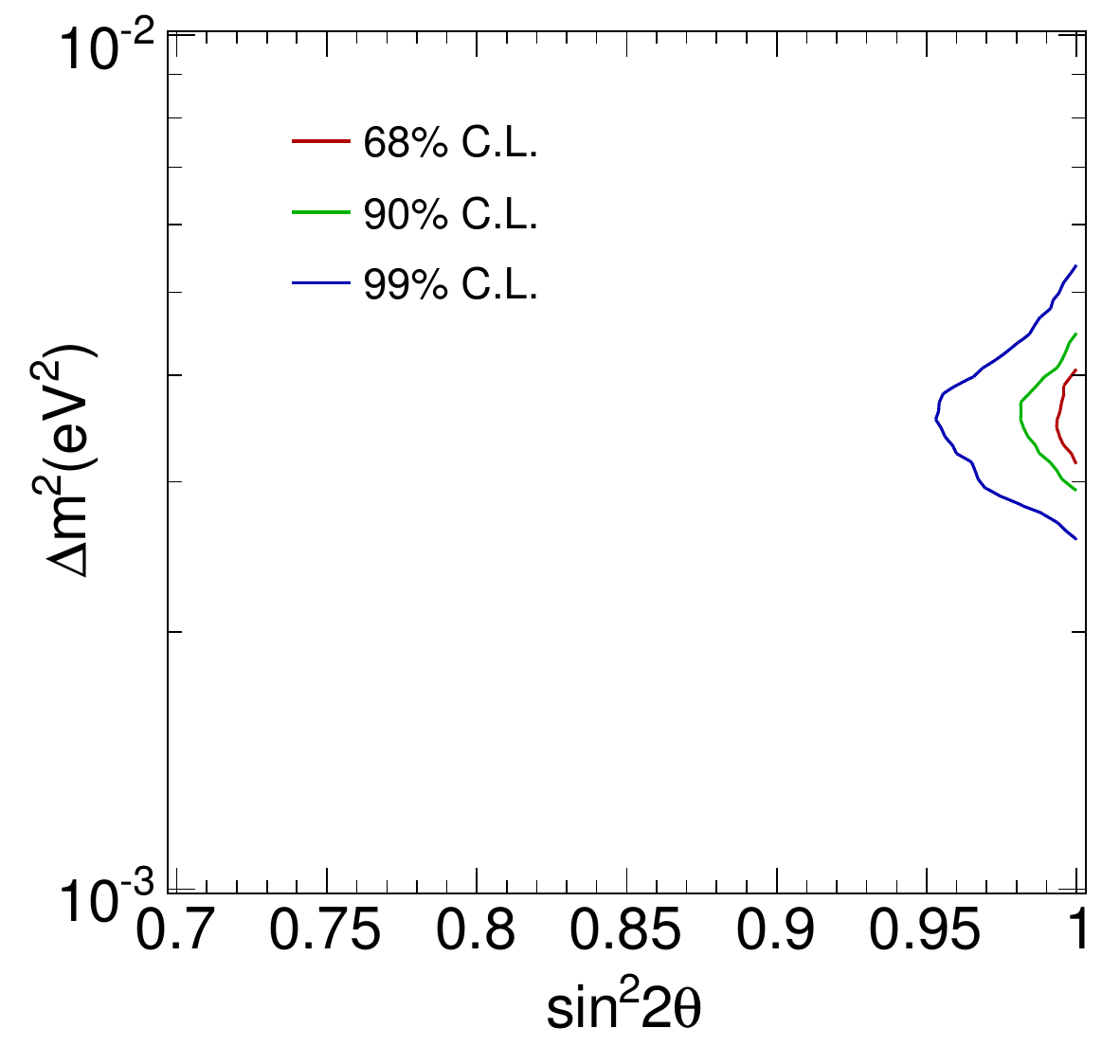}
\par\end{centering}

\caption{\label{fig:chisquare-contours-of-NC}$\chi^{2}$ contours of oscillations with NC
enhanced samples for (left) $\nu_{\mu}\to\nu_{\tau}$ (right) $\nu_{\mu}\to\nu_{s}$}
\end{figure}

Using the least chi-square method, a $\nu_{\mu}\to\nu_{\tau}$ oscillation fit 
using both CC and NC data samples is perfomed and the best-fit values
for the mixing parameters are obtained.  However, the goodness-of-fit
is worse when compared to the results of a fit based on CC samples only.  The 
minimum $\chi^2/ndf$ is 971.2/853 with a probability of 7.3\% as determined 
by toy Monte Carlo.  Figure~\ref{fig:chisquare-contours-of-NC} (left) shows the 
$\chi^2$ confidence contours. For the $\nu_{\mu}\to\nu_{s}$ oscillation, maximal 
mixing is still preferred by $\nu_{\mu}\to\nu_{s}$ oscillation as shown in
Fig.~\ref{fig:chisquare-contours-of-NC} (right).  However, the $\Delta m^{2}$
value is driven up a bit, $3.5\times10^{-3}{\mathrm eV}^{2}$. This is
because, as was argued in the previous sub-section, the survival probabilities
of $\nu_{\mu}\to\nu_{s}$ oscillations are suppressed, thus, in order
to fit the Super-K data, the best-fit value of $\Delta m^{2}$ is
increased to compensate for this suppression, which is also why the
constraint on mixing angle is tighter than the $\nu_{\mu}\to\nu_{\tau}$
case. The $\chi^2/ndf$ value at the best-fit point is 1023.6/853, which is much 
worse than $\nu_{\mu}\to\nu_{\tau}$ model. The $\Delta\chi^2$ of 52.4 corresponds 
to a 7.2$\sigma$ exclusion level for the pure $\nu_{\mu}\to\nu_{s}$ model.

\begin{figure}
\begin{centering}
\includegraphics[width=1.0\textwidth]{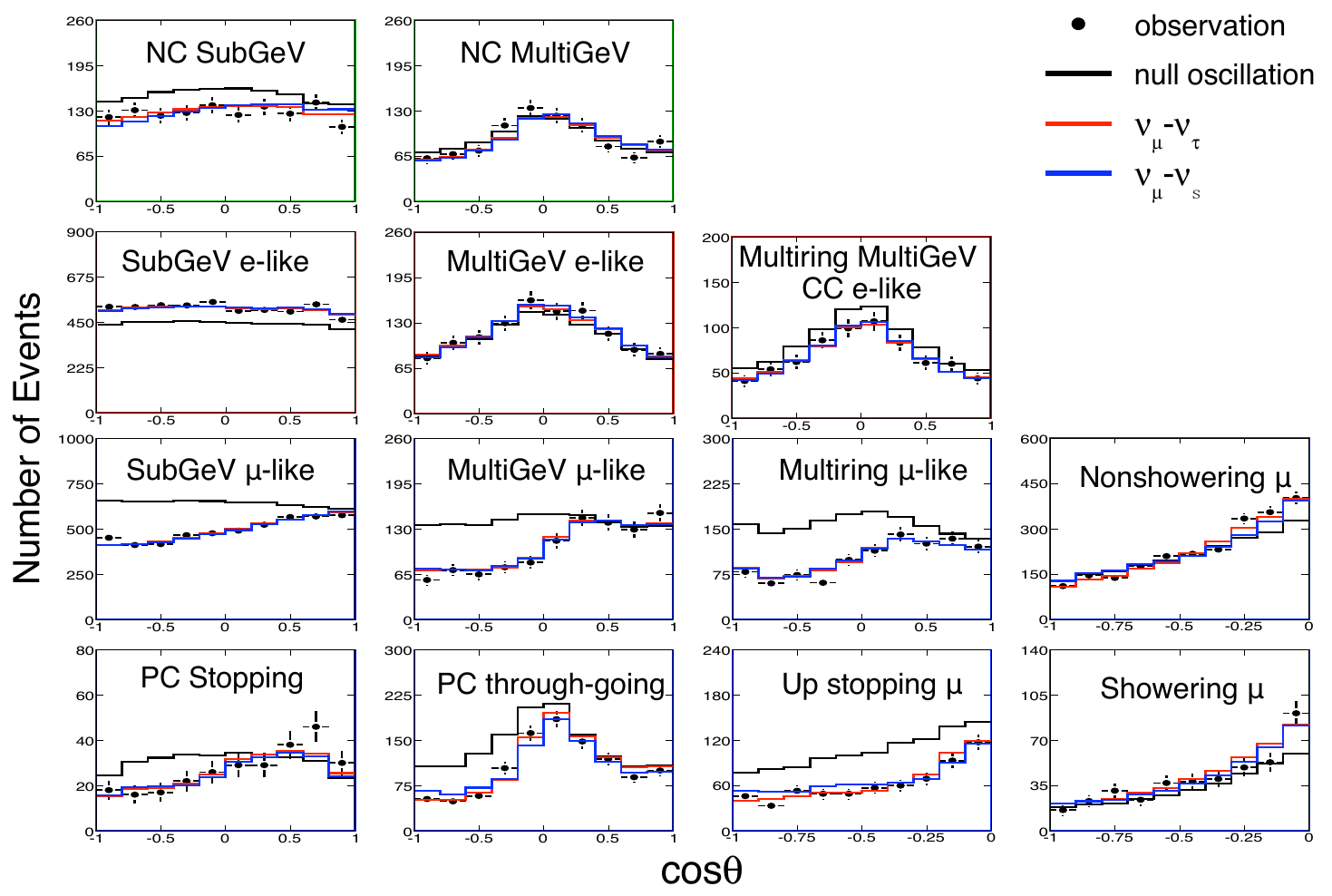}
\par\end{centering}

\caption{\label{fig:Comparison-of-the-zenith}Comparison of the best-fit zenith
angle distributions of $\nu_{\mu}\to\nu_{\tau}$ and $\nu_{\mu}\to\nu_{s}$
oscillations}
\end{figure}

\begin{figure}
\begin{centering}
\includegraphics[width=4.5in]{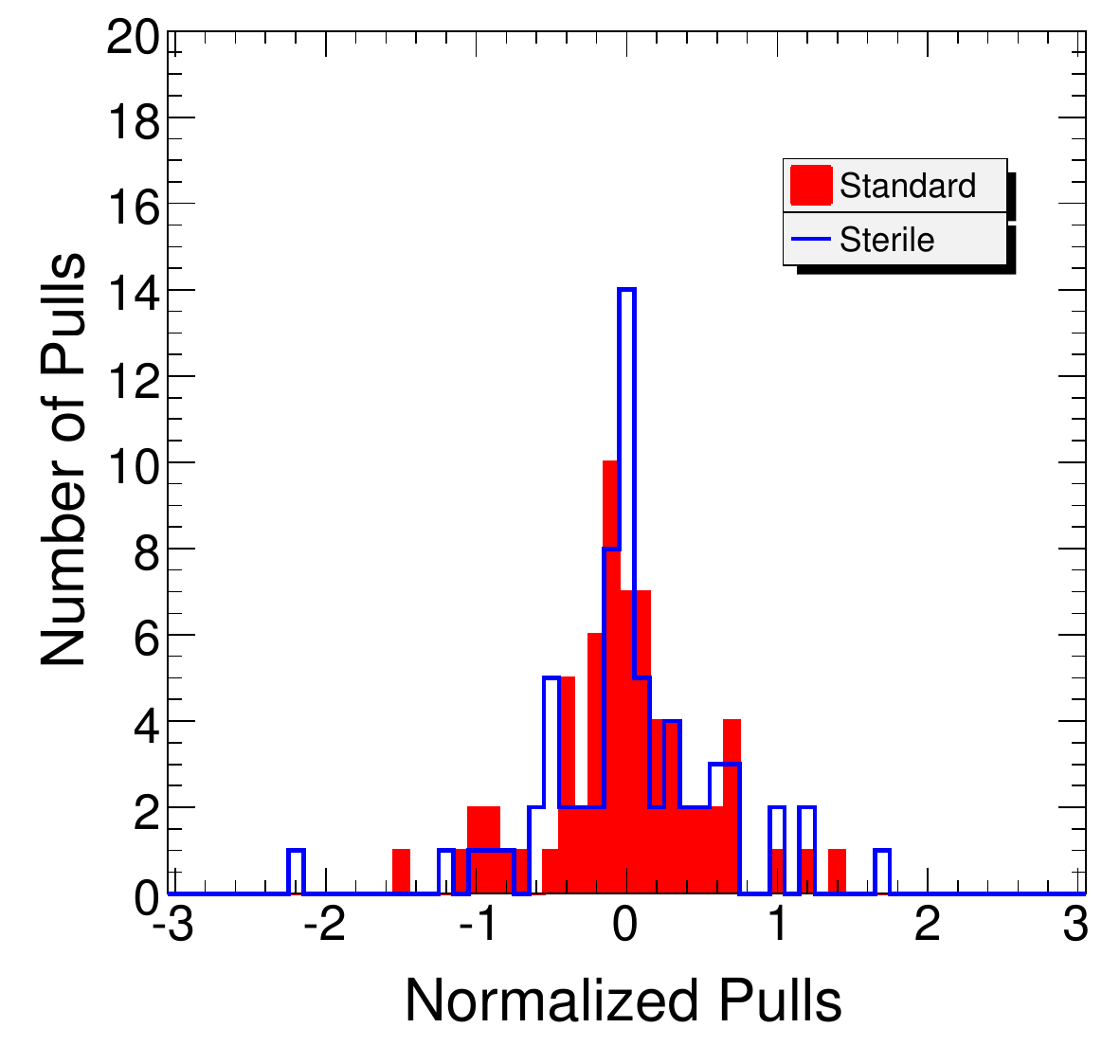}
\par\end{centering}

\caption{\label{fig:Comparison-of-the-pull-dist}Distributions of the normalized
pull terms of $\nu_{\mu}\to\nu_{\tau}$ and $\nu_{\mu}\to\nu_{s}$ oscillations}
\end{figure}

Figure~\ref{fig:Comparison-of-the-zenith} shows the comparison of
best-fit zenith angle distributions. For many bins of the data sample,
the $\nu_{\mu}\to\nu_{s}$ model reproduces SK observation as well as
the $\nu_{\mu}\to\nu_{\tau}$ model.  However, for sub-GeV neutral current
enhanced events, partially contained (PC) through-going events, upward $\mu$ 
stopping events and non-showering $\mu$ events, the $\nu_{\mu}\to\nu_{\tau}$ 
model fits the data better than the $\nu_{\mu}\to\nu_{s}$ model. 
Table~\ref{tab:Chi-square-difference-breakdown} shows the detailed $\chi^2$ 
differences for the different categories of events.  The $\Delta\chi^{2}$ 
from data bins is 38.3. And Fig.~\ref{fig:Comparison-of-the-pull-dist}
shows the comparison of the distributions of the pull terms. The pull
term distribution of the $\nu_{\mu}\to\nu_{s}$ oscillation is wider
than the $\nu_{\mu}\to\nu_{\tau}$ one, which is another sign that $\nu_{\mu}\to\nu_{\tau}$
oscillation is a better model accounting for the atmospheric neutrino
data. Pull terms contribute 14.1 to the $\chi^2$ difference out
of the total difference 52.4.

\begin{table}
\caption{\label{tab:Chi-square-difference-breakdown}Chi-square difference
breakdown (${\Delta\chi}^{2}=\chi_{\nu_{\mu}\to\nu_{s}}^{2}-\chi_{\nu_{\mu}\to\nu_{\tau}}^{2}$)}

\centering{}%
\begin{tabular}{|l|c|c|c|c|}
\hline 
Data Samples & Bins (SK-I+SK-II) & $\chi_{\nu_{\mu}\to\nu_{s}}^{2}$ & $\chi_{\nu_{\mu}\to\nu_{\tau}}^{2}$ & $\Delta\chi^{2}$\tabularnewline
\hline 
\hline 
Single ring sub-GeV $e$-like & 50+50 & 104.8 & 104.0 & 0.8\tabularnewline
\hline 
Single ring multi-GeV $e$-like & 50+50 & 108.6 & 110.7 & -2.1\tabularnewline
\hline 
Multi-ring multi-GeV CC $e$-like & 50+50 & 86.6 & 85.8 & 0.8\tabularnewline
\hline 
Single ring sub-GeV $\mu$-like & 50+50 & 104.9 & 106.1 & -1.2\tabularnewline
\hline 
Single ring multi-GeV $\mu$-like & 30+30 & 64.8 & 66.8 & -2.0\tabularnewline
\hline 
Multi-ring $\mu$-like & 40+40 & 79.3 & 75.5 & 3.8\tabularnewline
\hline 
NC-enhanced sub-GeV events & 10+10 & 19.5 & 14.5 & 5.0\tabularnewline
\hline 
NC-enhanced multi-GeV events & 40+40 & 105.7 & 104.5 & 1.2\tabularnewline
\hline 
PC stopping $\mu$ & 40+40 & 128.7 & 125.8 & 2.9\tabularnewline
\hline 
PC through-going $\mu$ & 40+40 & 114.4 & 102.1 & 12.3\tabularnewline
\hline 
Upward stopping $\mu$ & 10+10 & 21.1 & 14.1 & 7.0\tabularnewline
\hline 
Upward non-showering $\mu$ & 10+10 & 28.1 & 16.9 & 11.2\tabularnewline
\hline 
Upward showering $\mu$ & 10+10 & 24.5 & 25.0 & -1.5\tabularnewline
\hline 
TOTAL & 430+430 & 991.1 & 952.8 & 38.3\tabularnewline
\hline 
\end{tabular}
\end{table}

The $\nu_{\mu}\to\nu_{s}$ model can not reproduce sub-GeV neutral current
enhanced events as well as the $\nu_{\mu}\to\nu_{\tau}$ model. The
$\nu_{\mu}\to\nu_{\tau}$ model expects the data fairly flat which is
what is observed. However, for the $\nu_{\mu}\to\nu_{s}$ model, the
loss of neutral current events makes the expected distribution tilted
toward upward going bins where muon neutrinos oscillate into sterile
ones.

As is shown in Fig.~\ref{fig:Survival-probabilities},
around the atmospheric $\Delta m^{2}$ scale, the strongest matter
effect happen around 10~GeV. The typical energy of PC through-going
events, Up$\mu$ stopping events and non-showering $\mu$ events is
around this scale, which is why these events can differentiate the
two models better than other types of neutrino events. One can also see
that though neutral current enhanced events help to tell two models
apart, it is mainly the matter effect that distinguish two models
due to the advantages such as long baselines, high energies and high
matter densities that atmospheric neutrinos possess compared to long
baseline experiments.

\subsubsection{An Admixture Model}

Compared to $\nu_{\mu}\to\nu_{\tau}$ oscillation, the fact that the
$\nu_{\mu}\to\nu_{s}$ oscillation is excluded at 7.2$\sigma$ level
does not mean that there is no involvement of sterile neutrinos in
Super-K atmospheric neutrino oscillations. It is still possible that
a certain portion of muon neutrinos are oscillating into sterile neutrinos
in addition to the dominant $\nu_{\mu}\to\nu_{\tau}$ channel. 

In the case of 2+2 mass hierarchy,
according to G.~L.~Fogli {\it et. al.}~\cite{Fogli:2000ir},
based on the current experimental results, the four flavor mixing
can be transformed into two 2-flavor mixings by constructing two superposition
states of $\nu_{\tau}$ and $\nu_{s}$ in the following way:
\begin{eqnarray*}
\left(\begin{array}{c}
\nu_{+}\\
\nu_{-}
\end{array}\right)=\left(\begin{array}{cc}
\cos\xi & \sin\xi\\
-\sin\xi & \cos\xi
\end{array}\right)\left(\begin{array}{c}
\nu_{\tau}\\
\nu_{s}
\end{array}\right).
\end{eqnarray*}
For the atmospheric sector, the oscillation now is between $\nu_{\mu}$
and $\nu_{+}$which is a superposition state: $\cos\xi|\nu_{\tau}\rangle+\sin\xi|\nu_{s}\rangle$.
Thus, the portion of sterile neutrinos is $\sin^{2}\xi$. Accordingly,
the matter effect strength is weaken by a factor of $\sin^{2}\xi$.
Now, there are 3 parameters in this oscillation: the mixing angle between
$|\nu_{m_{3}}\rangle$ and $|\nu_{m_{4}}\rangle$, $\Delta m^{2}=m_{4}^{2}-m_{3}^{2}$
and the portion of sterile neutrinos $\sin^{2}\xi$.  Since the maximal
mixing is a very strong constraint, it is assumed in this analysis, 
{\it i.e.} $\sin^{2}2\theta_{atm}=1$.

\begin{figure}
\begin{centering}
\includegraphics[width=4in]{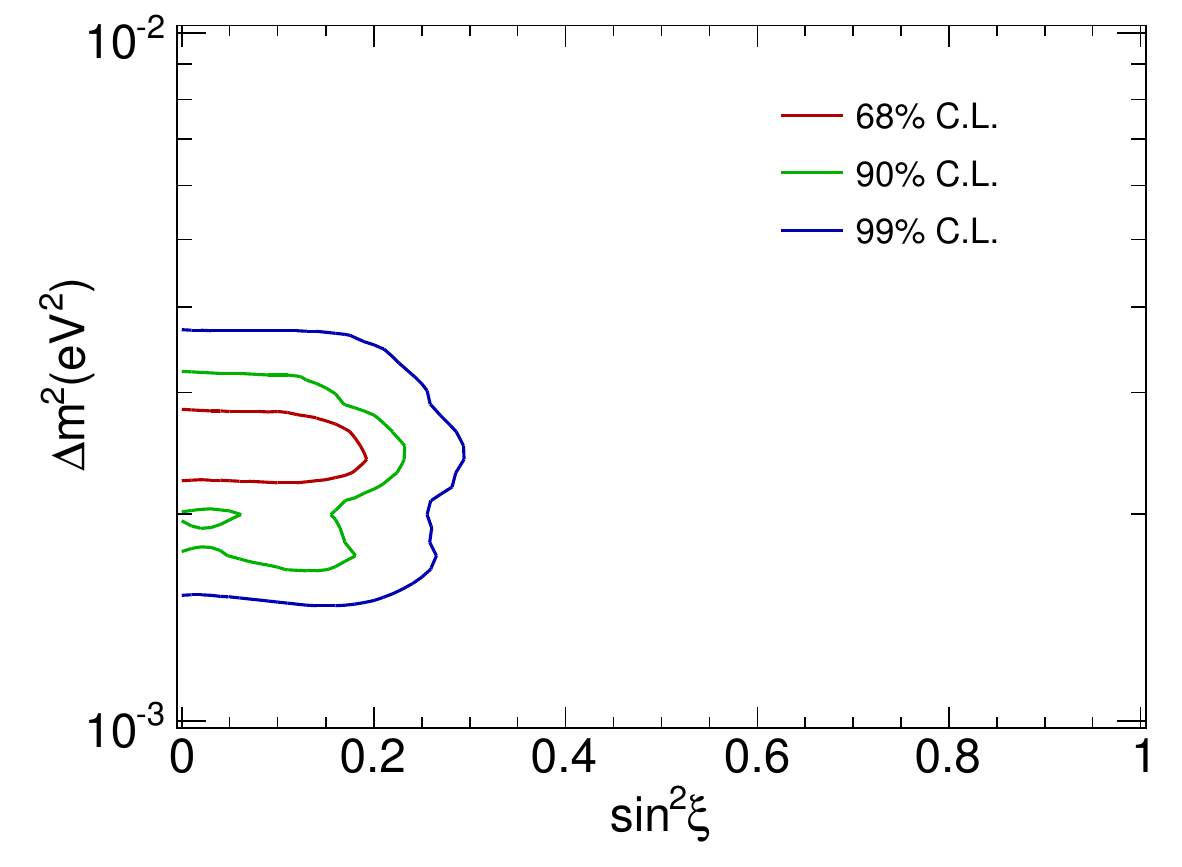}
\par\end{centering}
\caption{\label{fig:The-sterile-admixture}The $\nu_{s}$ admixture allowance }
\end{figure}

Using the same least chi-square framework, the best-fit is $\sin^{2}\xi=0$
which means Super-K data prefer no sterile neutrino involvement,
see Fig.~\ref{fig:The-sterile-admixture}. However, at 90\%~C.L., the data 
do allow a sterile admixture of 23\%.  Compared to the limit
of 67\% at 90\% C.L. set by Fogli~\emph{et~al} in Ref.~\cite{Fogli:2000ir}
using the published Super-K data, this limit using the complete SK-I
and SK-II data sets is much more stringent.

\subsection{The MiniBooNE $\nu_e$ and $\bar{\nu}_e$ Appearance Searches}

\subsubsection*{Description of the Experiment}

\begin{figure}
  \centerline{\includegraphics[width=\textwidth]{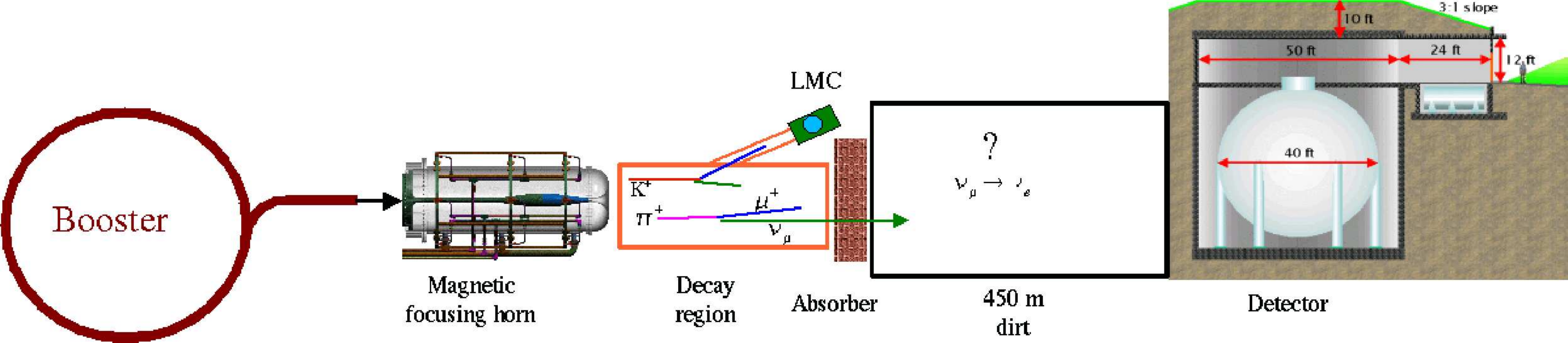}}
  \caption{\label{schematic} A schematic drawing of the MiniBooNE experiment.}
\end{figure}

A schematic drawing of the MiniBooNE experiment at FNAL is shown in
Fig.~\ref{schematic}.  The experiment is fed by protons with 8~GeV of kinetic
energy from the Fermilab Booster, which interact in a 71~cm long Be target
located at the upstream end of a magnetic focusing horn. The horn pulses with a
current of 174~kA and, depending on the polarity, either focuses $\pi^+$ and
$K^+$ and defocuses $\pi^-$ and $K^-$ to form a pure neutrino beam or focuses
$\pi^-$ and $K^-$ and defocuses $\pi^+$ and $K^+$ to form a somewhat pure
antineutrino beam.  The produced pions and kaons decay in a 50~m long pipe, and
a fraction of the neutrinos and antineutrinos \cite{AguilarArevalo:2008yp} interact in the
MiniBooNE detector, which is located 541~m downstream of the Be target. For the
MiniBooNE results presented here, a total of $6.5 \times 10^{20}$~POT (protons
on target) were collected in neutrino mode and $8.58 \times 10^{20}$~POT have
been collected so far in antineutrino mode.

\begin{figure}
  \centerline{\includegraphics[height=3.5in]{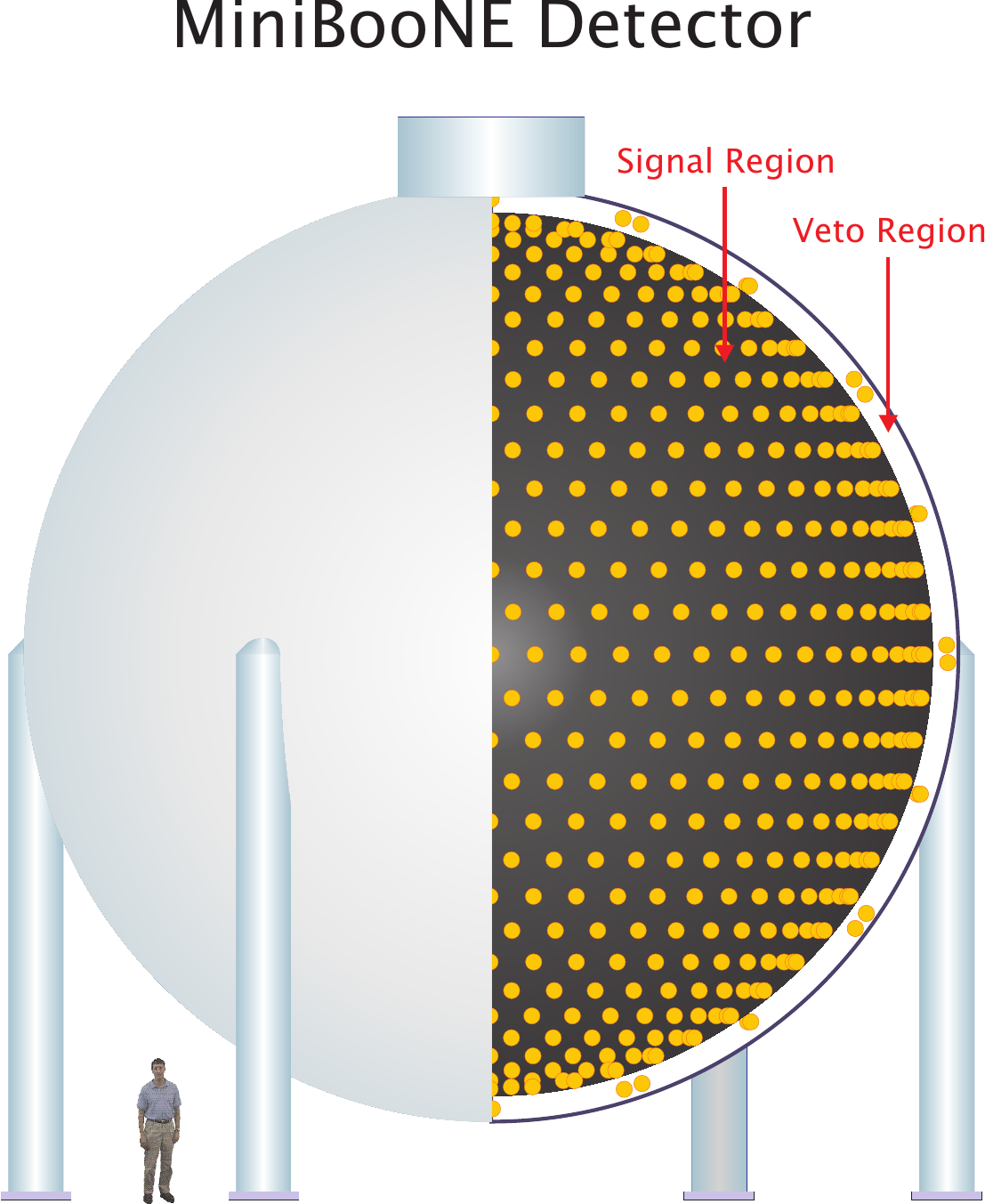}}
  \caption{\label{mb_schematic} A schematic drawing of the MiniBooNE detector.}
\end{figure}

The MiniBooNE detector \cite{AguilarArevalo:2008qa} consists of a spherical tank with a
diameter of 12.2~m, filled with approximately 800~tons of mineral oil ($CH_2$).
A schematic drawing of the MiniBooNE detector is shown in
Fig.~\ref{mb_schematic}.  There are a total of 1280 8~inch detector phototubes
(covering 10\% of the surface area) and 240 veto phototubes. The fiducial
volume has a 5~m radius and corresponds to approximately 450~tons. Only $\sim
2\%$ of the phototube channels failed so far over the course of the run.

\subsubsection*{Neutrino Oscillation Event Selection}

MiniBooNE searches for $\nu_\mu \rightarrow \nu_e$ oscillations by measuring
the rate of $\nu_e C \rightarrow e^- X$ (known as charged current quasi-elastic 
events or CCQE) and testing whether the
measured rate is consistent with the estimated background rate.  To select
candidate $\nu_e$ CCQE events, an initial selection is first applied: $>200$
tank hits, $<6$ veto hits, reconstructed time within the neutrino beam spill,
reconstructed vertex radius $<500$~cm, and visible energy $E_{vis}>140$~MeV.
It is then required that the event vertex reconstructed assuming an outgoing
electron and the track endpoint reconstructed assuming an outgoing muon occur
at radii $<500$~cm and $<488$~cm, respectively, to ensure good event
reconstruction and efficiency for possible muon decay electrons. Particle
identification (PID) cuts are then applied to reject muon and $\pi^0$ events.
Several improvements have been made to the neutrino oscillation data analysis
since the initial data was published \cite{AguilarArevalo:2008rc}, including an improved
background estimate, an additional fiducial volume cut that greatly reduces the
background from events produced outside the tank (dirt events), and an increase
in the data sample from $5.579 \times 10^{20}$~POT to $6.462 \times
10^{20}$~POT.  A total of 89,200 neutrino events pass the initial selection,
while 1069 events pass the complete event selection of the final analysis with
$E_\nu^{QE}>200$~MeV, where $E_\nu^{QE}$ is the reconstructed neutrino energy.

\subsubsection*{Neutrino Oscillation Signal and Background Reactions}

Table~\ref{signal_bkgd} shows the expected number of candidate $\nu_e$ CCQE
background events with $E_\nu^{QE}$ in the intervals 200--300~MeV,
300--475~MeV, and 475--1250~MeV, respectively, after the complete event
selection of the final analysis.  The background estimate includes antineutrino
events, representing $<2\%$ of the total.  The total expected backgrounds for
the three energy regions are $186.8 \pm 26.0$ events, $228.3 \pm 24.5$ events,
and $385.9 \pm 35.7$ events, respectively. For $\nu_\mu \rightarrow \nu_e$
oscillations at the best-fit LSND solution of $\Delta m^2 =1.2$~eV$^2$ and
$\sin^22\theta = 0.003$, the expected number of $\nu_e$ CCQE signal events for
the three energy regions are 7 events, 37 events, and 135 events, respectively.

\begin{table}
  \begin{center}
    \vspace{0.1in}
    \begin{ruledtabular}
    \begin{tabular}{lccc}
      Process&$200-300$~MeV&$300-475$~MeV&$475-1250$~MeV \\
      \hline
      $\nu_\mu$ CCQE&9.0&17.4&11.7 \\
      $\nu_\mu e \rightarrow \nu_\mu e$&6.1&4.3&6.4 \\
      NC $\pi^0$&103.5&77.8&71.2 \\
      NC $\Delta \rightarrow N \gamma$&19.5&47.5&19.4 \\
      Dirt Events&11.5&12.3&11.5 \\
      Other Events&18.4&7.3&16.8 \\
      \hline
      $\nu_e$ from $\mu$ Decay&13.6&44.5&153.5 \\
      $\nu_e$ from $K^+$ Decay&3.6&13.8&81.9 \\
      $\nu_e$ from $K^0_L$ Decay&1.6&3.4&13.5 \\
      \hline
      Total Background &$186.8 \pm 26.0$&$228.3 \pm 24.5$&$385.9 \pm 35.7$ \\
      \hline
      LSND Best-Fit Solution&$7 \pm 1$&$37 \pm 4$&$135 \pm 12$ \\
    \end{tabular}
  \end{ruledtabular}
  \end{center}
  \caption{\label{signal_bkgd} The expected number of events in the
    $200<E_\nu^{QE}<300$~MeV, $300<E_\nu^{QE}<475$~MeV, and $475<E_\nu^{QE}<1250$~MeV
    energy ranges from all of the significant backgrounds after the complete
    event selection of the final analysis.  Also shown are the expected number of
    $\nu_e$ CCQE signal events for two-neutrino oscillations at the LSND best-fit
    solution.}
\end{table}

\subsubsection*{MiniBooNE Neutrino Oscillation Results}

\begin{figure}
  \centerline{\includegraphics[width=0.8\textwidth]{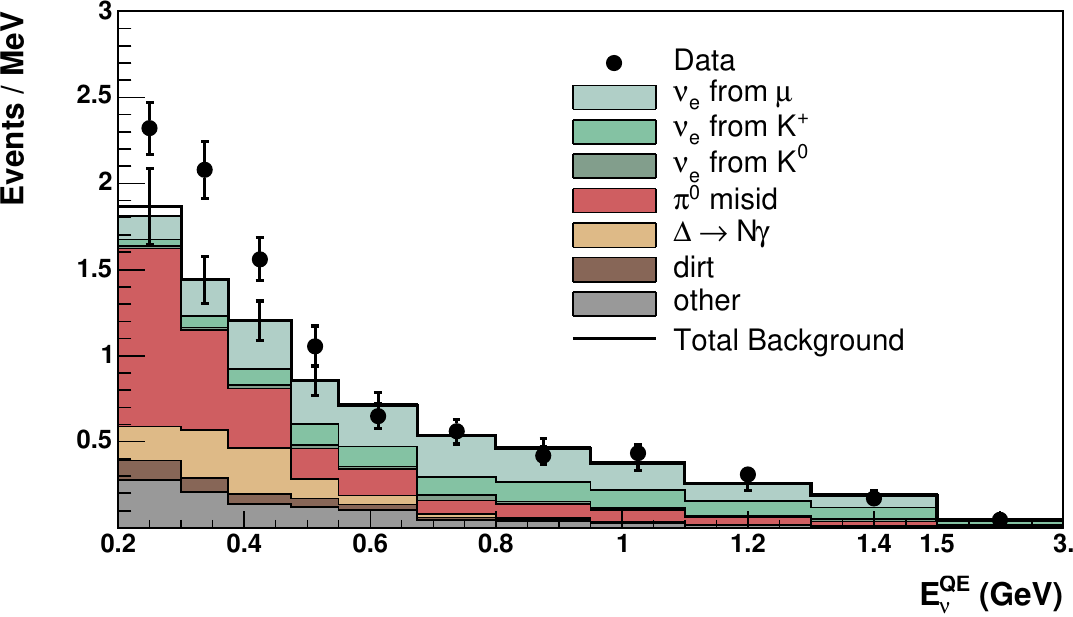}}
  \caption{\label{osc} The MiniBooNE reconstructed neutrino energy
    distribution for candidate $\nu_e$ data events (points with error bars)
    compared to the Monte Carlo simulation (histogram).}
\end{figure}

\begin{figure}
  \centerline{\includegraphics[width=0.8\textwidth]{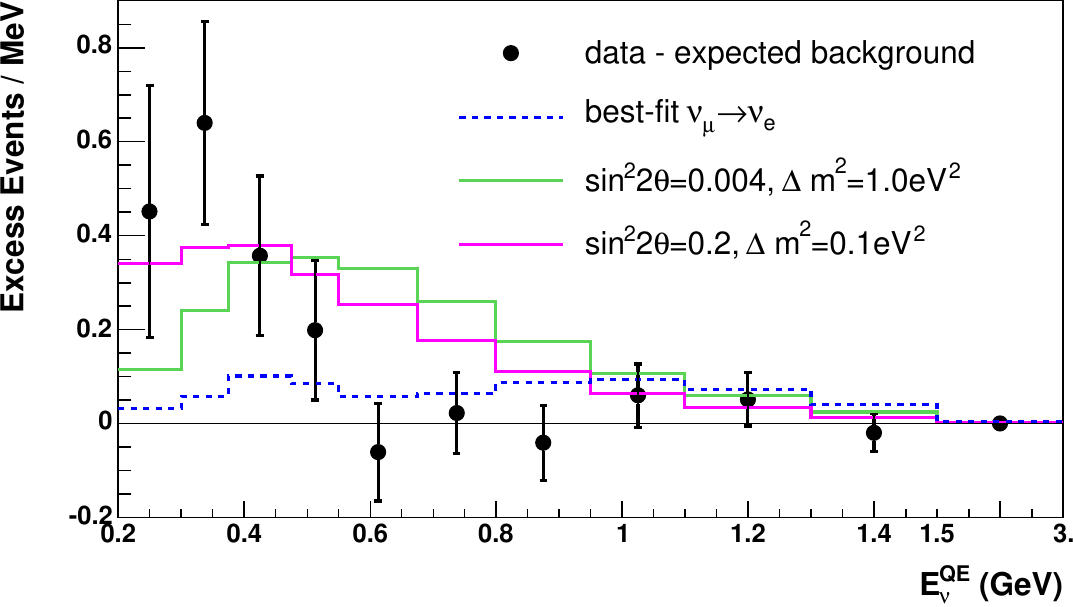}}
  \caption{\label{MB_neutrino_excess} The excess of $\nu_e$ candidate
    events observed by MiniBooNE in neutrino mode.}
\end{figure}

Fig.~\ref{osc} shows the reconstructed neutrino energy distribution for
candidate $\nu_e$ data events (points with error bars) compared to the MC
simulation (histogram) \cite{AguilarArevalo:2008rc}, while Fig.~\ref{MB_neutrino_excess} shows
the event excess as a function of reconstructed neutrino energy. Good agreement
between the data and the MC simulation is obtained for $E_\nu^{QE} > 475$~MeV;
however, an unexplained excess of electron-like events is observed for
$E_\nu^{QE} < 475$~MeV. As shown in Fig.~\ref{MB_neutrino_excess}, the
magnitude of the excess is very similar to what is expected from neutrino
oscillations based on the LSND signal. Although the shape of the excess is not
consistent with simple two-neutrino oscillations, more complicated oscillation
models with CP violation have shapes that may be consistent with the LSND
signal.

\begin{figure}
  \centerline{\includegraphics[width=0.8\textwidth,angle=0]{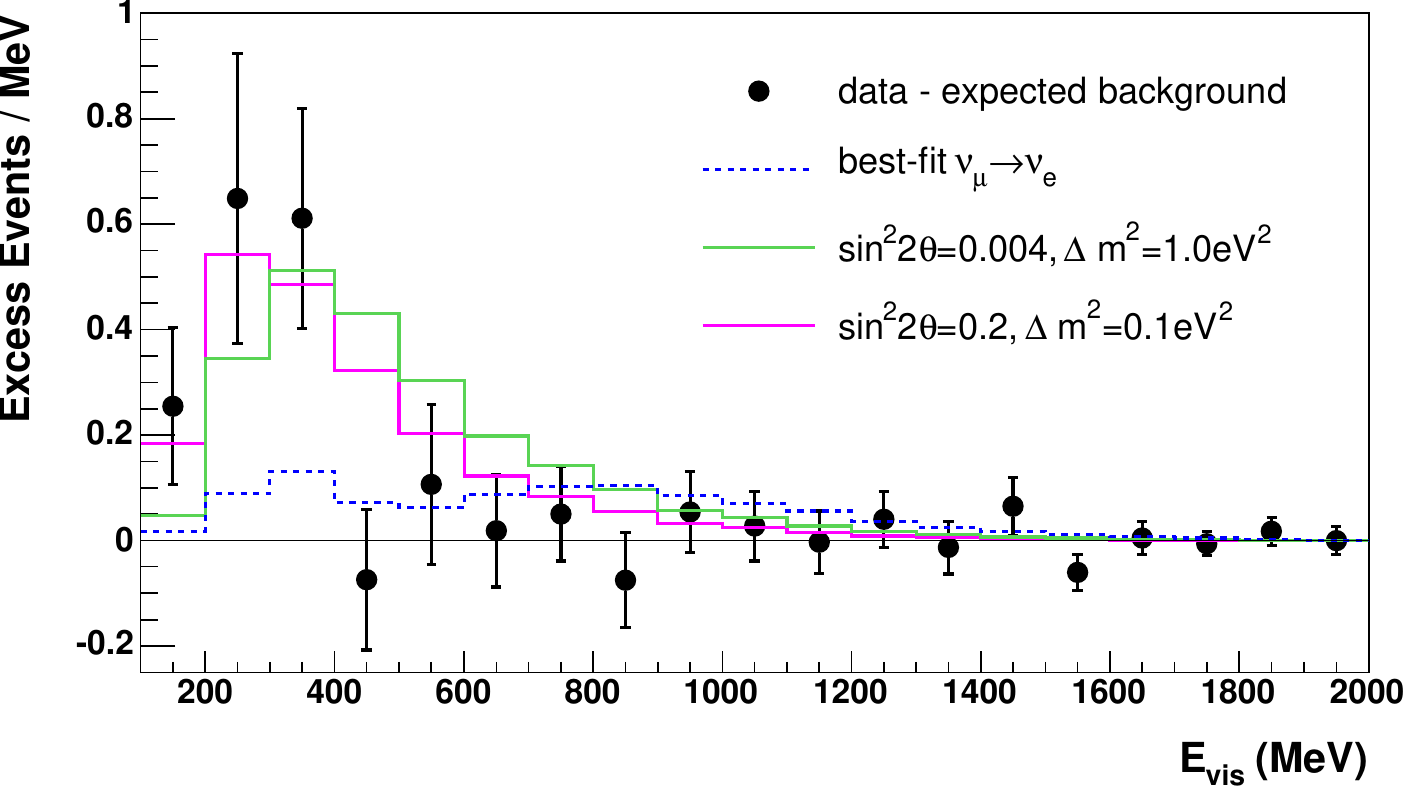}}
  \caption{The neutrino event excess as a function of $E_{vis}$ for
    $E_\nu^{QE} > 200$~MeV.  Also shown are the expectations from the best
    oscillation fit ($\sin^22\theta = 0.0017$, $\Delta m^2 = 3.14$~eV$^2$) and
    from neutrino oscillation parameters in the LSND allowed region. The error
    bars include both statistical and systematic errors.}
  \label{data_mc3}
\end{figure}

Table~\ref{mb_events} shows the number of data, background, and excess events
for different $E_\nu^{QE}$ ranges, together with the excess significance.  For
the final analysis, an excess of $128.8 \pm 20.4 \pm 38.3$ events is observed
for $200<E_\nu^{QE}<475$~MeV. For the entire $200<E_\nu^{QE}<1250$~MeV energy
region, the excess is $151.0 \pm 28.3 \pm 50.7$ events.  As shown in
Fig.~\ref{data_mc3}, the event excess occurs for $E_{vis} <400$~MeV, where
$E_{vis}$ is the visible energy.

\begin{table}
  \centering
  \parbox{10cm}{
  \begin{ruledtabular}
  \begin{tabular}{lc}
    Event Sample&Final Analysis \\
    \hline
    \multicolumn{2}{c}{$200-300$ MeV} \\
    \hline
    Data&232 \\
    Background&$186.8 \pm 13.7 \pm 22.1$ \\
    Excess&$45.2 \pm 13.7 \pm 22.1$ \\
    Significance&$1.7 \sigma$ \\
    \hline
    \multicolumn{2}{c}{$300-475$ MeV} \\
    \hline
    Data&312 \\
    Background&$228.3 \pm 15.1 \pm 19.3$ \\
    Excess&$83.7 \pm 15.1 \pm 19.3$ \\
    Significance&$3.4 \sigma$ \\
    \hline
    \multicolumn{2}{c}{$200-475$ MeV} \\
    \hline
    Data&544 \\
    Background&$415.2 \pm 20.4 \pm 38.3$ \\
    Excess&$128.8 \pm 20.4 \pm 38.3$ \\
    Significance&$3.0 \sigma$ \\
    \hline
    \multicolumn{2}{c}{$475-1250$ MeV} \\
    \hline
    Data&408 \\
    Background&$385.9 \pm 19.6 \pm 29.8$ \\
    Excess&$22.1 \pm 19.6 \pm 29.8$ \\
    Significance&$0.6 \sigma$ \\
  \end{tabular}
  \end{ruledtabular}}
  \caption{\label{mb_events} The number of data, background, and excess
    events for different $E_\nu^{QE}$ ranges, together with the significance of
    the excesses in neutrino mode.}
\end{table}

\begin{figure}
  \vspace{0.3cm}
  \centerline{\includegraphics[width=0.8\textwidth]{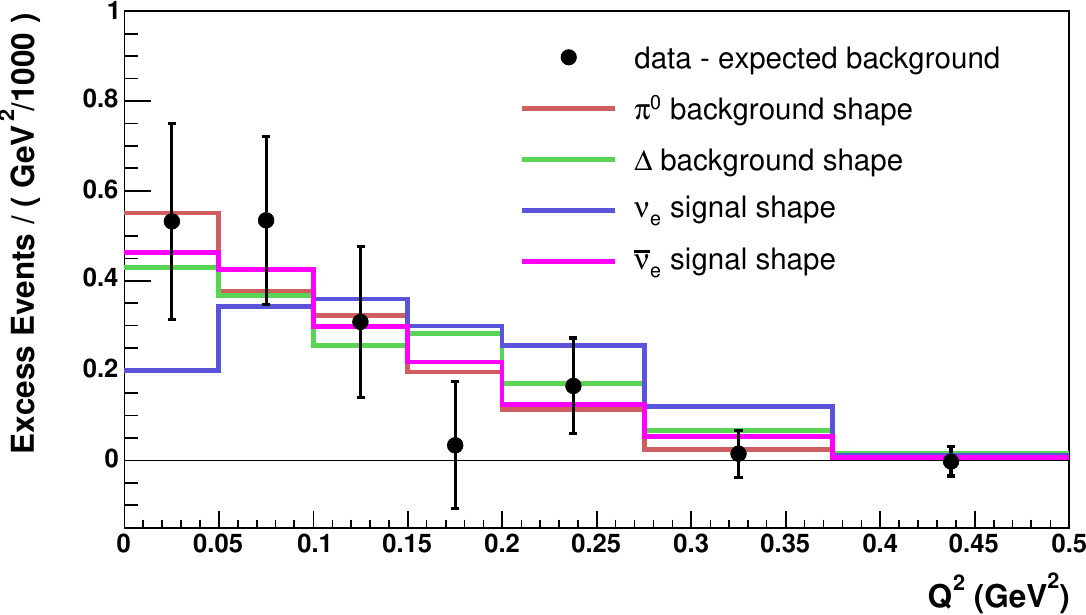}}
  \caption{The neutrino event excess as a function of $Q^2$ for $300 <
    E_\nu^{QE} < 475$~MeV.}
  \label{data_mc4}
\end{figure}

\begin{figure}
  \centerline{\includegraphics[width=0.8\textwidth]{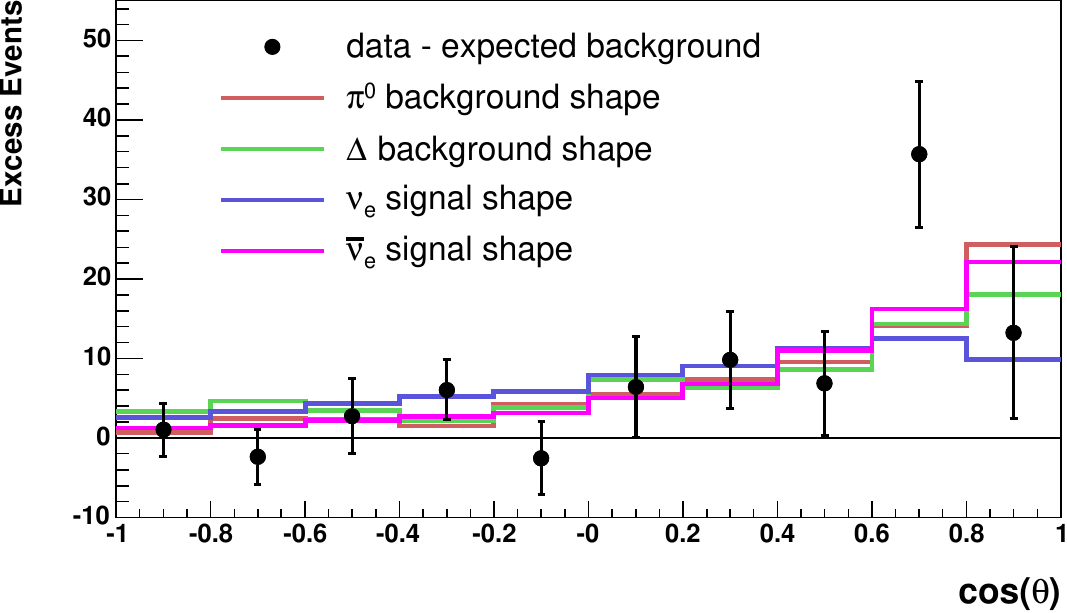}}
  \caption{The neutrino event excess as a function of $\cos (\theta)$ for
    $300 < E_\nu^{QE} < 475$~MeV.}
  \label{data_mc5}
\end{figure}

Figs.~\ref{data_mc4} and \ref{data_mc5} show the event excess as functions of
$Q^2$ and $\cos (\theta)$ for events in the $300 < E_\nu^{QE} < 475$~MeV range,
where $Q^2$ is determined from the energy and angle of the outgoing lepton and
$\theta$ is the angle between the beam direction and the reconstructed event
direction.  Also shown in the figures are the expected shapes from $\nu_e C
\rightarrow e^- X$ and $\bar \nu_e C \rightarrow e^+ X$ charged-current (CC)
scattering and from the NC $\pi^0$ and $\Delta \rightarrow N \gamma$ reactions,
which are representative of photon events produced by NC scattering.  The NC
scattering assumes the $\nu_\mu$ energy spectrum, while the CC scattering
assumes the transmutation of $\nu_\mu$ into $\nu_e$ and $\bar \nu_e$,
respectively.  As shown in Table~\ref{chisquare}, the $\chi^2$ values from
comparisons of the event excess to the expected shapes are acceptable for all
of the processes. However, any of the backgrounds in Table~\ref{chisquare}
would have to be increased by $>5 \sigma$ to explain the low-energy excess.

\begin{table}
  \begin{center}
    \begin{ruledtabular}
    \begin{tabular}{cccc}
      Process&$\chi^2(cos \theta)/9$ DF&$\chi^2(Q^2)/6$ DF&Factor Increase \\
      \hline
      NC $\pi^0$&13.46&2.18&2.0 $(6.8 \sigma)$ \\
      $\Delta \rightarrow N \gamma$&16.85&4.46&2.7 $(18.4 \sigma)$ \\
      $\nu_e C \rightarrow e^- X$&14.58&8.72&2.4 $(15.3 \sigma)$ \\
      $\bar \nu_e C \rightarrow e^+ X$&10.11&2.44&65.4 $(41.0 \sigma)$ \\
    \end{tabular}
    \end{ruledtabular}
  \end{center}
  \caption{\label{chisquare} The $\chi^2$ values from comparisons of the
    neutrino event excess $Q^2$ and $\cos (\theta)$ distributions for $300 <
    E_\nu^{QE} < 475$~MeV to the expected shapes from various NC and CC
    reactions. Also shown is the factor increase necessary for the estimated
    background for each process to explain the low-energy excess and the
    corresponding number of sigma.}
\end{table}

\subsubsection*{MiniBooNE Antineutrino Oscillation Results}

\begin{figure}
  \centering
  \includegraphics[width=0.45\textwidth]{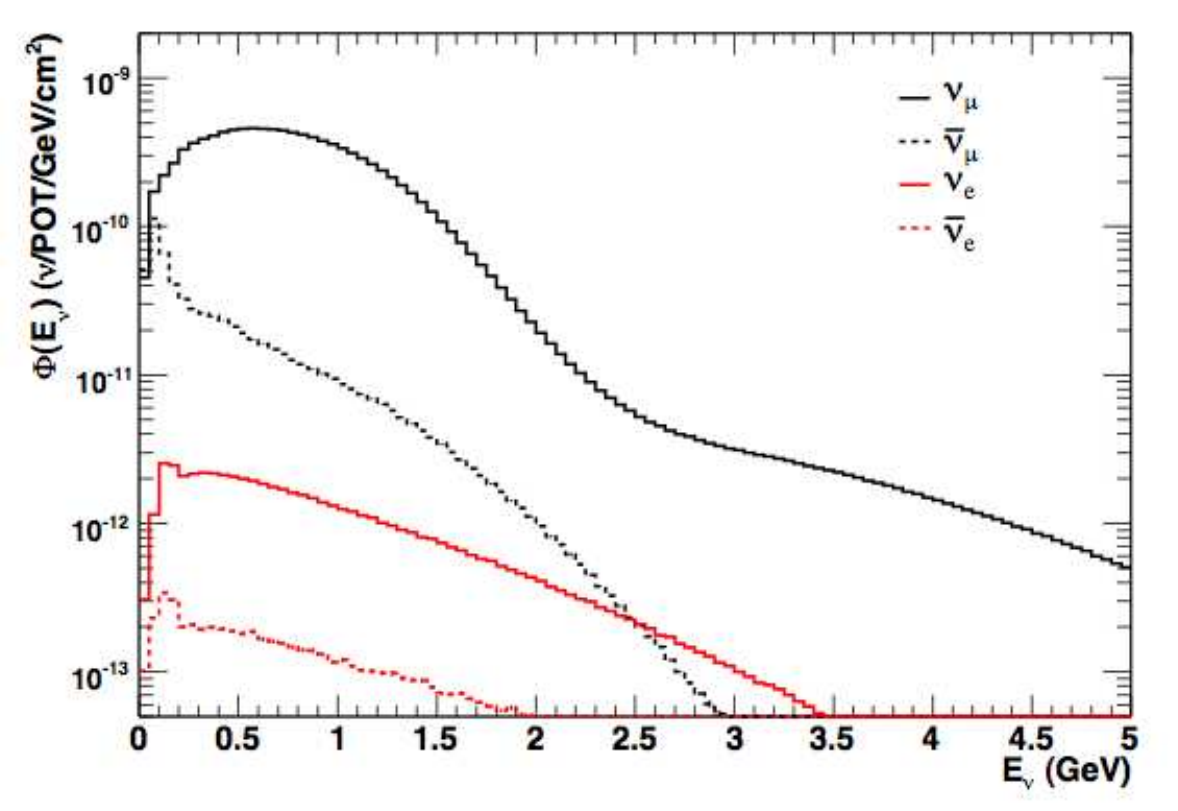}
  \includegraphics[width=0.45\textwidth]{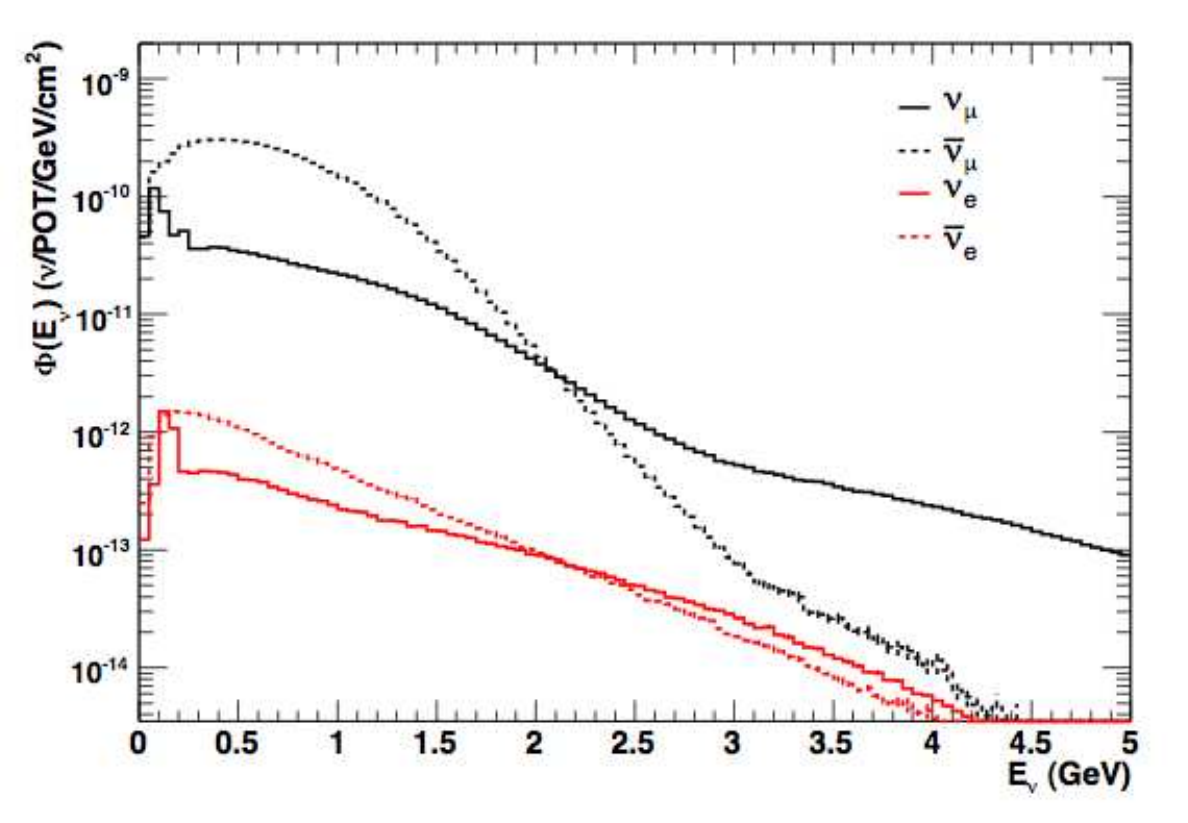}
  \caption{The estimated neutrino fluxes for neutrino mode (left plot) and
    antineutrino mode (right plot).
  \label{nu_flux}}
\end{figure}

\begin{figure}
  \centering
  \includegraphics[height=0.45\textwidth,angle=270]{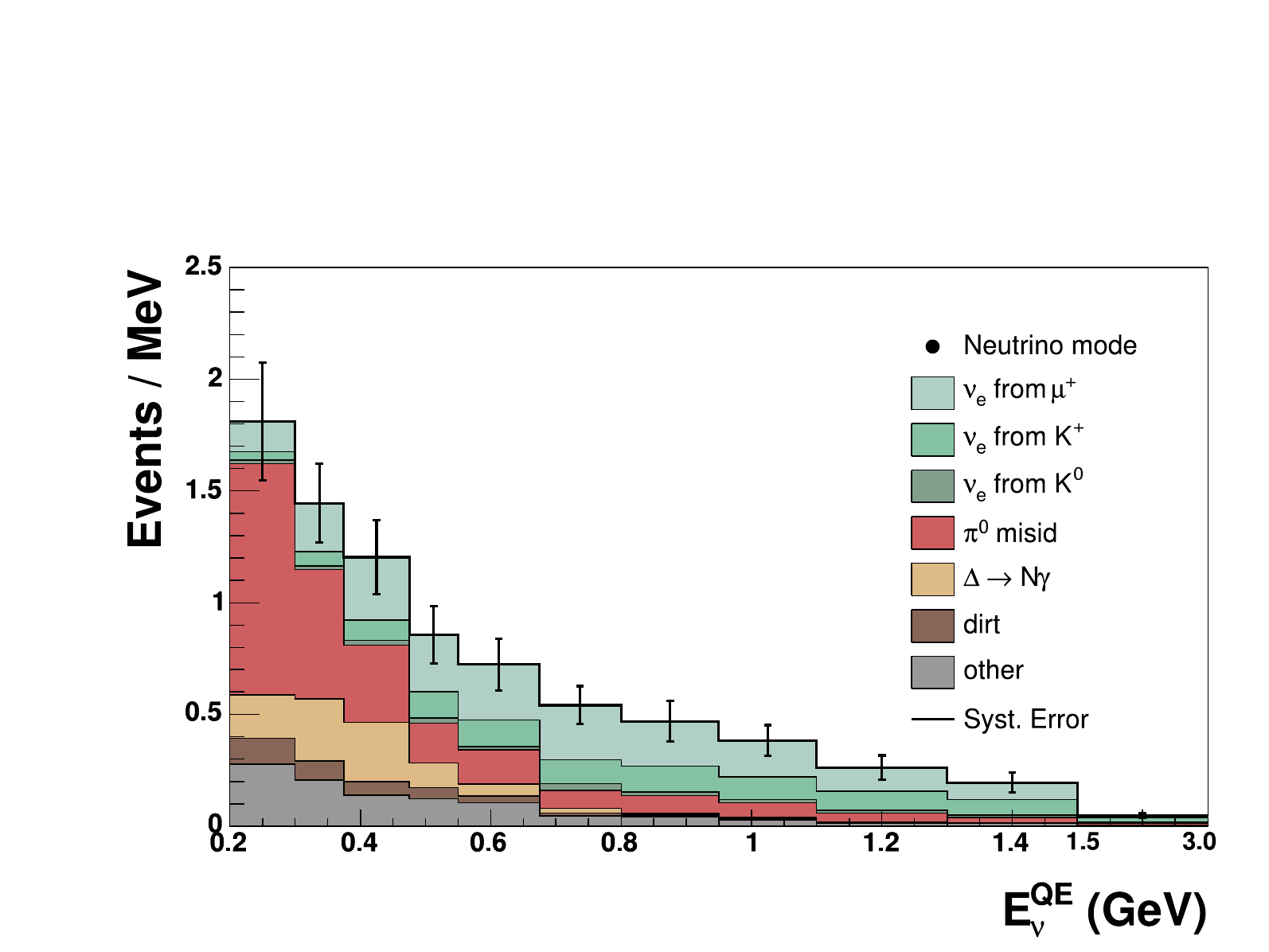}
  \hspace{0.3cm}
  \includegraphics[height=0.45\textwidth,angle=270]{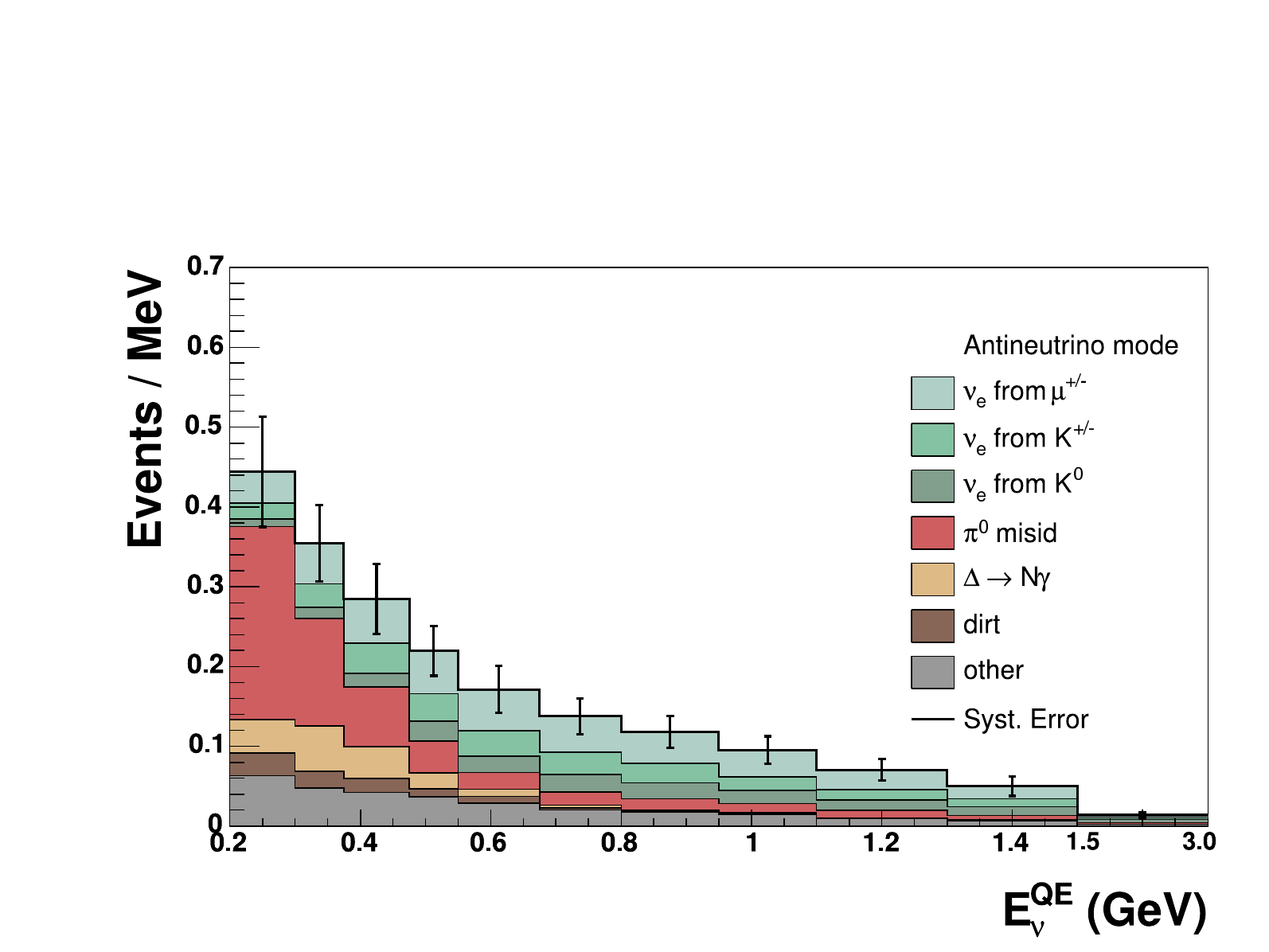}
  \caption{The estimated backgrounds for the neutrino oscillation search in
    neutrino mode (left plot) and antineutrino mode (right plot).  
    }
  \label{nu_bkgd}
\end{figure}

The same analysis that was used for the neutrino oscillation results is
employed for the antineutrino oscillation results~\cite{AguilarArevalo:2010wv}.
Fig.~\ref{nu_flux} shows the estimated neutrino fluxes for neutrino mode and
antineutrino mode, respectively. The fluxes are fairly similar (the intrinsic
electron-neutrino background is approximately 0.5\% for both modes of running),
although the wrong-sign contribution to the flux in antineutrino mode ($\sim
18\%$) is much larger than in neutrino mode ($\sim 6\%$). The average $\nu_e$
plus $\bar \nu_e$ energies are 0.96~GeV in neutrino mode and 0.77~GeV in
antineutrino mode, while the average $\nu_\mu$ plus $\bar \nu_\mu$ energies are
0.79~GeV in neutrino mode and 0.66~GeV in antineutrino mode.  Also, as shown in
Fig.~\ref{nu_bkgd}, the estimated backgrounds in the two modes are very
similar, especially at low energy.

\begin{figure}
  \centerline{\includegraphics[width=0.7\textwidth,angle=0]{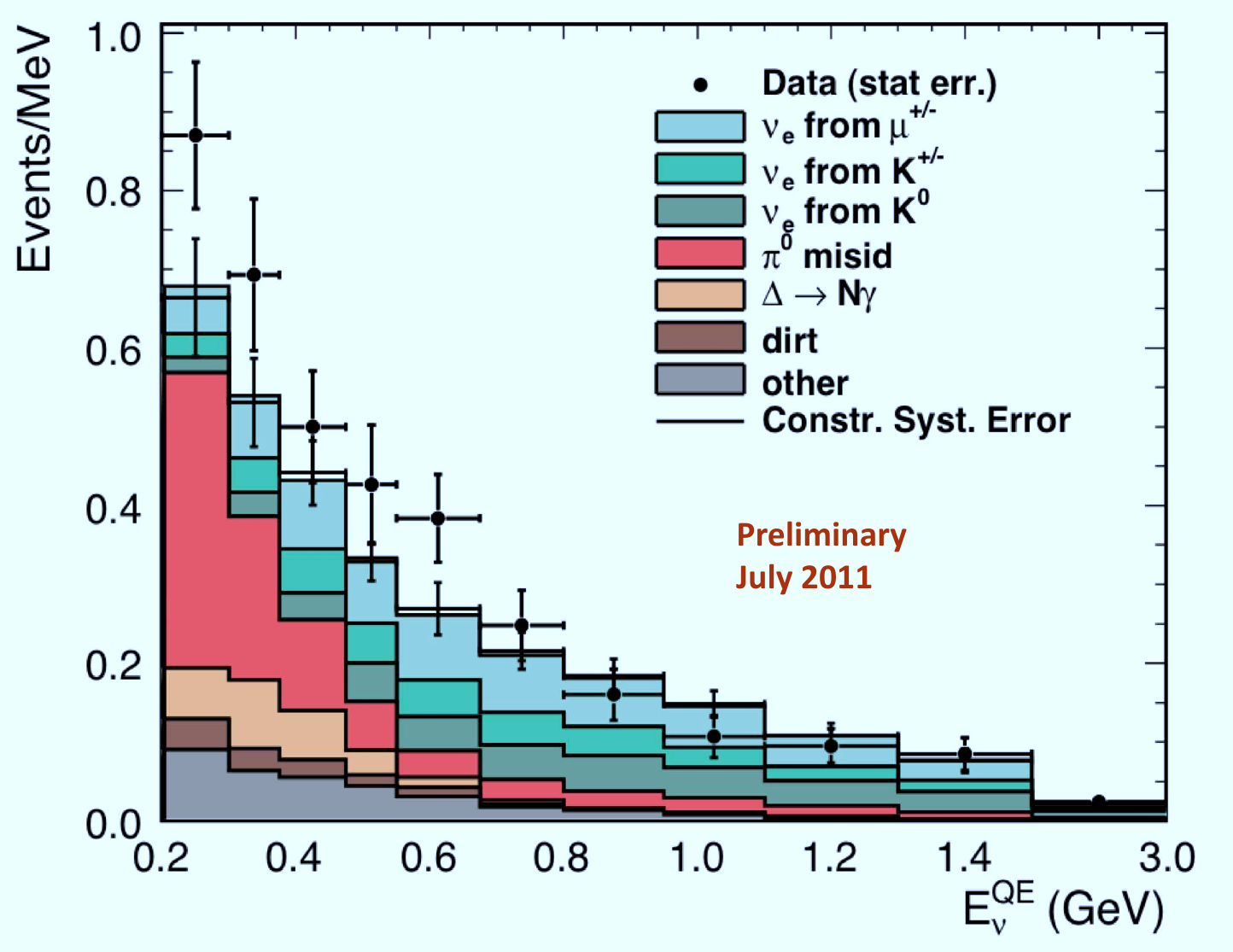}}
  \vspace{-0.3cm}
  \caption{\label{MB_antineutrino_data} The energy distributions of the
    MiniBooNE $\bar \nu_e$ candidate events and the expected background.}
\end{figure}

\begin{figure}
  \centerline{\includegraphics[width=0.7\textwidth,angle=0]{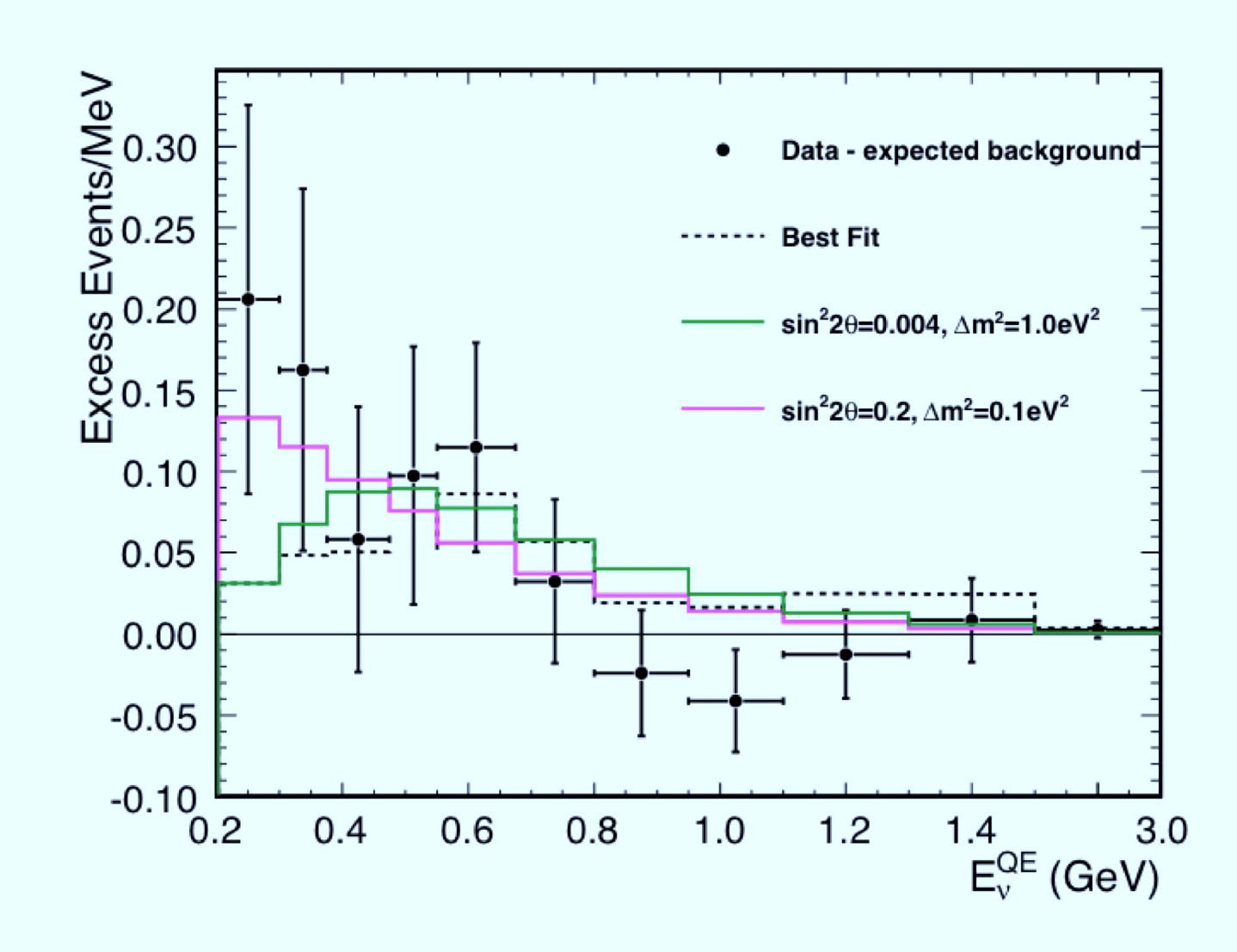}}
  \vspace{-0.5cm}
  \caption{\label{MB_antineutrino_excess} The excess of $\bar \nu_e$ candidate
    events observed by MiniBooNE in antineutrino mode.}
\end{figure}

\begin{figure}
  \centerline{\includegraphics[width=0.55\textwidth,angle=0]{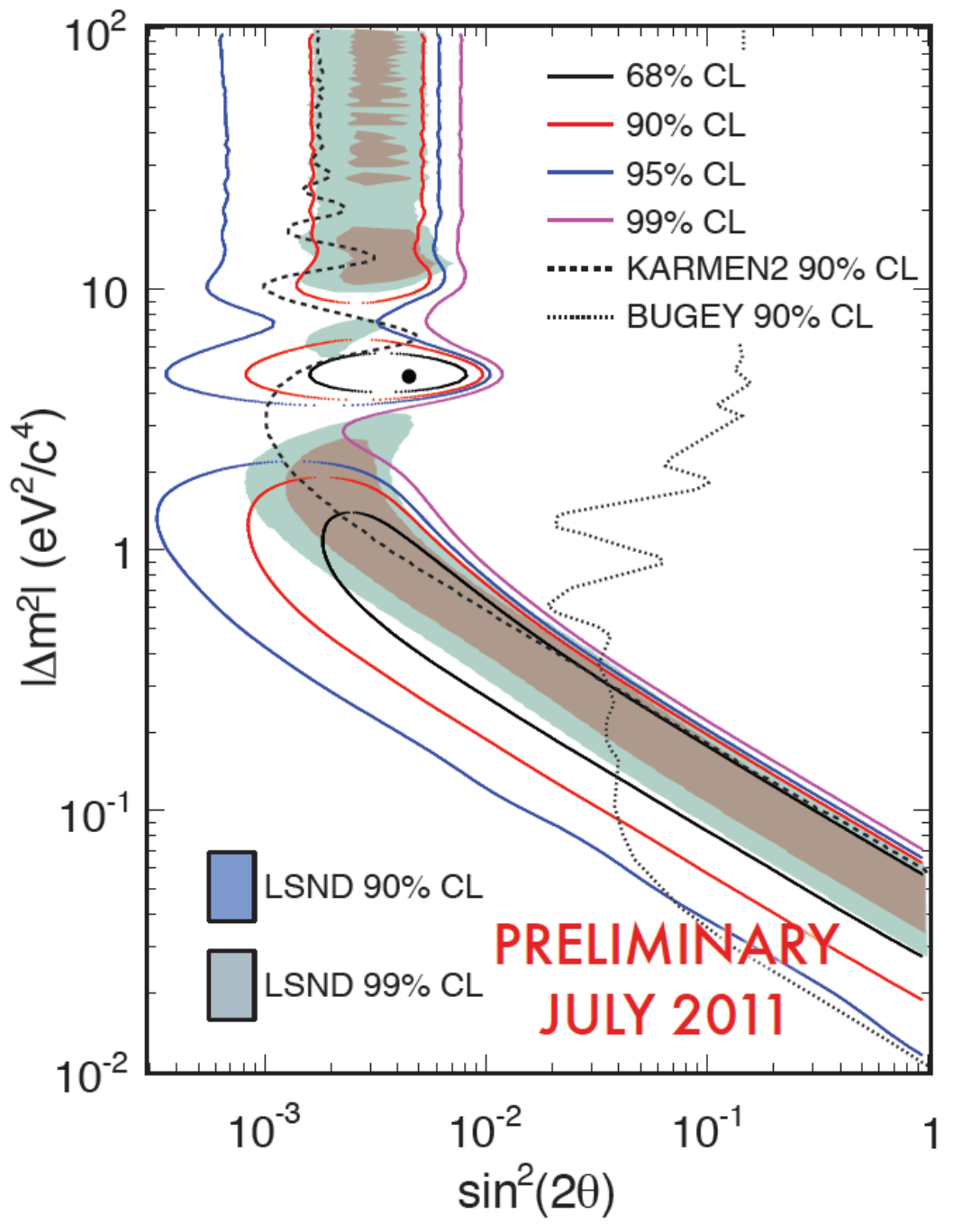}}
  \vspace{-0.3cm}
  \caption{\label{MB_allowed} The MiniBooNE oscillation allowed region in
    antineutrino mode agrees well with LSND.}
\end{figure}

\begin{figure}
 \centerline{\includegraphics[height=0.8\textwidth,angle=90]{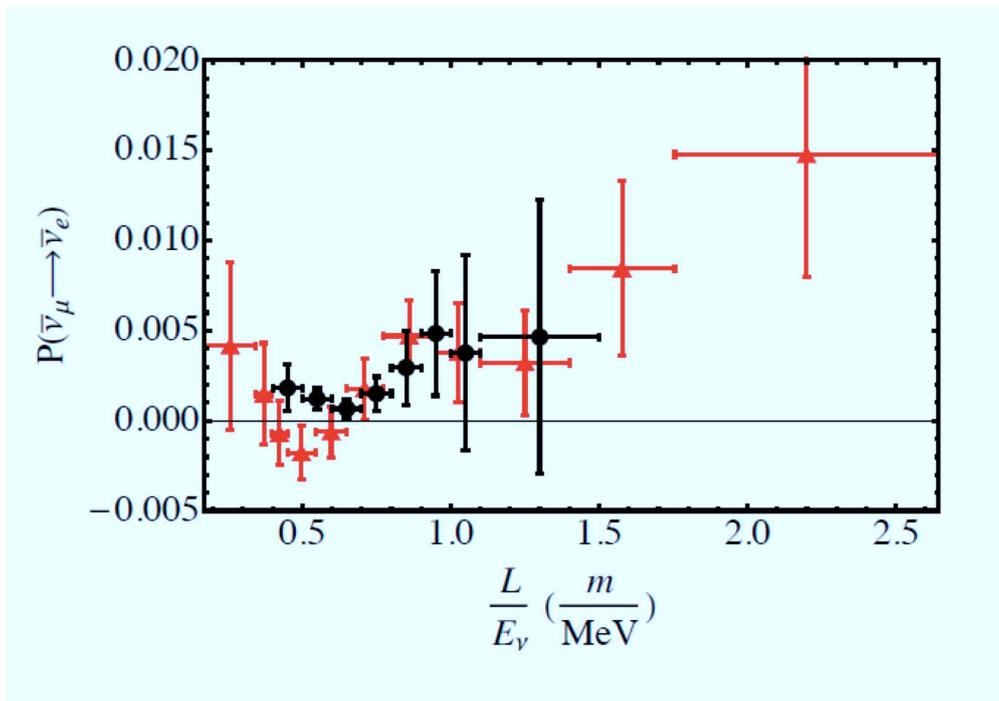}}
  \vspace{-0.5cm}
  \caption{\label{MB_LE} The MiniBooNE $L/E$ distribution in antineutrino
    mode (red data points) agrees well with LSND (black data points).}
\end{figure}

At present, with 8.58E20~POT in antineutrino mode, MiniBooNE observes an event
excess, $54.9 \pm 17.4 \pm 16.3$ events in the $200<E_\nu<1250$~MeV energy
range, which is consistent with the antineutrino oscillations suggested by the
LSND data~\cite{Aguilar:2001ty}.  Fig.~\ref{MB_antineutrino_data} shows the energy
distributions of the MiniBooNE $\bar \nu_e$ candidate events and the expected
background, while Fig.~\ref{MB_antineutrino_excess} shows the excess of $\bar
\nu_e$ candidate events in antineutrino mode. Figs.~\ref{MB_allowed} and
\ref{MB_LE} show that the MiniBooNE oscillation allowed region and $L/E$
distribution agree well with LSND.

\subsection{Disappearance Results from Accelerator Experiments}
\label{sec:disappearance}
\subsubsection{$\nu_{\mu}$ Disappearance from MiniBooNE and SciBooNE}

MiniBooNE performed an intitial search for $\nu_\mu$ and $\bar\nu_{\mu}$ disappearance 
in 2009~\cite {AguilarArevalo:2009yj}.  The search for $\nu_\mu$ disappearance was 
revisited with the inclusion of data from the SciBooNE 
detector~\cite{AguilarArevalo:2006se}, which operated concurently with MiniBooNE in 
the BNB for a period of time.  So far, no evidence for $\nu_\mu$ or $\bar\nu_{\mu}$ 
disappearance due to short-baseline oscillations has been found.  Limits have been 
placed on simple two neutrino mixing in the $\Delta m^{2}$ region of interest for 
sterile neutrino models. 

SciBooNE operated in the same beamline as MiniBooNE during 2007 and 2008, collecting 
data while the beam was operating in neutrino mode as well as antineutrino mode.  
SciBooNE was at a distance of 100~m from the target and consisted of three 
subdetectors.  The first part is the SciBar detector~\cite{Nitta:2004nt}, a finely 
segmented scintillation tracker that was relocated from the K2K neutrino beamline 
for cross section measurements at the neutrino energies available in the BNB.  
SciBar consists of 14,336 extruded polystyrene $(C_{8}H_{8})$ scintillator strips 
arranged horizontally and vertically in a $3 \times 3 \times 1.7$ ${\mathrm{m}^{3}}$ 
volume.  The 15~tons of the SciBar detector (10.6~tons fiducial volume)  provides 
the primary interaction target for SciBooNE.  The second sub-detector is an electron 
calorimeter (EC) constructed of 1~mm scintillating fibers embedded in lead foil and 
bundled into 64 modules.  The third subdetector is a muon range detector (MRD).  The 
MRD contains 12 iron plates sandwiched between scintillation counters, in 
alternating horizontal and vertical planes.  The iron plates are 5~cm thick and the 
counters 6~mm thick.  Photomultiplier tubes (PMTs) provide the readout for all three 
subdetectors.

\paragraph*{Neutrino Disappearance Event Selection}

A CCQE event in MiniBooNE is selected by identifying 
a single muon track and the muon's decay electron.  PMT timing information allows 
separating the neutrino interaction point from the decay electron.  PMT timing and 
charge response data are used to reconstruct muon location, energy, and direction.  
The number of hits in the outer (veto) region, beam timing information, and number 
of hits within the fiducial volume, are used to reject cosmic ray muons.  Particle 
ID cuts, based on event topology and reconstructions to alternate event hypotheses, 
are used to reject events that are not $\nu_\mu$ or $\bar{\nu}_{\mu}$ CCQE interactions.

Events that survive background rejection and particle ID cuts are binned in 
reconstructed neutrino energy, $E_\nu^{QE}$.  The reconstructed neutrino energy is 
computed assuming that all events in the final sample are CCQE, $\nu_l + N 
\rightarrow l + N'$.  This same assumption is applied to all events in data and 
Monte Carlo (MC), even though a sizeable fraction of the events are CC$1\pi$ or 
multi-nucleon knockout events misidentified as CCQE events.  16 data bins from 0 to 
1.9~GeV were used.  The first bin was from 0 to 400~MeV and the remaining bins were 
100~MeV wide each.

The MiniBooNE data set provided 190,454 events from $5.58 \times 10^{20}$ protons 
on target (POT) collected prior to SciBooNE operation (MiniBooNE neutrino mode Run~I).  
In addition, it included 29269 events from $0.83 \times 10^{20}$~POT taken 
concurrently with the SciBooNE run (MiniBooNE neutrino mode Run~II).  In 
antineutrino mode, there were 27,053 events in MiniBooNE's final sample, from $3.39 
\times 10^{20}$~POT.  The event rate as a function of POT is reduced by about a 
factor of five in antineutrino mode due to lower production and interaction rates.  

SciBooNE collected neutrino mode data for $0.99 \times 10^{20}$ POT.  
$\nu_\mu(\bar{\nu}_{\mu})$ CCQE interactions in SciBooNE are identified based on a 
single muon track~\cite{Nakajima:2010fp}.  The muon direction and energy is 
reconstructed based on the observed hits in each subdetector.  Events are 
subdivided based on whether the muon stopped in the SciBar fiducial volume, within 
the MRD, or exited the MRD.  Energy reconstruction was not possible for the MRD 
penetrated event sample.  After event selection cuts and subtracting the prediction 
for cosmic ray events, the SciBar stopped sample contained 13589 events, the MRD 
stopped sample contained 20236 events, and the MRD penetrated sample containted 3544 
events.  Data were binned in 16 equally spaced bins from 300~MeV to 1.9~GeV.

\paragraph*{Disappearance Signal and Background Reactions}

The MiniBooNE event samples were mostly $\nu_\mu(\bar\nu_{\mu})$ CCQE events, based 
on requiring two subevents (the initial interaction and the decay electron from the 
subsequent muon).  The neutrino-mode sample had an estimated 74\% CCQE purity, 
while the purity of the antineutrino mode sample was estimated at 70\%.  The 
antineutrino mode sample had roughly a 25\% $\nu_\mu$ content, while the 
$\bar{\nu}_{\mu}$ content of the neutrino mode sample was negligible.  The primary 
background was CC$1\pi$ events where the outgoing pion is unobserved due to 
absorption in the nucleus.

SciBooNE's SciBar stopped sample was estimated to consist of 51\% CCQE, 31\% 
CC$1\pi$, and the remainder CC or neutral current (NC) multi-pion events.  The MRD 
stopped sample had an estimated 52\% CCQE content and 34\% CC$1\pi$ content and 
the MRD penetrated sample had an estimated 57\% CCQE and 32\% CC$1\pi$ content.

Dominant uncertainties are the production of $\pi^{+}$ in the target ($\pi^{+}$ and 
$\pi^{-}$ for antineutrino mode) and CCQE and CC$1\pi$ cross sections for 
interactions in the detector.  Detector related uncertainties include light 
propagation and scattering in the mineral oil and PMT response.  SciBooNE specific 
systematic uncertainties include muon energy loss in the scintillator and iron 
plates, light attenuation in the wavelength shifting fibers used for SciBar readout, 
and PMT response.

MiniBooNE (and SciBooNE) can not distinguish between $\nu_\mu$ and $\bar{\nu}_{\mu}$ 
events on an event-by-event basis.  This is not a significant problem in neutrino 
mode, where the $\bar\nu_{\mu}$ contamination is small.  However, in antineutrino 
mode a sizeable fraction of the events are due to $\nu_\mu$ interactions.  In the 
oscillation analyses, neutrinos of the wrong sign ({\it i.e.}~antineutrino in neutrino 
mode or neutrino in antineutrino mode) are assumed to not oscillate, unless otherwise 
stated.  

A simple two neutrino oscillation model is used in the fits.  MiniBooNE was at a 
distance of 541~m from the target and SciBooNE was at a distance of 100~m from the 
target.   Mean neutrino flight distance, estimated from MC, was 76~m for SciBooNE 
events and 520~m for MiniBooNE events.  Oscillations at the SciBooNE location are 
non-negligible for some of the physics parameter space of interest.  Matter effects 
are assumed to be negligible.

\paragraph*{MiniBooNE $\nu_\mu$ and $\bar{\nu}_{\mu}$ Disappearance Results}

\begin{figure}[t]
\centering
\includegraphics[height=12cm]{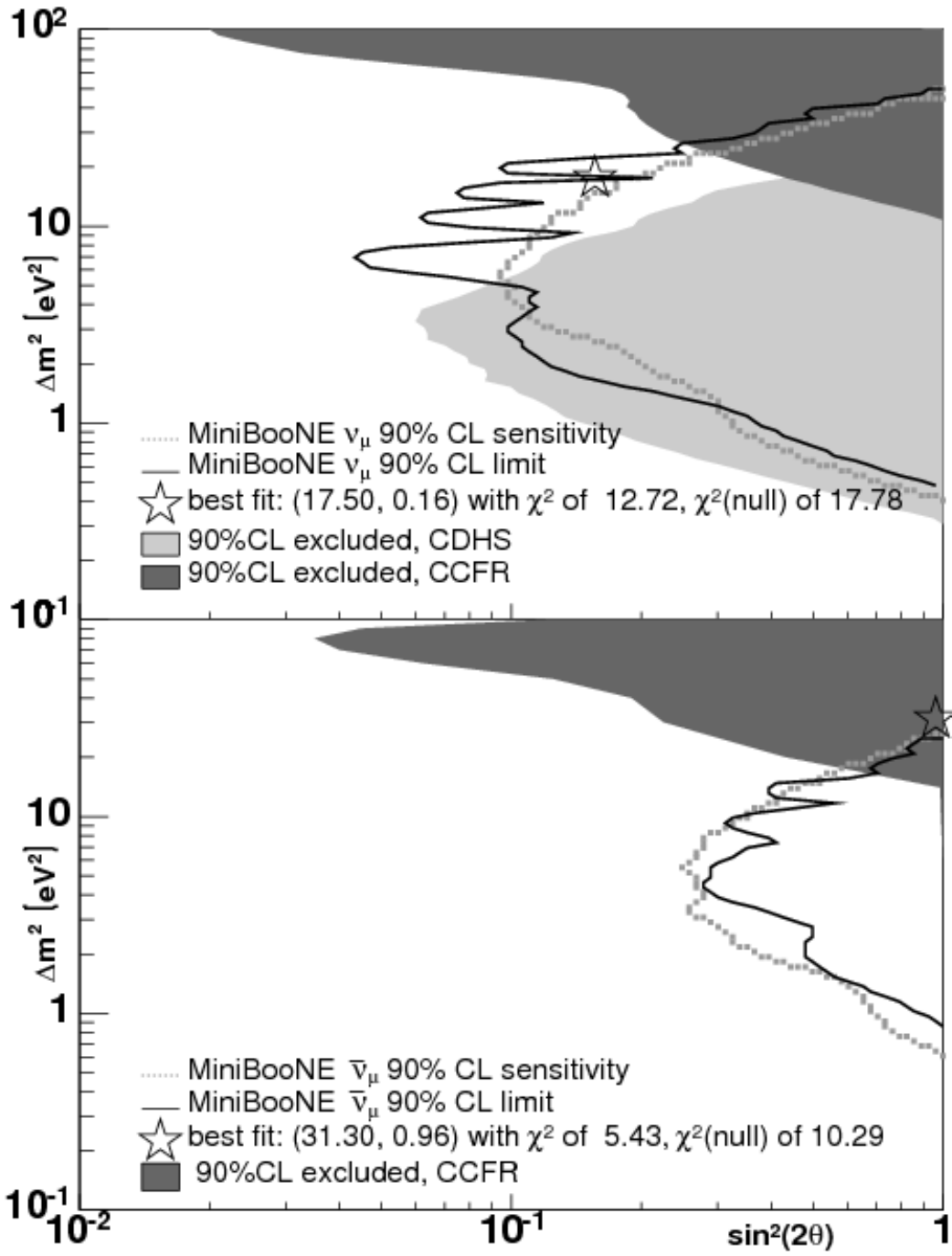}
\caption{$90\%$ CL sensitivity (dashed line) and limit (solid line) for MiniBooNE 
$\nu_\mu$ disappearance analysis (top) and $\bar\nu_{\mu}$ disappearance analysis 
(bottom).  Previous limits from CCFR (dark gray) and CDHSW (light gray) are also 
shown.}
\label{2009_disap}
\end{figure}

The 2009 MiniBooNE-only oscillation search was a shape-only test.  The MC prediction 
was normalized to data and the normalization uncertainty was removed from the error 
matrix.  A Pearson's $\chi^2$ test was used to determine confidence level (CL) 
boundaries.  A $16 \times 16$ error matrix accounted for bin-to-bin correlations in 
the uncertainties, which included flux, cross section, and detector systematic errors.  
The number of degrees of freedom (DOF) were determined with frequentist studies.  
The results of the analysis are shown in Fig.~\ref{2009_disap} (from 
Ref.~\cite{AguilarArevalo:2009yj}).  Also shown are the results from 
CCFR~\cite{Stockdale:1984cg,Stockdale:1984ce} ($\nu_\mu$ and $\bar\nu_{\mu}$) and 
CDHSW~\cite{Dydak:1983zq} ($\nu_\mu$).  More details on the CCFR and CDHSW results 
are provided below.

For the $\nu_\mu$ disappearance analysis, the $\chi^2$ for the null hypothesis (no 
disappearance) was 17.78.  Based on 16 DOF, this implies a $34\%$ probability.  The 
best fit point was at $\Delta m^2 = 17.5$~eV$^2$ and $\sin^2 2\theta = 0.16$.  With 
a $\chi^2$ of 12.72 for 13~DOF, this has a 47\% probability.

For the $\bar{\nu}_{\mu}$ disappearance analysis, only the $\bar{\nu}_{\mu}$ were 
assumed to oscillate in the fit.  The $\chi^2$ for the null hypothesis was 10.29 for 
16~DOF, a probability of 85.1\%.  The best fit point was at $\Delta m^2=31.32$~eV$^2$ 
and $\sin^2 2\theta = 0.96$, with a $\chi^2$ of 5.43 for 11~DOF, a 
99.5\% probability.  When neutrinos and antineutrinos were allowed to oscillate in 
the fit, each with the same mixing parameters, the best fit point was at $\Delta m^2 = 
31.32$~eV$^2$ and $\sin^2 2\theta = 0.44$, with a $\chi^2$ of 7.90, a 
99.2\% probability.

\paragraph*{Joint MiniBooNE/SciBooNE $\nu_\mu$ Disappearance Results}

Two complementary disappearance analyses were performed using the MiniBooNE and 
SciBooNE  $\nu_\mu$ CCQE event samples.  In the simultaneous fit method, data from 
both detectors were fit simultaneously to a two neutrino oscillation model.  In the 
spectrum fit method, SciBooNE data were used to determine energy dependent scale 
factors that were then applied to the MiniBooNE simulation.  For both methods, a 
likelihood ratio test was used for the test statistic.

The simultaneous fit method used a total of 48 bins; 16 bins for the SciBar stopped 
data, 16 bins for the MRD stopped data, and 16 bins for the MiniBooNE data (either 
the Run I or the Run II MiniBooNE data, which were handled separately).  The 
prediction in each bin was normalized by the ratio of the total number of observed 
events in the MRD stopped and SciBar stopped data sets divided by the total number 
of predicted events for these data sets.  The covariance matrix accounted for 
MiniBooNE uncertainties, SciBooNE uncertainties, as well as the correlations between 
the two detectors (a full 48 bin by 48 bin error matrix).

The spectrum fit method used a total of 32 bins; 16 bins for MiniBooNE Run I data, 
and 16 bins for MiniBooNE Run II data.  The data in each bin were renormalized based 
on the data to MC ratio for the SciBooNE data, similar to the spectrum fit method.  
In addition, energy-dependent scale factors were applied to the MiniBooNE data.  
These scale factors were determined from SciBooNE data and constrained some of the 
neutrino flux and cross section uncertainties.  Six scale factors, for six different 
true neutrino energy bins, were extracted by tuning the CC inclusive event 
predictions for all three SciBooNE event subsets.  Uncertainties in these scale 
factors were incorporated in the MiniBooNE $32 \times 32$ error matrix.

\begin{figure}[t]
\centering
\includegraphics[height=10cm]{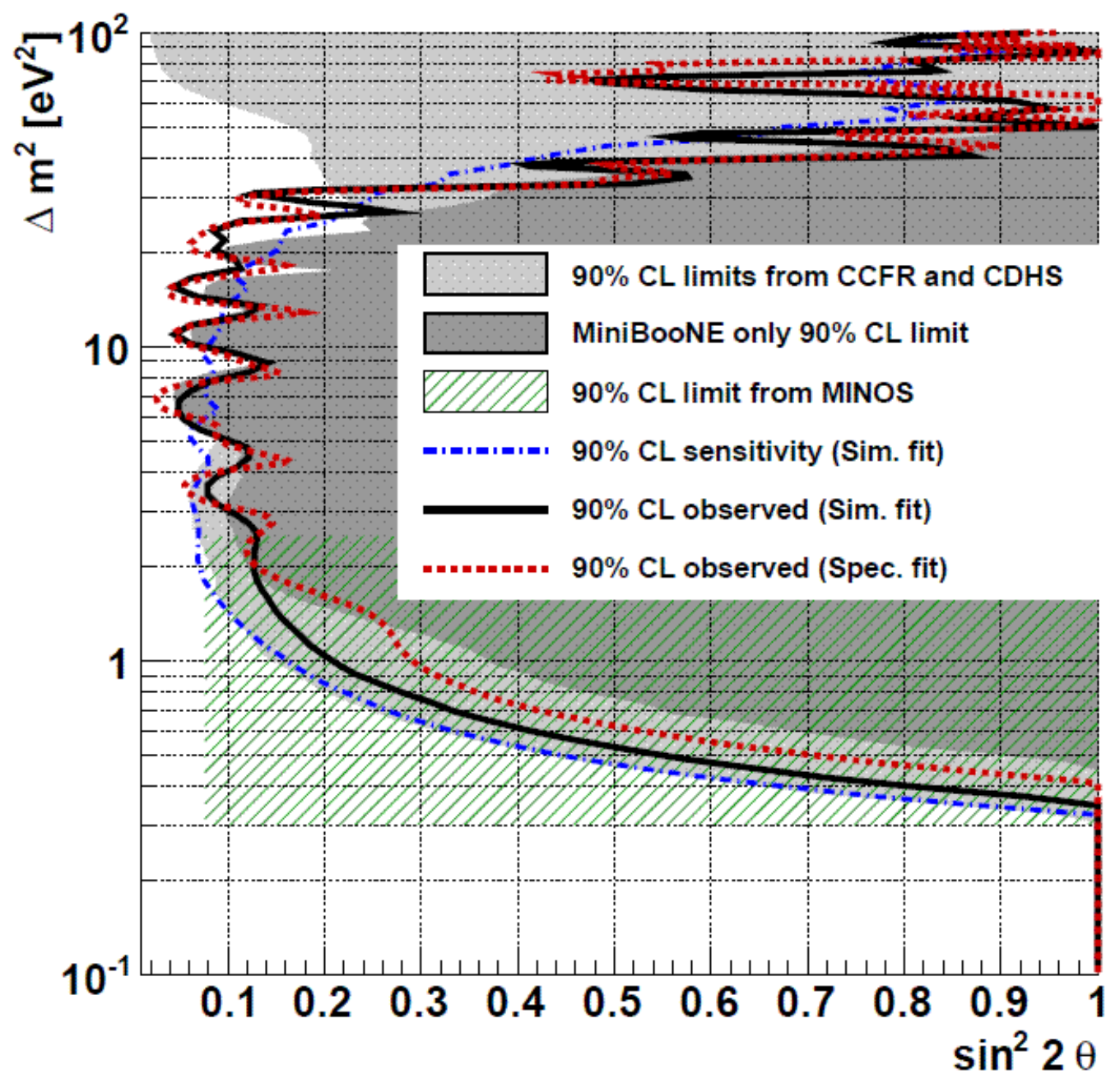}
\caption{$90\%$ CL sensitivity (dot-dash curve) and $90\%$ CL limit (solid black 
curve) from simultaneous MiniBooNE/SciBooNE fit, and $90\%$ CL limit from the 
spectrum fit method (red dashed curve).  Previous limits from CCFR, CDHSW, MINOS, 
and MiniBooNE are also shown.}
\label{jt_numu}
\end{figure}

The results of this analysis can be seen in Fig.~\ref{jt_numu} (from 
Ref.~\cite{Mahn:2011ea}).  Previous limits from 
CCFR~\cite{Stockdale:1984cg,Stockdale:1984ce}, CDHSW~\cite{Dydak:1983zq}, 
MINOS~\cite{Adamson:2011ku}, and MiniBooNE~\cite{AguilarArevalo:2009yj} are shown, 
for comparison.  No evidence for $\nu_\mu$ disappearance was found.  However, 
limits on $\sin^2 2\theta$ were improved, relative to previous results, in the 
region $10 \to 30$~eV$^2$.  Below $\Delta m^2=5$~eV$^2$, the 90\%~CL 
limit excludes less parameter space than expected from  the 90\%~CL sensitivity 
curve.

For the simultaneous fit, the $\chi^2$ for the null hypothesis was 45.1 for a $59\%$ 
probability (48 DOF).  Using MiniBooNE Run I data, the best fit point was at $\Delta 
m^{2}=43.7$~eV$^2$, $\sin^2 2\theta=0.60$, which had a $\chi^2$ of 39.5.  
The best fit point using Run II data had a $\chi^2$ of 41.5.  Combining the two 
MiniBooNE data run periods provided negligible improvement relative to the Run I 
data alone.  For the spectrum fit method, the $\chi^2$ for the null hypothesis was 
41.5 for a $12\%$ probability (32 DOF).  The best fit point was at 
$\Delta m^{2}=41.7$~eV$^2$, $\sin^2 2\theta=0.51$, which had a $\chi^2$ of 
35.6\@.  In Fig.~\ref{jt_numu}, the $90\%$ CL limit curve for the simultaneous fit is 
based on a $\Delta\chi^{2}$ of 9.34.  For the spectrum fit method, the $\Delta\chi^{2}$ 
value for the 90\%~CL limit curve is 8.41.

A joint search for $\bar\nu_{\mu}$ disappearance using data from MiniBooNE and 
SciBooNE is underway.  This new analysis will take advantage of neutrino cross 
section measurements from neutrino mode data, as well as constraints on neutrino 
background in the antineutrino beam.   In particular, the new analysis incorporates 
information learned from MiniBooNE's measurement of $\nu_\mu$ CCQE cross sections 
and CC$1\pi$ backgrounds in neutrino mode~\cite{AguilarArevalo:2010zc}.  The 
normalization of the neutrino contamination in the antineutrino beam is adjusted 
based on direct measurements~\cite{AguilarArevalo:2011sz,Grange:2011zi}.  A new 
$K^+$ constraint from SciBooNE~\cite{Cheng:2011wq} reduces the MC estimate for 
$K^+$ production in the beam and reduces its uncertainty.

The test statistic is a likelihood ratio for a two-neutrino oscillation fit.  
Normalization information is included (not a shape-only analysis).   MiniBooNE's 
antineutrino mode data from $7.4 \times 10^{20}$~POT and SciBooNE's antineutrino 
mode data from $1.53 \times 10^{20}$~POT will be used.  SciBooNE SciBar-stopped and 
MRD-stopped samples are combined.  $\nu_\mu$ are assumed to not oscillate, based on 
the results of the $\nu_\mu$ disappearance analyses discussed above.  MiniBooNE and 
SciBooNE CCQE event samples are in 21 bins, based on $E_\nu^{QE}$ from 300~MeV to 
1.9~GeV.  Both histograms are simultaneously fit to MC predictions.  A $42 \times 
42$ error matrix accounts for correlated systematic uncertainties.

The combination of improved constraints from MiniBooNE and SciBooNE, a MiniBooNE 
antineutrino dataset more than twice as large as that available for the 2009 
disappearance analysis, and various changes in the analysis methodology, will 
significantly improve our sensitivity to muon antineutrino disappearance.  Also 
planned is a joint MiniBooNE/SciBooNE analysis that will combine the neutrino mode 
and antineutrino mode data, further enhancing sensitivty to $\nu_\mu$ 
$(\bar\nu_{\mu})$ disappearance.

\subsubsection*{$\nu_\mu$ Disappearance Results from CDHSW}

The CDHSW (Cern Dortmund Heidelberg Saclay Warsaw) experiment performed a search 
for $\nu_\mu$ disappearance~\cite{Dydak:1983zq}.  19.2~GeV protons struck a 
beryllium target, producing a shower of mesons that subsequently entered a 52~m long 
decay tunnel.  The neutrino flux, produced by the decay of pions, kaons, and muons, 
peaked at about 1~GeV.  The mean energy of accepted neutrino events was about 3~GeV.  
No magnetic focusing of the meson beam was employed.  Two detectors operated 
simulataneously along the same beam line, enabling the reduction of systematic 
uncertrainties in flux and cross sections.

The far detector, the original CDHSW detector, was located 885 m from the proton 
target.  The near detector was located 130~m from the target.  Each detector was 
made up of combinations of three different types of modules.  The three module 
types consisted of layers of circular iron plates and scintillator hodoscopes 
sandwiched together, with PMTs for readout, but each module differed in the number 
of layers and the dimensions of the layers.  The iron plates were not magnetized.  
The far detector was made up of a total of 21 of these modules and the near detector 
consisted of a total of six modules.

Timing and event selection cuts were used to eliminate most of the cosmic ray 
backgorund and events from neutrino interactions that occured outside the detector.  
Events were classified based on whether they originated in a module of type~I or 
type~II, and ratios of events in the near detector to events in the far detector 
were determined separately for the two module types.  Events starting near the 
boundaries between modules were rejected.

Approximately 22,000 events passed selection cuts for the near detector, with a 
background contamination of 50 cosmic ray events.  For the far detector, these 
numbers were 3300 and 290.  Since the event range in the iron was a function of 
incident neutrino energy, events were binned according to this range projected onto 
the detector axis.

The far-to-near event ratios $(R_{data})$ for each bin were corrected based on the 
square of the distance to the target and the detector masses.  These ratios were 
also determined in MC that accounted for details of the beamline and the angular 
spread of the flux.  The binned values of $R_{data}/R_{MC}$ were then used in the 
disappearance analysis, based on two-neutrino mixing.  Systematic uncertainty in the 
ratio of event rates was estimated at $2.5\%$.  The data showed no significant signs 
of oscillations.  The $90\%$`CL result from CDHSW is shown in Figs.~\ref{2009_disap} 
and \ref{jt_numu}.

\subsubsection*{$\nu_\mu$ and $\bar\nu_{\mu}$ Disappearance Results from CCFR}

The CCFR experiment~\cite{Stockdale:1984cg,Stockdale:1984ce} performed searches for 
$\nu_\mu$ disappearance and $\bar\nu_{\mu}$ disappearance as well as a combined 
analysis where the $\nu_\mu$ and $\bar\nu_{\mu}$ were assumed to have identical 
oscillation parameters.  Two detectors were operated at different distances along 
Fermilab's narrow-band beam line.  Five different $\pi^+$ and $K^{+}$ momentum 
settings were used: 100, 140, 165, 200, and 250~${\mathrm{GeV}}/c$.  Mesons produced in 
the target were selected for sign and momentum by a series of dipole and quadrupole 
magents.  Subsequent neutrino energies ranged from 40 to 230~GeV.  For the 
$\bar\nu_{\mu}$ beam, a single momentum setting of 165~${\mathrm{GeV}}/c$ was used.

The detectors were located at 715 and 1116 m from the midpoint of a 352~m long decay 
pipe.  Each detector consisted of a calorimeter made up of steel plates instrumented 
with scintillation counters and spark chambers, followed by an iron toroidal muon s
pectrometer.  The near detector calorimeter contained 108 tons of steel plates while 
the far detector calorimeter contained 4444 tons.  Also, the near detector muon 
spectrometer had a single toroid while the far detector had three toroidal magnets.  
Muon neutrino and antineutrino CC interactions on iron produced a shower of hadrons 
measured by the iron calorimeter and a muon measured by the toroidal magnet(s).   The 
neutrino energy was determined by summing the measured hadron and muon energies.  For 
each beam setting, the events in each detector were divided into three energy bins.

Timing and event selection cuts were used to reduce background and to ensure that the 
same geometrical acceptance was realized for both detectors.  After final event 
selection, the neutrino data consisted of 33,700 events in the near detector and 
33,400 events in the far detector.  For antineutrino data,  the near detector sample 
contained 4960 events while the far detector sample contained 4600 events.

The oscillation analysis was based on the ratio of the number of events in each 
detector as a function of energy.    The mean neutrino energy for each bin depended on 
the beam setting, and was determined from MC simulation.  The ratios had to be 
corrected for differences between the detectors, including detector live times, veto 
counter dead times and reconstruction efficiencies.  Additionally, MC was used to 
account for the effects of the finite size of the decay pipe. A likelihood ratio 
method was used to set $90\%$~CL limits. 

CCFR results can be seen in Figs.~\ref{2009_disap} and \ref{jt_numu}.  For the 
$\nu_\mu$ disappearance analysis, the $\chi^2$ for the null hypothesis was 11.0 for 
15 DOF.   For the $\bar\nu_{\mu}$ disappearance analysis, the $\chi^2$ for the null 
hypothesis was 4.7 for 3 DOF.  For the combined oscillation analysis (not shown) 
$\nu_\mu$ and $\bar\nu_{\mu}$ data were combined and assumed to have identical mixing 
parameters.  In this case, the $\chi^2$ for the null hypothesis was 15.7 for 18~DOF.

\subsection{The Gallium Anomaly}
\label{sec:gallium}

The GALLEX \cite{Anselmann:1994ar, Hampel:1997fc,Kaether:2010ag} and SAGE
\cite{Abdurashitov:1996dp, Abdurashitov:1998ne, Abdurashitov:2005tb,Abdurashitov:2009tn} solar
neutrino detectors (see Refs.~\cite{Bilenky:1978nj, Bilenky:1987ty,
Bilenky:1998dt, GonzalezGarcia:2002dz, Giunti:2003qt, Maltoni:2004ei, Fogli:2005cq,
Strumia:2006db, Giunti-Kim-2007}) have been tested in so-called ``Gallium
radioactive source experiments'' which consist in the detection of electron
neutrinos produced by intense artificial ${}^{51}\text{Cr}$ and
${}^{37}\text{Ar}$ radioactive sources placed inside the detectors.

\begin{figure}[t!]
\null
\hfill
\includegraphics*[width=0.4\linewidth]{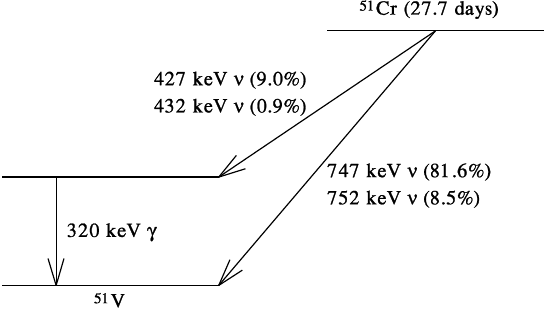}
\hfill
\includegraphics*[width=0.4\linewidth]{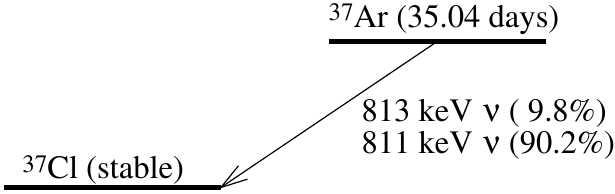}
\hfill
\null
\caption{\label{fig-ec}
Nuclear levels for the elecron capture processes in Eqs.~(\ref{Cr}) and
(\ref{Ar}).  Figures from Ref.~\protect\cite{Abdurashitov:1998ne} and
Ref.~\protect\cite{Abdurashitov:2005tb}.}
\end{figure}

The radioactive nuclei ${}^{51}\text{Cr}$ and ${}^{37}\text{Ar}$ decay through
electron capture,
\begin{align}
\null & \null
e^{-} + {}^{51}\text{Cr} \to {}^{51}\text{V} + \nu_{e}
\,,
\label{Cr}
\\
\null & \null
e^{-} + {}^{37}\text{Ar} \to {}^{37}\text{Cl} + \nu_{e}
\,,
\label{Ar}
\end{align}
emitting $\nu_{e}$ lines with the energies and branching ratios listed in
Tab.~\ref{tab-CrAr}.  The nuclear levels for the two processes are shown in
Fig.~\ref{fig-ec}.  The two energy lines for each nuclear transition are due to
electron capture from the K and L shells.

The neutrinos emitted by the radioactive sources have been detected through the
same reaction used for the detection of solar neutrinos \cite{Kuzmin-Ga-65}:
\begin{equation}
\nu_{e} + {}^{71}\text{Ga} \to {}^{71}\text{Ge} + e^{-}
\,,
\label{GaGe}
\end{equation}
which has the low neutrino energy threshold $
E_{\nu}^{\text{th}}({}^{71}\text{Ga}) = 0.233 \, \text{MeV} $.  The cross
sections of the $\nu_{e}$ lines emitted in ${}^{51}\text{Cr}$ and
${}^{37}\text{Ar}$ decay interpolated from Tab.~II of
Ref.~\cite{Bahcall:1997eg} are listed in Tab.~\ref{tab-CrAr}.

The ratios $R_{\text{B}}^{k}$ of measured and predicted ${}^{71}\text{Ge}$
production rates in the GALLEX ($k=\text{G1},\text{G1}$) and SAGE
($k=\text{S1},\text{S1}$) radioactive source experiments are listed in
Tab.~\ref{tab-exp}, together with the spatial characteristics of the
experiments.  The value of the average ratio is
\begin{equation}
R^{\text{Ga}}_{\text{B}}
=
0.86 \pm 0.05
\,.
\label{RGaB}
\end{equation}
Thus, the number of measured events is about $2.8\sigma$ smaller than the
prediction.  This is the ``Gallium anomaly'', which could be a manifestation of
short-baseline neutrino oscillations \cite{Bahcall:1994bq, Laveder:2007zz,
Giunti:2006bj, Giunti:2007xv, Acero:2007su, Giunti:2009zz, Giunti:2010wz, Gavrin:2010qj,
Giunti:2010zu}.

\begin{table}
\begin{center}
\begin{ruledtabular}
\begin{tabular}{l|cccc|cc}
&
\multicolumn{4}{c|}{${}^{51}\text{Cr}$}
&
\multicolumn{2}{c}{${}^{37}\text{Ar}$}
\\
\hline
$E_{\nu}\,[\text{keV}]$ & $ 747 $ & $ 752 $ & $ 427 $ & $ 432 $ & $ 811$ & $ 813$ \\
B.R. & $0.8163$ & $0.0849$ & $0.0895$ & $0.0093$ & $0.902$ & $0.098$ \\
$\sigma\,[10^{-46}\,\text{cm}^{2}]$ & $ 60.8 $ & $ 61.5 $ & $ 26.7 $ & $ 27.1 $ & $ 70.1$ & $70.3 $ \\
\end{tabular}
\end{ruledtabular}
\caption{\label{tab-CrAr}
  Energies ($E_{\nu}$), branching ratios (B.R.) and Gallium cross sections
  ($\sigma$) of the $\nu_{e}$ lines emitted in ${}^{51}\text{Cr}$ and
  ${}^{37}\text{Ar}$ decay through electron capture.  The cross sections are
  interpolated from Tab.~II of Ref.~\protect\cite{Bahcall:1997eg}.}
\end{center}
\end{table}

\begin{table}[t!]
\begin{center}
\begin{ruledtabular}
\begin{tabular}{l|cc|cc}
&
\multicolumn{2}{c|}{GALLEX}
&
\multicolumn{2}{c}{SAGE}
\\
\hline
k & G1 & G2 & S1 & S2
\\
source & ${}^{51}\text{Cr}$ & ${}^{51}\text{Cr}$ & ${}^{51}\text{Cr}$ & ${}^{37}\text{Ar}$ \\
$R_{\text{B}}^{k}$ & $ 0.953 \pm 0.11 $ & $ 0.812 {}^{+0.10}_{-0.11} $ & $ 0.95 \pm 0.12 $ & $ 0.791 \pm {}^{+0.084}_{-0.078} $ \\
$R_{\text{H}}^{k}$ & $ 0.84 {}^{+0.13}_{-0.12} $ & $ 0.71 {}^{+0.12}_{-0.11} $ & $ 0.84 {}^{+0.14}_{-0.13} $ & $ 0.70 \pm {}^{+0.10}_{-0.09} $ \\
radius [m] & \multicolumn{2}{c|}{$1.9$} & \multicolumn{2}{c}{$0.7$} \\
height [m] & \multicolumn{2}{c|}{$5.0$} & \multicolumn{2}{c}{$1.47$} \\
source height [m] & $2.7$ & $2.38$ & \multicolumn{2}{c}{$0.72$} \\
\end{tabular}
\end{ruledtabular}
\caption{ \label{tab-exp}
  Experiment index $k$, source type and ratios $R_{\text{B}}^{k}$ and
  $R_{\text{H}}^{k}$ of measured and predicted ${}^{71}\text{Ge}$ production
  rates in the GALLEX and SAGE radioactive source experiments.  We give also the
  radii and heights of the GALLEX and SAGE cylindrical detectors and the heights
  from the base of the detectors at which the radioactive sources were placed
  along the axes of the detectors.}
\end{center}
\end{table}

\begin{figure}[t!]
\null
\hfill
\includegraphics*[width=0.5\textwidth]{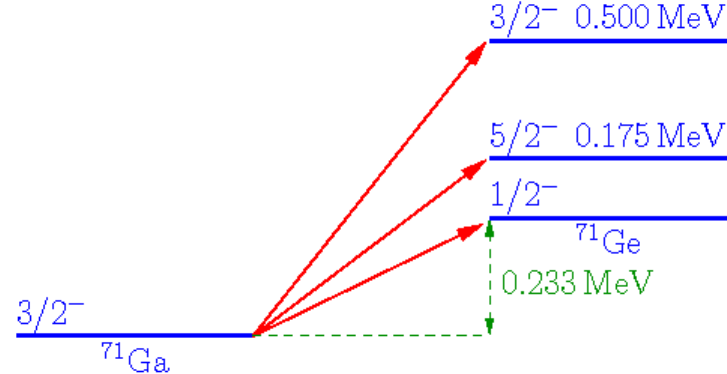}
\hfill
\null
\caption{ \label{fig-det} Nuclear levels for the Gallium detection process in
Eq.~(\ref{GaGe}).}
\end{figure}

\begin{figure}[t!]
\begin{center}
\includegraphics*[bb=5 11 571 571, width=0.6\linewidth]{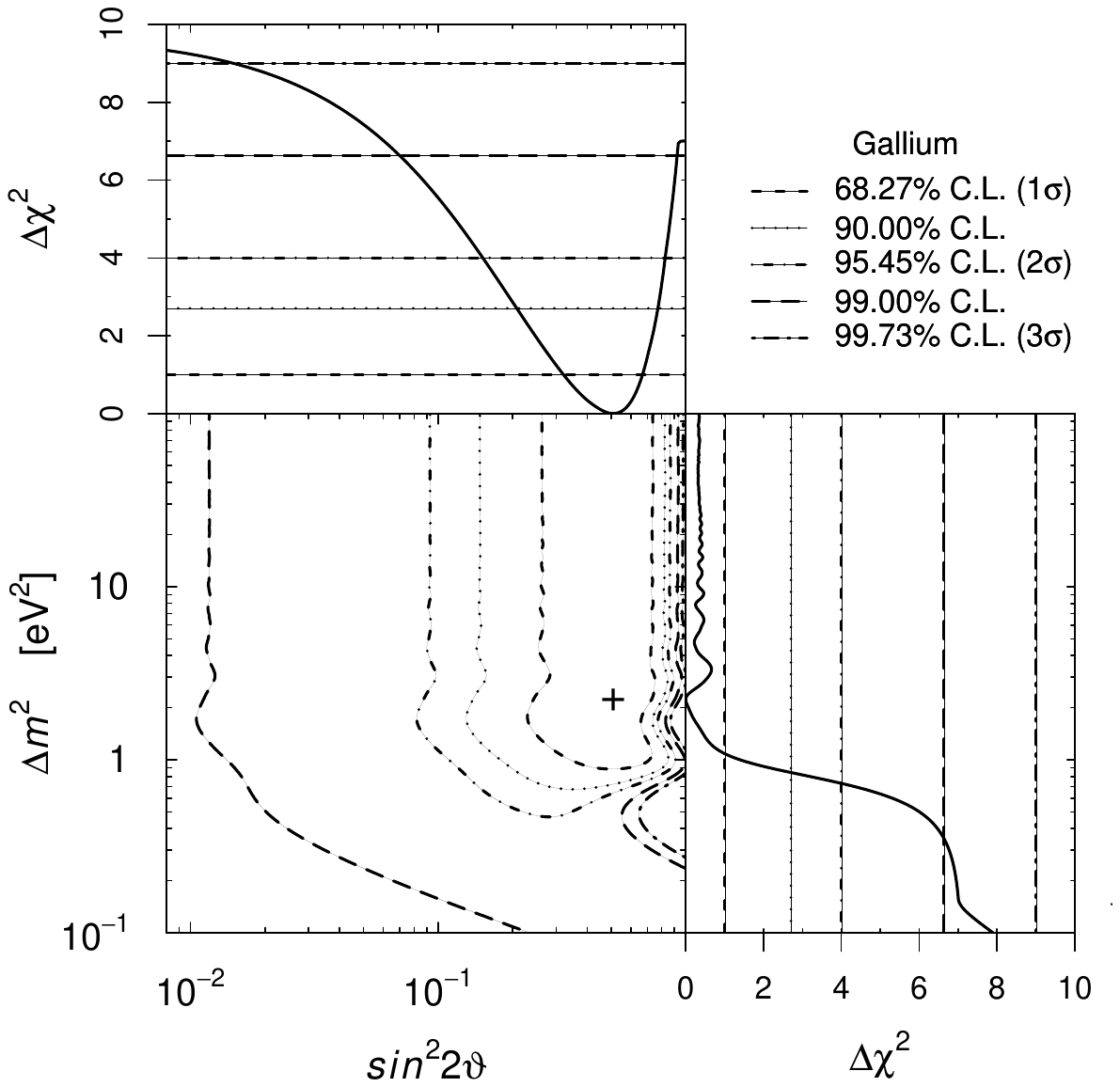}
\end{center}
\caption{ \label{fig-allowed}
Allowed regions in the $\sin^2 2\theta$--$\Delta{m}^{2}$ plane and marginal
$\Delta\chi^{2}$'s for $\sin^2 2\theta$ and $\Delta{m}^{2}$ obtained from
the combined fit of the results of the two GALLEX $^{51}\text{Cr}$
radioactive source experiments and the SAGE ${}^{51}\text{Cr}$ and
${}^{37}\text{Ar}$ radioactive source experiments \protect\cite{Giunti:2010zu}.
The best-fit point corresponding to $\chi^2_{\text{min}}$ is indicated by a
cross.  }
\end{figure}

As indicated by the ``B'' subscript, the ratios in Tab.~\ref{tab-exp} and
Eq.~(\ref{RGaB}) have been calculated with respect to the rate estimated using
the best-fit values of the cross section of the detection process (\ref{GaGe})
calculated by Bahcall \cite{Bahcall:1997eg},
\begin{align}
\sigma_{\text{B}}^{\text{bf}}({}^{51}\text{Cr})
=
\null & \null
\left( 58.1 {}^{+2.1}_{-1.6} \right) \times 10^{-46} \, \text{cm}^2
\,,
\label{007}
\\
\sigma_{\text{B}}^{\text{bf}}({}^{37}\text{Ar})
=
\null & \null
\left( 70.0 {}^{+4.9}_{-2.1} \right) \times 10^{-46} \, \text{cm}^2
\,.
\label{008}
\end{align}
The uncertainties of these cross sections are not taken into account in the
ratios in Tab.~\ref{tab-exp} and Eq.~(\ref{RGaB}).  These uncertainties are
large \cite{Hata:1995cw, Bahcall:1997eg, Haxton:1998uc}, because only the
cross section of the transition from the ground state of ${}^{71}\text{Ga}$ to
the ground state of ${}^{71}\text{Ge}$ is known with precision from the
measured rate of electron capture decay of ${}^{71}\text{Ge}$ to
${}^{71}\text{Ga}$.  Electron neutrinos produced by ${}^{51}\text{Cr}$ and
${}^{37}\text{Ar}$ radioactive sources can be absorbed also through transitions
from the ground state of ${}^{71}\text{Ga}$ to two excited states of
${}^{71}\text{Ge}$ at 175 keV and 500 keV (see Fig.~\ref{fig-det}), with cross
sections which are inferred using a nuclear model from $(p,n)$ measurements ($
p + {}^{71}\text{Ga} \to {}^{71}\text{Ge} + n $) \cite{Krofcheck:1985fg}.

Hence, at least part of the deficit of measured events with respect to the
prediction could be explained by an overestimation of the transitions to the
two excited states of ${}^{71}\text{Ge}$
\cite{Abdurashitov:2005tb,Fogli:2006fu,Abdurashitov:2009tn}.  However, since the
contribution of the transitions to the two excited states of ${}^{71}\text{Ge}$
is only 5\% \cite{Bahcall:1997eg}, even the complete absence of such
transitions would reduce the ratio of measured and predicted ${}^{71}\text{Ge}$
event rates to about $0.91\pm0.05$, leaving an anomaly of about $1.7\sigma$
\cite{Giunti:2010wz}.

The value and uncertainties of the detection cross section for the
${}^{51}\text{Cr}$ source have been calculated in a reliable way by Haxton in
Ref.~\cite{Haxton:1998uc} using the results of the $(p,n)$ measurements
\cite{Krofcheck:1985fg} and nuclear shell-model calculations.  His result is
\begin{equation}
\sigma_{\text{H}}({}^{51}\text{Cr}) = \left( 63.9 \pm 6.8 \right) \times 10^{-46} \, \text{cm}^2
\,.
\label{009}
\end{equation}
The mean value of this cross section is larger than the mean value of the
Bahcall cross section in Eq.~(\ref{007}).  This is due to a destructive
interference between the spin and spin-tensor matrix elements in the $(p,n)$
transition to the first excited states of ${}^{71}\text{Ge}$ according to
shell-model calculations.  Therefore, the contribution of the transition to the
first excited states of ${}^{71}\text{Ge}$ in $\nu_{e}$ detection is larger
than that inferred from the $(p,n)$ measurements, leading to a larger value for
the total cross section.

The ratios $R_{\text{H}}^{k}$ of measured and predicted ${}^{71}\text{Ge}$
production rates in the GALLEX and SAGE radioactive source experiments obtained
with the Haxton detection cross section and uncertainties \cite{Giunti:2010zu} are
listed in Tab.~\ref{tab-exp}.  The value of the average ratio is
\cite{Giunti:2010zu}
\begin{equation}
R^{\text{Ga}}_{\text{H}}
=
0.76 {}^{+0.09}_{-0.08}
\,.
\label{RGaH}
\end{equation}

The mean value of $R^{\text{Ga}}_{\text{H}}$ is smaller than that of
$R^{\text{Ga}}_{\text{B}}$, but the uncertainties of $R^{\text{Ga}}_{\text{H}}$
are rather large.  Taking both effects into account, the statistical
significance of the Gallium anomaly is about $2.7\sigma$, which is about equal
to the $2.8\sigma$ obtained with Eq.~(\ref{RGaB}).

A natural explanation of the Gallium anomaly is electron neutrino disappearance
due to short-baseline oscillations \cite{Bahcall:1994bq, Laveder:2007zz,
Giunti:2006bj,Giunti:2007xv,Acero:2007su,Giunti:2009zz,Giunti:2010wz,Gavrin:2010qj,Giunti:2010zu}
(another explanation based on quantum decoherence in neutrino oscillations has
been proposed in Ref.~\cite{Farzan:2008zv}).

Let us consider the electron neutrino survival probability
\begin{equation}
P_{\nu_{e}\to\nu_{e}}^{\text{SBL}}(L,E) = 1 - \sin^2 2\theta
\sin^2\!\left( \frac{ \Delta{m}^2 L }{ 4 E } \right)\,,
\label{026}
\end{equation}
where $\theta$ is the mixing angle, $\Delta{m}^2$ is the squared-mass
difference, $L$ is the neutrino path length and $E$ is the neutrino energy.
This survival probability is effective in short-baseline (SBL) experiments in
the framework of four-neutrino mixing schemes (see Refs.~\cite{Bilenky:1998dt,
Maltoni:2004ei, Strumia:2006db, GonzalezGarcia:2007ib}), which are the simplest
extensions of three-neutrino mixing schemes which can accommodate the two
measured small solar and atmospheric squared-mass differences
$\Delta{m}^2_{\text{SOL}} \simeq 8 \times 10^{-5} \, \text{eV}^2$ and
$\Delta{m}^2_{\text{ATM}} \simeq 2 \times 10^{-3} \, \text{eV}^2$ and one
larger squared-mass difference for short-baseline neutrino oscillations,
$\Delta{m}^2 \gtrsim 0.1 \, \text{eV}^2$.  The existence of a fourth massive
neutrino corresponds, in the flavor basis, to the existence of a sterile
neutrino $\nu_{s}$.

Figure~\ref{fig-allowed} shows the allowed regions in the
$\sin^{2}2\theta$--$\Delta{m}^{2}$ plane and the marginal $\Delta\chi^{2} =
\chi^2 - \chi^2_{\text{min}}$'s for $\sin^{2}2\theta$ and $\Delta{m}^{2}$,
from which one can infer the corresponding uncorrelated allowed intervals
\cite{Giunti:2010zu}.  The best-fit values of the oscillation parameters are
\begin{equation}
\sin^2 2\theta_{\text{bf}} = 0.50\,,
\quad
\Delta{m}^2_{\text{bf}} = 2.24 \, \text{eV}^2\,.
\label{028}
\end{equation}
The value of the likelihood ratio between the null hypothesis of no
oscillations and the oscillation hypothesis,
\begin{equation}
\frac{\mathcal{L}_{0}}{\mathcal{L}(\sin^2 2\theta_{\text{bf}},\Delta{m}^2_{\text{bf}})}
= 8 \times 10^{-3}\,,
\end{equation}
is in favor of the oscillation hypothesis.  It corresponds to $\Delta\chi^2 =
9.7$, which, with two degrees of freedom, disfavors the null hypothesis of no
oscillations at 99.23\% C.L.  ($2.7\sigma$).

In conclusion, the Gallium anomaly is statistically significant at a level of
about $2.7\sigma$.  The analysis of the data of the Gallium radioactive source
experiments in terms of neutrino oscillations indicates values of the
oscillation amplitude $\sin^{2}2\theta \gtrsim 0.07$ and squared-mass
difference $\Delta{m}^{2} \gtrsim 0.35 \, \text{eV}^2$ at 99\% C.L..

\subsection{The Reactor Antineutrino Anomaly}
\label{sec:reactor}

\subsubsection*{Antineutrinos from Reactors}

Nuclear reactors are very intense sources of neutrinos that have been used all
along the neutrino's history, from its discovery up to the most recent
oscillation studies. With an average energy of about 200~MeV released per
fission and 6 neutrinos produced along the $\beta$-decay chains of the fission
products, one expects about $2\times10^{20}$ $\nu/s$ emitted in a $4\pi$ solid
angle from a 1~GW reactor (thermal power). Since unstable fission products are
neutron-rich nuclei, all $\beta$-decays are of $\beta^-$ type and the neutrino
flux is actually pure electronic antineutrinos ($\bar\nu_e$).

The neutrino oscillation search at a reactor is always based on a disappearance
measurement, using the powerful inverse beta decay (IBD) detection process to
discriminate the neutrino signal from backgrounds. The observed neutrino
spectrum at a distance $L$ from a reactor is compared to the expected spectrum.
If a deficit is measured it can be interpreted in terms of the disappearance
probability which, in the two neutrino mixing approximation, reduces to 
\begin{eqnarray}
  P_{ee} &=& 1 - \sin^2 2\theta \, \sin^2 \left(\frac{\Delta m^2 L}{2E} \right) 
\end{eqnarray}
with $\Delta m^2$ the difference between the squared masses of the two
neutrino states and $\theta$ the mixing angle fixing the amplitude of the
oscillation.

Here, we will especially consider reactor antineutrino detector at short
distances below $100$~m from the reactor core, in particular ILL-Grenoble,
Goesgen, Rovno, Krasnoyarsk, Savannah River and Bugey~\cite{Kwon:1981ua,
Declais:1994su, Zacek:1986cu, Declais:1994ma, Afonin:1994, Kuvshinnikov:1991,
Vidyakin:1993, Vidyakin:1994ut, Greenwood:1996pb}. These experiments have
played an important role in the establishment of neutrino physics, and
escpecially neutrino oscillations, over the last fifty
years~\cite{Nakamura:2010zzi}.  Unlike modern long baseline reactor experiments
motivated by the measurement of the last unknown mixing angle
$\theta_{13}$~\cite{Ardellier:2006mn, Guo:2007ug, Ahn:2010vy}, which measure $P_{ee}$
by comparing the event rate and spectrum in two detectors at different
distances, the aforementioned short-baseline experiments can only employ one
detector and therefore depend on an accurate theoretical prediction for the
emitted $\bar\nu_e$ flux and spectrum to measure $P_{ee}$.

Until late 2010, all data from reactor neutrino experiments appeared to be
fully consistent with the mixing of $\nu_{e}$, $\nu_{\mu}$ and $\nu_{\tau}$
with three mass eigenstates, $\nu_1$, $\nu_2$ and $\nu_3$,  with the squared
mass differences $|\Delta m_{31}^2|\simeq2.4\,10^{-3}~{\mathrm eV}^2$ and $\Delta
m_{21}^2/|\Delta m_{31}^2|\simeq 0.032$.  The measured rate of $\bar\nu_e$ was
found to be in reasonable agreement with that predicted from the `old' reactor
antineutrino spectra~\cite{Schreckenbach:1985ep, Schreckenbach:1985ep,
VonFeilitzsch:1982jw, Hahn:1989zr}, though slightly lower than expected, with
the measured/expected ratio at $0.980\pm 0.024$, including recent revisions of
the neutron mean lifetime, $\tau_n=881.5~$s, in 2011~\cite{Nakamura:2010zzi}
(the cross section of the detection reaction of $\bar\nu_e$ on free protons
$\bar\nu_e+p\rightarrow e^{+}+n$ is inversely proportional to the neutron
lifetime). 

In preparation for the Double Chooz reactor experiment~\cite{Abe:2011fz},
the Saclay reactor neutrino group re-evaluated the specific reactor
antineutrino flux for $^{235}$U, $^{239}$Pu, $^{241}$Pu, and
$^{238}$U. In 2011, they reported their results~\cite{Mueller:2011nm},
which correspond to a flux that is a few percent higher than the previous
prediction. This also necessitates a reanalysis of
the ratio of observed event rate to predicted rate for 19 published
experiments at reactor--detector distances below $100$~m.

\subsubsection*{Reference antineutrino spectra}

The distribution of the fission products of uranium or plutonium isotopes
covers hundreds of nuclei, each of them contributing to $S_k(E)$ through
various $\beta$-decay chains. At the end the total antineutrino spectrum is a
sum of thousands of $\beta$-branches weighted by the branching ratio of each
transition and the fission yield of the parent nucleus. Despite the impressive
amount of data available in nuclear databases the {\it ab initio} calculation
of the emitted antineutrino spectrum is difficult. Moreover, when looking at
the detected spectrum through the IBD process, the 1.78~MeV threshold and the
quadratic energy dependence of the cross-section enhances the contribution of
transitions with large end-points ($E_0>4$~MeV). Systematic errors of the
nuclear data and the contribution of poorly known nuclei become a real
limitation for the high energy part of the antineutrino spectrum. Uncertainties
below the 10\% level seem to be out of reach with the \textit{ab initio}
approach, preventing any accurate oscillation analysis.

In order to circumvent this issue, measurements of total $\beta$-spectra of
fissile isotopes were performed in the 1980s at ILL
\cite{VonFeilitzsch:1982jw, Schreckenbach:1985ep, Hahn:1989zr}, a high flux
research reactor in Grenoble, France. Thin target foils of fissile isotopes
$^{235}$U, $^{239}$Pu and $^{241}$Pu  were exposed to the intense thermal
neutron flux of the reactor. A tiny part of the emitted electrons could exit
the core through a straight vacuum pipe to be detected by the high resolution
magnetic spectrometer BILL~\cite{BILL:1978}. The electron rates were recorded
by a pointwise measurement of the spectrum in magnetic field steps of 50~keV,
providing an excellent determination of the shape of the electron spectrum with
sub-percent statistical error. The published data were smoothed over 250~keV.
Except for the highest energy bins with poor statistics, the dominant error was
the absolute normalization, quoted around 3\% (90\% CL), with weak energy
dependence.

In principle the conversion of a $\beta$ spectrum into an antineutrino spectrum
can be done using the energy conservation between the two leptons
\begin{eqnarray}
  E_e+E_\nu &=& E_0
\end{eqnarray}
with $E_0$ the end-point of the $\beta$ transition. However this approach
requires to know the contribution of all single branches in the ILL sectra and
this information is not accessible from the integral measurement. Therefore a
specific conversion procedure was developped using a set of 30 `virtual'
$\beta$ branches, fitted on the data. The theoretical expression for the
electron spectrum of a virtual branch was of the form
\begin{eqnarray}
  \label{eq:BetaBranch}
  S_{virtual}(Z,A,E_e) & = & \underbrace{K}_{\text{Norm.}} \times
     \underbrace{\mathcal{F}(Z,A,E_e)}_{\text{Fermi function}} \times
     \underbrace{p_eE_e(E_e-E_0)^2}_{\text{Phase space}} \times
     \underbrace{\Big(1+\delta(Z,A,E_e)\Big)}_{\text{Correction}}
\end{eqnarray}
with $Z$ and $A$ the charge and atomic number of the parent nucleus and $E_0$
the end-point of the transition. The origin of each term is described by the
underbraces. The $\delta$ term contains the corrections to the Fermi theory. In
the ILL papers, it included the QED radiative corrections as calculated in
\cite{Sirlin:1967zza}. The $Z$ dependence comes from the Coulomb corrections.
Since a virtual branch is not connected to any real nucleus the choice of the
nuclear charge was described by the observed mean dependence of $Z$ on $E_0$ in
the nuclear databases
\begin{eqnarray}
  \label{eq:ZofE}
  Z(E_0) &=& 49.5 - 0.7 E_0 - 0.09 E_0^2,\qquad Z \leq 34
\end{eqnarray}
The $A$ dependence is weaker and linked to the determination of $Z$ through
global nuclear fits.

Once the sum of the 30 virtual branches is fitted to the electron data, each of
them is converted to an antineutrino branch by substituting $E_e$ by
$E_0-E_\nu$ in Eq.~(\ref{eq:BetaBranch}) and applying the correct radiative
corrections. The predicted antineutrino spectrum is the sum of all converted
branches. At the end of this procedure an extra correction term is implemented
in an effective way as
\begin{eqnarray}
  \label{eq:eff_delta}
  \Delta S_{branch}(E_\nu) &\simeq& 0.65\left(E_\nu-4.00\right)\,\%
\end{eqnarray}
This term is an approximation of the global effect of weak magnetism correction
and finite size Coulomb correction~\cite{Vogel:1983hi}.

\begin{figure}[h]
  \includegraphics[width=0.5\textwidth]{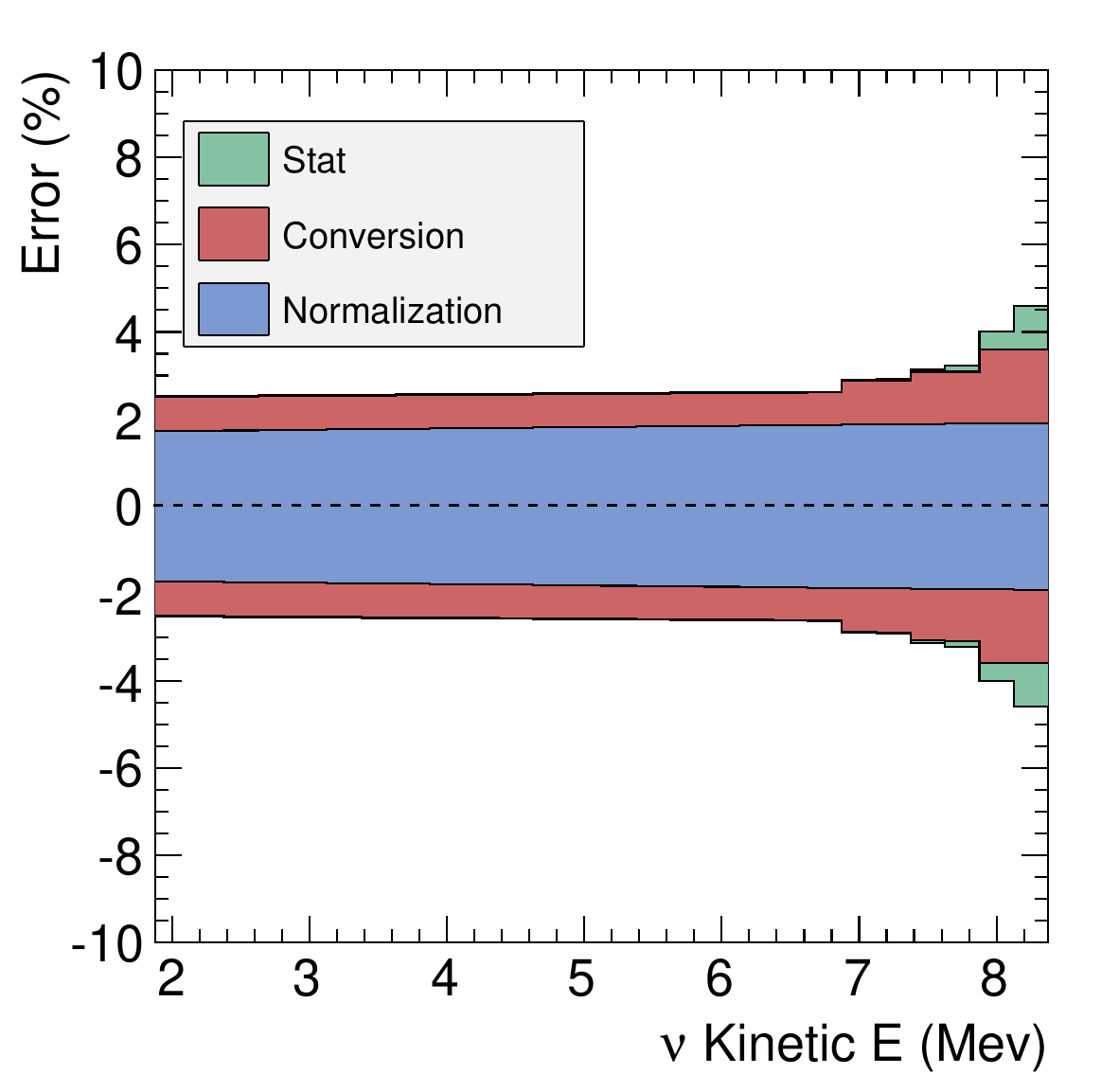}
  \caption{\label{fig:Stack_ILL_nu} Stack of errors of the $^{235}U$
    antineutrino spectrum as predicted in~\cite{Schreckenbach:1985ep}. Each new
    color is the quadratic sum of all previous contributions.}
\end{figure}

The final error of the conversion procedure was estimated to be 3--4\% (90\%
CL), to be added in quadrature with the electron calibration error which
directly propagates to the antineutrino prediction. Figure
\ref{fig:Stack_ILL_nu} shows the error stack for the case of $^{235}U$,
representative of the other isotopes.  From these reference spectra, the
expected antineutrino spectrum detected at a reactor can be computed.
All experiments performed at reactors since then relied on these reference
spectra to compute their predicted antineutrino spectrum.

\subsubsection*{New Reference Antineutrino Spectra}

Triggered by the need for an accurate prediction of the reactor antineutrino
flux for the first phase of the Double Chooz experiment, with a far detector
only, the determination of antineutrino reference spectra has been revisited
lately~\cite{Mueller:2011nm}.  In a first attempt a compilation of the most
recent nuclear data was performed for an up to date {\it ab initio} calculation
of the antineutrino fission spectra. The asset of this approach is the
knowledge of each individual $\beta$-branch, providing a perfect control of the
convertion between electron and antineutrino spectra. As a powerfull
cross-check, the sum of all the branches must match the very accurate electron
spectra measured at ILL. Despite the tremendous amount of nuclear data
available, this approach failed to meet the required accuracy of few \% for two
main reasons:

\begin{itemize}
  \item The majority of the $\beta$-decays is measured using $\beta$--$\gamma$
    coincidences , which are sensitive to the so-called pandemonium
    effect~\cite{Hardy:1983kh}. The net result is an experimental bias of the
    shape of the energy spectra, the high energy part being overestimated
    relative to the low energy part. New measurements are ongoing with dedicated
    experimental setups to correct for the pandemonium effect but in the case of
    the reference spectra many unstable nuclei have to be studied.
  \item As mentionned above, an important fraction of the detected neutrinos have
    a large energy ($> 4$~MeV). The associated $\beta$-transitions mostly come
    from very unstable nuclei with a large energy gap between the parent ground
    state and the nuclear levels of the daughter nucleus. Their decay scheme is
    often poorly known or even not measured at all.
\end{itemize}
A reference data set was constituted based on all fission products indexed in
the ENSDF database~\cite{ENSDF}. All nuclei measured separetely to correct for
the pandemonium effect were substituted when not in agreement with the ENSDF
data (67 nuclei from ~\cite{Tengblad:1989db} and 29 nuclei
from~\cite{Greenwood:1992}). A dedicated interface, BESTIOLE, reads the
relevant information of this set of almost 10000 $\beta$-branches and computes
their energy spectrum based on Eq.~(\ref{eq:BetaBranch}). Then the total beta
spectrum of one fissioning isotope is built as the sum of all fission fragment
spectra weighted by their activity. These activities are determined using a
simulation package called MCNP Utility for Reactor Evolution
(MURE~\cite{MURE}). Following this procedure the predicted fission spectrum is
about 90\% of the reference ILL $\beta$-spectra, as illustrated in figure 2 for
the $^{235}U$ isotope. The missing contribution is the image of all unmeasured
decays as well as the remaining experimental biases of the measurements. To
fill the gap one can invoke models of the decay scheme of missing fission
products. Reaching a good agreement with the ILL electron data remains
difficult with this approach.

\begin{figure}[h]
  \includegraphics[width=0.5\textwidth]{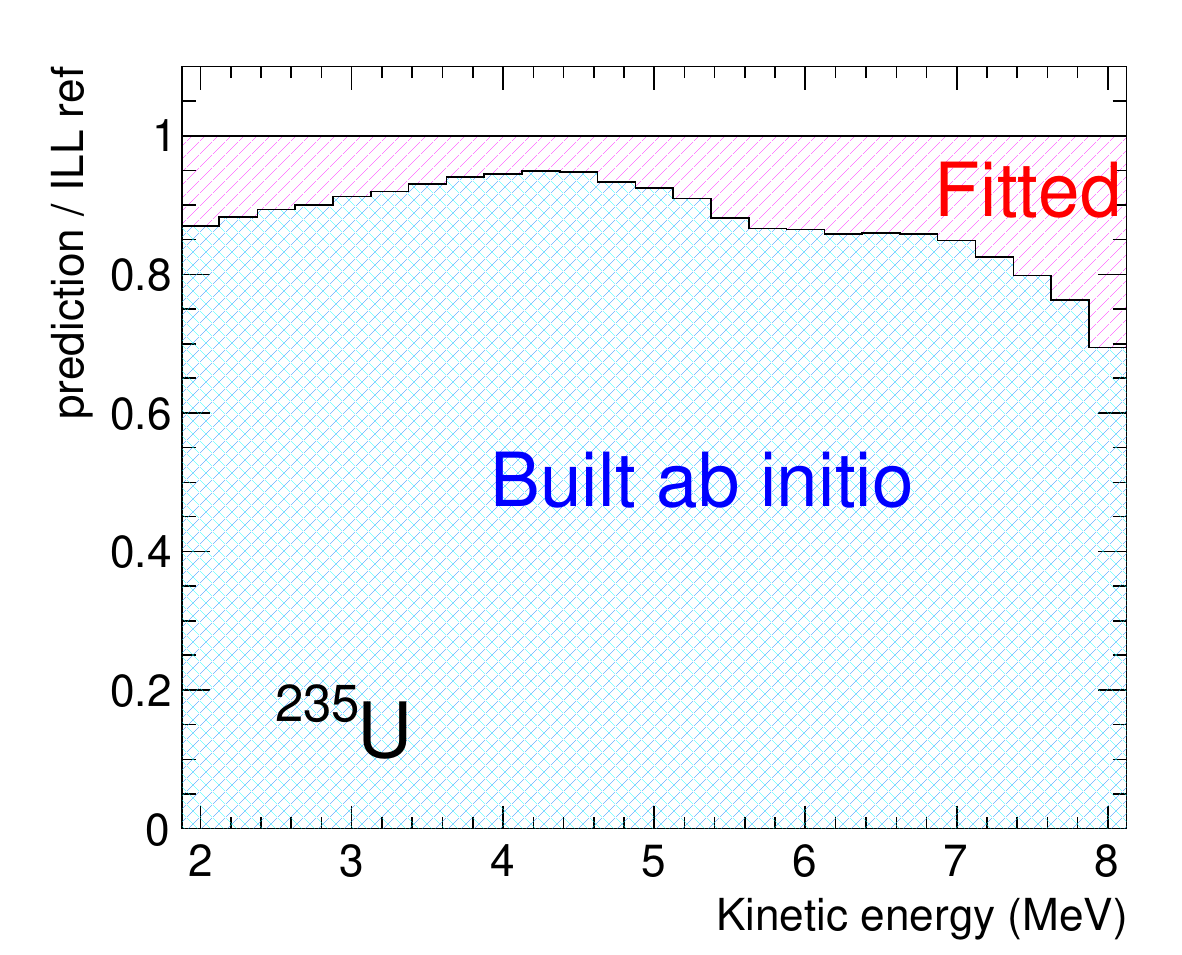}
  \caption{\label{fig:abinit_fit} The blue hatched area shows the contribution of
    the ENSDF + pandemonium corrected nuclei relative to the ILL reference data.
    The missing contribution coming from unknown nuclei and remaining systematic
    effects of nuclear databases (red hatched area) is fitted using a set of 5
    effective $\beta$-branches.}
\end{figure}

Another way to fill the gap is to fit the missing contribution in the electron
spectrum with few virtual branches. The same ILL procedure can be used except
that the virtual branches now rest on the base of physical transitions. This
mixed approach combines the assets of {\it ab initio} and virtual branches
methods: 

\begin{itemize}
  \item The prediction still matches accurately the reference electron data from
    the ILL measurements.
  \item 90\% of the spectrum is built with measured $\beta$-transitions with
    `true' distributions of end-point, branching ratios, nuclear charges, etc. This
    supresses the impact of the approximations associated with the use of virtual
    beta branches.
  \item All corrections to the Fermi theory are applied at the branch
    level, preserving the correspondence between the reference electron data and
    the predicted antineutrino spectrum.
\end{itemize}

\begin{figure}[h]
  \includegraphics[width=0.65\textwidth]{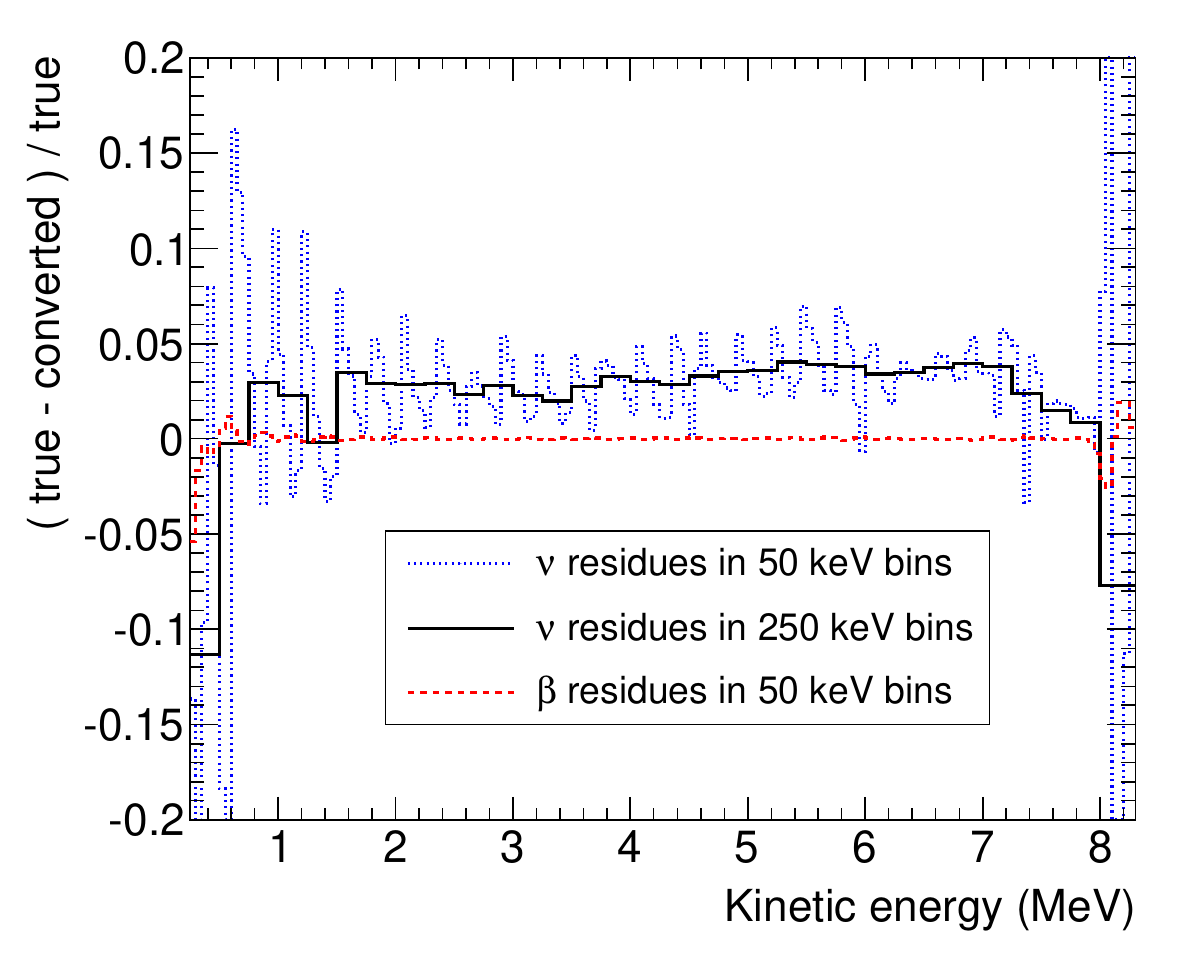}
  \caption{\label{fig:NumCheck} Independent cross-check of the new converted
    spectra based on known reference spectra (pure ENSDF database). Dashed red
    line: electron residuals after fitting with 30 virtual branches. Dotted blue
    line: relative difference between the reference antineutrino spectrum and the
    one converted according to the ILL procedure, in 50 keV bins. Smoothing out
    the residual oscillations in 250 keV bins (solid black line) exhibits a $3\%$
    normalization shift.}
\end{figure}

The new predicted antineutrino spectra are found about 3\% above the ILL
spectra. This effect is comparable for the 3 isotopes ($^{235}U$, $^{239}Pu$,
$^{241}Pu$) with little energy dependence. The origin and the amplitude of this
bias could be numerically studied in detail following a method initially
developped in \cite{Vogel:2007du}. A `true' electron spectrum is defined as the
sum of all measured branches (blue area in figure \ref{fig:abinit_fit}). Since
all the branches are known, the `true' antineutrino spectrum is perfectly
defined as well, with no uncertainty from the conversion. Applying the exact
same conversion procedure than in the eighties on this new electron reference
confirms the 3\% shift between the converted antineutrino spectrum and the
`true' spectrum (see figure \ref{fig:NumCheck}).

\begin{figure}[!h]
  \begin{minipage}{.5\textwidth}
    \centering \includegraphics[width=.95\textwidth]{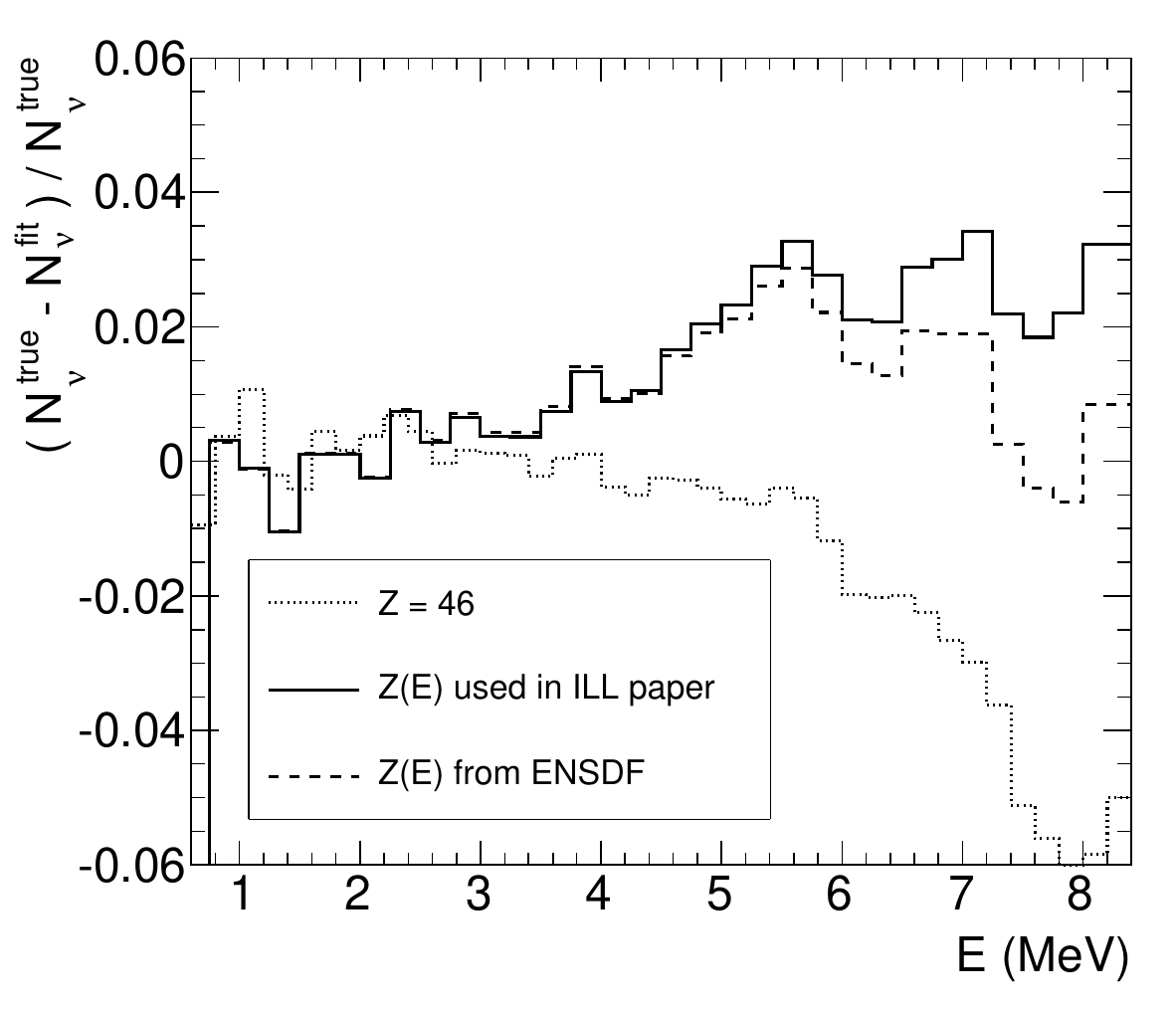}
  \end{minipage}\hfill
  \begin{minipage}{.5\textwidth}
    \centering \includegraphics[width=.90\textwidth]{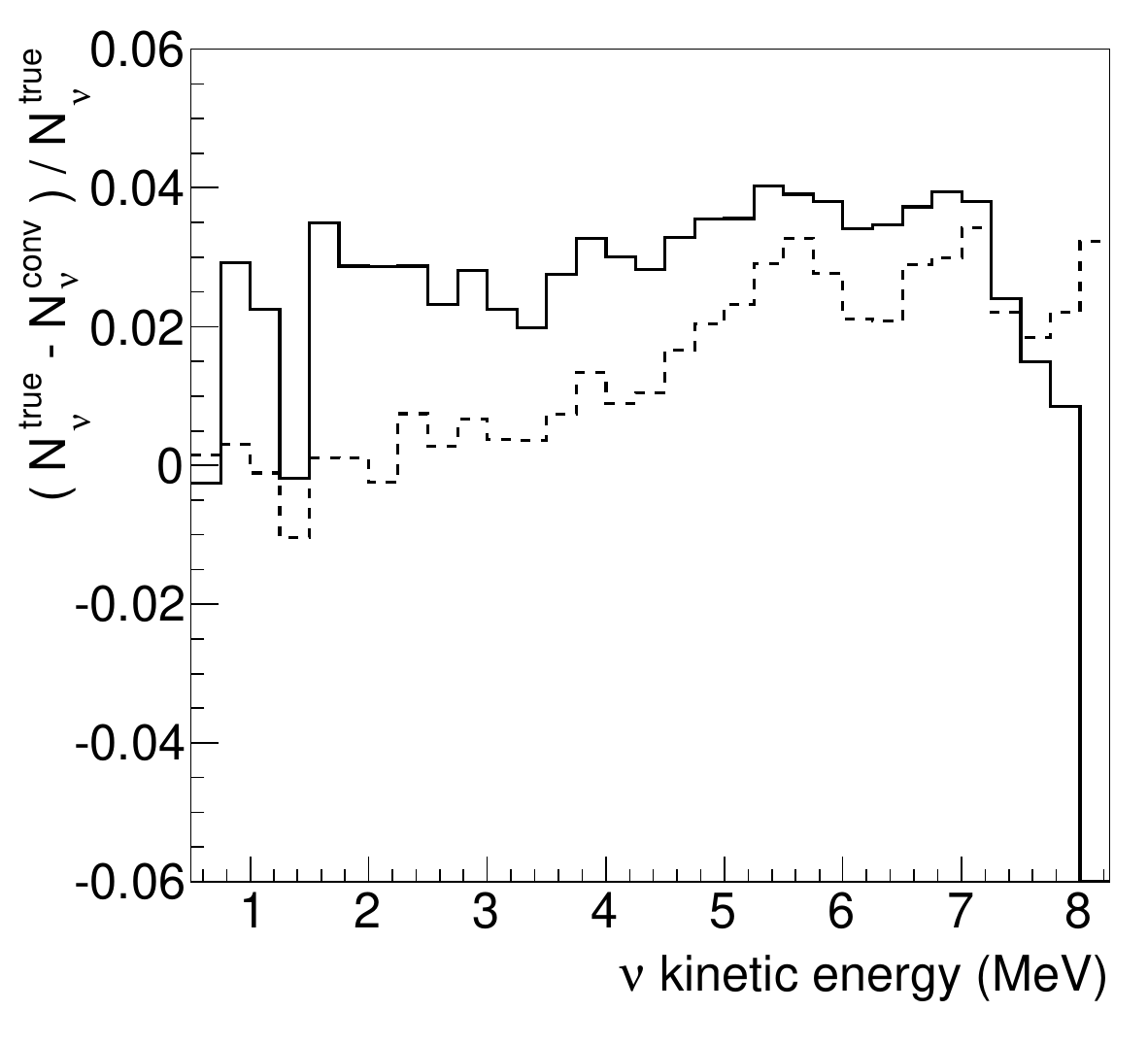}
  \end{minipage}\hfill
  \caption{Numerical tests of the conversion-induced deviations from a `true'
  spectrum built from a set of know branches (see text for details). Left:
  effect of various $Z(E)$ polynomials used in the formula of the virtual
  branches. Right: deviation of converted spectra with the effective correction
  of Eq(\ref{eq:eff_delta}) (solid line), or with the correction applied at the
  branch level.}
  \label{fig:biases}
\end{figure}

Further tests have shown that this global 3\% shift is actually a combination
of two effects. At high energy ($E>4$~MeV) the proper distribution of nuclear
charges, provided by the dominant contribution of the physical
$\beta$-branches, induces a 3\% increase of the predicted antineutrino
spectrum. On the low energy side it was shown that the effective linear
correction of Eq.~(\ref{eq:eff_delta}) was not accounting for the cancellations
operating between the numerous physical branches when the correction is applied
at the branch level (see figure \ref{fig:biases}).

Beyond the correction of these above biases, the uncertainty of the new fission
antineutrino spectra couldn't be reduced with respect to the initial
predictions. The normalisation of the ILL electron data, a dominant source of
error, is inherent to any conversion procedure using the electron reference.
Then a drawback of the extensive use of measured $\beta$-branches in the
mixed-approach is that it brings important constraints on the missing
contribution to reach the electron data. In particular the induced missing
shape can be difficult to fit with virtual branches, preventing a perfect match
with the electron reference. These electron residuals are  unfortunately
amplified as spurious oscillations in the predicted antineutrino spectrum
leading to comparable conversion uncertainties (see red curve in figure
\ref{fig:U235FinalComp}). Finally the correction of the weak magnetism effect
is calculated in a quite crude way and the same approximations are used since
the eighties.

\begin{figure}[h]
  \includegraphics[width=0.85\textwidth]{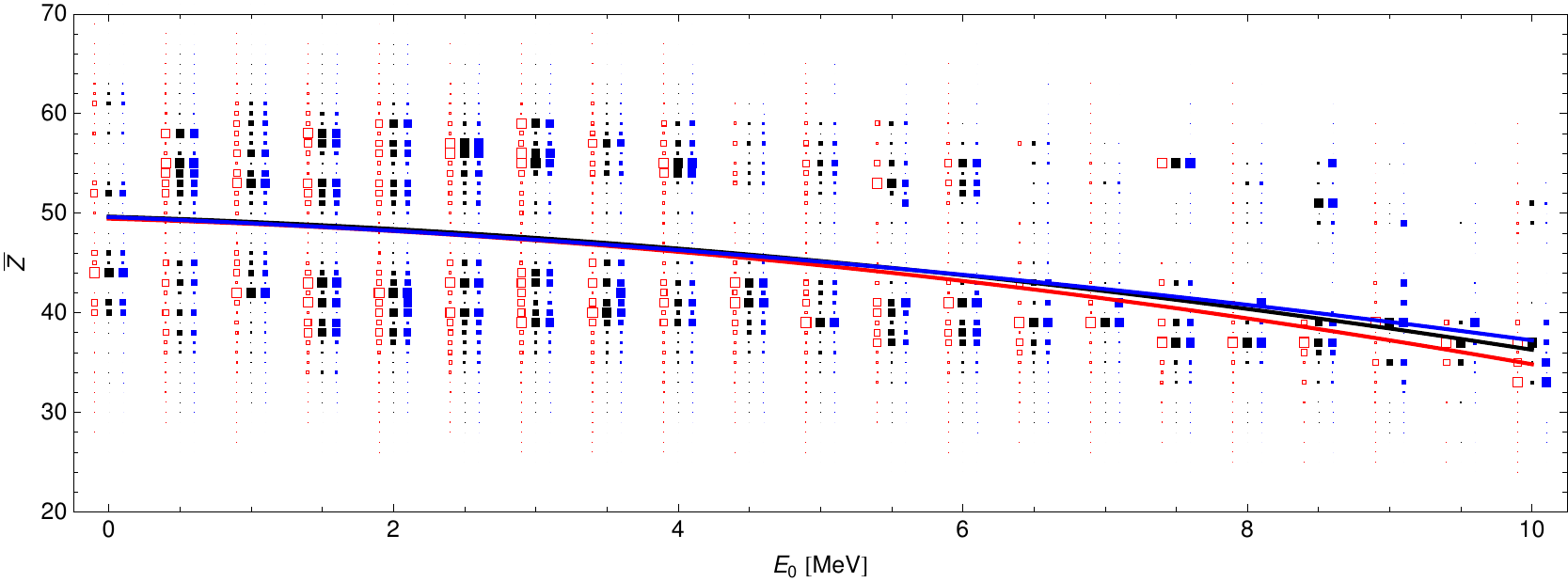}
  \caption{\label{fig:Zeff} The effective nuclear charge $\bar Z$ of the
    fission fragments of $^{235}$U as a function of $E_0$. The area of the each
    box is proportional to the contribution of that particular $Z$ to the fission
    yield in that energy bin. The lines are fits of quadratic polynomials: black
    -- ENSDF database, other colors illustrate the small sensitivity to different
    treatment of the missing isotopes.}
\end{figure}

\begin{figure}[h]
  \includegraphics[width=0.85\textwidth]{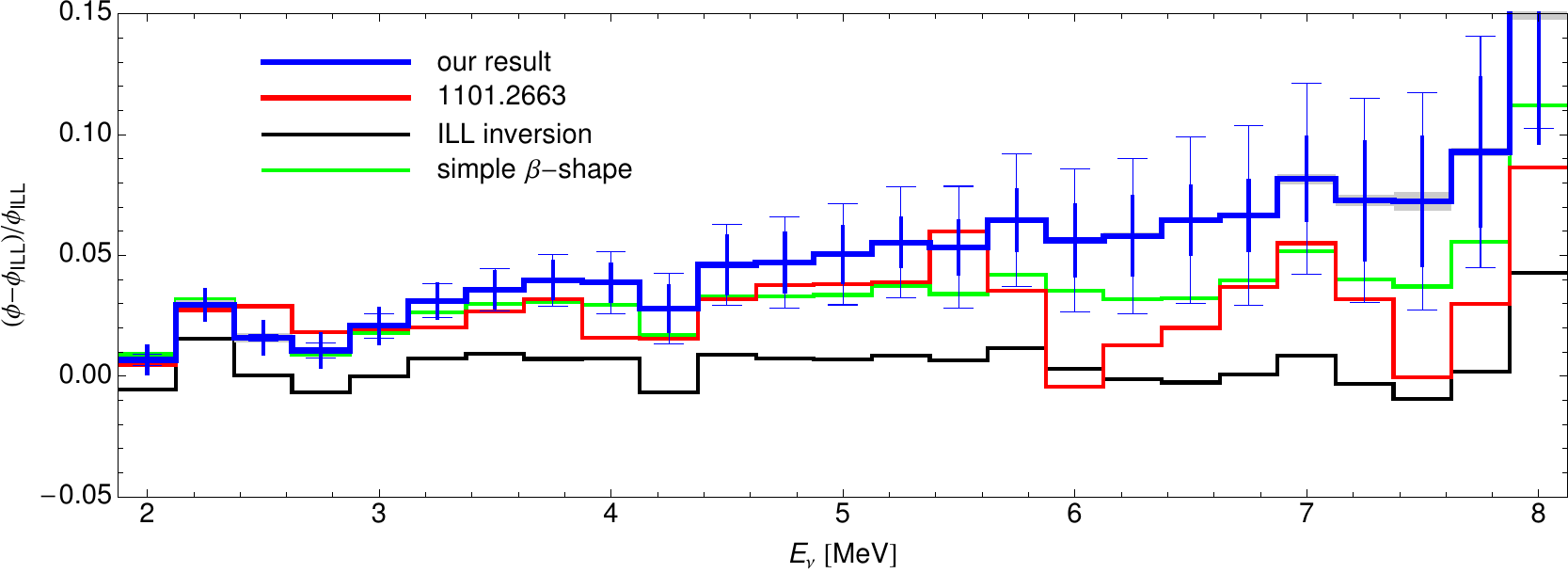}
  \caption{\label{fig:U235FinalComp} Comparison of different conversions of the
    ILL electron data for $^{235}U$. Black curve: cross-check of results from
    \cite{Schreckenbach:1985ep} following the same procedure. Red curve:
    results from \cite{Mueller:2011nm}. Green curve: results from
    \cite{Huber:2011wv} using the same description of $\beta$-decay as in
    \cite{Mueller:2011nm}.  Blue curve: Update of the results
    from~\cite{Mueller:2011nm}, including corrections to the Fermi theory as
    explained in the text.  The thin error bars show the theory errors from the
    effective nuclear charge $\bar Z$ and weak magnetism. The thick error bars
    are the statistical errors.}
\end{figure}

In the light of the above results, the initial conversion procedure of the ILL
data was revisited \cite{Huber:2011wv}. It was shown that a fit using only
virtual branches with a judicious choice of the effective nuclear charge could
provide results with minimum bias. A mean fit similar to Eq.~(\ref{eq:ZofE}) is
still used but the nuclear charge of all known branches is now weighted by its
contribution in the total spectrum, that is the associated fission yield. As
shown in figure \ref{fig:Zeff} the result is quite stable under various
assumptions for the weighting of poorly known nuclei. The bias illustrated in
the left plot of figure \ref{fig:biases} is corrected.

The second bias (right plot of figure \ref{fig:biases}) is again corrected by
implementing the corrections to the Fermi theory at the branch level rather
than using effective corrections as in Eq.~(\ref{eq:eff_delta}). Using the same
expression of these corrections than in \cite{Mueller:2011nm}, the two
independent new predictions are in very good agreement (figure
\ref{fig:U235FinalComp}) confirming the 3\% global shift. Note that the
spurious oscillations of the Mueller \textit{et al.}\ spectra are flattened out by
this new conversion because of the better zeroing of electron residuals.

\begin{figure}[h]
  \includegraphics[width=0.70\textwidth]{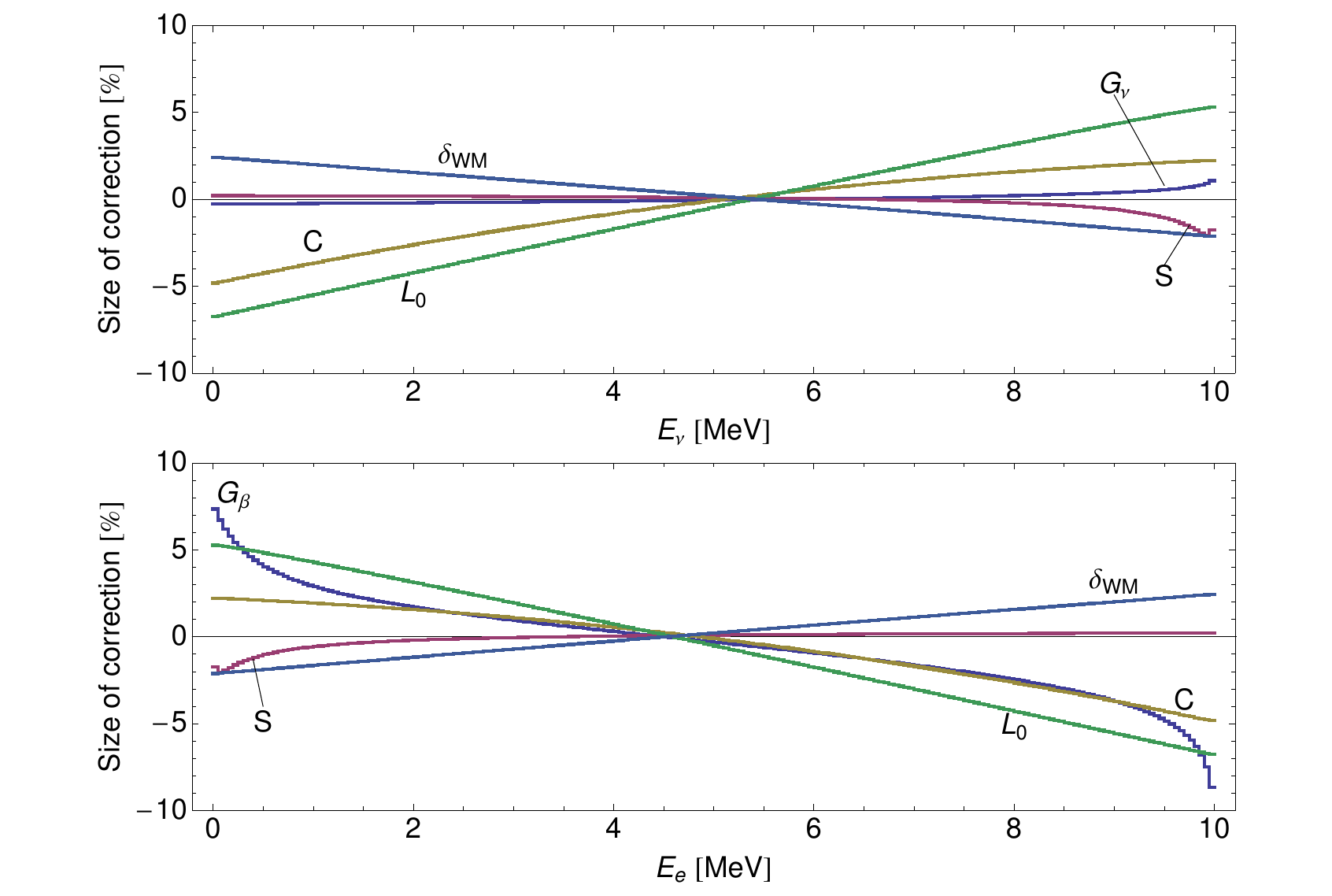}
  \caption{\label{fig:CorrSlopes} Shown is the relative size of the various
    corrections to the Fermi theory for a hypothetical $\beta$-decay with $Z=46$,
    $A=117$ and $E_0=10\,\mathrm{MeV}$. The upper panel shows the effect on the
    antineutrino spectrum, whereas the lower panels shows the effect on the
    $\beta$-spectrum. $\delta_{WM}$: weak magnetism correction; $L_0,C$: Coulomb
    and weak interaction finite size corrections; $S$ Screening correction;
    $G_{\nu,\beta}$: radiative corrections.}
\end{figure}

A detailed review of all corrections to the Fermi theory is provided in
\cite{Huber:2011wv} including finite size corrections, screening correction,
radiative corrections and weak magnetism. To a good approximation they all
appear as linear correction terms as illustrated in figure \ref{fig:CorrSlopes}
in the case of a 10 MeV end-point energy. This refined study of all corrections
leads to an extra increase of the predicted antineutrino spectra at high energy
as illustrated by the blue curve in figure \ref{fig:U235FinalComp}. The net
effect is between 1.0 and 1.4\% more detected antineutrinos depending on the
isotope (see table \ref{tab:ILLratio}).

The corrections of figure \ref{fig:CorrSlopes} are known with a good relative
accuracy except for the weak magnetism term. At the present time a universal
slope factor of about 0.5\% per MeV is assumed, neglecting any dependence on
nuclear structure~\cite{Vogel:1983hi}. Accurate calculation for every fission
product is out of reach. Using the conserved vector current hypothesis it is
possible to infer the weak magnetism correction from the electromagnetic decay
of isobaric analog states. Examples of the slope factors computed from the
available data are shown in table I of \cite{Huber:2011wv}. While most examples
are in reasonable agreement with the above universal slope, some nuclei with
large value of $\log ft$ have a very large slope factor. Moreover a review of
the nuclear databases \cite{SNAC:2011dl} shows that $\beta$-transitions with
$\log ft >7$ contribute between 15 and 30\% to the total spectrum. Still the
data on the weak magnetism slopes are scarce and none of them correspond to
fission products. At this stage it is difficult to conclude if the uncertainty
of the weak magnetism correction should be inflated or not. The prescription of
the ILL analysis, 100\%, coresponds to about 1\% of the detected neutrino rate.
The best constraints could actually come from shape analysis of the reactor
neutrino data themselves. The Bugey and Rovno data are accurate altough
detailed information on the detector response might be missing for such a
detailed shape analysis. The combination of the upcoming Daya Bay, Double Chooz
and Reno data should soon set stringent limits on the global slope factor.

The error budget of the predicted spectra remains again comparable to the first
ILL analysis. The normalization error of the electron data is a common
contribution. The uncertainties of the conversion by virtual branches have been
extensively studied and quantified based on the numerical approach illustrated
in figure \ref{fig:NumCheck}. The uncertainty induced by the weak magnetism
corrections is, faute de mieux, evaluated with the same 100\% relative error.
The final central values and errors are summarized in table
\ref{tab:NewSpectra}.

\subsubsection*{Off equilibrium effects}

For an accurate analysis of reactor antineutrino data, an extra correction to
the reference fission spectra has to be applied. It comes from the fact that
the ILL spectra were acquired after a relatively short irradiation time, between 12
hours and 1.8 days depending on the isotopes, whereas in a reactor experiment
the typical time scale is several months. A non-negligible fraction of the
fisison products have a life-time of several days. Therefore the antineutrinos
associated with their $\beta$-decay keep accumulating well after the `photograph
at 1 day' of the spectra taken at ILL. Very long-lived isotopes correspond to
nuclei close to the bottom of the nuclear valley of stability. Hence one
naively expects these $\beta$-transitions to contribute at low energy. For a
quantitative estimate of this effect the same simulations developed in
\cite{Mueller:2011nm} for the \textit{ab initio} calculation of antineutrino
spectra were used. The sensitivity to the nuclear ingredients is suppressed
because only the relative change between the ILL spectra and spectra of longer
irradiations at commercial reactors were computed. The corrections to be
applied are summarized in table \ref{tab:off_eq2}.

\begin{table}[h]
  \centering
  \begin{ruledtabular}
  \begin{tabular}{cccccc}
    \multicolumn{6}{c}{$^{235}$U} \\   
    \hline
      & 2.0 MeV & 2.5 MeV & 3.0 MeV & 3.5 MeV & 4.0 MeV \\ 
    \hline
    36 h    & 3.1 & 2.2 & 0.8 & 0.6 & 0.1 \\
    100 d   & 4.5 & 3.2 & 1.1 & 0.7 & 0.1 \\
    1E7 s   & 4.6 & 3.3 & 1.1 & 0.7 & 0.1 \\
    300 d   & 5.3 & 4.0 & 1.3 & 0.7 & 0.1 \\
    450 d   & 5.7 & 4.4 & 1.5 & 0.7 & 0.1 \\
    \hline
    \hline
    \multicolumn{6}{c}{$^{239}$Pu} \\   
    \hline
      & 2.0 MeV & 2.5 MeV & 3.0 MeV & 3.5 MeV & 4.0 MeV \\ 
    \hline
    100 d & 1.2 & 0.7  & 0.2 & $<0.1$  & $<0.1$ \\
    1E7 s & 1.3 & 0.7  & 0.2 & $<0.1$  & $<0.1$ \\
    300 d & 1.8 & 1.4  & 0.4 & $<0.1$ & $<0.1$ \\
    450 d & 2.1 & 1.7  & 0.5 & $<0.1$  & $<0.1$ \\
    \hline
    \hline
    \multicolumn{6}{c}{$^{241}$Pu} \\   
    \hline
      & 2.0 MeV & 2.5 MeV & 3.0 MeV & 3.5 MeV & 4.0 MeV \\ 
    \hline
    100 d & 1.0 & 0.5 & 0.2 & $<0.1$ & $<0.1$ \\
    1E7 s & 1.0 & 0.6 & 0.3 & $<0.1$ & $<0.1$ \\
    300 d & 1.6 & 1.1 & 0.4 & $<0.1$ & $<0.1$ \\
    450 d & 1.9 & 1.5 & 0.5 & $<0.1$ & $<0.1$ \\
  \end{tabular}
  \end{ruledtabular}
  \caption{Relative off-equilibrium correction (in \%) to be applied to the
    reference antineutrino spectra for several energy bins and several
    irradiation times significantly longer than the reference times (12~h for
    $^{235}$U, 36~h for $^{239}$Pu, and 43~h for $^{241}$Pu). Effects of neutron
    captures on fission products are included and computed using the simulation
    of a PWR fuel assembly with the MURE code.
    \label{tab:off_eq2}}
\end{table}

As expected they concern the low energy part of the detected spectrum and
vanish beyond 3.5 MeV. The corrections are larger for the $^{235}U$ spectrum
because its irradiation time, 12 h, is shorter than the others. The uncertainty
was estimated from the comparison between the results of MURE and
FISPAC~\cite{Fispact} codes as well as from the sensitivity to the simulated
core geometry. A safe 30\% relative error is recommended.

\subsubsection*{$^{238}U$ reference spectrum}

The $^{238}U$ isotope is contributing about 8\% of the total number of
fissions in a standard commercial reactor. These fissions are induced by fast
neutrons, therefore their associated $\beta$ spectrum could not be measured in
the purely thermal flux of ILL. A dedicated measurement in the fast neutron
flux of the FRMII reactor in Munich has been completed and should be published
in the coming months \cite{Haag}.

Meanwhile the \textit{ab initio} calculation developed
in~\cite{Mueller:2011nm} provides a useful prediction since the relatively
small contribution of $^{238}U$ can accomodate larger uncertainties in the
predicted antineutrino spectrum. An optimal set of $\beta$-branches was tuned
to match the ILL spectra of fissile isotopes as well as possible. The base of this data set
consists of the ENSDF branches corrected for the pandemonium effect as
described in the ``New Reference Antineutrino Spectra" Section. Missing $\beta$ emitters are taken
from the JENDL nuclear database~\cite{JENDL:2000} where they are calculated
using the gross-theory~\cite{GrossTheo:1971}. Finally the few remaining nuclei
were described using a model based on fits of the distributions of the
end-points and branching ratios in the ENSDF database, then extrapolated to the
exotic nuclei. 

\begin{figure}
  \includegraphics[width=0.50\textwidth]{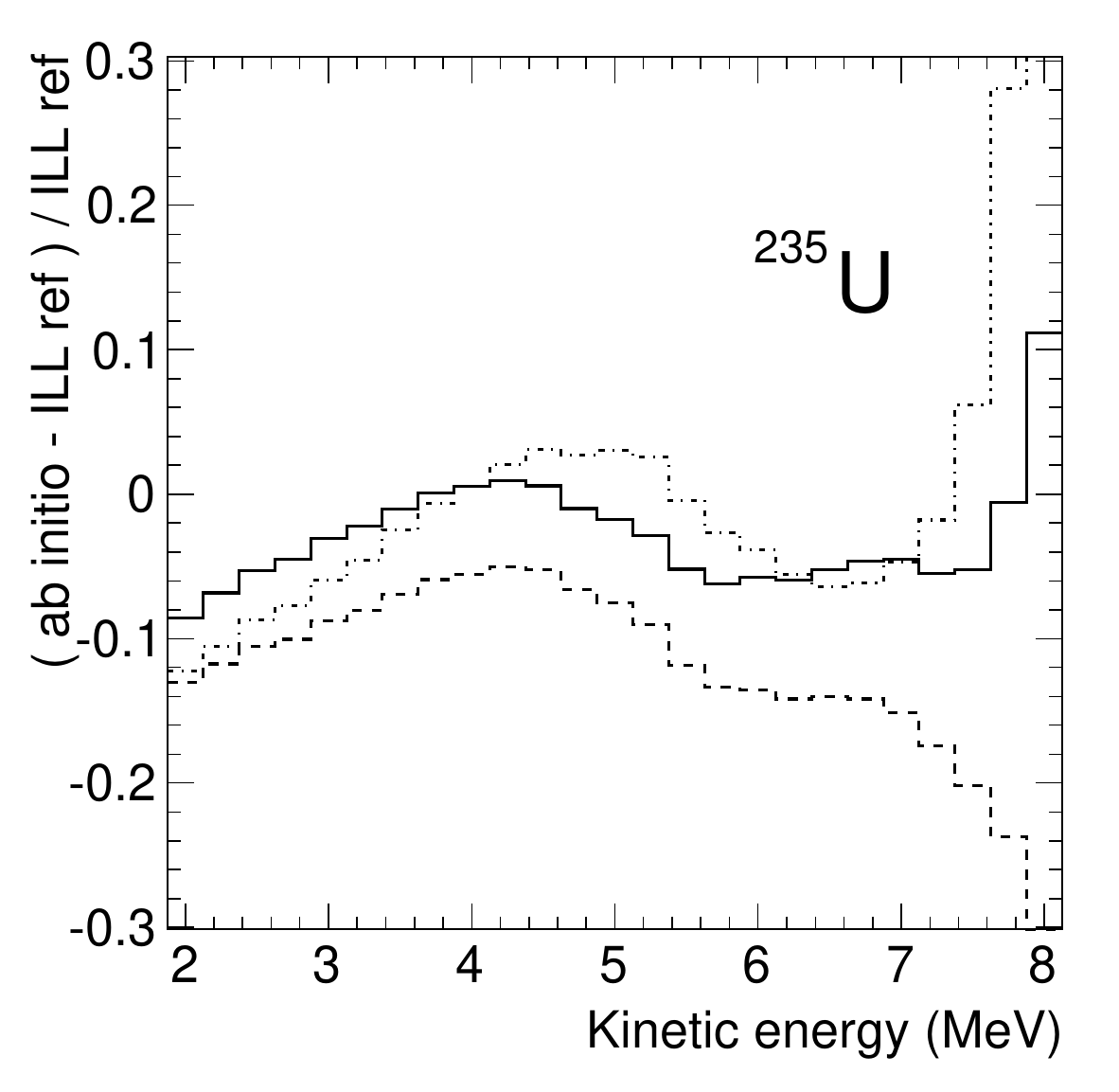}
  \caption{\label{fig:NuclDataSets} Relative difference between the {\it ab
    inito} calculation of the $^{235}U$ spectrum and the ILL $\beta$-spectrum.
    Dashed-dotted curve: ENSDF data only; dashed curve: some ENSDF data replaced by
    pandemonium corrected data; solid curve: unmeasured $\beta$ emitters are added
    on top of previous curve, using the gross-theory calculations of the JENDL
    nuclear database and few remaining exotic nuclei described by a simple model
    (see text).}
\end{figure}

The comparison of these various levels of ingredients with the reference
$^{235}U$ ILL data is shown in figure \ref{fig:NuclDataSets}. With the most
complete set of data one can see that the predicted spectrum agrees with the
reference at the $\pm 10\%$ level.

\begin{table}
  \begin{center}
  \medskip
  \begin{ruledtabular}
  \begin{tabular}{c|cccc|cccc}
    $E_\nu$ & \multicolumn{4}{c|}{$N_{\bar\nu}\,(\mathrm{fission}^{-1}\,\mathrm{MeV}^{-1})$ } & \multicolumn{4}{c}{$1\,\sigma$ errors (\%)} \\
    \hline
    (MeV) & $^{235}U$ & $^{239}Pu$ & $^{241}Pu$ & $^{238}U$ & $^{235}U$ & $^{239}Pu$ & $^{241}Pu$ & $^{238}U$ \\
    \hline
    $2.$ & $1.32$ & $1.08$ & $1.26$ & $1.48$ & $1.8$ & $2.6$ & $2.5$ & $14$ \\[0.3ex]
    $2.25$ & $1.12$ & $9.2\times 10^{-1}$ & $1.08$ & $1.30$ & $1.8$ & $2.6$ & $2.4$ & $14$ \\[0.3ex]
    $2.5$ & $9.15\times 10^{-1}$ & $7.19\times 10^{-1}$ & $8.94\times 10^{-1}$ & $1.15$ & $1.9$ & $2.5$ & $2.3$ & $14$ \\[0.3ex]
    $2.75$ & $7.70\times 10^{-1}$ & $6.20\times 10^{-1}$ & $7.77\times 10^{-1}$ & $1.00$ & $1.9$ & $2.6$ & $2.3$ & $14$ \\[0.3ex]
    $3.$ & $6.51\times 10^{-1}$ & $5.15\times 10^{-1}$ & $6.41\times 10^{-1}$ & $8.76\times 10^{-1}$ & $1.9$ & $2.9$ & $2.4$ & $14$ \\[0.3ex]
    $3.25$ & $5.53\times 10^{-1}$ & $3.98\times 10^{-1}$ & $5.36\times 10^{-1}$ & $7.59\times 10^{-1}$ & $2.0$ & $3.1$ & $2.5$ & $14$ \\[0.3ex]
    $3.5$ & $4.54\times 10^{-1}$ & $3.29\times 10^{-1}$ & $4.39\times 10^{-1}$ & $6.42\times 10^{-1}$ & $2.0$ & $3.3$ & $2.5$ & $14$ \\[0.3ex]
    $3.75$ & $3.64\times 10^{-1}$ & $2.61\times 10^{-1}$ & $3.46\times 10^{-1}$ & $5.39\times 10^{-1}$ & $2.1$ & $3.3$ & $^{+2.4}_{-2.5}$ & $14$ \\[0.3ex]
    $4.$ & $2.94\times 10^{-1}$ & $1.95\times 10^{-1}$ & $2.82\times 10^{-1}$ & $4.51\times 10^{-1}$ & $2.1$ & $3.4$ & $2.6$ & $14$ \\[0.3ex]
    $4.25$ & $2.30\times 10^{-1}$ & $1.57\times 10^{-1}$ & $2.20\times 10^{-1}$ & $3.67\times 10^{-1}$ & $2.3$ & $4.0$ & $2.9$ & $14$ \\[0.3ex]
    $4.5$ & $1.79\times 10^{-1}$ & $1.13\times 10^{-1}$ & $1.66\times 10^{-1}$ & $2.93\times 10^{-1}$ & $2.5$ & $4.9$ & $^{+3.3}_{-3.4}$ & $14$ \\[0.3ex]
    $4.75$ & $1.38\times 10^{-1}$ & $8.33\times 10^{-2}$ & $1.25\times 10^{-1}$ & $2.32\times 10^{-1}$ & $2.5$ & $5.0$ & $3.5$ & $14$ \\[0.3ex]
    $5.$ & $1.10\times 10^{-1}$ & $6.13\times 10^{-2}$ & $9.74\times 10^{-2}$ & $1.83\times 10^{-1}$ & $2.6$ & $4.7$ & $^{+3.3}_{-3.4}$ & $14$ \\[0.3ex]
    $5.25$ & $8.64\times 10^{-2}$ & $4.83\times 10^{-2}$ & $7.47\times 10^{-2}$ & $1.43\times 10^{-1}$ & $2.6$ & $5.1$ & $^{+3.4}_{-3.5}$ & $14$ \\[0.3ex]
    $5.5$ & $6.46\times 10^{-2}$ & $3.54\times 10^{-2}$ & $5.58\times 10^{-2}$ & $1.10\times 10^{-1}$ & $^{+2.7}_{-2.8}$ & $5.7$ & $^{+3.6}_{-3.8}$ & $18$ \\[0.3ex]
    $5.75$ & $5.10\times 10^{-2}$ & $2.92\times 10^{-2}$ & $4.11\times 10^{-2}$ & $8.35\times 10^{-2}$ & $^{+2.9}_{-3.0}$ & $6.4$ & $^{+4.2}_{-4.3}$ & $18$ \\[0.3ex]
    $6.$ & $3.89\times 10^{-2}$ & $1.92\times 10^{-2}$ & $3.05\times 10^{-2}$ & $6.21\times 10^{-2}$ & $^{+3.1}_{-3.2}$ & $8.5$ & $^{+4.9}_{-5.0}$ & $18$ \\[0.3ex]
    $6.25$ & $2.87\times 10^{-2}$ & $1.28\times 10^{-2}$ & $1.98\times 10^{-2}$ & $4.70\times 10^{-2}$ & $^{+3.3}_{-3.4}$ & $9.4$ & $^{+5.3}_{-5.5}$ & $18$ \\[0.3ex]
    $6.5$ & $2.17\times 10^{-2}$ & $9.98\times 10^{-3}$ & $1.54\times 10^{-2}$ & $3.58\times 10^{-2}$ & $^{+3.3}_{-3.5}$ & $^{+9.7}_{-9.8}$ & $^{+5.4}_{-5.7}$ & $18$ \\[0.3ex]
    $6.75$ & $1.61\times 10^{-2}$ & $7.54\times 10^{-3}$ & $1.09\times 10^{-2}$ & $2.71\times 10^{-2}$ & $^{+3.4}_{-3.7}$ & $11.$ & $^{+5.6}_{-5.9}$ & $18$ \\[0.3ex]
    $7.$ & $1.14\times 10^{-2}$ & $4.48\times 10^{-3}$ & $7.75\times 10^{-3}$ & $1.95\times 10^{-2}$ & $^{+3.6}_{-3.9}$ & $13.$ & $^{+5.8}_{-6.1}$ & $22$ \\[0.3ex]
    $7.25$ & $7.17\times 10^{-3}$ & $3.26\times 10^{-3}$ & $4.47\times 10^{-3}$ & $1.33\times 10^{-2}$ & $^{+4.1}_{-4.5}$ & $18.$ & $^{+7.2}_{-7.5}$ & $22$ \\[0.3ex]
    $7.5$ & $4.64\times 10^{-3}$ & $1.95\times 10^{-3}$ & $2.90\times 10^{-3}$ & $8.65\times 10^{-3}$ & $^{+4.3}_{-4.8}$ & $23.$ & $^{+8.4}_{-8.8}$ & $>22$ \\[0.3ex]
    $7.75$ & $2.97\times 10^{-3}$ & $8.47\times 10^{-4}$ & $1.78\times 10^{-3}$ & $6.01\times 10^{-3}$ & $^{+4.7}_{-5.2}$ & $27.$ & $^{+9.1}_{-9.4}$ & $>22$ \\[0.3ex]
    $8.$ & $1.62\times 10^{-3}$ & $5.87\times 10^{-4}$ & $1.06\times 10^{-3}$ & $3.84\times 10^{-3}$ & $^{+6.8}_{-7.2}$ & $29.$ & $^{+12.}_{-13.}$ & $>22$ \\[0.3ex]
  \end{tabular}
  \end{ruledtabular}
  \caption{\label{tab:NewSpectra} New reference spectra as predicted in
    \cite{Huber:2011wv} for the $^{235}U$, $^{239}Pu$ and $^{241}Pu$ isotopes and
    their relative error. These spectra are converted from the ILL electron data
    hence they correspond to 12\,h, 36\,h and 43\,h irradiation time
    respectively. Details on the error breakdown and related correlations can be
    found in \cite{Huber:2011wv}. The $^{238}$U antineutrino spectra is taken
    from the \textit{ab initio} calculation of~\cite{Mueller:2011nm}. A 10\%
    normalization error has been added to the error budget of this spectrum.}
  \vspace{0.6cm}
  \end{center}
\end{table}

Then this optimal data set is used to predict a $^{238}U$ spectrum. Again the
activity of each fission product is calculated with the evolution code MURE.
The case of an N4 commercial reactor operating for one year was simulated. After
such a long irradiation time the antineutrino spectrum has reached the
equilibrium. The results are summarized in table \ref{tab:NewSpectra}. The
central values are about 10\% higher than the previous prediction proposed in
\cite{Vogel:1980bk}. This discrepancy might be due to the larger amount of
nuclei taken into account in the most recent work. Nevertheless both results
are comparable within the uncertainty of the prediction, roughly estimated from
the deviation with respect to the ILL data and the sensitivity to the chosen
data set.

\subsubsection*{Summary of the new reactor antineutrino flux prediction}

In summary, a re-evaluation of the reference antineutrino spectra associated to
the fission of $^{235}U$, $^{239}Pu$ and $^{241}Pu$
isotopes~\cite{Mueller:2011nm} has revealed some systematic biases in the
previously published conversion of the ILL electron
data~\cite{VonFeilitzsch:1982jw, Schreckenbach:1985ep, Hahn:1989zr}. The net
result is a $\simeq+3\%$ shift in the predicted emitted spectra. The origin of
these biases were not in the principle of the conversion method but in the
approximate treatment of nuclear data and corrections to the Fermi theory. A
complementary work \cite{Huber:2011wv} confirmed the origin of the biases and
showed that an extra correction term should be added increasing further the
predicted antineutrino spectra at high energy. These most recent spectra are
the new reference used for the analysis of the reactor anomaly in the next
section. The prediction of the last isotope contributing to the neutrino flux
of reactors, $^{238}U$, is also updated by \textit{ab initio} calculations.

The new predicted spectra and their errors are presented in the summary
table~\ref{tab:NewSpectra}. The deviations with respect to the old reference
spectra are given in table~\ref{tab:ILLratio}

\begin{table}[h]
  \begin{center}
  \medskip
  \begin{ruledtabular}
  \begin{tabular}{ccccc}
    $(R_{new}-R_{ILL})/R_{ILL}$ & $^{235}U$ & $^{239}Pu$ & $^{241}Pu$ & $^{238}U$ \\
    \hline
    values from \cite{Mueller:2011nm} & 2.5 & 3.1 & 3.7 & 9.8 \\
    values from \cite{Huber:2011wv} & 3.7 & 4.2 & 4.7 & - \\
  \end{tabular}
  \end{ruledtabular}
  \caption{\label{tab:ILLratio} Relative change of the new predicted events
    rates with respect to the ILL reference (in~\%). The relative change of the
    emitted flux is always close to 3\%, dominated by the few first bins because
    the energy spectra are dropping fast.}
  \end{center}
\end{table}

\subsubsection*{New Predicted Cross Section per Fission}

Fission reactors release about $10^{20}\,\bar\nu_e\ {\mathrm GW}^{-1}{\mathrm s}^{-1}$,
which mainly come from the beta decays of the fission products of $^{235}$U,
$^{238}$U, $^{239}$Pu, and $^{241}$Pu.  The emitted antineutrino spectrum is
then given by: $S_{\mathrm tot} (E_\nu) =  \sum_{k} f_k S_{k}(E_\nu)$ where $f_k$
refers to the contribution of the main fissile nuclei to the total number of
fissions of the k$^{\mathrm th}$ branch, and $S_k$ to their corresponding neutrino
spectrum per fission. Antineutrino detection is achieved via the inverse
beta-decay (IBD) reaction $\bar{\nu}_e$ + $^1$H $\rightarrow$ $e^+$+n.
Experiments at baselines below $100$~m reported either the ratios (R) of the
measured to predicted cross section per fission, or the observed event rate to
the predicted rate.

The event rate at a detector is predicted based on the following formula
\begin{eqnarray}
  \label{eq:NuExp}
  N_\nu^{Pred} (s^{-1})&=& 
  \frac{1}{4\pi L^2} \,N_p\,\frac{P_{th}}{\left<E_f\right>}\, \sigma^{\mathrm pred}_{f} \,,
\end{eqnarray}
where the first term stands for the mean solid angle and $N_p$ is the number of
target protons for the inverse beta-decay process of detection. These two
detector-related quantities are usually known with very good accuracy. The last two
terms come from the reactor side. The ratio of $P_{th}$, the thermal power of the
reactor, over $\left<E_f\right>$, the mean energy per fission, provides the
mean number of fissions in the core. $P_{th}$ can be know at the subpercent
level in commercial reactors, somewhat less accurately at research reactors.
The mean energy per fission is computed as the average over the four main
fissionning isotopes, accounting for 99.5\% of the fissions
\begin{eqnarray}
  \label{eq:Ef}
  \left<E_f\right> = \sum_k \, \left<E_k\right>, \qquad
  k=^{235}\!U,^{238}\!U,^{239}\!Pu,^{241}\!Pu
\end{eqnarray}
It is accurately known from the nuclear databases and study of all decays and
neutron captures subsequent to a fission~\cite{Kopeikin:2004cn}. Finally the
dominant source of uncertainty and by far the most complex quantity to compute
is the mean cross-section per fission defined as
\begin{equation}
  \label{sigmaexp}
  \sigma^{\mathrm pred}_{f}   =  \int_{0}^{\infty}  S_{\mathrm tot} (E_\nu) \sigma _{\mathrm{V-A}}
  (E_\nu) dE_\nu =  \sum_{k} f_k \sigma^{\mathrm pred}_{f,k} , 
\end{equation}
where the $\sigma^{\mathrm pred}_{f,k}$ are the predicted cross sections
for each fissile isotope, $S_{\mathrm tot}$ is the model dependent reactor neutrino spectrum 
for a given average fuel composition ($f_k$) and $\sigma_{\mathrm{V-A}}$ is the theoretical
cross section of the IBD reaction:
\begin{equation}
  \sigma_{\mathrm{V-A}} (E_e)[{\mathrm cm}^2] = \frac{857 \times 10^{-43}}{\tau_n[{\mathrm s}]}\,
    p_e [{\mathrm MeV}]\, E_e [{\mathrm MeV}]\, (1+\delta_{\mathrm rec}+\delta_{\mathrm wm}+\delta_{\mathrm rad}),
  \label{sigVA}
\end{equation}
where $\delta_{\mathrm rec}$, $\delta_{\mathrm wm}$ and $\delta_{\mathrm rad}$ are
respectively the nucleon recoil, weak magnetism and radiative corretions to the
cross section (see~\cite{Mueller:2011nm,Mention:2011rk} for details).  The
fraction of fissions undergone by the $k^{th}$ isotope, $f_k$, can be computed
at the few percent level with reactor evolution codes (see for
instance~\cite{Jones:2011hi}), but their impact in the final error is well
reduced by the sum rule of the total thermal power, accurately known from
independent measurements

Accounting for new reactor antineutrino spectra~\cite{Huber:2011wv} the
normalization of predicted antineutrino rates, $\sigma^{\mathrm pred}_{f,k}$, is
shifted by +3.7\%, +4.2\%, +4.7\%, +9.8\% for k=$^{235}$U, $^{239}$Pu,
$^{241}$Pu, and $^{238}$U respectively.  In the case of $^{238}$U the
completeness of nuclear databases over the years largely explains the +9.8\%
shift from the reference computations~\cite{Mueller:2011nm}.

\begin{table}
  \begin{center}
  \medskip
  \begin{ruledtabular}
  \begin{tabular}{ccc}
           & old~\cite{Declais:1994ma} & Saclay/Huber~\cite{Mueller:2011nm,Huber:2011wv} \\
    \hline
    $\sigma^{\mathrm pred}_{f,^{235}{\mathrm U}}$    &  6.39$\pm$1.9\%  & 6.69$\pm$2.11\%  \\
    $\sigma^{\mathrm pred}_{f,^{239}{\mathrm Pu}}$   &  4.19$\pm$2.4\%  & 4.40$\pm$2.45\%   \\
    $\sigma^{\mathrm pred}_{f,^{238}{\mathrm U}}$    &  9.21$\pm$10\%   & 10.10$\pm$8.15\% \\
    $\sigma^{\mathrm pred}_{f,^{241}{\mathrm Pu}}$   &  5.73$\pm$2.1\%  & 6.03$\pm$2.15\%   \\
  \end{tabular}
  \end{ruledtabular}
  \caption{\label{tab:crosssec} Individual cross sections per fission per
    fissile isotope, $\sigma^{\mathrm pred}_{f,k}$ for the 'old' and the 'Saclay'
    predictions.}
  \end{center}
\end{table}

The new predicted cross section for any fuel composition
can be computed from Eq.~(\ref{sigmaexp}).
By default the new computation takes into account the so-called
off-equilibrium correction~\cite{Mueller:2011nm} of the antineutrino fluxes 
(increase in fluxes caused by the decay of long-lived
fission products). Individual cross sections per fission per fissile isotope,
$\sigma^{\mathrm pred}_{f,k}$ are simmarized in Table~\ref{tab:crosssec}.
These values are slightly different, by +1.25\% for the averaged
composition of Bugey-4~\cite{Declais:1994ma}, with respect to the original
publication of the reactor antineutrino anomaly~\cite{Mention:2011rk} because of the slight
upward shift of the antineutrino flux consecutive to the work of
Ref.~\cite{Huber:2011wv} (see the ``New Reference Antineutrino Spectra" Section for details).

\subsubsection*{Impact on past experimental results}

In the eighties and nineties, experiments were performed with detectors located
a few tens of meters from nuclear reactor cores at ILL, Goesgen, Rovno,
Krasnoyarsk, Bugey (phases~3 and~4) and Savannah River~\cite{Kwon:1981ua, 
Declais:1994su, Zacek:1986cu, Declais:1994ma, Afonin:1994, Kuvshinnikov:1991, 
Vidyakin:1993, Vidyakin:1994ut, Greenwood:1996pb}.  In the context of the
search of $\mathcal{O}(\text{eV})$ sterile neutrinos, these experiments, with
baselines below 100~m, have the advantage that they are not sensitive to a
possible $\theta_{13}$-, $\Delta m_{31}^2$-driven oscillation effect (unlike
the Palo Verde and CHOOZ experiments, for instance).

\begin{figure}
  \begin{center}
    \includegraphics[scale=0.55]{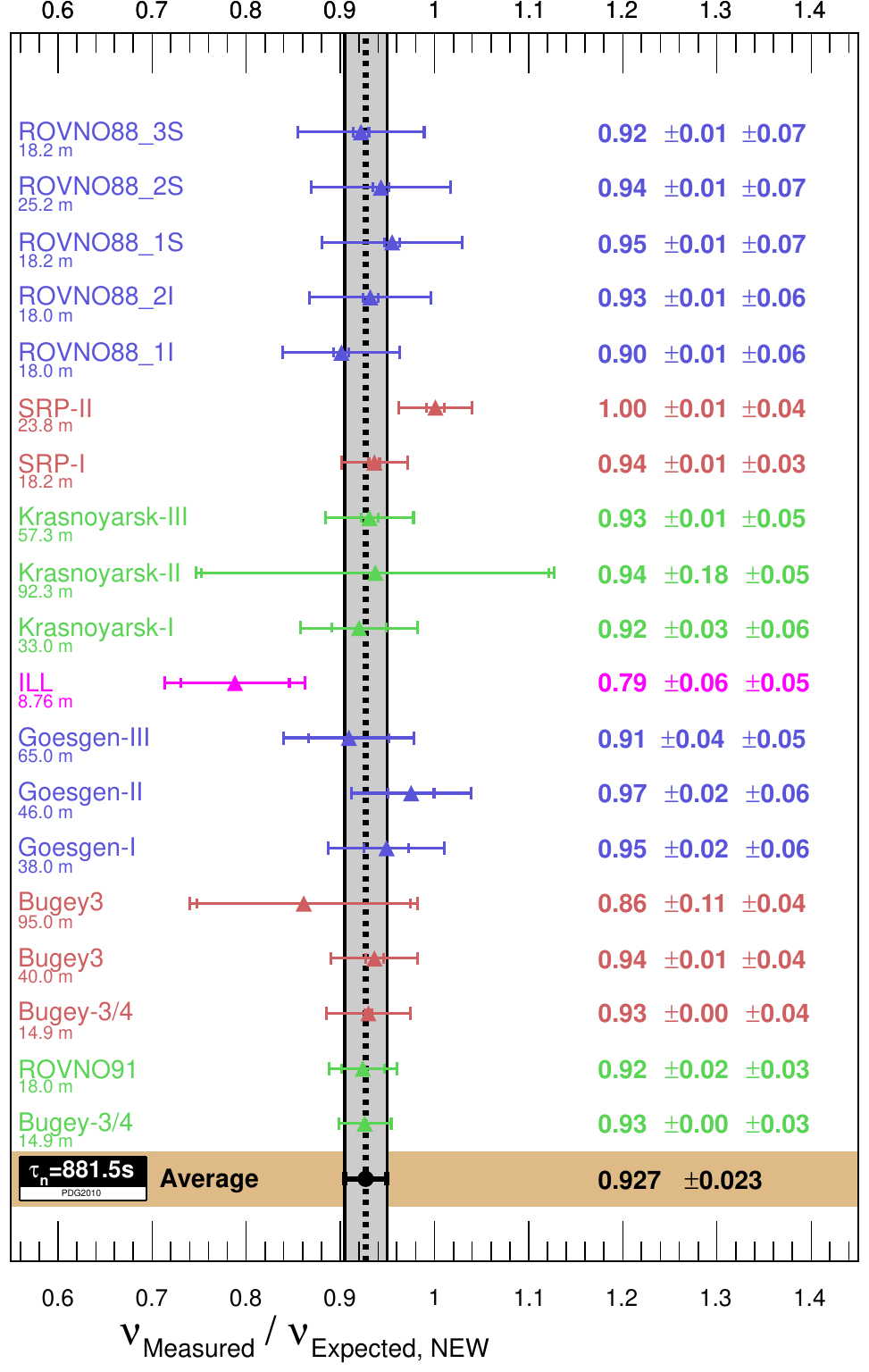}
    \caption{\label{raaplot} Weighted average (with correlations) of
      19~measurements of reactor neutrino experiments operating at short
      baselines. A summary of experiment details is given in
      Table~\ref{tab:other}.}
  \end{center}
\end{figure}

\begin{figure}
  \hspace*{-1cm}
  \includegraphics[scale=0.4,angle=90]{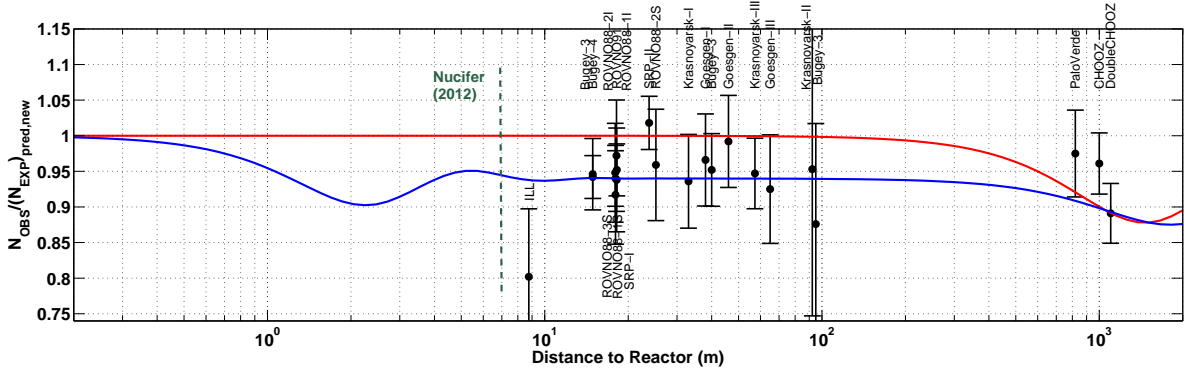}
  \caption{\label{raadistance} Short baseline reactor antineutrino anomaly.
    The experimental results are compared to the prediction without
    oscillation, taking into account the new antineutrino spectra, the
    corrections of the neutron mean lifetime, and the off-equilibrium effects.
    Published experimental errors and antineutrino spectra errors are added in
    quadrature. The mean averaged ratio including possible correlations is
    0.927$\pm$0.023.  As an illustration, the red line shows a 3 active
    neutrino mixing solution fitting the data, with
    $\sin^2(2\theta_{13})=0.15$.  The blue line displays a solution including a
    new neutrino mass state, such as $|\Delta m_{new,R}^2 | \gg$ 2 eV$^2$ and
    $\sin^2(2\theta_{new,R})$=0.12, as well as $\sin^2(2\theta_{13})=0.085$.}
\end{figure}

The ratios of observed event rates to predicted event rates (or cross
section per fission), $R=N_{\mathrm obs}/N_{\mathrm pred}$, are
summarized in Table~\ref{tab:other}. The observed event rates and
their associated errors are unchanged with respect to the
publications, the predicted rates are reevaluated
separately in each experimental case. One can observe a general systematic
shift more or less significantly below unity.  These reevaluations
unveil a new {\it reactor antineutrino anomaly}\footnote{http://irfu.cea.fr/en/Phocea/Vie\_des\_labos/Ast/ast\_visu.php?id\_ast=3045} \cite{Mention:2011rk},  clearly illustrated
in Fig.~\ref{raaplot} and in Fig.~\ref{raadistance} . In order to quantify the statistical
significance of the anomaly, one can compute the weighted average of the
ratios of expected over predicted rates, for all short baseline
reactor neutrino experiments (including their possible correlations). 

\begin{table*}
  \begin{center}
  \scalebox{.99}{%
  \begin{ruledtabular}
  \begin{tabular}{cccccccccccc}
     result   & Det. type & $\tau_n$ (s)& $^{235}$U & $^{239}$Pu &$^{238}$U & $^{241}$Pu& old  & new  & err(\%) & corr(\%) & L(m) \\
    \hline
    Bugey-4     & $^3$He+H$_2$O & 888.7 & 0.538 & 0.328 & 0.078 & 0.056 & 0.987 & 0.926 & 3.0  & 3.0 & 15\\
    ROVNO91     & $^3$He+H$_2$O & 888.6 & 0.614 & 0.274 & 0.074 & 0.038 & 0.985 & 0.924 & 3.9  & 3.0 & 18\\
    \hline
    Bugey-3-I   & $^6$Li-LS     & 889   & 0.538 & 0.328 & 0.078 & 0.056 & 0.988 & 0.930 & 4.8  & 4.8 & 15\\ 
    Bugey-3-II  & $^6$Li-LS     & 889   & 0.538 & 0.328 & 0.078 & 0.056 & 0.994 & 0.936 & 4.9  & 4.8 & 40\\ 
    Bugey-3-III & $^6$Li-LS     & 889   & 0.538 & 0.328 & 0.078 & 0.056 & 0.915 & 0.861 & 14.1 & 4.8 & 95\\
    \hline
    Goesgen-I   & $^3$He+LS     & 897   & 0.620 & 0.274 & 0.074 & 0.042 & 1.018 & 0.949 & 6.5  & 6.0 & 38\\
    Goesgen-II  & $^3$He+LS     & 897   & 0.584 & 0.298 & 0.068 & 0.050 & 1.045 & 0.975 & 6.5  & 6.0 & 45\\
    Goesgen-II  & $^3$He+LS     & 897   & 0.543 & 0.329 & 0.070 & 0.058 & 0.975 & 0.909 & 7.6  & 6.0 & 65\\
    ILL         & $^3$He+LS     & 889   & $\simeq1$ &--- &---   &---    & 0.832 & 0.7882 & 9.5  & 6.0 &  9\\
    \hline
    Krasn. I   & $^3$He+PE     & 899   & $\simeq1$ &--- &---   &---    & 1.013 & 0.920 & 5.8  & 4.9 & 33\\
    Krasn. II  & $^3$He+PE     & 899   & $\simeq1$ &--- &---   &---    & 1.031 & 0.937 & 20.3 & 4.9 & 92\\
    Krasn. III & $^3$He+PE     & 899   & $\simeq1$ &--- &---   &---    & 0.989 & 0.931 & 4.9  & 4.9 & 57\\
  \hline
    SRP I      & Gd-LS         & 887   & $\simeq1$ &--- &---   &---    & 0.987 & 0.936 & 3.7  & 3.7 & 18\\
    SRP II     & Gd-LS         & 887   & $\simeq1$ &--- &---   &---    & 1.055 & 1.001 & 3.8  & 3.7 & 24\\
  \hline
    ROVNO88-1I & $^3$He+PE     & 898.8 & 0.607 & 0.277 & 0.074 & 0.042 & 0.969 & 0.901 & 6.9  & 6.9 & 18\\ 
    ROVNO88-2I & $^3$He+PE     & 898.8 & 0.603 & 0.276 & 0.076 & 0.045 & 1.001 & 0.932 & 6.9  & 6.9 & 18\\
    ROVNO88-1S & Gd-LS         & 898.8 & 0.606 & 0.277 & 0.074 & 0.043 & 1.026 & 0.955 & 7.8  & 7.2 & 18\\
    ROVNO88-2S & Gd-LS         & 898.8 & 0.557 & 0.313 & 0.076 & 0.054 & 1.013 & 0.943 & 7.8  & 7.2 & 25\\
    ROVNO88-3S & Gd-LS        & 898.8 & 0.606 & 0.274 & 0.074 & 0.046 & 0.990 &0.922 & 7.2  & 7.2 & 18\\
  \end{tabular}
  \end{ruledtabular}}
  \end{center}
  \caption{\label{tab:other} $N_{\mathrm obs}/N_{\mathrm pred}$ ratios based on
    old and new spectra. Off-equilibrium corrections have been applied when justified.  
    The err column is the total error published by the
    collaborations including the error on $S_{\mathrm tot}$, 
    the corr column is the part of the error correlated among
    experiments (multiple-baseline or same detector).}
\end{table*}

In doing so, the authors of~\cite{Mention:2011rk} have considered the following
experimental rate information: Bugey-4 and Rovno91, the three Bugey-3
experiments, the three Goesgen experiments and the ILL experiment, the three
Krasnoyarsk experiments, the two Savannah River results (SRP), and the five
Rovno88 experiments.  $\overrightarrow{\text{R}}$~is the corresponding vector
of 19~ratios of observed to predicted event rates. A~2.0\% systematic
uncertainty was assumed, fully correlated among all 19~ratios, resulting from
the common normalization uncertainty of the beta-spectra measured
in~\cite{Schreckenbach:1985ep, VonFeilitzsch:1982jw, Hahn:1989zr}.  
In order to account for the potential experimental correlations,
the experimental errors of Bugey-4 and Rovno91, of the three Goesgen and the
ILL experiments, the three Krasnoyarsk experiments, the five Rovno88
experiments, and the two SRP results were fully correlated.  Also, the Rovno88
(1I~and~2I) results were fully correlated with Rovno91, and an arbitrary~50\%
correlation was added between the Rovno88 (1I~and~2I) and the Bugey-4
measurement.These latest correlations are motivated by the use of similar or
identical integral detectors.

In order to account for the non-gaussianity of the ratios $R$
a Monte Carlo simulation was developed to check this point, and it was found that
the ratios distribution is almost Gaussian, but with slightly longer
tails, which were taken into account in the calculations (in
contours that appear later, error bars are enlarged). With the old
antineutrino spectra the mean ratio is $\mu$=0.980$\pm$0.024. 

With the new antineutrino spectra, one obtains $\mu$=0.927$\pm$0.023, and the
fraction of simple Monte-Carlo experiments with $r \ge 1$ is 0.3\%,
corresponding to a $-2.9\,\sigma$ effect (while a simple calculation assuming
normality would lead to $-3.2\,\sigma$). Clearly the new spectra induce a
statistically significant deviation from the expectation. This motivates the definition
of an experimental cross section \mbox{$\sigma^{\mathrm ano,2012}_{f}=
0.927\times\sigma^{\mathrm pred,new}_{f}$~10$^{-43}$~cm$^2$/fission}. With the new
antineutrino spectra, the minimum $\chi^2$ for the data sample
is $\chi^2_{\mathrm min,data}=18.4$. The fraction of simple Monte-Carlo experiments
with $\chi^2_{\mathrm min}<\chi^2_{\mathrm min,data}$ is 50\%, showing that the
distribution of experimental ratios in $\overrightarrow{\text{R}}$ around the
mean value is representative given the correlations. 

Assuming the correctness of $\sigma^{\mathrm pred,new}_{f}$ the anomaly
could be explained by a common bias in all reactor neutrino
experiments. The measurements used different detection techniques
(scintillator counters and integral detectors). Neutrons were tagged
either by their capture in metal-loaded scintillator, or in
proportional counters, thus leading to two distinct systematics. As
far as the neutron detection efficiency calibration is concerned,
note that different types of radioactive sources emitting MeV or
sub-MeV neutrons were used (Am-Be, $^{252}$Cf, Sb-Pu, Pu-Be). It
should be mentioned that the Krasnoyarsk, ILL, and SRP experiments
operated with nuclear fuel such that the difference between the real
antineutrino spectrum and that of pure $^{235}$U was less than
1.5\%. They reported similar deficits to those observed at other
reactors operating with a mixed fuel. Hence the anomaly can be
associated neither with a single fissile isotope nor with a single
detection technique. All these elements argue against a trivial bias
in the experiments, but a detailed analysis of the most sensitive of
them, involving experts, would certainly improve the quantification of
the anomaly.

The other possible explanation of the anomaly is based on a real physical
effect and is detailed in the next section.  In that analysis, shape
information from the Bugey-3 and ILL published
data~\cite{Declais:1994su,Kwon:1981ua} is used.  From the analysis of the shape
of their energy spectra at different source-detector
distances~\cite{Declais:1994su,Zacek:1986cu}, the Goesgen and Bugey-3
measurements exclude oscillations with $0.06<\Delta m^2<1$~eV$^2$ for
$\sin^2(2\theta)>0.05$.  Bugey-3's 40~m/15~m ratio data
from~\cite{Declais:1994su} is used as it provides the best limit. As already
noted in Ref.~\cite{ILL95}, the data from ILL showed a spectral deformation
compatible with an oscillation pattern in their ratio of measured over
predicted events.  It should be mentioned that the parameters best fitting the
data reported by the authors of Ref.~\cite{ILL95} were $\Delta m^2=2.2$~eV$^2$
and $\sin^2(2\theta)=0.3$. A reanalysis of the data of Ref.~\cite{ILL95} was
carried out in order to include the ILL shape-only information in the analysis
of the reactor antineutrino anomaly.  The contour in Fig.~14 of
Ref.~\cite{Kwon:1981ua} was reproduced for the shape-only analysis (while for
the rate-only analysis discussed above, that of Ref.~\cite{ILL95} was
reproduced, excludeing the no-oscillation hypothesis at 2$\sigma$).

\subsubsection*{The fourth neutrino hypothesis (3+1 scenario)}

\paragraph{Reactor Rate-Only Analysis}

The reactor antineutrino anomaly could be explained through the
existence of a fourth non-standard neutrino, corresponding in the
flavor basis to a sterile neutrino $\nu_s$ (see~\cite{Nakamura:2010zzi} and
references therein) with a large $\Delta m_{\mathrm new}^2$ value.

\begin{figure}
  \begin{center}
  \includegraphics[scale=0.5]{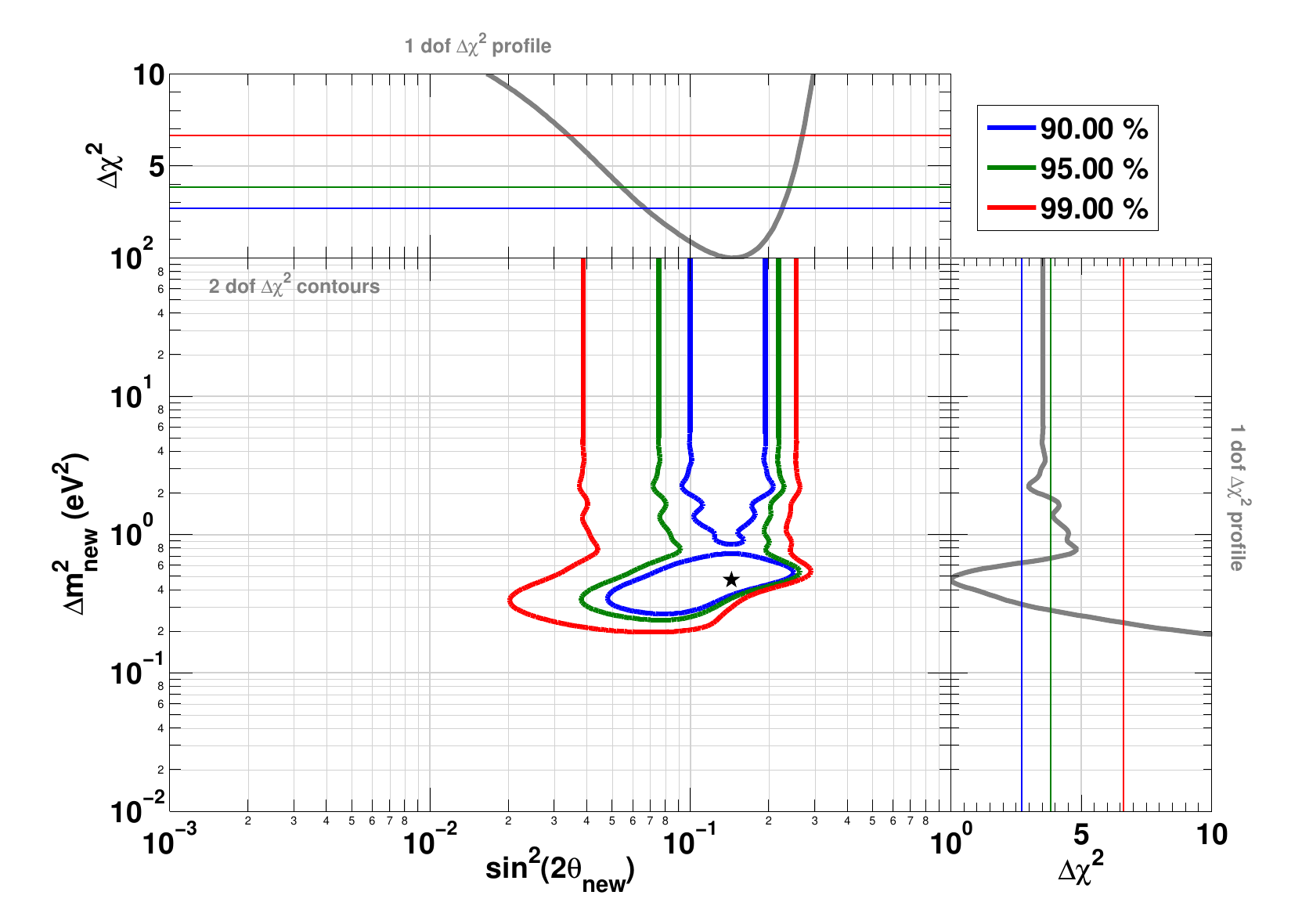}
  \caption{\label{f:sterile_r} Allowed  regions in the $\sin^2(2\theta_{\mathrm
    new})-\Delta m_{\mathrm new}^2$ plane obtained from the fit of the reactor
    neutrino data, without any energy spectra information, to the 3+1 neutrino
    hypothesis, with $\sin^2(2\theta_{13})=0$.  The best-fit point is indicated
    by a star.}
  \end{center}
\end{figure}

For simplicity the analysis presented here is restricted to the 3+1 four-neutrino
scheme in which there is a group of three active neutrino masses
separated from an isolated neutrino mass, such that $|\Delta m_{\mathrm
  new}^2 |\gg 10^{-2}$~eV$^2$. The latter would be responsible for
very short baseline reactor neutrino oscillations. For energies above
the IBD threshold and baselines below 100~m, the approximated oscillation 
formula
\begin{equation}
P_{ee} = 1 - \sin^2(2\theta_{\mathrm new})\sin^2\left(\frac{\Delta
    m_{\mathrm new}^2L}{4E_{\bar\nu_e}}\right)
\label{sterileproba}
\end{equation}
is adopted, where active neutrino oscillation effects are neglected at these
short baselines.  In such a framework the mixing angle is related to the $U$
matrix element by the relation:
\begin{equation}
\sin^2(2\theta_{\mathrm new}) =
4 |U_{e4}|^2 \left( 1 - |U_{e4}|^2 \right).
\end{equation}
One can us now fit the sterile neutrino hypothesis to the data (baselines
below 100~m) by minimizing the least-squares function 
\begin{equation}
\left( 
P_{ee}-\overrightarrow{\text{R}}
\right)^T 
W^{-1}
\left( 
P_{ee}-\overrightarrow{\text{R}}
\right), 
\end{equation}
assuming $\sin^2(2\theta_{13})=0$. Figure~\ref{f:sterile_r} provides the
results of the fit in the $\sin^2(2\theta_{\mathrm new})-\Delta m_{\mathrm new}^2$
plane, including only the reactor experiment rate information.  The fit to the
data indicates that $|\Delta m_{\mathrm new,R}^2 | > 0.2$~eV$^2$ (99\%) and
$\sin^2(2\theta_{\mathrm new,R}) \sim 0.14$. The best fit point is at $|\Delta
m_{\mathrm new,R}^2 | = 0.5$~eV$^2$ and $\sin^2(2\theta_{\mathrm new,R}) \sim 0.14$.
The no-oscillation analysis is excluded at 99.8\%, corresponding roughly to
3$\sigma$.

\paragraph{Reactor Rate+Shape Analysis}

\begin{figure}
  \begin{center}
  \includegraphics[scale=0.5]{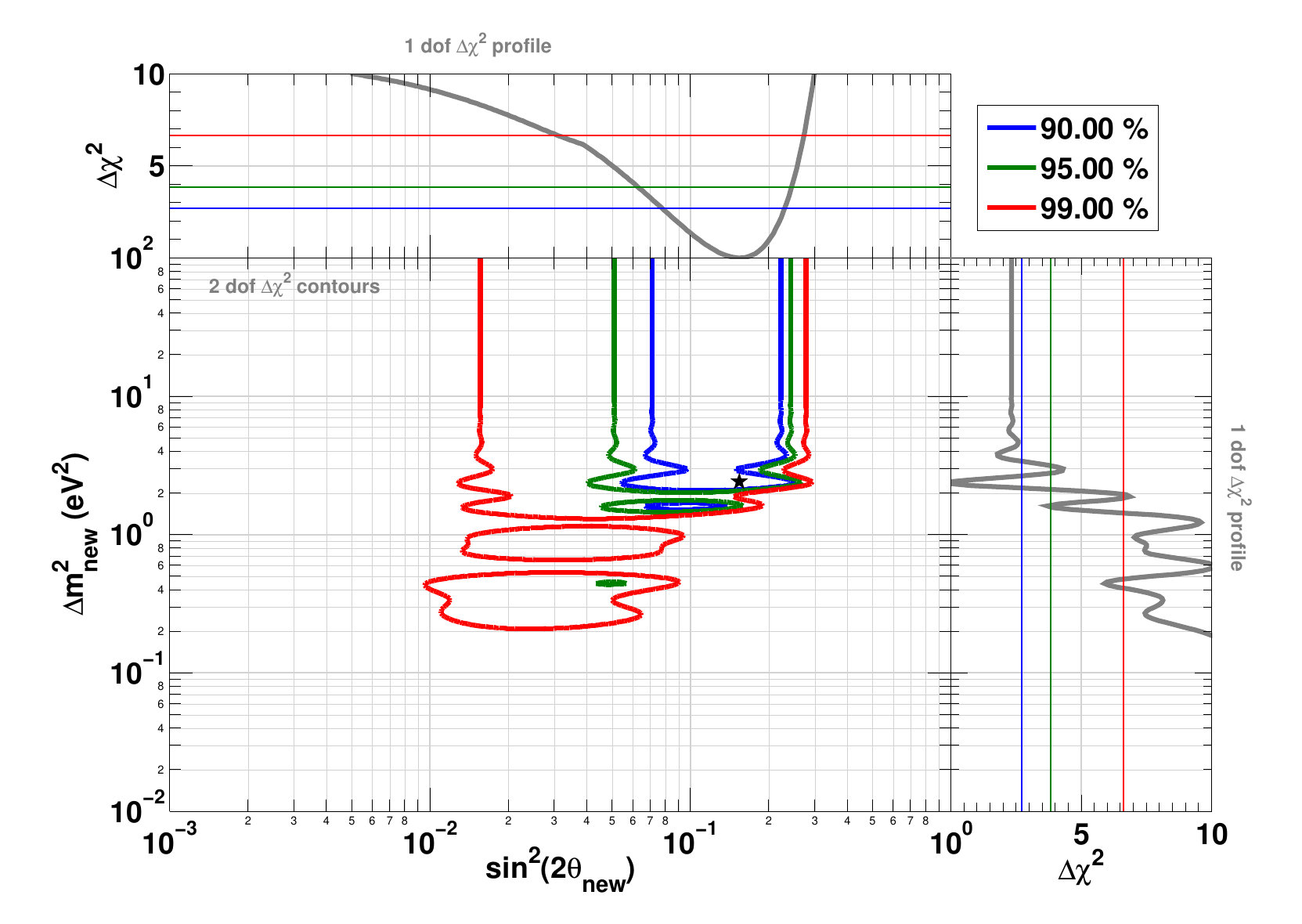}
  \caption{\label{f:sterile_rs} Allowed  regions in the $\sin^2(2\theta_{\mathrm
    new})-\Delta m_{\mathrm new}^2$ plane obtained from the fit of the reactor
    neutrino data, without ILL-shape information, but with the stringent
    oscillation constraint of Bugey-3 based on the 40~m/15~m ratios to the 3+1
    neutrino hypothesis, with $\sin^2(2\theta_{13})=0$.  The best-fit point is
    indicated by a star.}
  \end{center}
\end{figure}

The ILL experiment may have seen a hint of oscillation in their measured
positron energy spectrum~\cite{Kwon:1981ua,ILL95}, but Bugey-3's results do not
point to any significant spectral distortion more than 15~m away from the
antineutrino source. Hence, in a first approximation, hypothetical oscillations
could be seen as an energy-independent suppression of the $\bar\nu_e$ rate by a
factor of $\frac{1}{2}\sin^2(2\theta_{\mathrm new,R})$, thus leading to  $\Delta
m_{\mathrm new,R}^2 \gtrsim 1$ eV$^2$ and accounting for the Bugey-3 and Goesgen
shape analyses~\cite{Declais:1994su,Zacek:1986cu}. Considering the weighted
average of all reactor experiments one obtains an estimate of the mixing angle,
$\sin^2(2\theta_{\mathrm new,R}) \sim 0.15$. The ILL positron spectrum is thus in
agreement with the oscillation parameters found independently in the
re-analyses mainly based on rate information.  Because of the differences in
the systematic effects in the rate and shape analyses, this coincidence is in
favor of a true physical effect rather than an experimental anomaly.  Including
the finite spatial extension of the nuclear reactors and the ILL and Bugey-3
detectors, it is found that the small dimensions of the ILL nuclear core lead
to small corrections of the oscillation pattern imprinted on the positron
spectrum. However the large extension of the Bugey nuclear core is sufficient
to wash out most of the oscillation pattern at 15~m. This explains the absence
of shape distortion in the Bugey-3 experiment.  We now present results from a
fit of the sterile neutrino hypothesis to the data including both Bugey-3 and
ILL original results (no-oscillation reported). With respect to the rate only
parameters, the solutions at lower $|\Delta m_{\mathrm new,R+S}^2 |$ are now
disfavored at large mixing angle because they would have imprinted a strong
oscillation pattern in the energy spectra (or their ratio) measured at Bugey-3
and ILL. The best fit point is moved to $|\Delta m_{\mathrm new,R+S}^2| =
2.4$~eV$^2$ whereas the mixing angle remains almost unchanged,  at
$\sin^2(2\theta_{\mathrm new,R+S}) \sim 0.14$. The no-oscillation hypothesis is
excluded at 99.6\%, corresponding roughly to 2.9$\sigma$.
Figure~\ref{f:sterile_rs} provides the results of the fit in the
$\sin^2(2\theta_{\mathrm new})$--$\Delta m_{\mathrm new}^2$ plane, including both the
reactor experiment rate and shape (Bugey-3 and ILL) data. 

\subsubsection*{Combination of the Reactor and the Gallium Anomalies}

It is also possible to combine the results on the reactor antineutrino anomaly
with the results on the gallium anomaly (see
Section~\ref{sec:gallium})~\cite{Giunti:2010wz}.
The goal is to quantify the compatibility of the reactor and the gallium data. 

For the reanalysis of the Gallex and Sage calibration runs with $^{51}$Cr and
$^{37}$Ar radioactive sources emitting $\sim 1$~MeV electron
neutrinos~\cite{Anselmann:1994ar, Hampel:1997fc, Kaether:2010ag,
Abdurashitov:1996dp, Abdurashitov:1998ne, Abdurashitov:2005tb,
Abdurashitov:2009tn}, the methodology developed in
Ref.~\cite{Giunti:2010zu,Giunti:2010wz} is used. However, in the analysis shown
here, possible correlations between these four measurements are included.
Details are given in Ref.~\cite{Mention:2011rk}. This has the effect of being
slightly more conservative, with the no-oscillation hypothesis disfavored at
97.7\%~C.L., instead of 98\%~C.L.\ in Ref.~\cite{Giunti:2010wz}. Gallex and
Sage observed an average deficit of \mbox{$R_G=0.86\pm 0.06\,(1\sigma)$}.  The
best fit point is at $|\Delta m_{\mathrm gallium}^2 | = 2.4$~eV$^2$ (poorly
defined) whereas the mixing angle is found to be $\sin^2(2\theta_{\mathrm gallium})
\sim 0.27\pm0.13$. Note that the best fit values are very close to those
obtained by the analysis of the rate+shape reactor data.

\begin{figure}
  \begin{center}
    \includegraphics[scale=0.5]{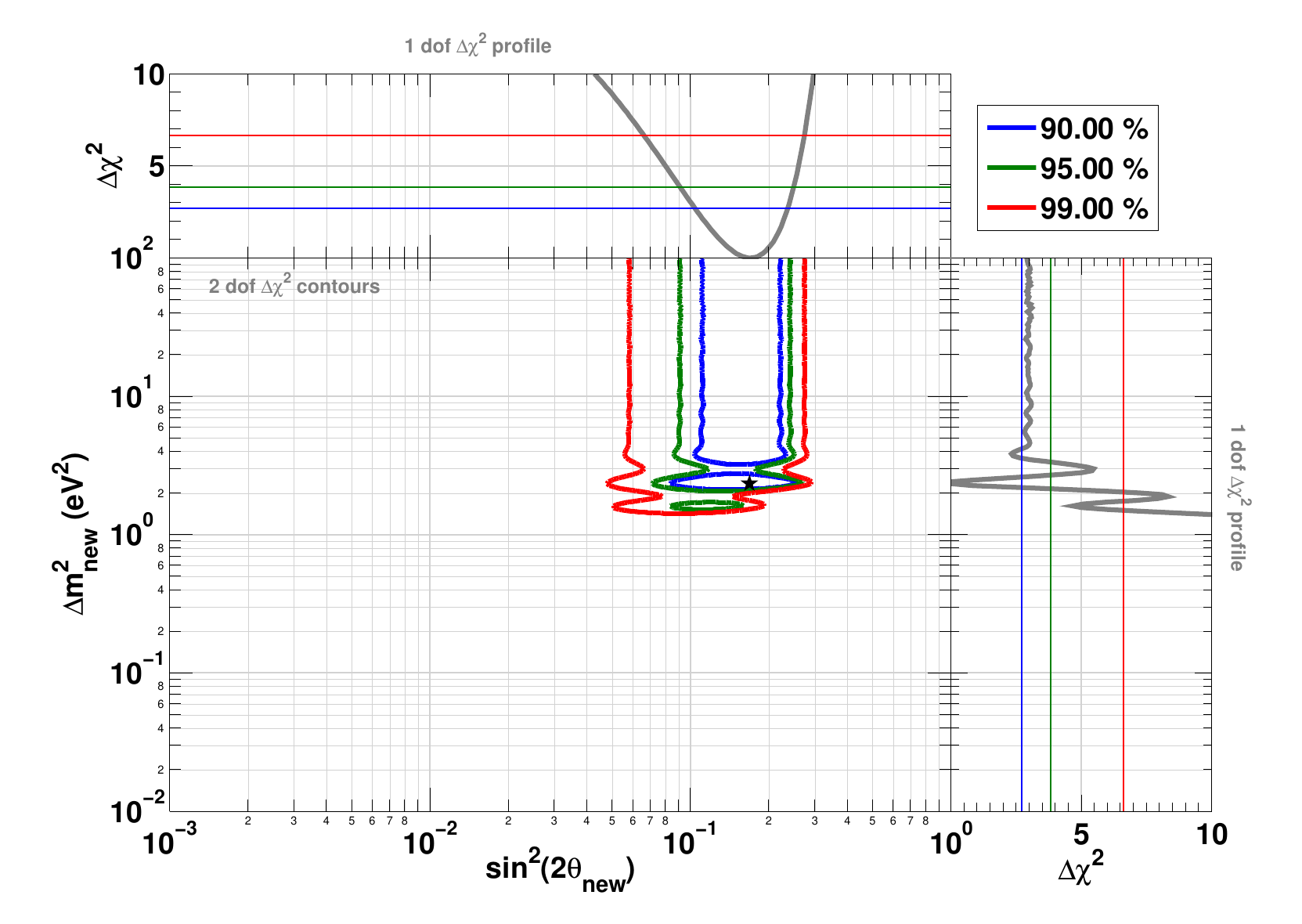}
    \caption{\label{raacontour} Allowed regions in the $\sin^2(2\theta_{\mathrm
      new})$--$\Delta m_{\mathrm new}^2$ plane from the combination of reactor neutrino
      experiments, the Gallex and Sage calibration sources experiments, and the ILL
      and Bugey-3-energy spectra.  The data are well fitted by the 3+1 neutrino
      hypothesis, while the no-oscillation hypothesis is disfavored at
      99.97\%~C.L (3.6 $\sigma$).}
  \end{center}
\end{figure}

Combing both the reactor and the gallium data, The no-oscillation hypothesis is
disfavored at 99.97\%~C.L (3.6 $\sigma$).  Allowed regions in the
$\sin^2(2\theta_{\mathrm new})$--$\Delta m_{\mathrm new}^2$ plane are displayed in
Fig.~\ref{raacontour}, together with the marginal $\Delta\chi^2$ profiles for
$|\Delta m_{\mathrm new}^2 |$ and $\sin^2(2\theta_{\mathrm new})$. The combined fit
leads to the following constraints on oscillation parameters: $|\Delta m_{\mathrm
new}^2|>1.5$~eV$^2$ (99\%~C.L.) and $\sin^2(2\theta_{\mathrm new})=0.17\pm0.04$ (1
$\sigma$).  The most probable $|\Delta m_{\mathrm new}^2 |$ is now rather better
defined with respect to what has been published in Ref.~\cite{Mention:2011rk},
at $|\Delta m_{\mathrm new}^2 | = 2.3\pm0.1$~eV$^2$.

\subsubsection*{Summary of the reactor antineutrino anomaly}

The impact of the new reactor antineutrino spectra has been extensively studied
in Ref.~\cite{Mention:2011rk}. The increase of the expected antineutrino rate
by about~4.5\% combined with revised values of the antineutrino cross section
significantly decreased the normalized ratio of observed to expected event
rates in all previous reactor experiments performed over the last 30~years at
distances below 100~m~\cite{Kwon:1981ua, Declais:1994su, Zacek:1986cu,
Declais:1994ma, Afonin:1994, Kuvshinnikov:1991, Vidyakin:1993, Vidyakin:1994ut,
Greenwood:1996pb}.  The new average ratio, updated early 2012, is now $0.927\pm
0.023$, leading to an enhancement of reactor antineutrino anomaly, now
significant at the $3\sigma$ confidence level.  The best fit point is at
$|\Delta m_{\mathrm new,R+S}^2 | = 2.4$~eV$^2$ whereas the mixing angle is  at
$\sin^2(2\theta_{\mathrm new,R+S}) \sim 0.14$. 

This deficit could still be due to some unknown in the reactor physics, but it
can also be analyzed in terms of a suppression of the $\bar\nu_e$ rate at short
distance as could be expected from a sterile neutrino, beyond the standard
model, with a large $|\Delta m_{\mathrm new}^2| \gg |\Delta m_{31}^2|$.  Note that
hints of such results were already present at the ILL neutrino experiment
in~1981~\cite{ILL95}.

Considering the reactor $\bar\nu_e$ anomaly and the gallium $\nu_e$ source
experiments~\cite{Anselmann:1994ar, Hampel:1997fc, Kaether:2010ag,
Abdurashitov:1996dp, Abdurashitov:1998ne, Abdurashitov:2005tb,
Abdurashitov:2009tn, Giunti:2010wz} together, it is interesting to note that in
both cases (neutrinos and antineutrinos), comparable deficits are observed at
a similar $L/E$.  Furthermore it turns out that each experiment fitted
separately leads to similar values of $\sin^2(2\theta_{\mathrm new})$ and similar
lower bounds for $|\Delta m_{\mathrm new}^2|$, but without a strong significance. A
combined global fit of gallium data and of short-baseline reactor data, taking
into account the reevaluation of the reactor results discussed here, as well as
the existing correlations, leads to a solution for a new neutrino oscillation,
such that $|\Delta m_{\mathrm new}^2|>1.5$~eV$^2$ (99\%~C.L.) and
$\sin^2(2\theta_{\mathrm new})=0.17\pm 0.04$ (1$\sigma$), disfavoring the
no-oscillation case at 99.97\%~C.L (3.6 $\sigma$).  The most probable $|\Delta
m_{\mathrm new}^2 |$ is now at $|\Delta m_{\mathrm new}^2 | = 2.3\pm0.1$~eV$^2$. This
hypothesis should be checked against systematical effects, either in the
prediction of the reactor antineutrino spectra or in the experimental results.

\subsection{Limit on Disappearance Derived from KARMEN and LSND $\nu_e$-Carbon Cross Sections}
\label{sec:nue-Carbon}

The $\nu_e$-carbon cross section data from the KARMEN~\cite{Armbruster:1998uk,Bodmann:1994py} 
and LSND~\cite{Auerbach:2001hz} experiments have been interpreted  within the context of 
electron neutrino oscillations at high $\Delta m^2$, leading to the most stringent limit on 
electron-flavor disappearance relevant to sterile neutrinos~\cite{Conrad:2011ce}.  Both 
experiments measured the cross-section for the 2-body interaction $\nu_e +^{12}{\mathrm C} 
\rightarrow ^{12}{\mathrm N}_{gs} + e^-$.  The neutrino energy can be reconstructed by measuring 
the outgoing visible energy of the electron and accounting for the 17.3 MeV $Q$-value, 
allowing a measurement of the cross section versus neutrino energy.  KARMEN and LSND were 
located at 17.7~m and 29.8~m respectively from the neutrino source.  The neutrino flux 
normalization is known to 10\% \cite{Burman:1996gt,Burman:1989dq}.  Thus, the consistency of 
the two cross section measurements, as a function of antineutrino energy, sets strong limits 
on $\nu_e$ oscillations.

\begin{figure}[t]
\begin{center}
{\includegraphics[width=0.7\textwidth]{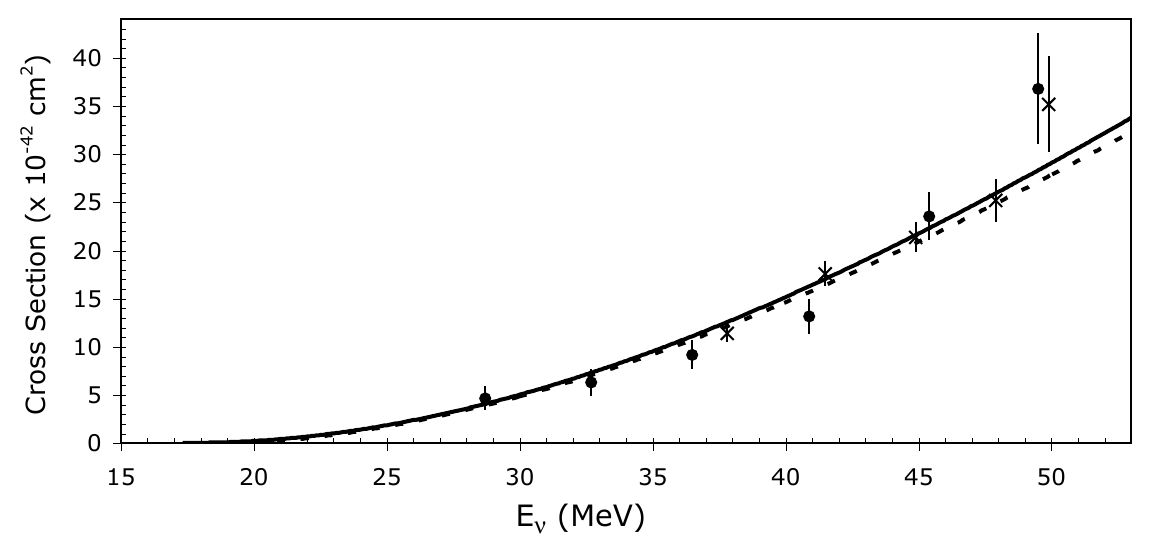}} 
\end{center}
\caption{The KARMEN (points) and LSND (crosses) measured cross sections with statistical 
errors for $\nu_e +^{12}{\mathrm C}\rightarrow ^{12}{\mathrm N}_{gs} + e^-$ compared to the 
theoretical prediction of Fukugita~\cite{Fukugita:1988hg} (solid line), based on the EPT model, 
and Kolbe~\cite{Kolbe:1999au} (dashed line), based on the CRPA model. 
\label{xsecs} }
\end{figure}

Fig.~\ref{xsecs} shows the KARMEN and LSND energy-dependent $\nu_e +^{12}{\mathrm C} \rightarrow 
^{12}{\mathrm N}_{gs} + e^-$ cross sections~\cite{Armbruster:1998uk,Bodmann:1994py,Auerbach:2001hz}.  
Table~\ref{fluxavetab} reports the corresponding flux-averaged cross sections measured by 
KARMEN, LSND and the LANL E225 experiment \cite{Krakauer:1991rf}, which was located 9~m from a 
decay-at-rest source.  Unfortunately, E225 did not publish energy-binned cross section 
measurements, and so is  not included in this analysis.  The agreement between all three 
experiments is excellent.

Predictions for the cross section, also shown in Fig.~\ref{xsecs}, come from Fukugita,
{\it et al.}~\cite{Fukugita:1988hg} and by Kolbe {\it et al.}~\cite{Kolbe:1999au}.  Models 
follow a $(E_\nu-Q)^2$ form, where $Q=17.3$ MeV because the interaction is an allowed 
transition ($0^+$ ($^{12}$C) $\rightarrow$ $1^+$ ($^{12}$N).  The calculations are approached 
in two models: the ``elementary particle model'' (EPT), with an associated 12\% normalization 
uncertainty~\cite{Fukugita:1988hg,Donnelly:1978tz,Mintz:1988zz}; and the ``continuum random 
phase approximation'' (CRPA) approach~\cite{Kolbe:1999au}.    Kolbe,  Langanke, and Vogel 
have provided a discussion of the e relative merits of EPT versus CRPA models for describing 
$\nu_e +^{12}{\mathrm C} \rightarrow ^{12}{\mathrm N}_{gs} + e^-$~\cite{Kolbe:1997jx}. However, from 
a strictly phenomenological point of view, both EPT and CRPA models fit the data well and 
serve as good predictions for tests of oscillations.

\begin{table}[tbp] \centering
\begin{tabular}
[c]{l|c|c}%
\hline
Experiment (Dist.)& Flux-Averaged Cross Section & Ref. \\ \hline
KARMEN (17.7 m)& $(9.1\pm 0.5\pm 0.8) \times 10^{-42}$ cm$^2$ & \cite{Armbruster:1998uk,Bodmann:1994py} \\
LSND (29.8 m) & $(8.9\pm 0.3 \pm 0.9) \times 10^{-42}$ cm$^2$ & \cite{Auerbach:2001hz} \\
E225 (9 .0 m)& $(1.05 \pm 0.10 \pm 0.10) \times 10^{-41}$ cm$^2$ & \cite{Krakauer:1991rf} \\ \hline\hline
Prediction & Flux-Averaged Cross Section & Ref. \\ \hline
Fukugita {\it et al.} & $9.2 \times 10^{-42}$ cm$^2$ & \cite{Fukugita:1988hg}  \\
Mintz {\it et al.}&  $8.0 \times 10^{-42}$ cm$^2$ & \cite{Mintz:1988zz} \\
Donnelly & $9.4 \times 10^{-42}$ cm$^2$  & \cite{Donnelly:1978tz} \\
Kolbe {\it et al.} &  $8.9 \times 10^{-42}$ cm$^2$ & \cite{Kolbe:1999au} \\ \hline
\end{tabular}
\caption{Top: Flux-averaged $\nu_e + ^{12}{\mathrm C} \rightarrow e^+ + ^{12}{\mathrm N}_{gs}$ cross 
section measurements with statistical and systematic error. Bottom:  Flux-averaged predictions 
from EPT (Fukugita, Mintz and Donnelly) and CRPA (Kolbe) models. Flux-average cross section 
values are equivalent to those for a neutrino of 35 MeV energy. \label{fluxavetab}}
\end{table}

The analysis compares the  LSND and KARMEN data with respect to the Fukugita and Kolbe 
predictions to set a limit on $\nu_e \rightarrow \nu_s$ oscillations.  For each set of 
oscillation paramaters, ($\Delta m^2$ and $\sin^2 2\theta_{ee}$), one calculates a combined 
$\chi^2$ for LSND and KARMEN with respect to the oscillation-modified prediction using the 
statistical error for each data point.  Three pull terms incorporate systematic uncertainties: 
\begin{enumerate}
\item{The correlated normalization error.  As noted in the KARMEN 
paper~\cite{Armbruster:1998uk,Bodmann:1994py}, LSND and KARMEN have a common 7\% systematic error 
on the neutrino flux normalization from the flux simulation~\cite{Burman:1996gt,Burman:1989dq}.  
The pull term combines this in quadrature with the 12\% systematic error on the normalization for 
the Fukugita prediction.}   
\item{The uncorrelated LSND experimental uncertainty of 7\%~\cite{Auerbach:2001hz}.}
\item{The uncorrelated KARMEN  experiemental uncertainty of 
5\%~\cite{Armbruster:1998uk,Bodmann:1994py}.}
\end{enumerate}
To obtain the 90\% CL allowed regions, one marginalizes over the three normalization pull 
parameters and use a $\Delta \chi^2 > 4.61$ requirement for the two-degrees-of-freedom 
(associated with $\Delta m^2$ and $\sin^2 2\theta_{ee}$) excluded region.

\begin{figure}[t]
\begin{center}
{\includegraphics[width=0.5\textwidth]{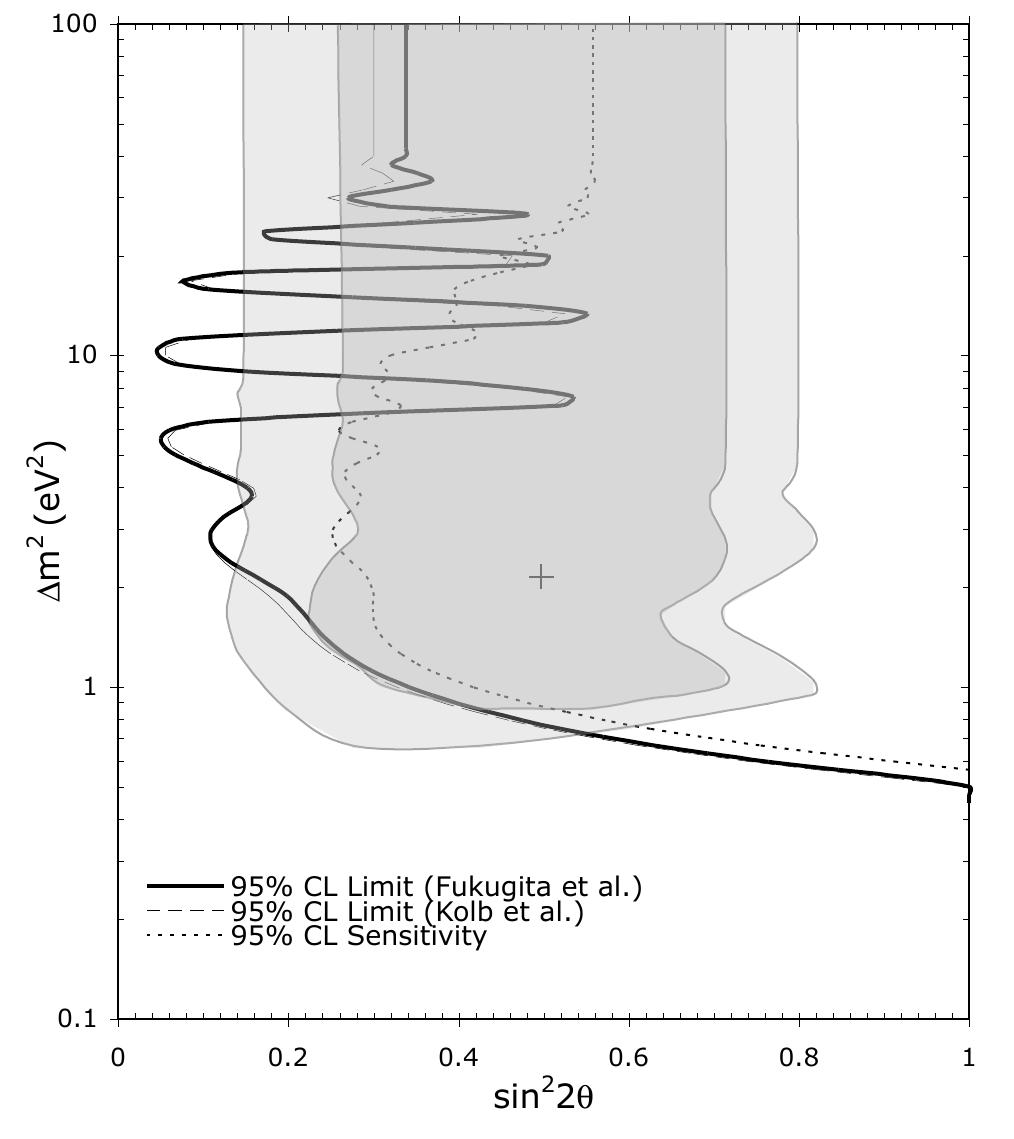}} 
\end{center}
\caption{The 95\% $\nu_e$ disappearance limit from the Fukugita (EPT) fit (solid, black line) 
compared to the predicted sensitivity (dotted line).  Also shown is the 68\% (darker, shaded 
region) and 90\% (lighter, shaded region) contours from the Gallium anomaly.  The dashed line is 
the Kolbe (CRPA) fit.  \label{contour} }
\end{figure}

The result is a 95\% CL exclusion region, although the fit without oscillations has a $\Delta 
\chi^2$ probability of 91.5\% and is only excluded at the 1.7$\sigma$ level.  The result is 
shown in Fig.~\ref{contour}.  The best fit  is at $\Delta m^2 = 7.49 \pm 0.39$ eV$^2$ and 
$\sin^2 2\theta_{ee} = 0.290 \pm 0.115$.  This is true regardless of whether the Fukugita 
(EPT) and Kolbe (CRPA) predictions are used.  Agreement between the results using these two 
predictions indicates that there is no substantial systematic effect from the underlying cross 
section model. 

Fig.~\ref{contour} compares this limit to the gallium anomaly described in 
Section~\ref{sec:gallium}, showing significant disagreement with this signal across substantial 
parameter space.  CPT conservation requires that $\nu_e$ and $\bar \nu_e$ disappearance 
should occur at the same rate.   Assuming that CPT is conserved allows comparison of this limit 
to the antineutrino reactor anomaly of Section~\ref{sec:reactor}, where it addresses a modest 
region of the allowed space.

\subsection{Constraints from the MINOS Long-Baseline Experiment}
\label{subsec:MINOS}

MINOS probes active to sterile neutrino mixing by measuring the rate of
neutral-current (NC) events at two locations, over a baseline of
\unit[735]{km}. Because NC cross-sections are identical among the three active
flavors, NC event rates are unaffected by standard neutrino mixing. However,
oscillations into a sterile non-interacting neutrino flavor would result in an
energy-dependent depletion of NC events at the far site. Furthermore, evidence
of disappearance of charged-current (CC) $\nu_\mu$ events at higher neutrino
energies ($> \unit[10]{\mathrm{GeV}}$), for which oscillations driven by $\Delta
\mathrm{m}^2_{\mathrm{atm}}$ oscillations are suppressed, could be an indication of
oscillations into sterile neutrinos driven by a large mass-square difference
$\sim\unit[1]{\mathrm{eV^2}}$.

MINOS measures neutrinos from the NuMI beam using two detectors: the
\unit[980]{ton} (\unit[27]{ton} fiducial) Near Detector (ND), located
\unit[1.04]{km} downstream of the beam target at Fermilab; and the
\unit[5.4]{kton} (\unit[4.0]{kton} fiducial) Far Detector (FD), placed
\unit[735]{km} downstream of the target in the Soudan Underground Laboratory,
in Minnesota~\cite{Michael:2008bc}. The energy resolution function for
neutrino-induced hadronic showers is approximately $56\%/\sqrt{E}$.  The
results reported were obtained using an exposure of $7.07\times10^{20}$~protons
on target taken exclusively with a beam configuration for which the peak
neutrino event energy is \unit[3.3]{GeV}. The NuMI beam includes a 1.3\%
($\nu_e+\bar{\nu}_e$) contamination arising from the decay of muons originating
in kaon and pion decays.  

\begin{figure}[!h]
\includegraphics[width=0.75\linewidth]{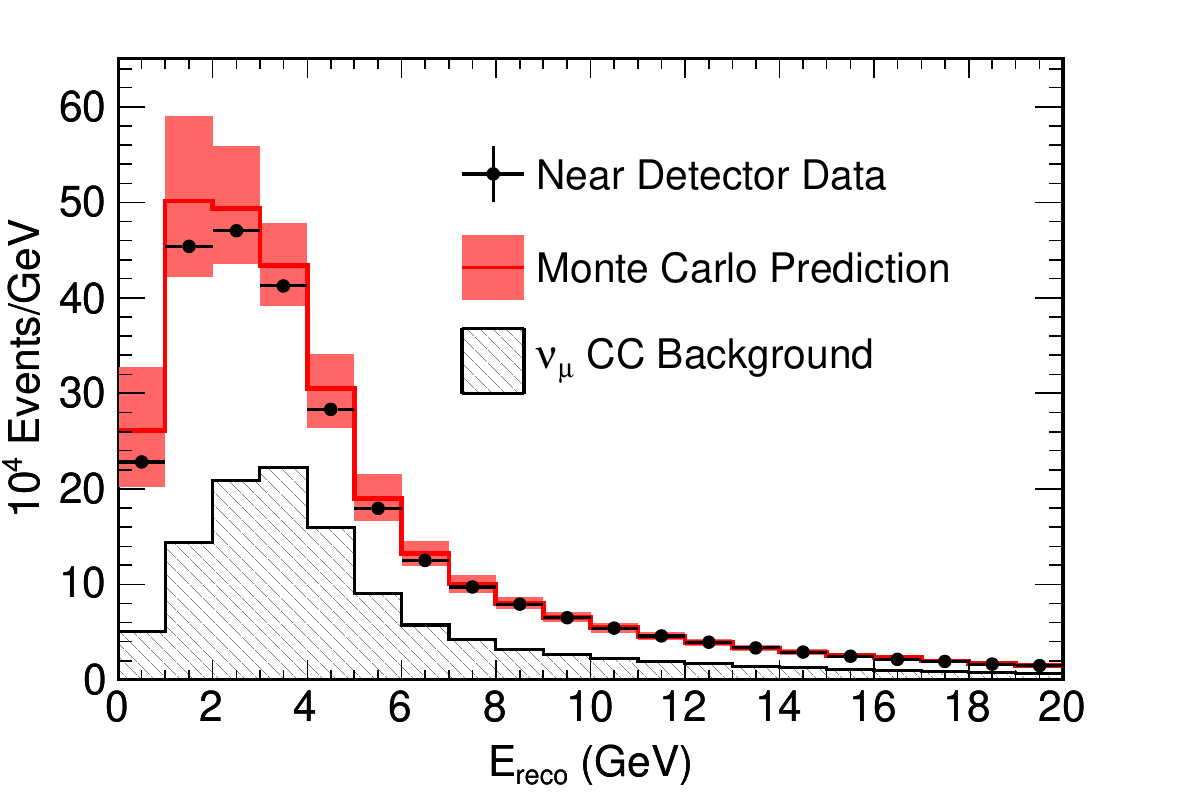}
\caption{ Reconstructed energy spectrum of NC-selected events in the ND compared to the Monte Carlo prediction shown with \unit[1]{$\sigma$} systematic errors (shaded band). Also displayed is the simulation of the background from misidentified CC events (hatched histogram).}
\label{fig:NDSpectrum}
\end{figure}

\begin{figure}[!h]
\includegraphics[width=0.75\linewidth]{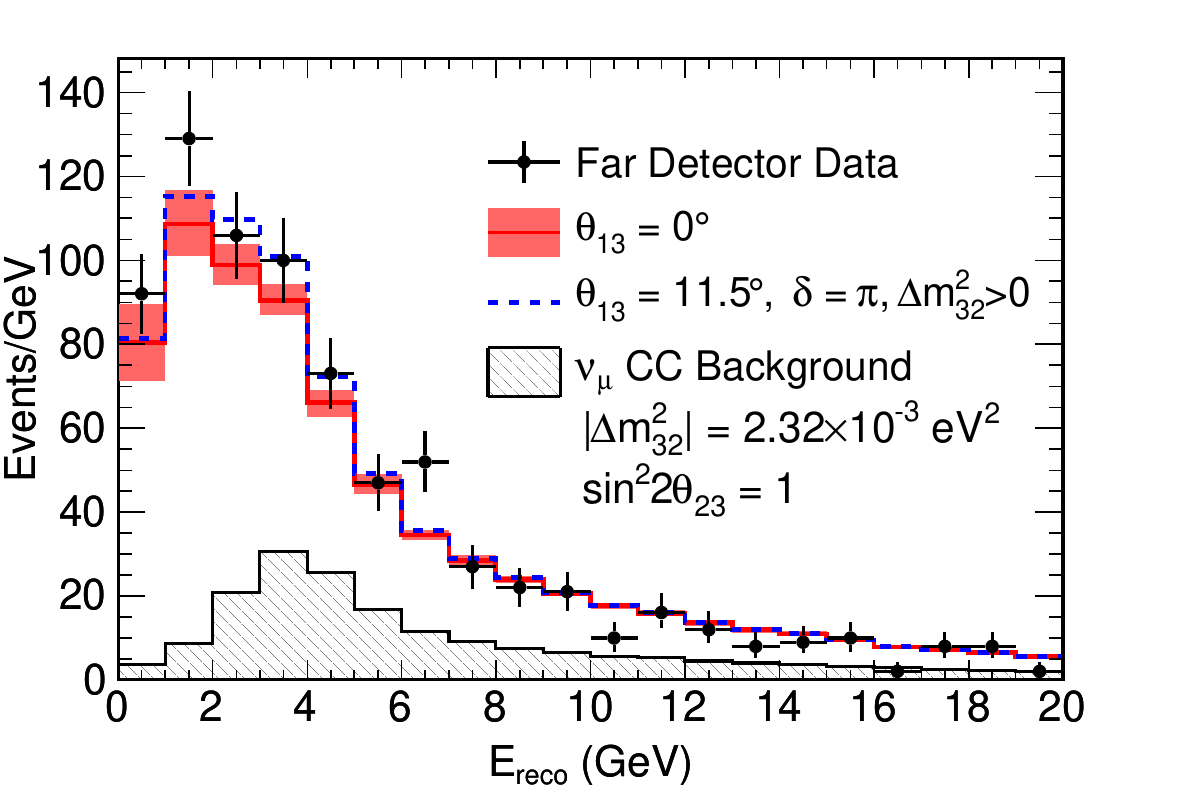}
\caption{Reconstructed energy spectrum of NC-selected events in the FD compared with predictions for standard three-flavor mixing with and without $\nu_e$~appearance at the MINOS 90\% C.L. limit~\cite{Adamson:2010uj}.\\ \\}
\label{fig:FDSpectrum} 
\end{figure}

In the MINOS detectors, NC interactions give rise to events with a short
diffuse hadronic shower and either small or no tracks, whereas CC events
typically display a long muon track accompanied by hadronic activity at the
event vertex.  The separation between NC and CC events proceeds through
selection criteria based on topological variables: events crossing fewer than
47 planes for which no track is reconstructed are selected as NC; events
crossing fewer than 47 planes that contain a track are classified as NC only if
the track extends less than 6 planes beyond the shower. These selections result
in an NC-selected sample with 89\% efficiency and 61\% purity. Highly inelastic
$\nu_{_\mu}$~and $\bar{\nu}_\mu$~CC events, where the muon track is not
distinguishable from the hadronic shower, are the main source of background for
the NC-selected spectrum. Furthermore, the analysis classifies 97\% of
$\nu_e$-induced CC events as NC, requiring the possibility of
$\nu_e$~appearance to be considered when extracting results.  The predicted NC
energy spectrum in the FD is obtained using the ND data. An estimate of the
ratio of events in the FD and ND as a function of reconstructed energy,
$E_{\mathrm{reco}}$, is calculated from Monte Carlo simulations. The ratio is
multiplied by the observed ND energy spectrum to produce the predicted FD
spectrum. To avoid biases, the analysis selections and procedures were
determined prior to examining the FD data, following the precepts of a blind
analysis. Figures~\ref{fig:NDSpectrum} and~\ref{fig:FDSpectrum} show the
reconstructed energy spectra in each detector. The events not selected as
NC-like are fed to the kNN-based selection method used by the MINOS $\nu_\mu$
CC disappearance analysis~\cite{Adamson:2011ig} and the predicted CC FD energy
spectrum is obtained with the same extrapolation methodology used for the NC
prediction.

The NC selection procedures identify 802 NC interaction candidates in the FD,
with $754\pm28\mathrm{(stat)}\pm{37}\mathrm{(syst)}$ events expected from standard
three-flavor mixing (assuming $\theta_{13}=0^\circ$)~\cite{Adamson:2011ku}. The
NC-like reconstructed energy spectrum is shown on Fig.~\ref{fig:FDSpectrum}.

The agreement between the observed and predicted NC spectra is quantified using
the statistic $R\equiv\frac{N_{\mathrm{data}}-B_{\mathrm{CC}}}{S_{\mathrm{NC}}}$, where
$N_{\mathrm{data}}$ is the observed number of events, $B_{\mathrm{CC}}$ is the
predicted CC background from all flavors, and  $S_{\mathrm{NC}}$ is the expected
number of NC interactions. The values of $N_{\mathrm{data}}$, $S_{\mathrm{NC}}$ and
contributions to $B_{\mathrm{CC}}$ for various reconstructed energy ranges are
shown in Table~\ref{tab:nums}. 

\begin{table}[h!]
\begin{center}
\begin{tabular}{llcccc}
\hline
$E_{\mathrm{reco}}$ (GeV)      & $N_{\mathrm{Data}}$ & ~~$S_{\mathrm{NC}}$~~ & ~~$B^{\nu_\mu}_{\mathrm{CC}}$~~ & ~~$B^{\nu_\tau}_{\mathrm{CC}}$~~ & ~~$B^{\nu_e}_{\mathrm{CC}}$~~ \bigstrut \\ \hline
$0-3$    & 327 & 248.4 & 33.2 & 3.2 & 3.1~(21.5) \bigstrut  \\
$3-120$  & 475 & 269.6 & 156.0 & 9.2 & 31.2~(53.8) \bigstrut \\
\hline
$0-3$ &\multicolumn{5}{l}{$R=1.16\pm0.07\pm0.08-0.08(\nu_e)$} \bigstrut\\
$3-120$ &\multicolumn{5}{l}{$R=1.02\pm0.08\pm0.06-0.08(\nu_e)$} \bigstrut\\
$0-120$ &\multicolumn{5}{l}{$R=1.09\pm0.06\pm0.05-0.08(\nu_e)$} \bigstrut \\
\hline
\end{tabular}\\
\end{center}
\caption{Values of the $R$ statistic and its components for several reconstructed energy ranges.  The numbers shown in parentheses include $\nu_e$~appearance with $\theta_{13}=11.5^\circ$ and $\delta_{CP}=\pi$.  The displayed uncertainties are statistical, systematic, and the uncertainty associated with $\nu_e$~appearance.} 
\label{tab:nums}
\end{table}

The values of $R$ for each energy range show no evidence of a depletion in the
NC rate at the FD, supporting the hypothesis that standard three-flavor
oscillations explain the data. 

\begin{figure}[h!]
\includegraphics[width=0.75\linewidth]{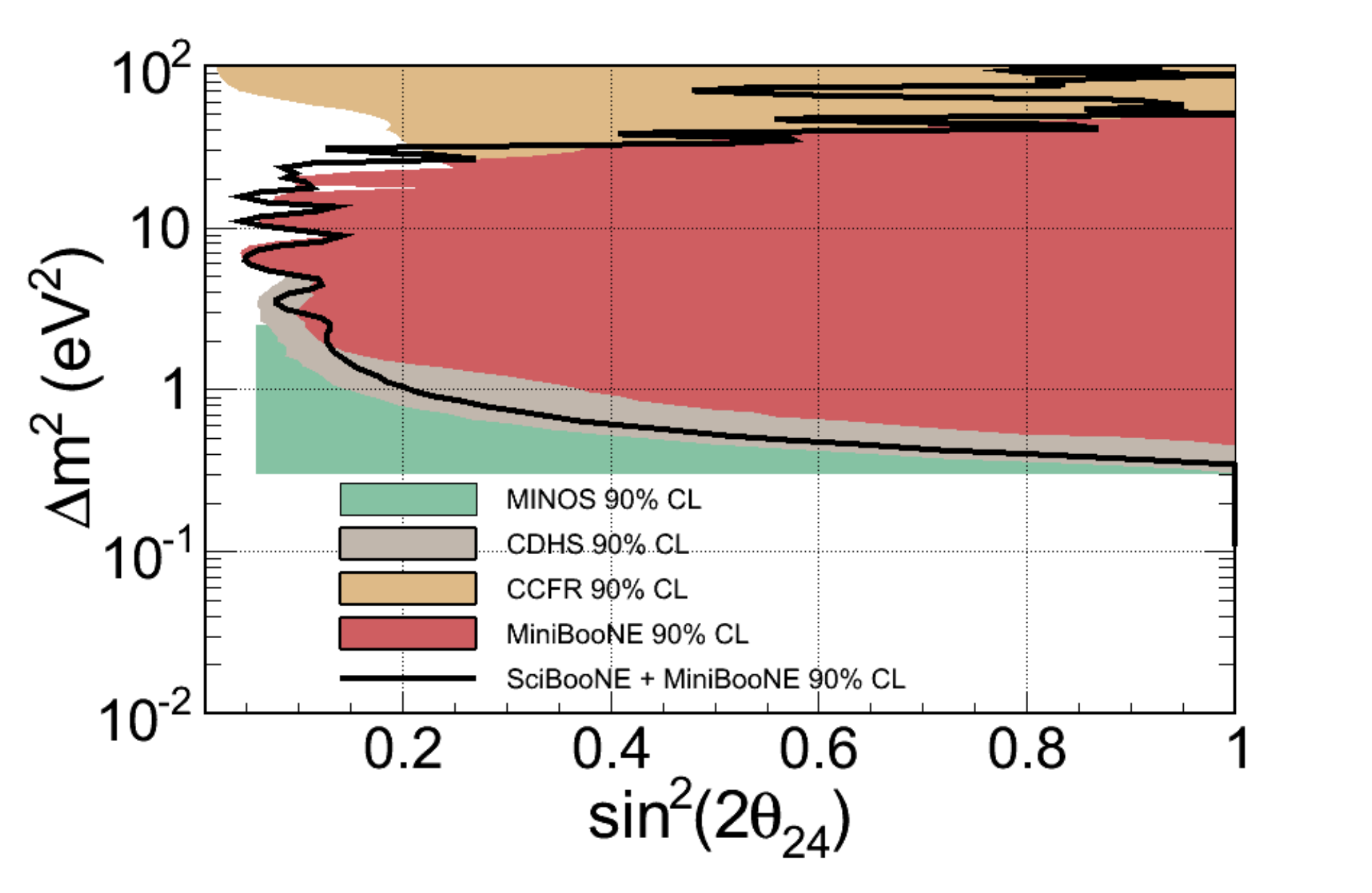}
\caption{MINOS exclusion compared to MiniBooNE, CDHS, CCFR $\nu_\mu$ disappearance results. The MINOS 90\%CL excluded region is shown in green. Following the discussion in the text, the bound is assumed to be valid for $0.3~<~\Delta m_{43}^2~<~\unit[2.5]{eV^2}$. }
\label{fig:theta24} 
\end{figure}

The data are compared with a neutrino oscillation model that allows admixture
with one sterile neutrino. In this model, an additional mass scale $\Delta
m_{43}^2$ with magnitude $\mathcal{O}$(\unit[1]{eV$^{2}$}) is introduced, along
with the assumption that no oscillation-induced change of the neutrino event
rate is measurable at the ND site, but rapid oscillations are predicted at the
FD location. Based on the magnitude of the systematic uncertainties at the ND,
this approximation is assumed to be valid for $0.3<\Delta
m_{43}^2<\unit[2.5]{eV^2}$. A detailed description of this model is provided in
Ref.~\cite{Adamson:2010wi} and references therein. 
Both the NC-selected energy spectrum shown in Fig.~\ref{fig:FDSpectrum} and the
CC-selected spectrum in the FD data are used in the fits to the oscillation
models.  Limits on the sterile mixing angles  $\theta_{34} <
26^\circ\,(37^\circ)$ and $\theta_{24} < 7^\circ\,(8^\circ)$ are obtained at
the 90\% C.L..  The numbers in parentheses represent the limits extracted for
maximal $\nu_e$~appearance. The latter result is presently the most stringent
constraint on $\theta_{24}$ for $\Delta m_{43}^2\sim\unit[1]{eV^2}$. A
comparison of the MINOS result with other disappearance results is shown in
Fig.~\ref{fig:theta24}.

The coupling between active and sterile neutrinos may also be quantified in terms of the fraction of disappearing $\nu_\mu$~that oscillate into $\nu_s$, $f_{s}~\equiv~P_{\nu_\mu\rightarrow\nu_s}/(1-P_{\nu_\mu\rightarrow\nu_\mu})$, where the $P_{\nu_\mu\rightarrow\nu_x}$ refer to neutrino oscillation probabilities. MINOS places the most stringent constraint to date on this quantity, $f_{s} < 0.22\,(0.40)$ at the 90\%~C.L., where the number in parentheses denotes the limit assuming $\nu_e$~appearance.

A recent paper by Hernandez and Smirnov~\cite{Hernandez:2011rs} analyzed the
MINOS sterile neutrino search results in detail and points out that for large
enough values of $\Delta m_{43}^2$, oscillations at the ND, even if not visible
due to the size of systematic uncertainties on the ND energy spectrum, could
affect the FD predicted spectrum and reduce the strength of the constraint on
$\theta_{24}$ for $\Delta m_{43}^2>\unit[1]{eV^2}$. Moreover, in the same paper
and also in a paper by Giunti and Laveder~\cite{Giunti:2011hn}, it is pointed
out that the combination of the MINOS bound on $\theta_{24}$ with the Bugey
bound on $\theta_{14}$ largely rules out the LSND signal region for $0.3<\Delta
m_{43}^2<\unit[1]{eV^2}$. Motivated by these remarks, MINOS is working on the
inclusion of ND oscillations into the sterile mixing models and will update the
analysis with reviewed results during 2012. These models are already being
employed in the computation of sterile oscillation sensitivities for the MINOS+
project, as described in Appendix Section~\ref{subapp:MINOS+}.
The results described above use exclusively NuMI data taken in neutrino running
mode, which include a small 7\% antineutrino contamination flatly distributed
at higher neutrino energies. MINOS also plans to run the sterile neutrino
analysis on the $3.3\times10^{20}$~POT of antineutrino data already collected.
The main challenges for this analysis arise from limited statistics and
feed-down NC interactions from the higher energy neutrino component, which
accounts for 58\% of the beam composition when the NuMI horns focus $\pi^-$ and
$K^-$. Results of this analysis are also expected during 2012.

\subsection{Conclusion}

The LSND and MiniBooNE antineutrino experiments observe excesses of events
consistent with $\bar \nu_\mu \rightarrow \bar \nu_e$ oscillations at a $\Delta
m^2 \sim 1$ eV$^2$.  Although no excess is observed by the KARMEN experiment,
which was very similar to LSND, a joint analysis of the two experiments still
favors an allowed oscillation region with $\Delta m^2 < 2$~eV$^2$. The
MiniBooNE neutrino experiment observes an excess of events at low energy, which
may be consistent with the antineutrino experiments with the inclusion of CP
violation. A joint analysis of the MiniBooNE neutrino and antineutrino runs by
the MiniBooNE collaboration is underway.

Further potential evidence for the existence of eV-scale sterile neutrinos
comes from non-accelerator experiments. The Gallium anomaly---an event deficit
observed in experiments with neutrinos from intense radioactive sources---has a
statistical significance of $2.7\sigma$. The reactor antineutrino anomaly---a
deficit of the measured reactor antineutrino flux compared to new and improved
flux predictions---has a signficance of $3\sigma$.

There are thus hints for the existence of sterile neutrinos from several
different experiments, employing different neutrino sources and detector
technologies, though none of them can claim a discovery.  On the other hand, a
number of other short-baseline oscillation searches have yielded null results,
which, at least in the simplest sterile neutrino scenarios, are in some tension
with the positive hints (see section~\ref{sec:global} for a discussion of
global fits).

\clearpage
\newcommand{\boss}[2]{\ensuremath{\rlap{\kern-2.5pt\ensuremath{\overset{\scriptscriptstyle(-)}{\phantom{#1}}}}{\ensuremath{{#1}_{#2}}}}}

\section{Global Picture}
\label{sec:global}

As we have seen in the previous sections, a number of (yet inconclusive)
experimental results cannot be explained within the standard three-flavor
framework and, when interpreted in terms of neutrino oscillations, seem to
require at least an additional neutrino with a mass at the eV
scale~\cite{Aguilar:2001ty, AguilarArevalo:2010wv,Mention:2011rk}. Such
neutrinos cannot participate in the weak interactions due to collider
constraints, and are therefore called ``sterile'' neutrinos. Apart from
those hints for sterile neutrinos several data set exist which do not show any
evidence for neutrino oscillations at the eV scale and the important
question has to be addressed, whether a consistent description of all data
is possible if sterile neutrinos are assumed, see \cite{Goswami:1995yq, 
GomezCadenas:1995sj, Bilenky:1996rw, Okada:1996kw, Barger:1997yd, Barger:1998bn, 
Bilenky:1998ne, Bilenky:1999ny, Barger:2000ch, Fogli:2000ir, Peres:2000ic, 
Grimus:2001mn, GonzalezGarcia:2001uy, Maltoni:2001mt, Maltoni:2001bc, 
Maltoni:2002xd, Strumia:2002fw} for a list of early references. The following sections 
\ref{sec:giunti-global}, \ref{sec:KMS}, \ref{sec:Karagiorgi} review
phenomenological studies and global fits taking into account all relevant
data~\cite{Giunti:2011gz,Giunti:2011hn,Giunti:2011cp, Maltoni:2007zf,
Akhmedov:2010vy, Kopp:2011qd, Karagiorgi:2006jf,Karagiorgi:2009nb,
Karagiorgi:2011ut}. In Section~\ref{sec:beta} possible implications
of sterile neutrinos for $\beta$-decay and neutrinoless double $\beta$-decay
are discussed. A brief summary is given in Section~\ref{sec:summary}.

\subsection{3+1 Global Fit of Short-Baseline Neutrino Oscillation Data}
\label{sec:giunti-global}


\bigskip

In this Section we review the results of the global fit of short-baseline
neutrino oscillation data in the framework of 3+1 neutrino mixing presented
in Refs.~\cite{Giunti:2011gz,Giunti:2011hn,Giunti:2011cp}.

Short-baseline (SBL) neutrino oscillations are generated by a squared-mass difference
$\Delta{m}^2_{\text{SBL}} \gtrsim 0.1 \, \text{eV}^2$,
which is much larger than
the two measured solar (SOL) and atmospheric (ATM) squared-mass differences
$
\Delta{m}^2_{\text{SOL}}
=
(7.6 \pm 0.2) \times 10^{-5} \, \text{eV}^2
$
\cite{Abe:2010hy}
and
$
\Delta{m}^2_{\text{ATM}}
=
2.32 {}^{+0.12}_{-0.08} \times 10^{-3} \, \text{eV}^2
$
\cite{Adamson:2011ig}.
The minimal neutrino mixing schemes which can provide a third squared-mass difference
for short-baseline neutrino oscillations
require the introduction of a sterile neutrino $\nu_{s}$
(see Refs.~\cite{Bilenky:1998dt,Maltoni:2004ei,Strumia:2006db,GonzalezGarcia:2007ib}).
Hierarchical
3+1 neutrino mixing is a perturbation of the standard three-neutrino mixing in which
the three active neutrinos
$\nu_{e}$,
$\nu_{\mu}$,
$\nu_{\tau}$
are mainly composed of three massive neutrinos
$\nu_1$,
$\nu_2$,
$\nu_3$
with light masses
$m_1$,
$m_2$,
$m_3$,
such that
$
\Delta{m}^2_{\text{SOL}}
=
\Delta{m}^2_{21}
$
and
$
\Delta{m}^2_{\text{ATM}}
=
|\Delta{m}^2_{31}|
\simeq
|\Delta{m}^2_{32}|
$,
with the standard notation
$\Delta{m}^2_{kj} \equiv m_{k}^2 - m_{j}^2$
(see Ref.~\cite{Giunti-Kim-2007}).
The sterile neutrino is mainly composed of a heavy neutrino
$\nu_{4}$
with mass $m_{4}$
such that
$\Delta{m}^2_{\text{SBL}} = \Delta{m}^2_{41}$
and
\begin{equation}
m_{1}
\,,\,
m_{2}
\,,\,
m_{3}
\ll
m_{4}
\quad
\Rightarrow
\quad
m_{4} \simeq \sqrt{\Delta{m}^2_{41}}
\,.
\label{hierarchy}
\end{equation}
Under these hypotheses, the effects of active-sterile neutrino mixing in 
solar~\cite{Giunti:2009xz,Palazzo:2011rj} and 
atmospheric~\cite{Choubey:2007ji,Razzaque:2011ab,Gandhi:2011jg,Barger:2011rc}
neutrino experiments are small, but should be revealed sooner or later.

In 3+1 neutrino mixing, the effective flavor transition and survival 
probabilities in short-baseline neutrino oscillation experiments are given 
by (see Refs.~\cite{Bilenky:1998dt,Maltoni:2004ei,Strumia:2006db,
GonzalezGarcia:2007ib})
\begin{align}
\null & \null
P_{\boss{\nu}{\alpha}\to\boss{\nu}{\beta}}^{\text{SBL}}
=
\sin^{2} 2\theta_{\alpha\beta}
\sin^{2}\left( \frac{\Delta{m}^2_{41} L}{4E} \right)
\qquad
(\alpha\neq\beta)
\,,
\label{trans}
\\
\null & \null
P_{\boss{\nu}{\alpha}\to\boss{\nu}{\alpha}}^{\text{SBL}}
=
1
-
\sin^{2} 2\theta_{\alpha\alpha}
\sin^{2}\left( \frac{\Delta{m}^2_{41} L}{4E} \right)
\,,
\label{survi}
\end{align}
for
$\alpha,\beta=e,\mu,\tau,s$,
with the transition amplitudes
\begin{align}
\null & \null
\sin^{2} 2\theta_{\alpha\beta}
=
4 |U_{\alpha4}|^2 |U_{\beta4}|^2
\,,
\label{transsin}
\\
\null & \null
\sin^{2} 2\theta_{\alpha\alpha}
=
4 |U_{\alpha4}|^2 \left( 1 - |U_{\alpha4}|^2 \right)
\,.
\label{survisin}
\end{align}

The hierarchical 3+1 scheme may be compatible with the results of standard 
cosmological $\Lambda$CDM analyses of the Cosmic Microwave Background and 
Large-Scale Structures data, which constrain the three light neutrino masses
to be much smaller than 1~eV~\cite{Fogli:2008ig,Reid:2009nq,Thomas:2009ae,GonzalezGarcia:2010un}
and indicate that one or two sterile neutrinos may have been thermalized in 
the early Universe~\cite{Hamann:2010bk,Giusarma:2011ex,Kristiansen:2011mp,Hou:2011ec,GonzalezMorales:2011ty,Hamann:2011ge,Archidiacono:2011gq,Hamann:2011hu},
with a upper limit of the order of 1 eV on their masses.  Also Big-Bang 
Nucleosynthesis data~\cite{Cyburt:2004yc,Izotov:2010ca} are compatible with 
the existence of sterile neutrinos, with the indication however that the 
thermalization of more than one sterile neutrino is 
disfavored~\cite{Mangano:2011ar,Hamann:2011ge}.

We made a global fit of the following sets of data:

\begin{itemize}

\item
The short-baseline $\boss{\nu}{\mu}\to\boss{\nu}{e}$ data of the
LSND~\cite{Aguilar:2001ty}, KARMEN~\cite{Armbruster:2002mp},
NOMAD~\cite{Astier:2003gs} and MiniBooNE neutrino~\cite{AguilarArevalo:2008rc} 
and antineutrino~\cite{AguilarArevalo:2010wv,Zimmerman-PANIC2011,
Djurcic-NUFACT2011} experiments.

\item
The short-baseline $\bar\nu_{e}$ disappearance data
of the
Bugey-3 \cite{Declais:1994su},
Bugey-4 \cite{Declais:1994ma},
ROVNO91 \cite{Kuvshinnikov:1990ry},
Gosgen \cite{Zacek:1986cu},
ILL \cite{Hoummada:1995zz}
and
Krasnoyarsk \cite{Vidyakin:1990iz}
reactor antineutrino experiments,
taking into account the new calculation of the reactor $\bar\nu_{e}$ flux
\cite{Mueller:2011nm,Huber:2011wv}
which indicates a small $\bar\nu_{e}$ disappearance
(the reactor antineutrino anomaly \cite{Mention:2011rk}),
and the KamLAND \cite{:2008ee} bound on $|U_{e4}|^2$
(see Ref.~\cite{Giunti:2010uj}).

\item
The short-baseline $\nu_{\mu}$ disappearance data
of the CDHSW experiment
\cite{Dydak:1983zq},
the constraints on $|U_{\mu4}|^2$ obtained in Ref.~\cite{Maltoni:2007zf}
from the analysis of
the data of
atmospheric neutrino oscillation experiments,
and
the bound on $|U_{\mu4}|^2$ obtained from MINOS 
neutral-current data \cite{Adamson:2011ku}
(see Refs.~\cite{Hernandez:2011rs,Giunti:2011hn}).

\item
The
data of Gallium radioactive source experiments
(GALLEX
\cite{Anselmann:1994ar,Hampel:1997fc,Kaether:2010ag}
and
SAGE
\cite{Abdurashitov:1996dp,Abdurashitov:1998ne,Abdurashitov:2005tb,Abdurashitov:2009tn})
which indicate a $\nu_{e}$ disappearance
(the Gallium neutrino anomaly
\cite{Bahcall:1994bq,Laveder:2007zz,Giunti:2006bj,Giunti:2007xv,Acero:2007su,
Giunti:2009zz,Giunti:2010wz,Gavrin:2010qj,Giunti:2010zu,Mention:2011rk}).
We analyze the Gallium data according to Ref.~\cite{Giunti:2010zu}.

\item
The
$\nu_{e} + {}^{12}\text{C} \to {}^{12}\text{N}_{\text{g.s.}} + e^{-}$
scattering data of the
KARMEN \cite{Bodmann:1994py,Armbruster:1998uk}
and
LSND \cite{Auerbach:2001hz}
experiments,
which constrain the short-baseline $\nu_{e}$ disappearance
\cite{Conrad:2011ce}.

\end{itemize}

\begin{figure*}[t!]
\begin{center}
\begin{tabular}{c}
\includegraphics*[width=0.9\textwidth]{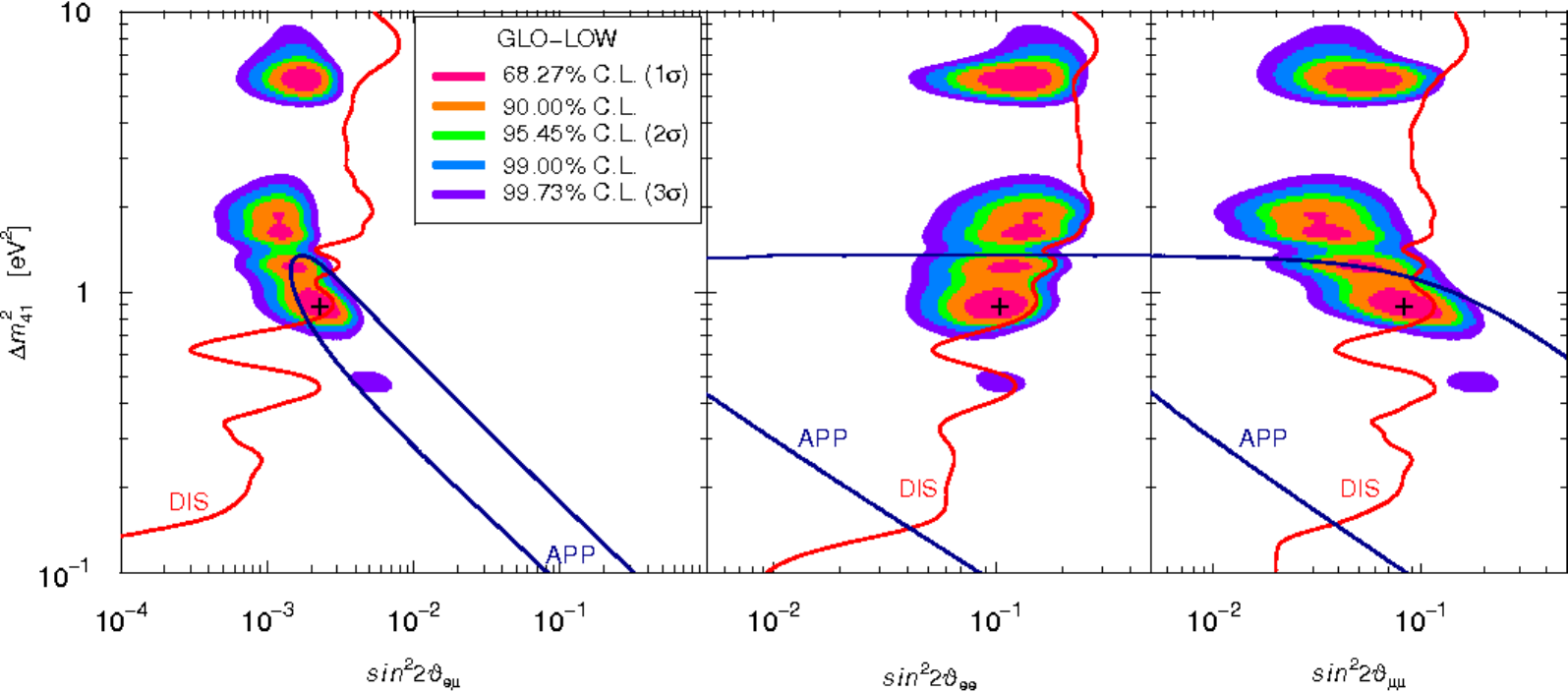}
\\
\includegraphics*[width=0.9\textwidth]{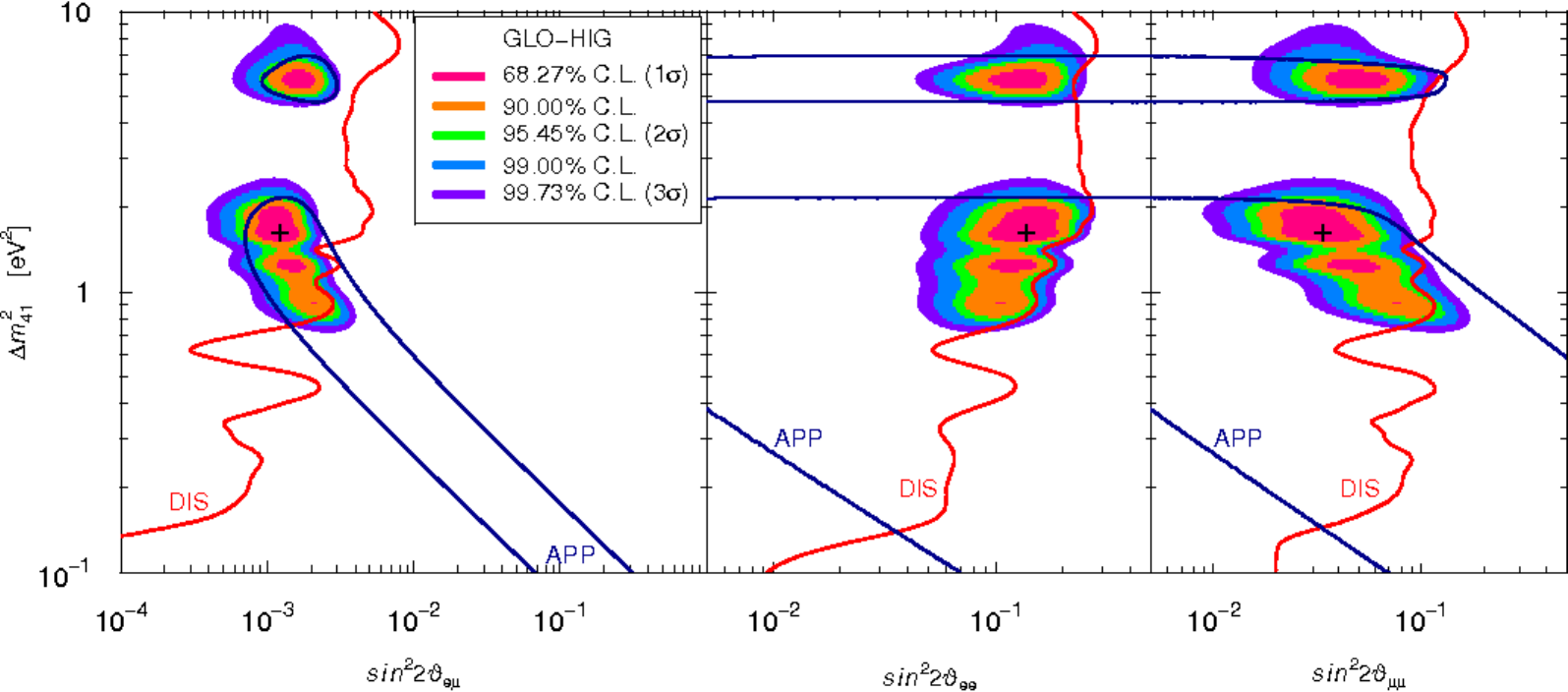}
\end{tabular}
\end{center}
\caption{ \label{con-glo}
Allowed regions in the
$\sin^{2}2\theta_{e\mu}$--$\Delta{m}^2_{41}$,
$\sin^{2}2\theta_{ee}$--$\Delta{m}^2_{41}$ and
$\sin^{2}2\theta_{\mu\mu}$--$\Delta{m}^2_{41}$
planes obtained
\protect\cite{Giunti:2011cp}
from the GLO-LOW and GLO-HIG
global analyses of short-baseline neutrino oscillation data
(see Tab~\ref{tab-bef}).
The best-fit points are indicated by crosses (see Table.~\ref{tab-bef}).
The thick solid blue lines with the label APP
show the $3\sigma$ allowed regions obtained from the analysis of
$\protect\boss{\nu}{\mu}\to\protect\boss{\nu}{e}$
appearance data.
The thick solid red lines with the label DIS
show the $3\sigma$ allowed regions obtained from the analysis of
disappearance data.
}
\end{figure*}

Figure~\ref{con-glo} shows the allowed regions in the
$\sin^{2}2\theta_{e\mu}$--$\Delta{m}^2_{41}$,
$\sin^{2}2\theta_{ee}$--$\Delta{m}^2_{41}$ and
$\sin^{2}2\theta_{\mu\mu}$--$\Delta{m}^2_{41}$
planes obtained from the global analysis of short-baseline neutrino oscillation data.
We considered values of
$\Delta{m}^2_{41}$
smaller than $10\,\text{eV}^2$,
since larger values are strongly incompatible with
the cosmological constraints on neutrino masses\cite{Hamann:2010bk,Giusarma:2011ex,
Kristiansen:2011mp,Hou:2011ec,GonzalezMorales:2011ty,Hamann:2011ge,Archidiacono:2011gq,
Hamann:2011hu}.
We made two global analyses named
GLO-LOW and GLO-HIG, respectively,
with and without
the three MiniBooNE electron neutrino and antineutrino bins
with reconstructed neutrino energy smaller than $475 \, \text{MeV}$,
which have an excess of events called the "MiniBooNE low-energy anomaly".
The best-fit values of the oscillation parameters are listed in
Tab.~\ref{tab-bef}.

\begin{table*}[t!]
\begin{center}
\setlength{\tabcolsep}{5pt}
\begin{tabular}{cccc}
&
&
GLO-LOW
&
GLO-HIG
\\
\hline
 No Osc. &$\chi^{2}_{\text{min}}$ &195.1 &178.1 \\
 &NDF &147 &141 \\
 &GoF &0.49\% &1.9\% \\
\hline 3+1 &$\chi^{2}_{\text{min}}$ &152.4 &137.5 \\
 &NDF &144 &138 \\
 &GoF &30\% &50\% \\
 &$\Delta{m}^2_{41} [\text{eV}^2]$ &0.9 &1.6 \\
 &$|U_{e4}|^2$ &0.027 &0.036 \\
 &$|U_{\mu4}|^2$ &0.021 &0.0084 \\
 &$\sin^22\theta_{e\mu}$ &0.0023 &0.0012 \\
 &$\sin^22\theta_{ee}$ &0.10 &0.14 \\
 &$\sin^22\theta_{\mu\mu}$ &0.083 &0.034 \\
\hline PG &$\Delta\chi^{2}_{\text{min}}$ &18.8 &11.6 \\
 &NDF &2 &2 \\
 &GoF &0.008\% &0.3\% \\
\hline
\end{tabular}
\caption{ \label{tab-bef}
Values of $\chi^{2}$, number of degrees of freedom (NDF), goodness-of-fit (GoF)
and best-fit values of the 3+1 oscillation parameters obtained~\cite{Giunti:2011cp}
from the global fit with (GLO-LOW) and without (GLO-HIG)
the MiniBooNE electron neutrino and antineutrino data
with reconstructed neutrino energy smaller than $475 \, \text{MeV}$.
The first three lines correspond to the case of no oscillations (No Osc.).
The following nine lines correspond to the case of 3+1 mixing.
The last three lines give the parameter goodness-of-fit (PG) \protect\cite{Maltoni:2003cu}.
}
\end{center}
\end{table*}

Comparing the GLO-LOW and GLO-HIG parts of Fig.~\ref{con-glo} and Tab.~\ref{tab-bef}
one can see that the inclusion of the fit of
the MiniBooNE low-energy data
favors small values of
$\Delta{m}^2_{41}$.
This fact has been noted and explained in Ref.~\cite{Giunti:2011hn}.
Hence,
the results of the GLO-LOW analysis are more attractive than those of the GLO-HIG in view of
a better compatibility with
cosmological constraints on the neutrino masses.

The results of both the
GLO-LOW and GLO-HIG analyses confirm
the well known tension between appearance and disappearance data
\cite{Bilenky:1996rw,Bilenky:1999ny,Maltoni:2002xd,Sorel:2003hf,Maltoni:2004ei,
Karagiorgi:2006jf,Maltoni:2007zf,Akhmedov:2010vy,Karagiorgi:2009nb,Giunti:2010uj,
Kopp:2011qd,Giunti:2011gz,Giunti:2011hn,Karagiorgi:2011ut}.
In the GLO-LOW analysis the
0.008\%
parameter goodness-of-fit shows that the tension is very strong.
Nevertheless,
we do not fully discard this solution here,
because we are not aware of an a-priori argument which allows to
exclude from the analysis
the MiniBooNE low-energy data.
The exclusion a-posteriori motivated by the results of the fit
may be hazardous,
taking also into account the nice value of the global
goodness-of-fit
(30\%)
and the above-mentioned preference for small values of $\Delta{m}^2_{41}$ in agreement 
with the same preference of the cosmological data.
Considering the GLO-HIG analysis, the 0.3\%
appearance-disappearance
parameter goodness-of-fit
is not dramatically low and the fit cannot be rejected,
also taking into account the pleasant
50\%
value of the global goodness-of-fit.

\subsection{3+1 and 3+2 Fits of Short-Baseline Experiments}
\label{sec:KMS}


\bigskip

Here we report the results on global fits assuming one or two sterile
neutrinos obtained in Ref.~\cite{Kopp:2011qd}, updated with the latest
constraints on $\nu_\mu$ mixing with eV states from MINOS NC data
\cite{Adamson:2011ku}.  We first discuss the implications of the new reactor
antineutrino fluxes (see section~\ref{sec:oscillation} and
refs.~\cite{Mueller:2011nm, Huber:2011wv}) and later we present the fit
results of the global data.

\subsubsection{Short-baseline reactor experiments}

Following~\cite{Mention:2011rk}, we have analyzed data from several short
baseline ($L \lesssim 100$) reactor experiments.  In particular, we include
full spectral data from the Bugey3 detectors~\cite{Declais:1994su} at distances
of 15, 40 and 95~m from the reactor core, and we take into account the total
event rates measured at Bugey4~\cite{Declais:1994ma},
ROVNO~\cite{Kuvshinnikov:1990ry}, Krasnoyarsk~\cite{Vidyakin:1987ue},
ILL~\cite{Kwon:1981ua}, and G\"osgen~\cite{Zacek:1986cu} (see Table~II of
\cite{Mention:2011rk} for a summary of these measurements). We also include the
Chooz~\cite{Apollonio:2002gd} and Palo Verde~\cite{Boehm:2001ik} experiments at
$L \simeq 1$~km. We use the neutrino fluxes from the isotopes $^{235}$U,
$^{239}$Pu, $^{238}$U, $^{241}$Pu obtained in \cite{Mueller:2011nm} and we
include the uncertainty on the integrated flux for each isotope given in
Table~I of \cite{Mention:2011rk}, correlated between all experiments. For
further technical details see~\cite{Schwetz:2011qt}.

\begin{figure} \centering 
  \includegraphics[width=0.85\textwidth]{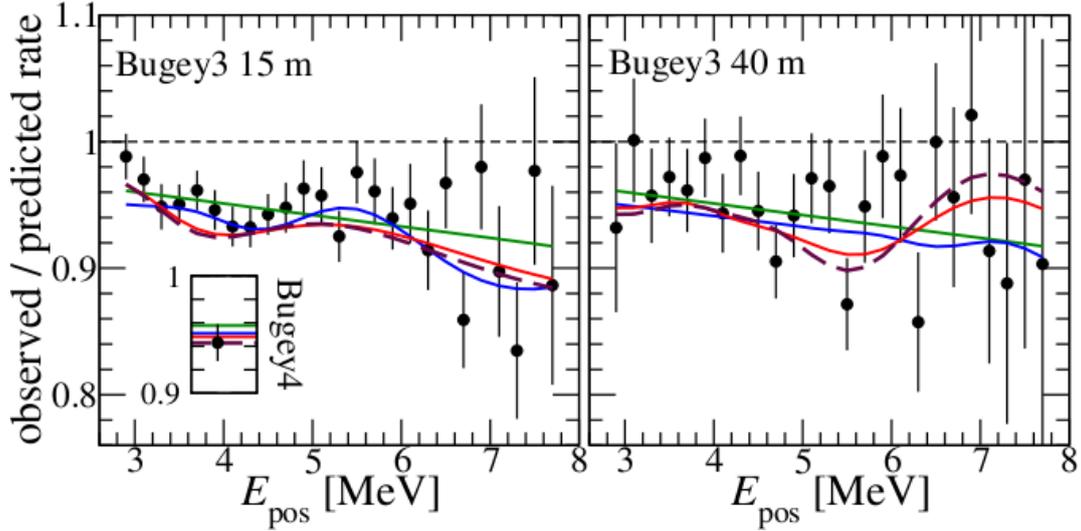}
  \caption{Comparison of sterile neutrino models to data from the Bugey
    reactor neutrino experiment. The data points show the ratio of the observed
    event spectra to the predicted ones in the absence of oscillations. The
    prediction is based on the new (higher) reactor antineutrino fluxes
    from~\cite{Mueller:2011nm}. The inset shows the ratio of the total rate
    measured in Bugey4 compared to the prediction.  Bugey3 error bars are
    statistical only, whereas the error on the Bugey4 rate includes statistics
    and systematics and is dominated by the latter. The green solid curve shows
    the prediction for the no oscillation hypothesis, the blue solid and red
    solid curves correspond to the 3+1 and 3+2 best fit points for SBL reactor
    data (Table~\ref{tab:react-bfp}), and the dashed curve corresponds to the
    3+2 best fit point of global SBL data from~Table II of~\cite{Kopp:2011qd}.}
  \label{fig:spect-react}
\end{figure}

Even though the 3+1 (3+2) framework has a large number of free parameters, SBL
reactor experiments are only sensitive to two (four) independent parameters.
These parameters are the mass-squared differences $\Dmq_{41}$ and $\Dmq_{51}$
between the eV-scale sterile neutrinos and the light neutrinos, and the
elements $|U_{e4}|$ and $|U_{e5}|$ of the leptonic mixing matrix, which
describe the mixing of the electron neutrino flavor with the heavy neutrino
mass states $\nu_4$ and $\nu_5$. Obviously, for the 3+1 case, only $\nu_4$ is
present. The best fit points for the two scenarios are summarized in
Table~\ref{tab:react-bfp} and illustrated in figure~\ref{fig:spect-react}, which
are both taken from ref.~\cite{Kopp:2011qd}. The lines in
figure~\ref{fig:spect-react} show the best fit theoretical prediction for the
no oscillation case (green) and for the 3+1 (blue) and 3+2 (red) models.  Note
that, even for no oscillations, the prediction may deviate from 1 due to
nuisance parameters included in the fit to parametrize systematic
uncertainties. The fit is dominated by Bugey3 spectral data at 15~m and 40~m
and the precise rate measurement from Bugey4.

\begin{figure} \centering
  \includegraphics[width=0.7\textwidth]{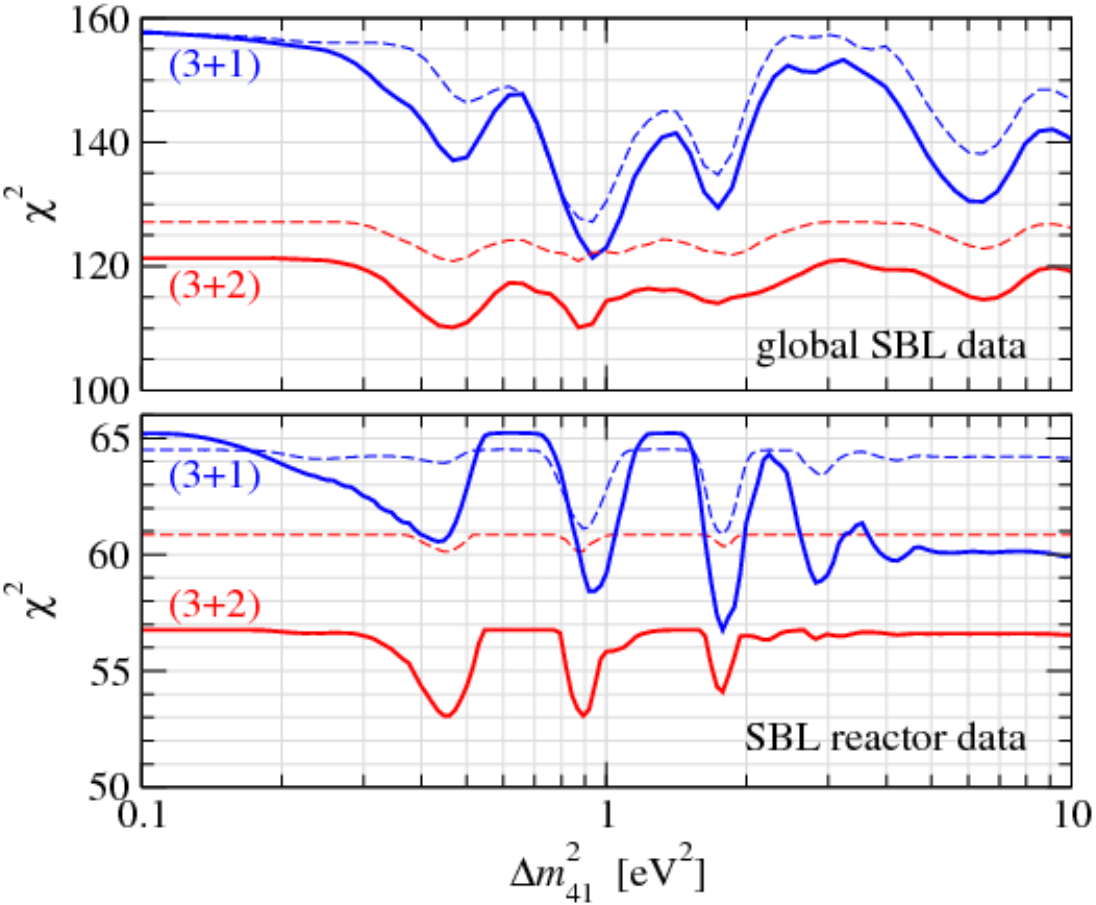}
  \caption{$\chi^2$ from global SBL data (upper panel) and from SBL
    reactor data alone (lower panel) for the 3+1 (blue) and 3+2 (red)
    scenarios. Dashed (solid) curves were computed using the
    old~\cite{Schreckenbach:1985ep} (new~\cite{Mueller:2011nm})
    reactor $\bar\nu_e$ flux prediction. All undisplayed parameters
    are minimized over.  The total number of data points is 137 (84)
    for the global (reactor) analysis.}
  \label{fig:chisq-dmq}
\end{figure}

\begin{table}[b!] \centering
  \begin{ruledtabular}
    \begin{tabular}{l@{\quad}ccccc}
      & $\Dmq_{41}$ [$\eVq$] & $|U_{e4}|$ 
      & $\Dmq_{51}$ [$\eVq$] & $|U_{e5}|$ & $\chi^2$/dof \\
      \hline      
      3+1 & 1.78 & 0.151 &      &       & 50.1/67\\
      3+2 & 0.46 & 0.108 & 0.89 & 0.124 & 46.5/65\\
    \end{tabular}
  \end{ruledtabular}
  \caption{Best fit points for the 3+1 and 3+2 scenarios from reactor
    antineutrino data alone. The total number of data points is 69 (Bugey3
    spectra plus 9 SBL rate measurements; we have omitted data from
    Chooz and Palo Verde, which are not very sensitive to the model
    parameters, but would dilute the $\chi^2$ by introducing 15
    additional data points). For no oscillations we have $\chi^2 /
    \text{dof} = 59.0/69$.}
  \label{tab:react-bfp}
\end{table}

To study the sterile neutrino masses favored by short baseline reactor neutrino
data, we show in the lower part of Fig.~\ref{fig:chisq-dmq} the $\chi^2$ of the
SBL reactor fit as a function of $\Dmq_{41}$. We compare the 3+1 (blue) and 3+2
(red) models, and we also compare results obtained using the new (slod) and old
(dashed) reactor flux predictions. Especially for the new fluxes, we find a
clear preference for sterile neutrino oscillations: the $\Delta\chi^2$ between
the no oscillation hypothesis and the 3+1 best fit point is 8.5, which implies
that the no oscillation case is disfavored at about 98.6\%~CL (2~dof). In the
3+2 case the no oscillation hypothesis is disfavored compared to the 3+2 best
fit point with $\Delta\chi^2 = 12.1$, or 98.3\%~CL (4~dof). In contrast, with
the old flux predictions the improvement of the fit is not significant, with a
$\Delta\chi^2$ between the best fit points and the no oscillation case of only
3.6 and 4.4 for the 3+1 and 3+2 hypotheses, respectively.

\subsubsection{Global analysis of SBL data}

We now move on to a combined analysis of short-baseline reactor neutrino data
together with other short-baseline data, including the LSND and MiniBooNE
excesses. LSND has provided evidence for $\bar\nu_\mu\to\bar\nu_e$
transitions~\cite{Aguilar:2001ty}, and MiniBooNE has reported an excess of
events in the same channel, consistent with the LSND
signal~\cite{AguilarArevalo:2010wv}. (In a new preliminary data release by the
MiniBooNE collaboration~\cite{MiniBooNE:2011prelim}, the significant of this
excess has decreased, but it is still consistent with the LSND signal.) This
hint for oscillations is however not confirmed by a MiniBooNE search in the
$\nu_\mu\to\nu_e$ channel~\cite{AguilarArevalo:2007it}, where the data in the
energy range sensitive to oscillations is consistent with the background
expectation.  These results, together with the reactor antineutrino anomaly,
seem to suggest an explanation involving CP violation in order to reconcile the
differing results from neutrino and antineutrino searches.

An explanation of the LSND and MiniBooNE anomalies in terms of sterile neutrino
oscillations requires mixing of the sterile neutrinos with both electron and
muon neutrinos. Their mixing with electron neutrinos is probed also by reactor
experiments, which lead to a tight constraint if the previous reactor flux
prediction~\cite{Schreckenbach:1985ep} is used. With the new reactor fluxes
from~\cite{Mueller:2011nm}, however, lead to a preference for active--sterile
neutrino mixing at th 98\% confidence level, hence the interesting question
arises whether a consistent description of the global data on SBL oscillations
(including LSND/MiniBooNE) becomes now possible. To answer this question we
perform a fit by including, in addition to the reactor searches for $\bar\nu_e$
disappearance, the LSND~\cite{Aguilar:2001ty} and
MiniBooNE~\cite{AguilarArevalo:2010wv,AguilarArevalo:2007it} results, as well
as additional constraints from the appearance experiments
KARMEN~\cite{Armbruster:2002mp} and NOMAD~\cite{Astier:2001yj}, from the
$\nu_\mu$ disappearance search in CDHS~\cite{Dydak:1983zq}, from atmospheric
neutrinos, and from the MINOS search for active neutrino
disappearance~\cite{Adamson:2010wi,Adamson:2011ku}.  Technical details of our
analysis can be found in \cite{Maltoni:2007zf, Akhmedov:2010vy} and references
therein.  For our MINOS simulation and for the combination of the different
analyses, we have used the GLoBES package~\cite{Huber:2004ka,Huber:2007ji}.
For a recent discussion of MINOS data in the context of sterile neutrinos
see also Ref.~\cite{Bhattacharya:2011ee}.

\begin{figure} \centering 
  \includegraphics[width=0.6\textwidth]{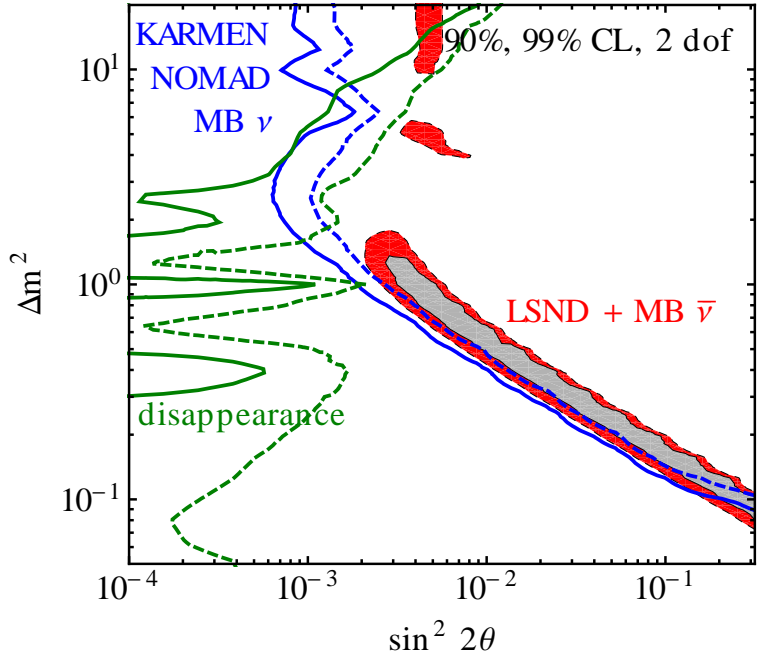}
  \caption{Global constraints on sterile neutrinos in the 3+1 model.
    We show the allowed regions at 90\% and 99\%~CL from a combined
    analysis of the LSND~\cite{Aguilar:2001ty} and MiniBooNE
    antineutrino~\cite{AguilarArevalo:2010wv} signals (filled
    regions), as well as the constraints from the null results of
    KARMEN~\cite{Armbruster:2002mp}, NOMAD~\cite{Astier:2001yj}
    and MiniBooNE neutrino~\cite{AguilarArevalo:2007it} appearance
    searches (blue contour).  The limit from disappearance experiments
    (green contours) includes data from CDHS~\cite{Dydak:1983zq},
    atmospheric neutrinos, MINOS, and from the SBL reactor experiments.
    For the latter, we have used the new reactor flux predictions
    from~\cite{Mueller:2011nm}, but we have checked that the results,
    especially regarding consistency with LSND and MiniBooNE $\bar\nu$
    data, are qualitatively unchanged when the old reactor fluxes
    are used \cite{Kopp:2011qd}.}
  \label{fig:regions-3p1}
\end{figure}

In the 3+1 scheme, short-baseline oscillations depend on the three
parameters $\Dmq_{41}$, $|U_{e4}|$, and $|U_{\mu 4}|$. Since short baseline
oscillations in this case can be effectively reduced to a two-flavor problem
in all experiments considered here, it is not possible to obtain CP
violation. Therefore, oscillations involving one sterile neutrino are not
capable of reconciling the different results for neutrino (MiniBooNE) and
antineutrino (LSND and MiniBooNE) appearance searches.
Fig.~\ref{fig:regions-3p1} compares the allowed regions from LSND and
MiniBooNE antineutrino data to the constraints from the other experiments
in the 3+1 model.  Note that, even though reactor analyses using the new
flux prediction prefer non-zero $U_{e4}$, no closed regions appear for the
disappearance bound (solid curve), since $\sin^22\theta_\text{SBL} = 4
|U_{e4}|^2 |U_{\mu 4}|^2$ can still become zero if $U_{\mu 4} = 0$. We find
that the parameter region favored by LSND and MiniBooNE antineutrino data
is ruled out by other experiments at the 99\% confidence level. The strength
of this exclusion is almost the same for the new~\cite{Mueller:2011nm} and
old~\cite{Schreckenbach:1985ep} reactor fluxes.\footnote{Let us comment on
the comparison of Fig.~\ref{fig:regions-3p1} based on \cite{Kopp:2011qd} to the
GLO-HIG analysis from \cite{Giunti:2011cp} shown in the lower left panel of
Fig.~\ref{con-glo}. The appearance region shown in Fig.~\ref{con-glo}
extends to lower values than the LSND+MB$\bar\nu$ region in
Fig.~\ref{fig:regions-3p1}, since it includes other appearance data (shown as
blue curves in Fig.~\ref{fig:regions-3p1}) which push the appearance region to
lower values of $\sin^22\theta$. There are also differences in the limits
from disappearance data, especially around $\Delta m^2 \sim 1$~eV$^2$.
Presumably the origin of those differences is related to a different
treatement of spectral data from the Bugey reactor experiment.}

Using the parameter goodness of fit test developed in~\cite{Maltoni:2003cu}, we
find compatibility of the LSND+MiniBooNE($\bar\nu$) signal with the rest of the
data only at the level of $\text{few} \times 10^{-6}$, with $\chi^2_\text{PG} =
25.8 (27.3)$ for new (old) reactor fluxes, see Table~\ref{tab:PG}. Hence we
conclude that the 3+1 scenario does not provide a satisfactory description of
the data despite the new hint coming from reactors.

\begin{table} \centering
  \begin{ruledtabular}
    \begin{tabular}{cccccccccc}
            & $|\Delta m_{41}^2|$ & $|U_{e4}|$ & $|U_{\mu 4}|$
            & $|\Delta m_{51}^2|$ & $|U_{e5}|$ & $|U_{\mu 5}|$
            & $\delta/\pi$ & $\chi^2$/dof \\
      \hline
      (3+1) &(0.48)&(0.14)&(0.23)&      &      &      &       &(255.5/252)\\
      3+2   & 1.10 & 0.14 & 0.11 & 0.82 & 0.13 & 0.12 & -0.31 & 245.2/247 \\
      1+3+1 & 0.48 & 0.13 & 0.12 & 0.90 & 0.15 & 0.15 &  0.62 & 241.6/247
    \end{tabular}

  \end{ruledtabular}
  \caption{Parameter values and $\chi^2$ at the global best fit points
    for 3+1, 3+2 and 1+3+1 oscillations. The mass squared differences
    are given in units of $\eVq$. Note that the 3+1 model, in spite of
    its seemingly unoffending $\chi^2$/dof value, does not provide a
    satisfactory fit to the data, as demonstrated by the parameter
    goodness of fit test (see text and Table~\ref{tab:PG}). We therefore
    show the corresponding best fit point in parentheses.}
  \label{tab:global-bfp}
\end{table}

Moving on to the 3+2 model, we note that in this case, SBL experiments depend
on the seven parameters listed in Table~\ref{tab:global-bfp}. In addition to
the two mass-squared differences and the moduli of the mixing matrix elements,
there is now also a physical complex phase, $\delta \equiv \arg(U_{\mu 4}
U_{e4}^* U_{\mu 5}^* U_{e5})$. This phase leads to CP violation in SBL
oscillations~\cite{Maltoni:2007zf, Karagiorgi:2006jf}, allowing to reconcile
differing neutrino and antineutrino results from MiniBooNE, LSND, and
short-baseline reactor experiments.  Table~\ref{tab:global-bfp} shows the
parameter values at the global best fit point and the corresponding $\chi^2$
value. Changing from the previous to the new reactor flux calculations the
$\chi^2$ decreases by 10.5 units, indicating a significant improvement of the
description of the data, see also upper panel of Fig.~\ref{fig:chisq-dmq}. From
that figure follows also that going from 3+1 to 3+2 leads to a significant
improvement of the fit with the new reactor fluxes, which was not the case with
the old ones. The $\chi^2$ improves by 10.3 units, which means that 3+1 is
disfavored at the 96.4\%~CL (4~dof) with respect to 3+2, compared to
$\Delta\chi^2 = 5.5$ (76\%~CL) for the old fluxes.

\begin{figure} \centering 
  \includegraphics[width=0.9\textwidth]{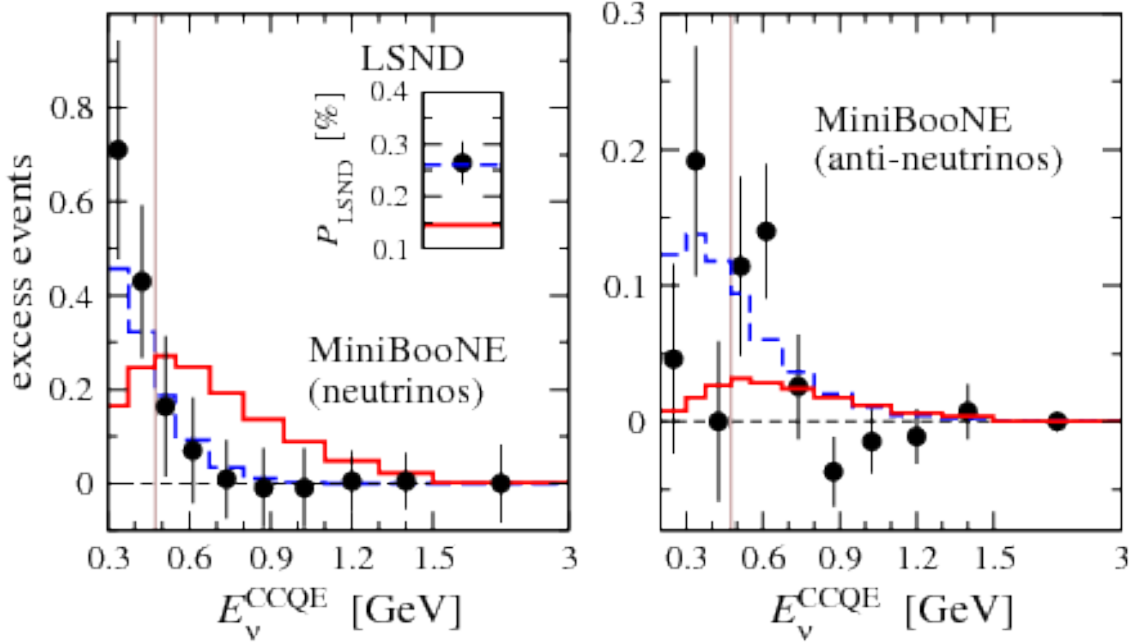}
  \caption{Predicted spectra for MiniBooNE data and the transition
    probability for LSND (inset).  Solid histograms refer to the 3+2 global
    best fit point from Table.~\ref{tab:global-bfp}.  Dashed histograms
    correspond to the best fit of appearance data only (LSND, MiniBooNE
    $\nu/\bar\nu$, KARMEN, NOMAD). For MiniBooNE we fit only data above
    475~MeV.}
  \label{fig:spect-MB}
\end{figure}

In Fig.~\ref{fig:spect-react} we show the prediction for the Bugey
spectra at the 3+2 global best fit point as dashed curves. Clearly
they are very similar to the best fit of reactor data
only. Fig.~\ref{fig:spect-MB} shows the predicted spectra for
MiniBooNE neutrino and antineutrino data, as well as the LSND
$\bar\nu_\mu\to\bar\nu_e$ transition probability. In this case some
tension in the fit is apparent and the description of the data is
clearly worse at the global best fit point than a fit to appearance
data only (dashed histograms). The main difference to the results
presented in \cite{Kopp:2011qd} is the inclusion of MINOS NC
results. Note that MiniBooNE observes an event excess in the lower
part of the spectrum. This excess can be explained if only appearance
data are considered, but not in the global analysis including
disappearance searches~\cite{Maltoni:2007zf}. Therefore, we
follow~\cite{AguilarArevalo:2007it} and assume an alternative
explanation for this excess, {\it e.g.}~\cite{Hill:2010zy}.

\begin{figure} \centering 
  \includegraphics[width=0.6\textwidth]{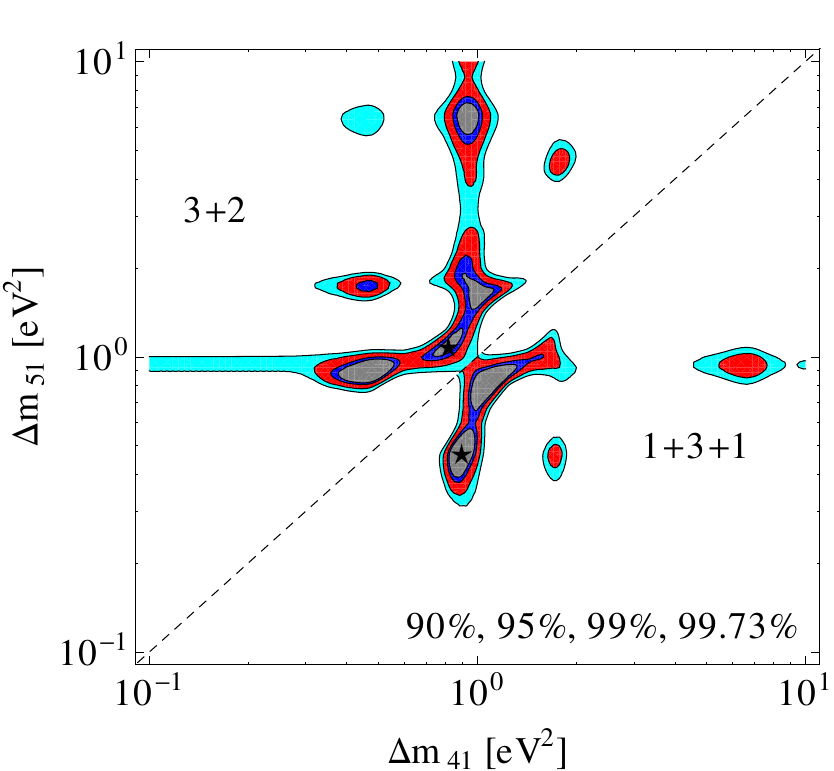}
  \caption{The globally preferred regions for the neutrino mass
    squared differences $\Dmq_{41}$ and $\Dmq_{51}$ in the 3+2 (upper
    left) and 1+3+1 (lower right) scenarios.}
  \label{fig:regions-3p2}
\end{figure}

In Table~\ref{tab:PG} we show the compatibility of the
LSND/MiniBooNE($\bar\nu$) signal with the rest of the data, as well as
the compatibility of appearance and disappearance searches using the
PG test from~\cite{Maltoni:2003cu}. Although the compatibility
improves drastically when changing from old to new reactor fluxes, the
PG is still well below 0.1\% for 3+2. This indicates that some tension
between data sets remains. The reason for the reduced PG values
compared to the ones obtained in \cite{Kopp:2011qd} comes from the
inclusion of MINOS NC data. Note that within the 3+1 scenario there
is an internal tension between LSND and MiniBooNE neutrino data
worsening the fit of the appearance data set. This in turn can lead to
an ``improved'' PG value in the comparision between appearance and
disappearance data, since the tension in the fit is hidden in the
appearance data set and does not contribute to the PG. Therefore the
apparently better PG value for 3+1 compared to 3+2 has to be
interpreted with care. We considered also a ``1+3+1'' scenario, in
which one of the sterile mass eigenstates is lighter than the three
active ones and the other is heavier~\cite{Goswami:2007kv}. As can be
seen from Tables~\ref{tab:global-bfp} and~\ref{tab:PG} the fit of
1+3+1 is slightly better than 3+2, with $\Delta\chi^2 = 13.9$ between
3+1 and 1+3+1 (99.2\%~CL for 4~dof). However, due to the larger total
mass in neutrinos, a 1+3+1 ordering might be in more tension with
cosmology than a 3+2 scheme~\cite{GonzalezGarcia:2010un,
Hamann:2010bk, Giusarma:2011ex}. Fig.~\ref{fig:regions-3p2} shows
the allowed regions for the two eV-scale mass-squared differences for
the 3+2 and 1+3+1 schemes.

\begin{table} \centering
  \begin{ruledtabular}
    \begin{tabular}{ccccc}
       & \multicolumn{2}{c}{LSND+MB($\bar\nu$) vs rest}
       & \multicolumn{2}{c}{appearance\ vs disapp.} \\
       & old & new & old & new \\
       \hline
       $\chi^2_\text{PG,3+1}$/dof & 27.3/2 & 25.8/2 & 15.7/2 & 14.2/2 \\
       PG$_\text{3+1}$            & $1.2 \times 10^{-6}$ & $2.5 \times 10^{-6}$
                                  & $3.9 \times 10^{-4}$ & $8.2 \times 10^{-4}$ \\
       \hline
       $\chi^2_\text{PG,3+2}$/dof & 30.0/5 & 24.8/5 & 24.7/4 & 19.5/4 \\
       PG$_\text{3+2}$            & $1.5 \times 10^{-5}$ & $1.5 \times 10^{-4}$
                                  & $5.7 \times 10^{-5}$ & $6.1 \times 10^{-4}$ \\
       \hline
       $\chi^2_\text{PG,1+3+1}$/dof & 24.9/5 & 21.2/5 & 19.6/4 & 10.7/4 \\
       PG$_\text{1+3+1}$          & $1.5 \times 10^{-4}$ & $7.5 \times 10^{-4}$
                                  & $6.0 \times 10^{-3}$ & $3.1 \times 10^{-2}$
    \end{tabular}

  \end{ruledtabular}
  \caption{Compatibility of different data sets according to the parameter
    goodness of fit (PG) test~\cite{Maltoni:2003cu} for the 3+1, 3+2
    and 1+3+1 oscillation scenarios. Results are shown for both the
    old~\cite{Schreckenbach:1985ep} and the new~\cite{Mueller:2011nm}
    reactor antineutrino flux predictions.}
  \label{tab:PG}
\end{table}

%
Let us comment briefly on other signatures of eV-scale sterile neutrinos. We
have checked the fit of solar neutrino data and the KamLAND reactor experiment,
and found excellent agreement. The effect of non-zero $U_{e4}$ and $U_{e5}$ for
these data are similar to the one of $U_{e3}$ in the standard three-active
neutrino case, and hence the 3+2 best fit point mimics a non-zero $U_{e3}$
close to the preferred value of these data, see~\cite{Schwetz:2011qt,
GonzalezGarcia:2010er,prep}.  Our best-fit points also fall in the range of
parameter values required to explain the Gallium anomaly, a small $\nu_e$
deficit observed in the neutrino flux from intense radioactive sources by
radiochemical neutrino detectors~\cite{Acero:2007su}.  Finally, sterile
neutrinos may manifest themselves in cosmology.  Recent
studies~\cite{GonzalezGarcia:2010un, Hamann:2010bk, Giusarma:2011ex} indicate a
slight preference for extra radiation content in the universe (mainly from CMB
measurements), favoring the existence of light sterile neutrinos. On the other
hand, Big-Bang nucleosynthesis constrains the number of extra neutrino species
to be $< 1.2$ at the 95\%~confidence level~\cite{Mangano:2011ar} and is
therefore problematic in schemes with two or more sterile neutrinos. Moreover,
global fits to cosmological data constrain the sum of the neutrino masses to be
$\leq 0.7$ to 1.5~eV at 95\%~CL~\cite{GonzalezGarcia:2010un, Hamann:2010bk,
Giusarma:2011ex}, depending on the used data, whereas our 3+2 best fit point
leads to $\sum m_\nu \approx 1.7$~eV. Hence, sterile neutrino explanations of
short-baseline oscillation data are in tension with cosmology, or, if
confirmed, would indicate a deviation from the standard cosmological picture.
For instance, a mechanism that suppresses their production in the early
universe may be required (see {\it e.g.}~\cite{Bento:2001xi, Dolgov:2004jw}).

In conclusion, a global fit to short-baseline oscillation searches assuming two
sterile neutrinos improves significantly when new predictions for the reactor
neutrino flux are taken into account, although tension remains. We are thus
facing an intriguing accumulation of hints for the existence of sterile
neutrinos at the eV scale, and a confirmation of these hints in the future
could certainly be considered a major discovery.

\subsection{Discussion of the LSND and MiniBooNE Results}
\label{sec:Karagiorgi}


At present, the evidence for oscillations at $\Delta m^2_{41}\sim1$
eV$^2$ and $\sin^22\theta_{\mu e}\sim0.005$ from the LSND experiment
\cite{Athanassopoulos:1996jb,Athanassopoulos:1997pv,Aguilar:2001ty}
remains the most significant hint for the existence of light sterile
neutrinos, at $3.8\sigma$. The $3.0\sigma$ excess of $\nu_e$ observed
by MiniBooNE, referred to as the ``MiniBooNE low energy excess'', is
only marginally consistent with a $\nu_{\mu}\rightarrow\nu_e$ single
sterile neutrino oscillation signature, and, under that
interpretation, it corresponds to a significantly different $L/E$
distribution than that suggested by the LSND excess
\cite{AguilarArevalo:2007it,AguilarArevalo:2008rc}. As a result, the
MiniBooNE neutrino and LSND results have been shown to be
significantly incompatible \cite{Maltoni:2007zf,Karagiorgi:2009nb}
under the single sterile neutrino oscillation assumption.

Instead, past attempts to simultaneously interpret the MiniBooNE
neutrino and LSND antineutrino excesses as signals induced by sterile
neutrino oscillations lean on CP-violation
\cite{Maltoni:2007zf,Karagiorgi:2009nb} and the assumption that more
than one mostly-sterile mass eigenstates exist. This scenario will be
discussed in Sec.~\ref{3p2}. Generally, one must resort to models
which induce differences between neutrino and antineutrino oscillation
probabilities and/or more complex $L/E$ dependence in the observable
oscillation probabilities. There are other scenarios besides
CP-violating sterile neutrino oscillations which can induce such
differences, and those include models with altered neutrino dispersion
relations \cite{Hollenberg:2009ak}, extra dimensions
\cite{Hollenberg:2009ws}, CPT-violation
\cite{Diaz:2011ia,Diaz:2011tx,AguilarArevalo:2011yi}, and non-standard
neutrino interactions
\cite{Akhmedov:2010vy,Akhmedov:2011zz,Nelson:2007yq,Karagiorgi:2011ut}. At
the same!  time, there have been several attempts to interpret the
MiniBooNE low energy excess independently of oscillations. The
possibility of mis-estimated background has been considered but has
been ruled out by independent {\it in situ} measurements at MiniBooNE
\cite{AguilarArevalo:2008rc}, in particular for the case of
neutral-current single-photon production, which makes up the most
dominant background to the MiniBooNE $\nu_e$ appearance search at low
energy. However, the possibility of a new background, such as that
contributed by anomaly-mediated single-photon production
\cite{Harvey:2007rd}, remains a viable interpretation for the
MiniBooNE low energy excess. Viable theoretical interpretations
include heavy sterile neutrino decay
\cite{Gninenko:2011xa,Gninenko:2009yf,Gninenko:2010pr,Gninenko:2009ks,Gninenko:2011hb,Dib:2011jh}.

While it is unclear whether the MiniBooNE neutrino and LSND antineutrino
excesses are a manifestation of the same underlying scenario, be it
CP-violating light sterile neutrino oscillations, or some other physics,
there are two independent measurements which are in direct agreement with
LSND under the single sterile neutrino oscillation hypothesis. The first one
is the appearance measurement performed by MiniBooNE during antineutrino
running \cite{AguilarArevalo:2010wv}. While the results of that measurement
are statistics-limited, the small excess observed has an oscillations-like
$L/E$ dependence which agrees with that of LSND, and favors oscillations
over the null hypothesis at the 91\% CL \cite{Zimmerman:2011hy}.

The second measurement in support of the LSND single sterile neutrino
oscillation interpretation comes from the electron antineutrino
disappearance sector, and is what has been described in previous sections as
the ``reactor antineutrino anomaly'' \cite{Mention:2011rk}. This new finding
has shed new light to light sterile neutrino oscillations. Until 2011,
electron neutrino and antineutrino disappearance at $\Delta m^2\sim1$ eV$^2$
had been thought to be constrained to less than 10\% by short-baseline
reactor antineutrino experiments including Bugey \cite{Declais:1994su} and
CHOOZ \cite{Apollonio:2002gd}. This level of constraint on $\bar{\nu}_e$
disappearance, when combined with equally stringent $\nu_{\mu}$
disappearance limits from MiniBooNE/SciBooNE \cite{Mahn:2011ea}, CDHS
\cite{Dydak:1983zq}, CCFR \cite{Stockdale:1984cg}, MINOS
\cite{Adamson:2010wi}, K2K and atmospheric neutrino experiments
\cite{Maltoni:2004ei}, reduces observable appearance to a level well below
what! is necessary to account for the LSND excess. Instead, the now
anomalous measurements performed at ILL, Bugey, ROVNO, CHOOZ, and others
\cite{Mention:2011rk}, can be interpreted as---and are in support
of---oscillations at a $\Delta m^2$ consistent with that from LSND, $\Delta
m^2\sim 1$ eV$^2$, and relatively small disappearance amplitude
$\sin^22\theta_{ee}\sim0.1$, consistent with a sterile neutrino hypothesis.

Finally, there has been a long-standing piece of evidence for much larger $\nu_e$ 
disappearance at very short baselines from the Gallex and SAGE calibration source 
experiments. This evidence now seems excluded by a comparison of KARMEN and LSND 
$\nu_e$ cross section measurements to theoretical predictions \cite{Conrad:2011ce}. 

The following two sections present global oscillation fit results from the 
perspective of CPT-conserving light sterile neutrino oscillations. The conclusions 
are drawn from Refs.~\cite{Karagiorgi:2011ut,Kopp:2011qd,Giunti:2011gz}. In 
summary, the present light sterile neutrino oscillation fits reveal that neutrino 
and antineutrino oscillation signatures which may be responsible for the LSND, 
MiniBooNE, and reactor antineutrino anomalies are incompatible with other, 
primarily neutrino, null results. The neutrino and antineutrino data sets show 
differences which cannot be explained away by just CP violation. That observation 
has prompted the consideration of even less standard scenarios, beyond the 
``reference picture'' of CPT-conserving 3+N oscillations \cite{everett}.

\subsubsection{(3+1) Oscillations}
It has already been established that the MiniBooNE antineutrino and LSND
excesses are highly compatible with each other under the single sterile
neutrino oscillation scenario, referred to as the (3+1) scenario
\cite{Maltoni:2007zf,Karagiorgi:2009nb}. It is, however, also necessary to
consider how those excesses compare with other available data sets. This has
been done by several authors in the past
\cite{Giunti:2010uj,Maltoni:2007zf,Karagiorgi:2009nb,Karagiorgi:2011ut},
which find that the high compatibility persists even when one considers all
other short-baseline {\it antineutrino} data sets, including the KARMEN
$\bar{\nu}_{\mu}\rightarrow\bar{\nu}_e$ appearance \cite{Armbruster:2002mp}
and the Bugey and CHOOZ $\bar{\nu}_e$ disappearance data sets, with a
combined fit favoring oscillations at the $>99$\% confidence level. The
compatibility drops drastically, however, when one folds in constraints from
$\nu_{\mu}$ disappearance experiments, as well as additional $\nu_e$
appearance constraints from NOMAD \cite{Astier:2003gs} and the MiniBooNE
$\nu_{\mu}\rightarrow\nu_e$ search. It should be noted that, under a CPT
conserving (3+1) scenario, neutrino and antineutrino oscillations are
identical, and therefore constraints from neutrino data sets are directly
applicable to the global fits considered under this scenario. The
compatibility found in a joint fit using most to all available experimental
constraints is found to be at the $\sim1$\% or less level
\cite{Karagiorgi:2009nb,Karagiorgi:2011ut,Giunti:2011hn}, even after
accounting for the reactor antineutrino anomaly.

\subsubsection{(3+2) Oscillations}
\label{3p2} The high compatibility seen in antineutrino-only fits in a (3+1)
hypothesis is instructive, and the apparent differences seen among neutrino
and antineutrino appearance signals make the (3+2) oscillation scenario
interesting because of the offered possibility of CP violation. Combined
analyses of most to all {\it appearance} data sets under the (3+2) scenario
have been performed by several independent groups (see,
{\it e.g.}~\cite{Karagiorgi:2009nb,Maltoni:2007zf}), and consistently yield high
compatibility when large CP violation is invoked. Furthermore,
appearance-only fits yield oscillation parameters which predict excesses of
low energy events at MiniBooNE in both neutrino and antineutrino running,
which can account for the observed excesses. However, when disappearance
data sets all included in the fits, the compatibility once again reduces
significantly, and one finds tension among neutrino and antineutrino data
sets. It is worth noting that the reactor anomaly is responsible for the
significant reduction in tension found previously between neutrino and
antineutrino data sets, or appearance and disappearance data sets; however,
some tension remains, such that in order to make (3+2) CP-violating models
viable, one or more experimental data sets other than those with
antineutrino signals must be rejected \cite{Kopp:2011qd,Giunti:2011gz}.

\subsection{Impact of Sterile Neutrinos for Absolute Neutrino Mass Measurments}
\label{sec:beta}

\subsubsection{Impact for $\beta$-Decay and Neutrinoless $\beta\beta$-Decay}


\bigskip

In this section we investigate possible signals of eV sterile neutrinos as
indicated by the experiments discussed above in searches for the absolute
neutrino mass in $\beta$-decay and neutrinoless double $\beta$-decay, based
on the GLO-LOW and GLO-HIG analyses from section~\ref{sec:giunti-global} of
short-baseline neutrino oscillation data.

The effective electron neutrino mass $m_{\beta}$ in $\beta$-decay experiments is given by
\cite{Shrock:1980vy,McKellar:1980cn,Kobzarev:1980nk,Vissani:2000ci}
(other approaches are discussed in Refs.~\cite{Farzan:2001cj,Studnik:2001hs,Farzan:2002zq})
\begin{equation}
m_{\beta}^2
=
\sum_{k} |U_{ek}|^2 m_{k}^2
\,.
\label{029}
\end{equation}
The most accurate measurements of $m_{\beta}$
have been obtained in the
Mainz \cite{Kraus:2004zw}
and
Troitsk \cite{Lobashev:2003kt}
experiments,
whose combined upper bound is
\cite{Giunti:2010wz}
\begin{equation}
m_{\beta} \leq 1.8 \, \text{eV}
\qquad
(\text{Mainz+Troitsk, 95\% C.L.})
\,.
\label{034}
\end{equation}

In the hierarchical 3+1 scheme
we have the lower bound
\begin{equation}
m_{\beta}
\geq
|U_{e4}| \sqrt{\Delta{m}^2_{41}}
\equiv
m_{\beta}^{(4)}
\,.
\label{mb4}
\end{equation}
Therefore,
from the analysis of short-baseline neutrino oscillation data we
can derive predictions for the possibility of observing a neutrino mass effect
in the KATRIN experiment \cite{Fraenkle:2011uu},
which is under construction and scheduled to start in 2012,
and in other possible future experiments.

Let us however note that the effective electron neutrino mass
in Eq.~(\ref{029}) has been derived assuming that all the neutrino masses are smaller
than the experimental energy resolution
(see Ref.~\cite{Giunti-Kim-2007}).
If $m_{4}$ is of the order of 1 eV,
the approximation is acceptable for the interpretation of the result of the
Mainz and Troitsk
experiments, which had, respectively,
energy resolutions of
4.8 eV and 3.5 eV \cite{Weinheimer:2009cb}.
On the other hand,
the energy resolution of the KATRIN experiment will be
0.93 eV near the end-point of the energy spectrum of the electron emitted in Tritium decay,
at
$T=Q$,
where $T$ is the kinetic energy of the electron
and
$Q=18.574\,\text{keV}$
is the $Q$-value of the decay.
If the value of $m_{4}$ is larger than the energy resolution of the experiment,
its effect on the measured electron spectrum
cannot be summarized by one effective quantity,
because the Kurie function $K(T)$ is given by
\begin{align}
\null & \null
\frac{K^2(T)}{Q-T}
=
\sqrt{(Q-T)^2-\widetilde{m}_{\beta}^2}
-
|U_{e4}|^2
(Q-T)
\nonumber
\\
\null & \null
+
|U_{e4}|^2
\sqrt{(Q-T)^2-m_{4}^{2}}
\
\theta(Q-T-m_{4})
\,,
\label{kurie}
\end{align}
where
$
\widetilde{m}_{\beta}^2
=
\sum_{k=1}^{3} |U_{ek}|^2 m_{k}^2
$
is the contribution of the three neutrino masses much smaller than 1 eV
and $\theta$ is the Heaviside step function.

\begin{table*}[t!]
\begin{center}
\setlength{\tabcolsep}{5pt}
\begin{tabular}{ccc}
&
GLO-LOW
&
GLO-HIG
\\
\hline
$m_{4}$
&
\begin{tabular}{c}
$ 0.91 - 2.5 $
\\
$ 0.88 - 2.5 $
\\
$ 0.85 - 2.8 $
\end{tabular}
&
\begin{tabular}{c}
$ 1.2 - 2.4 $
\\
$ 0.91 - 2.5 $
\\
$ 0.87 - 2.8 $
\end{tabular}
\\
\hline
$|U_{e4}|^2$
&
\begin{tabular}{c}
$ 0.02 - 0.04 $
\\
$ 0.02 - 0.06 $
\\
$ 0.01 - 0.07 $
\end{tabular}
&
\begin{tabular}{c}
$ 0.03 - 0.05 $
\\
$ 0.02 - 0.06 $
\\
$ 0.01 - 0.07 $
\end{tabular}
\\
\hline
$m_{\beta}^{(4)}$
&
\begin{tabular}{c}
$ 0.14 - 0.49 $
\\
$ 0.12 - 0.56 $
\\
$ 0.10 - 0.65 $
\end{tabular}
&
\begin{tabular}{c}
$ 0.21 - 0.45 $
\\
$ 0.13 - 0.56 $
\\
$ 0.11 - 0.63 $
\end{tabular}
\\
\hline
$m_{\beta\beta}^{(4)}$
&
\begin{tabular}{c}
$ 0.020 - 0.10 $
\\
$ 0.015 - 0.13 $
\\
$ 0.011 - 0.16 $
\end{tabular}
&
\begin{tabular}{c}
$ 0.035 - 0.09 $
\\
$ 0.018 - 0.12 $
\\
$ 0.013 - 0.16 $
\end{tabular}
\\
\hline
\end{tabular}
\caption{ \label{tab-rng}
Allowed
$1\sigma$,
$2\sigma$ and
$3\sigma$
ranges of
$m_{4}=\sqrt{\Delta{m}^2_{41}}$,
$|U_{e4}|^2$,
$m_{\beta}^{(4)}=|U_{e4}|\sqrt{\Delta{m}^2_{41}}$
and
$m_{\beta\beta}^{(4)}=|U_{e4}|^2\sqrt{\Delta{m}^2_{41}}$
obtained
\protect\cite{Giunti:2011cp}
from
the global fit with (GLO-LOW) and without (GLO-HIG)
the MiniBooNE electron neutrino and antineutrino data
with reconstructed neutrino energy smaller than $475 \, \text{MeV}$.
Masses are given in eV.
}
\end{center}
\end{table*}

\begin{figure}[t!]
\begin{center}
\includegraphics*[width=0.5\linewidth]{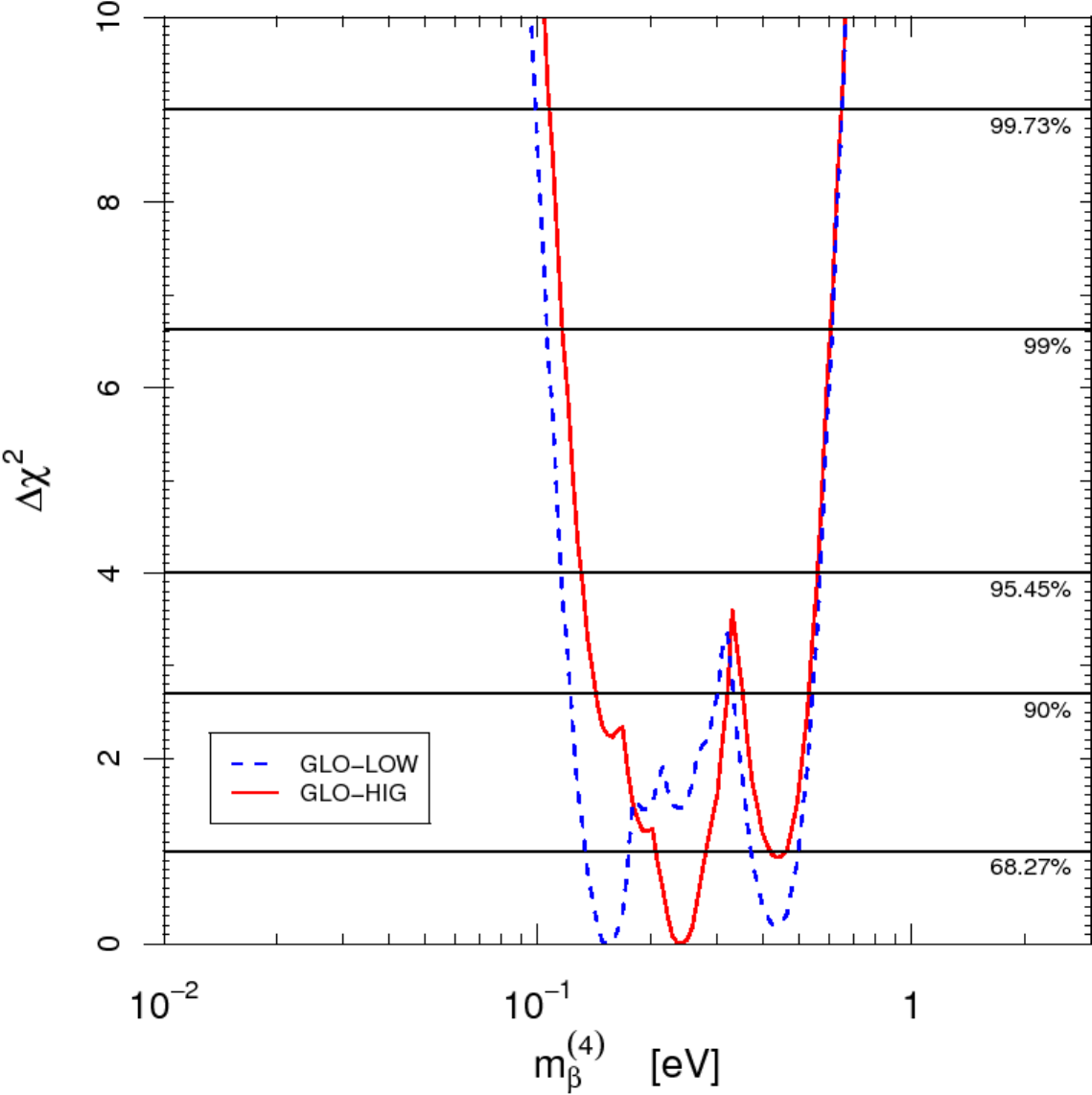}
\end{center}
\caption{ \label{mbt-chi-plt}
Marginal
$\Delta\chi^2 = \chi^2 - \chi^2_{\text{min}}$
as a function of the contribution
$m_{\beta}^{(4)} = |U_{e4}| \sqrt{\Delta{m}^2_{41}}$
to the effective $\beta$-decay electron-neutrino mass $m_{\beta}$
obtained
\protect\cite{Giunti:2011cp}
from
the GLO-LOW and GLO-HIG global fits.
}
\end{figure}

In the following we discuss the predictions obtained from the GLO-LOW and
GLO-HIG analyses from section~\ref{sec:giunti-global} of short-baseline
neutrino oscillation data for both the contribution $m_{\beta}^{(4)}$ to the
effective mass in $\beta$-decay and the distortion of the Kurie function due
to $m_{4}$. Figure~\ref{mbt-chi-plt} shows the marginal $\Delta\chi^2 =
\chi^2 - \chi^2_{\text{min}}$ as a function of $m_{\beta}^{(4)}$, which
gives the $1\sigma$, $2\sigma$ and $3\sigma$ allowed ranges of $m_{4}$,
$|U_{e4}|^2$ and $m_{\beta}^{(4)}$ listed in Tab.~\ref{tab-rng}. As one can
see from Fig.~\ref{mbt-chi-plt} and Tab.~\ref{tab-rng}, the results of both
the GLO-LOW and GLO-HIG analyses favor values of $m_{\beta}^{(4)}$ between
about 0.1 and 0.7 eV. This is promising for the perspectives of future
experiments.

\begin{figure}[t!]
\begin{center}
\includegraphics*[width=0.5\linewidth]{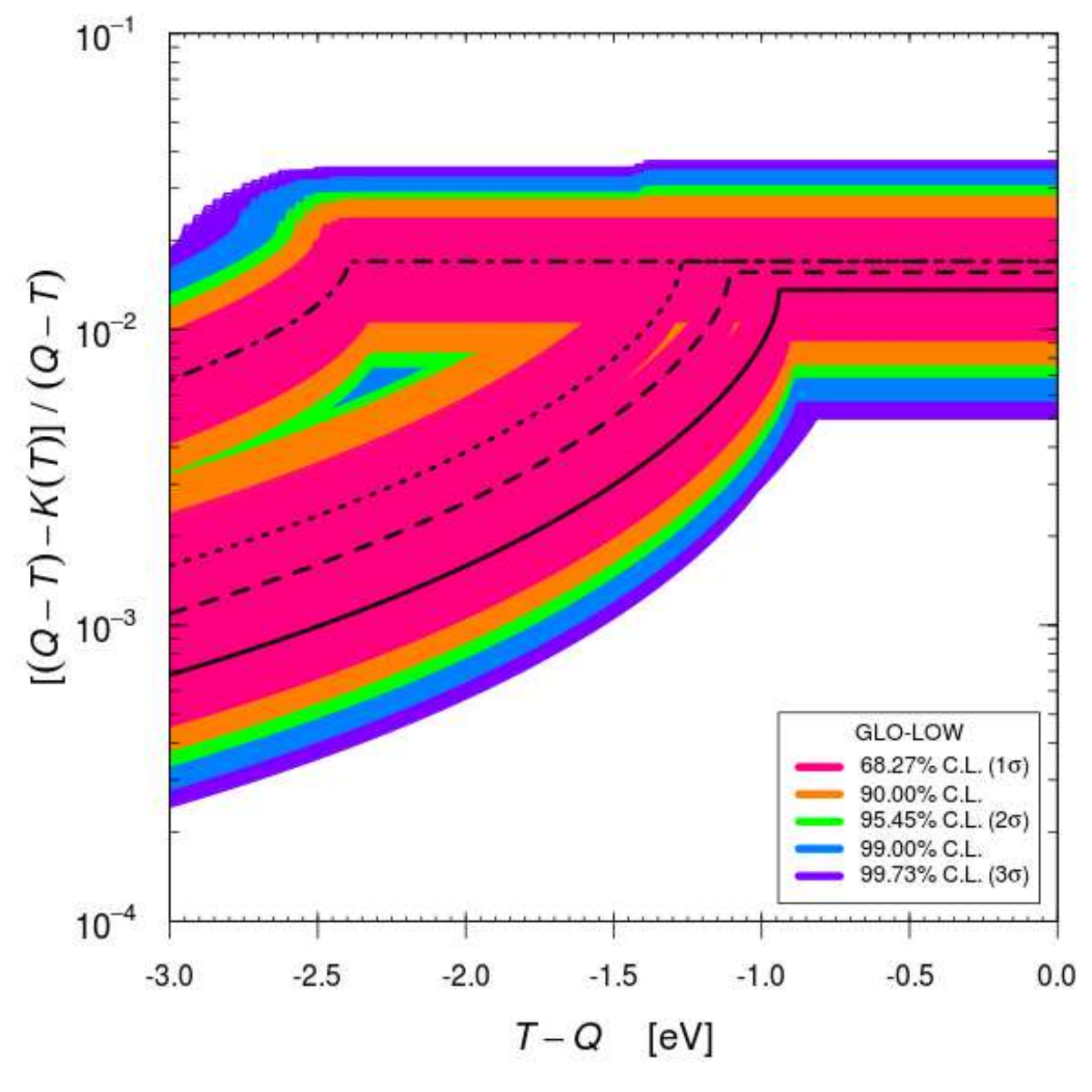}
\end{center}
\caption{ \label{plt-kur-dif-low}
Bands of the relative deviation of the Kurie plot in $\beta$-decay
corresponding to the allowed regions in the
$\sin^{2}2\theta_{ee}$--$\Delta{m}^2_{41}$
plane in Fig.~\ref{con-glo},
obtained
\protect\cite{Giunti:2011cp}
from the GLO-LOW global analysis of short-baseline neutrino oscillation data
(see Tab~\ref{tab-bef}).
The black solid line corresponds to the best-fit point
($m_{4} = \protect0.94 \, \text{eV}$
and
$|U_{e4}|^2 = \protect0.027$).
The dashed, dotted and dash-dotted lines correspond,
respectively,
to the local minima at
($m_{4} = \protect1.11 \, \text{eV}$,
$|U_{e4}|^2 = \protect0.03$),
($m_{4} = \protect1.27 \, \text{eV}$,
$|U_{e4}|^2 = \protect0.035$) and
($m_{4} = \protect2.40 \, \text{eV}$,
$|U_{e4}|^2 = \protect0.033$).
}
\end{figure}

\begin{figure}[t!]
\begin{center}
\includegraphics*[width=0.5\linewidth]{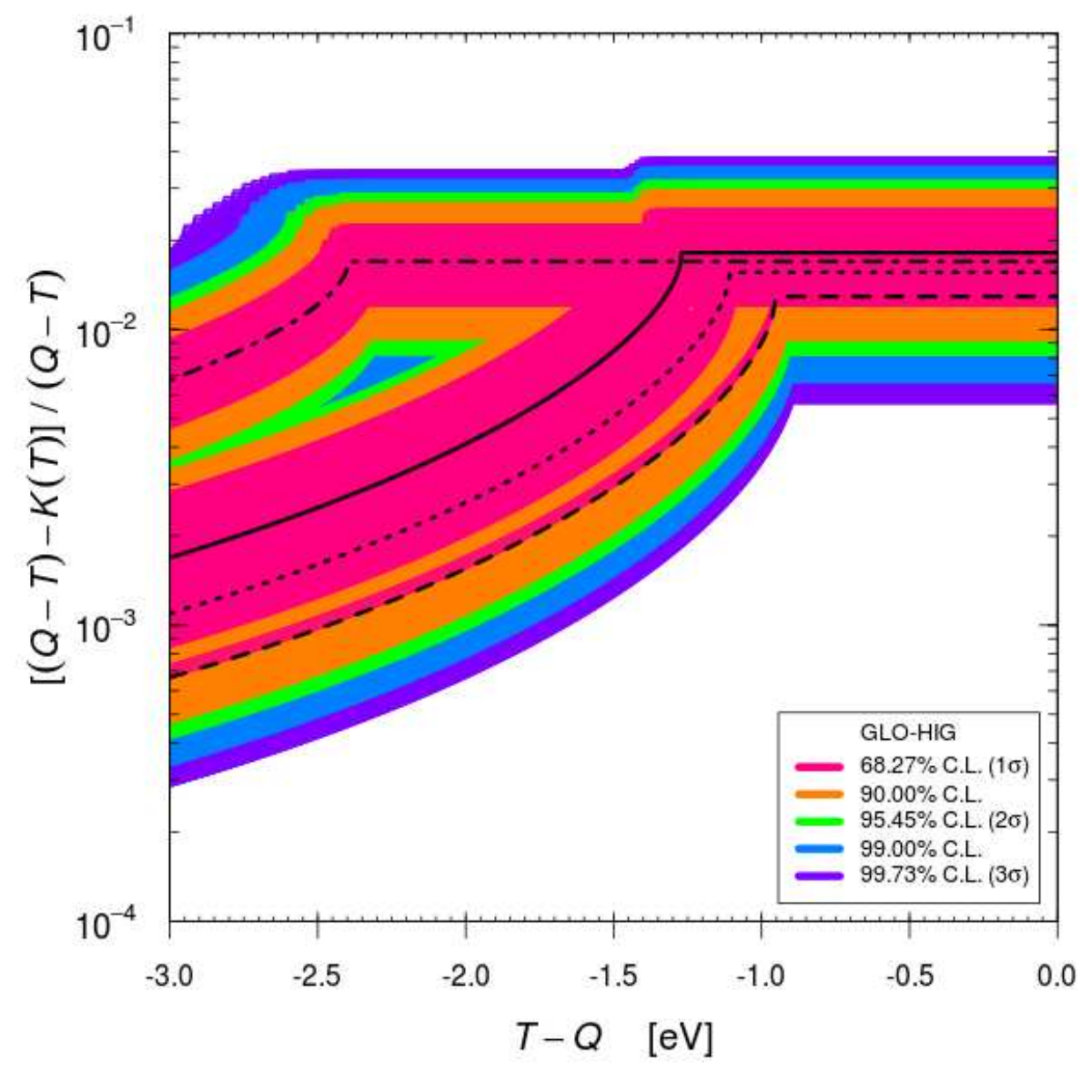}
\end{center}
\caption{ \label{plt-kur-dif-hig}
Bands of the relative deviation of the Kurie plot in $\beta$-decay
corresponding to the allowed regions in the
$\sin^{2}2\theta_{ee}$--$\Delta{m}^2_{41}$
plane in Fig.~\ref{con-glo},
obtained
\protect\cite{Giunti:2011cp}
from the GLO-HIG global analysis of short-baseline neutrino oscillation data
(see Tab~\ref{tab-bef}).
The black solid line corresponds to the best-fit point
($m_{4} = \protect1.27 \, \text{eV}$
and
$|U_{e4}|^2 = \protect0.036$).
The dashed, dotted and dash-dotted lines correspond,
respectively,
to the local minima at
($m_{4} = \protect0.95 \, \text{eV}$,
$|U_{e4}|^2 = \protect0.027$),
($m_{4} = \protect1.11 \, \text{eV}$,
$|U_{e4}|^2 = \protect0.031$) and
($m_{4} = \protect2.40 \, \text{eV}$,
$|U_{e4}|^2 = \protect0.033$).
}
\end{figure}

In Figures~\ref{plt-kur-dif-low} and \ref{plt-kur-dif-hig}
we show the relative deviation of the Kurie function
from the massless case
($K(T) = Q - T$)
obtained in the GLO-LOW and GLO-HIG analyses,
neglecting the contribution of
$\widetilde{m}_{\beta}$
in Eq.~(\ref{kurie}).
For $T > Q - m_{4}$
the deviation is constant, because the Kurie function in Eq.~(\ref{kurie})
reduces to
\begin{equation}
K(T)
=
(Q-T) \sqrt{1 - |U_{e4}|^2}
\,.
\label{kurie-1}
\end{equation}
For
$T = Q - m_{4}$
there is a kink and
for $T < Q - m_{4}$
the Kurie function depends on both
$m_{4}$ and $|U_{e4}|^2$,
as given by Eq.~(\ref{kurie}).

From Figs.~\ref{plt-kur-dif-low} and \ref{plt-kur-dif-hig} one can see that
high precision will be needed in order to see the effect of $m_{4}$ and
measure $|U_{e4}|^2$, which is the only parameter which determines the
deviation of $K(T)$ from the massless Kurie function near the end point, for
$T > Q - m_{4}$. If the mixing parameters are near the best-fit point of the
GLO-LOW analysis, a precision of about one percent will be needed within 1
eV from the end-point of the spectrum. Finding the effect of $m_{4}$ farther
from the end-point, for $T < Q - m_{4}$ is more difficult, because the
relative deviation of the Kurie function can be as small as about $10^{-3}$.
The GLO-HIG analysis prefers slightly larger values of $m_{4}$, but the
discovery of an effect in $\beta$-decay will require a similar precision.


If massive neutrinos are Majorana particles,
neutrinoless double-$\beta$ decay is possible,
with a decay rate proportional to the effective Majorana mass
(see Refs.~\cite{Elliott:2002xe,Bilenky:2002aw,Elliott:2004hr,Avignone:2007fu,Giunti-Kim-2007,Rodejohann:2011mu})
\begin{equation}
m_{\beta\beta}
=
\left| \sum_{k} U_{ek}^2 m_{k} \right|
\,.
\label{050}
\end{equation}
The results of the analysis of short-baseline oscillation data
allow us to calculate the contribution of the heaviest massive neutrino $\nu_{4}$
to $m_{\beta\beta}$, which is given by
\begin{equation}
m_{\beta\beta}^{(4)}
=
|U_{e4}|^2 \sqrt{\Delta{m}^2_{41}}
\,,
\label{m2b4}
\end{equation}
taking into account the mass hierarchy in Eq.~(\ref{hierarchy}).
If there are no unlikely cancellations among the contributions of
$m_{1}$,
$m_{2}$,
$m_{3}$
and that of
$m_{4}$
\cite{Giunti:1999jw}
(possible cancellations are discussed in Refs.\cite{Goswami:2005ng,Li:2011ss}),
the value of
$m_{\beta\beta}^{(4)}$
is a lower bound for the effective neutrino mass
which could be observed
in future
neutrinoless double-$\beta$ decay experiments
(see the review in Ref.~\cite{GomezCadenas:2011it}).

\begin{figure}[t!]
\begin{center}
\includegraphics*[width=0.5\linewidth]{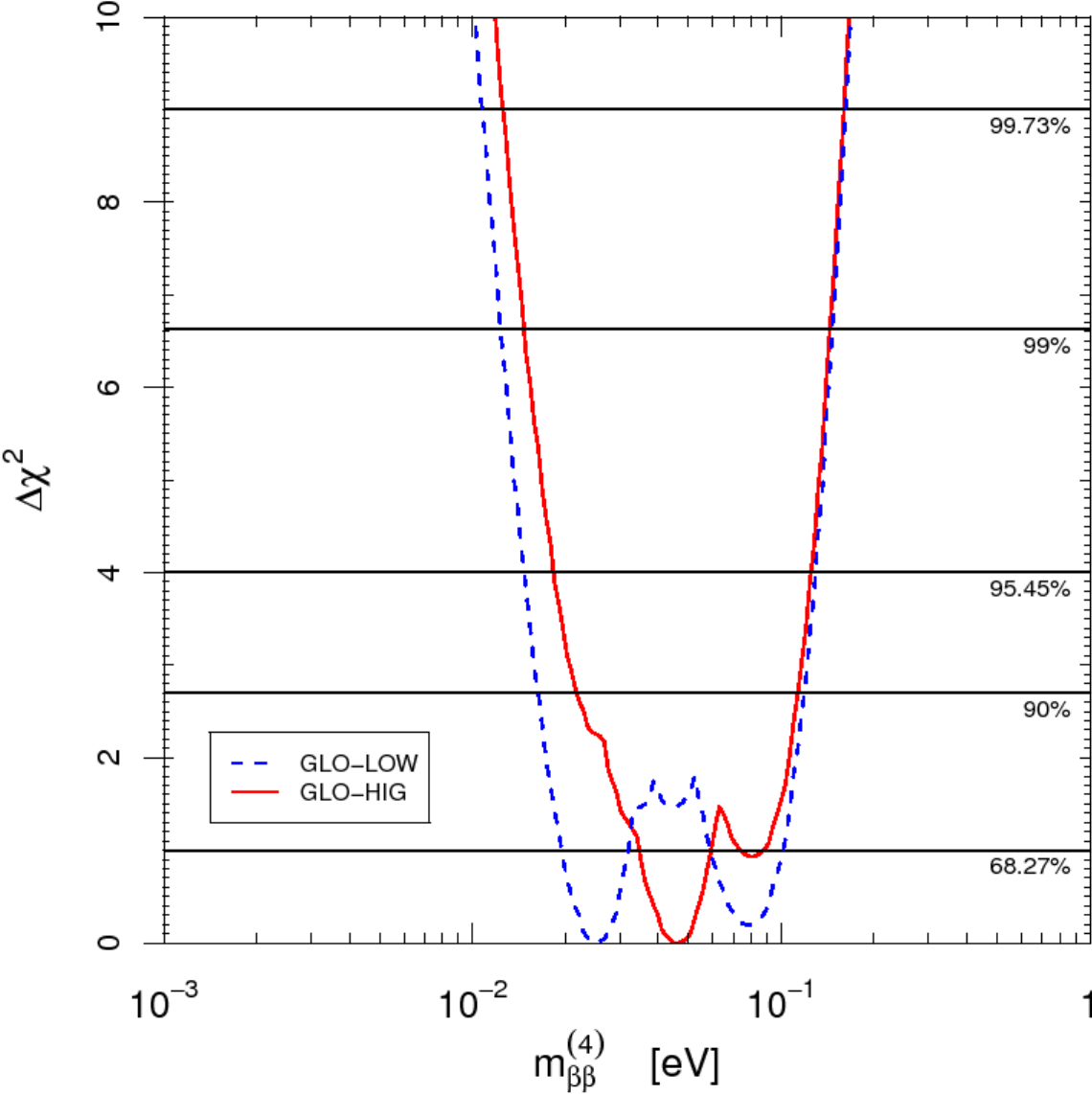}
\end{center}
\caption{ \label{mbb-chi-plt-hig}
Marginal
$\Delta\chi^2 = \chi^2 - \chi^2_{\text{min}}$
as a function of the contribution
$m_{\beta\beta}^{(4)} = |U_{e4}|^2 \sqrt{\Delta{m}^2_{41}}$
to the effective neutrinoless double-$\beta$ decay Majorana mass $m_{\beta\beta}$
obtained
\protect\cite{Giunti:2011cp}
from
the GLO-LOW and GLO-HIG global fits.
}
\end{figure}

\begin{figure}[t!]
\begin{center}
\includegraphics*[width=0.45\linewidth]{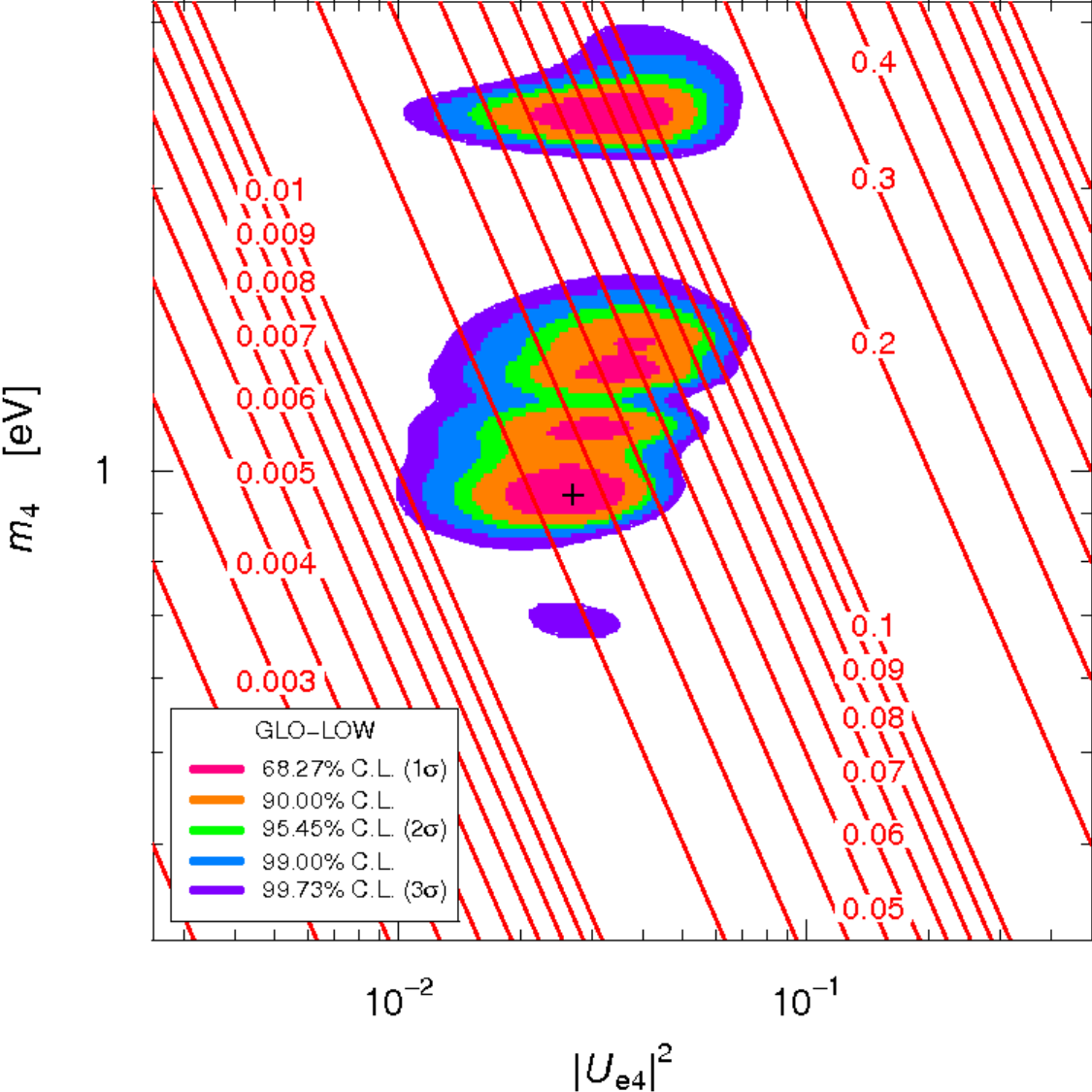}
\includegraphics*[width=0.45\linewidth]{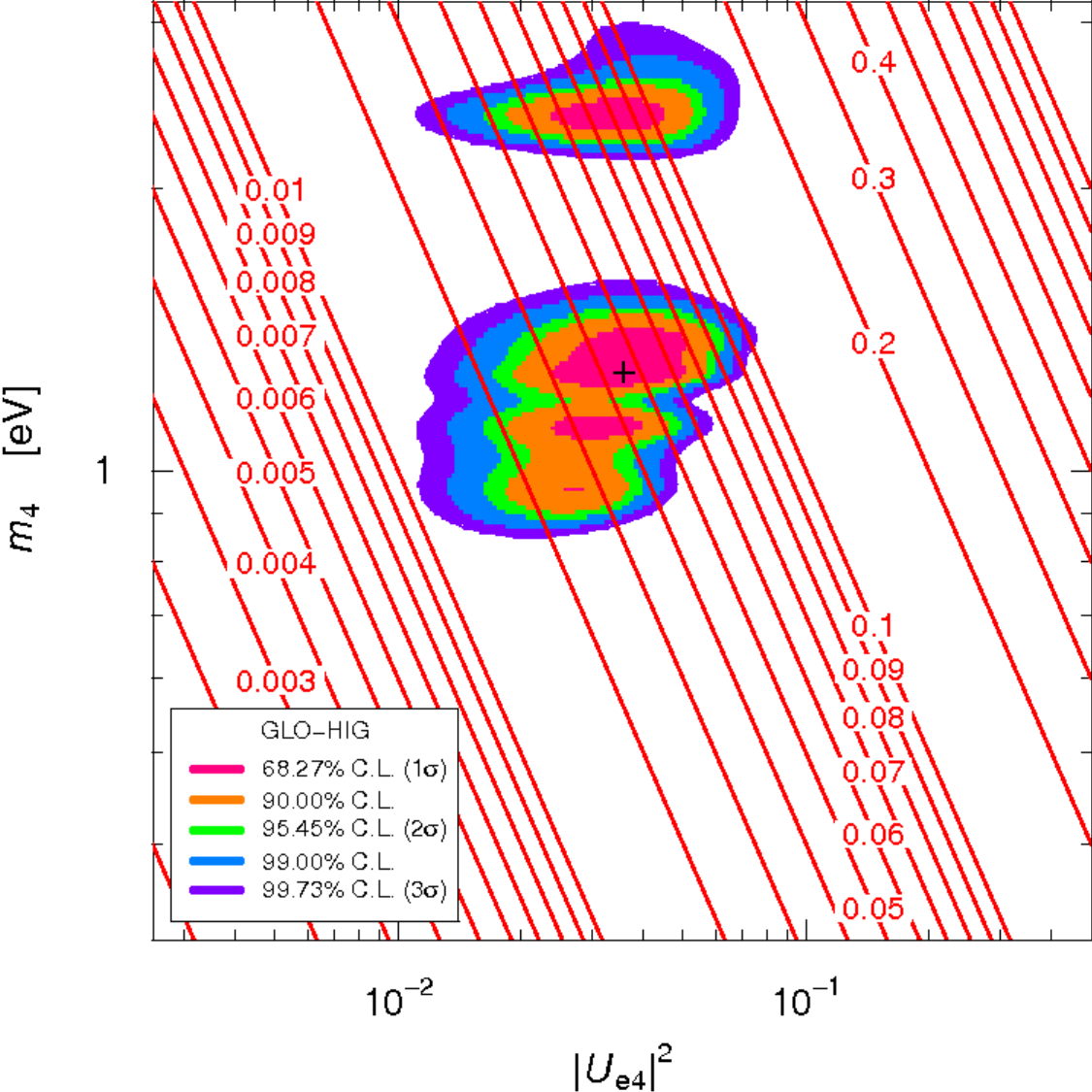}
\end{center}
\caption{ \label{con-img-see-mbb}
Allowed regions in the
$|U_{e4}|^2$--$m_{4}$
plane obtained
\protect\cite{Giunti:2011cp}
from the GLO-LOW (left) and GLO-HIG (right) 
global analysis of short-baseline neutrino oscillation data
(see Tab~\ref{tab-bef}).
The best-fit point is indicated by a cross (see Table.~\ref{tab-bef}).
The red lines have the indicated constant value of
$m_{\beta\beta}^{(4)} = |U_{e4}|^2 \sqrt{\Delta{m}^2_{41}}$.
}
\end{figure}

Figure~\ref{mbb-chi-plt-hig}
shows the marginal
$\Delta\chi^2$
as a function of
$m_{\beta\beta}^{(4)}$,
which gives the
$1\sigma$,
$2\sigma$ and
$3\sigma$
allowed ranges of $m_{\beta\beta}^{(4)}$
listed in Tab.~\ref{tab-rng}.
The predictions for $m_{\beta\beta}^{(4)}$
obtained from global GLO-LOW and GLO-HIG agree in indicating a $3\sigma$
allowed range between about 0.01 and 0.1 eV.
The connection of the value of $m_{\beta\beta}^{(4)}$
with the allowed regions for the oscillation parameters
is clarified in Fig.~\ref{con-img-see-mbb},
where we show the allowed regions in the
$|U_{e4}|^2$--$m_{4}$
plane obtained
from the GLO-LOW and GLOW-HIG analyses,
together with lines of constant $m_{\beta\beta}^{(4)}$.
One can see that if the oscillation parameters are close to the best-fit 
point of the GLO-LOW analysis, at
$m_{4} = 0.94 \, \text{eV}$,
which is favored by cosmological data,
the value of $m_{\beta\beta}^{(4)}$
is about 0.02--0.03~eV.
In order to have a large value of $m_{\beta\beta}^{(4)}$,
around 0.1~eV,
the oscillation parameters must lie in the large-$m_{4}$ region at
$m_{4} \simeq 2.40 \, \text{eV}$,
or on the large-$|U_{e4}|^2$ border
of the allowed region at
$m_{4} \simeq 1.27 \, \text{eV}$.

\subsubsection{On Neutrinoless double-$\beta$ decay with Sterile Neutrinos}


\bigskip

\begin{figure}[t!]
\begin{center}
\includegraphics*[width=0.45\linewidth]{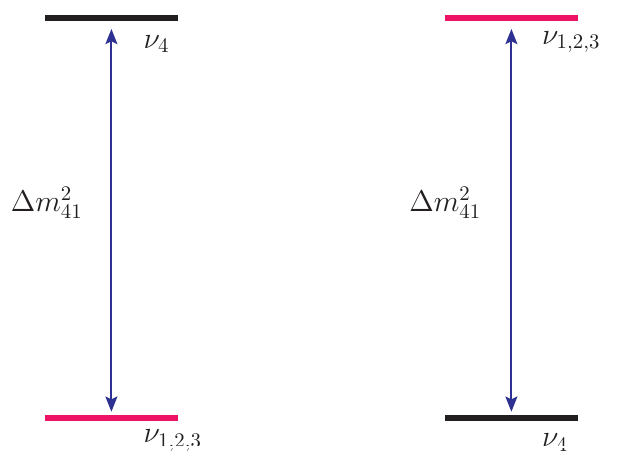}
\end{center}
\caption{ \label{mass-ordering-4} Description of the two possible
mass orderings in the 3+1 case (left) and 1+3 case (right). Here
$\nu_{1,2,3}$ are the mostly active neutrinos whose mass splitting can
be either normal or inverted.}
\end{figure}

Let us discuss in more detail how light sterile neutrinos may
significantly affect neutrinoless double-$\beta$ decay. Needless to
say, one needs to assume for the following discussion that neutrinos
are Majorana particles, and that no other of the many proposed
particle physics scenarios other than light neutrino exchange
contribute to the process (see \cite{Rodejohann:2011mu} for a recent
review).  
The contributions of sterile neutrinos to the
effective mass $m_{\beta\beta}$ depend on the sterile
neutrino masses, the active-sterile mixings and the neutrino mass
ordering. In the presence of one sterile neutrino
$\nu_4$, there are two typical mass orderings between active and
sterile neutrinos, which are depicted in Fig.~\ref{mass-ordering-4}.
\begin{figure}[t!]
\begin{center}
\includegraphics*[width=0.7\linewidth]{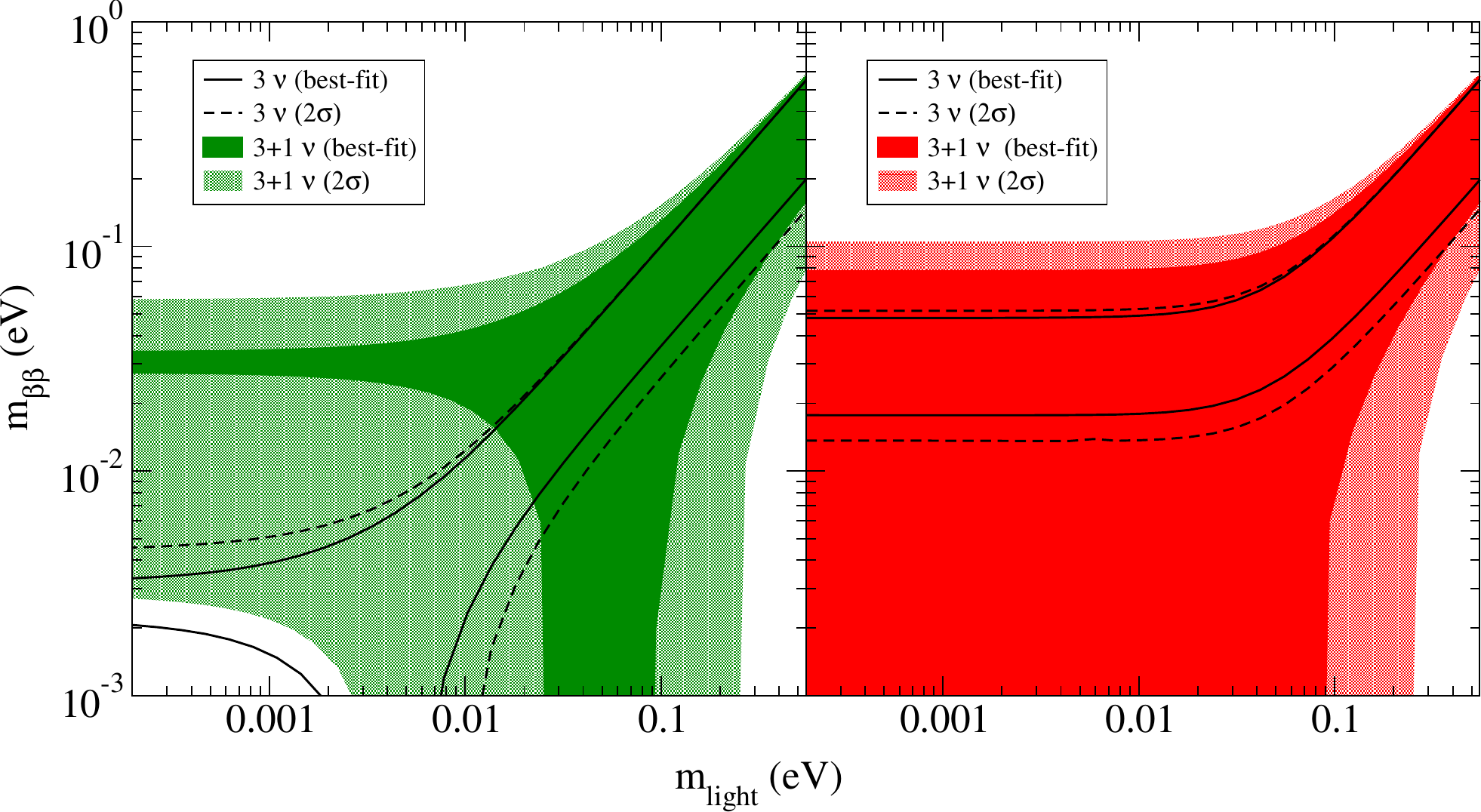}
\includegraphics*[width=0.7\linewidth]{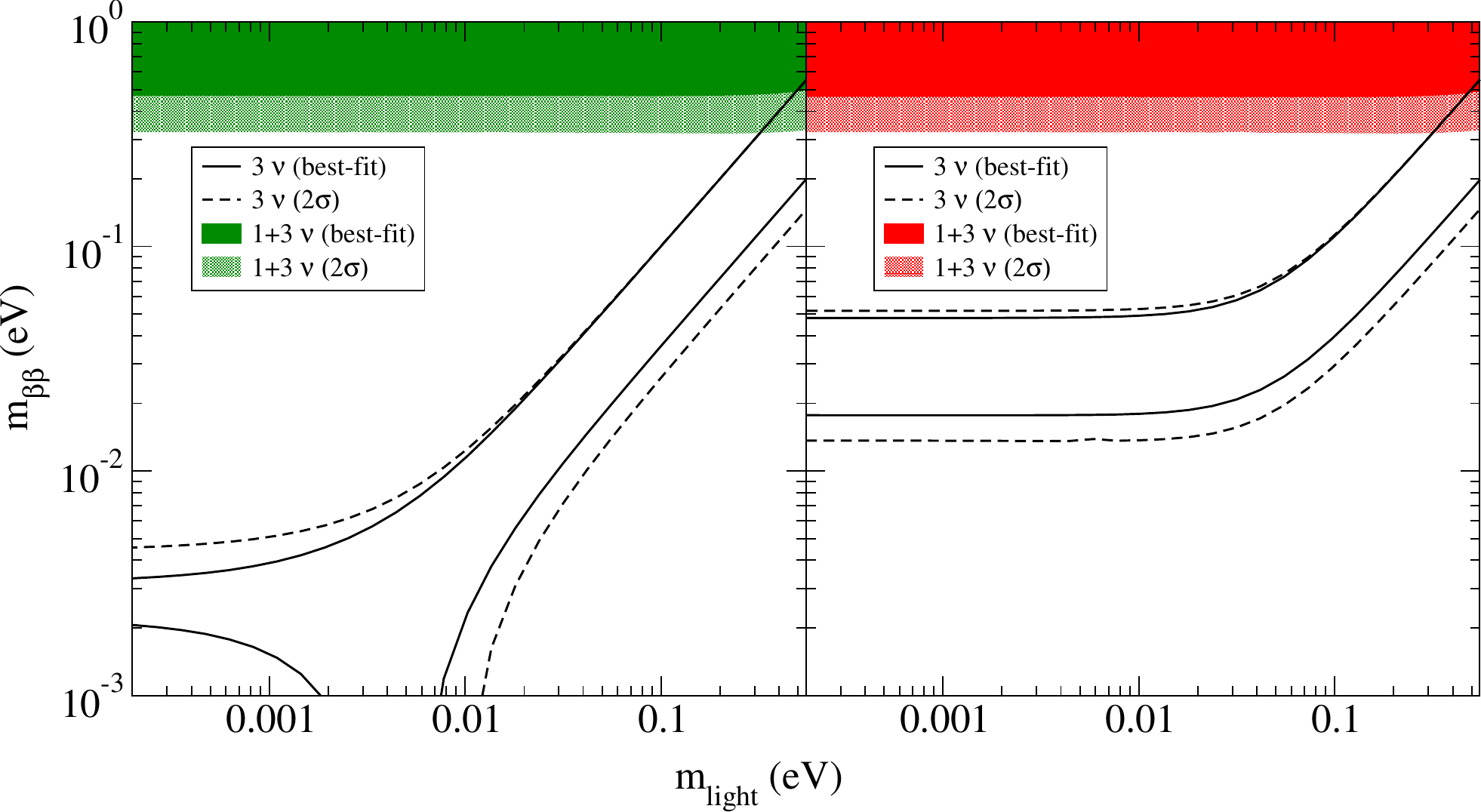}
\end{center}
\caption{\label{fig:mee_4nu} The allowed ranges in the
$m_{\beta\beta} -m_{\mathrm light}$ parameter space, both in the
standard three-neutrino picture (unshaded regions) and with one
sterile neutrino (shaded regions), for the 3+1 (bottom) and 1+3
(top) cases.}
\end{figure}
The effective mass can be written as 
\begin{equation}
m_{\beta\beta} = | \underbrace{|U_{e1}|^2  \, m_1 + |U_{e2}|^2  \, m_2 \, e^{ i
\alpha} + |U_{e3}^2|  \, m_3 \, e^{i \beta} 
}_{\displaystyle m_{\beta\beta}^{\mathrm (3)}} + 
\underbrace{|U_{e4}|^2  \, m_4 \, e^{ i \gamma}}_{\displaystyle m_{\beta\beta}^{\mathrm (4)}} | \, ,
\end{equation} 
where $\gamma$ denotes the additional Majorana CP phase\footnote{In
most parameterizations of scenarios with sterile neutrinos both Dirac and
Majorana phases appear in the effective mass, and care has to be taken
to introduce only the Majorana phases, see {\it e.g.}~\cite{Rodejohann:2011vc}.}, and
$m^{(3)}_{\beta\beta}$ is the active neutrino contribution to
$m_{\beta\beta}$. Generalization to 2 or more sterile neutrinos is
straightforward. 
We illustrate in Fig.~\ref{fig:mee_4nu} the allowed ranges of
$m_{\beta\beta}$ as a function of the lightest mass $m_{\mathrm light}$,
using data from Refs.~\cite{Schwetz:2011qt,Kopp:2011qd}.  
It is important to recall the standard three neutrino picture, namely that in the normal
hierarchy ($m_1 = 0$) $|m_{\beta\beta}^{(3, \mathrm NH)}|$ can vanish and
that in the inverted hierarchy ($m_3=0$) 
$|m_{\beta\beta}^{(3, \mathrm IH)}|$ cannot vanish, having a typical value of
0.02 eV. In the typically considered case of 3+1 ordering, let us take
for illustration the  parameters $\Delta m^2_{41} \simeq 1$ eV$^2$ and
$|U_{e 
4}| \simeq 0.15$. One finds with $m_4 \simeq \sqrt{\Delta m^2_{41}} \gg m_{1,2,3}$ that 
\be 
|m_{\beta\beta}^{(4)}| \simeq \sqrt{\Delta m^2_{41}} \, |U_{e4}|^2 \simeq 0.02~{\mathrm eV} \left\{ 
\begin{array}{c}
\gg | m_{\beta\beta}^{(3, \mathrm NH)}|\\
\simeq | m_{\beta\beta}^{(3, \mathrm IH)}|
\end{array} 
\right. .
\ee 
Therefore, if the active neutrinos are normally ordered, the effective
mass {\it cannot} vanish anymore, whereas it {\it can} vanish when they are
inversely ordered. Hence, the usual standard phenomenology has been
completely turned around \cite{Barry:2011wb,Li:2011ss}. 
 The effective mass can also be zero in the regime
where the active neutrinos are quasi-degenerate (lightest mass above 
$0.1~{\mathrm eV}$). This feature is of particular interest if future
neutrinoless double-$\beta$ decay experiments measure a tiny
effective mass smaller than the usual lower bound on the effective
mass (cf.~the solid and dashed lines in Fig.~\ref{fig:mee_4nu}) and
the neutrino mass hierarchy is confirmed to be inverted from  
long-baseline neutrino oscillations. In such a case one needs
additional sources to cancel the active neutrino contributions to
$m_{\beta\beta}$, and the sterile neutrino hypothesis would be an
attractive explanation for this inconsistency.

If active neutrinos are heavier than the sterile one, {\it i.e.}, the 1+3
case (somewhat disfavored by cosmological bounds on the sum of
neutrino masses) shown in  Fig.~\ref{mass-ordering-4}, three active
neutrinos are quasi-degenerate and the effective mass is approximately
given by 
\begin{equation} \label{eq:meffQD}
 m_{\beta\beta} \simeq \sqrt{\Delta m^2_{41}} \, \sqrt{1 - \sin^2 2 \theta_{12} \, \sin^2
\alpha/2} \, ,
\end{equation}
no matter whether the active neutrino mass spectrum is normal 
($m_3>m_2>m_1$) or inverted ($m_2>m_1>m_3$). For values of $\Delta
m^2_{41}$ around 1 eV$^2$ this puts already constraints on the
Majorana phase $\alpha$.

\begin{figure}[t!]
\begin{center}
\includegraphics*[width=0.75\linewidth]{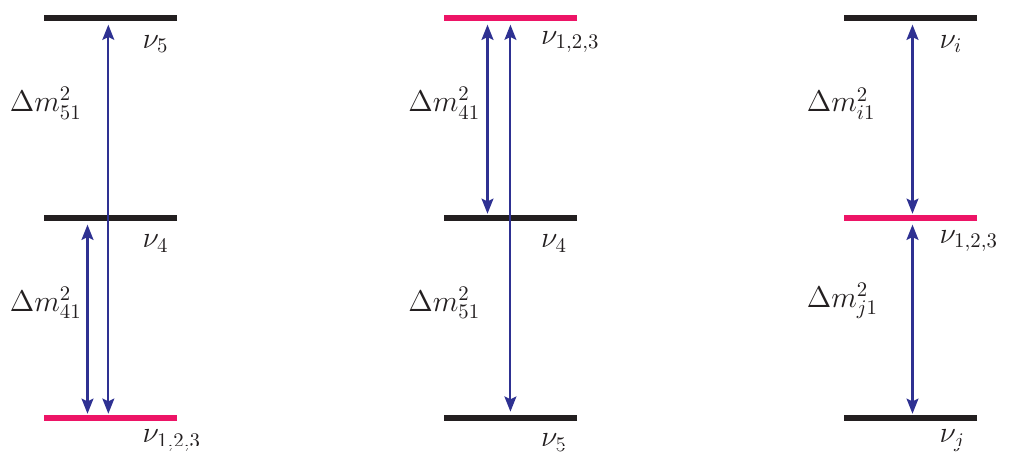}
\end{center}
\caption{ \label{mass-ordering-5} Description of three possible mass
hierarchies in the 3+2 case (left), the 2+3 case (middle) and 1+3+1
case (right).}
\end{figure}

\begin{figure}[h!]
\begin{center}
\includegraphics*[width=0.7\linewidth]{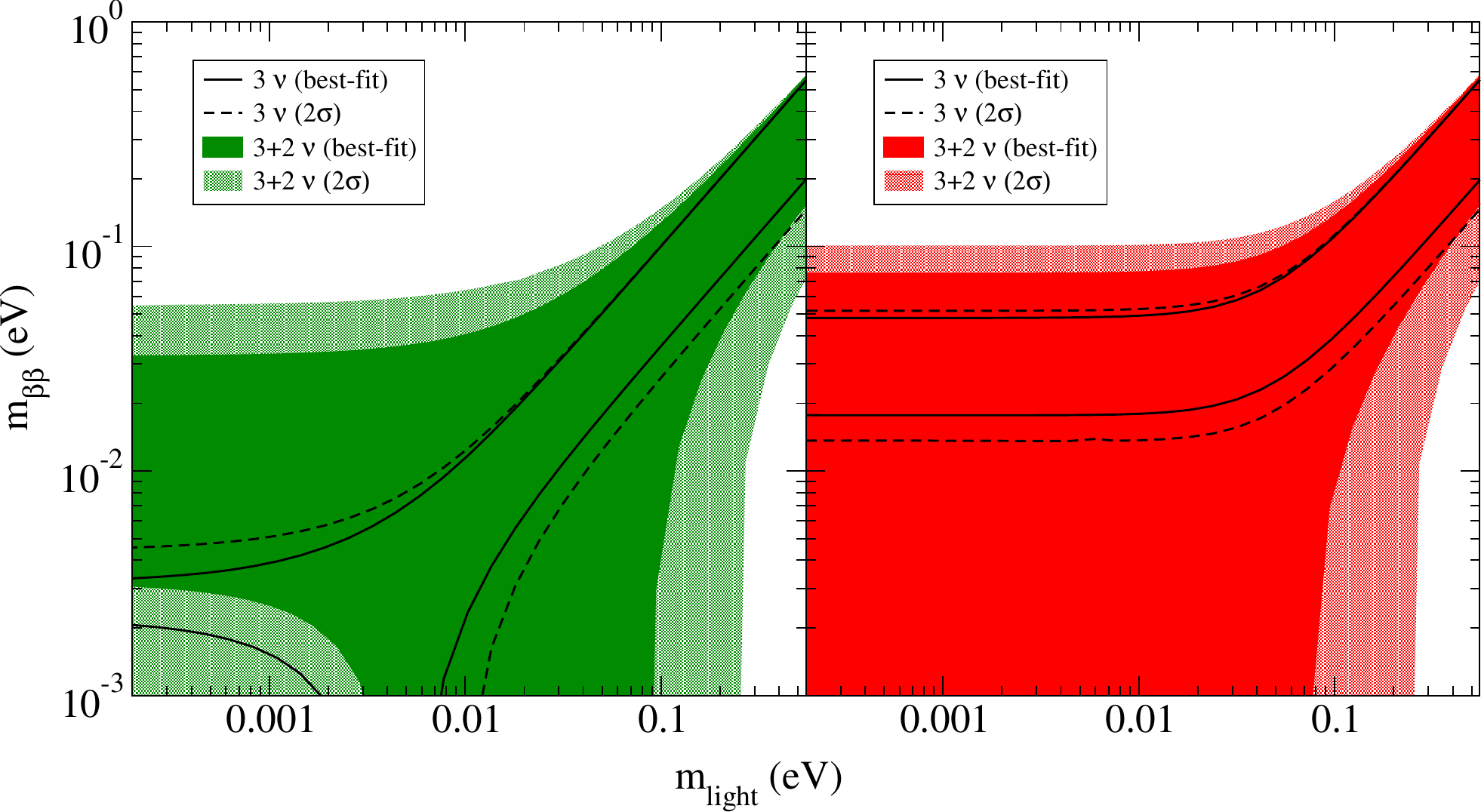}
\end{center}
\caption{\label{fig:mee_5nu} The allowed ranges of $m_{\beta\beta}
-m_{\mathrm light}$ in the 3+2 case with normal mass ordering (left) and
inverted mass ordering (right).}
\end{figure}

The situation is more complicated in the five neutrino case,
since there are in principle three classes of neutrino mass spectra:
$m_5>m_4 >  m_{1,2,3}$ (3+2 case), $m_{1,2,3} > m_4 > m_5$ (2+3
case), and $m_i > m_{1,2,3} > m_j$ (1+3+1 case)
\cite{Goswami:2007kv}, see Fig.~\ref{mass-ordering-5}. 
The latter two cases generally lead to a large sum of neutrino masses,
because the active neutrinos are quasi-degenerate and there are
in addition eV-scale sterile neutrinos. They induce more tension with
cosmological constraints and we therefore only show in
Fig.~\ref{fig:mee_5nu} the allowed ranges of $m_{\beta\beta}-m_{\mathrm
light}$ in the interesting 3+2 case. Similar to the above discussion for the
3+1 case above, the presence of two sterile neutrinos allows
$m_{\beta\beta}$ to take smaller values in the hierarchical region for
the normal ordering, compared to the standard three neutrino case.

Let us remark here that for the other interesting sterile neutrino
case discussed in the White Paper, keV-scale neutrinos with mixing of
order $10^{-4}$ with active neutrinos, the contribution to
neutrinoless double-$\beta$ decay is negligible. In addition, if these
keV-scale particles are part of the seesaw mechanism, they imply one
essentially massless active neutrino, {\it i.e.}~$m_1 = 0$ or $m_3=0$.

\subsection{Sterile Neutrinos and IceCube}
\label{sec:IceCube}


\bigskip

Neutrino experiments offer opportunities for new discoveries and,
occasionally, total surprises. Examples of new physics include sterile
neutrinos and additional degrees of freedom in the energy density of the
Universe. Even though it is premature to motivate future facilities on the
basis of present indications (which include some hints from short-baseline
experiments \cite{AguilarArevalo:2010wv} and reactor data
\cite{Mueller:2011nm}), recent developments underscore the possibility of
unexpected discoveries, supporting the construction of neutrino facilities
with the widest science reach. IceCube is such a facility
\cite{Halzen:2010yj}. IceCube measures the flux of atmospheric muons and
neutrinos with high statistics in a high-energy range that has not been
previously explored. The procedure is simply to compare data with
expectations based on the extrapolation of lower-energy data. Any deviation
observed is an opportunity for discovery in astrophysics or the physics of
the neutrinos themselves.

The potential of IceCube to detect eV-mass sterile neutrinos has been
recognized for some time \cite{Nunokawa:2003ep,Choubey:2007ji}. The
problem has been revisited \cite{Kopp:2011qd,Razzaque:2011ab,Barger:2011rc}
with the measurement of the atmospheric neutrino flux in the energy range
100\,GeV--400\,TeV with unprecedented statistics, taken when IceCube was
half complete \cite{Abbasi:2011jx}.  The signature for sterile neutrinos is the
disappearance of $\nu_{\mu} (\bar\nu_{\mu})$ resulting from the fact that
they mix with the sterile neutrino during propagation between production and
detection. Simply stated, some of the time they propagate in the sterile
state with direct impact on matter effects when propagating through the
Earth. As a result resonant oscillations occur in the atmospheric neutrino
beam at characteristic zenith angles and energies in the TeV range for
$\delta\!m^2 \sim 1 eV^2$. Because the sterile neutrino acts as an
intermediary between muon and electron neutrinos, an additional signature is
the appearance of an excess of $\nu_{e} (\bar\nu_{e})$ with a characteristic
energy and zenith dependence.

Given the high statistics, the focus is on the systematics of the
experiment. The systematic issues are being studied with renewed emphasis;
up to now, the priority has been the search for neutrino sources beyond the
atmosphere, analyses where the simulation of the background is done using
the data themselves.

\begin{figure}
\begin{center}
\includegraphics[width=1.0\textwidth]{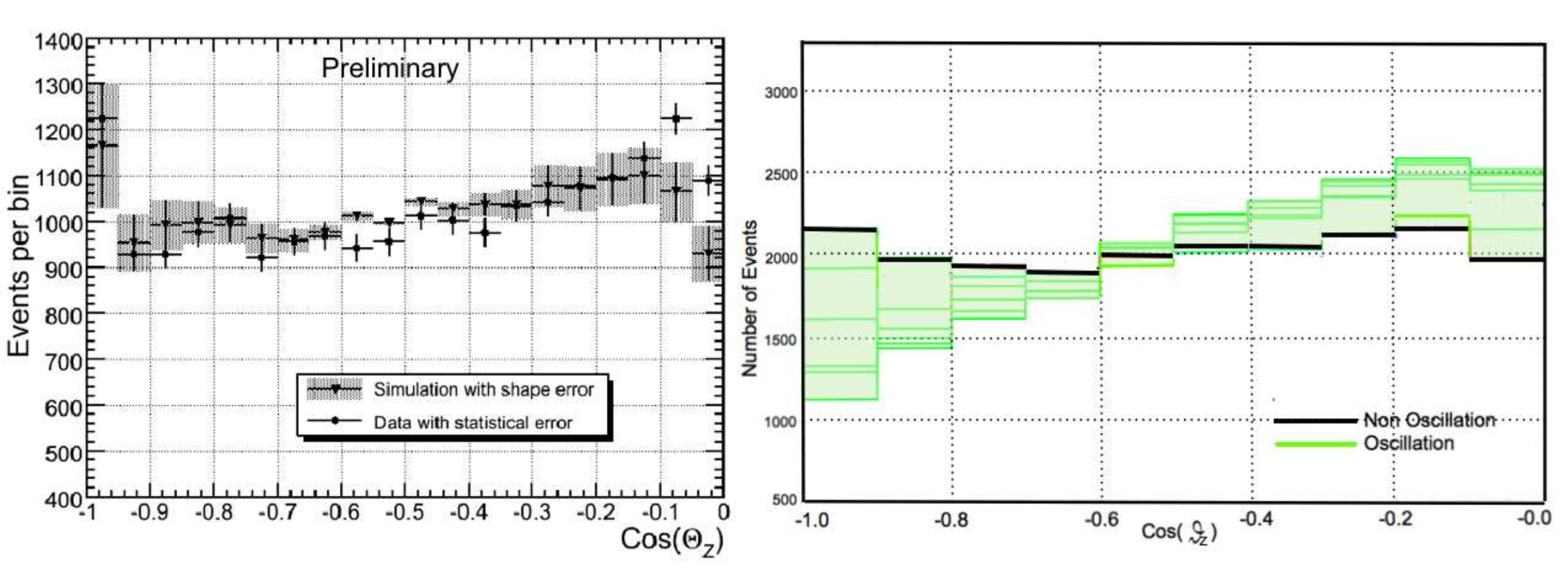}
\caption{Zenith distribution cos$\theta_z$ of atmospheric neutrinos from \cite{Abbasi:2010ie}. \textit{On the left}:  Simulation has been normalized to the data.  Error bars on the data are statistical only.  Error bars on the simulation indicate the uncertainty in the shape of the predicted distribution only.  \textit{On the right}:  Illustration of the impact of adding a sterile neutrino into the conventional 3-flavor scheme on the cos$\theta_z$ distribution.  For this particular model (green lines), $\Delta m_{41}^2=0.4$ $\mathrm{eV}^2$, $\mathrm{sin}^2\theta_{24}=0.1$, and $\mathrm{sin}^2\theta_{34}$ is varied from 0 to 0.5 in increments of 0.1.  Green and black lines include standard atmospheric neutrino oscillations and detector acceptance.  (credit: A. Esmaili, Universidade Estadual de Campinas, Brazil). }
\label{cos_zen}
\end{center}
\end{figure}

{\bf Zenith Distribution.}
Figure~\ref{cos_zen} compares the predicted atmospheric $\nu_\mu$ (plus
$\bar{\nu}_{\mu}$) flux with data taken while IceCube operated with 40 out
of its final 86 strings.  Since the uncertainty in the normalization is
rather large \cite{Abbasi:2010ie}, the shape of the zenith distribution will be
more sensitive to the resonant oscillation.  In Fig.~\ref{cos_zen}, the
predicted distribution, normalized to the event rate, is compared with the
data.  Uncertainties on the shape of the predicted flux are shown.  These
shape uncertainties reflect uncertainty in the relative ratio of pions to
kaons produced by the cosmic ray flux, uncertainties in the spectral shape
and composition of the cosmic ray flux, and uncertainties in the simulation
of digital optical module sensitivity and ice properties \cite{Abbasi:2011jx,
Abbasi:2010ie}.  Some of these will be reduced as the final detector and its
irreducible systematic uncertainties are better understood.

Several possibilities have been explored to resolve the apparent mismatch
between data and simulation in the near horizontal region, including the
possibilities discussed in \cite{Abbasi:2010ie, Abbasi:2010rd}: regional and seasonal
variations in atmospheric neutrino production, simulation of ice properties
and photon propagation in the ice, and inadequate simulation of atmospheric
muon backgrounds including various composition models of cosmic rays.  

While these effects have some impact on the zenith distribution, we have
found that an important change arose from improvements in the modeling of
the rock layer below the detector and the ice/rock boundary, as well as the
assignment of event weights according to model simulations.  These event
weights account for the probability of a simulated neutrino to survive
propagation through the Earth and interact in or near the detector.
Preliminary testing of improved simulation for the 59-string detector
configuration indicates that these inadequacies in the modeling of the
detector were a major contributor to the disagreement between the 40-string
simulation and data in reference~\cite{Abbasi:2010ie}.  Atmospheric neutrino
analyses and searches for a diffuse flux of astrophysical neutrinos, with
data taken by the 59-string detector, are currently underway and are using
this improved simulation.

\begin{figure}
\begin{center}
\includegraphics[width=0.5\textwidth]{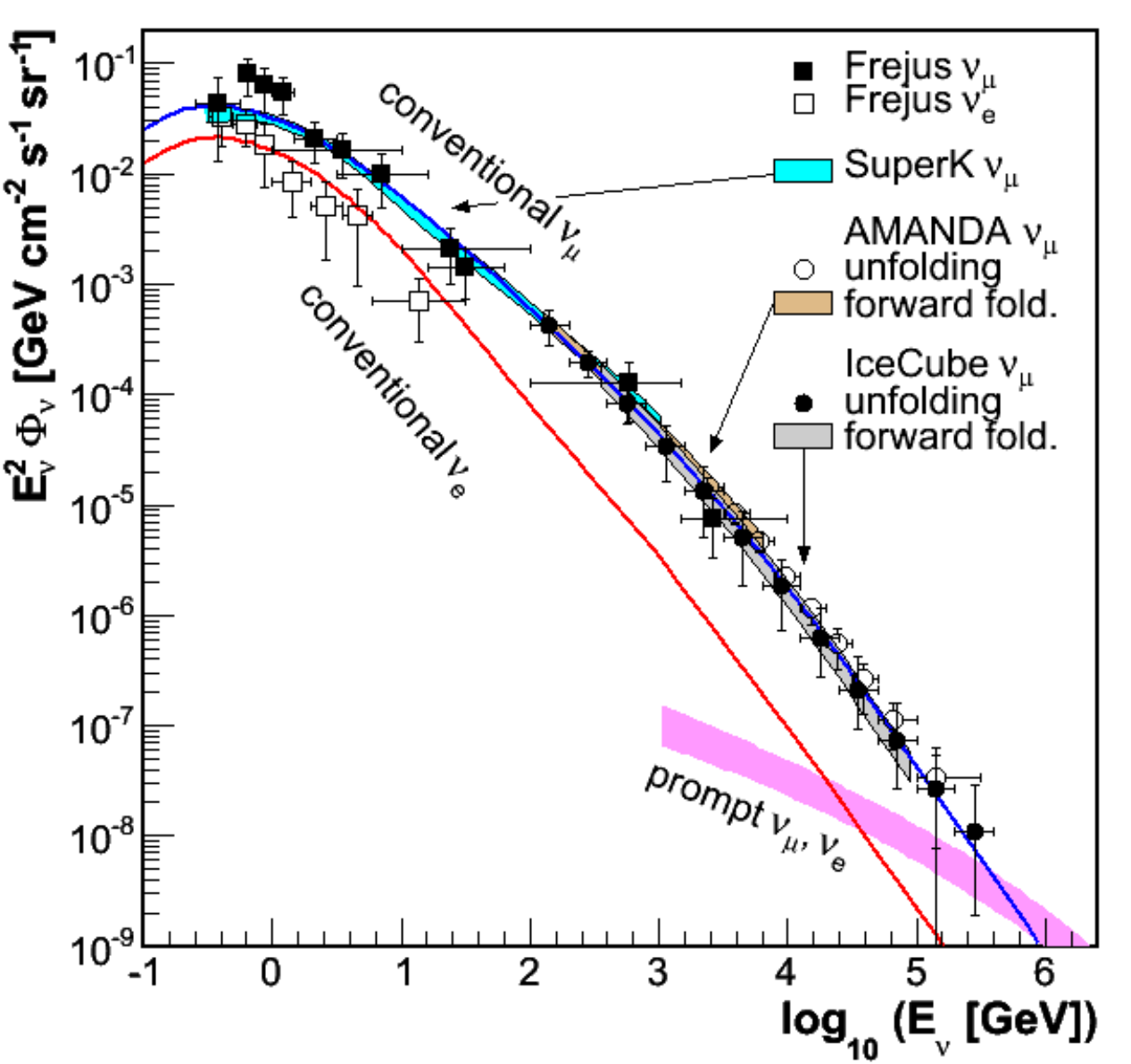}
\caption{Measurements of the atmospheric neutrino energy spectrum; the Fr\'{e}jus results \cite{Daum:1994bf}, SuperK \cite{GonzalezGarcia:2006ay}, 
AMANDA forward-folding analysis \cite{Abbasi:2009nfa} and unfolding analysis \cite{Abbasi:2010qv}, IceCube (40 strings) forward folding analysis \cite{Abbasi:2011jx} and unfolding analysis \cite{Abbasi:2010ie}.  All measurements include the sum of neutrinos and antineutrinos.  The expectations for conventional $\nu_\mu$ and $\nu_e$ flux are from \cite{Barr:2004br}.  The prompt flux is from \cite{Enberg:2008te}.}
\label{compare}
\end{center}
\end{figure}

{\bf Energy Spectrum.}
Figure~\ref{compare} compares the results of a variety of atmospheric
neutrino measurements in the GeV -- 400~TeV energy range.  For IceCube's
spectrum unfolding measurement \cite{Abbasi:2010ie}, events in the $90^ \circ $ to
$97^ \circ $ zenith region were not used.  However, in Fig.~\ref{compare},
this result has been scaled to the flux for the entire hemisphere, to allow
a more direct comparison to the other results.  IceCube's forward folding
result from reference~\cite{Abbasi:2011jx} is also shown.  Both of these results
are affected by the simulation issues discussed above.  New analyses using
improved simulation are underway using data taken while IceCube operated in
the 59-string configuration.

{\bf Conclusion.} IceCube has been stably operating with a completed
detector since May 2011. Atmospheric neutrinos are collected at the rate of
\textit{O}($10^5$) per year.  Our ability to constrain a variety of sterile
neutrino models is only limited by systematic uncertainties.  The
calibration of the detector in its final configuration will be superior to
that of configurations operated during construction. A new generation of
analysis tools are resulting in improved effective area and angular and
energy resolution.

\subsection{Sterile Neutrinos and Dark Matter Searchs}
\label{sec:SNDarkMatter}

Recently, it has been noted that the existence of light sterile neutrinos
would have important consequences for dark matter searches. On the one hand,
as we will outline in section~\ref{sec:snu-dm-icecube},
MSW-enhanced transitions between active and sterile neutrino flavors would
have a substantial impact on searches for neutrinos from dark matter annihilation in
the Sun~\cite{Esmaili:2012ut,Arguelles:2012cf}. On the other hand, if sterile
neutrinos possess, in addition to their mixing with the active neutrinos, also
some new gauge interactions, they can lead to very interesting signals in
direct dark matter searches, which can either be used to discover or constrain
such sterile neutrino models, but may also be confused with dark matter
signals~\cite{Pospelov:2011ha,Harnik:2012ni,Pospelov:2012gm}.
We will discuss sterile neutrinos in direct dark matter detector in
section~\ref{sec:snu-dm-direct}.

\subsubsection{Sterile neutrinos and indirect dark matter search in IceCube}
\label{sec:snu-dm-icecube}

In section~\ref{sec:IceCube} and ref.~\cite{Razzaque:2011ab}, it
has been argued that transitions between high-energy active and sterile
neutrinos can be enhanced by new MSW resonances. In refs.~\cite{Esmaili:2012ut,
Arguelles:2012cf}, it was then pointed out that for high-energy ($\gg$~GeV)
neutrinos propagating out of the Sun, these resonances can lead to almost
complete conversion of certain flavors into sterile states. 
This is particularly relevant for neutrinos produced when dark matter is captured
by the Sun and annihilates into standard model particles at its center (see
{\it e.g.}~\cite{Gould:1991hx, Cirelli:2005gh, Blennow:2007tw}). Such high-energy
neutrinos from the Sun provide a unique opportunity to search for dark matter,
and are actively exploited to that end by the IceCube and Super-Kamiokande
collaborations~\cite{Abbasi:2009uz,IceCube:2011aj,Desai:2004pq}.
If part of the high-energy neutrino flux from dark matter annihilation is converted
into undetectable sterile neutrinos, these limits will be weakened. Moreover,
if the existence of sterile neutrinos should be established in the future,
but their parameters still remain uncertain, this uncertainty would constitute
a significant systematic uncertainty for indirect dark matter searches using
high-energy neutrinos from the Sun.

\begin{figure}
  \begin{center}
    \begin{tabular}{c@{}c@{}c}
      \includegraphics[width=5.5cm]{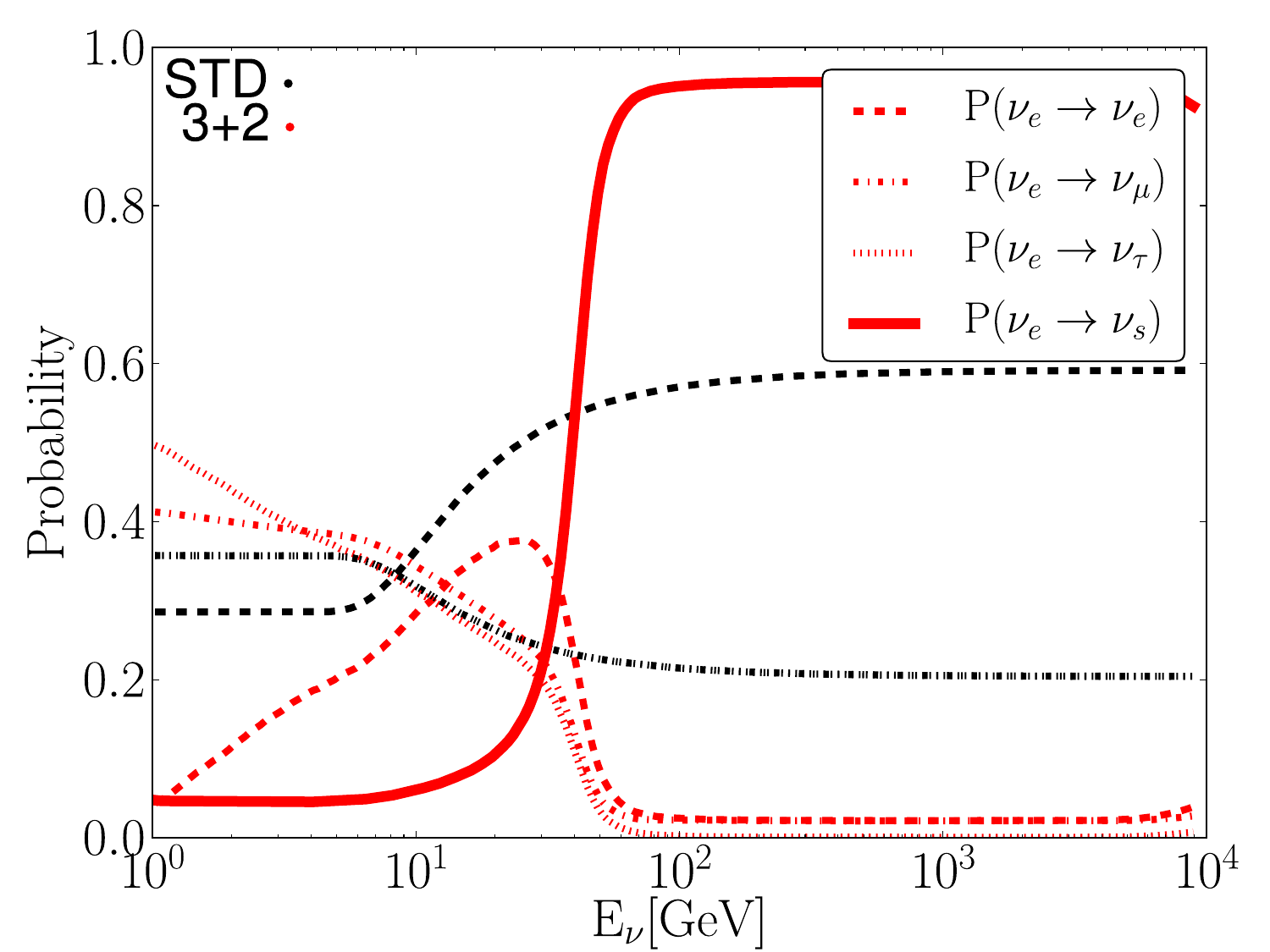} &
      \includegraphics[width=5.5cm]{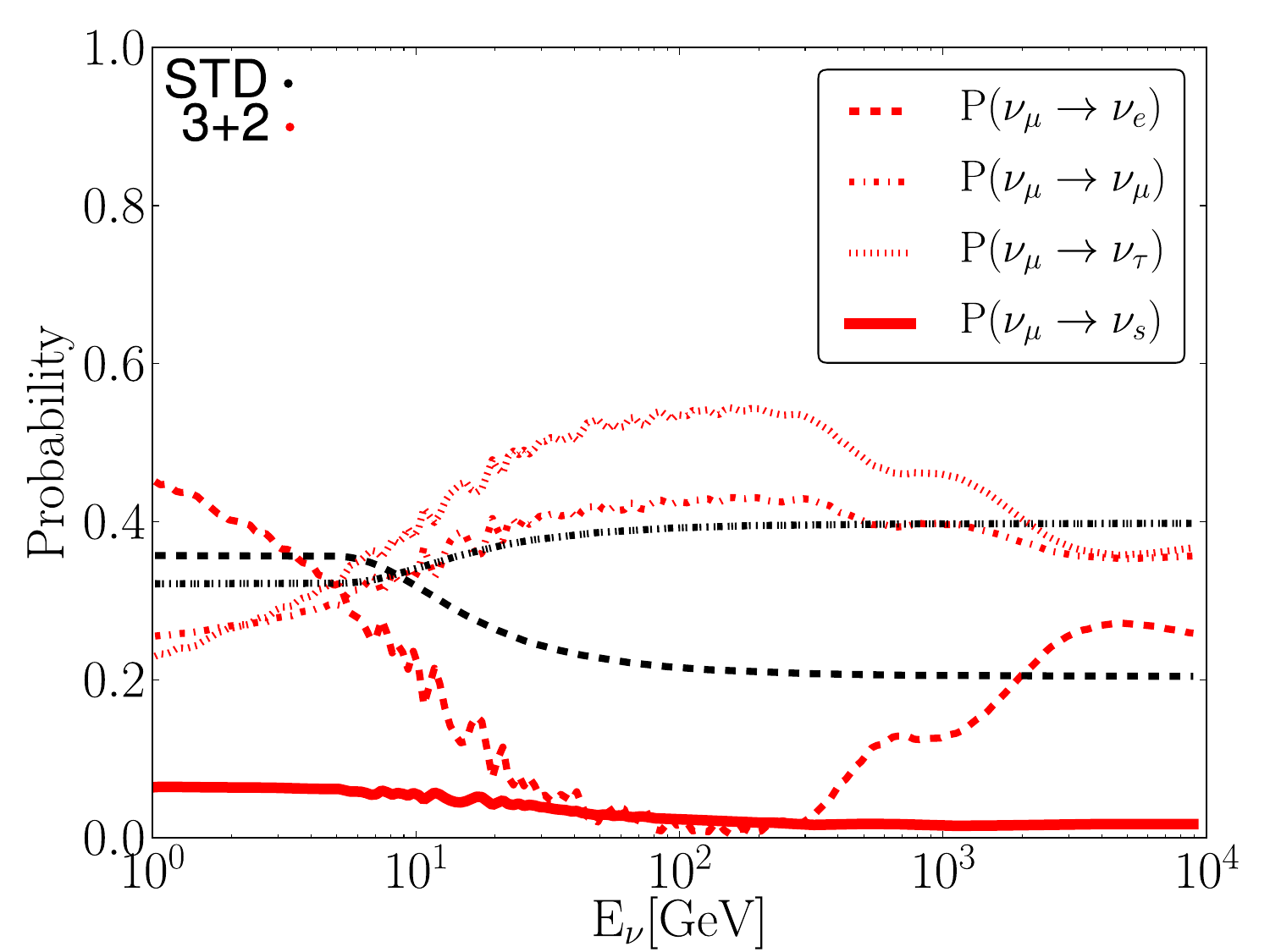} &
      \includegraphics[width=5.5cm]{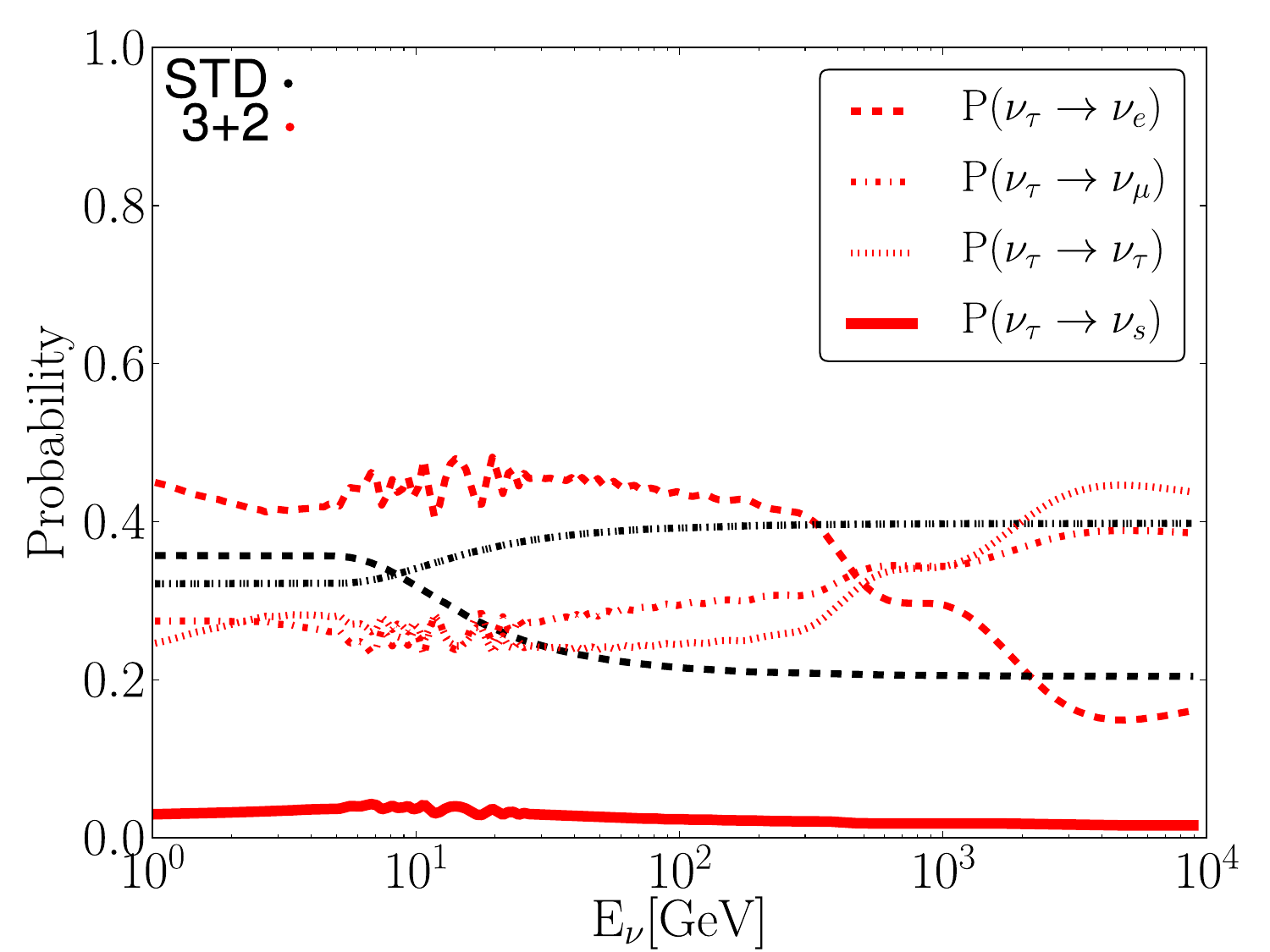} \\
      \includegraphics[width=5.5cm]{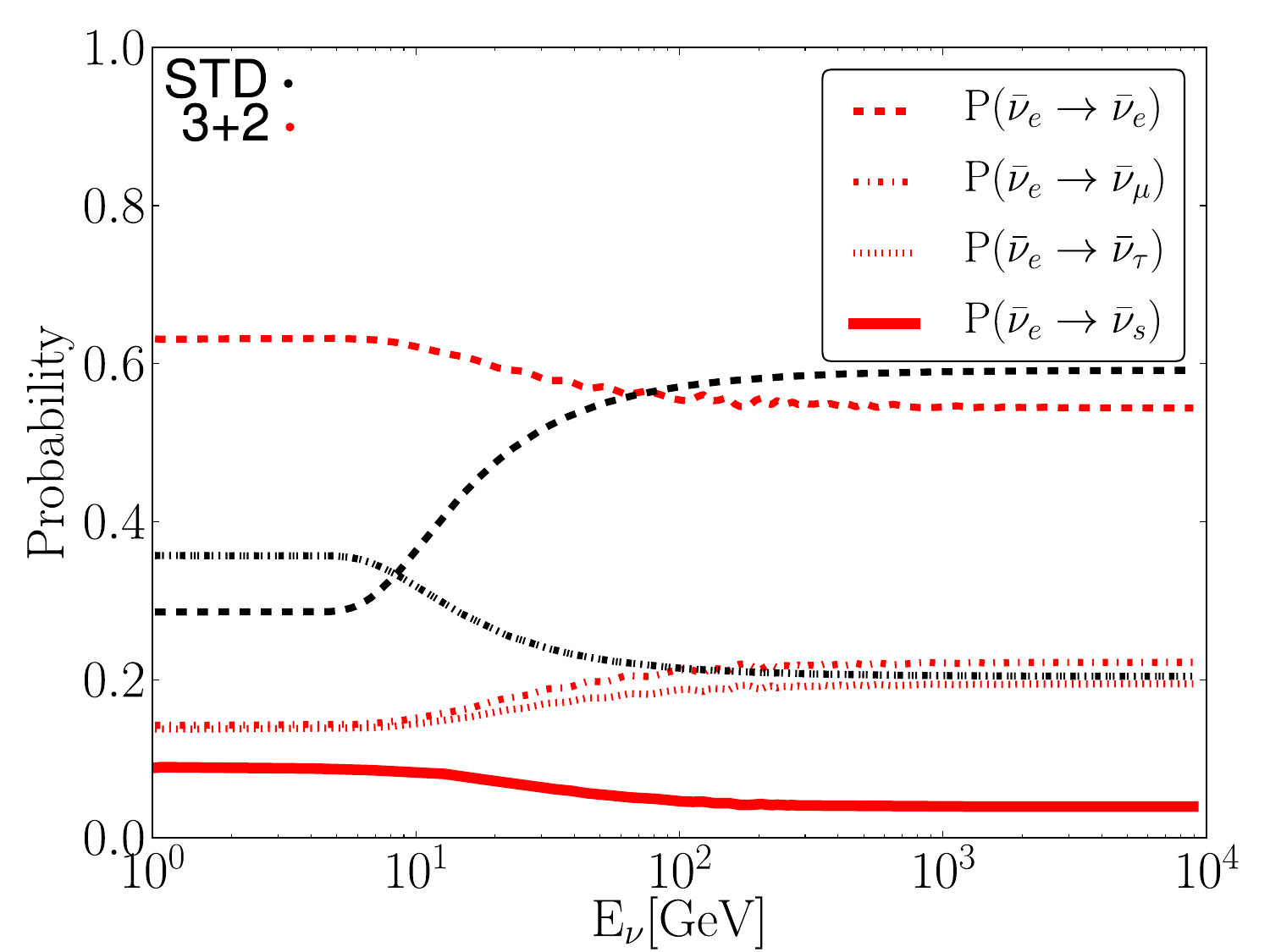} &
      \includegraphics[width=5.5cm]{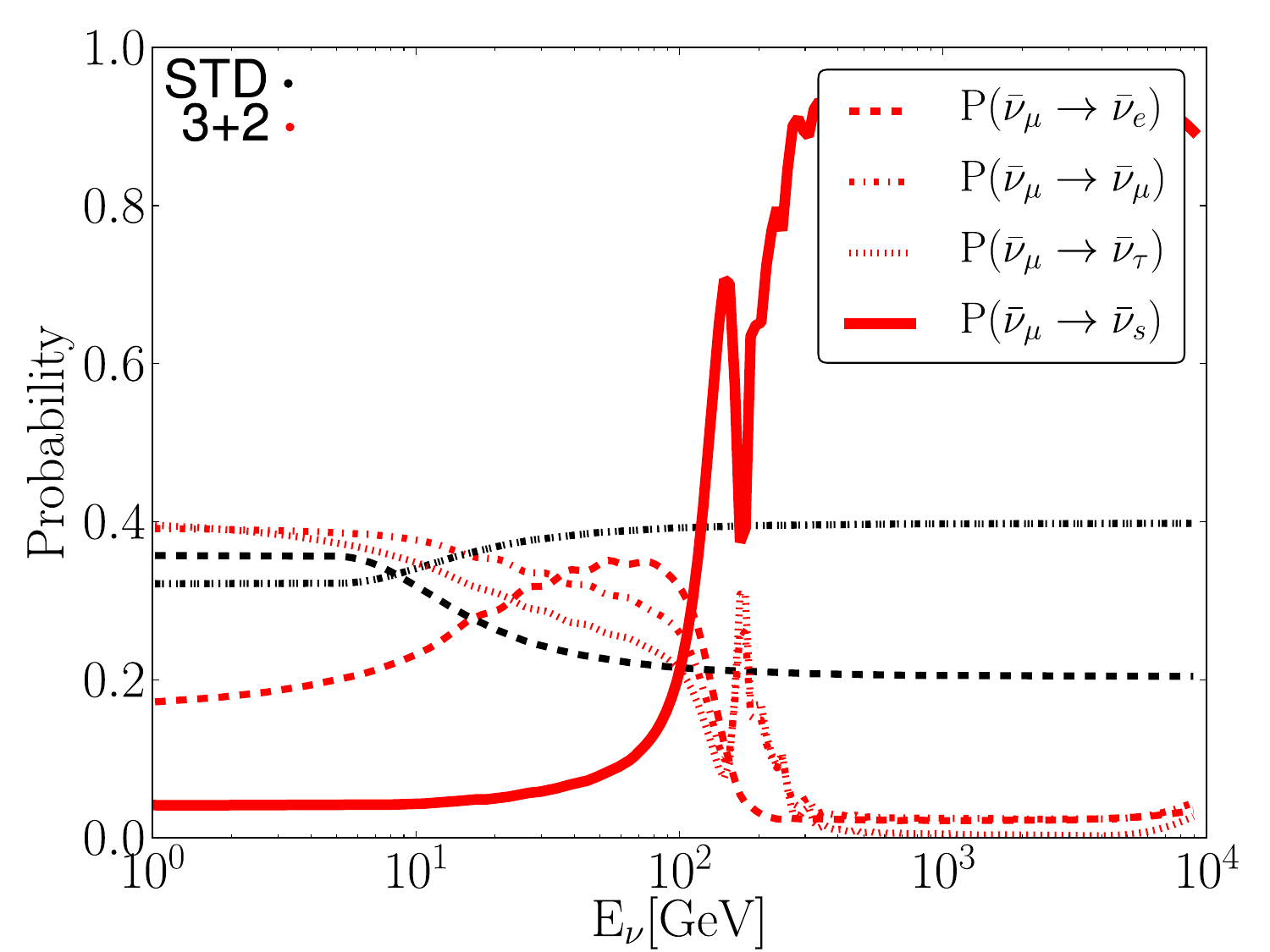} &
      \includegraphics[width=5.5cm]{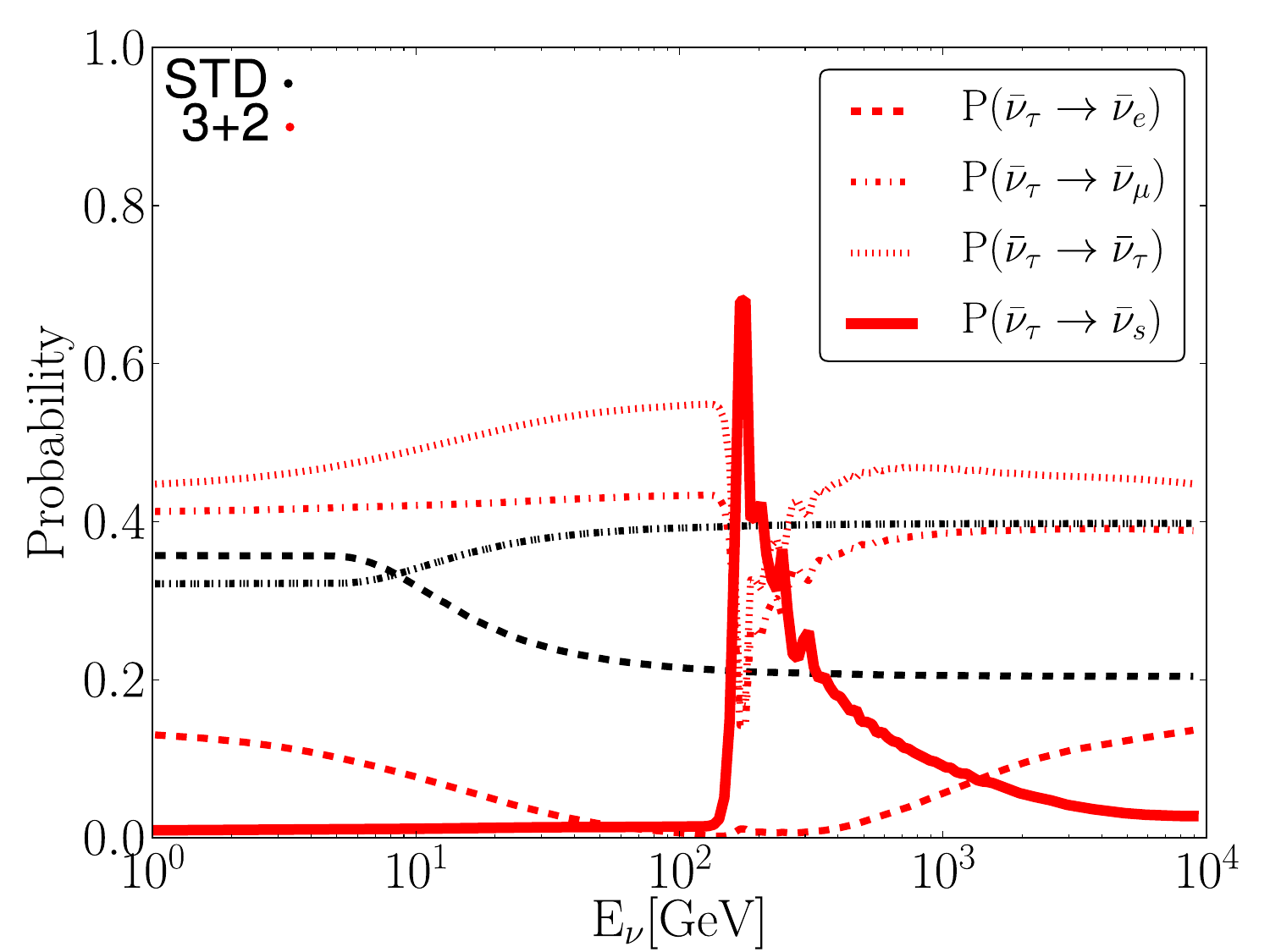} \\
    \end{tabular}
  \end{center}
  \caption{Flavor transition probabilities in the Sun as a function of the neutrino
    energy for an initial $\nu_e$ (left), an initial $\nu_\mu$ (center), and an
    initial $\nu_\tau$ (right). The top plots are for neutrinos, the bottom
    ones are for antineutrinos. Black lines are for standard three-flavor
    oscillation, red lines are for a ``$3+2$'' model with two sterile
    neutrinos, and with mixing parameters determined from the global fit in
    section~\ref{sec:KMS}.
    Absorption and $\tau$ regeneration effects are neglected in these plots.
    Note that the black dotted lines ($\nu_x \to \nu_\tau$ in the SM) and the
    black dot-dashed lines ($\nu_x \to \nu_\mu$ in the SM) lie on top of each
    other since $\nu_\mu$--$\nu_\tau$ mixing is assumed to be maximal.  Plots
    taken from ref.~\cite{Arguelles:2012cf}, see~\cite{Esmaili:2012ut} for a
    similar study.}
  \label{fig:snu-dm-prob-3+2}
\end{figure}

To illustrate the impact of active-to-sterile oscillations on neutrinos from
dark matter annihilation, we show in fig.~\ref{fig:snu-dm-prob-3+2} the flavor
transition probabilities in the Sun as a function of energy (thick red lines).
Here, a scenario with two sterile neutrinos, and with mixing parameters
determined from the global fit in section~\ref{sec:KMS} has been assumed.  We 
see that, for heavy dark matter whose
annihilation leads to very hard neutrino spectra, almost all the $\nu_e$,
$\bar\nu_\mu$, and a significant fraction of $\bar\nu_\tau$ could be converted
into sterile states. Note that, for sterile neutrino models with different
parameter sets, the effects could be even larger.

To study how active--sterile flavor transitions would affect the actual limits
that a neutrino telescope can set on the dark matter parameter space, it is useful
to perform a complete simulation of neutrino propagation from the center of
the Sun to a terrestrial detector, and to compute the ratios of expected signal
events in a sterile neutrino model and in the standard oscillation framework.
In doing so, it is important to include not only oscillations, but also
neutrino absorption, scattering, and the generation of secondary neutrinos
from $\tau$ lepton decay (see~\cite{Arguelles:2012cf} for details on the
simulations used here).

\begin{figure}[tbp]
  \centering
  \includegraphics[width=10cm]{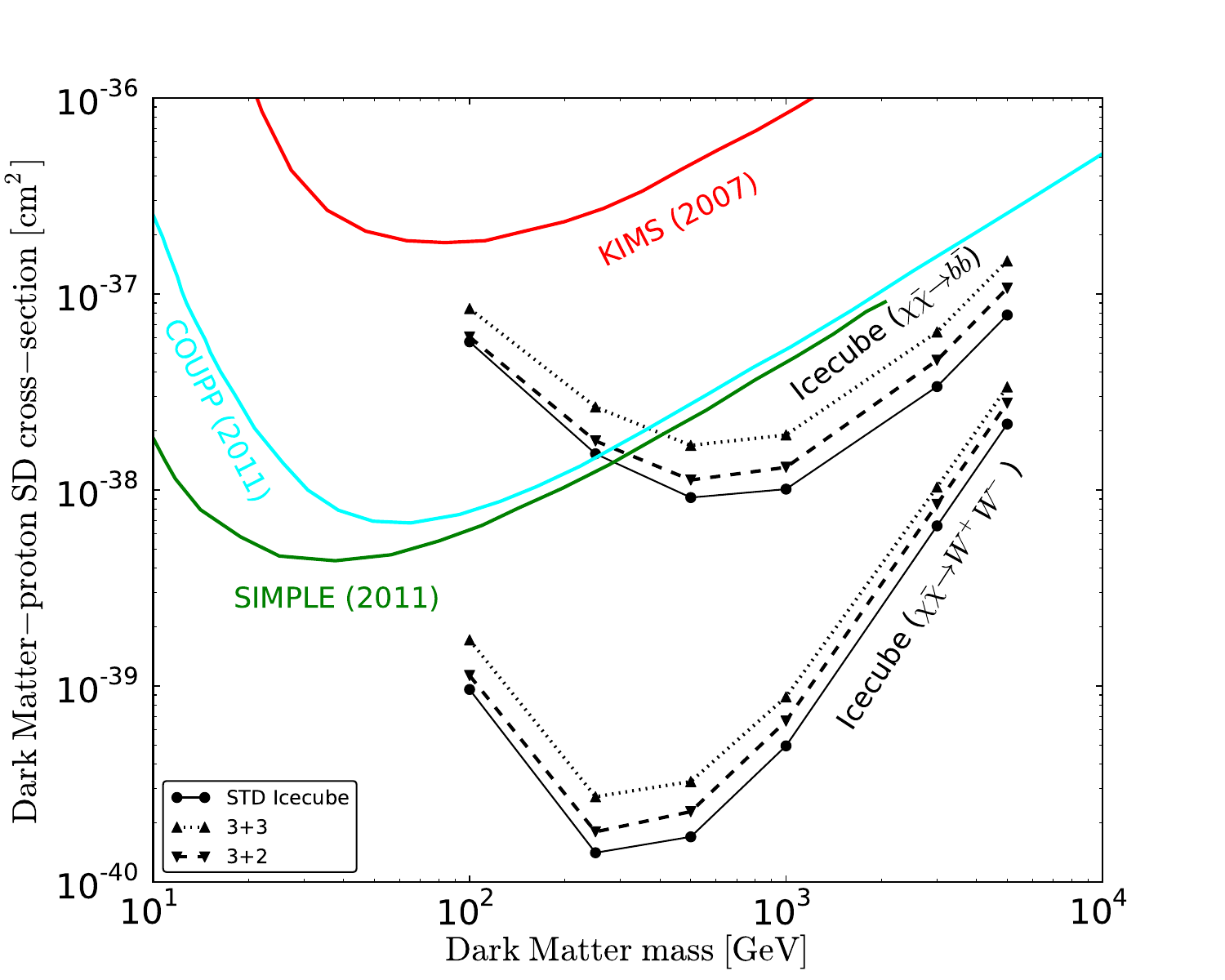}
  \caption{IceCube limits on spin-dependent dark matter-proton
    scattering~\cite{IceCube:2011ae,IceCube:2011aj} in scenarios with (black
    dashed and dotted lines) and without (black solid lines) sterile neutrinos
    compared to data from direct detection experiments (colored
    lines)~\cite{Lee:2007qn, Behnke:2010xt, Felizardo:2011uw}. We
    see that for the $3+2$ scenario which provides the best fit to short
    baseline neutrino oscillation data, the limits are only moderately
    weakened. A $3+3$ toy model, on the other hand, illustrates that larger
    modifications are possible. Plot based on ref.~\cite{Arguelles:2012cf}.}
  \label{fig:snu-dm-IC-limit}
\end{figure}

Two illustrative examples are shown in fig.~\ref{fig:snu-dm-IC-limit}, where we
compare the IceCube limit on the spin-dependent dark matter--nucleon scattering
cross sections obtained under the assumption of standard 3-flavor oscillations
to the limits that would be obtained if the best fit $3+2$ model from 
section~\ref{sec:KMS} is realized in nature
(dashed black lines). We also show the results that would be obtained in a toy
model with three sterile neutrinos, each of which mixes with only one of the
active neutrinos, and with the corresponding mixing angles being equal (black
dotted lines, see ref.~\cite{Arguelles:2012cf} for details). We show results
for two different dark matter annihilation channels, with $b\bar{b}$ and $W^+
W^-$ final states, respectively.  For comparison, we also show results from
various direct dark matter searches. We observe that IceCube limits are
weakened by an $\mathcal{O}(1)$ factor in the $3+2$ model, and by roughly a
factor 2 in the $3+3$ model.

\begin{figure}
  \begin{center}
    \includegraphics[width=12cm]{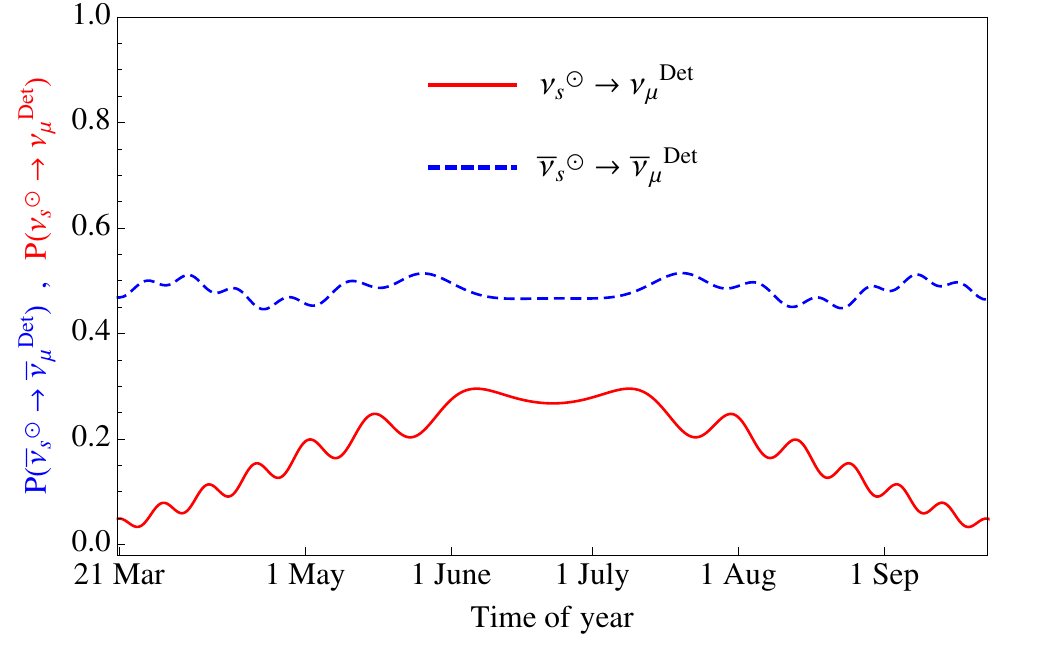}
  \end{center}
  \caption{Seasonal variation of the oscillation probability for neutrinos
  from the dark matter annihilation channel $\chi \bar{\chi} \to \nu_s \bar{\nu}_s$.
  A neutrino energy of 400~GeV was assumed here and the sterile neutrino mixing
  parameters are $\Delta m_{41}^2 = 1$~eV$^2$, $\sin^2\theta_{14} = \sin^2\theta_{24}
  = \sin^2\theta_{34} = 0.008$. Plot taken from~\cite{Esmaili:2012ut}.}
  \label{fig:snu-dm-IC-seasonal}
\end{figure}

Another interesting annihilation channel for dark matter particles in the
presence of sterile neutrinos is $\chi \bar{\chi} \to \nu_s \bar{\nu}_s$, which
leads to monochromatic sterile neutrinos. For the low mass DM particles
($m_\chi \lesssim 100$~GeV) this channel of annihilation is almost inaccessible
to IceCube. However, thanks to the MSW resonant flavor conversion mentioned in
this section, for the $m_\chi \gtrsim 100$~GeV, the sterile neutrinos
propagating to the surface of the Sun could convert to active neutrino which
can be detected in neutrino telescopes.  Moreover, the annual variation of the
Earth--Sun distance as well as MSW-enhanced oscillation effects inside the
Earth can lead to a seasonal variation in the count rate~\cite{Esmaili:2009ks,
Esmaili:2010wa, Esmaili:2012ut}.  As an illustrative example,
Fig.~\ref{fig:snu-dm-IC-seasonal} shows the oscillation probabilities $\nu_s
(\bar{\nu}_s)\to \nu_\mu (\bar{\nu}_\mu)$ from the production point in the
center of Sun to the IceCube detector at the South Pole. As can be seen, the
MSW indued conversion results in a nonzero, and time-dependent, oscillation
probability.

\subsubsection{Sterile neutrinos and direct dark matter searches}
\label{sec:snu-dm-direct}

It has been recognized for a long time that dark matter detectors sensitive to
$\mathcal{O}(\text{keV})$ electron recoils and nuclear recoils will become
sensitive to solar neutrino interactions once they reach the ton
scale~\cite{Gutlein:2010tq}. In extensions of the standard model, low-energy
neutrino interactions can be considerably enhanced by new physics, making dark
matter detectors unique tools for probing such types of new
physics~\cite{Pospelov:2011ha, Harnik:2012ni, Pospelov:2012gm}.

Particularly appealing models involve light ($\ll 1$~MeV) sterile neutrinos,
which are singlets under the standard model gauge group, but are charged under
a new $U(1)'$ gauge group.  (Instead of introducing a new gauge group one could
also postulate that neutrinos carry a large magnetic moment,
see~\cite{Harnik:2012ni} for details on this possibility.) The $U(1)$' gauge
boson $A'$ is also assumed to be light ($M_{A'} \ll 1$~GeV) and very weakly
coupled to standard model particles to avoid constraints. The new gauge group
could, for instance, correspond to gauged baryon number~\cite{Pospelov:2011ha,
Harnik:2012ni, Pospelov:2012gm} so that also standard model quarks are charged
under it.  This model would lead to sterile neutrino--nucleus scattering in a
dark matter detector, whereas sterile neutrino--electron scattering would be
absent at tree level. Alternatively, the coupling of standard model particles
to the $U(1)'$ gauge boson could be exclusively through a kinetic mixing term
of the form $-\frac{1}{2} \epsilon F_{\mu\nu} F'^{\mu\nu}$, where $F_{\mu\nu}$
and $F'^{\mu\nu}$ are the field strength tensors of the photon and the $A'$
boson, respectively. In this case, sterile neutrinos can scatter both on
electrons and on nuclei in a detector.

The differential $A'$-mediated sterile neutrino--electron scattering cross
section is given by
\begin{align}
  \frac{d\sigma_{A'}(\nu e \to \nu e)}{dE_r} =
      \frac{g_{\nu_s}'^2 g_e'^2 m_e}{4 \pi p_\nu^2 (M_{A'}^2 + 2 E_r m_e)^2}
      \big[ 2 E_\nu^2 + E_r^2 - 2 E_r E_\nu - E_r m_e - m_\nu^2 \big] \,,
  \label{eq:snu-dm-dsigmadEr}
\end{align}
where $g_e'$ and $g_{\nu_s}'$ are the $A'$ couplings to sterile neutrinos and
electrons, respectively, $E_\nu$, $p_\nu$ and $m_\nu$ are the neutrino energy,
momentum and mass, respectively, $E_r$ is the observed electron recoil energy,
and $m_e$ is the electron mass.  We notice that, for small gauge boson mass
$A'$, the denominator of the gauge boson propagator is dominated by the
momentum transfer, so that the cross section for low-$E_r$ recoils drops as
$1/E_r^2$. Thus, significant scattering rates can be obtained in low-threshold
dark matter detectors, whereas dedicated neutrino experiments, which have
energy thresholds of at least few hundred keV, would be insensitive (see
refs.~\cite{Pospelov:2011ha, Harnik:2012ni, Pospelov:2012gm} for detailed
studies of constraints on the models discussed here).
Eq.~\eqref{eq:snu-dm-dsigmadEr} can be readily generalized to the case of
sterile neutrino--nucleus scattering by replacing $g_e'$ by the $A'$ coupling
to nucleons and $m_e$ by the nuclear mass, and by including a coherence factor
$A^2$ (where $A$ is the nucleon number of the target nucleus) and a nuclear
form factor $F^2(E_r)$.

Sterile neutrinos can be produced by oscillations of solar neutrinos or,
if they are too heavy to oscillate, by incoherent production of the heavy mass
eigenstate due to its (mxing angle-suppressed) coupling to the $W$.

\begin{figure}
  \begin{center}
    \includegraphics[width=8cm]{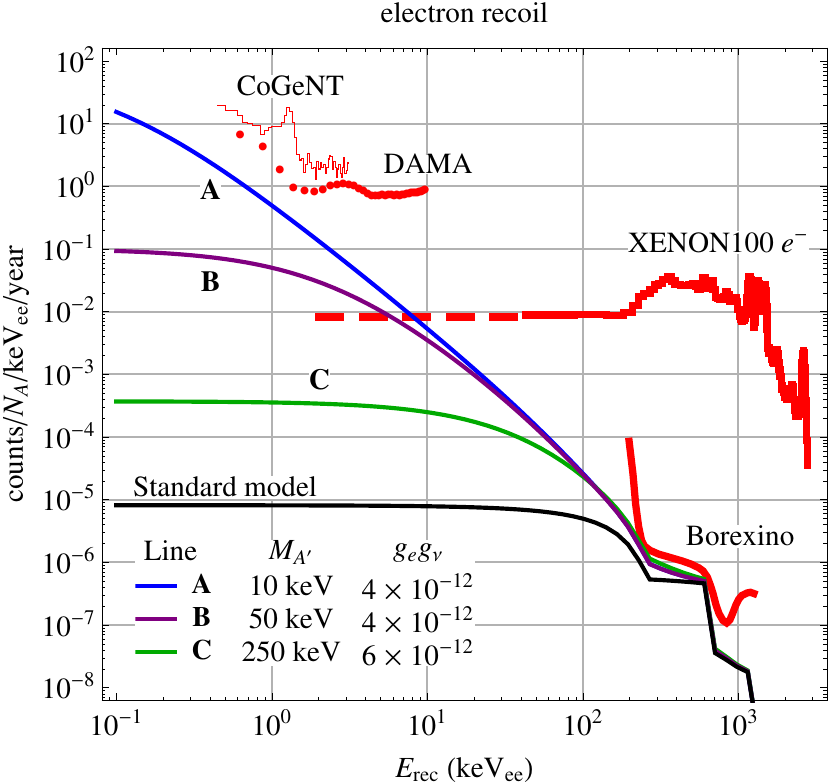}
    \includegraphics[width=8cm]{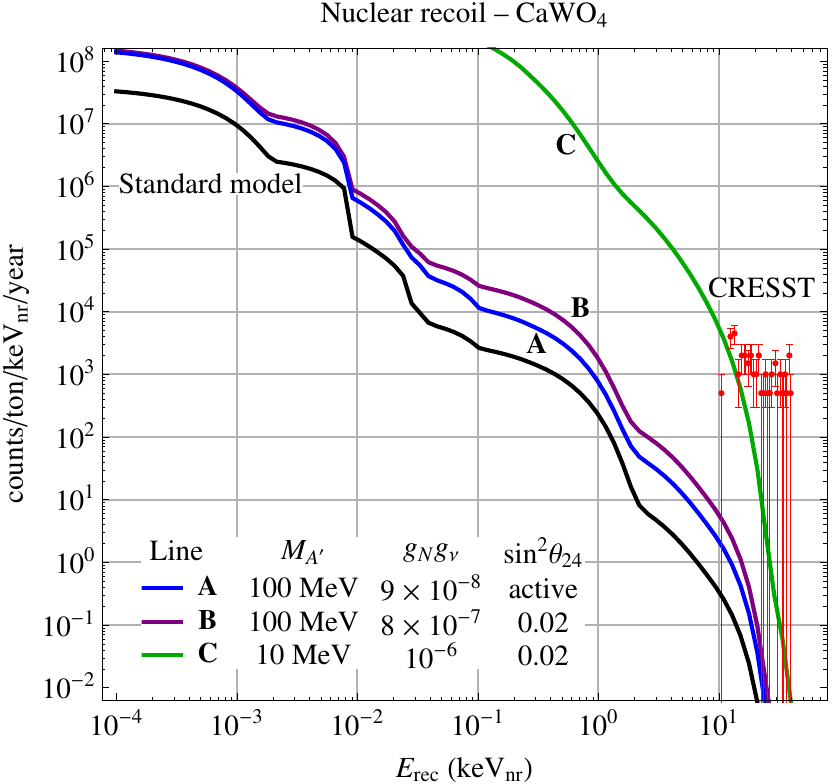}
  \end{center}
  \caption{Expected event spectra in a dark matter detector from
    scattering of solar neutrinos on electrons (left) and on nuclei (right, for
    the specific case of a CaWO$_4$ detector) mediated by a light new gauge
    boson with couplings $g_e$ ($g_N$) to electrons (nucleons) and $g_\nu$ to
    neutrinos. The different colored curves correspond to different new physics
    models and parameter points: In the left-hand plot, a model with one
    sterile neutrino charged under a $U(1)'$ gauge group was assumed, and the
    coupling of the $A'$ boson to standard model particles was assumed to occur
    through kinetic mixing with the photon. In the right-hand plot, curve A is
    for a model with only active neutrinos and with scattering mediated by a
    $U(1)_{B-L}$ (baryon number minus lepton number) boson, and curves B and C
    are for a model with gauged baryon number, and a sterile neutrino charged
    under it.  The black curves shows the standard model rate, and the red
    curves and data points show the observed electron recoil rates in
    XENON-100~\cite{Aprile:2011vb}, Borexino~\cite{Bellini:2011rx},
    CoGeNT~\cite{Aalseth:2011wp}, and DAMA~\cite{Bernabei:2008yi}. (Note that
    CoGeNT and DAMA cannot distinguish nuclear recoils from electron recoils,
    so their data can be interpreted as either.) Plots based on
    ref.~\cite{Harnik:2012ni}.}
  \label{fig:snu-dm-rate}
\end{figure}

In fig.~\ref{fig:snu-dm-rate}, we show examples of neutrino--electron and
neutrino--nucleus scattering rates in models with light new gauge boson.  The
model parameters we have chosen for the different colored curves are still
allowed by current bounds~\cite{Harnik:2012ni}.  The black curves show the
standard model rates, whereas in red we show the data from various dark matter
experiments and from Borexino. Note that for electron recoils in Xenon-100, it
is not entirely clear where the detection threshold is (electron recoils are
not important for Xenon-100's dark matter search except as a background), and
this uncertainty is roughly indicated by a dashed line in
fig.~\ref{fig:snu-dm-rate}.  The kinks in the solar neutrino event spectra
shown in the plots arise from the kinematic edges of the various nuclear
reactions contributing to the solar neutrino spectrum.  We see that a
substantial enhancement of the event rate compared to the standard model is
easily possible, making present and future dark matter detectors very sensitive
to models with sterile neutrinos and light new gauge bosons. We also see that
it is even possible to \emph{explain} some of the anomalous event excesses
observed by some dark matter experiments (CoGeNT, CRESST) in terms of sterile
neutrino interactions. On the other hand, there is also a danger that a sterile
neutrino signal is misinterpreted as a dark matter signal.

One of the most robust signatures of dark matter in a low-threshold detector
is an annual modulation of the nuclear recoil rate~\cite{Freese:1987wu},
which is induced by the varying velocity of the Earth with respect to the
Mily Way's dark matter halo throughout the year. Depending on the dark matter
mass, the scattering rate is expected to peak either in early June or in early
December. Indeed, the DAMA experiment reports a statistically significant anual
modulation in the observed event rate peaking in summer~\cite{Bernabei:2008yi,
Bernabei:2010mq} (see, however, \cite{Kudryavtsev:2009gd, Ralston:2010bd,
Nygren:2011xu, Blum:2011jf, Bernabei:2012wp} for a discussion of systematic
effects that could cause this modulation). In view of this, it is interesting
to observe that the sterile neutrino scattering rate in the models introduced
above can \emph{also} show annual modulation.

There are several physical effects that can lead to this.  First, the
ellipticity of the Earth's orbit around the Sun imply variations in the
Earth--Sun distance, which in turn lead to a stronger geometric suppression of
the flux in summer than in winter, with the extrema of the modulation occuring
around January 3rd and July 4th, respectively. This phase differs by about one
month from the one expected from heavy dark matter, and by almost half a year
from the phase observed in DAMA, which is consistent with light dark matter
scattering.

\begin{figure}
  \begin{center}
    \includegraphics[width=0.9\textwidth]{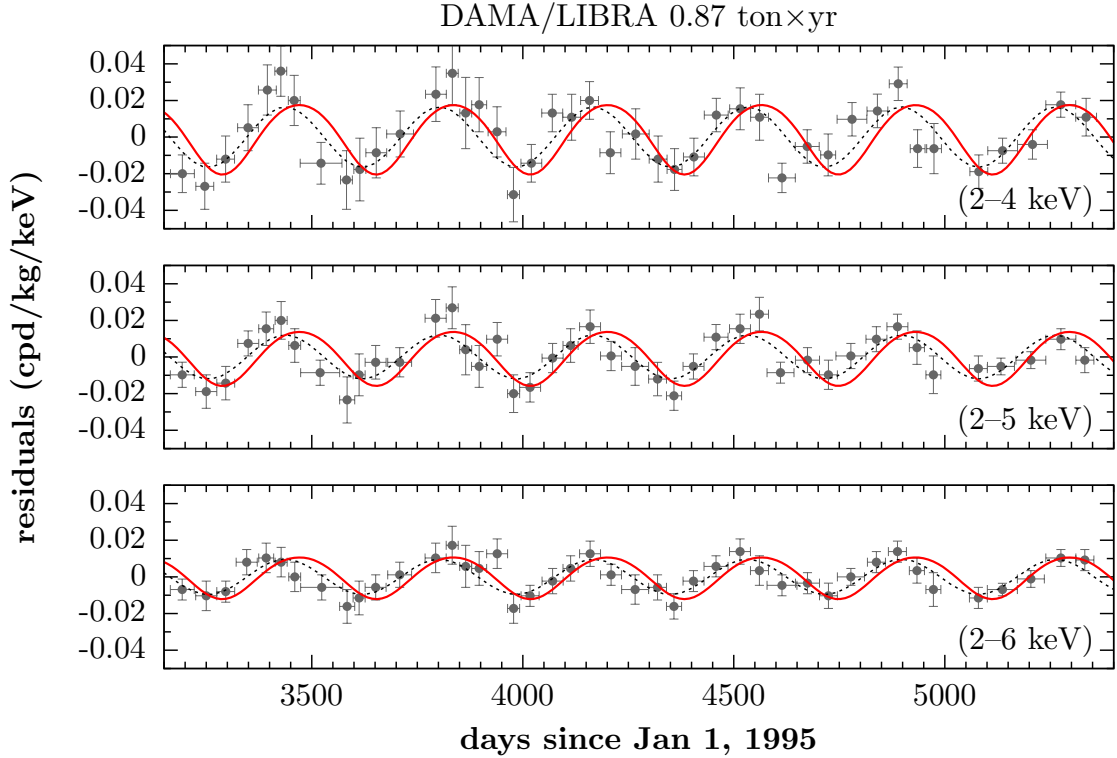}
  \end{center}
  \caption{Comparison of DAMA/LIBRA data~\cite{Bernabei:2010mq} (data points
    with error bars) to the predictions of a model with a sterile neutrino
    charged under gauge baryon number (solid red line). Also shown is a simple
    sinusoidal fit to the data, assuming a period of one year and a phase as
    expected from the scattering of light dark matter. Plot taken from
    ref.~\cite{Pospelov:2012gm}.}
  \label{fig:snu-dm-dama}
\end{figure}

The modulation of the sterile neutrino flux can be \emph{reversed} if
oscillations are taken into account~\cite{Pospelov:2011ha, Harnik:2012ni,
Pospelov:2012gm}.  Consider in particular scenarios where sterile neutrinos are
produced through vacuum oscillations of active neutrinos, with an oscillation
length somewhat smaller than one astronomical unit (this requires a $\Delta m^2
\sim 10^{-10}$~eV$^2$~\cite{Harnik:2012ni}). In this case, the oscillation
probability can increase with increasing Earth--Sun distance, and this can
overcompensate for the larger geometric suppression at larger distance. An
illustration of such a scenario is shown in fig.~\ref{fig:snu-dm-dama}, where
the predicted residual sterile neutrino scattering rate in DAMA (after
subtracting the time-averaged rate) in several energy bins is shown as a
function of time and compared to the experimental data
from~\cite{Bernabei:2010mq}.  We observe that the observed modulation amplitude
can be well reproduced, whereas the phase differs from the observed one by
about one month (on average roughly one time bin).

Another possible sources of modulation in the sterile neutrino flux are Earth matter
effects~\cite{Harnik:2012ni}.  The new gauge boson mediating the sterile
neutrinos scattering can also lead to new MSW matter effects if ordinary matter
carries some net $U(1)'$ charge (as is the case, for instance, in the gauged
baryon number model mentioned above).  In this case, active--sterile
oscillations in the Earth can lead to a day--night asymmetry in the flux, and
since nights are shorter in summer than in winter, this will also induce an
annual modulation. If the new interaction of sterile neutrinos with matter is
\emph{very} strong, it is even imaginable that the Earth becomes partially
opaque to sterile neutrinos.

In conclusion, if the ``hidden sector'' of particle physics contains at least
one sterile neutrino \emph{and} a new light gauge boson, coupled only weakly to
standard model particles, an interesting and rich phenomenology arises in dark
matter detectors. In particular, sterile neutrinos from the Sun can lead to
strongly enhanced electron and/or nuclear recoil rates, and they can even lead
to a temporal modulation of this signal not too different from what is expected
from dark matter.  Therefore, sterile neutrino recoils can on the one hand be
used to detect or constrain such models, but they may also be problematic if
they can be confused with a dark matter signal.

\subsection{Brief Summary}
\label{sec:summary}

Several anomalies have been presented in Section~\ref{sec:oscillation} In 
this Section we present studies coming to the conclusion that, even though 
the three neutrino scheme is unable to explain the data, augmenting the
standard model by one (``3+1'') sterile neutrino with a mass at the eV
scale is not sufficient to well describe all data (Sections
\ref{sec:giunti-global}, \ref{sec:KMS}, \ref{sec:Karagiorgi}). Adding
two eV-scale sterile neutrinos (``3+2'') is in better agreement with the
data though severe tension remains even in this case.

We can summarize the situation in the following way:
\begin{itemize}

\item Three anomalies affect the oscillation data: LSND (about 3.3
  $\sigma$), The Reactor Antineutrino Anomaly (about 3 $\sigma$), and the
  Gallium Anomaly (about 2.5 to 3 $\sigma$).

\item
Despite several hints pointing towards sterile neutrinos around the eV
scale, no fully consistent picture has emerged so-far. New
experimental data are mandatory to understand the origin of the
anomalies.

\item
One important reason for the tension in the global data including the LSND
evidence is the non-observation of $\nu_\mu$ disappearance at the eV-scale,
which is a generic prediction of an explanation of the LSND signal in terms
of sterile neutrino oscillations.

\item
If the LSND signal is discarded the hints for sterile neutrinos from the
reactor and Gallium anomalies remain. These hints are related to $\bar\nu_e
(\nu_e)$ disappearance and do not require $\nu_\mu$ disappearance.

\item
The $\Delta m^2$ values implied by global fits are in tension with
constraints on the sum of neutrino masses from cosmology, assuming a
standard cosmological scenario.

\item
The presence of additional neutrino mass states can affect the
interpretation of beta decay and neutrinoless double beta decay
experiments. This is studied in subsection~\ref{sec:beta}.

\item
The impact of an eV sterile neutrino on atmospheric neutrino data 
in Ice Cube is presented in subsection~\ref{sec:IceCube}.

\item
Finally, the existance of light sterile neutrinos can impact existing and 
future dark matter searches.  This is discussed in 
subsection~\ref{sec:SNDarkMatter}.

\end{itemize}


\clearpage
\section{Requirements for Future Measurements}
\label{sec:requirements}

\subsection{Historical Precedent}

As the community proceeds forward in answering the question to the existence of sterile neutrinos (or equivalent mechanism which would be consistent with the observed phenomena), a natural question rises to the surface: {\it What criteria would need to be met for the community to be convinced of their existence?}  Clearly, a mere statistically significant signal may not be sufficient, as even the present situation seems to suggest.  Rather, a robust sequence of measurements, both testing the anomalies with as little model dependence as possible (re-doing the measurements but with definitive outcomes) and testing the anomalies in different ways (different L and E) to be able to best interpret the results. 

Fortunately, we may draw on historical precedent to help guide us.  Let us examine, for example, the acceptance of neutrino oscillations as the proper physical phenomenon behind solar neutrino measurements.  Such a conclusion was not attained immediately by a single measurement, but was rather the culmination of almost thirty years of scientific investigation.  Let us take the case of solar neutrinos as a concrete example.  The progression over time suggests improvements not just in statistical accuracy, but in approach as well.  Initial measurements with Ray David Jr. at Homestake indicated a discrepancy in the observed rate only.  Follow-up experiments with SAGE, GALLEX and GNO, occurring at a slightly lower energy threshold and different targets, continued to confirm the discrepancy.  However, all these experiments were confined to a single channel (charged current scattering), with little spectral information.  In the advent of experiments such as Kamiokande and Super-Kamiokande, real-time measurements with an alternate channel (elastic scattering) provided stronger evidence of neutrino oscillations.  With the appearance of the Sudbury Neutrino Observatory (SNO), one finally had multiple channels accessible within the same experiment (CC, ES, and NC), effectively making SNO both a disappearance {\it and} an appearance experiment.  The last nail in the coffin came from an orthogonal confirmation using reactor neutrinos.  KamLAND was able to confirm that the mechanism responsible for these observations was indeed neutrino oscillations, and further improve on the determination of the parameters.  To date, neutrino oscillations are confirmed at the 13 sigma level.  In truth neutrino oscillations and mass stand as the first and only definitive evidence in contradiction to the standard model of nuclear and particle physics, yet the neutrino oscillation model is now so well established that many view it as a part of a revised standard model.

If we now look at the sterile neutrino sector, we see a similar history unfolding.  The LSND experiment stands perhaps as the first possible evidence, showing an excess $\bar{\nu}_e$ signal indicative of the existence of one or more sterile neutrinos.  The nature of the signal, however, is limited to rate (with some energy information).  The MiniBooNE experiment does not see direct evidence in $\nu_e$ appearance, though a low energy excess in their data could potentially accommodate the existence of sterile neutrinos.  Their antineutrino data, though statistically limited, appears to show consistency with the original LSND signal. The latest results from a re-analysis of the neutrino reactor data, which reflects a rate dependence as well, appears consistent with the sterile oscillation picture.  Finally, calibration data from Gallex and Sage also suggest a deficit which hints at the existence of sterile neutrinos.  

To proceed forward with a program whose goal is to establish or refute the existence of sterile neutrinos, multiple and possibly orthogonal approaches in the same spirit as employed for neutrino oscillations is warranted.  Current and future short-baseline experiments, such as MiniBooNE, BooNE, MicroBooNE, and possible follow-ons to MicroBooNE, readily provide measurements of the energy and rate dependence using charged current reactions.  There will also be in the near future an expansion of the reactor measurements to provide a charged current measurement of the length dependence of the oscillation formula.  

New on the scene is the emergence of source-based measurements, whereby strong radioactive isotopes can be generated to provide a mono-chromatic neutrino source within a single detector.  Such experiments provide a test of the {\em length} dependence of the oscillation formula.  For large detectors, one can explore this for charged current and elastic scattering interactions.  If one is able to push to very low thresholds, neutral current coherent scattering may also be accessible, providing the most direct test of sterile neutrino disappearance.  More importantly, just as was done in the solar sector, multiple observations with several different channels will provide robustness to the sterile neutrino interpretation.  Experiments in cosmology and direct neutrino mass measurements may finally provide the final orthogonal check on this physics.

\begin{table}[htdp]
\caption{List of past experiments able to probe neutrino oscillations from the solar sector $(\theta_{12})$.  Experiments broken down in terms of incoming flux (solar or reactor), how they are sensitive to oscillations (total rate, energy spectrum, or both) and the reaction channel probed (either charged current (CC), neutral current (NC), or elastic scattering (ES)).}
\begin{center}
\begin{tabular}{|l|c|c|c|}
\hline
Experiment(s) & Source & Sensitivity to Oscillations & Channel\\
\hline
Homestake & Solar & Total Rate & CC \\
SAGE/GALLEX/GNO & Solar & Total Rate & CC\\
Kamiokande/Super-Kamiokande & Solar, Rate & Energy & ES \\
SNO & Solar & Energy, Rate & CC, ES, NC \\
KamLAND & Reactor & Energy, Rate & CC \\
Borexino & Solar & Energy, Rate & ES, CC \\
\hline
\end{tabular}
\end{center}
\label{tab:solar}
\end{table}%

\begin{table}[htdp]
\caption{List of past and future planned experiments able to probe oscillations to sterile neutrinos.  Experiments broken down in terms of type of experiment (decay-at-rest, short baseline, reactor, etc.) how they are sensitive to oscillations (total rate, energy spectrum, and/or length dependence) and the reaction channel probed (either charged current (CC), neutral current (NC), or elastic scattering (ES)). Past experiments are denoted by $\dagger$.\label{tab:sterile}}
\begin{center}
\begin{tabular}{|l|l|c|l|c|}
\hline 
Experiment(s) & Source & Type & Sensitivity to Oscillations & Channel \\
\hline
LSND$^\dagger$ & Decay-at-rest & $\bar{\nu}_\mu \rightarrow \bar{\nu}_e$ & Total Rate, Energy & CC \\
MiniBooNE$^\dagger$ & Short baseline & $\bar{\nu}_\mu \rightarrow \bar{\nu}_e$ & Total Rate, Energy & CC \\
Reactor measurements$^\dagger$ & Reactor & $\bar{\nu}_e$ dis. & Total Rate & CC \\
Gallium Anomaly$^\dagger$  & EC Source & $\nu_e$ dis. & Total Rate & CC \\
\hline \hline
Future Decay-at-Rest  & & & & \\
({\it OscSNS, Super-K}) & Decay-at-rest & $\bar{\nu}_{\mu} \rightarrow \bar{\nu}_e$ & Total Rate, Energy & CC \\
 & & $\nu$ dis. & Total Rate & NC \\
\hline
 Future Short Baseline & & & & \\
 ({\it $\mu$BooNE, BooNE, NESSiE, LArLAr}) & Short baseline & 
 $\boss{\nu}{\mu} \rightarrow \boss{\nu}{e}$ & Total Rate, Energy & CC \\
 & & $\boss{\nu}{\mu}$ dis. & Total Rate, Energy & CC \\
 ({\it VLENF}) & Short baseline & $\boss{\nu}{e} \rightarrow \boss{\nu}{\mu}$ & Total Rate, Energy & CC \\
 & & $\boss{\nu}{e},\,\boss{\nu}{\mu}$ dis. & Total Rate, Energy & CC, NC \\
\hline
Future Reactor Measurements & & & & \\
({\it Nucifer, SCRAAM, Stereo}) & Reactor & $\bar{\nu}_e$ dis. & Total Rate, Length & CC \\
\hline
Future Source Experiments & & & & \\
({\it Borexino, Ce-LAND, Daya Bay}) & $\beta^-$ Source & $\bar{\nu}_e$ dis. & Total Rate, Length & CC \\
({\it Borexino, SNO+Cr}) & EC Source & $\nu_e$ dis. & Total Rate, Length & ES \\
({\it LENS, Baksan}) & EC Source & $\nu_e$ dis. & Total Rate, Length & CC \\
({\it  RICOCHET}) & EC Source & $\nu_e$ dis. & Total Rate, Length & NC \\
\hline
\end{tabular}
\end{center}
\end{table}%

\subsection{Requirements for a Future Sterile Program}

\begin{itemize}

\item {\bf\emph{Multiple Approaches:}}\\  Given the potential implications of sterile neutrinos, it is important to confirm their existence in multiple (preferably orthogonal) approaches.  A few of such approaches are already being sought by the experimental community, such as short/long baseline experiments, reactor experiments, neutrino source experiments, cosmology, and beta decay.  Support on existing efforts should continue and new efforts should continue to be pursued.  Both experiments which repeat existing experiments but with technologies that allow for definitive measurements, and experiments which assume an L/E model but vary L and E will be needed to identity and interpret the results.

\item{\bf\emph{For Decay-at-Rest experiments:  A stopped pion source}}\\  A stopped pion neutrino source combined with a $\nu_e$ detection via inverse $\beta$-decay at a distance of about 30\,m is required. A key component will be to reduce both cosmogenic as well as beam-induced backgrounds. The reduction of cosmogenic backgrounds can be achieved by either a pulsed beam or sufficient overburden or a combination thereof. The reduction of beam-induced backgrounds requires the detector to not be downstream of the target.  This approach takes advantage of the philosophy of reducing model dependencies by repeating measurements but with an improved experimental setup.

\item{\bf\emph{For Short Baseline Experiments:  Two Detector Approach}}\\  An experiment's sensitivity to sterile neutrinos depends highly on the ability to model the flux and cross-section for neutrinos on various targets and to definitively identify outgoing particles.  Although improvements on understanding neutrino cross-sections is improving, it is likely not sufficient to definitely determine whether new physics may be at play.  A two-detector approach, already adopted by a number of experiments, will significantly increase the sensitivity to sterile neutrinos and should be adopted by any future sterile neutrino search.  An important feature of these new two detector approaches is to take advantage of precision detection techniques, such as Liquid Argon TPCs, in particular because of their ability to distinguish electrons from photons.  This approach takes advantage of the philosophy of reducing model dependencies by repeating measurements but with an improved experimental setup.

\item{\bf\emph {For Reactor Experiments:  Emphasis on Short Baseline Measurements}}\\  Recent re-analysis of reactor neutrino data indicates the possibility of an eV scale sterile neutrino at the 3$\sigma$ level.  Given the energy involved in such experiments, any sterile neutrino search would benefit greatly from measurements at very short baselines.  A program that constrains/measures the flux of neutrinos at very short (and possibly variable) baselines may provide further evidence and guidance for sterile neutrino oscillations. 

\item{\bf\emph {Source-Based Experiments: Development of Radioactive Sources}}\\  Source-based initiatives add a complementary approach to the reactor and short baseline program both because of the purity of their source and because of their probe of the length dependence (as opposed to energy dependence) of their signal.  As such, these programs are less susceptible to flux and cross-section uncertainties.  However, the production of megacurie neutrino sources depends on reactor facilities for irradiation, and many, that have been used in the past, are no longer in operation.  One the other hand, the kilocurie antineutrino sources come from fission fragments and can only be produced in coutries with existing fuel reprocessing.  Further, short half-lives and regulatory issues may make it difficult to transport these sources, especially across international borders.  In regions like North America, where work on large scale source technology is just starting, a robust R\&D program will be needed to mount such experiments.  We strongly encourage the development of such sources ($^{51}$Cr, $^{37}$Ar, $^{144}$Ce, etc.) for use in sterile neutrino experiments.

\item{\bf\emph {Source-Based Experiments: Multiple Channels}}\\  Source-based initiatives also benefit from multiple channel approaches (elastic, neutral current, and charged current scattering).  These channels may provide the necessary \emph{orthogonal} measurement to positively claim the existence of sterile neutrinos.  We strongly encourage that multiple channels for source-based initiatives be pursued.

\item{\bf\emph {Theoretical Framework: Global Analysis Approach}}\\  Given the complexity of data available, we encourage the development of a coordinated phenomenological approach in analyzing existing neutrino data (from both current and future experiments).  Such an approach should not be limited to fitting for a possible signal, but also the backgrounds and systematic uncertainties associated with each measurement.  Experiments should provide comprehensive information about their response function for use by the theoretical community.  We also encourage that \emph{all data}, including cosmological and beta decay measurements, also be incorporated in these fits.

\end{itemize}

\clearpage


%
\begin{appendix}
\section{Future Experiments}
\label{sec:future_exps}
This appendix contains a series of contributed descriptions of possible future experiments
and experimental approaches that may be sensitive to sterile neutrino oscillations.  The
projects described are in various stages of development from under construction to
conceptual.  

\begin{itemize}
\item{The first eight sections (\ref{LENS-Sterile} to \ref{SNO+Cr}) describe experiments using
radioactive neutrino sourecs.}

\item{Sections \ref{Yasuda} through \ref{NIST_Reactor} describe experiments using nuclear reactors 
as their neutrino source.}

\item{The experiments described in Sections \ref{OscSNS} and \ref{LSND_reloaded} are using stopped
pion beams to make a direct test of LSND.}

\item{Section~\ref{sec:Kaons} discusses the use of kaon decay at rest beams.}

\item{Sections \ref{subapp:MINOS+} to \ref{VLENF} cover experiments using decay in flight neutrino
beams.}

\item{And finally, Section~\ref{Ghoshal} describes the sensitivity of possible future
atmospheric neutrino detectors.}
\end{itemize}

\clearpage
\subsection{LENS-Sterile\footnote{Proposed by the LENS Collaboration \\ 
http://www.phys.vt.edu/\~kimballton/lens/public/collab/}}
\label{LENS-Sterile}

The LENS (for Low Energy Neutrino Spectroscopy) detector is intended to observe solar neutrinos in real time and as a function of energy.  In particular, LENS is designed to study the lowest energy neutrino from proton-proton (pp) fusion.  The unique capabilities of LENS not only open a definitive window on solar neutrinos, but also create opportunities for fundamental measurements of neutrino properties.  In particular, LENS, when paired with a mono-energetic radioactive neutrino source, like $^{51}$Cr, could undertake a meaningful search for large $\Delta m^2$ ($\mathcal{O}\sim 1$eV$^2$) sterile neutrinos~\cite{Grieb:2006mp}.

The design of the typical neutrino oscillation experiment is greatly complicated by the long distance required for the oscillation probability to reach the first maximum.  The relatively short baseline experiments LSND~\cite{Aguilar:2001ty} and MiniBooNE~\cite{AguilarArevalo:2008qa} are most sensitive to $\Delta m^2$ of about 1~eV$^2$.  For LSND this was achieved with a mean neutrino energy of $\sim$30~MeV and a baseline of ~30~m, while MiniBooNE used a mean neutrino energy of $\sim$500~MeV and a baseline of $\sim$500~m.  It then follows that an oscillation search using neutrinos with energies in the few hundred keV range, typical of radioactive decays, would require a baseline of only a few tenths of a meter to cover the same $\Delta m^2$ range.  Such a compact experiment would, for the first time, allow for the possibility of fully reconstructing the oscillation pattern and could potentially observe oscillation minima and maxima beyond the first.  

\subsubsection{LENS Technology}
Making an energy spectrum measurement on low energy neutrinos requires a low threshold charged current process and the capability to reject the large background from radioactive decays.  LENS uses the neutrino induced transition of $^{115}$In to an excited state of $^{115}$Sn:
\begin{eqnarray}
^{115}{\rm{In}} + \nu_e & \rightarrow & ^{115}{\rm{Sn}}^* + e^-\ (E = E_\nu - 114\ {\rm{keV}}) \ \ \ (\tau_{\rm{Sn}^*} = 4.76\ \mu s)\\
& &\hspace{0.8cm}\raisebox{0.52em}{$\mid$}\!\negthickspace\rightarrow\ ^{115}{\rm{Sn}} + \gamma\ (498\ {\rm{keV}}) + \gamma\ (116\ {\rm{keV}}) \nonumber 
\end{eqnarray}
to detect low energy neutrinos ($E_{\nu} > 114$~keV) and measure their energy.  The primary interaction and secondary $\gamma$ cascade make a triple coincidence, correlated in both time and space.  The detection medium in LENS is liquid scintillator (LS), chemically doped with natural indium which is 95.7\% $^{115}In$. To exploit the spatial correlation, the detector volume is segmented into 7.5~cm cubic cells by clear foils.  The foils have a lower index of refraction than the LS and so the scintillation light produced in a cell is channeled, by internal reflection, in the directions of the 6 cell faces.  The channeled light is read-out at the edge of the detector by photomultiplier tubes (PMTs).  This segmentation, know as the scintillation lattice, allows the position of each event to be determined in three dimensions with the precision of an individual cell.  The combination of the 4.76~$\mu$s mean delay and the spatial correlation of the primary electron and the two de-excitation $\gamma$s of known energies provides a sharp tag for neutrino interactions.  Radioactive backgrounds, including events from the $\beta$-decay of $^{115}$In, are rejected by a large factor ($10^{11}$ in the pp solar neutrino energy range) with an efficiency for neutrino interactions of greater than 85\% above 500~keV.    

\subsubsection{MCi Neutrino Source}
Radioactive neutrino sources (as opposed to antineutrino sources) involve either $\beta^+$-decay or electron capture.  Electron capture decays produce mono-energetic neutrinos, a unique property in all of experimental neutrino physics.  Mono-energetic neutrinos allow for a precise determination of $L/E$ and provide the means for a sharp signal tag.  Several electron capture sources have been proposed for the calibration of radiochemical solar neutrino experiments including $^{65}$Zn~\cite{Alvarez:1973}, $^{51}$Cr~\cite{Raghavan:1978}, $^{152}$Eu~\cite{Cribier:1987}, and $^{37}$Ar~\cite{Haxton:1988ee}.  Of the proposed source only $^{51}$Cr and $^{37}$Ar have ever been used.  Production of an $^{37}$Ar source requires a large fast breeder reactor, the reactor used to make the only MCi-scale $^{37}$Ar source, is no longer running, that leaves $^{51}$Cr as the best source candidate.  

$^{51}$Cr has a half-life of 27.7~days.  90.1\% of the time it decays to the ground state of $^{51}$V and emits a 751~keV neutrino while 9.9\% of the time it decays to the first excited state of $^{50}$V and emits a 413~keV neutrino followed by 320~keV $\gamma$.  $^{50}$Cr has a relatively high average thermal neutron capture cross section of 17.9~b which makes the large scale production of $^{51}$Cr straightforward.  Natural Cr is primarily $^{52}$Cr (83.8\%) and  contains 4.35\% $^{50}$Cr.  The isotope $^{53}$Cr, which is present in natural chromium at 9.5\%, has an average thermal neutron of 18.7~b, so when natural chromium is irradiated, $^{53}$Cr absorbs 2.5 neutrons to every one captured on $^{50}$Cr.  This significantly reduces the $^{51}$Cr yield.  Therefore enriched $^{50}$Cr must be used to eliminate $^{53}$Cr and to produce a compact target for irradiation.  The chromium enrichment for both Sage and Gallex was done by gas centrifuge at the Kurchatov Institute~\cite{Popov:1995} in Russia.  The material used by Gallex was enriched to 38.6\% in $^{50}$Cr while the Sage target was enriched to 92\%.  In both cases the remaining $^{53}$Cr content was less than 1\%.   

\subsubsection{LENS-Sterile Measurement}
\begin{figure}
\centerline{
\includegraphics[width=0.8\textwidth]{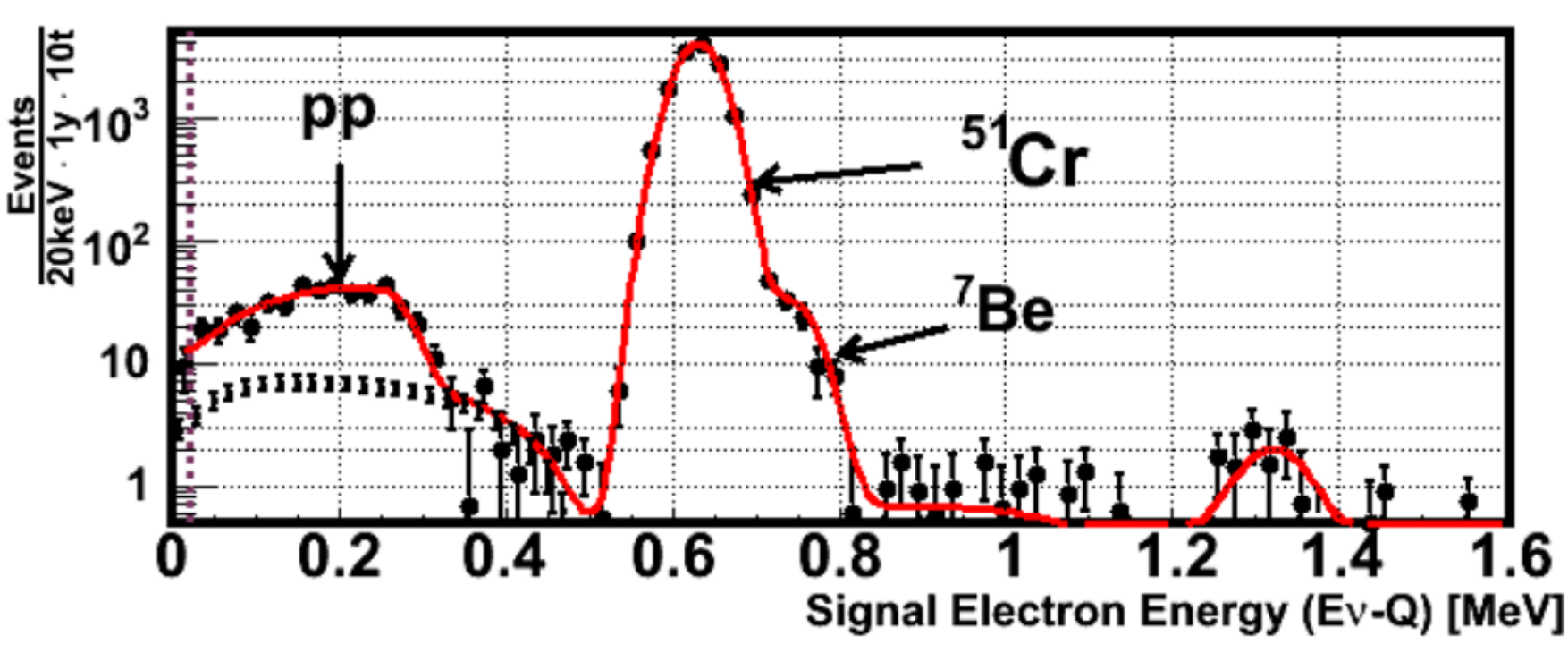}
}
\caption{\label{lens_source_spectrum}This log scale plot shows the 751~keV $^{51}$Cr neutrino peak superimposed on the solar neutrino spectrum.  The $^{51}$Cr signal represents 4 runs of a 10~MCi source each lasting 100~days, while the solar rate corresponds to 400~days.}
\end{figure}

The LENS-Sterile concept calls for placing a multi-MCi source in the center of the LENS detector, and counting the source neutrino interactions as a function of distance from the source, or radius.  By far the largest background under the 751~keV $^{51}$Cr peak comes from the $^7$Be solar neutrinos.  Figure~\ref{lens_source_spectrum} shows the $^{51}$Cr signal superimposed on the solar background.  The solar neutrino rate is down by two order of magnitude compared to the $^{51}$Cr signal.  In addition, will be well measured in LENS from the extended source-free running.  As our measurement is of a rate for a fixed energy, as a function of radius in a single detector, all other potential sources of systematic error generally cancel to first order.  Therefore the sensitivity is limited by statistics and should improve with additional source running.

\begin{figure}
\centerline{
\includegraphics[width=0.6\textwidth]{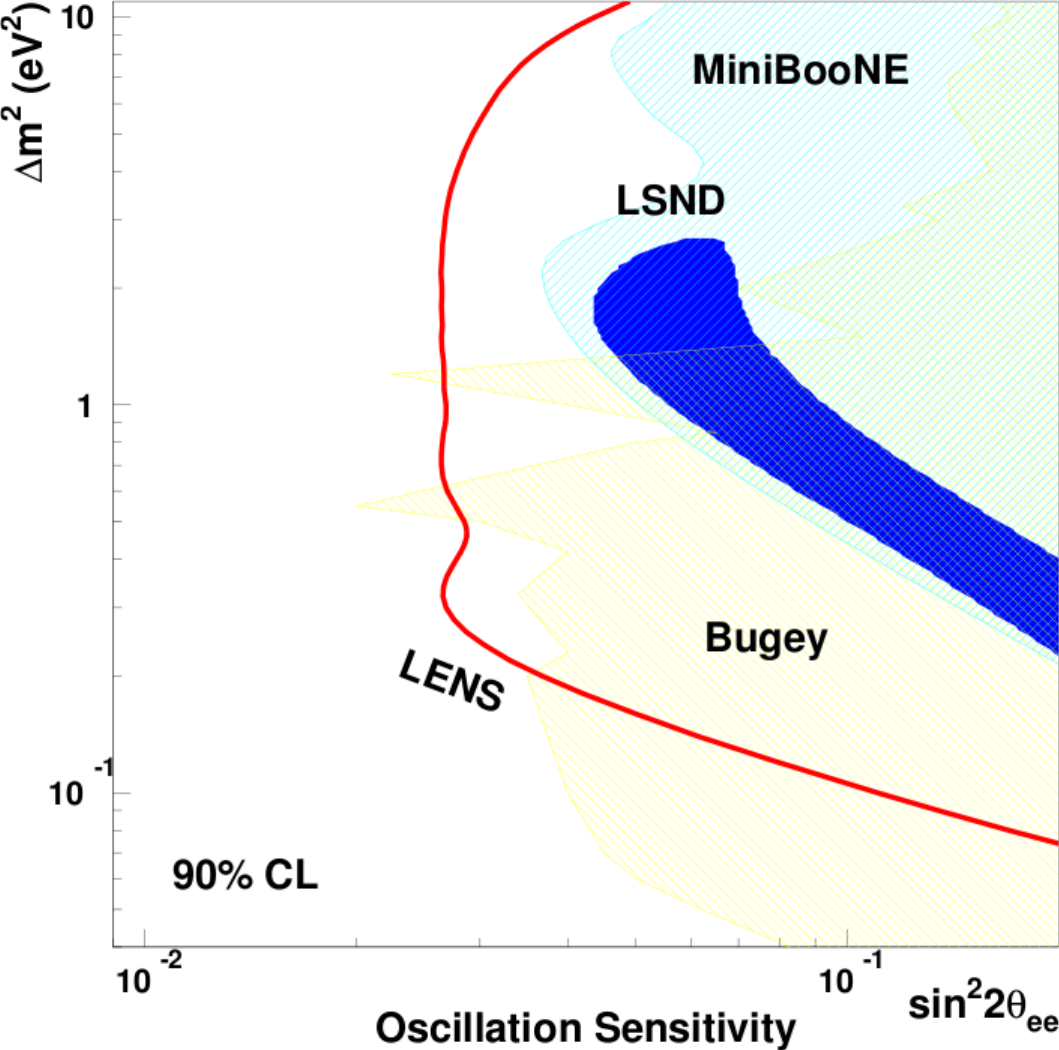}
}
\caption{\label{lens_sense}Sensitivity of the LENS $\nu_e$ disappearance measurement with an exposure of 4$\times$100~days to a $^{51}$Cr source with an initial activity of 10~MCi.  The LSND allowed region~\cite{Aguilar:2001ty}, the MiniBooNE appearance limit~\cite{AguilarArevalo:2007it}, and the $\nu_e$ disappearance exclusion region from  Bugey~\cite{Declais:1994su} are also shown.  The LSND allowed region and MiniBooNE limit have been converted to $\sin^2 2\theta_{ee}$ by assuming equal overlap of the $\nu_e$ and $\nu_{\mu}$ floavor statesa with the fourth mass eigenstate.}
\end{figure}

The sensitivity calculations are based on a 10~MCi initial source strength and assume 4 runs of 100~days each.  This preliminary calculation uses no rate information, and depends only on the variation in neutrino interaction rate as a function of distance from the source~\footnote{This is a conservative estimate of the sensitivity.  On the one hand, absolute rate information can not be used, because the nuclear matrix element for the $^{115}$In-$^{115}$Sn$^*$ system is not known.  On the other hand, the points of zero oscillation probability in the detector are known for a given $\Delta m^2$ hypothesis, and the unoscillated interaction rate can be determined at these points.  With this approach the $^{51}$Cr neutrino rate can be used to determine the $^{115}$In-neutrino matrix element to better than 5\%, independent of possible oscillation, allowing a precision measurement of the absolute solar neutrino flux.}.  Figure~\ref{lens_sense} shows the sensitivity of LENS-Sterile to $\nu_e$ disappearance in the 0.1 to 10~eV$^2$ range, compared to the existing limit from the short baseline reactor neutrino experiment Bugey~\cite{Declais:1994su}.  The LSND allowed region~\cite{Aguilar:2001ty} and the MiniBooNE~\cite{AguilarArevalo:2007it} limit are shown converted into $\sin^2\theta_{ee}$ by assuming equal overlap of the $\nu_e$ and $\nu_{\mu}$ flavor states with the fourth mass eigenstate.

\clearpage
\subsection{RICOCHET: Coherent Scattering and Oscillometry Measurements with Low-temperature
Bolometers\footnote{Proposed by 
                    K.~Scholberg (Duke University),
		    G.~Karagiorgi, M.~Shaevitz (Columbia University),
		    M.~Pyle (UC Berkeley),
		    E.~Figueroa-Feliciano, J.~Formaggio, J.~Conrad, K.~Palladino, J.~Spitz,
		    L.~Winslow, A.~Anderson, N.~Guerrero (MIT)}}

If the existence of sterile neutrinos is confirmed, the observation would be a clear manifestation
of physics beyond the standard model. As such, the observation should be tested by making the least number of assumptions with regards to the underlying physics.  Neutral current coherent scattering provides such a test, since it explicitly probes active-to-sterile neutrino oscillations.  Such a role should be seen as complementary to other measurement programs, including charged current and elastic scattering channels currently being pursued.

Coherent scattering offers distinct advantages compared to other techniques in disentangling the signature of sterile neutrinos.  First and foremost, coherent scattering off nuclei is a neutral current process.  Thus, any observation of an oscillation structure would indicate mixing solely to non-active neutrinos.  Other methods, such as neutrino-electron scattering, must disentangle the mixing to sterile neutrinos from mixing to active neutrinos.  The technique becomes even more powerful when combined with low energy mono-energetic sources.  Oscillations to neutrinos at the eV mass scale would manifest themselves over the length of a few meters (for $\sim 1$ MeV neutrino energies).  The signature would be quite difficult to mimic with typical backgrounds.  Finally, the cross-section for the process is greatly enhanced thanks to the coherent nature of the reaction.

Neutrino-nucleus interactions which are coherent in character have the advantage of scaling as $A^2$, where $A$ is the mass number of the target nucleus.  For a target nucleus with atomic number $Z$ and neutron number $N$, the cross-section as a function of recoil kinetic energy is given by the expression~\cite{Freedman:1977xn}:

\begin{equation}
\frac{d\sigma(\nu A \rightarrow \nu A)}{dT}  = \frac{G_F^2}{4\pi} M_A Q_W^2 (1 - \frac{M_A T}{2 E_\nu^2}) F(q^2)^2
\end{equation}

\noindent where $G_F$ is the Fermi coupling constant, $M_A$ is the mass of the nucleus, $F(q^2)$ is the nuclear form factor, and $Q_W$ is the weak charge, defined by the relation:

\begin{equation}
Q_W = N - Z(1-4\weakangle)
\end{equation}

In our study, we will mainly consider mono-energetic electron capture sources, all of which have neutrino energies below 1 MeV.  The maximum momentum transfer for such sources is $|q_{\rm max}| \le 2 E_\nu \ll 2$ MeV.  Since the form factor $F(q^2) \rightarrow 1$ for cases where the scale of the momentum probe is much larger than the size of the nucleus, we can safely ignore this correction factor for our analysis.

The maximum kinetic energy imparted on the nuclear recoil depends on the neutrino energy and the mass of the recoil target:

\begin{equation}
T_{\rm max} \le \frac{E_\nu}{1+\frac{M_A}{2E_\nu}}
\end{equation}

For a silicon target at 1 MeV, that implies a maximum kinetic energy of about 50 eV.  For a germanium target the maximum kinetic energy would be around 20 eV. Such low kinetic energies are why detection of the process has been so elusive to date.  Any detector hoping to detect such a signal with sufficient statistics must achieve as low a recoil threshold as possible.

\subsubsection*{The Source}

Oscillometry-based measurements benefit greatly from the use of mono-energetic neutrino sources, since it reduces the measurement to a pure flux-versus-distance analysis.  Low energy electron capture sources provide the most effective and clean source of such neutrinos available to date~\cite{Haxton:1988ee}.  Historically, two such high intensity source have been produced for neutrino studies:  a \isotope{Cr}{51} source, used by the SAGE and GALLEX experiments~\cite{Abdurashitov:1998ne,Anselmann:1994ar}, and an \isotope{Ar}{37} gaseous source used in conjunction with the SAGE experiment~\cite{Abdurashitov:2005tb}. 

The \isotope{Ar}{37} source is perhaps the most ideal with respect to a future coherent-scattering measurement, for a number of reasons:

\begin{itemize}
\item \isotope{Ar}{37} produces a very high-energy, near mono-energetic neutrino (90.2\% at 811 keV, 9.8\% at 813 keV).
\item With the exception of inner bremsstrahlung photons, almost all the energy is carried away by neutrinos, facilitating shielding and enabling the source to be extremely compact.
\item Extremely high production yield per reactor target.
\end{itemize}

The SAGE collaboration successfully produced such a source with a total activity of about 400 kCi to be used in conjunction with their gallium solar neutrino detector.  The source was also very compact, extending 14 cm in length and 8 cm in diameter, including shielding~\cite{Barsanov:2007fe}.  Further reduction in size might be possible, even with increased activity, making \isotope{Ar}{37} an ideal portable neutrino source.  

Despite its clear advantages as a source and its historical precedent, production of such sources is less than ideal.  Far less complex to produce is \isotope{Cr}{51}, which requires only thermal neutrons capturing on \isotope{Cr}{50}.  However, as a source, the high energy gamma produced from the decay of the excited state of \isotope{V}{51} imposes more shielding requirements.  As such, intense  \isotope{Cr}{51} may be less ideal for this investigation, but still worth considering given the advantages in producing the required activity.  

\subsubsection*{The Detector}
\label{sec:detector}

The detector requirements for this experiment are extremely challenging.  Due to the low energy of the neutrinos ($\le1$~MeV), the recoil energy deposited in the target is in the order of tens of eV, while the minimum mass needed is hundreds of kilograms. Methods of determining the energy deposition from particle interactions in a target include measuring the ionization, the scintillation, and/or the phonon excitations in the material. For nuclear recoils of tens of eV, the fraction of the energy deposited by the scattering event that produces free or conduction band electrons (the quenching factor) is unknown at these energies, and is expected to be very low (could be zero for some materials). Thus any readout scheme involving ionization channels will be at a severe disadvantage. Similar uncertainties hold for the scintillation yield from nuclear recoils at these energies. An additional problem for both ionization and scintillation readout is that the energy required to create a single electron, electron-hole pair, or scintillation photon from a nuclear recoil in most liquid or solid targets is a few eV for ionization and tens of eV for scintillation. Thus, even if any quanta were produced, Poisson statistics would make the measurement of the energy of any given recoil event fairly poor. We have therefore focused our attention to the measurement of phonons created in the interaction. With mean energies of the order of $\mu$eV, thermal phonons provide high statistics at 10 eV and sample the full energy of the recoil with no quenching effects.

The threshold for a bolometer is a function of its baseline energy resolution. A dimensionless measure of the sensitivity of a resistive thermometer at a temperature $T$ and resistance $R$ is the quantity $\alpha$, defined as $\alpha \equiv \frac{T}{R}\frac{dR}{dT}$. The energy resolution of a TES bolometer is approximately~\cite{Irwin:1995ie}
\begin{equation}
\Delta E_{\rm rms} = \sigma_{E} \approx \sqrt{\frac{4 k_{B}T^{2} C_{\rm tot}}{\alpha} \sqrt{\frac{\beta+1}{2}}},
\end{equation}
where $k_{B}$ is the Boltzmann constant, $C_{\rm tot}$ is the total heat capacity of the bolometer, and $\beta$ is the exponent of the temperature dependance of the thermal conductivity between the bolometer and the refrigerator. To unambiguously detect events above the noise from the detector, we set the experimental threshold to 7.5 $\sigma_{E}$. For a 10~eV threshold, we then need a detector with $\sigma_{E} < 1.33$~eV, or expressed in terms of the full width at half maximum, $\Delta E_{\rm FWHM} = 2\sqrt{2 \ln 2} \, \sigma_{E} < 3.14$~eV.

Assuming a conduction path to the cold bath of the refrigerator dominated by Kapitza resistance, $\beta = 4$, and with a temperature $T=15$~mK, a 10~eV threshold could be attained with a heat capacity $C_{\rm tot} \leq 200$~pJ/K.  However, this model is not complete, as it assumes a perfectly isothermal bolometer. In practice, the various internal heat capacity systems of the bolometer are decoupled from each other through internal conductances, and thermalization times of each separate heat capacity must also be taken into account. These internal decouplings introduce various sources of noise, degrading the energy resolution of the bolometer and consequently requiring a smaller heat capacity to attain the desired threshold. 

For this study, we will assume the excess heat capacity is negligible, and optimize the heat capacity of the thermometer $C_{\rm TES}$, the electron-phonon coupling $G_{\rm ep}$, and the Si heat capacity $C_{\rm Si}$ to obtain the desired 10~eV threshold with the highest possible target mass. We make the following assumptions:

\begin{itemize}
\item Each detector is a Si cube ranging in mass from 20--100~g. The heat capacity is determined from Debye theory.
\item The conductance between the Si and the cold bath, $G_{\rm pb}$, can be engineered to give a desired value. The value is chosen to give a thermal impulse response time of 50~ms as measured by the thermometer readout. 
\item The thermometer is a Mo/Au TES bilayer with a superconducting transition engineered to a specific temperature between 10--100~mK. Mo/Au TES X-ray detectors have achieved resolutions of $\Delta E_{\rm FWHM}=2$~eV~\cite{Bandler:2008fv}.
\item The TES heat capacity and electron-phonon coupling are taken from the literature and are a function of the chosen volume of the TES and the temperature.
\end{itemize}

Given these general assumptions, several combinations of detector mass and transition temperature were tested for both Si and Ge targets, scaling the TES volume to obtain the best energy resolution, following the theoretical framework of~\cite{FigueroaFeliciano:2006bc}. Due to practical limitations in refrigeration and considering the readout necessary for the size of the experiment, we focus on a transition temperature of 15~mK, with the refrigerator base temperature at 7.5~mK. At this temperature, a low-energy threshold of 10~eV can be obtained with bolometers with 50~g of Si or 20~g of Ge. A 50~g Si target sees about twice the rate of neutrino coherent recoil events as a 20~g Ge target when both have a 10~eV threshold. 

\begin{figure}[htbp]
\begin{center}
\includegraphics[width=0.7\columnwidth,keepaspectratio=true]{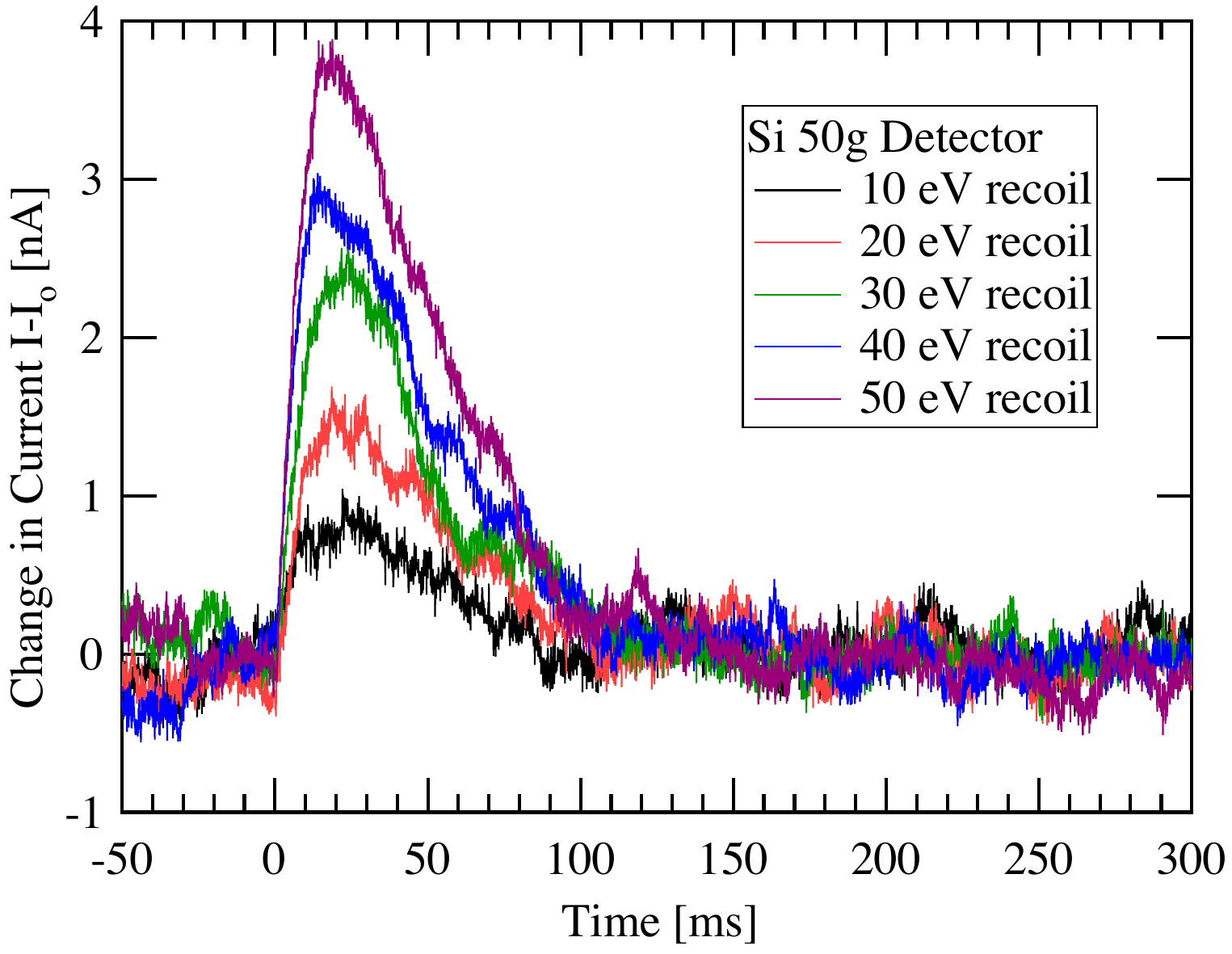} 
\caption{Simulated current readout for 10--50~eV recoils using the model parameters in this study. The current has been multiplied by -1 to make the pulses positive. Noise sources modeled are: the phonon noise between the target to the bath, the internal thermal fluctuation noise between the target and the TES thermometer, the Johnson noise from the TES and its bias resistor, and the electronics noise. The modeled 10~eV pulses are clearly separated from the noise.}
\label{fig:pulses}
\end{center}
\end{figure}

Based on the performance of such low threshold detectors, it is possible envision constructing a highly modular coherent neutrino experiment with a total target mass of 500 kg.  Multiplexing readout schemes for transition-edge sensors are now a mature technology being developed for many astronomical applications, for example~\cite{Reintsema:2009er}, and 10,000 channel systems with time constants similar to this application are already in operation~\cite{Bintley:gz}. Schemes for even larger multiplexing gains are in development~\cite{Niemack:2010gs}. Given the slow time constants of this application, a 10,000 channel multiplexer design carries a fairly low risk and would allow 500~kg of Si to be instrumented.

\begin{figure}[htbp]
\begin{center}
\includegraphics[width=0.45\columnwidth,keepaspectratio=true]{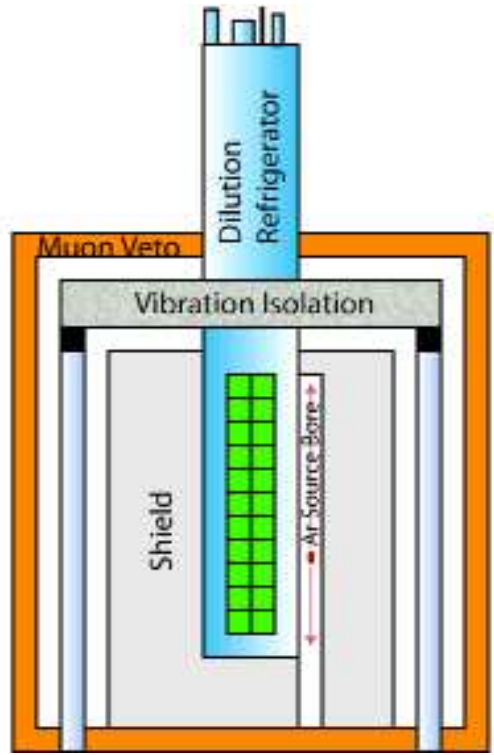} 
\caption{Conceptual schematic of the experimental setup for a bolometric measurement of coherent scattering from a high-intensity \isotope{Ar}{37} neutrino source.  An array of 10,000 Si bolometers is arranged in a column of dimensions  0.42 (dia.)~$\times$~2.0~(length)~meters (shown in green) inside a dilution refrigerator suspended from a vibration isolation mount.  Each Si bolometer has a mass of 50 g for a total active mass of 500~kg. Appropriate passive or active shielding surrounds the refrigerator. A cylindrical bore in the shield allows the \isotope{Ar}{37} source, mounted on a translation mechanism, to be moved to different positions along the side of the array. Periodic movement of the source throughout the measurement sequence allows each detector to sample multiple baselines, enables cross-calibration among detectors, and aids in background subtraction. The minimum distance from the source to a bolometer is assumed to be $\sim$10~cm.}
\label{fig:array}
\end{center}
\end{figure}

A concept for a 500~kg payload is shown in Fig.~\ref{fig:array}. The 10,000 Si bolometers are arranged in a column of dimensions 0.42 (dia.)~$\times$~2.0~(length)~meters inside a dilution refrigerator suspended from a vibration isolation mount.  Passive or active shielding surrounds the refrigerator. A cylindrical bore, perhaps 10~cm or less in diameter, is removed from the shield and allows the \isotope{Ar}{37} source, mounted on a radio-pure translation mechanism, to be moved to different positions along the side of the array. Periodic movement of the source throughout the measurement sequence allows each detector to sample multiple baselines, enables cross-calibration among detectors, and aids in background subtraction. 

\subsubsection*{Sensitivity and Outlook}

\begin{figure*}[!h]
\begin{center}
\begin{tabular}{c c}
\subfigure{\label{fig:resultsSiAr}\includegraphics[width=0.5\columnwidth]{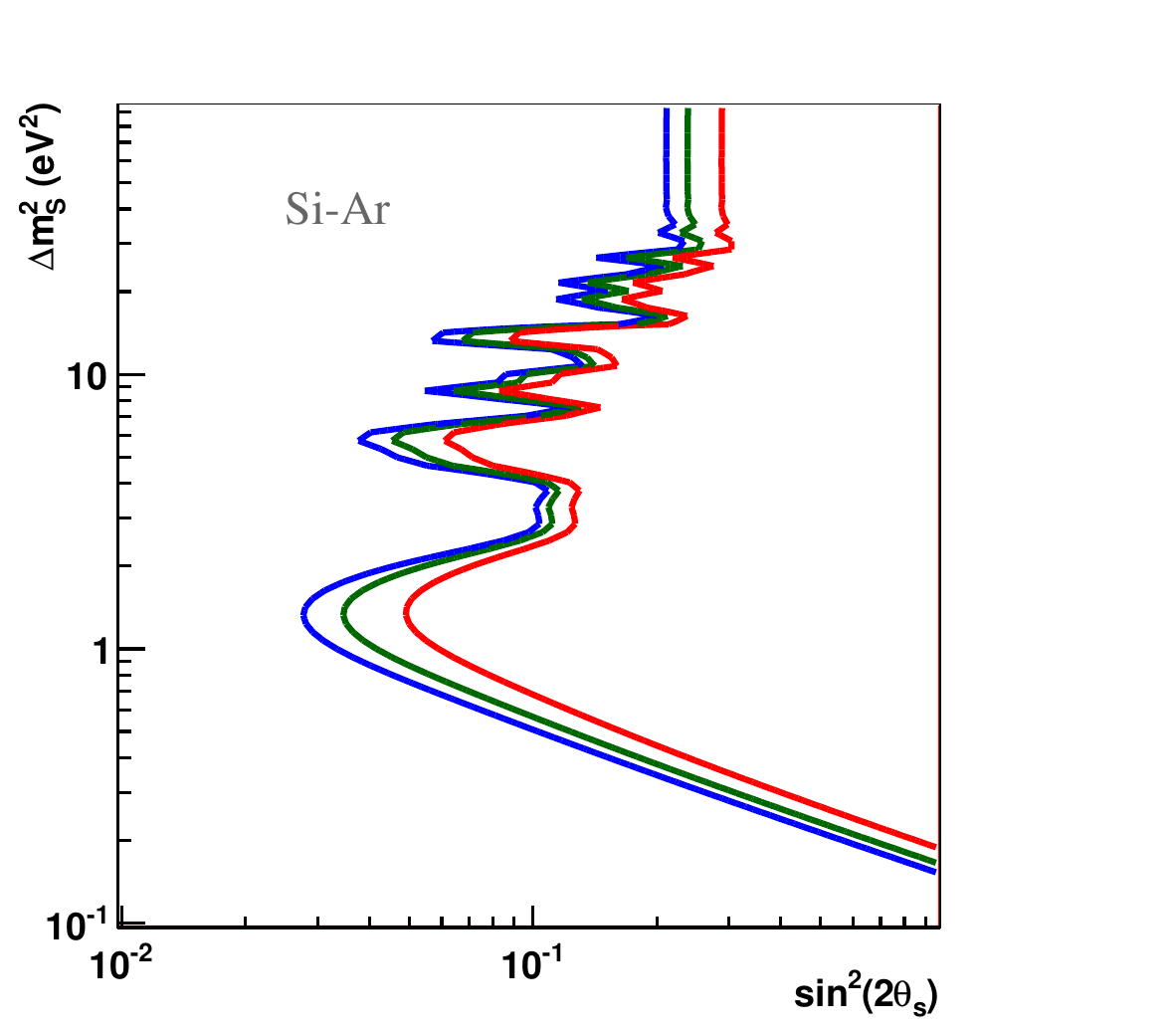}} &
\subfigure{\label{fig:thesignal}\includegraphics[width=0.5\columnwidth]{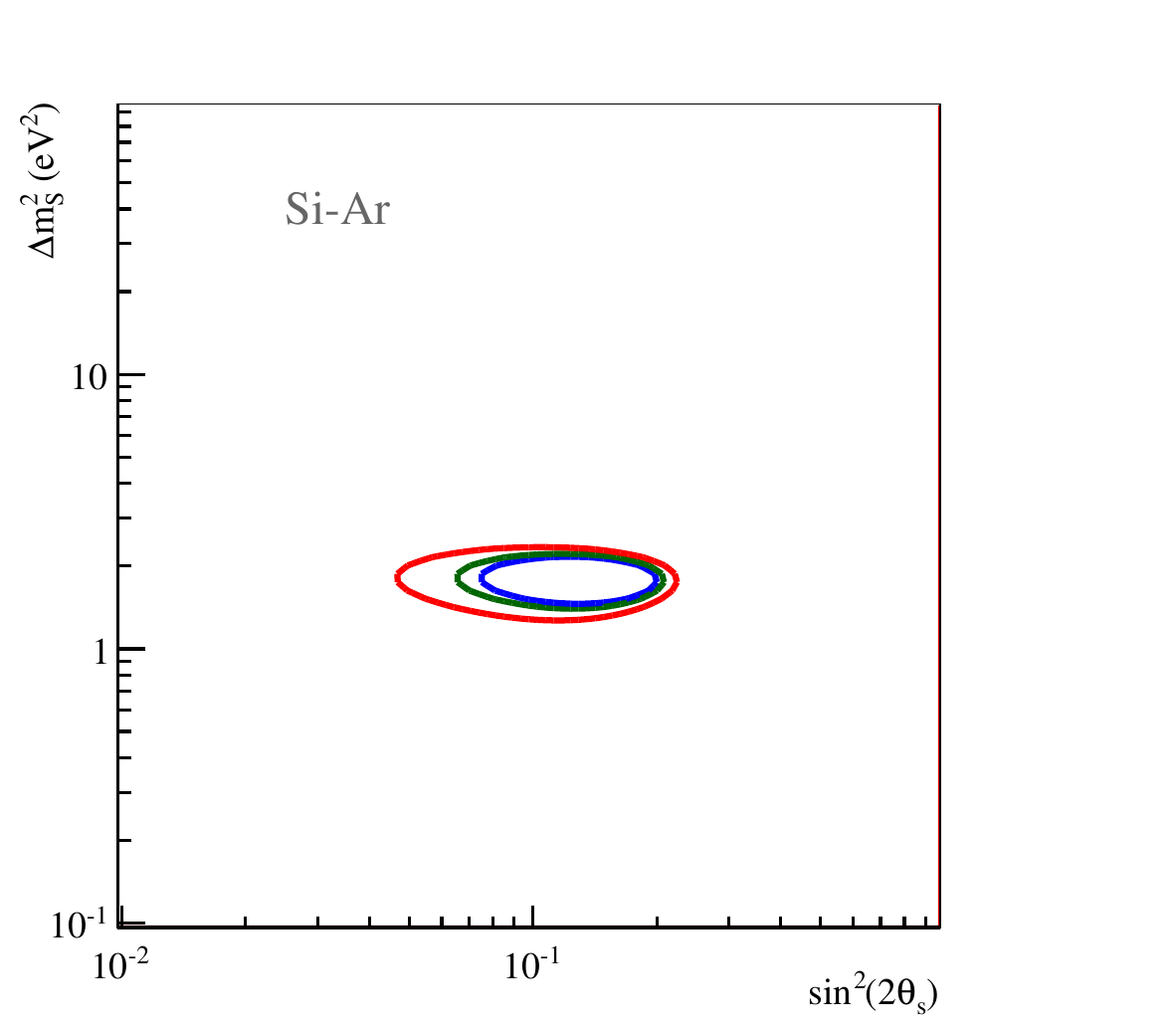}}
\end{tabular}
\caption{Left: Likelihood contours for a 300-day run on a 500 kg Si array exposed to a 5 MCi \isotope{Ar}{37} source, using both shape and rate information. Confidence levels in all plots are shown at 90\% (blue), 95\% (green), and 99\% (red). Statistical and systematic errors are included in the signal analysis. Right: Likelihood contour curves for a simulated neutrino oscillation signal ($\Delta m_S^2$ = 1.5 eV$^2$, $\sin{(2\theta_S)^2} = 0.15$) after 300 days of data taking.  Contour levels are shown at 90\% (blue), 95\% (green), and 99\% (red). }
\end{center}
\end{figure*}

For a monoenergetic source, the oscillation signal is all encoded within the spatial distribution of events.  A deviation from the expected $r^{-2}$ dependence could constitute a possible oscillation signal.  We use simulated data from a mock experiment to determine the potential sensitivity to sterile neutrinos.  We consider a compact 5 MCi \isotope{Ar}{37} source to be used in conjunction with a 500~kg silicon array.  We consider a total exposure of 300 days in order to extract both signal and background rates. Results for a 500 kg Si detector array are shown in Fig.~\ref{fig:resultsSiAr}. The distortion caused by a non-zero sterile mixing is statistically distinguishable in the measured distance profile.  For the bulk of the region of $\Delta m_S^2 = 1-10$ eV$^2$ and $\sin{(2\theta_S)}^2 \ge 0.08$, typically preferred from the reactor data, is ruled out at the 90\% C.L.  If the best fit solution from the reactor anomaly is viable, then the measurement should be detectable at the 99\% C.L. (see Fig.~\ref{fig:thesignal}). It is possible to also conduct a shape-only analysis.  Most of the sensitivity to sterile oscillations is retained for $\Delta m_S^2$ masses below 10 eV$^2$.  

We have outlined the possibility of probing the existence of sterile neutrinos using coherent scattering on a bolometric array.  Such a method could provide the most direct test of oscillations to sterile neutrinos.  With the advent of low threshold detectors and the use of intense neutrino sources, such an experiment appears feasible with our current technology.  This program (known as RICOCHET) is also very complimentary to any existing dark matter search.  Even in the absence of sterile neutrinos, the experiment as described in this letter can make other important measurements.  Most prominently, such an experiment may constitute the first observation of coherent scattering.  For a 500 kg detector, it should be able to make a $\simeq 5\%$ measurement on the overall cross-section, pending on the absolute calibration of the efficiency.  For an isoscalar target, such as silicon, this provides a direct measurement of the weak mixing angle at momentum transfer as low as 1~MeV. 

Sensitivity to the neutrino oscillation parameter space indicative of sterile neutrinos ($\Delta m^2\sim 1~\mathrm{eV}^2$) can also be provided with a dark matter style detector in combination with a decay-at-rest pion/muon source~\cite{Anderson:2011bi,Anderson:2012pn}. Reference~\cite{Anderson:2012pn} calls for a germanium- or argon-based device placed within a few tens of meters of a DAE$\delta$ALUS-style~\cite{Conrad:2009mh} cyclotron configured with two proton targets at different baselines from the single detector. The two targets allow the composition of the neutrino beam to be studied as a function of distance without the complication of having to move detectors or instrument multiple devices.

The disappearance of neutrinos interacting via the neutral current is a strict probe of active-to-sterile oscillations as there are no complicating contributions from active-to-active$^{'}$ disappearance. That is, the disappearance of neutrinos via neutral current coherent neutrino-nucleus scattering could definitively establish the existence of the sterile neutrino, especially when considered in combination with charged-current-based searches. Depending on the detector technology and run scenario, such an experiment could supply sensitivity to the LSND best-fit mass splitting at the 3-5$\sigma$ level and large regions of the LSND and reactor anomaly allowed regions. Furthermore, the configuration described could produce a first observation of the as-yet-unseen coherent scattering process as well as improve on non-standard neutrino interaction limits by over an order of magnitude in some cases~\cite{Anderson:2011bi}.

\clearpage
\subsection{Very Short Baseline $\nu_e \rightarrow \nu_x$ Oscillation Search 
with a Dual Metallic Ga Target at Baksan and a $^{51}$Cr Neutrino Source
\footnote{Proposed by: 
          B.~T.~Cleveland, R.~G.~H.~Robertson (University of Washington),
   	  H.~Ejiri (Osaka University),
   	  S.~R.~Elliott (Los Alamos National Laboratory),
   	  V.~N.~Gavrin, V.~V.~Gorbachev, D.~S.~Gorbunov, E.~P.~Veretenkin, A.~V.~Kalikhov, 
	  T.~V.~Ibragimova (Institute for Nuclear Research of the Russian Academy of Sciences),
   	  W.~Haxton (University of California, Berkeley),
   	  R.~Henning, J.~F.~Wilkerson (University of North Carolina, Chapel Hill), 
   	  V.~A.~Matveev (Joint Institute for Nuclear Research, Dubna)
   	  J.~S.~Nico (National Institute of Standards and Technology)
   	  A.~Suzuki (Tohoku University)
}}

There have been four experiments in which very intense
neutrino-emitting radioactive sources, either $^{51}$Cr
\cite{Abdurashitov:1998ne,Kaether:2010ag} or $^{37}$Ar \cite{Abdurashitov:2005tb}, have irradiated
a Ga target.  The weighted-average result of these experiments,
expressed as the ratio $R$ of the measured neutrino capture rate to
the expected rate, based on the measured source intensity and the
known neutrino capture cross section, is $R = 0.87 \pm 0.05$
\cite{Abdurashitov:2009tn}, considerably less than the expected value of unity.

\begin{figure}[b]
\centering
\includegraphics[width=0.5\textwidth]{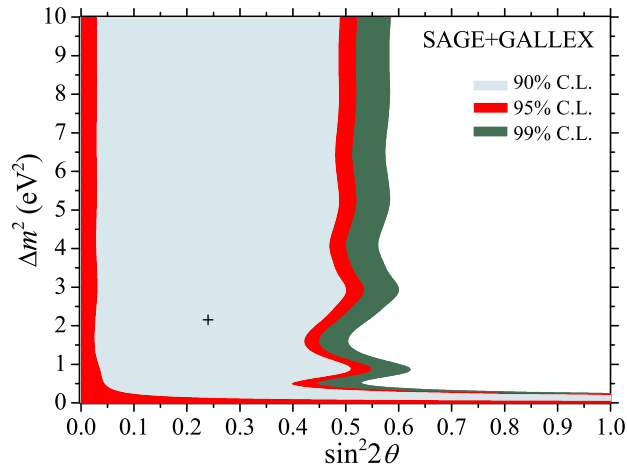}
\caption{Region of allowed mixing parameters inferred from gallium
source experiments assuming oscillations to a sterile neutrino.  The
plus sign at $\Delta m^2=2.15$~eV$^2$ and $\sin^2 2\theta=0.24$
indicates the best-fit point.}

\label{sage-gallex}
\end{figure}

If we assume that this unexpectedly low value of $R$ is the result of
transitions from active to sterile neutrinos, then the region of
allowed oscillation parameters inferred from these Ga source
experiments is shown in Fig.~\ref{sage-gallex}.

Sterile neutrinos are, however, not the only possible explanation for
this lower than expected result.  It could also be simply the result
of a large statistical fluctuation, an effect that is unexpected, but
which conceivably could have occurred.  Another explanation, namely
overestimation of the cross section for neutrino capture to the
lowest two excited states in $^{71}$Ge, has been essentially ruled
out by a recent experiment \cite{Frekers:2011zz}.  A final possible
explanation could be an error in one of the efficiency factors that
enters into the calculation of the rate, such as the extraction
efficiency or the counting efficiency.  As discussed in \cite{Abdurashitov:2009tn},
however, many ancillary experiments have been performed which give us
great confidence in these measured efficiencies.

The interpretation of the Ga source experiments in terms of
oscillations to a sterile neutrino with $\Delta m^2 \approx
1$~eV$^2$, as well as the agreement of these results with the reactor
experiments Bugey, Chooz, and G\"osgen and the accelerator
experiments LSND and MiniBooNE is considered in detail in
Ref.~\cite{Giunti:2006bj,Acero:2007su,Giunti:2009zz,Giunti:2010wz}.

\begin{figure}
\centering
\includegraphics[width=0.4\textwidth]{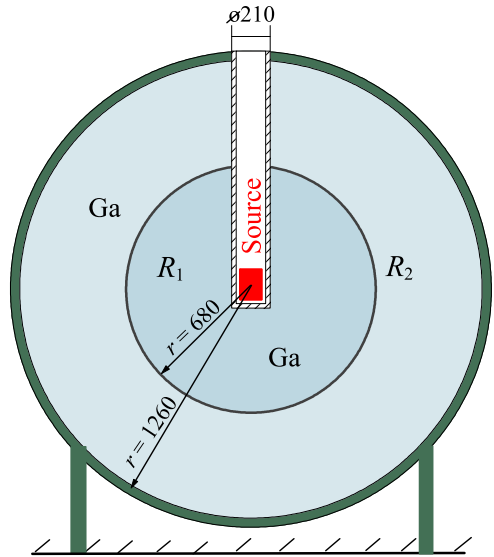}
\caption{Schematic drawing of proposed neutrino source experiment.
$R_1$ and $R_2$ are the ratios of measured capture rate to predicted
rate in the absence of oscillations in the inner and outer zones,
respectively.  Outer radii $r$ of the two zones and diameter $\phi$
of source reentrant tube are given in mm.}

\label{src_scheme}
\end{figure}

To further investigate the sterile neutrino explanation for the Ga
source experiments, we intend to pursue an improved version of these
measurements which will now be described.  In outline, our plan, as
schematically pictured in Fig.~\ref{src_scheme}, is to place a
$^{51}$Cr source with initial activity of 3~MCi at the center of a
50-tonne target of liquid Ga metal that is divided into two
concentric spherical zones, an inner 8-tonne zone and an outer
42-tonne zone.  If oscillations to sterile neutrinos do not occur,
then at the beginning of irradiation there is a mean of 65~atoms of
$^{71}$Ge produced by the source per day in each zone.  After an
exposure period of a few days, the Ga in each zone is transferred to
reaction vessels and the $^{71}$Ge atoms produced by neutrino
capture are extracted.  These steps are the same as were used in our
prior source experiments \cite{Abdurashitov:1998ne,Abdurashitov:2005tb} and are well tested.
Finally, the Ge atoms are placed in proportional counters and their
number is determined by counting the Auger electrons released in the
transition back to $^{71}$Ge, which occurs with a half life of
11.4~days.  A series of exposures is made, each of a few days
duration, with the $^{71}$Ge atoms from each zone measured in
individual counters.

These extraction and counting procedures have been used in the SAGE
experiment for the last 20~years \cite{Abdurashitov:2009tn} and are all very well
understood.  A Monte Carlo simulation of the entire experiment --~ten
extractions each with a 9-day exposure --~that uses typical values of
extraction efficiency, counter efficiency, counter background rates,
and includes the known solar neutrino rate, indicates that the rate
in each zone can be measured with a statistical uncertainty of
$\pm3.7$\%.  From our prior experience in measurement of the solar
neutrino flux we expect a systematic uncertainty of $\pm2.6$\%,
leading to a total uncertainty, statistical plus systematic, of
$\pm4.5$\%.

Some further details on various aspects of this experiment will now
be given.

$^{51}$Cr will be produced by the capture of thermal neutrons on
the stable isotope $^{50}$Cr, whose content in natural Cr is
4.35\%.  To make the volume of the final source as small as possible,
the first step in source production will be to use centrifuge
technology to produce 3.5~kg of enriched $^{50}$Cr in the form of
highly-purified chromium trioxide.  Several facilities exist in
Russia where this task can be carried out and enrichment in
$^{50}$Cr up to 97\% is achievable.

The isotopically enriched Cr will be converted to metal by
electrolysis and pressed into 81 metallic Cr rods with diameter
8.5~mm and length 95~mm whose total mass is 3050~grams.  These
$^{50}$Cr rods will be placed in 27~cells in the central neutron
trap of the research reactor SM--3 in Dimitrovgrad, where the flux is
$5.0 \times 10^{14}$ neutrons/(cm$^2$-sec), and irradiated for 54
effective days (63 calendar days).  At the end of irradiation the
reactor engineers calculate that the specific activity of the Cr will
be 1003~Ci/g, and its total activity will be 3.06~MCi.

\begin{figure}
\centering
\includegraphics[width=0.6\textwidth]{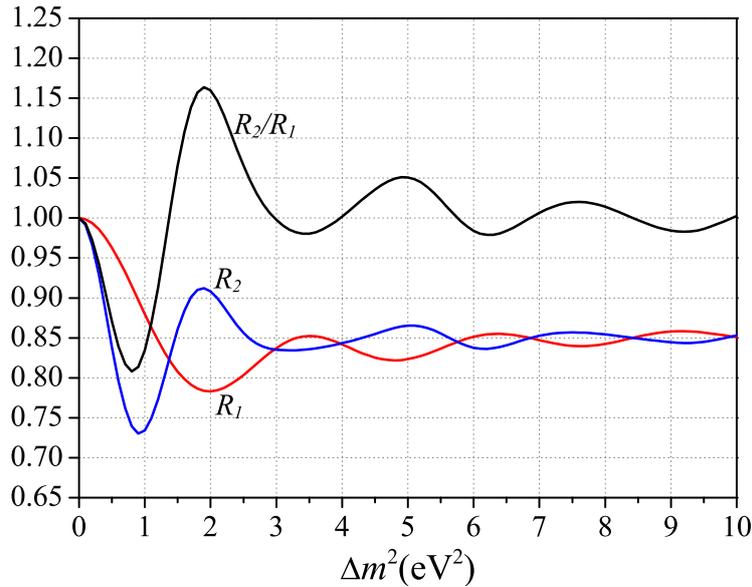}
\caption{Ratio of measured capture rate to predicted rate in inner
zone ($R_1$), in outer zone ($R_2$), and their ratio $R_2/R_1$ as a
function of $\Delta m^2$ for the case of $\sin^2 2\theta = 0.3$.  The
outer shell of the outer zone is assumed to be a cylinder, not a
sphere.}

\label{rate_ratio}
\end{figure}

The Cr rods will be removed from the reactor and, in a hot cell,
transfered to stainless steel tubes which are placed within a
stainless steel can.  The can is welded closed, put into a transport
cask, and brought to the Baksan Neutrino Observatory in the northern
Caucasus mountains of Russia.  The gallium irradiation will take
place in the SAGE laboratory where the overhead shielding is
equivalent to 4700~m of water and the measured muon flux is $(3.03
\pm 0.10) \times 10^{-9}$/(cm$^2$-s).

The inner compartment of the vessel for irradiation of Ga will be
made from Ti or from a thin inert plastic and the outer compartment
will be plastic-lined steel.  The distance that the neutrinos travel
through the Ga in both the inner and outer zones is 55~cm.  (To
simplify construction, the outer shell of the outer zone will
actually be a cylinder, not a sphere, but that has only a 2\% effect
on the neutrino capture rate and almost negligibly changes the
sensitivity of the experiment to oscillations.)

We intend to follow the same experimental procedures as were used in
the two previous SAGE source measurements~\cite{Abdurashitov:1998ne,Abdurashitov:2005tb}.  Each
exposure of the Ga to the source will begin by placing the source at
the center of the inner zone of Ga.  After an exposure period of
9~days, the source will be removed from the tank and moved to a
calorimeter, which will measure its intensity.  During this time the
Ga in each zone will be pumped to reaction vessels where the
$^{71}$Ge atoms produced by neutrino capture will be extracted.
When the extraction is completed, which requires about 12~hours, the
Ga will be pumped back into the 2-zone vessel, the source will be
again placed at its center, and the next irradiation will begin.

The activity of the source will be measured by calorimetry and other
methods.  Each $^{51}$Cr decay emits an average of 36~keV of
thermal energy, {\it i.e.}, the energy that is not carried away by
neutrinos.  A 3--MCi source will thus emit about 650~W of heat.  The
source activity will be measured in a calorimeter several times while
it decays and again, when all measurements are finished, the activity
will be determined by measuring the amount of accumulated $^{51}$V
-- the product of $^{51}$Cr decay.

\begin{figure*}
\centering
\hspace{1em}
\includegraphics*[width=0.45\textwidth]{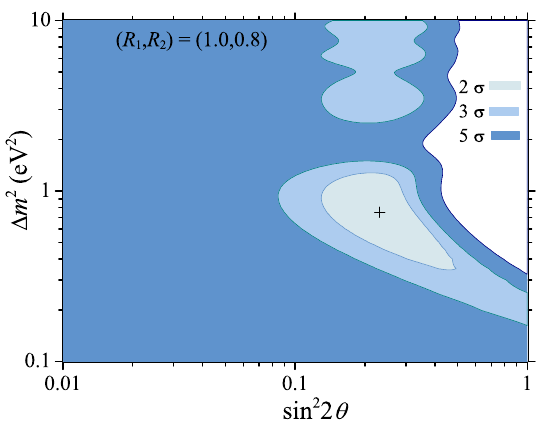}
\hspace{1em}%
\includegraphics*[width=0.45\textwidth]{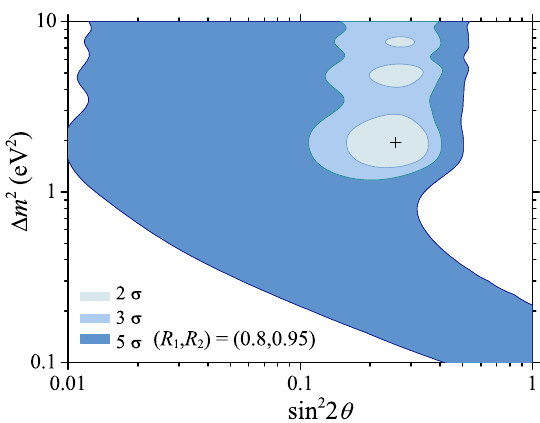}
\caption{Regions of allowed mixing parameters for two possible
outcomes of new two-zone Ga experiment.  $R_1$ and $R_2$ are the
ratios of measured rate to predicted rate in the absence of
oscillations in the inner and outer zones, respectively.  Plus sign
indicates the best-fit point.}

\label{regions}
\end{figure*}

Because many isotopes have high neutron capture cross sections and
their reaction products emit high-energy gamma rays great care will
be taken during the production and handling of the Cr rods to
minimize the introduction of chemical impurities.  The source will be
scanned with a semiconductor Ge detector to look for
$\gamma$-emitting radioactive impurities, as was done in the first
SAGE $^{51}$Cr experiment.

If oscillations to a sterile neutrino are occurring with mass-squared
difference of $\Delta m^2$ and mixing parameter $\sin^2 2\theta$,
then the rates in the outer and inner zones of gallium will be
different and their ratio, for the specific case of $\sin^2 2\theta =
0.3$, will be as shown in Fig.~\ref{rate_ratio}.

To see how our new 2-zone experiment may shed light on transitions to
sterile neutrinos, let us consider several possible outcomes and the
inferences therefrom:

\begin{itemize}

\item If the ratio of rates $R$ in the two zones are statistically
incompatible, such as $R_1=1.00 \pm 0.05$ in the inner zone and
$R_2=0.80 \pm 0.04$ in the outer zone, or vice versa, then it can be
evidence that transitions to sterile neutrinos are occurring.  We
show in Fig.~\ref{regions} the allowed neutrino mixing parameters for
two such possible outcomes of the experiment.

\item If the ratio of rates $R$ in both zones are statistically
compatible, then the inferences will depend on what the average value
may be.  For example, average values of $R$ much below 0.92 are also
evidence of transitions to sterile neutrinos.

\end{itemize}

\begin{figure}
\centering
\includegraphics[width=0.6\hsize]{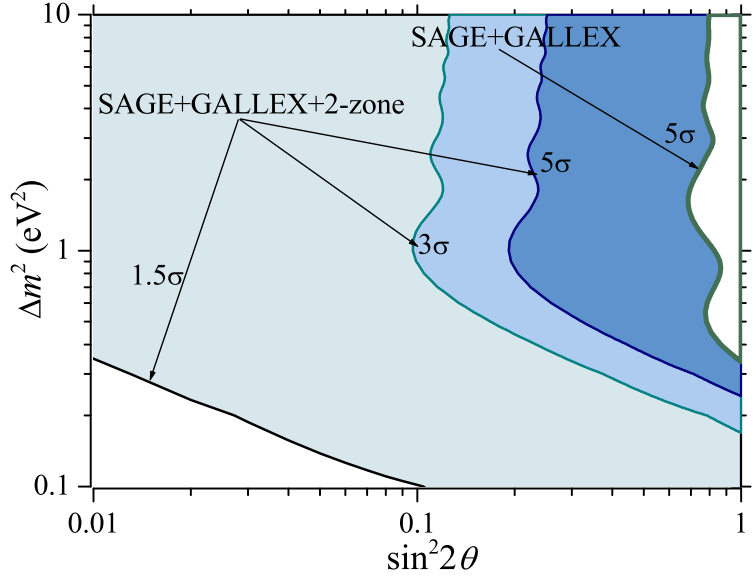}
\caption{Region of mixing parameters to which proposed new
$^{51}$Cr source experiment, in combination with previous source
experiments, is sensitive with various levels of confidence.  The
white region in upper-right corner has been already excluded with
$5\sigma$ confidence by previous SAGE and GALLEX source experiments.
As indicated, the new 2-zone experiment has the capability to greatly
expand this exclusion region with high confidence.}

\label{SIGMA5log}
\end{figure}

Figure~\ref{SIGMA5log} shows the region in $\Delta m^2 - \sin^2
2\theta$ space to which the new 3~MCi $^{51}$Cr experiment will be
sensitive.  Our proposed experiment has several significant
advantages over other methods of investigation of oscillations with
parameters $\Delta m^2 \approx 1$~eV$^2$:

\begin{itemize}

\item The neutrino spectrum is extremely well known.  90\% of the
decays of the $^{51}$Cr source give a neutrino with energy 750~keV
and 10\% an energy of 430~keV.  This nearly-monochromatic energy is
important for oscillation experiments because the energy occurs in
the denominator of the transition probability $P$ in the form
\begin{equation}
P(\nu_e \rightarrow \nu_e) = 1 - \sin^2(2\theta) \sin^2[1.27\Delta m^2(\text{eV}^2) \frac{L}{E_{\nu}}]
\end{equation}
where $\theta$ is the mixing angle, $L$ is the distance in m from the
point in the source where neutrino emission occurs to the point in
the target where this neutrino is captured, and $E_{\nu}$ is the
neutrino energy in MeV.  In the two-zone experiment the source is
very compact, with typical linear dimension of 10~cm, and $L$ is
only about 1~m.  As a result, the ripples of oscillation are strongly
manifested and are not averaged out when $\Delta m^2 \approx
1$~eV$^2$.

\item The cross section for neutrino capture on $^{71}$Ga is $5.5
\times 10^{-45}$ cm$^2$ and is known to an uncertainty of 2--3\%
\cite{Frekers:2011zz}.  Further, the density of gallium at its melting
temperature of 29.8$^{\circ }$C is 6.095~g/cm$^3$.  These factors ensure
that the neutrino capture rate will be very high and can be measured
with good statistical accuracy.

\item The main background will be from solar neutrinos, whose flux
has been measured with the SAGE detector for many years and is now
known to an uncertainty of 4.7\% \cite{Abdurashitov:2009tn}.  The extremely high
intensity of the source, $\sim$3 MCi, should provide a production
rate in the detector that will exceed the rate from the Sun by
several factors of ten.

\item The activity of the neutrino source can be measured in several
ways leading to a total uncertainty on the source emission rate as
low as 0.5\% (see, {\it e.g.}, Ref.~\cite{Abdurashitov:2005tb}).

\item The statistical uncertainty can be reduced by re-irradiating
the Cr and repeating the entire experiment.

\end{itemize}

In contrast, experiments with reactor and accelerator neutrinos
suffer from several disadvantages.  The neutrino energy $E_{\nu}$ is
distributed over a wide spectrum and the dimensions $L$ of the
sources and detectors are on the scale of several meters.  Other
disadvantages of a reactor or accelerator experiment are that the
knowledge of the neutrino flux incident on the target is usually
significantly worse than with a neutrino source and that, with some
targets, there are appreciable uncertainties in the cross section for
neutrino interaction.

To summarize, we propose a new experiment in which a very-intense
$^{51}$Cr source irradiates a target of Ga metal that is divided
into two zones and the neutrino capture rate in each zone is
independently measured.  If there is either a significant difference
between the capture rates in the two zones, or the average rate in
both zones is considerably below the expected rate, then there is
evidence of nonstandard neutrino properties.

The proposed experiment has the potential to test neutrino
oscillation transitions with mass-squared difference $\Delta
m^2>0.5$~eV$^2$ and mixing angle $\theta$ such that $\sin^2 2\theta >
0.1$.  This capability exists because the experiment uses a compact
nearly-monochromatic neutrino source with well-known activity, the
dense target of Ga metal provides a high interaction rate, and the
special target geometry makes it possible to study the dependence of
the rate on the distance to the source.

\clearpage
\subsection{Proposed search of sterile neutrinos with the Borexino detector 
using neutrino and antineutrino sources\footnote{Proposed by the Borexino 
Collaboration \\ 
({\tt http://borex.lngs.infn.it}).}}

The solar neutrino detector Borexino is perfectly suited to host a source 
experiment able to shed light on the many intriguing experimental hints, 
accumulated so far, pointing to the possible existence of a sterile neutrino at 
the few ev2 mass scale. The extreme radiopurity achieved in the liquid 
scintillator acting as detection medium, witnessed by the successful precise 
$^7$Be solar neutrino detection~\cite{Bellini:2011rx}, and the thorough 
understanding of the detector performances gained throughout almost five years 
of data taking (and several calibration campaigns), make specifically Borexino 
the ideal choice for the sterile neutrino experimental investigation. 
Preliminary studies indicate, in particular, that Borexino could be a well 
suited location both for an external neutrino source experiment (with an 
ultimate background of about 50~eV/day mainly due to the irreducible 
contribution of solar neutrinos) and for an internal antineutrino source test 
(in which case a virtually zero background can be achieved, 
{\it e.g.}~10~cpy/100~tons~\cite{Bellini:2010hy}, due to reactor and geophysical 
antineutrinos).

Neutrino source experiments have been part of the Borexino program since the 
very beginning of the project. In the 1991 Borexino 
proposal~\cite{Borexino_proposal}, source experiments were explicitly mentioned, 
at that time mainly to perform neutrino magnetic moment search and for 
calibration purposes. During the detector construction, a small tunnel was built 
right below the water tank, providing a location as close as 8.25~m to the 
scintillator inner vessel center. This tunnel was maintained clean and sealed, 
as other parts of the Hall C, during the underground maintenance works done in 
2004-2005, so it is ready to be used.

\begin{figure}[b]
\includegraphics[width=0.6\textwidth]{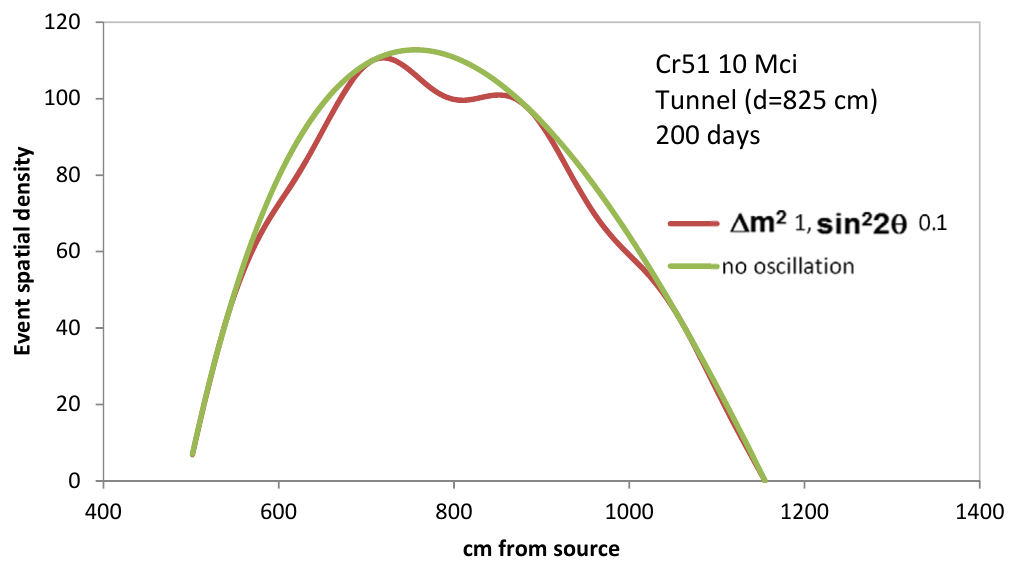}
\caption{
Distribution of the events as a function of the distance from the external 
$^{51}$Cr monochromatic neutrino, obtained with $\Delta m^2=1\ {\rm eV}^2$ and 
$\sin^2(2\theta)=0.1$.
}
\label{Borexino1}
\end{figure}

This location, together with other possible locations inside the detector which 
have been also investigated, is very appropriate for a sterile neutrino search 
with the current Borexino set-up. The distance between the source and the 
detector, and the size of the detector itself, are of the order of few meters, 
thus comparable with the expected oscillation length, that at the mass scale of 
$\Delta m^2=1\ {\rm eV}^2$ is also of this magnitude: therefore a possible 
striking signature of the occurrence of the oscillation involving the putative 
sterile neutrino would be the development across the detector of a clearly 
visible ``spatial wave" of detected neutrino events.

Actually, evidence of neutrino oscillations might be obtained in Borexino by 
exploiting jointly the standard disappearance technique, in which the total 
number of expected events without oscillations is compared with the observed 
rate, and the observation of the spatial waves in the distribution of the 
registered events.  An example of these spatial waves is shown in 
Fig.~\ref{Borexino1} for the case  of a source located externally to the 
detector. The position of each event in Borexino can be reconstructed with a 
precision of about 12~cm at 1~MeV energy, and it scales as $1/\sqrt{E}$. For 
monochromatic sources, like $^{51}$Cr, this automatically ensures a good 
sensitivity; in case of a continuous energy anti-nu source, like 
$^{90}$Sr-$^{90}$Y, the complication of the non-monochromaticity is 
counterbalanced by the zero-background characteristic of the measurement and by 
the higher cross section: both features taken together ensure even superior 
performances for the anti-nu source.  Preliminary evaluations show that in theT.~V.~Ibragimova 
range above 0.02-0.1~eV$^2$ in $\Delta m^2$ (depending upon the energy of the 
emitted neutrinos) and for $\sin^2(2\theta)\ge 0.04$ the sensitivity to 
oscillations into a sterile neutrino should be good with a reasonable neutrino 
or antineutrino source, also thanks to the additional handle of the search of 
the spatial waves.

\begin{figure}[t]
\includegraphics[width=0.49\textwidth]{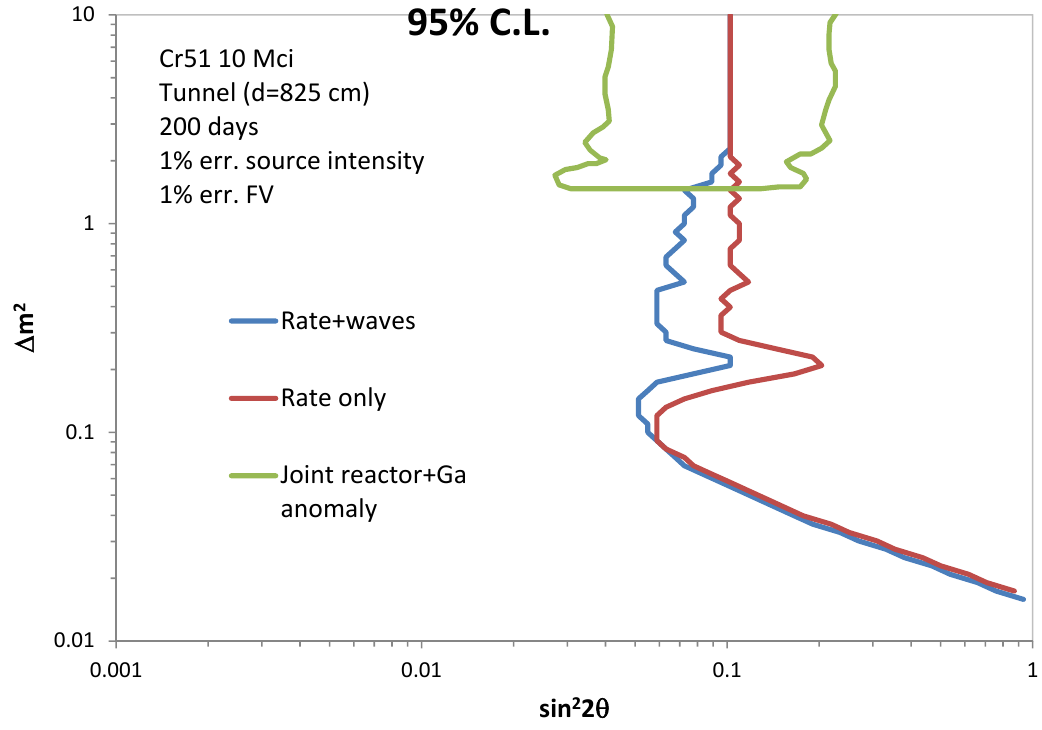}
\includegraphics[width=0.49\textwidth]{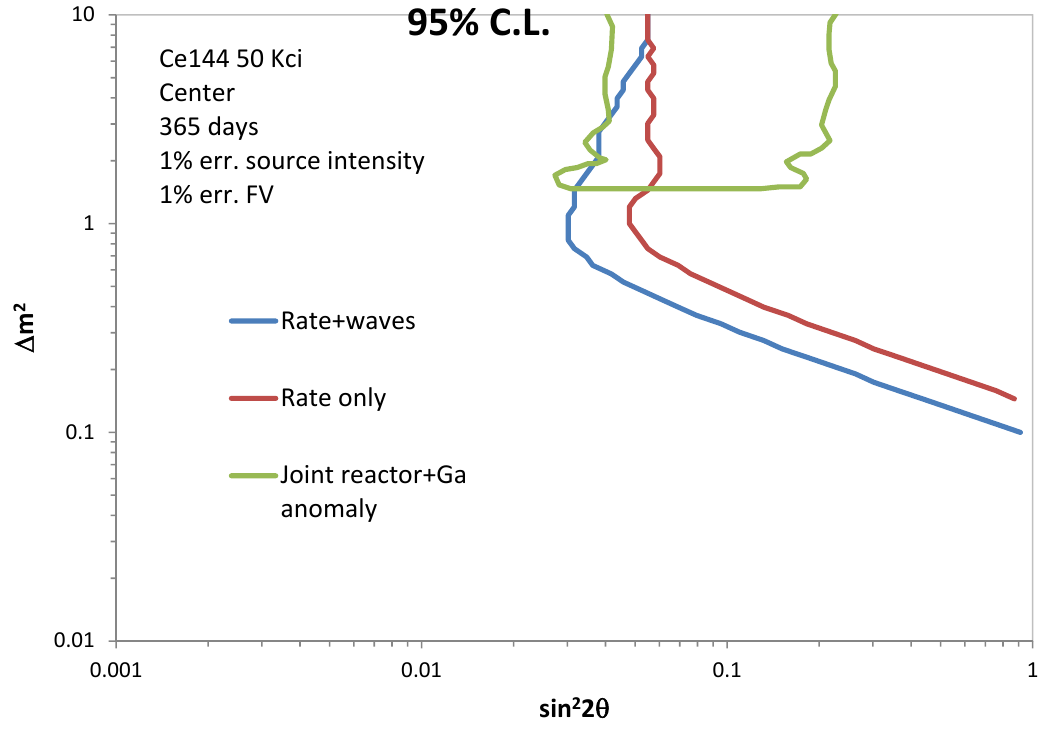}
\caption{
Exclusion plots (both rate only and rate+waves) at the 95\% C.L. illustrating 
the sensitivity reach of the experiment performed with (left) a $^{51}$Cr external 
source of intensity equal to 10~MCi and (right) a $^{144}$Ce source deployed in 
the center of Borexino.  The green contour represents the joint 
Gallium-reactor anomaly.  The significant reach of $^{144}$Ce experiment is clearly 
shown by the overlap of the sensitivity contours with practically the entire joint 
Gallium-reactor anomaly region.
}
\label{Borexino2}
\end{figure}

Fig.~\ref{Borexino2} (left) shows the exclusion plots that might be obtained at 
95\%~C.L., with a 10~MCi $^{51}$Cr source located externally at 8.25~m from the 
centre of the detector. The red line is the curve stemming from the rate-only 
analysis, while the blue contour is obtained through the more complete analysis 
involving also the spatial information; from the comparison of the two it can be 
appreciated clearly the boost to the sensitivity implied by the wave analysis, 
in the region where the spatial waves would be characterized by oscillation 
lengths well matched to the size of the detector. 

In order to highlight the physics reach of a test accomplished with an external 
source, the figure displays also the 95\% contour resulting from the joint 
analysis of the Gallium and reactor anomalies: though not conclusive, an 
external test performed with a sufficiently strong neutrino source will be 
extremely significant, since it will start to address a sizable portion of the 
oscillation parameter region of interest.

Fig.~\ref{Borexino2} (right) shows the sensitivity for the other scenario being 
considered, {\it i.e.}~an antineutrino source deployed in the center of the detector.  
As anticipated above, several characteristics denote the antineutrino source as 
a most powerful choice with respect to the neutrino option: the correlated 
nature of the event detection (a positron-neutron correlation), which makes the 
background irrelevant and allows the possibility to deploy the source directly 
in the scintillator; furthermore, the higher energy spectrum with respect to 
that of $^{51}$Cr implies an higher cross section, and thus for the same number 
of events a lower source intensity, that hence in this respect can be more 
manageable.

The global superior characteristic of the antineutrino scenario is also 
witnessed by the sensitivity features illustrated in Fig 3, where the 95\% 
sensitivity contour covers the entire 95\% joint gallium-reactor region. 
Therefore the internal antineutrino source test appears to be very conclusive, 
being able to disprove or confirm convincingly the sterile neutrino hypothesis. 

It should be pointed out that also in Fig.~\ref{Borexino2} (right) both the rate only 
and the rate + waves analysis are reported, highlighting the zone where the 
powerful tool of the waves detection across the detector is effective to enhance 
the experimental sensitivity. This region is located at higher $\Delta m^2$ with 
respect to the similar "wave" region in the Fig.~\ref{Borexino2} (left), as 
consequence of the different (higher) energy range of the antineutrino source, 
and this is an advantage to better cover the gallium-reactor zone.  

In both the preliminary calculations whose results are illustrated in 
Fig.~\ref{Borexino2}, the two main errors affecting the test 
are considered, {\it e.g.}~the uncertainty on the intensity of the source and the 
error on the fiducial volume, both assumed to amount to 1\%. For the fiducial 
volume such an uncertainty is well within the Borexino capability, as determined 
through numerous calibration efforts. On the other hand, the ambitious 1\% error 
on the source intensity will require an important calibration effort to be 
carried out after its preparation through several complementary measurement 
methods. Anyhow, it should be stressed that in the region of the parameter space 
where the sensitivity relies mostly on the spatial distribution of the detected 
events the degree to which both errors affect the measure is drastically 
attenuated, since the shape of the waves does not depend on the precision of the 
source activity nor on the knowledge of the fiducial volume. 

Moreover, in the antineutrino source scenario it can be conceived to exploit 
the full volume to perform the measurement, due to the positron-neutron 
correlated tag, and thus the fiducial volume error can be ignored, further 
increasing the sensitivity region of the test with respect to 
Fig.~\ref{Borexino2} (right). 

From the general framework described so far it stems that our proposed 
experimental program contemplates the construction and deployment of two 
sources, a mono-chromatic $^{51}$Cr neutrino external source and an 
antineutrino internal source¸ being the $^{90}$Sr-$^{90}$Y ``historical" 
option, cited in the Borexino proposal, likely replaced by the 
$^{144}$Ce-$^{144}$Pr material, emerged recently as the most attractive and 
practical anti-nu solution.

For the former source the path was traced in the 90's by the pioneering work of 
the Gallex/GNO collaboration, which calibrated the Gallium detector twice with 
a $^{51}$Cr neutrino source~\cite{Cribier:1996cq}. Another similar effort was 
successfully accomplished by the SAGE collaboration~\cite{Abdurashitov:1996dp}.

\begin{figure}[t]
\includegraphics[height=0.7\textwidth,angle=270]{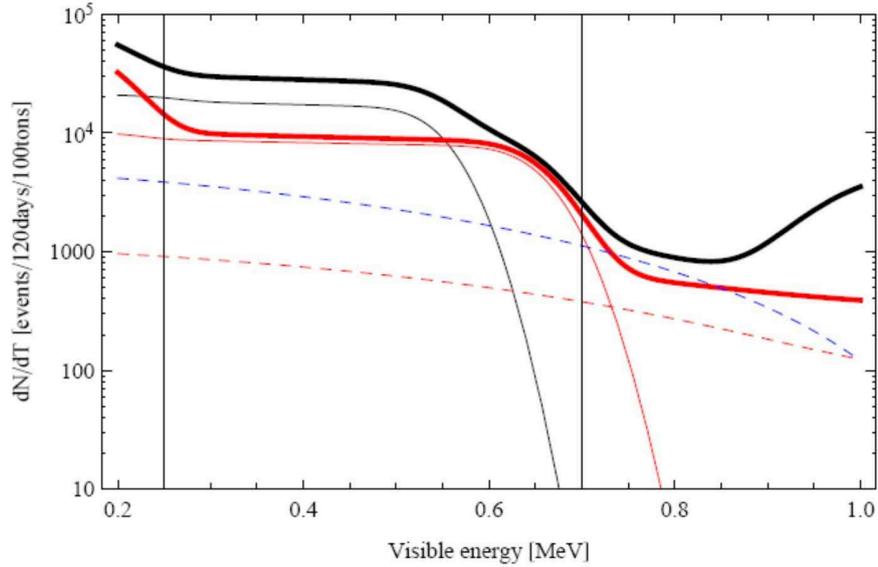}
\caption{
The expected spectra for: $^7$Be neutrinos (red-thin), $^{51}$Cr events 
(black-thin), $^{210}$Bi at the level of 20~cpd/100~t (blue-dashed), CNO 
neutrinos (red-dashed). The red thick line is the solar neutrino total, while 
the black thick line is the $^{51}$Cr signal + background. The rise above 
0.8~MeV is due to cosmogenic $^{11}$C background. The $^{51}$Cr events are 
computed for a 5~MCi source (initial activity) with 120~days of data taking.
}
\label{Borexino4}
\end{figure}

The $^{51}$Cr neutrino source decays by electron capture and emits monochromatic 
neutrinos of energy 746~keV, with a branching ratio of about 90\%. The spectrum 
of the $^{51}$Cr events is shown in Fig.~\ref{Borexino4} together with the main 
expected background. All curves in Fig.~\ref{Borexino4} are normalized to 
120~days of data taking. The $^{51}$Cr technology is known and the enriched 
$^{50}$Cr material is still available at CEA Saclay. Work is in progress to 
understand where a new irradiation could be done, and at which cost. The 
$^{51}$Cr life-time is 39.96~days, so a $^{51}$Cr run might be done with little 
impact on the Borexino solar neutrino program. 

On the other hand, sizable quantities of both materials ($^{90}$Sr-$^{90}$Y and 
$^{144}$Ce-$^{144}$Pr) being considered for the anti-nu test are available 
world-wide as fission products; we are currently investigating the possibility 
to procure in Russia the source whose strength would be of the order of 50~kCi.

The antineutrinos interact in the scintillator via the usual inverse beta decay 
on proton, yielding a prompt signal from the positron and a delayed coincidence 
from the neutron capture. Therefore, as mentioned before, the background is in 
this case virtually zero, though only the part of the neutrino spectrum above 
the 1.8 ~MeV kinematic threshold contributes to the signal. The end-point of the 
spectrum of the $^{144}$Ce-$^{144}$Pr is about 3~MeV, higher than the 2.2~MeV 
featured by $^{90}$Sr-$^{90}$Y, therefore the former originates more events, 
further strengthening the case of its superior features.

Contrary to the external neutrino source, the deployment of the anti-nu source 
in the center of the detector will require a major technical effort and also 
important hardware changes, therefore it is envisioned after the end of the 
Borexino solar phase.

\clearpage
\subsection{Ce-LAND: A proposed search for a fourth neutrino with a PBq 
            antineutrino source\footnote{
	    Proposed by  M.~Cribier, M.~Fechner, Th.~Lasserre, A.~Letourneau, 
	    D.~Lhuillier, G.~Mention, D.~Franco, V.~Kornoukhov, S.~Schoenert, 
	    M.~Vivier}}

As reported in this white paper several observed anomalies in neutrino
oscillation data can be explained by a hypothetical fourth neutrino
separated from the three standard neutrinos by a squared mass
difference of a few eV$^2$.  

We propose in the following an unambiguous search for this
fourth neutrino by using a 1.85~PBq (50 kCi) $^{144}$Ce or $^{106}$Ru
antineutrino $\beta$-source deployed at
the center of a kilo-ton scale detector such as
Borexino~\cite{Alimonti:2000xc}, KamLAND~\cite{Decowski:2008zz}, or
SNO+~\cite{Chen:2005yi}.

Antineutrino detection will be made via the inverse beta-decay (IBD) reaction
$\bar{\nu}_e$ + p $\rightarrow$ $e^+$+n. The delayed coincidence between
detection of the positron and the neutron capture gamma rays will
allow for a nearly background free experiment. The small size
($\sim$10~g) of the source compared to the size a nuclear reactor
core, may allow the observation of an energy-dependent oscillating
pattern in event spatial distribution that would unambiguously
determine neutrino mass differences and mixing angles.

This project is called
Ce-LAND\footnote{http://irfu.cea.fr/Phocea/Vie\_des\_labos/Ast/ast\_sstechnique.php?id\_ast=3139}
  for Cerium in a Liquid Antineutrino Detector~\cite{Cribier:2011fv}.  

\begin{figure}[!h]
\begin{center}
\includegraphics[scale=0.4]{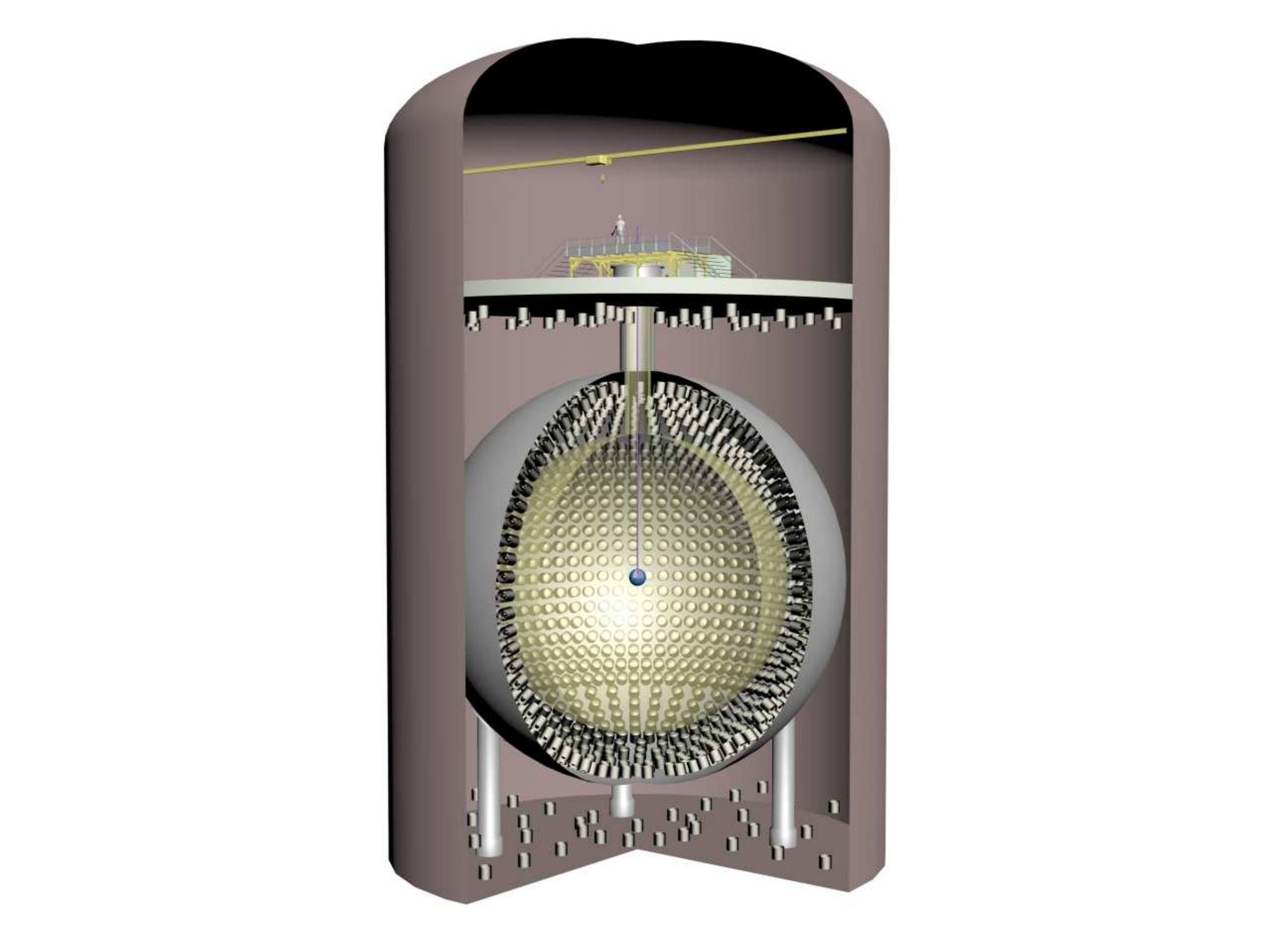}
\caption{\label{f:celand} 
Sketch of the Ce-LAND experiment: a 15 g $^{144}$Ce antineutrino source and
its 80 cm diameter shielding deployed at the center of a large liquid scintillator
detector.}
\end{center}
\end{figure}

\subsubsection*{Experimental Concept}
Large liquid scintillator (LS) detectors, called~{\rm
  LLSD} hereafter, share key features well suited to search for an
eV-scale fourth $\nu$ ($\nubar$) state mixing with $\nu_e$ ($\nuebar$). The
active mass is composed of about thousand tons of ultra-pure LS
contained in a nylon or acrylic vessel. 
The scintillation light is detected via thousands of photomultipliers
uniformly distributed on a stainless steel spherical vessel.
In Borexino and KamLAND the target is surrounded by mineral oil or scintillator contained in a stainless
steel vessel. This buffer is enclosed in a water tank instrumented by
photomultipliers detecting the Cherenkov light radiated by cosmic
muons. 
In the following we study the deployment of a $\nu$ source of energy
spectrum $\mathcal{S}(E_{\nu})$, mean lifetime  $\tau$, and initial
activity $\mathcal{A}_0$, encapsulated inside a thick tungsten (W) and
copper (Cu) shielding sphere, at the center of a~{\rm LLSD}. We consider a
running time $t_e$ with a fully efficient detector. 
The theoretical expected number of interactions at a radius R and energy $E_{\nu} $ can be written: 
\begin{equation}
\frac{d^2N (R,E_{\nu})}{dR dE_{\nu}} = \mathcal{A} _0  \cdot n
\cdot \sigma(E_{\nu}) \cdot \mathcal{S}(E_{\nu})  \cdot \mathcal{P} (R,E_{\nu})
\int _{0}^{t_e} e^{-t/\tau}dt , 
\label{nevt}
\end{equation}
where $n$ is the density of free protons in the target for inverse beta decay,  
$\sigma$ is the cross section. 
$\mathcal{P} (R,E_{\nu})$ is the 2-$\nu$ oscillation survival probability, defined as:
\begin{equation}
\mathcal{P} (R,E_{\nu}) = 1-\sin^2(2\theta_{new})\cdot\sin^2\left(1.27\frac{\Delta
m_{new}^2[{\mathrm eV}^2]R[{\mathrm m}]}{E_{\nu}[{\mathrm MeV}]}\right), 
\label{prob}
\end{equation}
where $\Delta m_{\mathrm new}^2$ and~$\theta_{\mathrm new}$ are the new oscillation
parameters relating $\nu_e$ to the fourth $\nu$. 
In our simulations we assume a 15 cm vertex resolution
and a 5\% energy resolution.
In the no-oscillation scenario we expect a constant $\nu$ rate in
concentric shells of equal thickness (see Eq.~\ref{nevt}).
At 2~MeV the oscillation length is 2.5~m for $\Delta m_{\mathrm new}^2$= 2
eV$^2$, proportional to $1/\Delta m_{\mathrm new}^2$ (see Eq.~\ref{prob}). 
A definitive test of the reactor antineutrino anomaly, independent of
the knowledge of the source activity, would be the observation of the
oscillation pattern as a function of the $\nu$ interaction radius
and possibly the $\nu$ energy. 

\subsubsection*{Limitations of {\large $\nu_{\lowercase{e}}$} Sources}
Intense man-made $\nu$ sources were used for the calibration of
solar-$\nu$ experiments. In the nineties, $^{51}$Cr ($\sim$750~keV,
$\mathcal{A}_0 \sim$MCi) and $^{37}$Ar (814~keV,
$\mathcal{A}_0$=0.4~MCi) were used as a check of the radiochemical
experiments Gallex and Sage~\cite{Cribier:1996cq, Abdurashitov:1998ne, Abdurashitov:2005tb}.
There are two options for deploying $\nu$ sources in LS:
monochromatic $\nu_e$ emitters, like $^{51}$Cr or $^{37}$Ar, or
$\nuebar$ emitters with a continuous $\beta$-spectrum.
In the first case, the signature is provided by $\nu_e$ elastic
scattering off electrons in the LS molecules. This signature
can be mimicked by Compton scattering induced by radioactive and
cosmogenic background, or by Solar-$\nu$  interactions. The
constraints of an experiment with $\nu_e$ impose the use of
a very high activity source (5-10~MCi) outside of the detector
target. In the second option, $\nuebar$ are detected via inverse beta
decay. Its signature, provided by the $e^+$-n delayed
coincidence, offers an efficient rejection of the mentioned
background. For this reason, we focus our studies on $\nuebar$ sources.

\subsubsection*{Choice and Production of {\large $\nuebar$} Sources}
A suitable $\bar{\nu}_e$ source must have $Q_\beta>$1.8~MeV (the IBD
threshold) and a lifetime that is long enough ($\gtrsim$1~month) to allow for
production and transportation to the detector. For individual nuclei,
these two requirements are contradictory so we expect candidate
sources to involve a long-lived low-$Q$ nucleus that decays to a short-lived
high-$Q$ nucleus. We identified four such pairs
$^{144}$Ce-$^{144}$Pr  ($Q_\beta$(Pr)=2.996~MeV), 
$^{106}$Ru-$^{106}$Rh ($Q_\beta$(Rh)=3.54~MeV), 
$^{90}$Sr-$^{90}$Y ($Q_\beta$(Y)=2.28~MeV), 
and $^{42}$Ar-$^{42}$K  ($Q_\beta$ (K)=3.52~MeV), some of them also
reported in~\cite{Kornoukhov:1994zq}.

\begin{figure}[!b]
\begin{center}
\includegraphics[scale=1.]{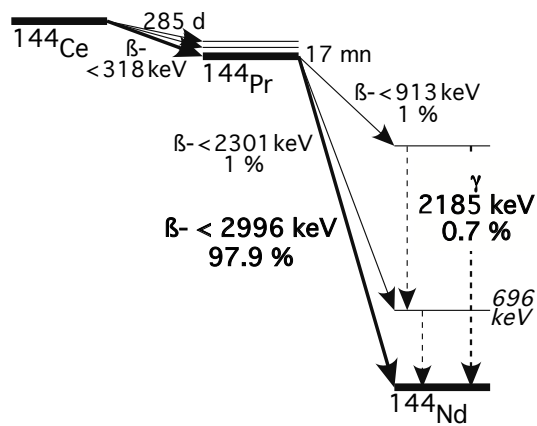}
\caption{\label{f:cepr144} 
In this simplify decay scheme the relatively long lifetime $^{144}$Ce decay
into the short life $^{144}$Pr decaying itself mainly through an energetic
$\beta$ extending up to 2996 keV.
}
\end{center}
\end{figure}

The first three are common fission products from nuclear reactors that
can be extracted from spent fuel rods. While not minimizing the
difficulty of doing this, the nuclear industry does have the
technology to produce sources of the appropriate intensity, at the ppm
purity level. In fact, 10~kCi $^{90}Sr$ sources have been produced and
used industrially for heat generation. Delays obtaining
authorizations for transportation and deployment of the source into
an underground laboratory should be addressed at the start of
the project. 

For this white paper, we concentrate on the $^{144}$Ce source
(Fig. \ref{f:cepr144}) because its $Q_\beta$ is greater than that of
$^{90}$Sr and because it is easier to extract chemically than
$^{106}$Ru. We note also that it has a very low production rate of
high-energy $\gamma$ rays ($>1MeV$) from which the $\nuebar$ detector
must be shielded to limit background events.  Finally cerium is present in
fission products of uranium and plutonium at the level of a few percent.

\begin{center}
\begin{table*}[htb!]
\medskip
\scalebox{.75}{
\begin{tabular}{c|c|c|l|l|l|l|c|c}
  \hline\hline
 Source & F.Y. $^{235}$U/$^{239}$Pu  & $t_{1/2}$ & 1$^{\mathrm st}$
 $\beta ^-$ (keV)& 2$^{\mathrm nd}$ $\beta ^-$ (keV) &  $I_{\gamma>1MeV}$ &$ I_{\gamma>2MeV}$&
 W/kCi & kCi/4\,10$^4$ int./y\\
\hline\hline
\hline
 \multirow{3}{*}{$^{144}$Ce-$^{144}$Pr} &
 \multirow{3}{*}{5.2\%/3.7\%} & \multirow{3}{*}{285 d}  & 318 (76\%) &
 &  & &  \multirow{3}{*}{7.47} & \multirow{3}{*}{43.7}\\
 & & & 184 (20\%) &2996 (99\%) & 1380 (0.007\%) &  2185  (0.7\%) & & \\
 & & & 238 (4\%) & 810 (1\%) & 1489 (0.28\%)& & & \\
\hline
\hline
 \multirow{4}{*}{$^{106}$Ru-$^{106}$Rh} &  \multirow{4}{*}{0.5\%/4.3\%} &
  \multirow{4}{*}{373 d}  &
 \multirow{4}{*}{39.4 (100\%)} & 3540 (78\%)
 & 1050 (1.6\%) &  & \multirow{4}{*}{8.40} &\multirow{4}{*}{23.0}\\
 & & & & 3050 (8\%)  &1128-1194 (0.47 \%) & 2112 (0.04\%)  & & \\
 & & & & 2410 (10\%) & 1496-1562 (0.19 \%) & 2366 (0.03\%) & & \\
 & & & & 2000 (2\%)  & 1766-1988 (0.09 \%) & 3400 (0.016\%) & & \\
\hline\hline
\end{tabular}
}
\caption{\label{tab:sources} Features of $^{144}$Ce-$^{144}$Pr
  and $^{106}$Ru-$^{106}$Rh pairs, extracted from spent nuclear
  fuel. F.Y. are the fission yields of $^{144}$Ce and
  $^{106}$Ru, $t_{1/2}$, their half-lives. $\beta$-end-points
  are given for 1$^{\mathrm st}$ and 2$^{\mathrm nd}$ nucleus of each
  pair. The $I_{\gamma}$'s are the branching ratio of gammas $\gamma$
  rays per beta-decay above 1 and 2~MeV.
The two last columns are the heat produced/kCi and the activity required to get 40,000 events/year.
}
\end{table*}
\end{center}

\subsubsection*{The  {\large $\nuebar$} source and its signal}
We now focus on the unique oscillation signature induced by
an eV-scale sterile $\nu$ at the center of a LLSD. 
\begin{figure}[!t]
\begin{center}
\includegraphics[scale=0.5]{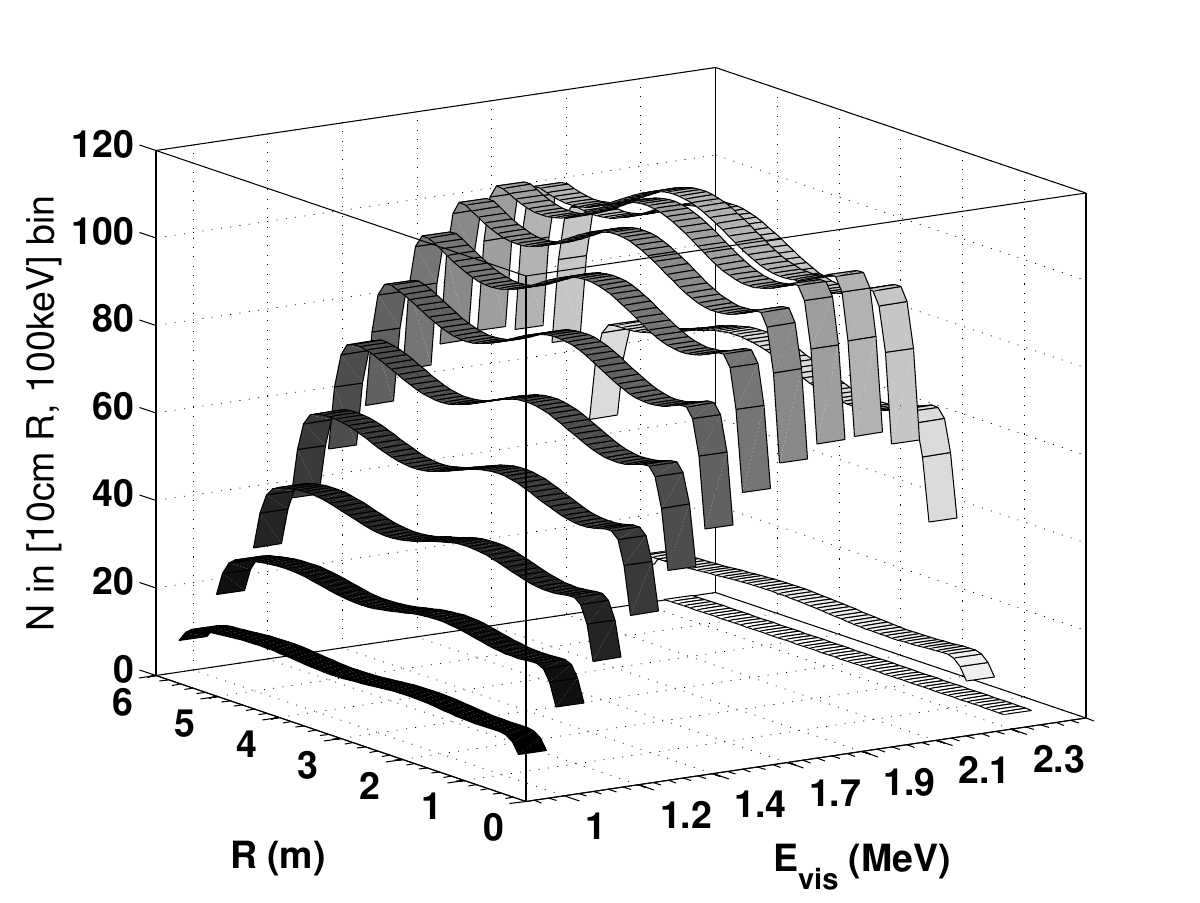}
\caption{\label{f:signal} Advantage of $\nuebar$ sources
  providing both R and E$_{vis}$ oscillation patterns. IBD rate for a 50~kCi
  $^{144}$Ce source deployed at a center of a LLSD, in 10 cm radius bins
  and 100~keV bins of visible energy, E$_{vis}$=E$_e$+2m$_e$. In one
  year, 38,000 $\nuebar$ interact between 1.5~m  and 6~m radius, for
  $\Delta m_{\mathrm new}^2=2$~eV$^2$ and $\sin^2(2\theta_{\mathrm
    new})=0.1$. 
}
\end{center}
\end{figure}
For $^{144}$Ce-$^{144}$Pr, 1.85 PBq (50 kCi) source is needed to reach 40,000
interactions in one year in a LLSD, between 1.5 and 6~m away from the
source (n$_H$=5.3 $\cdot$ 10$^{28}$ m$^{-3}$).
This is realized with 14~g of $^{144}$Ce, whereas the total mass of all
cerium isotopes is~$\sim$1.5 kg, for an extraction from selected fission products.
The compactness of the source, $<$4~cm, is small
enough to be considered as a point-like source for $\Delta m_{\mathrm new}^2$~eV$^2$ oscillation
searches. This source initially releases $\sim$300 W, and it could
be cooled either by convective exchanges with the LS, or via
conduction though an ultrapure copper cold finger connecting the
massive passive shield to a low temperature bath.
$\beta^-$-decay induced $\nuebar$ are detected through IBD. The cross section is
$\sigma(E_e) = 0.956 \,  10^{-43}\times p_e E_e \,\,
\mathrm{cm}^2$, where $p_e$ and $E_e$ are the momentum
and energy (MeV) of the detected $e^+$, neglecting
recoil, weak magnetism, and radiative corrections (\%-level correction). 
The $e^+$ promptly deposits its kinetic energy in the LS and annihilates emitting two 511~keV $\gamma$-rays,
yielding a prompt event, with a visible energy of E$_e$=
E$_\nu$-($m_n$-$m_p$)~MeV; the emitted keV neutron is captured on a free proton
with a mean time of a few hundred microseconds, followed by the emission of a
2.2~MeV deexcitation $\gamma$-ray providing a delayed coincidence
event. The expected oscillation signal for $\Delta m_{\mathrm new}^2=2$~eV$^2$
and $\sin^2(2\theta_{\mathrm new})=0.1$, is shown on Fig.~\ref{f:signal}. 
LLSD are thus well suited to search for an eV-scale fourth $\nu$ state. 
Note that a study of signals of a $^{90}$Sr MCi source external of a
LLSD was done in~\cite{Ianni:1999nk}.

\subsubsection*{Backgrounds}
The space-time coincidence signature of IBD
events ensure an almost background-free detection. Backgrounds are
of two types, those induced by the environment or detector, and those
due to the source and its shielding. 
 
The main concern is accidental coincidences between a prompt
(E$>$0.9 MeV) and a delayed energy depositions (E$>$2.0 MeV) occurring within a
time window taken as three neutron capture lifetimes on hydrogen 
(equivalent to about 772 $\mu$sec), and within a volume of 10~m$^3$
(both positions are reconstructed, this last cut leading to a background
rejection of a factor 100).
\begin{figure}[!t]
\begin{center}
\includegraphics[scale=0.5]{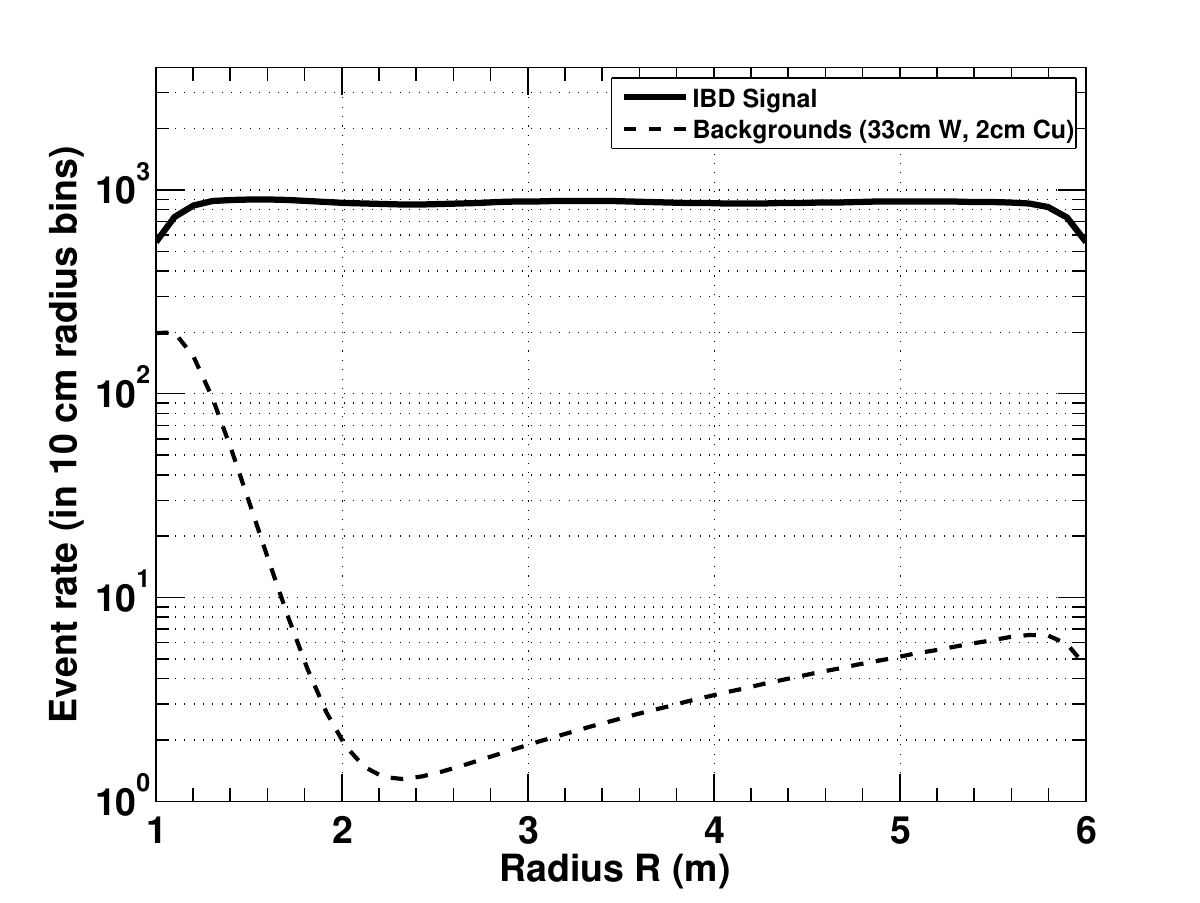} 
\caption{\label{f:signalbkg} 
Signal rate of a 50~kCi $^{144}$Ce deployed for 1 year at a center of
a LLSD (plain line),
compared to the sum of all identified background rates (dashed line), as a
function of the detector radius in 10 cm concentric bins.
A shield made of 33 cm of W and 2 cm of Cu attenuates the backgrounds,
dominated primarily by $^{144}$Ce $\gamma$ lines, then by external
bremsstrahlung of the $^{144}$Ce $\beta$-decay electrons slowing down
in the Cerium material.}
\end{center}
\end{figure}
The main source of detector backgrounds originates from
accidental coincidences, fast neutrons, and the long-lived muon induced
isotopes $^9$Li/$^8$He and scales with $R^2$ when using concentric $R$-bins. 
These components have been measured in-situ for the Borexino
geo-$\nu$ searches~\cite{Bellini:2010hy}, at 0.14$\pm$0.02
counts/day/100 tons. Being conservative we increase it to
10 counts/day/100 tons in our simulation.

Geologic $\nuebar$ arising from the decay of radioactive isotopes of
Uranium/Thorium in the Earth have been measured at a rate of a few 
 events/(100 ton.year) in KamLAND~\cite{Inoue:2010} and Borexino~\cite{Bellini:2010hy}. 
Reactor $\nuebar$ emitted by the $\beta$-decays of the fission
products in the nuclear cores have been measured in KamLAND at a rate of
$\sim$10~events/(100 ton.year) in the energy range of interest~\cite{Inoue:2010}. 
We use a rate of 20~events/(100 ton.year), which is negligible with
respect to the $\nuebar$ rate from a kCi source. 

The most dangerous source background originates from the energetic
$\gamma$ produced by the decay through excited states of $^{144}$Pr (Table~\ref{tab:sources}).  
We approximate $\gamma$ ray attenuation in a shield of 33~cm of W and
2~cm of Cu with an exponential attenuation
law accounting for Compton scattering and photoelectric effect.
The intensity of 2185~keV $\gamma$ rays is decreased by a factor
$<10^{-12}$ ($\lambda _W \sim$1.2~cm)~\cite{Nist}, to reach a tolerable rate. 

The energy spectrum of external bremsstrahlung photons in the
cerium is estimated with a simulation using the cross
section of~\cite{Koch:1959zz}. Results were confirmed with a
GEANT4~\cite{Agostinelli:2002hh} simulation. The number of photons above a prompt
signal threshold of 0.9 MeV is $6.5 \cdot 10^{-3}$ photons per
$\beta$ decay, and $10^{-4}$ photon per $\beta$ decay $>$2.0~MeV. 

An important remaining background source could be the W shield
itself. Activities at the level of ten to hundreds mBq/kg have been
reported. We anticipate the need of an external layer of ultrapure
copper, set to 2~cm in our simulation. It allows one to achieve the
radiopurity and material compatibility requirements. Assuming a
$\sim$4 tons shield we consider a prompt and delayed event rates of
50~Hz and 25~Hz, respectively. 
The escaping $\gamma$ are attenuated in the LS, assuming
a 20~cm attenuation length~\cite{Nist}. Beyond a distance of 1.5~m from the
source the backgrounds become negligible.
Any of the bremsstrahlung photons or shielding backgrounds can
account for either the prompt or delayed event, depending on their energy. 
The sum of the backgrounds integrated over their energy spectrum is
shown on Fig.~\ref{f:signalbkg}, supporting the case of kCi $\nuebar$
source versus MCi $\nu_e$ source for which solar-$\nu$'s become an
irreducible background.
A light doping of the LS with gadolinium or an oil buffer surrounding the shielding would further
suppress backgrounds; finally, non-source backgrounds could be
measured in-situ during a blank run with an empty shielding.

\subsubsection*{Sensitivity}
\label{sensitivity}
We now assess the sensitivity of an experiment with a 50~kCi $^{144}$Ce
source running for 1~year. 
\begin{figure}[bh!]
\begin{center}
\includegraphics[scale=0.5]{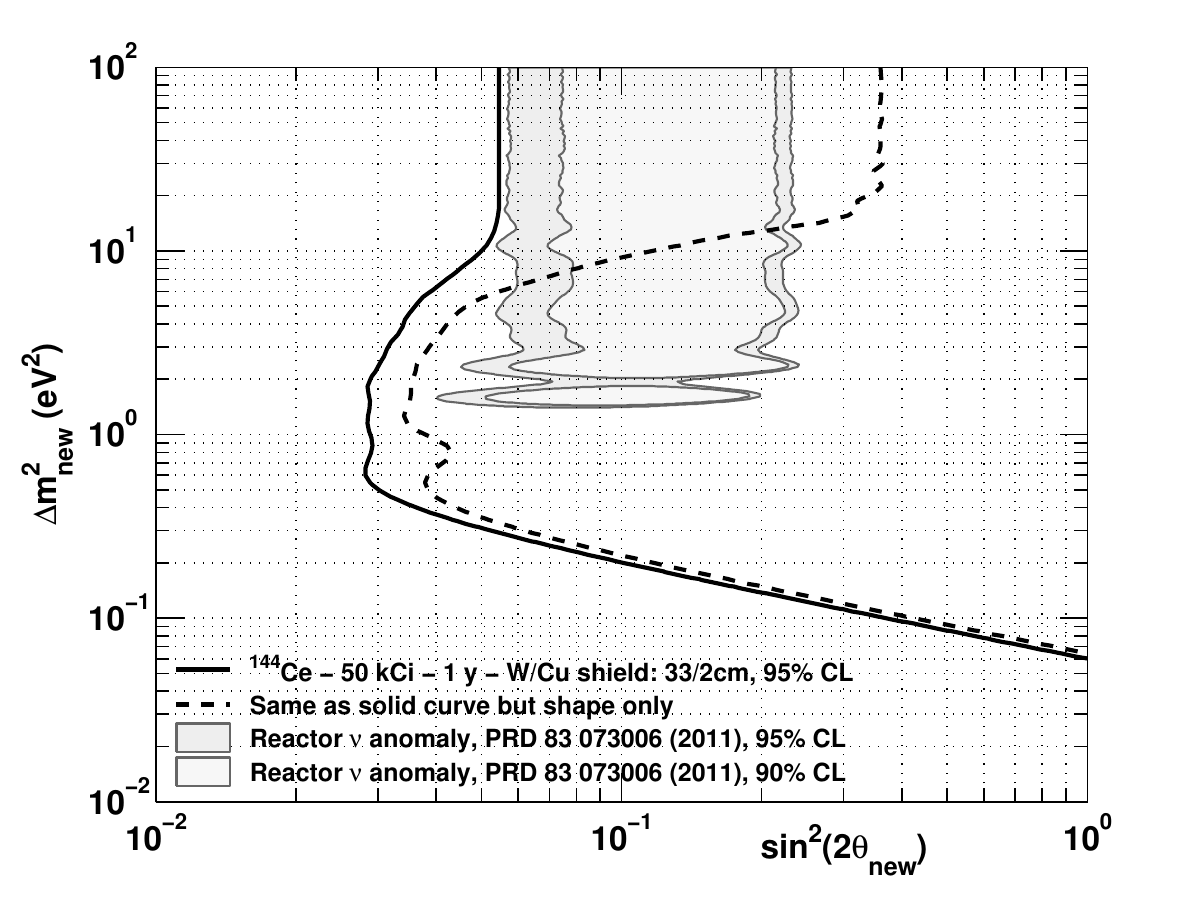}
\caption{\label{f:sensitivity} 95\% C.L. exclusion limits of the 50
  kCi.y $^{144}$Ce experiment obtained in the $\Delta m_{\mathrm new}^2$
  and $\sin^2(2\theta_{\mathrm new})$ plane (2~dof). Our result (plain
  and dashed lines with and w/o knowledge of source activity) is
  compared to the 90\% and 95\% C.L. inclusion domains given by the combination
  of reactor neutrino experiments, Gallex and Sage calibration sources
  experiments, as described in Fig. 8 of~\cite{Mention:2011rk}  (gray
  areas).}
\end{center}
\end{figure} 
With the shield described above and using events between
1.5~m and 6~m, the background is negligible. With $\Delta
m_{\mathrm new}^2=2$~eV$^2$ and $\sin^2(2\theta _{\mathrm new})=0.1$, the interaction 
rate decreases from 40,000 to 38,000 per year. 
The 95\% C.L. sensitivity is extracted through the following function:
\begin{equation}
\label{chi2estimator}
\chi^2 = \sum_i \sum_j
\frac{\left( N_{\mathrm obs}^{i,j}-(1+\alpha)N_{\mathrm exp}^{i,j}\right)^2}{N_{\mathrm
    exp}^{i,j}(1+\sigma_b^2 N_{\mathrm exp}^{i,j})} + \left(
  \frac{\alpha}{\sigma_N}\right)^2,
\end{equation}
where $N_{\mathrm obs}^{i,j}$ are the simulated data in the no-oscillation case and
 $N_{\mathrm exp}^{i,j}$ the expectations for a given oscillation scenario, in each energy $E_i$ and radius $R_j$ bin. 
$\sigma_b$~is a 2\% fully uncorrelated systematic error, accounting
for a fiducial volume uncertainty of 1\% in a calibrated
detector, as well as for ($e^+$, n)  space-time coincidence detection
efficiencies uncertainties at the sub-percent level. 
$\sigma_N$ is a normalization error of 1\%,
describing for the source activity uncertainty (from calorimetric
measurement, see~\cite{Kornoukhov:1997cv}), and $\alpha$ is the associated
nuisance parameter. 
Fig.~\ref{f:sensitivity} clearly shows that 50~kCi of $^{144}$Ce
allows us to probe most of the reactor antineutrino anomaly parameter
space~\cite{Mention:2011rk} at 95\% C.L. An analysis assuming no knowledge on
the source activity shows that the oscillatory behavior can be established for $\Delta m_{\mathrm new}^2 <10$~eV$^2$.
Fig.~\ref{f:sensitivity5sigma} provides the same analysis as described
before, but the confidence level is now 5$\sigma$. This illustrates
the potential of the Ce-LAND project to definitively test the reactor
and gallium neutrino anomalies~\cite{Mention:2011rk}.  
\begin{figure}[bh!]
\begin{center}
\includegraphics[scale=0.5]{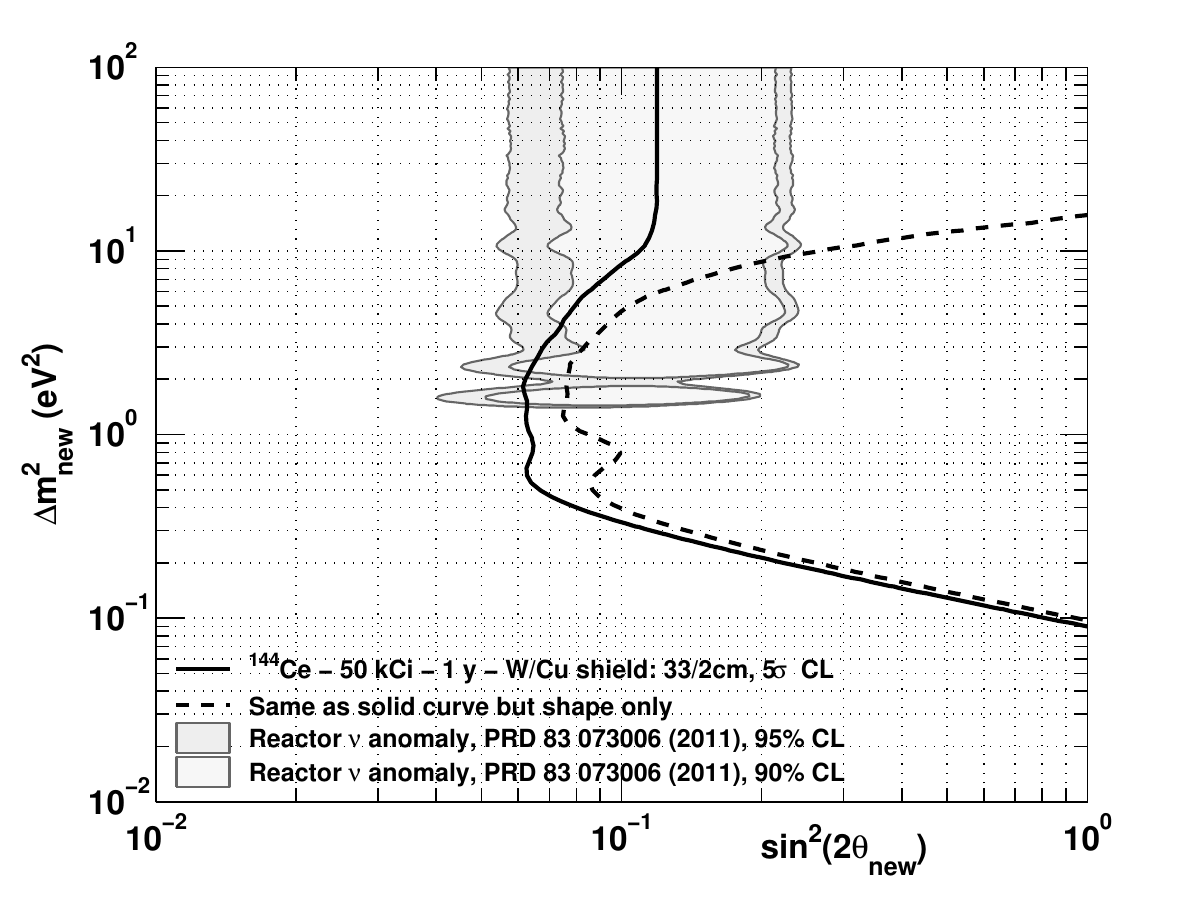}
\caption{\label{f:sensitivity5sigma} 5$\sigma$ exclusion limits of the 50
  kCi.y $^{144}$Ce experiment obtained in the $\Delta m_{\mathrm new}^2$
  and $\sin^2(2\theta_{\mathrm new})$ plane (2~dof).}
\end{center}
\end{figure} 
We note that a 5~kCi source would be enough to test the anomaly at
90\%~C.L. This is illustrated on Fig.~\ref{f:sensitivity5kci}
\begin{figure}[bh!]
\begin{center}
\includegraphics[scale=0.5]{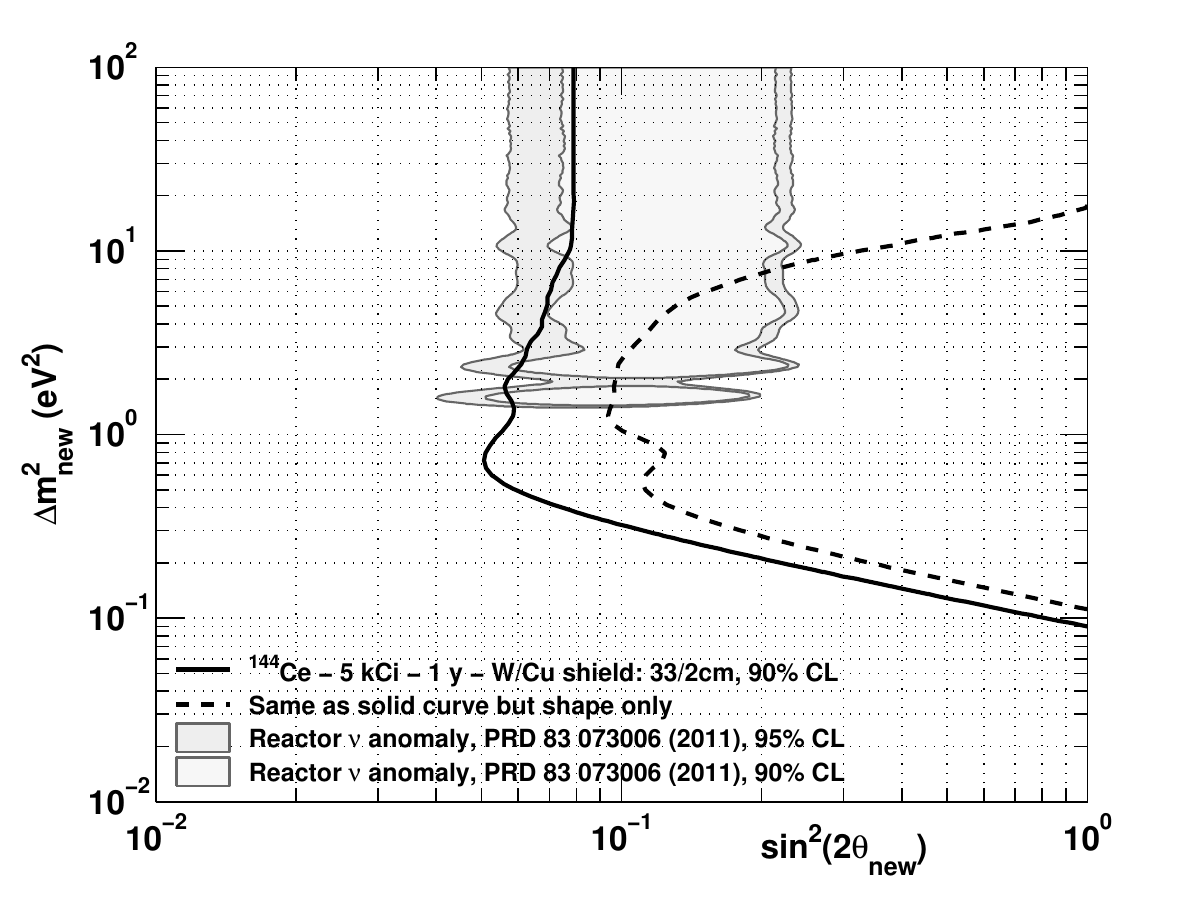}
\caption{\label{f:sensitivity5kci} 90\% C.L. exclusion limits of the 5
  kCi.y $^{144}$Ce experiment obtained in the $\Delta m_{\mathrm new}^2$
  and $\sin^2(2\theta_{\mathrm new})$ plane (2~dof). The reactor
  antineutrino anomaly can be tested, but it requires the knowledge of
the source activity at the percent level. The low statistics does not
allow a powerful shape-only analysis at low mixing angle values.}
\end{center}
\end{figure} 
To conclude we simulate an oscillation signal induced by a fourth
neutrino state mixing with the electron antineutrino, assuming $\Delta m_{\mathrm
  new}^2$=2.35 $eV^2$ and $\sin^2(2\theta_{\mathrm new})=0.1$. Results are
displayed in Fig.~\ref{f:sensitivity50kci_osc}. A precise measurement
of the mixing parameters is obtained at the 3$\sigma$ level.
\begin{figure}[bh!]
\begin{center}
\includegraphics[scale=0.5]{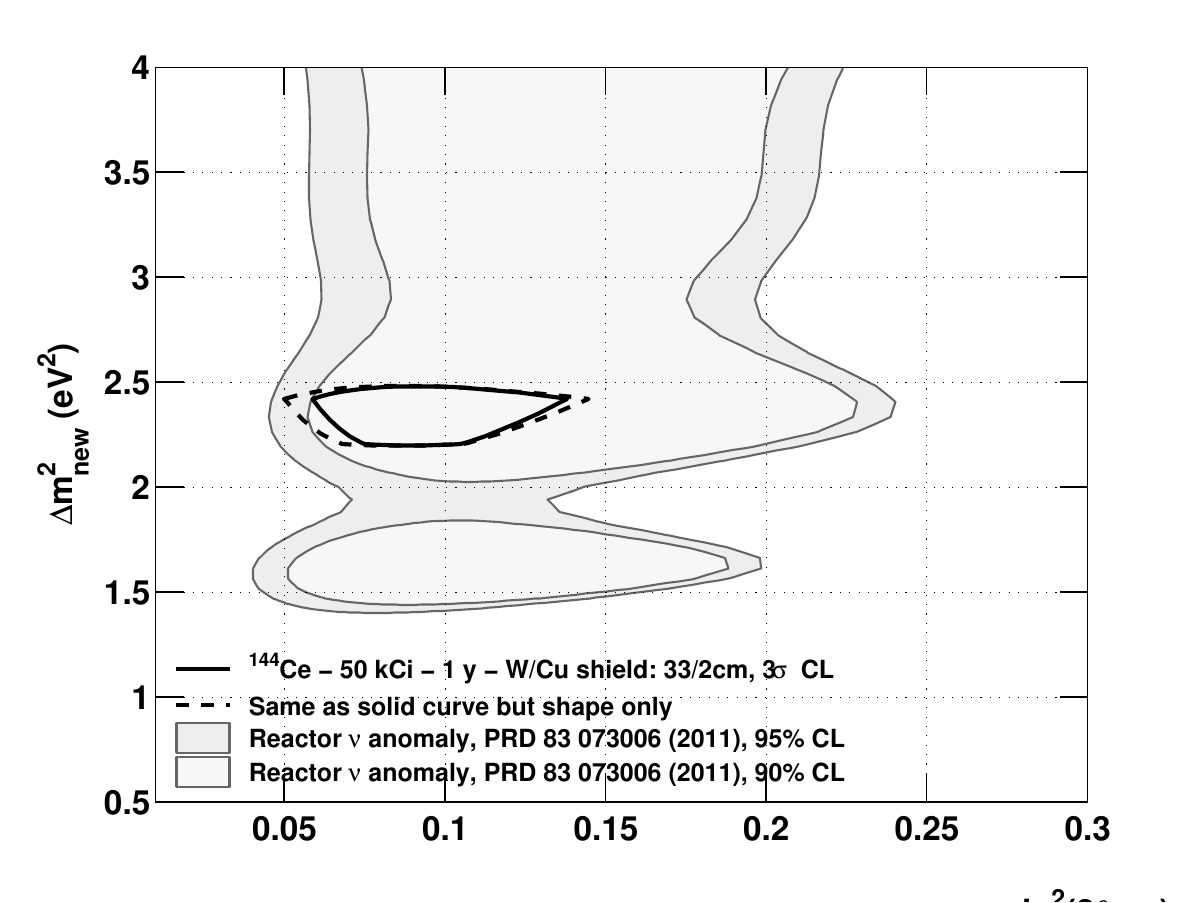}
\caption{\label{f:sensitivity50kci_osc} 3$\sigma$ contours obtained with a 50
  kCi.y $^{144}$Ce experiment in the $\Delta m_{\mathrm new}^2$
  and $\sin^2(2\theta_{\mathrm new})$ plane (2~dof). The true oscillation
  parameters simulated are as follows: $\Delta m_{\mathrm
  new}^2$=2.35 $eV^2$ and $\sin^2(2\theta_{\mathrm new})=0.1$.}
\end{center}
\end{figure} 

\clearpage
\subsection{Search for Sterile Neutrinos with a Radioactive Source at Daya
Bay\footnote{Proposed by: D.A.~Dwyer, P.~Vogel (Caltech), K.~M.~Heeger, 
B.R.~Littlejohn (University of Wisconsin)}}
\label{DYB}

Data from a variety of short-baseline experiments as well as astrophysical observations and
cosmology favor the existence of additional neutrino mass states  beyond the 3 active species in
the standard model. Most recently, a re-analysis of short-baseline reactor neutrino experiments
found a 3\% deficit between the predicted antineutrino flux and observations
\cite{Mention:2011rk}. This has been interpreted as indication for the existence of at least one
sterile neutrino. with a mass splitting of $\sim 1{\rm eV}^2$ \cite{Giunti:2011hn}. The possible
implications of additional sterile neutrino states would be profound and change the paradigm of
the standard model of Particle Physics. As a result, great interest has developed in testing the hypothesis of sterile neutrinos and providing a definitive resolution to the question if sterile neutrinos exist \cite{SNAC, FNAL}.  

We propose to use the far site detector complex of the Daya Bay reactor experiment together with a compact PBq $\overline{\nu}_e$ source as a location to search for sterile neutrinos with $\geq$~eV mass~\cite{Dwyer:2011xs}. The Daya Bay reactor experiment is located at the Daya Bay Nuclear Power Plant near Shenzhen, China and designed to make a high-precision measurement of the neutrino mixing angle $\theta_{13}$ using antineutrinos from the Day Bay reactor complex \cite{Guo:2007ug}.  The experiment has three underground sites, two at short distances from the reactors ($\sim$400~m) with two $\overline{\nu}_e$  detectors each, and one at a further baseline ($\sim$1.7~km) with four $\overline{\nu}_e$ detectors. The far site detector complex of the Daya Bay reactor experiment houses four 20-ton antineutrino detectors with a separation of 6~m. 

When combined with a compact radioactive $\overline{\nu}_e$ source the Daya Bay far detectors provide a unique setup for the study of  neutrino oscillation with multiple detectors over baselines ranging from 1.5-8~m. The geometric arrangement of the four
identical Daya Bay detectors and the flexibility to place the $\overline{\nu}_{e}$ source at multiple locations outside the antineutrino detectors and inside the water
pool allows for additional control of experimental systematics. Daya Bay's unique feature of being able to use multiple detectors and multiple possible source positions will allow us to cross-check any results. In addition, the water pool surrounding the four far-site Daya Bay detectors provides natural shielding and source cooling minimizing technical complications resulting from a hot, radioactive source.  As a source we propose to use a heavily shielded, 18.5 PBq $^{144}$Ce source approximately 16~cm in diameter (${\Delta}Q=$ 2.996~MeV). 

\begin{figure}[h]
\includegraphics[width=0.85\textwidth,angle=0]{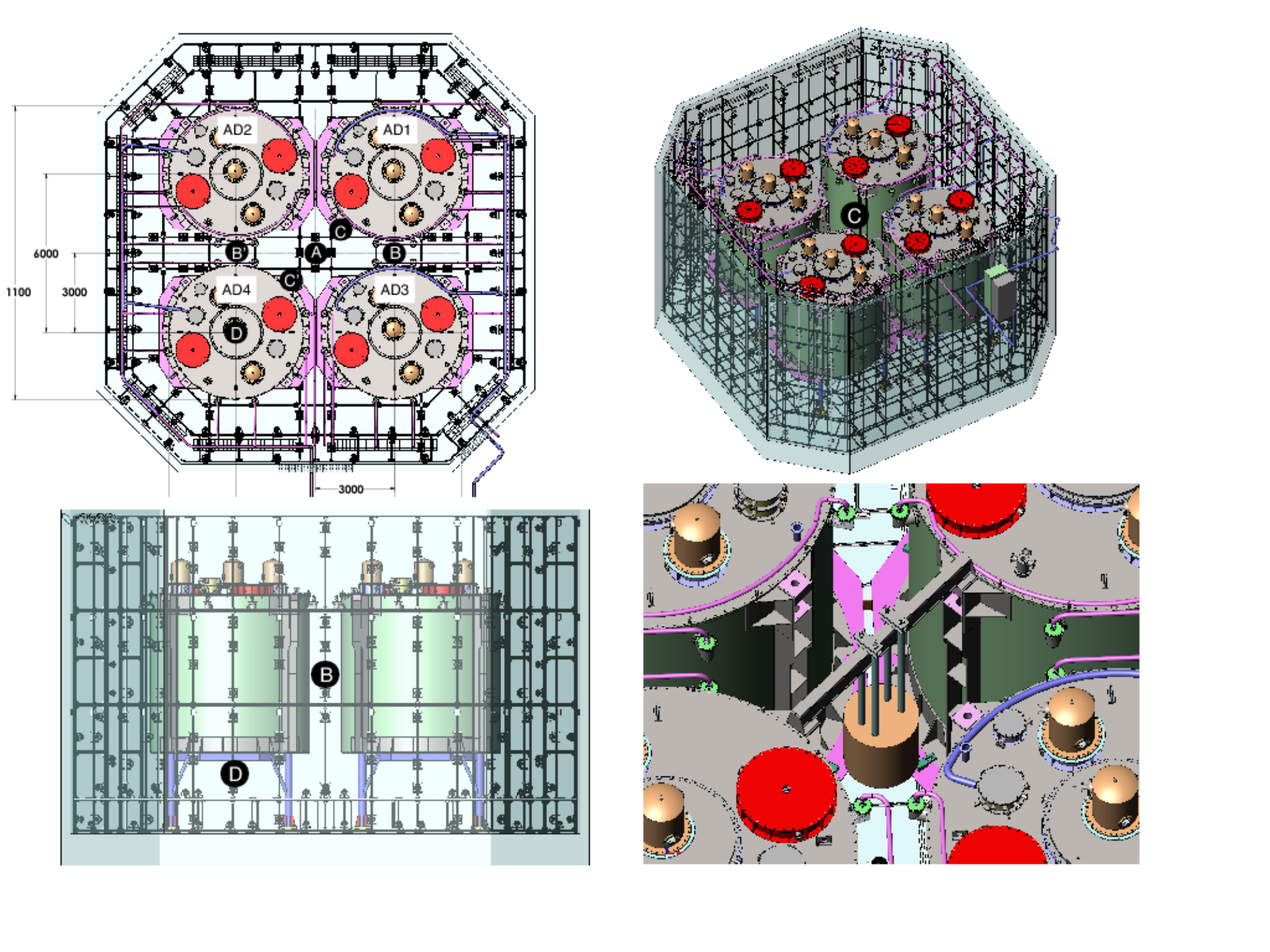}
\caption{Model of the four antineutrino detectors (AD1-4) in the Daya Bay Far Hall. Left: Top and side view of the
 Far Hall with water pool (octagonal frame), four antineutrino detectors
  (grey cylinders) on their support stands (pink), and water cosmic ray veto photomultipliers and support structure (small
  black features).  {\em A}, {\em B}, {\em C}, and {\em D} mark
  possible antineutrino source locations. Positions {\em B'} and {\em C'} are symmetric to  {\em B} and {\em C} and can be used as cross-checks and for systematic studies. Detector dimensions are given in mm. Right: ISO view of the Far Hall.  {\em A}, {\em B}, {\em C} are all at half-height of the antineutrino detectors, {\em D} is directly below it.  Right bottom: Illustration of suspending a source between the detectors in the water pool.
  Figures adapted from Ref.~\cite{Guo:2007ug, DayaBayTAUP2011}
 .}
\label{fig:FarHallfig2}
\end{figure} 

 This experimental setup can probe sterile neutrino oscillations most powerfully by measuring spectral distortions of the energy and baseline spectrum.  If the source's $\overline{\nu}_e$ rate normalization is well-measured, further information can be provided by measuring total rate deficits.  The dominant background of this experiment,  reactor $\overline{\nu}_e$, will be measured to less than 1\% in rate and spectra by the near-site detectors.  In addition, the detector systematics of all detectors will be well-understood after 3 years of dedicated $\theta_{13}$ running, minimizing expected detector-related systematics. 

The proposed Daya Bay sterile neutrino experiment can probe the 0.3-10 eV$^2$ mass splitting range to a sensitivity of as low as sin$^2$2$\theta_{new}<$0.04 at 95\% CL.  The experiment will be sensitive at 95\% CL to most of the 95\% CL allowed sterile neutrino parameter space suggested by the reactor neutrino anomaly, MiniBooNE, LSND, and the Gallium experiments.  In one year, the 3+1 sterile neutrino hypothesis can be tested at essentially the full suggested range of the parameters $\Delta m^2_{\rm new}$ and $\sin^2 2\theta_{\rm new}$ (90\% C.L.).  

\begin{figure}[htb]
\includegraphics[width=.9\textwidth,angle=0]{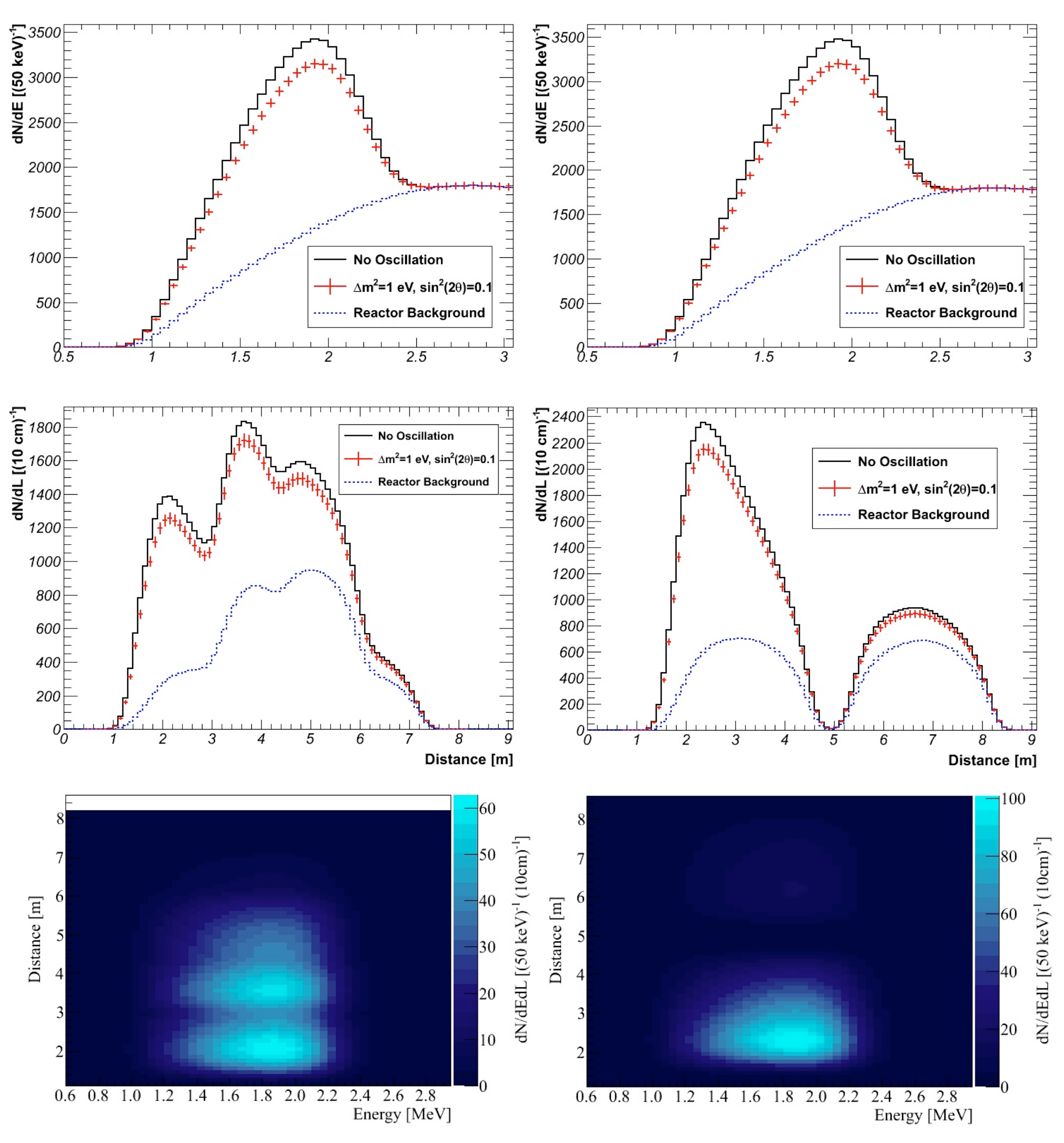}
\caption{Energy and position dependence of the event rates in the antineutrino detectors.  The bottom row shows the 2-dimensional distributions of event rate versus energy and distance from source.  Top and middle rows are the 1-dimensions projections of expected events versus energy (top) and distance from source (middle) for the case of no oscillation (black histogram), the observed event rate in case of $\overline{\nu}_{e} \rightarrow \nu_{s} $ oscillation (red points), and the reactor $\overline{\nu}_{e}$ background (blue dots). Left panels correspond to source position $C$ while right panels are for source position $B$. }
\label{fig:AnalysisPlotfig4}
\end{figure} 

\begin{figure}[h]
\includegraphics[width=.9\textwidth,angle=0]{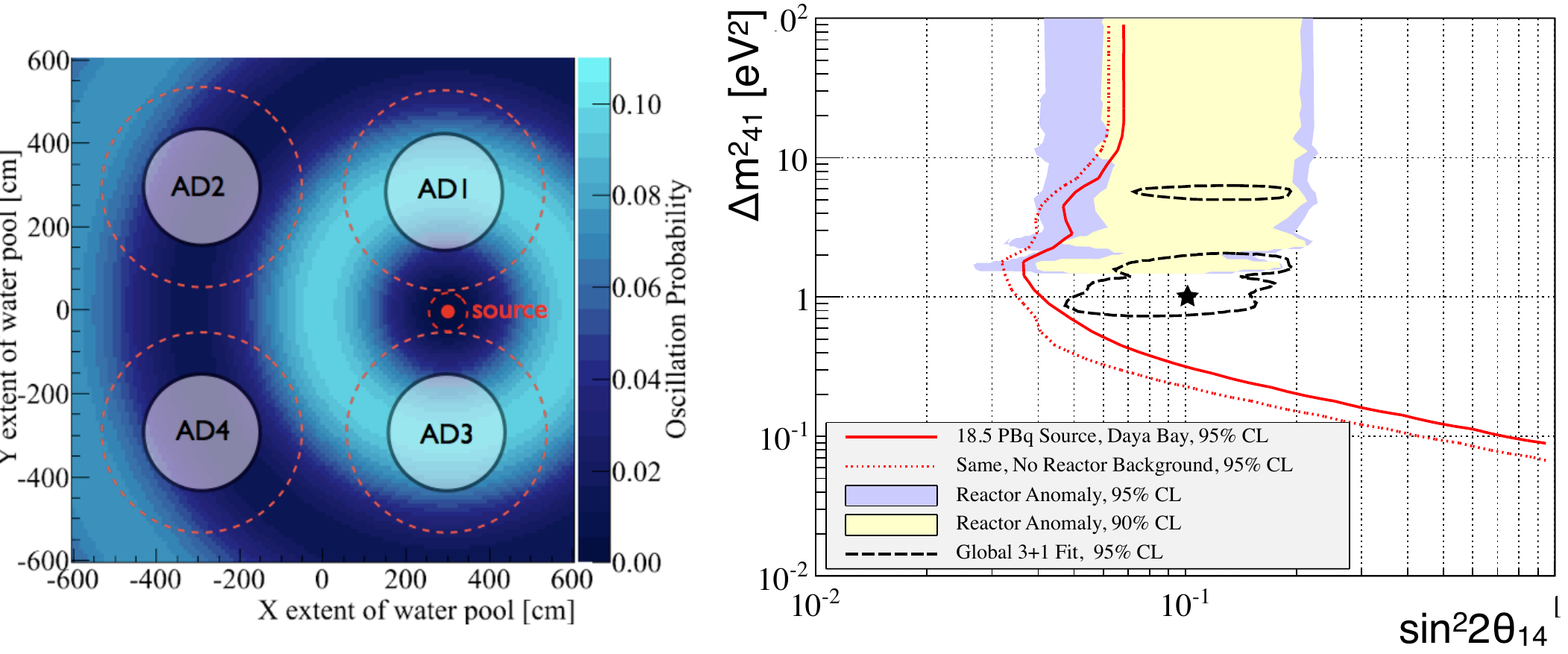}
\caption{Left: 
Illustration of sterile neutrino oscillation in the Daya Bay Far Hall and top view of the geometric arrangement of the antineutrino detectors and the $\overline{\nu}_{e}$ source. Source at position {\em B}. Figure shows an overlay of the positions of the active regions of the antineutrino detectors with the disappearance probability for $\overline{\nu}_{e} \rightarrow \nu_{s}$ oscillation with $\Delta m^2_{41}$=1~eV$^2$ and sin$^22\theta_{14}$=0.1 into sterile species.  Active regions of the source and detectors are shown in solid red and grey color respectively, while the physical outer dimensions of the source and detectors are indicated by the dashed lines. 
Right:  Sensitivity of a $\overline{\nu}_{e}$ search at Daya Bay to
  the oscillation parameters $\Delta m^2_{41}$ and sin$^22\theta_{14}$
  assuming a 500~kCi $^{144}$Ce source at position $B$ in the Daya Bay
  Far Hall. We show the 95\% C.L. sensitivity of the Daya Bay source
  experiment with reactor background (red solid) and without (red
  dashed), the 90\% and 95\% C.L. preferred regions of the reactor
  anomaly (shaded yellow and blue)~\cite{Mention:2011rk}, and the 95\%
  best-fit region from a 3+1 global fit to all neutrino data (dashed
  black)~\cite{Giunti:2011hn}. The parameter space to the left and above
  the Daya Bay sensitivity curve will be excluded at 95\% C.L.}
\label{fig:Sensfig6}
\end{figure} 

 In order to realize such an experiment,  R\&D towards the development of a PBq $\overline{\nu}_e$ source is necessary.  The process of selectively harvesting fission products from spent nuclear fuel has been developed in the nuclear reprocessing industry, and will need to be tailored to remove $^{144}$Ce with high efficiency and purity from a small number of spent fuel assemblies.  The necessary R\&D and development work can be conducted in the years ahead during the $\theta_{13}$ measurement at Daya Bay.

\clearpage
\subsection{SNO+Cr\footnote{Proposed by P.~Huber and J.M.~Link(Virginia
Tech)}}
\label{SNO+Cr}
As was noted in Section~\ref{LENS-Sterile}, combining a low-energy solar neutrino 
detector sensitive with an electron capture neutrino source like $^{51}$Cr, can result in
an experiment for sensitive sterile neutrino search.  There are currently three detectors
in operation or under construction that are designed to study solar neutrinos down
around the 861~MeV $^7$Be neutrino peaks -- which is only a little more energetic than 
the 751~KeV $^{51}$Cr neutrino.  They are Borexino~\cite{Arpesella:2008mt}, 
KamLAND~\cite{Decowski:2008zz}, and SNO+~\cite{Chen:2005yi}.  The best way to use a 
radioactive neutrino source is to put it in the middle of the detector, which will 
typically result in a factor of 3 or more neutrino interactions than for a source placed 
outside the detector.  Of the thee active solar neutrino detectors, only SNO+ has a neck 
wide enough accommodate an appropriately shielded neutrino source.  Hence this concept 
shall be referred to as SNO+Cr.

Unlike the LENS-Sterile proposal discussed in Section~\ref{LENS-Sterile}, where the
neutrinos are detected through a pure charged current process, when a $^{51}$Cr source 
is used with an undoped liquid scintillator (LS) detector the reaction channel is 
neutrino-electron elastic scattering.  While on the positive side, the elastic scattering 
interaction rate in LS is two orders of magnitude larger than the $^{115}$In interaction 
rate in LENS, on the negative side, there is no coincidence tag to eliminate backgrounds.  
So care must be taken in preparing detector and source materials to reduce radioactive
backgrounds inside the detector.

The source is modeled as a 4.5~cm diameter by 4.5~cm tall bundle of Cr rods embedded in
the center of a 16~cm radius sphere of tungsten alloy.  The tungsten shield reduces the
rate of gamma-rays from $^{51}$Cr decay to a few Hz at its surface.  In this analysis it 
is assumed that impurities, that have lead to MeV gamma-rays in past sources, have been
eliminated prior to source irradiation.  Shielding of only 5~cm is sufficient to reduce the 
rate of 320~keV gammas, present in 10\% of all $^{51}$Cr decays, to the Hz level.  
Instead the gamma rate at the surface of a 16~cm shielded is dominated by the internal 
bremstrahlung (IB) gammas which are present in about a part in 2000 
decays~\cite{Mutterer:1973zza}.  Only a tiny fraction of gammas in the high energy tail 
of the IB spectrum pass through the shielded, but they, nonetheless, drive the shield 
design. 

\begin{figure}[b!]
\centerline{
\includegraphics[width=\textwidth]{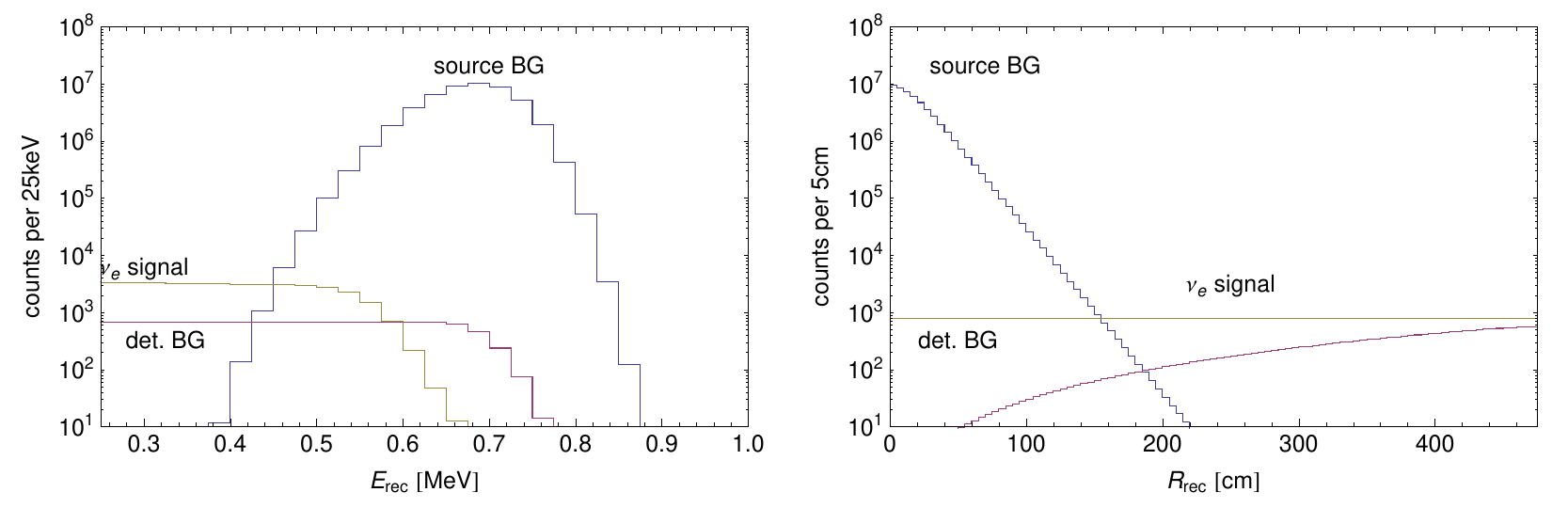}
}
\caption{\label{SNO+rates}Plots of signal events, source background and detector
background as a function of energy (left) and radius (right).}
\end{figure}

It should be noted that solar neutrinos interactions may constitute a significant fraction
of the detector background.  Fortunately, the detector background, which is present during 
source-free running will be measured directly to high precision.  On the other hand, the
source background is only present when the source is active, and, assuming that it is
dominated by gamma-rays from $^{51}$Cr decay, it decays away at the same rate as the Cr
neutrino flux.   

\begin{figure}
\centerline{
\includegraphics[width=0.7\textwidth]{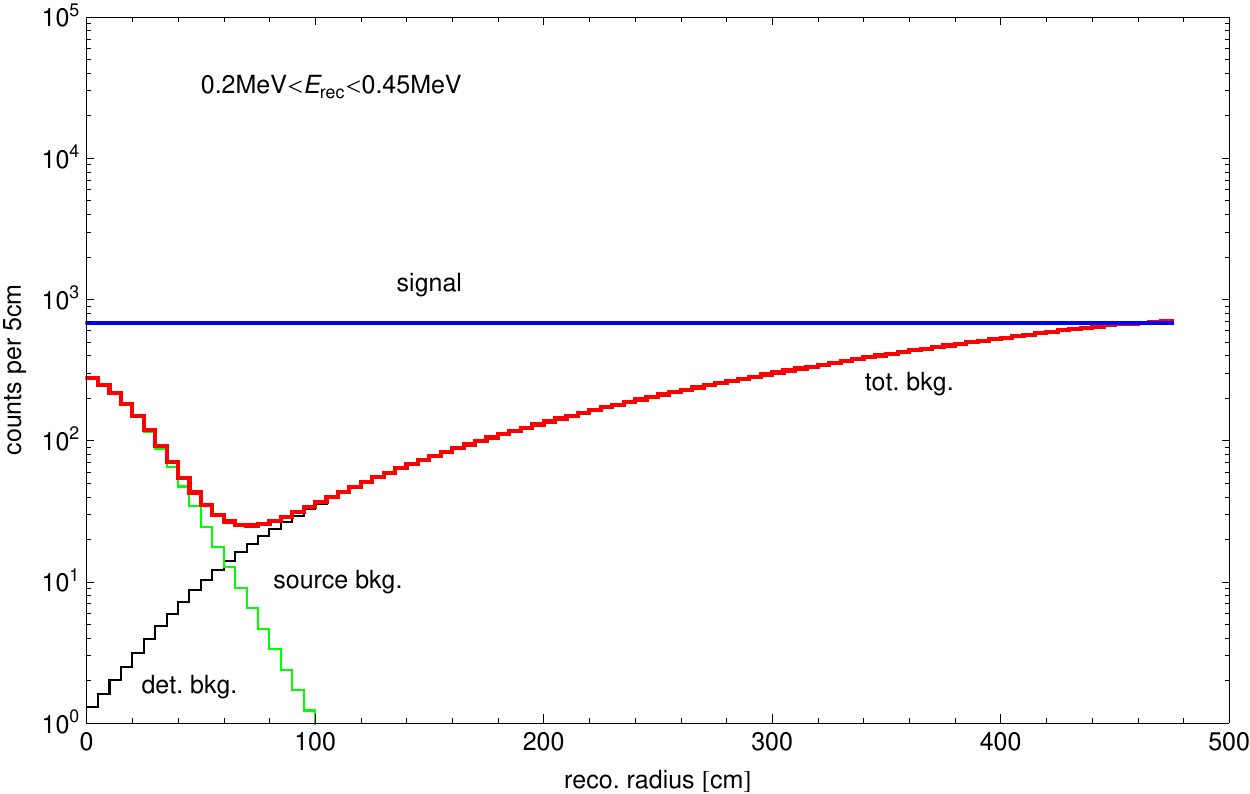}
}
\caption{\label{rate-vs-radius}Signal, source background and detector background event
rates after a reconstructed energy cut and plotted as a function of radius in the 
detector.}
\end{figure}

The left side of figure~\ref{SNO+rates} shows the resolution smeared energy spectrum of 
the elastic scattering signal events, the source gamma-ray background and the projected 
detector background for SNO+\footnote{Personal communication with Gabriel Orebi Gann}. 
It shows that the the source background is only significant at energies above 450~keV 
and therefore can be eliminated with an energy cut..  The right side of 
figure~\ref{SNO+rates} shows the event rates as a function of detector radius.  Without 
oscillations the signal rate is flat as a function of radius.  The detector background, 
which is distributed uniformly through the detector fiducial volume, grows as $R^3$, 
while the source background falls off rapidly as the product of $R^2$ and an exponential 
for attenuation. Figure~\ref{rate-vs-radius} shows the rates as a function of detector
radius with an energy cut applied (200~keV$<E_{rec}<$450~keV).  With the energy cut, 
signal events outnumber background events through out the entire fiducial volume.

\begin{figure}
\centerline{
\includegraphics[width=0.9\textwidth]{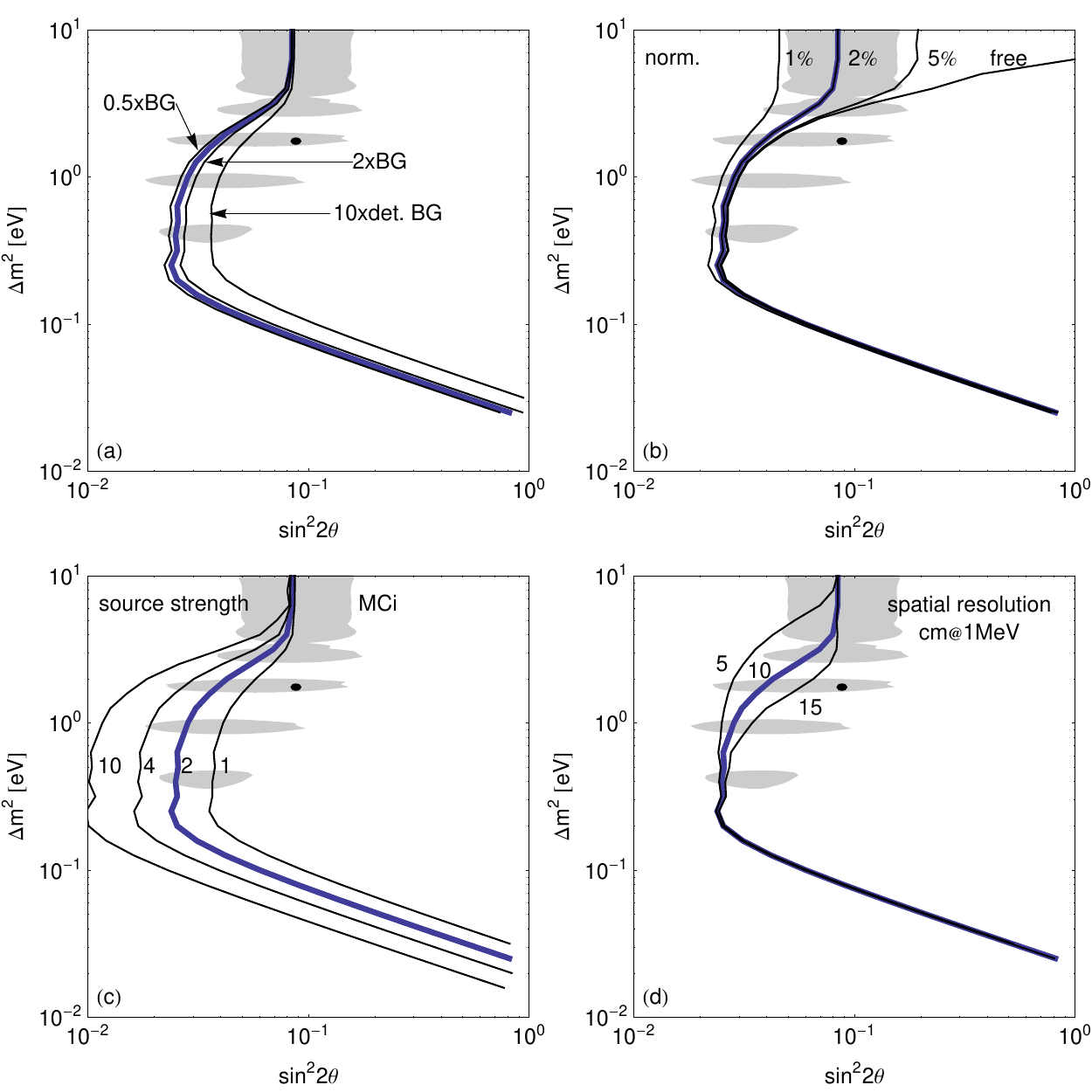}
}
\caption{\label{SNO+Sensitivity}The 90\%CL sensitivity of SNO+ to sterile neutrinos as a
function of $\sin^2 2\theta$ and $\Delta m^2$.  The heavy curve corresponds to the
default assumptions. Plot (a) Shows the sensitivity as the background from
radioactivity in the scintillator is varied; in (b) the uncertainty on
the source normalization is varied;  in (c) the source strength is
varied; and in (d) the detector spatial resolution is varied.  The gray
region is the 90\% CL sterile allowed region from Ref.~\cite{Kopp:2011qd}.}
\end{figure}

The experiment's sensitivity is calculated using GLoBES~\cite{Huber:2004ka}. A single 
75~day run is assumed.  In the baseline scenario the source strength is 2~MCi, the source
strength normalization uncertainty is 2\%, and the detector spatial resolution is 10~cm
at 1~MeV.  Figure~\ref{SNO+Sensitivity} shows the sensitivity of the baseline scenario
and with a series of variations on the baseline including, changes to the background
rate (Fig~\ref{SNO+Sensitivity}a), the source strength normalization
(Fig~\ref{SNO+Sensitivity}b), the source strength (Fig~\ref{SNO+Sensitivity}c), and the 
detector spatial resolution (Fig~\ref{SNO+Sensitivity}d).  The background variation
shows that the sensitivity does not have a strong dependence on the background rate.  
Even increasing backgrounds by a factor of 10, only lowers the sensitivity by a half. 
With the baseline spatial resolution, the signal strength normalization uncertainty 
primarily effects sensitivity for $\Delta m^2$ greater than 2~eV$^2$.  For larger 
$\Delta m^2$ the oscillation length is too short to be resolved and sensitivity comes 
from the ability to resolve a deficit in the total rate.  The sensitivity is strongly
dependent on the source strength showing that the experiment is statistically limited.  
Therefore, the reach in $\sin^2 2\theta$ can be improved by either increasing the 
source strength or by adding additional runs. Finally, we see that the detector spatial
resolution determines the dividing point in $\Delta m^2$ between the sensitivity from 
the resolved rate variation in radius or from a total rate deficit.  Better spatial
resolution would allow a direct observation of oscillometry over a greater range in
$\Delta m^2$.

\clearpage
\subsection{Reactors with a small core\footnote{Contributed by Osamu Yasuda (Tokyo
Metropolitan University)}}
\label{Yasuda}

\begin{figure}[h]
\includegraphics[width=0.9\textwidth]{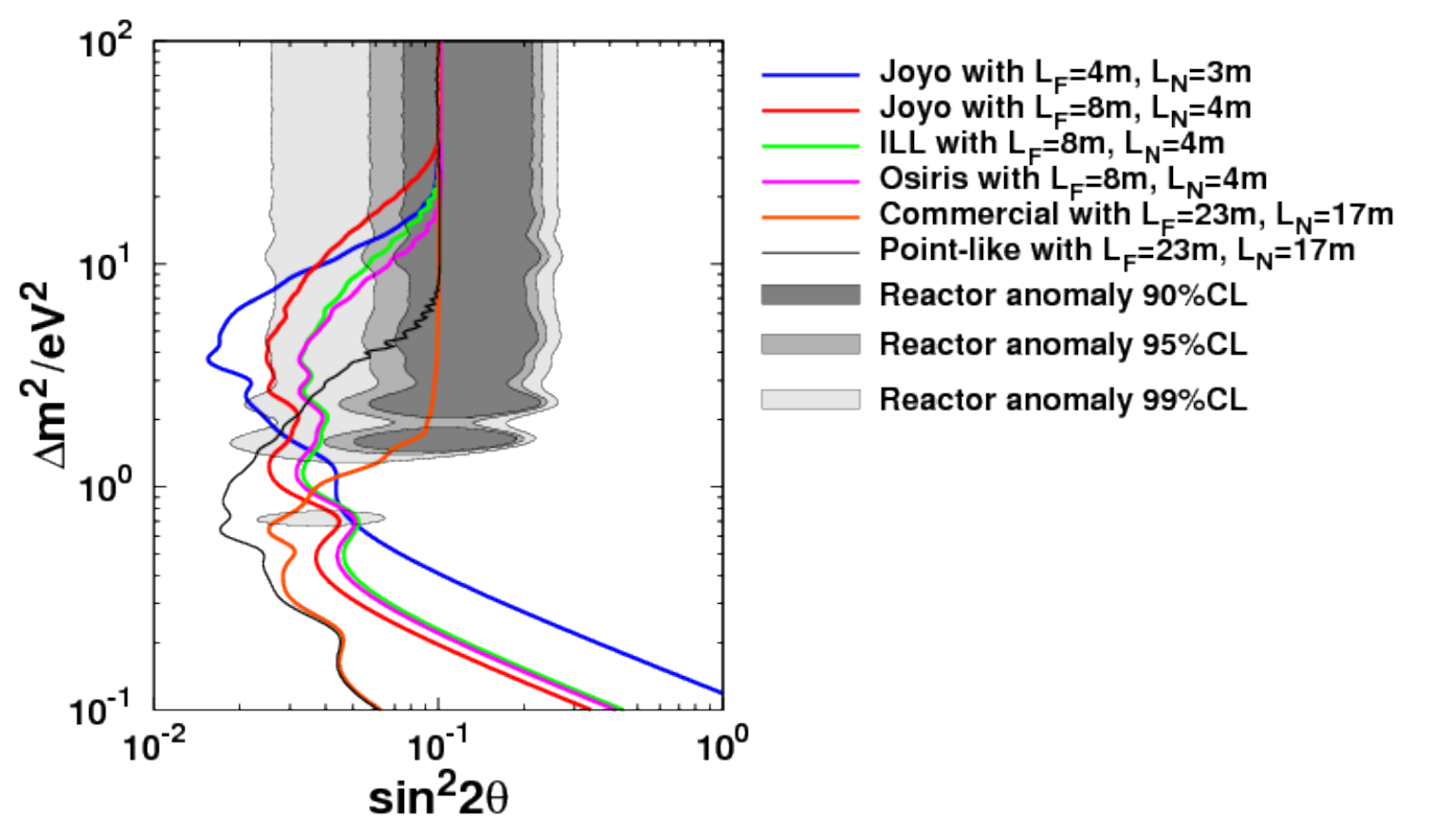}
\caption{
The sensitivity to $\sin^22\theta$
of each reactor with the two detectors at its
optimum baseline lengths.
The region given in Ref.\,\cite{Mention:2011rk}
from the combination of the reactor neutrino experiments, Gallex and
Sage calibration sources experiments, the MiniBooNE reanalysis of 
Ref.\,\cite{Giunti:2010wz}, and the ILL-energy spectrum
distortion is also shown
as a shaded area for comparison.}
\label{Future-Experiments-Reactors}
\end{figure}

In Refs.\,\cite{Yasuda:2011np,Yasuda:2011wk} the
sensitivity to the mixing angle of sterile neutrino
oscillations at very short baseline reactor oscillation
experiments is examined by a spectrum analysis.\footnote{
See, {\it e.g.}, Refs.\,\cite{nucifer,panda}
which discuss the same issue from other points of view,
and Refs.\,\cite{Sugiyama:2005ir} (the published
version), \cite{Latimer:2007qe,deGouvea:2008qk} for
earlier works on search for sterile neutrinos at a
reactor.}  The
assumptions are that the experiment has two identical
detectors whose size and efficiency are exactly the same
as those used at the Bugey
experiment\,\cite{Declais:1994su} and $\chi^2$ is
optimized with respect to the baseline lengths of the two
detectors.  In the case of a commercial reactor, the
sensitivity is lost above 1eV$^2$ due to the smearing of
the finite core size.  In the case of a research reactor
with a small core (such as Joyo\,\cite{joyo} with MK-III
upgrade\,\cite{joyomk3}, ILL\,\cite{ill},
Osiris\,\cite{osiris}, PIK\,\cite{Derbin:2012kf}), on the other hand, one obtains the
sensitivity as good as a several $\times 10^{-2}$ for 1
eV$^2$$\lesssim\Delta m^2_{41} \lesssim 10 $eV$^2$ if the
detectors are located at $L_N=$ 4m and $L_F=$ 8m
(See Fig.~\ref{Future-Experiments-Reactors}).  To turn
this idea into reality, one has to put detectors at a
location very near to a research reactor and has to veto
potentially huge backgrounds from the reactor.

\clearpage
\subsection{SCRAAM: A reactor experiment to rapidly probe the Reactor Antineutrino Anomaly}
\label{sec:SCRAAM}

Here we describe the proposed Southern California Reactor Antineutrino Anomaly Monitor (SCRAAM) experiment. The primary objective of SCRAAM is to experimentally probe the existence of sterile neutrinos in the $\Delta m^2_{14} \approx 1$~eV$^2$ region allowed by the Reactor Antineutrino Anomaly (RAA)~\cite{Mention:2011rk} using carefully chosen reactor antineutrino sources. The project would begin by performing a rapid experiment at a large Pressurized Water Reactor (PWR) (the San Onofre Nuclear Generating Station (SONGS) in Southern California). A follow-up effort at a compact research reactor (most likely the Advanced Test Reactor (ATR) at the Idaho National Laboratory) would broaden the $\Delta m^2_{14}$ sensitivity of SCRAAM, while also allowing verification of the SONGS measurement at one or more different baselines. The secondary objective of the project is to use the high precision reactor antineutrino spectrum measurement central to the sterile oscillation search to also improve reactor antineutrino flux predictions~\cite{Huber:2011wv}.

Notably, much of the reactor-sector contribution to the RAA excluded phase space arises from comparison of rate measurements with flux and background predictions, as is also true of the MiniBOONE and radiochemical experiments. Any experiment seeking to test these anomalies and have the potential to unequivocally prove the existence of a sterile neutrino must use a more conclusive experimental technique. The oscillation phenomena at the heart of the problem provides such a probe: a sterile oscillation should have a distinctive pattern that varies with both neutrino energy and baseline. A direct oscillation pattern measurement can be rapidly and inexpensively provided by well sited reactor antineutrino measurements at short (10-25m) baselines. This experimental approach has several advantages: the high antineutrino flux close to a reactor allows a rapid measurement using a  small, inexpensive detector, and the broad energy range spanned by the reactor antineutrino emissions (2-10 MeV) allows a sterile oscillation search directly via spectral distortion. Finally, a high precision reactor spectrum measurement is a natural outcome of such an effort, which could help constraint the magnitude of corrections to flux predictions.

\begin{figure}
\includegraphics[width=0.5\textwidth]{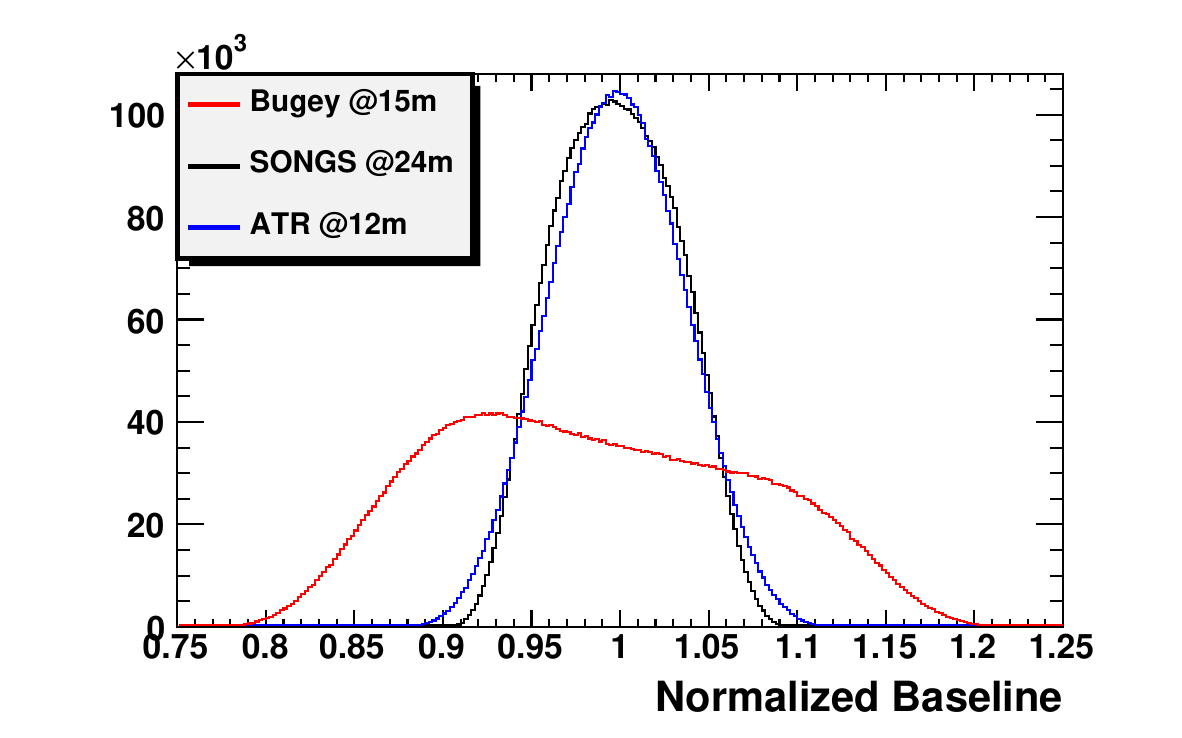}
\caption{\label{fig:baseline} Normalized baseline distributions for the proposed reactor deployment sites and that of the Bugey3 experiment. The wide Bugey3 distribution, where the baseline is only a few times the size of the reactor, is marked asymmetric, largely due to the difference in solid angle from the front to back of the core. }
\end{figure}

\begin{figure}
\includegraphics[width=0.8\textwidth]{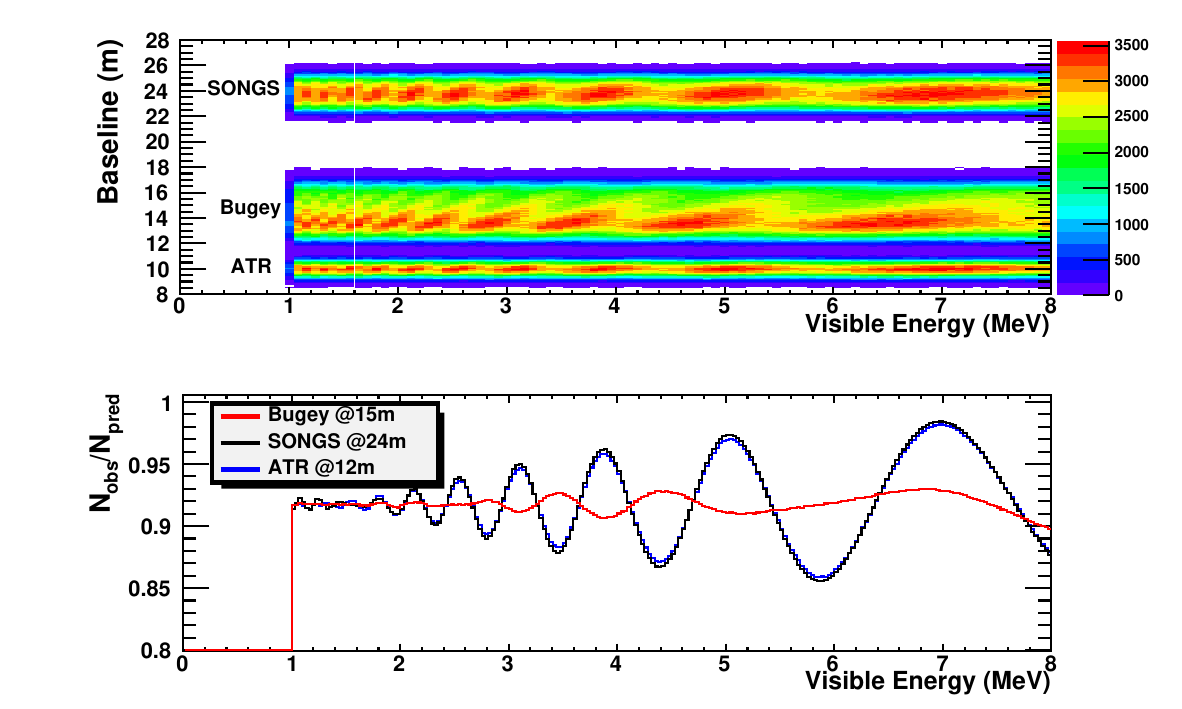}
\caption{\label{fig:L_v_E} The predicted pattern of a sterile oscillation with
$sin^2(2\theta_{14}) = 0.15$ and $\Delta m^2_{14}/\bar L = 0.1$~eV$^2$m$^{-1}$ ({\it i.e.}~$\Delta m^2_{14} = 2.4$~eV$^2$ for the SONGS baseline of $24$~m) for each of the three reactor sites considered. The top panel shows the variation of the $\bar \nu_e$ survival probability with $E$ and $L$, while the bottom panel shows the projection onto the energy axis - the experimentally accessible quantity. For clarity, the ATR pattern in the top panel is displaced to $\bar L=10$.}
\end{figure}

The spread in both $E$, {\it e.g.}~from detector energy resolution, and $L$, due to the spatial extent of the reactor core and detector, must be considered when planning such a measurement. Relevant parameters are summarized in Table~\ref{tab:reactors} for the SONGS and ATR reactors and for the Bugey3 measurement conducted at $15$~m~\cite{Declais:1994su}.  The relative baseline distribution for these three reactor sites is shown in Fig.~\ref{fig:baseline}. The effect the baseline spread has upon the ability to resolve an oscillation pattern is demonstrated in Fig.~\ref{fig:L_v_E}. Here, the electron antineutrino survival probability is displayed as a function of both baseline and energy for the RAA best fit sterile oscillation parameter values for each reactor site. It is apparent that the large baseline spread of the Bugey3 experiment substantially reduces the oscillation pattern contrast compared to that accessible at the SONGS and ATR sites.

\begin{table}[b]
\begin{center}
\begin{tabular}{l|c|c|c|c|c} \hline
Reactor &Baseline &Core & $\Delta$L/L &Power &Antineutrino Flux\\
\hline
Bugey3 &$15$~m &$\diameter2.5$~m x $2.5$~m &$\approx30$\% &$2800$~MW$_{th}$ &$\approx2\times10^{17}~$m$^{-2}$s$^{-1}$\\
SONGS &$24$~m &$\diameter3.0$~m x $3.8$~m &$\approx10$\% &$3400$~MW$_{th}$ &$\approx1\times10^{17}~$m$^{-2}$s$^{-1}$\\
ATR &$12$~m &$\diameter1.2$~m x $1.2$~m &$\approx10$\% &$150$~MW$_{th}$ &$\approx2\times10^{16}~$m$^{-2}$s$^{-1}$\\
\hline\end{tabular}
\caption{\label{tab:reactors} Reactor parameters of the two appropriate sites identified, as well as those for the $15$~m Bugey3 measurement.}
\end{center}
\end{table}

Besides providing sensitivity to the physics of interest, the two identified reactor sites have many other desirable features. In the case of SONGS, our group has almost a decade of operational experience at this site to draw on, including many antineutrino detector deployments and measurements~\cite{Bowden:2006hu,Bernstein:2008tj,Bowden:2008gu}. We maintain an excellent relationship with the operator and are confident of being able to obtain ready access to the site for this effort. Finally, the below-ground ``tendon gallery'' deployment location provides a convenient $30$~m.w.e. overburden. A preliminary survey of the ATR site has yielded potential deployment sites with $10-20$~m.w.e. overburden at $\approx 12$~m baseline, as well as the possibility to vary that baseline. Also, the $150$~MW$_{th}$ power of the ATR is relatively high for a research reactor, the core is primarily $^{235}$U, and it's $60$~day on, $30$~day off cycle provide ample opportunity for background measurement. 

Drawing upon our own recent experience, and that of the wider neutrino oscillation physics community, we believe that a relatively compact, inexpensive, and high precision liquid scintillator antineutrino detector could be rapidly designed and built for this experiment. The key features required are: high detection efficiency; $\approx 4\%$ precision on the absolute value of that efficiency, including reactor power uncertainties; and good energy resolution. Reactor antineutrinos would be detected via the inverse beta decay interaction: ($p(\bar \nu_e,e^+)n)$. Due to the kinematics of the interaction, the final state positron energy is closely related to the antineutrino energy. Near-coincident detection of the final state neutron, typically via Gd-doping, provides powerful background rejection. An efficiency of $50\%$ for a $1.5$~ton active mass would yield a detection rate of $\approx4,500$/day at SONGS and provide a high statistics spectrum measurement within $\approx6$~months of running. Given the width of the baseline distributions of the proposed sites, a readily achievable resolution of $10\%/\sqrt{E}$ ($E$ in MeV) will not limit sensitivity. We estimate the component costs of such a device to be $<\$1$M.

To achieve these performance goals, we envisage a cylindrical detector with PMTs at either end. Application of a Teflon coating to the detector walls will provide a highly diffuse reflective surface. This configuration will provide excellent light collection performance and uniformity across the entire detector volume, while fitting inside the relatively narrow confines of a tendon gallery. We have performed preliminary Monte Carlo simulations of this design which indicate encouraging performance. Detector calibration will be a critical part of this experiment, because knowledge of the absolute energy scale and its stability will be central to an oscillation analysis and a high precision spectrum measurement. This will be achieved via regular calibrations with small sealed gamma ray and neutron sources deployed via guide tubes within the detector volume. 
 
\begin{figure}
\includegraphics[width=0.9\textwidth]{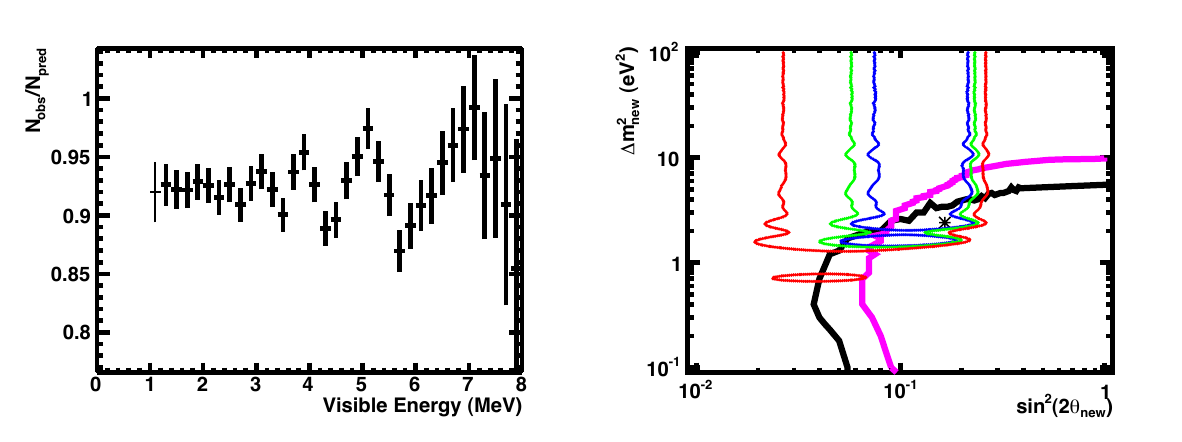}
\caption{\label{fig:excl} LEFT: Simulated spectral data for the RAA best fit parameters after $150$~days at SONGS. The error bars shown include both statistical and systematic errors.
RIGHT: 99\% C.L. SCRAAM shape-only sensitivity at SONGS (solid black) and the ATR (solid purple). Also shown are the 90\%, 95\% and 99\% allowed contours of the RAA (dashed). The RAA best-fit point is indicated with a star.}
\end{figure}

Using simulated data, we have estimated the sensitivity of SCRAAM to the oscillation parameters of a sterile neutrino flavor using shape information only. For SONGS, we assume $150$~days of reactor on operation, $45$~days of reactor off operation, a flat background spectrum with signal to background of $8/1$ (a reasonable factor 2 improvement over ÒSONGS1Ó), and an energy scale uncertainty of $1.5\%$. The expected spectral distortion that would be observed for the RAA best fit parameters and the $99\%$ exclusion region in the oscillation parameter space are shown in Fig.~\ref{fig:excl}. For the ATR deployment we assume $300$~days of reactor on operation, $150$~days of reactor off operation, and a similar signal to background as SONGS. It can be seen that SONGS alone excludes much of the RAA phase space around 1eV2, including the best fit region. It is notable that this measurement alone covers the RAA phase space that is consistent with astrophysical indications of a sterile neutrino with mass $<1$~eV. The ATR deployment extends the range of $\Delta m^2$ covered by approximately a factor of 2, which is to be expected give that it is placed at half the baseline. Combined, these two measurements cover a broad portion of the RAA phase space and provide a crosscheck of the Bugey3 result that provides much of the RAA exclusion at low $\Delta m^2$ values. Finally, we note that additional sensitivity could probably be gained by performing a combined analysis of the SONGS and ATR results, by performing the ATR measurement at two baselines, or by using position resolution in the detector at the ATR to probe multiple baselines. Recent work~\cite{Yasuda:2011np} suggests that sensitivity to lower $sin^2(2\theta_{14})$ values could be achieved in this way, so long as relative detector normalization is well controlled. 

In summary, SCRAAM can provide a direct sterile neutrino exclusion measurement at the source of one of the hints of their existence: nuclear reactors. Deployment of a compact antineutrino detector at the locations identified will probe a large fraction the sterile oscillation parameter phase space allowed by the ÒReactor Antineutrino AnomalyÓ. Because of the advantageous features of the reactor sites we have identified, this experiment can proceed rapidly and at relatively modest cost.

This work performed under the auspices of the U.S. Department of Energy by Lawrence Livermore National Laboratory under Contract DE-AC52-07NA27344. LLNL-PROC-520315

\clearpage
\subsection{Nucifer: a Small Detector for Short-Distance Reactor Electron
Antineutrino Studies\footnote{
            Proposed by J.~Gaffiot, D.~Lhuillier, Th.~Lasserre, A.~Cucoanes, 
	    A.~Letourneau, G.~Mention, M.~Fallot, A.~Porta, R.~Granelli, L.~Giot, 
	    F.~Yermia, M.~Cribier, J.~L.~Sida,  M.~Fechner, M.~Vivier}}

\subsubsection{Introduction}
The Nucifer~\cite{2010ITNS...57.2732P} project firstly aims at testing a small
electron-antineutrino detector in order to apply this technology to non proliferation.
Installing such a detector a few 10 meters from a reactor core
allows to monitor its thermal power and to estimate its plutonium content.
The design of the detector emphasizes compactness and simplicity,
while keeping efficient background rejection.

As reported in Section~\ref{sec:reactor}, recent work
on reactor neutrino spetra~\cite{Mueller:2011nm} has lead to
the surprising re-evaluation of the expected neutrino counting rate
by an amount of 3 \%. This has triggered the re-analysis of
previous short-distance reactor neutrino experiments~\cite{Mention:2011rk},
showing a significant discrepancy (about 3 $\sigma$, see~\ref{sec:oscillation})
between measured and expected neutrino counting rates.
Among various hypotheses such as experimental bias, a new oscillation
to a sterile neutrino state, invisible to detectors has been proposed.

The Nucifer experiment is going to take data at the Osiris research reactor
in 2012/13, at only 7 meters from the core.
This unique configuration, short distance and the compactness of Osiris core
reduces the dispersion of neutrino paths.
If an eV$^2$ sterile neutrino exists (as discussed in Section~\ref{sec:theory}),
the measured energy spectrum could appear distorted according to the neutrino 
oscillation hypothesis.  The shape-only analysis of the Nucifer data will be the first
unambiguous test of the existence of a sterile neutrino, although the complete area
of the reactor anomaly contour can not be covered by this experiment alone.

The installation of the detector is now completed and 1 year of data taking
is foreseen at the French Osiris research reactor.

\subsubsection{Description}
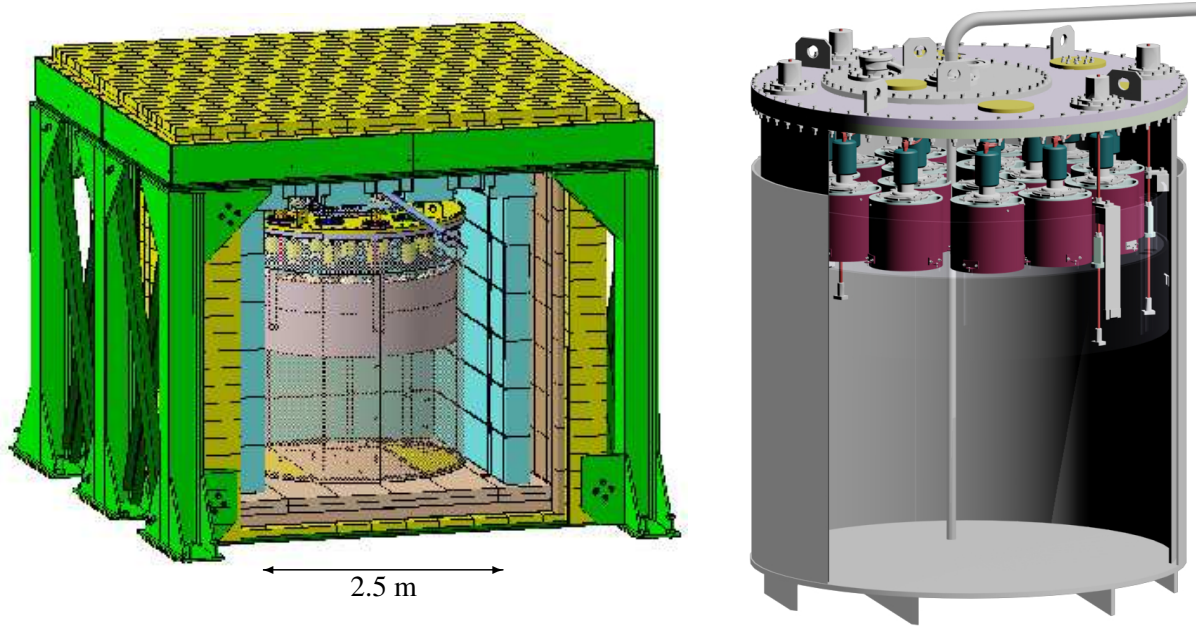
\begin{figure}
 \begin{minipage}{.55\linewidth}
  \begin{center}
   \includegraphics[width=\linewidth]{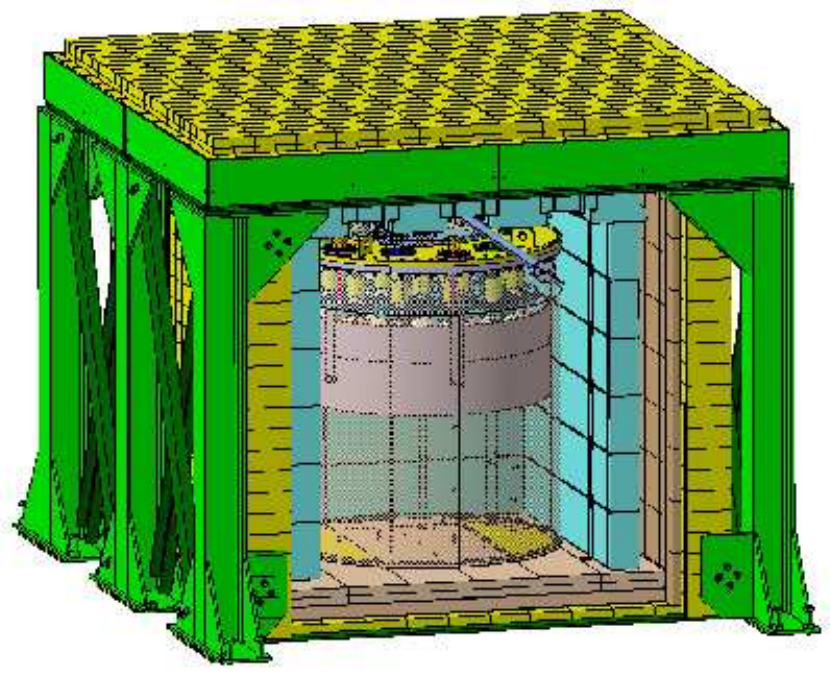}
    \begin{picture}(0,0)
    \color{black}
	\put(-30,15){\vector(1,0){90}}
	\put(60,15){\vector(-1,0){90}}
	\put(0,5){ 2.5~m }
    \end{picture}

  \end{center} 
 \end{minipage}\hfill
 \begin{minipage}{.45\linewidth}
  \centering \includegraphics[width=.8\linewidth]{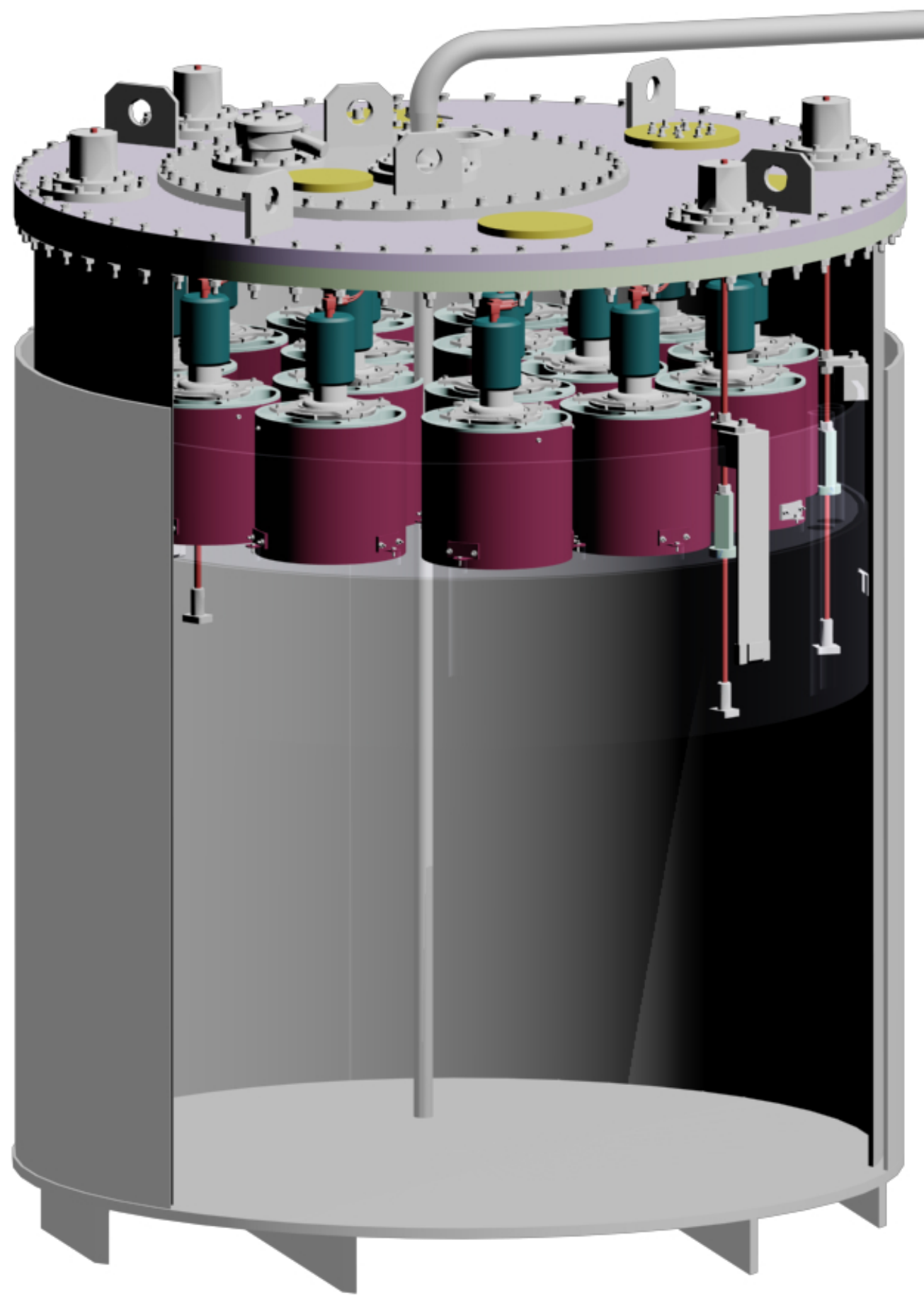}
 \end{minipage}\hfill
 \caption{Left: Nucifer with shieldings, including 10 cm of lead, 14 cm of boron doped plastic 
          and $\mu$ veto Right: Nucifer tank topped by the acrylic buffer and the 16 PMTs}
 \label{fig:nucifer}
\end{figure}

The Nucifer~\cite{2010ITNS...57.2732P} detector target is a stainless steel vessel
of 1.8 m in height, and 1.2 m in diameter (see Figure~\ref{fig:nucifer})
filled with about 0.85 m$^{3}$ of Gd-doped liquid scintillator
(EJ335 from Eljen technology). The internal surface of the vessel is coated
with Teflon to ensure the compatibility with the liquid scintillator and
to increase the light diffusionthe light reflections (diffusive and specular).
All mechanical parts, in particular welding materials, are low radioactive materials
and their radiopurity was controlled with a High purity Germanium in our low
background laboratory~\cite{Fechner:2011wy}. The photodetection system is based
on 16 large (8 inches in diameter, R5912) photomultipliers (PMTs) from Hamamatsu,
providing a large dynamic of light detection from the single photoelectron to
few hundreds of photoelectrons and ensuring an efficient light collection.

PMTs are coupled to a 250~mm thick acrylic vessel placed at the top
of the target vessel. This so-called acrylic buffer aims at ensuring
the uniformity of the response in the whole target volume while
reducing the light generated by the intrinsic PMT radioactivity
in the scintillator. 80 liters of mineral oil are used to ensure
the optical coupling between the PMTs and the acrylic.

The Data Acquisition system (DAQ) is based on the VMEbus and the LabView
software, and allows a dead time below 1\% at 1.5 kHz trigger rate
and a remote control of the DAQ. The physic trigger is basically
a threshold on an analogic summation of all PMTs' signals. The time and the charge
of each PMT are recorded with commercial CAEN Time to Digital Converter (TDC)
and Charge to Digital Converter (QDC) modules. The collection of the total
and late charges of the pulse is used off line for the Pulse Shape Discrimination 
analysis.

During operation, a set of 5 LEDs injects light in the detector.
One of them is a low intensity diode to monitor the gain of PMTs
at the level of the single photoelectron and the rest are
high intensity diodes with light Teflon diffuser to monitor
the liquid stability and the linearity of the response of the detector.

Four sources are used for the absolute calibration of the detector:
$^{137}$Cs, $^{22}$Na, $^{60}$Co and Am-Be, providing an energy deposition
in the range 0.7-5.5~MeV.  A vertical calibration tube located
at the center of the target volume allows the introduction
of the sources along the target central axis.
Moreover the Am-Be source provides tagged neutrons with a 4.4~MeV
gamma-ray that is used to calibrate the neutron detection efficiency.

\subsubsection{Background}
The experimental challenge is to operate such a small detector in a high background 
environment.  Due to the proximity to the reactor, Nucifer will operate under a high 
flux of gamma-rays, including high energy gamma-rays from neutron capture on metals. 
A remnant flux of thermal neutrons has also been mesured on site. Osiris is a pool 
reactor, and only 7.0~meters separate the core center from the detector center, which 
means only 2.9 meters of water and 2.0~meters of concrete as shielding. To reduce the 
accidental background in the detector due to random coincidences of external gamma-rays 
and neutrons, the detector is enclosed within a shielding composed of one layer of 
10~cm of lead and one layer of 14~cm of boron-doped polyethylene. As part of the 
gamma-rays come directly from the core, a lead wall of 10 cm is added between the 
concrete wall of the reactor pool and the detector. Using this shielding, the accidental
background rate is estimated to be of the same level than the neutrino signal.
It will be measured online using the reverse and shifted coincidence windows
between the prompt and the delayed signal.

Moreover, the detector is close to the surface, with only 5 meters equivalent water
above, so the muon flux is only divided by a factor 2.7 compared to surface and
significant rapid neutron flux is expected, leading to a correlated event rate
few times above the expected neutrino rate.
To reduce this background, a 5-sided muon veto will allow us to veto the acquisition
during 200 $\mu$s after each muon, with few percent dead time.
Since muon-induced rapid neutron could be produced
outside of the muon veto, we will use Pulse Shape Discrimination (PSD)
to reject remaining correlated background.

Our PSD analysis  is based on the comparison of the total charge of the event
and the charge of its tail, given by our two QDCs.
Test on a small 10 $\text{cm}^3$ cell filled with the final EJ335 liquid
from Eljen Technology shows a great power of separation
between rapid neutron induced proton recoil and gamma induced electron recoil.
To preserve the final liquid, tests in lab were performed with Nucifer filled
by a homemade liquid, non loaded, based on linear alkyl-benzene (LAB). It was
shown that loss of separation power in a large diffusive tank are limited,
thanks to our good light collection. We finally expect a rejection of correlated background
higher than a factor 10, leading to a signal over correlated background ratio higher than 4.

\subsubsection{Short baseline oscillation study}
In the scheme of a new oscillation at short distance, we calculated the performances
of Nucifer to measure a spectral distortion of the signal.

The Osiris reactor has a thermal power of 70~MW, with a compact core of 
$57\times57\times60$~cm$^{3}$,
and an enrichment in $^{235}$U of almost 20\%.  With preliminary plans, with estimate the 
core center to detector center distance to 7.0~meter.  A Monte-Carlo simulation allow us 
to calculate the solid angle and propagate neutrinos from core to detector. The neutrino 
energy is drawn in the neutrino spectra of uraniums and plutoniums, and the Nucifer GEANT4 
energy resolution is used to draw the visible energy.  We add also a 5\% uncertainty on 
the whole energy scale.  The neutron efficiency is largely dependent of background, and 
according to our simulation it could be from 12\% with a neutron energy cut at 6~MeV to 
60\% with a cut at 4~MeV.  We use here an efficiency of 40\% corresponding to a cut at 
5~MeV.  So 400 neutrino interactions are expected in the tank with only 330 above
threshold on visible energy at 2~MeV.  Since our PSD measurement shown a great power of 
correlated background rejection, we suppose that the dominant background is accidental, 
with an expected signal over background ratio of 1. We use here an exponential shape at 
low energy, below 3~MeV, and a flat shape at high energy.

\begin{figure}
\begin{minipage}{.6\linewidth}
\centering
\includegraphics[width=\linewidth]{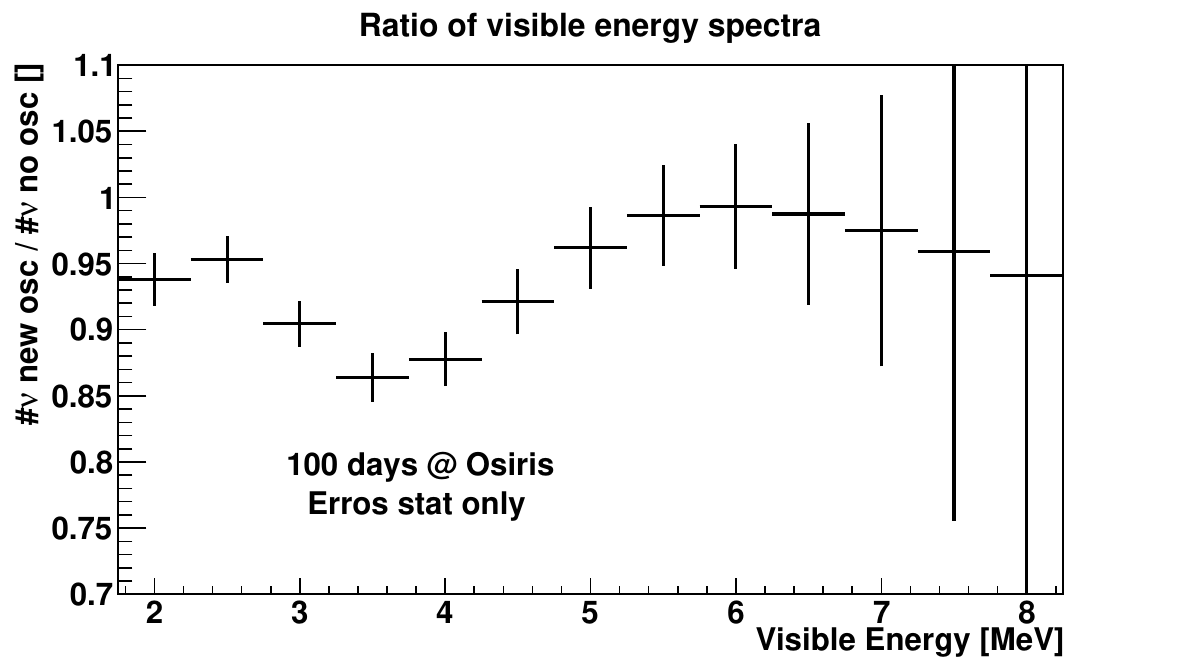}
\caption{Expected distortion of neutrino spectrum in Nucifer with a new oscillation controlled 
          by $\Delta m_{new}^{2}=2.3$ and $\theta_{new}=0.17$}
\label{fig:spectra}
\end{minipage}\hfill
\begin{minipage}{.05\linewidth}
\end{minipage}\hfill
\begin{minipage}{.35\linewidth}
 \centering \includegraphics[width=\linewidth]{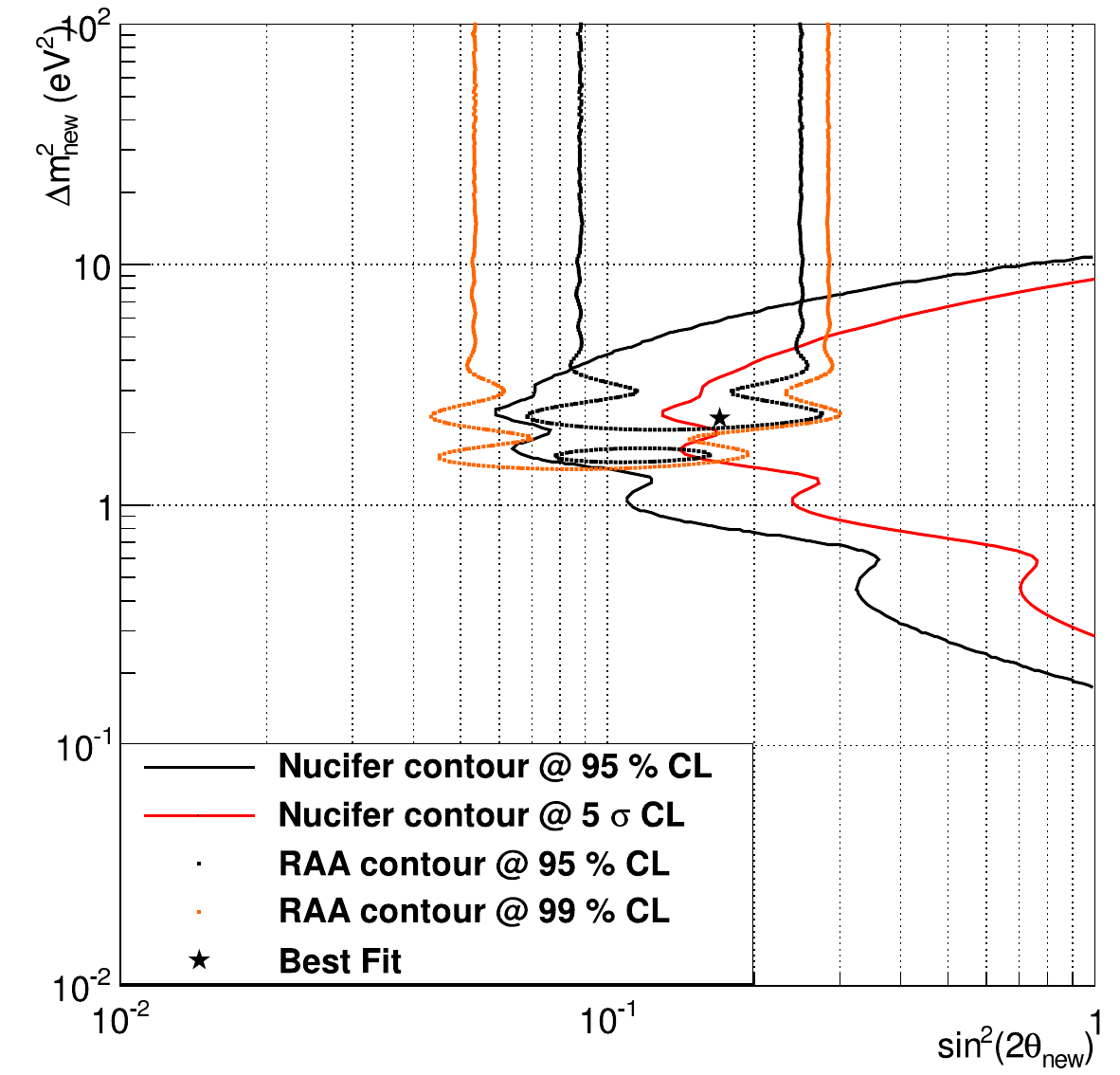}
 \caption{Discovery potential of Nucifer compared to Reactor Antineutrino Anomaly}
 \label{fig:contour}
\end{minipage}\hfill
\end{figure}

Figure \ref{fig:spectra}) shows the result for 100 days of running (corresponding to 
$\sim$6~months with reactor off periods) with $\Delta m_{new}^{2}$ and $\theta_{new}$ 
best fit values taken in Ref~\cite{Mention:2011rk}.  We see a distortion, but its 
significance will be decreased by accounting for the full set of systematic errors.
Figure~\ref{fig:contour} shows the expected contour with 300 days of data, corresponding
to about 1~year on site.

To go further, precise power history, controlled at the percent level, will be used,
and dedicated simulation of the Osiris core
is under development to control the variations of the fissions barycenter.
The measurement of the detector position in respect with the core will allow an uncertainty
on the neutrino propagation length at the five centimeters level.
The tank will be weighted after and before filling with weigh sensors,
allowing a precision in liquid mass below one percent.
The hydrogen fraction in the liquid scintillator has been mesured
by a BASF laboratory, which finally provides a proton number at the percent level.
The proton number in the tank will thus be controlled at the percent level.

\subsubsection{Conclusions}
The Nucifer detector will take data in 2012/2013, and due to its closeness
with the compact Osiris research reactor core, it will be the first reactor experiment
testing the eV$^2$ sterile neutrino hypothesis at reactors.
The sensitivity contour shows interesting prospects for the test of the best fit parameters
pointed out by the combination of the gallium and reactor anomaly (Ref~\cite{Mention:2011rk}).
Detection efficiency and background rejection will dominate
the final systematics and thus performances.

\clearpage
\subsection{Stereo Experiment\footnote{Proposed by 
     D.~Lhuillier, A.~Collin, M.~Cribier, Th.~Lasserre, A.~Letourneau, G.~Mention, 
     J.~Gaffiot (CEA-Saclay, DSM/Irfu),
     A.~Cucoanes, F.~Yermia (CNRS-IN2P3, Subatech),
     D.~Duchesneau, H.~Pessard (LAPP, Université de Savoie, CNRS/IN2P3).}}
\label{sec:stereo}

\subsubsection*{Principle}
The combination of the reactor anomaly with the neutrino deficit already observed
in Gallium experiments favors the hypothesis of a sterile neutrino at the 3.6
$\sigma$ level (see section \ref{sec:reactor}). The most probable values of the new oscillation parameters are
\begin{eqnarray}
\label{eq:RaaBestFit}
|\Delta m_{\mathrm new}^2 | &=& 2.3\pm 0.1~{\mathrm eV} ^2 \qquad
\sin^2(2\theta_{\mathrm new})=0.17\pm 0.04 \,\, (1\sigma)
\end{eqnarray}
So far the sterile neutrino hypothesis relies mainly on a missing contribution in the integrated  observed rate. An unambiguous signal of a new neutrino must show an oscillation pattern related to the quite large $\Delta m^2_{new}$. At the typical few MeV energy scale of reactor antineutrinos, the expected oscillation length is of the order 1 m. Hence to avoid a complete smearing of the oscillation signal the uncertainty of the event-by-event baseline should be below 1 m. On the source side the compact core of a research reactor can meet this specification. On the detector side, standard vertex resolutions are well below 1 m.

\begin{figure}[h]
\includegraphics[width=0.65\textwidth]{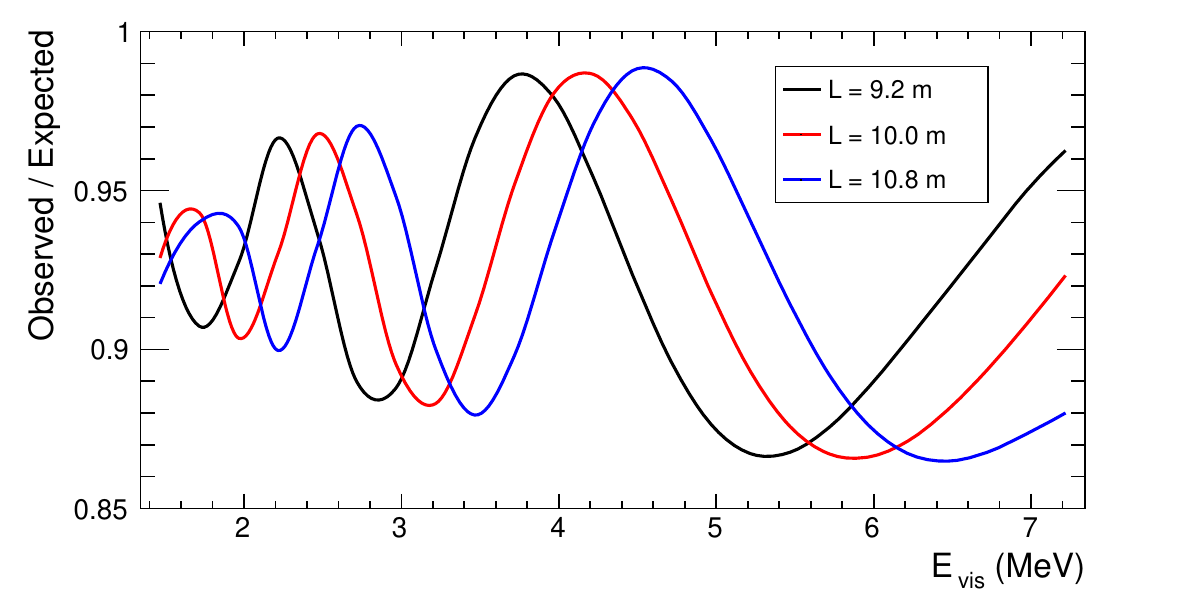}
\caption{\label{fig:OscilinDet} Ratio of expected/observed neutrino spectra for three of the 5 baseline bins of the Stereo detector. Each bin is 40 cm wide.}
\end{figure}
The proposed experiment is an antineutrino spectrum measurement at short distance from a research reactor, using the standard beta-inverse process in a liquid scintillator (LS) doped with Gd
\begin{eqnarray}
\bar{\nu}_e + p \rightarrow e^+ + n
\end{eqnarray}
The prompt positron signal gives the antineutrino energy. The emitted neutron is captured mainly on Gd with a capture time $\simeq 30 \mu s$. This delayed signal is used to reject backgrounds. Along the axis pointing toward the core, the detector is built long enough to see the phase of the new oscillation in the energy spectrum changing along this axis. To prove the existence of a sterile neutrino, emphasis is put on the shape-only analysis, looking for the relative deformation of the spectrum along the detector with as little as possible normalization input. Figure \ref{fig:OscilinDet} illustrates the expected oscillation pattern of the antineutrino energy spectrum in three baseline bins of the Stereo detector. All detection effects presented in the Inner Detector Section are included and the best fit parameters of Eq.(\ref{eq:RaaBestFit}) are used. The oscillation patterns and their phase variation are clearly measurable.

\subsubsection*{Inner Detector}
\begin{figure}[h]
\includegraphics[width=0.5\textwidth]{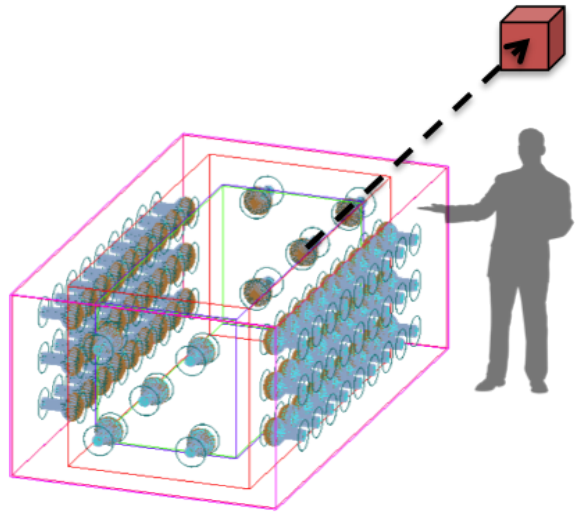}
\caption{\label{fig:StereoSetup} Illustration of the proposed setup (shielding not represented). The longest detector axis points toward the core of the reactor.}
\end{figure}

The proposed setup, based on experience from the Double Chooz \cite{Ardellier:2006mn} and Nucifer experiments, is illustrated in figure \ref{fig:StereoSetup}. The target liquid scintillator, doped with Gd, is contained in a 8 mm thick acrylic vessel. The section of the target is about $1\,m\times 1\,m$ and the length is $2\,m$. The target vessel is immersed in another LS, not doped with Gd, contained in a second acrylic vessel. This 15 cm thick outer layer collects part of the energy of $\gamma$-rays escaping from the target, reducing the low energy tail of the detector response. In the current version, 64 PMTs of 8 inches diameter are distributed across the lateral surfaces of the outer acrylic box. The roof and bottom planes are covered with diffusive white Teflon. Fixed optical coupling of the photomultipliers with thick lateral acrylic walls can be used to provide mechanical support and serve as a buffer layer for a more uniform detector response.

\subsubsection*{Backgrounds}
At the shallow depths considered here, the main source of background is induced by the cosmic rays. When interacting in the surrounding material, they can generate fast neutrons. In the liquid scintillator these fast neutrons will induce recoils of protons. Despite a typical quenching factor of about 5, these protons can mimic a prompt signal above the detection threshold. Then the neutron will keep loosing energy and will be captured, generating the same delayed signal than in the case of an antineutrino interaction. The production of multiple neutrons induced by cosmic rays contributes to the same effect. This correlated background can be efficiently suppressed by the use of the pulse shape discrimination (PSD) properties of liquid scintillators to separate proton recoils from the seeked positron signals. Measurements performed in the 1 $m^3$ target of Nucifer has shown that the volume effects do not reduce significantly the PSD. Muons interacting in the detector can also produce long-lived $\beta$-n emitters like $^9Li$ which mimic the $e^+,n$ emission of the IBD process. This signal passes all the antineutrino selection cuts and it can't be rejected  by a long veto after every muon because of dead time issue. Still the final contribution of muon-induced background is accurately measured during reactor off periods, comparable in length to on periods at research reactors. The expected muon rate in Stereo is few 100 Hz, leading to an acceptable few percent deadtime with a 100 $\mu s$ veto (3 times the neutron capture time in Gd doped LS).

\begin{figure}[h]
\includegraphics[width=0.5\textwidth]{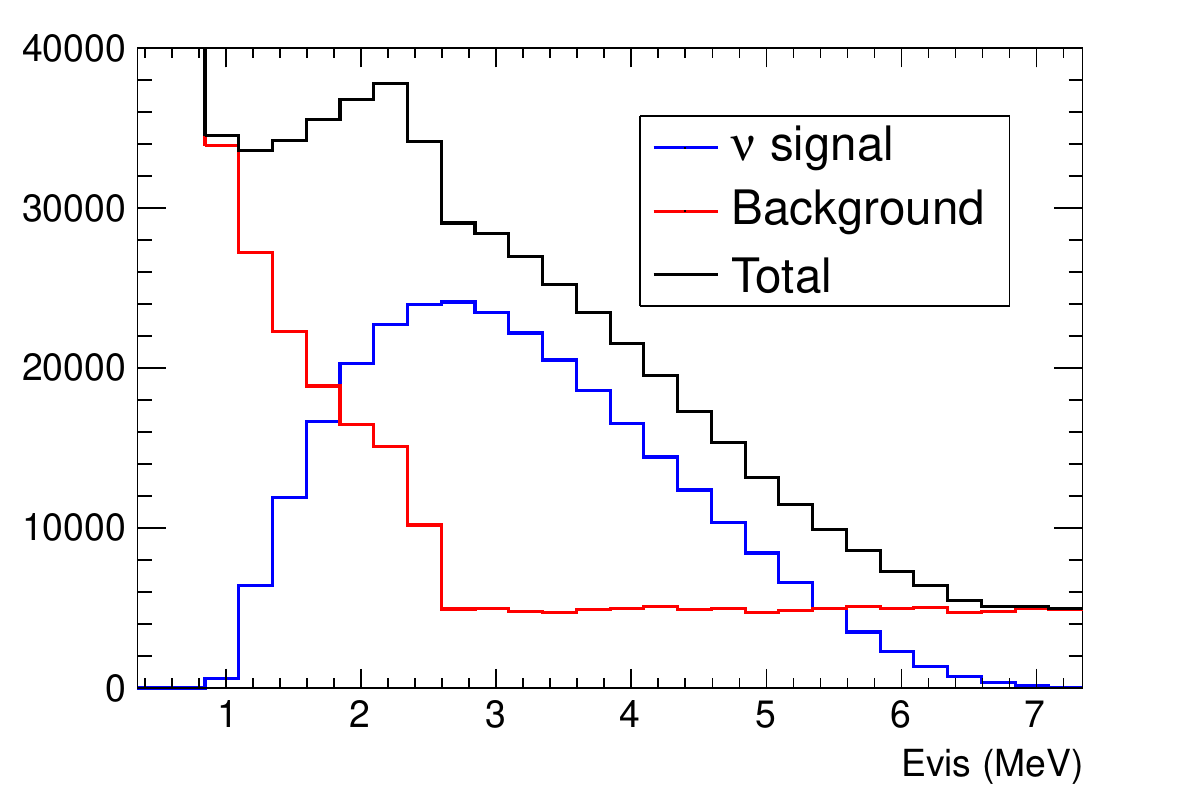}
\caption{\label{fig:SoverB} Example of expected spectra for S/B=2 in the range of visible energy [2-7] MeV.}
\end{figure}
Accidental background is induced by the ambiant radioactivity. This source is limited by the use of radiopure materials. At very short baseline, thermal neutrons and high energy $\gamma$s emitted by the core itself can also reach the casemate of the detector. By definition this background is correlated with the operation of the reactor and can't be measured during the reactor off periods. It must be suppressed by heavy shieldings isolating the inner detector from the outside. The final geometry has to be optimized depending on the background measured on site. The default shielding structure considered here is based on the Nucifer study at the Osiris site: 
\begin{itemize}
\item The first layer around the target is an active muon veto made of plastic scintillator or water for Cerenkov light detection.
\item If thick enough the muon veto can thermalize and absorb most neutrons. An extra polyethylene layer doped with Boron can be added.
\item An external layer of $\approx 10$ cm of lead suppresses the $\gamma$ background.
\end{itemize}
The mass of the setup is dominated by the lead shielding, estimated at about 45 tons for the Stereo experiment. The total mass is about 60 tons for a foot print of $4.5\,m \times 3.5\,m$ and a total heigth of $2\,m$.

The ILL experiment performed in the eigthies \cite{Kwon:1981ua} has shown that the particle flux from the reactor did not induce correlated background in the detector and that the accidental background was efficiently rejected by the shieldings. The remaining contribution of accidentals can always be measured online using shifted coincidence windows. The Nucifer experiment will provide valuable inputs on background rejection in the next few months for the Osiris site. Meanwhile, for the following studies we assume a background level similar to the published ILL data \cite{Kwon:1981ua}. The figure \ref{fig:SoverB} illustrates the crude model of prompt background spectrum, with a fast decrease at low energy and an almost flat behavior at high energy. With a detection threshold set at 2 MeV (visible energy) the ratio of the total expected antineutrino signal over the total background is about 1.5.

\subsubsection*{Performances}
The Stereo setup is simulated with a GEANT4 software adapted from  the simulation of the Double Chooz experiment \cite{Ardellier:2006mn}. It includes a detailed optical model of ligth emission by the liquid scintillator as well as its propagation to the PMT's. The response to each photo-electron (p.e.) is simulated according to the measured response of the 8" PMTs. The analogic total pulse of each PMT is then built from the sum of all single p.e. The arrival time of each PMT pulse is determined via a simulated constant fraction discriminator. The total charge is integrated in a 150 ns gate.

\begin{figure}[h]
\includegraphics[width=0.7\textwidth]{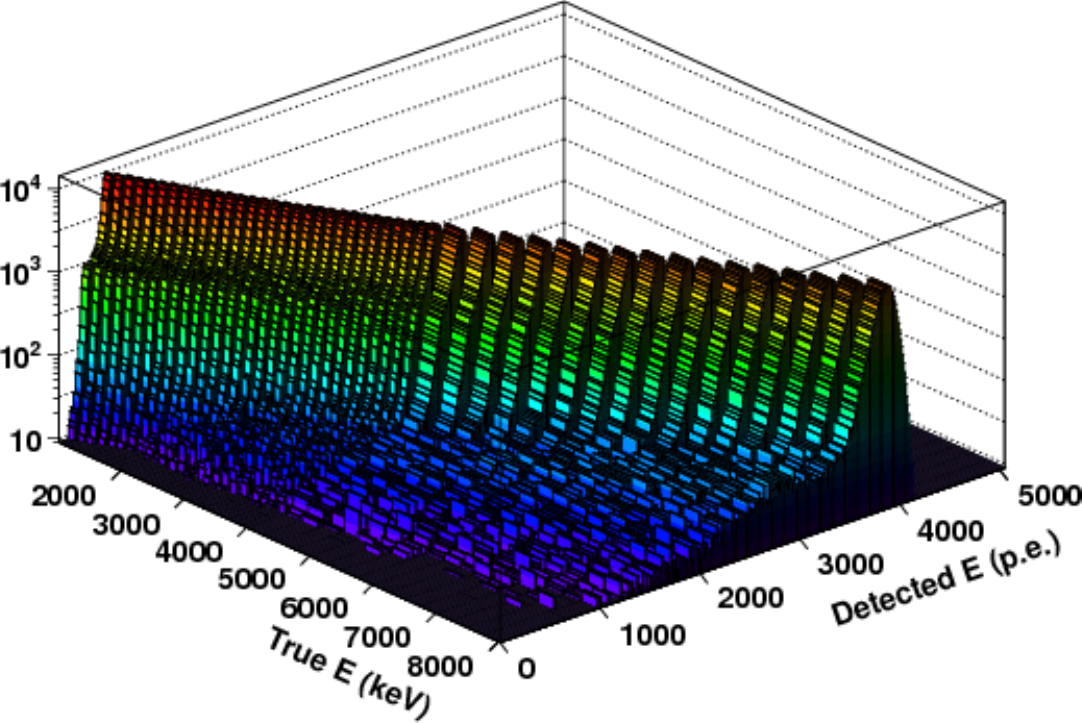}
\caption{\label{fig:TransMat} Transition matrix describing the energy response of the detector.}
\end{figure}
A vertex resolution of 15 cm is assumed, based on the arrival time of the first photons on the PMT's of opposite sides of the detector. This parameter is not critical since the extension of the reactor core itself already generate a smearing of the antineutrino baseline of about 30 cm ($1 \sigma$). For the oscillation analysis the antineutrino events are collected in bins of $\simeq$ 40 cm width along the detector-core axis.

The energy response is described by a transition matrix connecting any incident antineutrino energy with a reconstructed energy projected in bins (fig.\ref{fig:TransMat}). The more important energy loss of events close to the edges of the target induces some spectral distorsion along the detector axis. To account for this effect a specific transition matrix is computed for each detector slice considered in the oscillation analysis (see the Sensitivity Section). A symmetry with respect to the middle vertical plane of the detector is assumed.

\begin{figure}[h]
\includegraphics[width=0.5\textwidth]{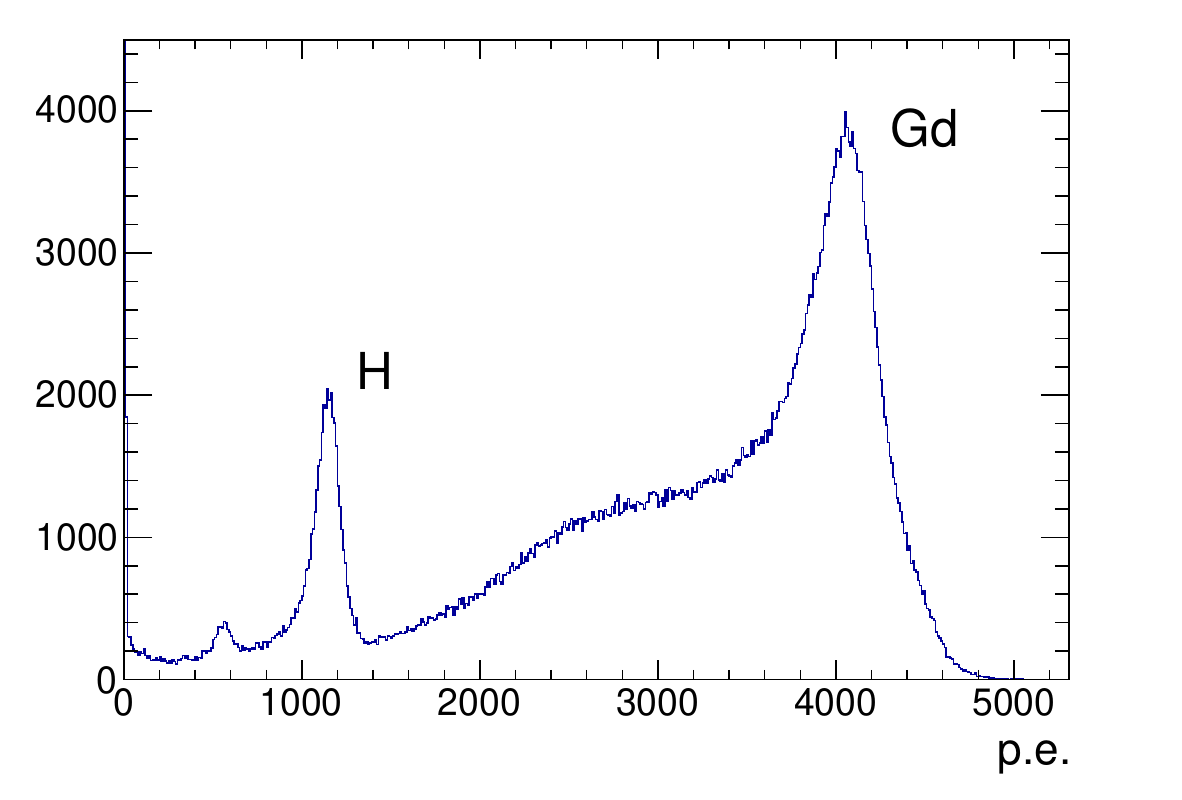}
\caption{\label{fig:nCaptPlot} Visible energy spectrum for neutron capture with initial vertex uniformly distributed in the target volume. The energy scale is about 500 p.e. per MeV.}
\end{figure}
The signal of the neutron capture is the emission of $\gamma$ rays. About 3 $\gamma$s and a total energy of $\simeq$8 MeV for the capture on Gadolinium, a single 2.2 MeV $\gamma$ for the capture on hydrogen. Figure \ref{fig:nCaptPlot} shows the simulated response to neutron capture. Assuming a cut of 6 MeV for neutron capture candidates, the predicted detection efficiency is 58\%.

\subsubsection*{Sensitivity}
The contours expected from the Stereo experiment are computed using the full simulated detector response described above. Two candidate sites are tested: the ILL research reactor in Grenoble, France and the Osiris research reactor in Saclay, France. The relevant parameters of each site are summarized in table \ref{tab:sites}. 
\begin{table}[h]
\centering
\begin{tabular}{|c|c|c|}
\hline
\hline
 & ILL & Osiris \\
\hline
Power & 58 MW & 70 MW \\ 
\hline
Core size & $\phi=28 cm,\,h=80 cm$ & $61\times 61 \times 63 cm$ \\ 
\hline
Nuclear fuel & 93\% U5 enriched & 20\% U5 enriched \\ 
\hline
Det-Core distance & 8 m & 10 m \\ 
\hline
Overburden & $\simeq 5$ m.w.e. & $\simeq 5$ m.w.e. \\ 
\hline
Expected $\bar{\nu}_e$/day & 708 & 1043 \\ 
\hline
\hline
\end{tabular}
\caption{\label{tab:sites} Characteristics of the ILL and Osiris sites.}
\end{table}
The detection threshold is set at 2 MeV on the visible energy, equivalent to a 2.78 MeV threshold on the antineutrino energy. The background model of figure \ref{fig:SoverB} is used with an signal/background ratio of 1.5. The exposure time is 300 full days, corresponding to about 2 years running at the reactor when accounting for the commissioning and the reactor off periods.

To evaluate the sensitivity to the new oscillation we introduce the following $\chi^2$, built according to annex A of \cite{Huber:2003pm}
\begin{eqnarray}
\label{eq:chi2}
\chi^2(\Delta m^2_{new}, \theta_{new}) &=& \sum_{l,i} \left(\dfrac{O_{l,i}(\Delta m^2, \sin^2\theta)-T_{l,i}}{\sigma^2(O_{l,i})}\right) 
+\sum_l \left(\dfrac{\beta_l}{\sigma_\beta}\right)^2
+\sum_i \left(\dfrac{\alpha_i}{\sigma_{\alpha_i}}\right)^2 \\
& & +\left(\dfrac{\alpha_{WM}}{\sigma_{WM}}\right)^2 \nonumber 
\end{eqnarray}
with
\begin{eqnarray}
T_{l,i} &=& \left[1+\alpha_{l}+\alpha_i+\alpha_{WM}*(E_i-1)\right]\,N_{l,i}+\beta_l\,\left.\dfrac{dN_{l,i}}{d\beta_l}\right|_{\beta_l=0}
\end{eqnarray}
$O_{l,i}(\Delta m^2, \sin^2\theta)$ is the expected event rate for the mass square splitting $\Delta m^2_{new}$ and the mixing angle $\theta_{new}$. $T_{l,i}$ is the theoretical prediction without oscillation and $N_{l,i}$ the expected number of events in the $i^{th}$ energy bin (250 keV bin width) and $l^{th}$ baseline bin in the detector. The default energy binning is 250 keV, adapted to the simulated detector resolution. Similar sensitivity is achieved with 500 keV bins. With a 2 m long target 5 baseline bins of $\simeq 40$~cm are used in the analysis. These detector slices are perpendicular to the detector-reactor axis which can differ significantly from the detector axis, depending on the site.

\begin{figure}[!h]
\begin{minipage}{.5\textwidth}
\centering \includegraphics[width=0.95\textwidth]{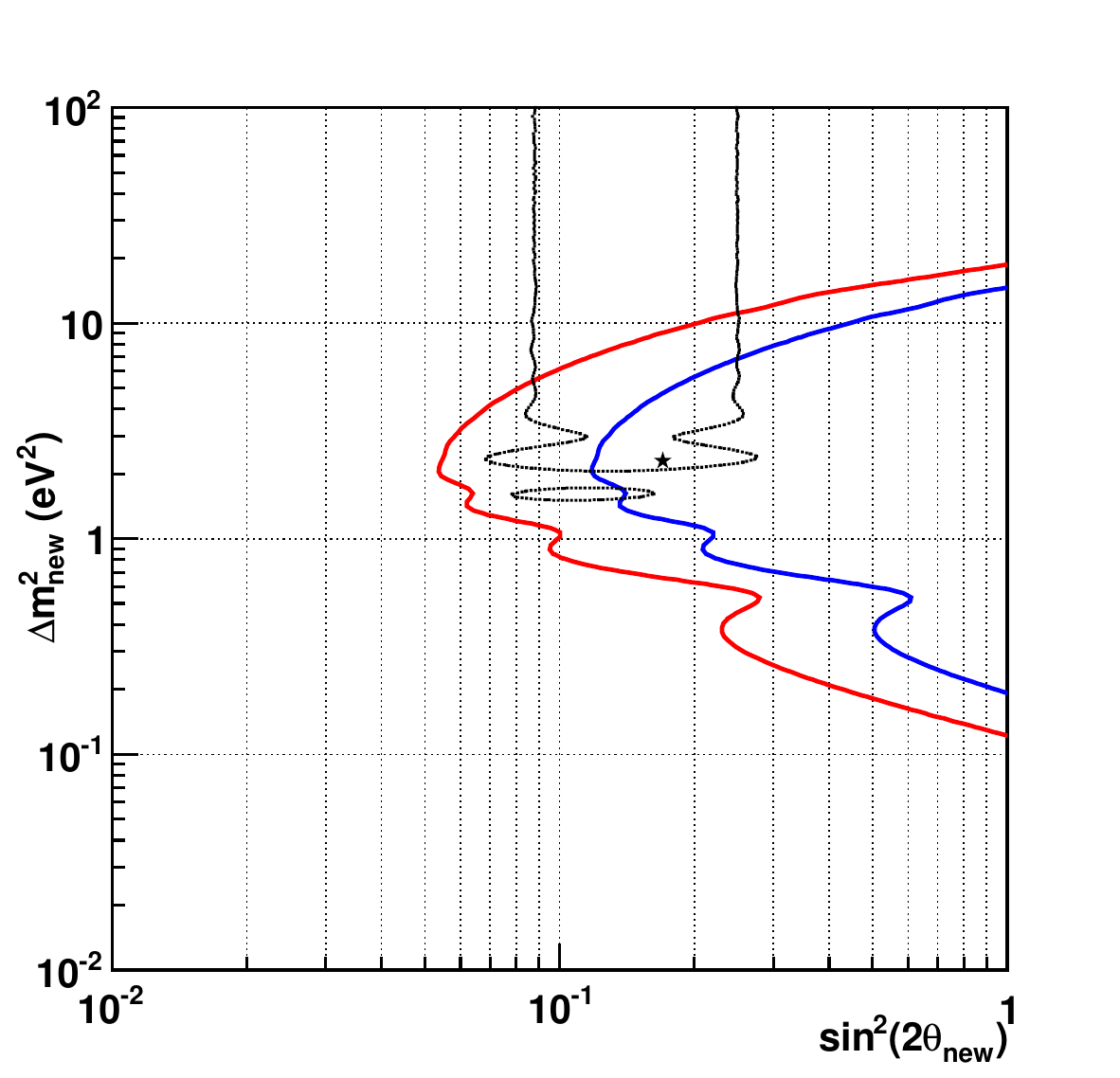}
\end{minipage}\hfill
\begin{minipage}{.5\textwidth}
\centering \includegraphics[width=0.95\textwidth]{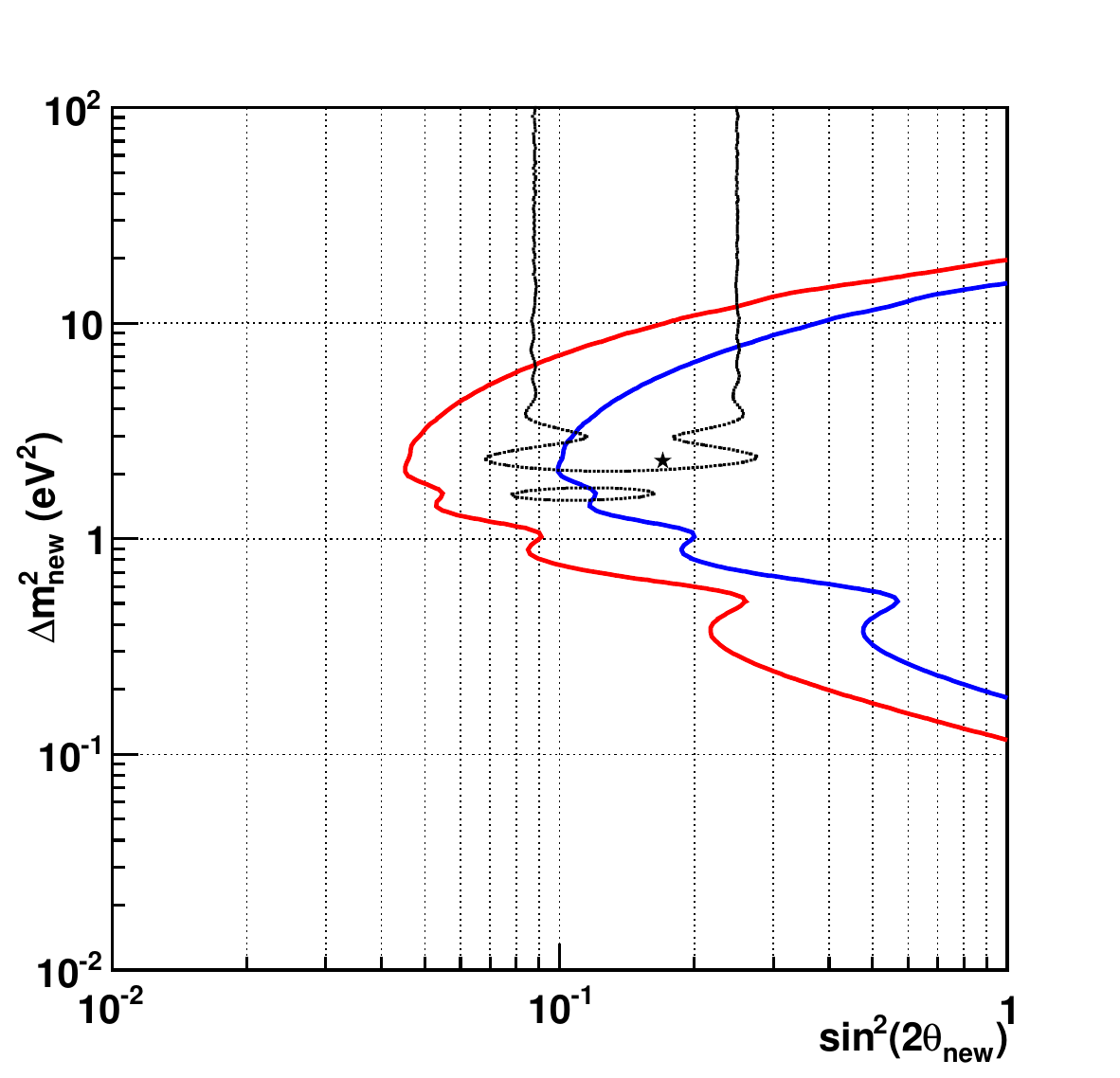}
\end{minipage}\hfill
\caption{\label{fig:Contours} Contours of the Stereo experiment for the ILL site (left) and Osiris site (right). The contours are obtained for 300 full days of data, 58\% detection efficiency, 2~MeV visible energy threshold and signal/background=1.5} 
\end{figure}
For each set of oscillation parameters, the $\chi^2$-function has to be minimized with respect to the parameters $\alpha_l,\,\alpha_i,\,\alpha_{WM}$, and $\beta_l$ modeling the systematical errors. $\alpha_l$ are free normalization parameters for the $L$ baseline bins. The absence of associated pull term provides a shape-only analysis. Their minimization absorbs the variation of rate with $L^2$, as well as any mean effect of the oscillation. The parameters $\alpha_i$ refer to bin-to-bin uncorrelated error uncertainty on the shape of the expected energy spectrum. They are dominated by the electron to antineutrino conversion error and the statistical error. The $\alpha_i$ are taken from \cite{Huber:2011wv}. The uncertainty on the weak magnetism correction is another independent source of shape uncertainty, correlated between energy bins but not equivalent to a simple normalisation error. It is described by a linear correction with $\alpha_{WM}$ a slope parameter. The size of the uncertainty used here, $\sigma(\alpha_{WM})$=1\%, corresponds to a $\sim$3\% change of the total predicted rate. This is the dominant theoretical error beyond normalization for the predicted neutrino spectra. The energy scale uncertainty in the two detectors is taken into account by the parameter $\beta_l$ with $\sigma_\beta=2\%$ and
\begin{eqnarray}
\left.\dfrac{dN_{l,i}}{d\beta_l}\right|_{\beta_l=0} &=& \dfrac{ E_i^+(N_{l,i} + N_{l,i}^+)- E_i^-(N_{l,i}^- + N_{l,i}) }{ 2*(E_i^+-E_i^-) }
\end{eqnarray}
The $+(-)$ exponents refer to the values at the upper (lower) edge of the energy bin.

Finally $\sigma(O_{l,i}) = \sqrt{N_{l,i}+bi+bi}$, taking into account the subtracted background contribution $b_i$ and assuming comparable time dedicated to background measurement and data. The same background model (rate and spectrum shape) is used in all baseline bins.

This formalism provides a quite complete treatment of all experimental systematics. The contours are illustrated in figure \ref{fig:Contours}. The 99\% contour of the experiment covers the 99\% contour of the reactor anomaly in the $\Delta^2m_{new}$ area of interest. The key points of the experiment are to insure a good background rejection and to minimize the energy losses outside the target volume. This study uses background conditions already achieved by previous experiments. With a 15~cm $\gamma$-catcher layer, the simulation shows that the edge effects in the first and last baseline bins are not critical. For instance, we checked that using the central transition matrix for all baseline bins doesn't modify significantly the contours in the reactor anomaly area. We aslo checked numerically that the impact of the other parameters (bin size, bin-to bin uncorreleted errors of the prediction, weak magnetism correction, energy scale) was not dominant.

This experiment proposes a powerfull test of the reactor anomaly using a well known technology. The best fit of the reactor anomaly is tested well beyond the 5 sigma level.

\clearpage

\subsection{A Very Short-Baseline Study of Reactor Antineutrinos at the National Institute of Standards and
Technology Center for Neutron Research\footnote{Proposed by 
          H.~P.~Mumm (National Institute of Standards and Technology) and
          K.~M.~Heeger(University of Wisconsin)}}
\label{NIST_Reactor}

Data from a variety of short-baseline experiments, astrophysical observations, and evidence from
cosmology are beginning to favor the existence of additional neutrino mass states beyond the three
active species in the standard model of Particle Physics (SM). Most recently, a re-analysis of short-baseline reactor neutrino experiments  has revealed a discrepancy between observations and the predicted antineutrino flux of about 3\%~\cite{Mention:2011rk}. This can be interpreted as an indication of the existence of at least one sterile neutrino with a mass splitting of $\sim 1{\rm eV}^2$ relative to the SM neutrinos~\cite{Giunti:2011hn}. The possible implications of additional sterile neutrino states would be profound.  This would be the first direct observation of a particle beyond the SM, and have a significant impact on our understanding of the Universe.  As a result, great interest has developed in carrying out a definitive test of the sterile neutrino hypothesis with the suggested mass splitting~\cite{SNAC,FNAL} as well as in making a precision determination of the absolute reactor neutrino flux.

Various approaches have been proposed to address this situation~\cite{FNAL,Dwyer:2011xs}.  One possibility is to measure the reactor antineutrino flux from compact research reactors at distances comparable to the expected oscillation length of sterile neutrinos ($\sim$2-3\,m). A measurement of the rate and shape of the antineutrino spectrum as a function of distance will reveal the signature of sterile neutrinos in the form of both an overall rate deficit and a spectral distortion.  Though the finite dimensions of the reactor core as well as the spread in the position reconstruction of events inside the detector will smear the spectral and rate signatures of sterile neutrinos, compact cores with dimensions of $\sim$ 1~m and demonstrated detector technology will allow a direct observation of a spatial deviation from $1/r^2$. Furthermore, Highly Enriched Uranium (HEU) research reactors may allow a more accurate absolute prediction of the reactor neutrino flux from thermal power data and fuel simulations compared to commercial reactors. This in turn should result in a better comparison of the absolute measured antineutrino flux  with the flux prediction from the reactor.  In this context the major challenge of antineutrino measurements near research reactors is expected to come from the lack of overburden and the need to operate the detectors close to the core. (Overburden and underground locations typically help reduce environmental and cosmogenic backgrounds in low-energy neutrino measurements.) In addition, fast neutron backgrounds associated with the reactor and adjacent experiments will contribute significantly to the ambient backgrounds near the reactor. In spite of these challenges, recent development of antineutrino detectors in the context of non-proliferation and nuclear verification efforts have demonstrated the feasibility of antineutrino detection in such a situation.

Measurement of the reactor antineutrinos in a typical detector utilizes the inverse beta decay reaction yielding a prompt signal followed by a neutron capture tens of microseconds later. This detection principle has been used for decades in many reactor neutrino experiments and a variety of methods  are available ({\it e.g.}~choice of scintillator and doping isotope) for optimization of the detector design in a specific background  environment and detector geometry. The delayed coincidence allows for a significant reduction in accidental backgrounds from natural radioactivity and neutron capture gamma backgrounds. 

\begin{figure}[t]
        \includegraphics[width=3.43in]{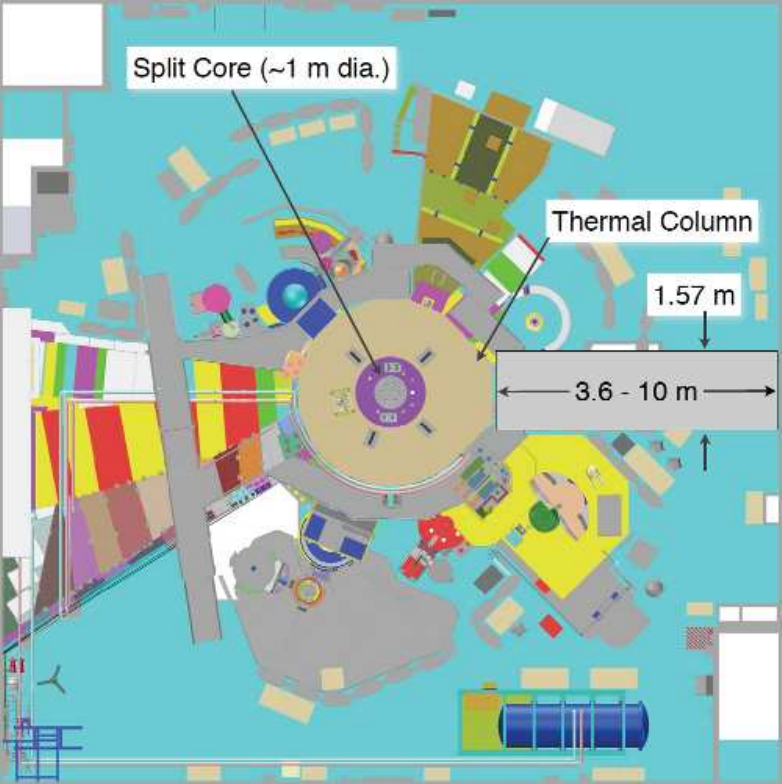} \hspace{.2in}
        \includegraphics[width=2.46in]{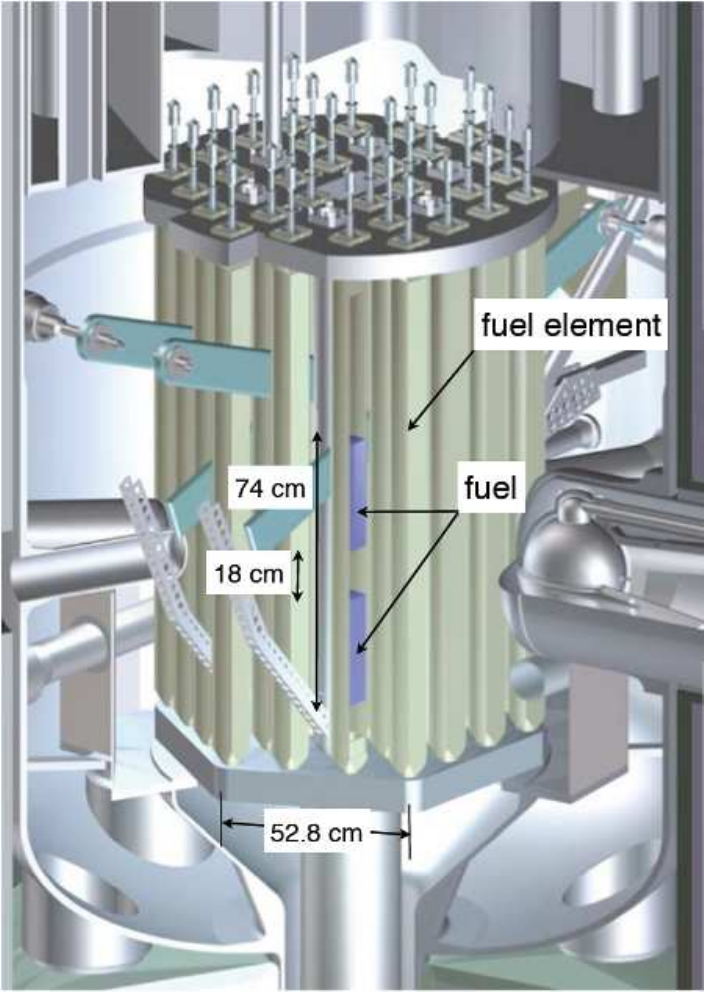}
        \caption{Left: The layout the NBSR confinement building with the reactor and adjacent instruments~\cite{Nist}. The confinement building allows the placement of a detector at distances of 3.6 to 10\,m from the reactor core. A several-ton antineutrino detector can be located in and moved through the region indicated in grey.   
     Right: Model of the NIST reactor core~\cite{Nist}.  The purple region indicates location of fuel within an element.}
        \label{fig:reactor}
\end{figure}

The National Institute of Standards and Technology (NIST)~\cite{Nist} operates a 20~MW D$_2$O-moderated research reactor at the NIST Center for Neutron Research (NCNR)~\cite{NISTCNR}.  The National Bureau of Standards Reactor (NBSR) is a split core geometry with 18~cm vertical separation between fuel segments and overall dimensions of roughly 1~m diameter  by 74~cm in height. See Figure~\ref{fig:reactor}. The primary purpose of the reactor is to produce neutrons for a broad program of fundamental and applied neutron science at NIST~\cite{NISTCNRexpansion}. The reactor operates in 38-day cycles with approximately 10-day refueling periods which provide an ideal opportunity for detailed studies of natural backgrounds. During every refueling 4 of the 30 fuel elements are replaced while others are repositioned.  Overall operating time at 20~MW averages 250 days per year.  Furthermore, an engineering model-based MCNP simulation package has been developed and validated at NIST to quantitatively predict fast neutron fluxes at instruments inside the facility. Finally, detailed information on the operation of the reactor is expected to be available to an experiment at the NIST facility.  

The reactor is inside a confinement building (Figure~\ref{fig:reactor}) that houses seven thermal neutron beams supporting a variety of applied and fundamental neutron experiments. Space and infrastructure include a full suite of standard utilities, an overhead crane, truck access, and floor loading capacity are suitable to support the installation and operation of an antineutrino detector of several tons. The space next to the biological shield indicated in Fig.~\ref{fig:reactor} is designated the ``thermal column".  This experimental area is designed to produce a high flux of thermal neutrons through successive moderation in a heavy water tank (currently filled with light water) and graphite blocks.  It is consequently the area of the confinement building that has the lowest fast neutron flux.  Fast neutron backgrounds in this area are expected to be dominated by scattering from adjacent instruments.  A volume measuring 10\,m long by 1.57\,m wide by 2.5\,m tall centered at roughly core height and at a minimum baseline of 3.6~m from the reactor core is potentially available, though floor level access to rear areas of the confinement building must be maintained.  In practice this suggests a smaller detector that is moved through the region indicated by grey in Fig.~\ref{fig:reactor}.
 
\begin{table}[htbp]
   \centering
   \begin{tabular}{l|ll} 
\hline
reactor type && D$_2$O, HEU \\
power &&  20 MW$_{th}$\\
fuel cycle && 38 days \\ 
number of fuel elements && 30 (load 4 fresh fuel elements in every cycle) \\
fuel composition && 93\% $^{235}\rm U_{3}\rm O_{8}+\rm Al$\\ 
average operating days per year   \hspace{.1in}&& 250 \\
core geometry && split core, hexagonal \\ 
core dimensions && overall height: 74\,cm (including 18\,cm central gap) \\
&& hexagonal sides: 52.8\,cm\\
&& overall diameter: $\sim$1\,m\\ 
\hline
   \end{tabular}
   \caption{Characteristics of the NIST research reactor, NBSR.}
   \label{tab:booktabs}
\end{table}

A variety of existing neutron spectrometers and the expertise of the NIST group would allow a thorough characterization of the fast neutron background in the vicinity of the thermal column and a validation of the shielding design for a potential detector package. Along with these measurements this background can be simulated once additional support is available. The NIST reactor is currently undergoing an upgrade and precise data will become available once the reactor is operational again in mid-2012.  Measurements  of the neutron background taken for health physics surveys can be used to estimate backgrounds in the near term. It is clear that substantial shielding will be required for the envisioned surface detector. However, based on initial estimates, cost, and structural requirements would not be expected to be prohibitive. 

The NIST Center for Neutron Research appears to be an attractive location for a possible short baseline reactor antineutrino experiment.  A more detailed study of the sensitivity and shielding requirements of a several ton liquid-scintillator  experiment is currently underway.

\clearpage
\subsection{OscSNS: A Precision Neutrino Oscillation Experiment at the SNS\footnote{
Proposed by: 
S.~Habib, I.~Stancu (University of Alabama),
M.~Yeh (Brookhaven National Laboratory),
R.~Svoboda (University of California, Davis),
B.~Osmanov, H.~Ray (University of Florida, Gainesville),
R.~Tayloe (Indiana University, Bloomington),
G.~T.~Garvey, W.~Huelsnitz, W.~C.~Louis, G.~B.~Mills,
Z.~Pavlovic, R.~Van~de~Water, D.~H.~White (Los Alamos National Laboratory),
R.~Imlay (Louisiana State University),
B.~P.~Roe (University of Michigan),
M.~Chen,  (Oak Ridge National Laboratory),
Y.~Efremenko (University of Tennessee),
F.~T.~Avignone (University of South Carolina),
J.~M.~Link (Virginia Tech).
}}
\label{OscSNS}
 
The issue of whether or not light sterile neutrinos ($m_{\nu_s} \sim 1$ eV/c$^2$) exist has received 
very considerable attention recently. Such exotic particles, first invoked to explain the 
$\bar \nu_\mu \rightarrow \bar \nu_e$ appearance signal observed by LSND, require a neutrino mass of 
$\sim 1$ eV/c$^2$, which is far above the mass scale associated with the active neutrinos. This unexpected
appearance is ascribed a process that proceeds through a sterile neutrino. More recently, lower than 
expected neutrino event rates from calibrated radioactive sources and nuclear reactors can also be 
explained by the existence of a sterile neutrino with mass $\sim 1$ eV/c$^2$. No experiment 
directly contradicts the existence of such a sterile neutrino, though there is tension between limits on
$\nu_\mu$ disappearance and the LSND observation. Fits to the world neutrino and antineutrino 
data are consistent with one
or two light sterile neutrinos at this $\sim 1$ eV/c$^2$ mass scale. It is crucial to establish if such 
totally unexpected light sterile neutrinos exist. The Spallation Neutron Source (SNS), located at Oak Ridge 
National Laboratory and built to herald a new era in neutron research, provides the opportunity to make 
that critical determination. The SNS with 1MW beam power is a prodigious source of neutrinos from $\pi^+$ 
and $\mu^+$ decay at rest and has a duty factor more than 100 times shorter than that of LAMPF. This much 
smaller duty factor not only reduces backgrounds, it also allows the $\nu_\mu$ induced events from $\pi^+$ 
decay to be separated from the $\nu_e$ and $\bar \nu_\mu$ events from $\mu^+$ decay. The 
monoenergetic 29.8 MeV
$\nu_\mu$ can be used to investigate the existence of light sterile neutrinos via the neutral-current 
reaction $\nu_\mu C \rightarrow \nu_\mu C^*$(15.11 MeV), which has the same cross section for all active 
neutrinos but is zero for sterile neutrinos. An oscillation or suppression of this reaction would be 
direct evidence for sterile neutrinos. OscSNS can obviously carry out a unique and decisive test of the 
LSND appearance $\bar \nu_\mu \rightarrow \bar \nu_e$ signal that has not been executed to date. The 
existence and properties of light sterile neutrino is central to understanding the creation of the heaviest 
elements and verifying our understanding of the nuclear reactions involved in nuclear reactors. An OscSNS 
detector is based on the LSND and MiniBooNE detectors and can be built for $\sim\$12$M (or $\sim\$8$M 
if the MiniBooNE oil and phototubes are reused and $\sim\$5$M if the tank size is reduced).

Observations of neutrino oscillations, and therefore neutrino mass, have been made by solar-neutrino 
experiments at a $\Delta m^2 \sim 8 \times 10^{-5}$ eV$^2$, and by atmospheric-neutrino experiments at 
a $\Delta m^2 \sim 3 \times 10^{-3}$ eV$^2$, where $\Delta m^2$ is the difference in squared mass of the 
two mass eigenstates contributing to the oscillations. In addition to these observations, the LSND 
experiment, which took data at Los Alamos National Laboratory (LANSCE) for six years from 1993 to 1998, 
obtained evidence for $\bar \nu_\mu \rightarrow \bar \nu_e$ oscillations at a $\Delta m^2 \sim 1$ eV$^2$. 
Oscillations at the mass splittings seen by LSND do not fit with well-established oscillation observations 
from solar and atmospheric experiments. The standard model, with only three flavors of neutrinos, cannot 
accommodate all three observations. Confirmation of LSND-style oscillations would require further non-trivial 
extensions to the standard model.

\begin{figure}
\centering
\includegraphics[height=0.9\textwidth,angle=90,clip=true]{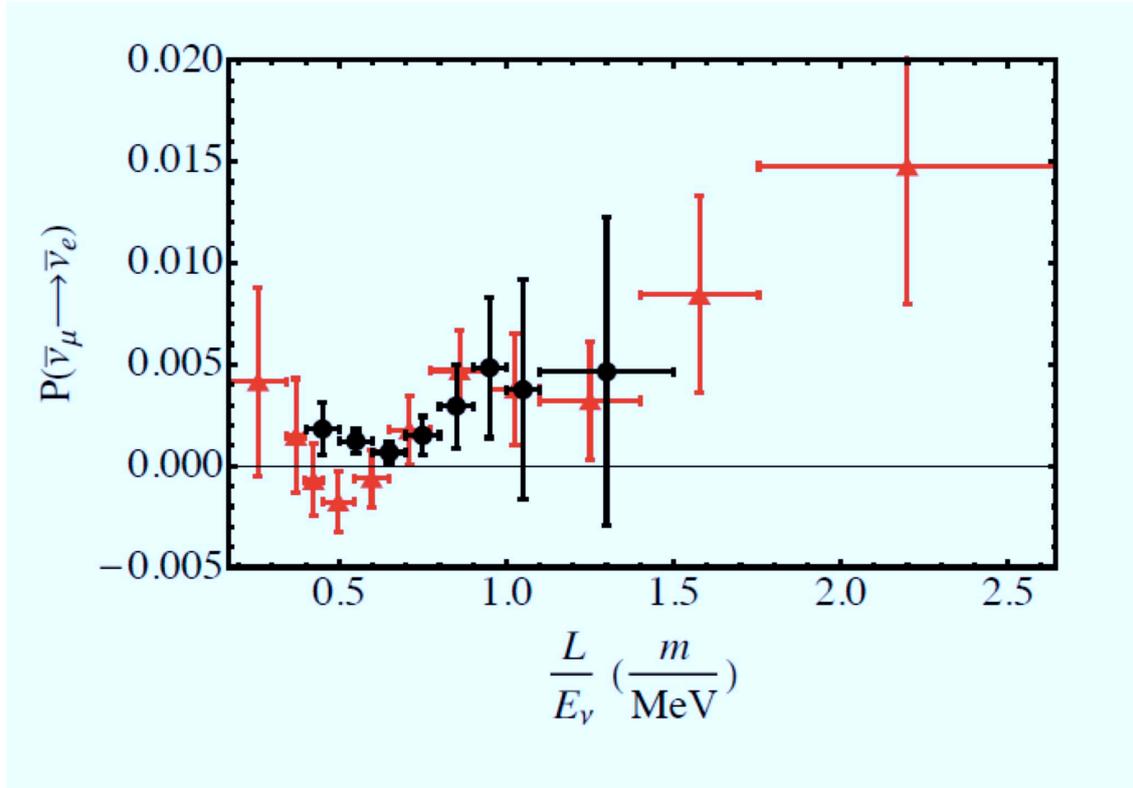}
\caption{
The probability of $\bar \nu_e$ appearing in a $\bar \nu_\mu$ beam as a function of the 
neutrino proper time.
}
\label{L_E}
\end{figure}

\begin{figure}
\centering
\includegraphics[width=11cm,angle=90,clip=true]{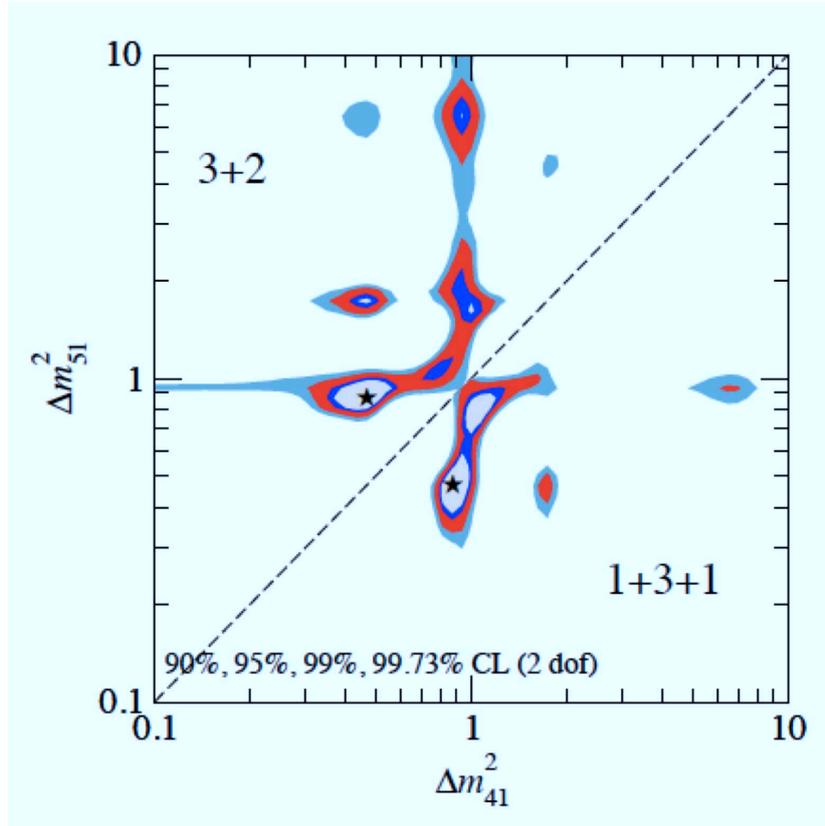}
\caption{
A global fit to the world neutrino plus antineutrino data indicates that the world data fit 
reasonably well to a 3+2 model with three active neutrinos plus two sterile neutrinos.
}
\label{global}
\end{figure}

The MiniBooNE experiment at Fermilab, designed to search for $\nu_\mu \rightarrow \nu_e$ and
$\bar \nu_\mu \rightarrow \bar \nu_e$ oscillations and to further explore the LSND neutrino oscillation 
evidence, has presented neutrino and antineutrino oscillation results. Combining these results, MiniBooNE 
observes a 3.7$\sigma$ excess of events in the 200-1250 MeV oscillation energy range. Many of the beyond 
the standard model explanations of this excess involve sterile neutrinos, which would have a huge impact 
on astrophysics, supernovae neutrino bursts, and the creation of the heaviest elements. Figure \ref{L_E} 
shows the 
L/E (neutrino proper time) dependence of $\bar \nu_\mu \rightarrow \bar \nu_e$ from LSND and MiniBooNE. 
The correspondence between the two experiments is striking. Furthermore, Figure \ref{global} shows fits to the world 
neutrino plus antineutrino data that indicate that the world data fit reasonably well to a 3+2 model with
three active neutrinos plus two sterile neutrinos.

The SNS offers many advantages for neutrino oscillation physics, including known neutrino spectra, well 
understood neutrino cross sections (uncertainties less than a few percent), low duty cycle for cosmic ray 
background rejection, low beam-induced neutrino background, and a very high neutrino rate of $>10^{15}$/s 
from the decay of stopped pions and muons in the Hg beam dump. Stopped pions produce 29.8 MeV monoenergetic 
$\nu_\mu$ from $\pi^+ \rightarrow \mu^+ \nu_\mu$ decay, while stopped muons produce $\bar \nu_\mu$ and 
$\nu_e$ up to the 52.8 MeV endpoint from $\mu^+ \rightarrow e^+ \nu_e \bar \nu_\mu$ decay. Note that 
$\pi^-$ and $\mu^-$ mostly capture in Hg before they have a chance to decay, so that hardly any neutrinos 
are produced from either $\pi^- \rightarrow \mu^- \bar \nu_\mu$ or $\mu^- \rightarrow e^- \bar \nu_e \nu_\mu$
decay. The rapid capture of the negatively charged meson in the Hg environment is an advantage over the LSND 
experiment, where the production target was more open with greater possibility of $\pi^-$ decay in flight and 
the resulting $\mu^-$ decaying to $e^- \bar \nu_e \nu_\mu$.

The SNS neutrino flux is ideal for probing $\bar \nu_\mu \rightarrow \bar \nu_e$ and
$\nu_\mu \rightarrow \nu_e$ appearance,
as well as $\nu_\mu$ disappearance into sterile neutrinos. The appearance searches
both have a two-fold coincidence for the rejection of background. For $\bar \nu_\mu \rightarrow \bar \nu_e$
appearance, the signal is an $e^+$ in coincidence with a 2.2 MeV $\gamma$: $\bar \nu_e p \rightarrow e^+ n$,
followed by $n p \rightarrow D \gamma$. For $\nu_\mu \rightarrow \nu_e$ appearance, the signal is an $e^-$ 
in coincidence with an $e^+$ from the $\beta$ decay of the ground state of $^{12}N$: $\nu_e~^{12}C \rightarrow
e^-~^{12}N_{gs}$, followed by $^{12}N_{gs} \rightarrow ~^{12}C e^+ \nu_e$. The disappearance search will detect 
the prompt 15.11 MeV $\gamma$ from the neutral-current reaction $\nu_\mu C \rightarrow \nu_\mu C^*$(15.11). 
This reaction has been measured by the KARMEN experiment, which has determined a cross section that is 
consistent with theoretical expectations. However, the KARMEN result was measured in a sample of 86 events, 
and carries a 20\% total error. OscSNS will be able to greatly improve upon the statistical and systematic 
uncertainties of this measurement. If OscSNS observes an event rate from this neutral-current reaction that 
is less than expected, or if the event rate displays a sinusoidal dependence with distance, then this will be 
evidence for $\nu_\mu$ oscillations into sterile neutrinos.

\begin{figure}
\centering
\includegraphics[height=\textwidth,angle=90,clip=true]{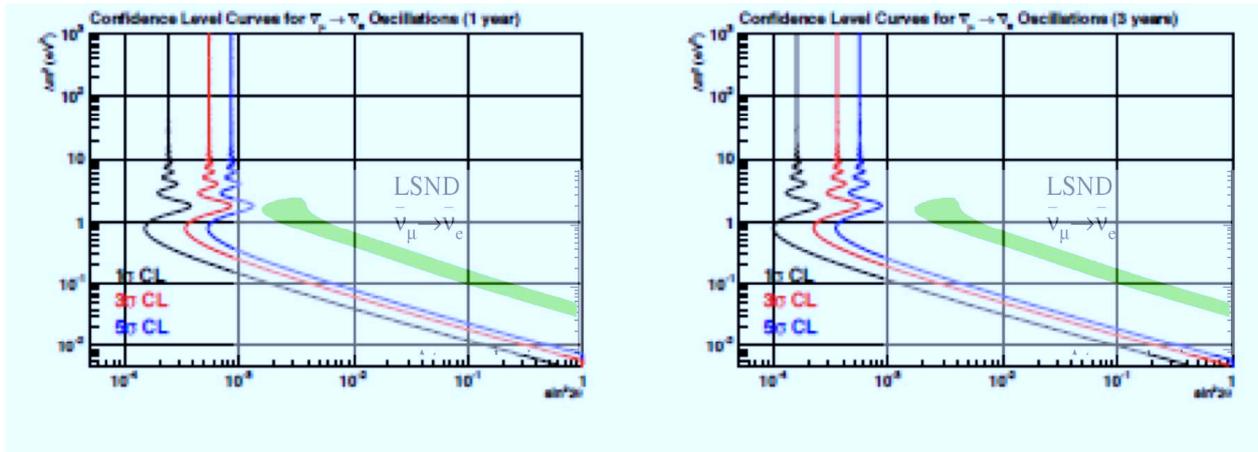}
\caption{
The OscSNS sensitivity curves for the simulated sensitivity to $\bar \nu_\mu \rightarrow \bar \nu_e$  
oscillations after one (left) and three (right) years of operation. Note that it has more than 5$\sigma$
sensitivity to the LSND result in 1 year.
}
\label{app}
\end{figure}

\begin{figure}
\centering
\includegraphics[height=\textwidth,angle=90,clip=true]{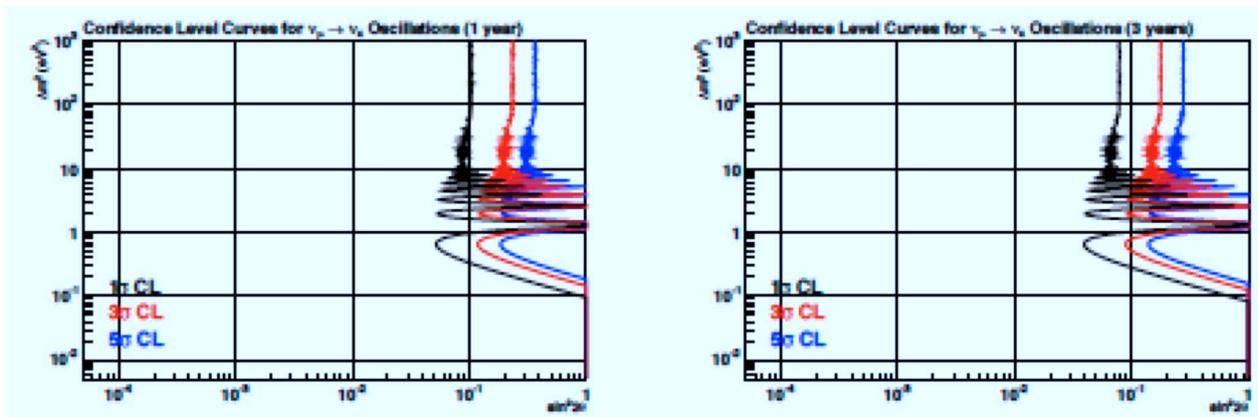}
\caption{
The OscSNS sensitivity curves for $\nu_\mu$ disappearance for 1 and 3 years, respectively.  
}
\label{disap}
\end{figure}

In addition to the neutrino oscillation searches, OscSNS will also make precision cross section measurements of 
$\nu_e C \rightarrow e^- N$ scattering and $\nu e^- \rightarrow \nu e^-$ elastic scattering. The former reaction 
has a well-understood cross section and can be used to normalize the total neutrino flux, while the latter 
reaction, involving $\nu_\mu$, $\nu_e$, and $\bar \nu_\mu$, will allow a precision measurement of 
$\sin^2 \theta_W$.
We have simulated many of the processes mentioned above, and
in Figures \ref{app} and \ref{disap} we present some of the most relevant results.

\clearpage
\subsection{LSND Reloaded\footnote{Proposed by Sanjib K. Agarwalla and Patrick Huber
(Virginia Tech).}}
\label{LSND_reloaded}

\begin{figure}[b]
\includegraphics[width=0.55\columnwidth]{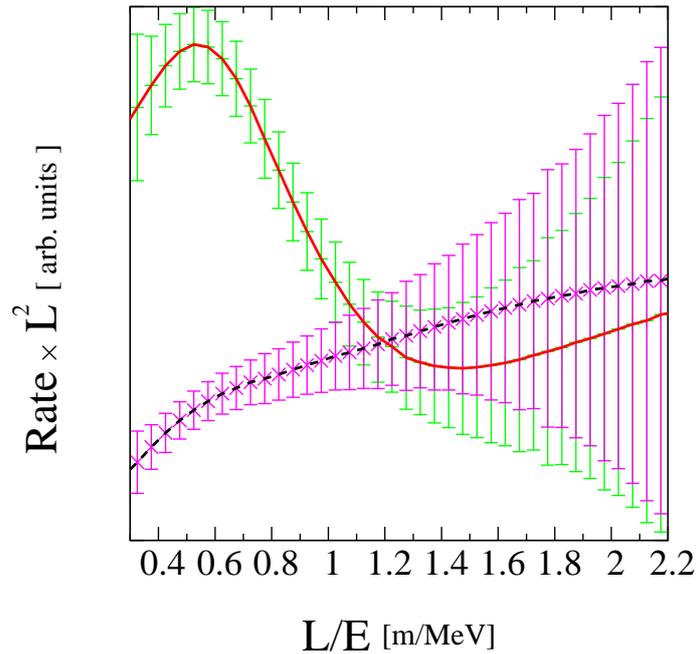}
\caption{\label{fig:LSND_baseline} The signal event rate after one year weighted with
  $L^2$ as function of the reconstructed baseline divided by
  reconstructed neutrino energy $L/E$, shown as solid line. The dashed
  line shows the background weighted with $L^2$. The error bars show
  the statistical errors only.  The oscillation signal is computed for
  $\sin^22\theta=10^{-3}$ and $\Delta m^2=2\,\mathrm{eV}^2$. Figure and caption taken from~\cite{Agarwalla:2010zu}.}
\end{figure}

This note is based on Ref.~\cite{Agarwalla:2010zu} and more details
can be found in there. The basic proposal is to repeat
LSND~\cite{Athanassopoulos:1995iw,Athanassopoulos:1997er,Aguilar:2001ty}
using a stopped pion neutrino source and Super-Kamiokande doped with
Gadolinium~\cite{Beacom:2003nk,Watanabe:2008ru,Dazeley:2008xk} as
detector.  In this configuration both neutrino production and
detection employ the same mechanism as LSND did. The crucial
difference is that the cosmogenic background rate in Super-Kamiokande
is negligible and the enormous size of the detector allows to map out
the $L/E$-dependence of the effect over the range of
$0.3-2.2\,\mathrm{m}\,\mathrm{MeV}^{-1}$. The $\bar\nu_e$ signal for a
typical choice of oscillation parameters is shown as red line in
figure~\ref{fig:LSND_baseline}. At the same time any beam related
background scales as $L^3/L^{-2}=L$ shown as black dashed line in
figure~\ref{fig:LSND_baseline}, which is quite distinct form the signal.

Super-Kamiokande has a fiducial mass of $22.5\,\mathrm{kt}$ compared
to around $120\,\mathrm{t}$ in LSND and thus a relatively modest beam
power can yield significant event rates, for details of the proton
source, see ~\cite{Conrad:2009mh,Alonso:2010fs}. If we assume 300\,kW beam power we obtain the following
 event rates per year of
operation are, whereas the background event
rate due to beam contamination is $765$.
\begin{table}[h!]
\begin{tabular}{|c|cccc|}
\hline
$\Delta m^2$ $[\mathrm{eV}^2]$&0.1&1&10&100\\
\hline
signal&29&1605&1232&1314\\
\hline
\end{tabular}
\caption{\label{tab:events} Number of signal events after one year for 
  $\sin^22\theta=10^{-3}$ including efficiency and energy resolution. Table and caption taken from ~\cite{Agarwalla:2010zu}.}
\end{table}
Secondly, the large rock overburden of approximately
$2,700\,\mathrm{mwe}$, compared to $120\,\mathrm{mwe}$ in LSND,
reduces cosmic ray induced backgrounds to negligible
levels~\cite{Conrad:2009mh,Alonso:2010fs}.

\begin{figure}[t!]
\includegraphics[width=0.55\columnwidth]{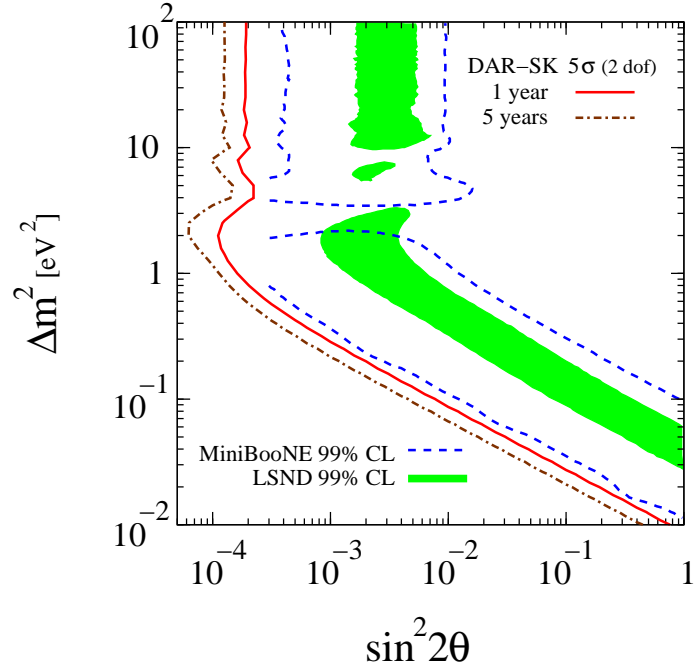}
\caption{\label{fig:sens} Sensitivity limit to sterile neutrino
  oscillation at the $5\,\sigma$ confidence level, shown as red, solid
  line, labeled DAR-SK. This limit corresponds to a one year run. The
  green/gray shaded region is the LSND allowed region at $99\%$
  confidence level, whereas the dashed line is the MiniBooNE
  antineutrino run allowed region at $99\%$ confidence
  level~\cite{AguilarArevalo:2010wv}. Figure and caption taken from~\cite{Agarwalla:2010zu}. }
\end{figure}

A detailed $\chi^2$ analysis is presented in
Ref.~\cite{Agarwalla:2010zu} which does include a 5\% systematic
error. The resulting $5\,\sigma$ sensitivity in comparison to the
allowed regions from LSND and MiniBooNE is shown in
figure~\ref{fig:sens}. A one year run at 300\,kW beam power provides a
significant test of LSND and MiniBooNE, or equivalently a 5 year run
at 60\,kW beam power. Spreading out the luminosity over several years
is possible because cosmogenic backgrounds are very small compared to
beam backgrounds.

In summary, the experiment proposed in Ref.~\cite{Agarwalla:2010zu}
will allow to conclusively test the LSND and MiniBooNE claims for
$\bar\nu_\mu\rightarrow\bar\nu_e$ oscillations with more than
$5\,\sigma$ significance. Due to the ability to observe a wide range
of $L/E$ it will be straightforward to disentangle backgrounds and the
signal as well as to identify the underlying physical mechanism for
flavor transitions. Since both the production and detection mechanism
as well as the energy and baseline are identical to the ones used in
LSND, this will be the final test of LSND.

\clearpage
\subsection{Kaon Decay-at-Rest for a Sterile Neutrino Search~\footnote{Proposed by J.~Spitz (MIT)}}
\label{sec:Kaons}

Monoenergetic muon neutrinos from positive charged kaon decay-at-rest can be used as a source for an electron neutrino appearance search in probing the neutrino oscillation parameter space near $\Delta m^2\sim 1~\mathrm{eV}^2$~\cite{Spitz:2012gp}. The charged kaon decays to a muon and a 235.5~MeV muon neutrino ($K^{+} \rightarrow  \mu^+ \nu_{\mu}$) about 64\% of the time . Electron neutrinos, originating from these muon neutrinos ($\nu_\mu \rightarrow \nu_e$), can be searched for in a narrow reconstructed energy window around 235.5~MeV. The signature of an electron neutrino event is the charged current, quasi-elastic interaction $\nu_e n \rightarrow e^- p$. The analogous muon neutrino charged current interaction as well as electron neutrino events, coming from the three body kaon decays $K^{+} \rightarrow \pi^0 e^+ \nu_{e}$ (BR=5.1\%) and $K^{0}_L \rightarrow \pi^\pm e^\mp \nu_{e}$ (BR=40.6\%), are the background for this appearance search. However, these event classes can also be used to perform the flux, cross section, and \textit{in-situ} background measurements necessary to extract the oscillation probability from an observed electron neutrino signal.

\subsubsection*{Experimental description}
\begin{figure}
\includegraphics[width=0.7\textwidth]{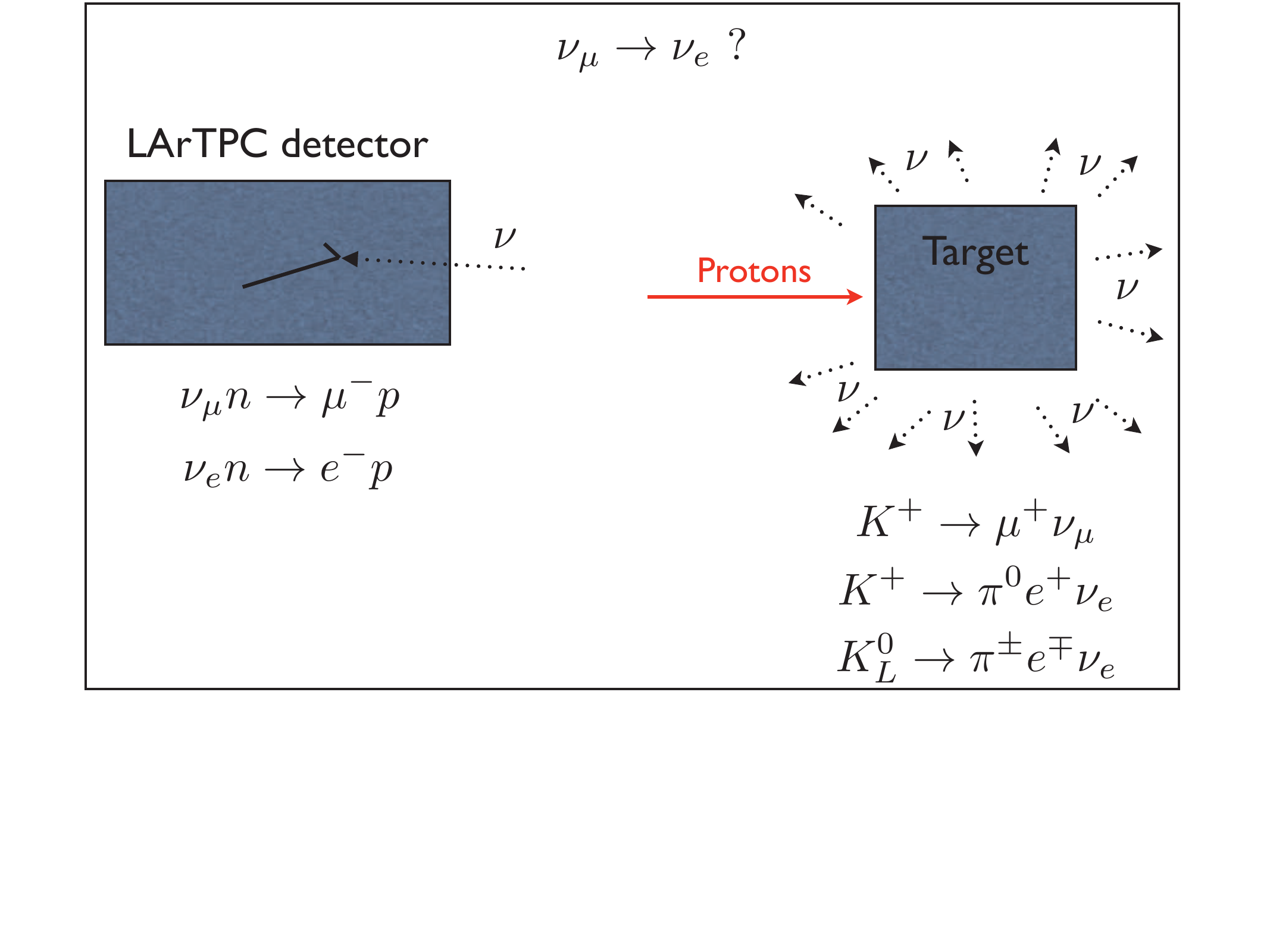}
\vspace{-2.0cm}
\caption{\label{fig:schematic} A kaon decay-at-rest based experimental design for probing the sterile neutrino parameter space. The relevant kaon decay modes and neutrino interactions are shown on the right and left, respectively.}
\end{figure}

This experimental design calls for an intense source of charged kaon decays. A large copper block is placed just downstream of a powerful $\gtrsim$3~GeV kinetic energy proton beam. Kaons created in primary proton-copper and secondary interactions quickly come to rest and subsequently decay. A LArTPC-based detector is envisioned opposite the primary proton beam's direction at a baseline of 160~m from the isotopic neutrino source in order to search for electron neutrino appearance events near a reconstructed energy of 235.5~MeV. A schematic of the experimental design can be seen in Fig.~\ref{fig:schematic}. The baseline is chosen in consideration of sensitivity to LSND and the $1/r^2$ dependence of the flux. The oscillation maximum at the LSND best fit $\Delta m^2=1.2~\mathrm{eV}^2$ occurs at about 240~m for 235.5~MeV neutrinos. 

The Booster Neutrino Beam (BNB) at Fermilab, currently featuring an 8~GeV kinetic energy beam providing $\sim3 \times 10^{20}$~protons on target (POT) every year, has been considered as a source for such an experiment--although many proton beam facilities around the world can be used. The precise calorimetric reconstruction, background identification, and three dimensional imaging capabilities of LArTPCs make the technology optimal for this design~\cite{Rubbia:2004tz}.

\subsubsection*{Flux, nuclear effects, background and detector resolution}
The signal-inducing monoenergetic muon neutrino and all-energy background electron neutrino flux are determined using the Geant4 simulation package~\cite{Agostinelli:2002hh}. Both decay-at-rest and decay-in-flight kaons (charged and neutral) are studied. The simulation predicts a monoenergetic $\nu_\mu/$proton yield of 0.038.

Nuclear effects and detector resolution act to significantly smear the reconstructed energy of signal and background events, making the nominally distinctive monoenergetic signal more difficult to identify. The NuWro neutrino event generator and its argon spectral function implementation are employed for simulating neutrino events on an argon nucleus~\cite{Juszczak:2009qa,Golan:2012wx}. Intra-nuclear interactions, often resulting in multiple ejected nucleons, are simulated as well. 

The dominant background contribution to the electron neutrino appearance search comes from three body charged kaon decay at rest ($K^{+} \rightarrow \pi^0 e^+ \nu_{e}$) which produces electron neutrinos with maximum $\sim$226~MeV energy. The concept described here can be thought of as similar to that of a neutrinoless double beta decay search with a signal sought near the endpoint of a well predicted and measured background spectrum. There are also background contributions from three body kaon decay-in-flight with neutrino energies up to and exceeding 235.5~MeV.  

The background-free identification and precise calorimetric reconstruction of electron neutrino events is vital to this experimental concept. Along with differentiating muon and electron events, calorimetrically reconstructing the outgoing proton(s) and electron-induced electromagnetic shower are important for distinguishing the monoenergetic signal from background. The reconstruction resolution assumptions employed for the sensitivity estimates discussed later are consistent with currently available LArTPC-based measurements~\cite{Ankowski:2008aa,Arneodo:2006ug}. 

\begin{figure}
\includegraphics[width=0.7\textwidth]{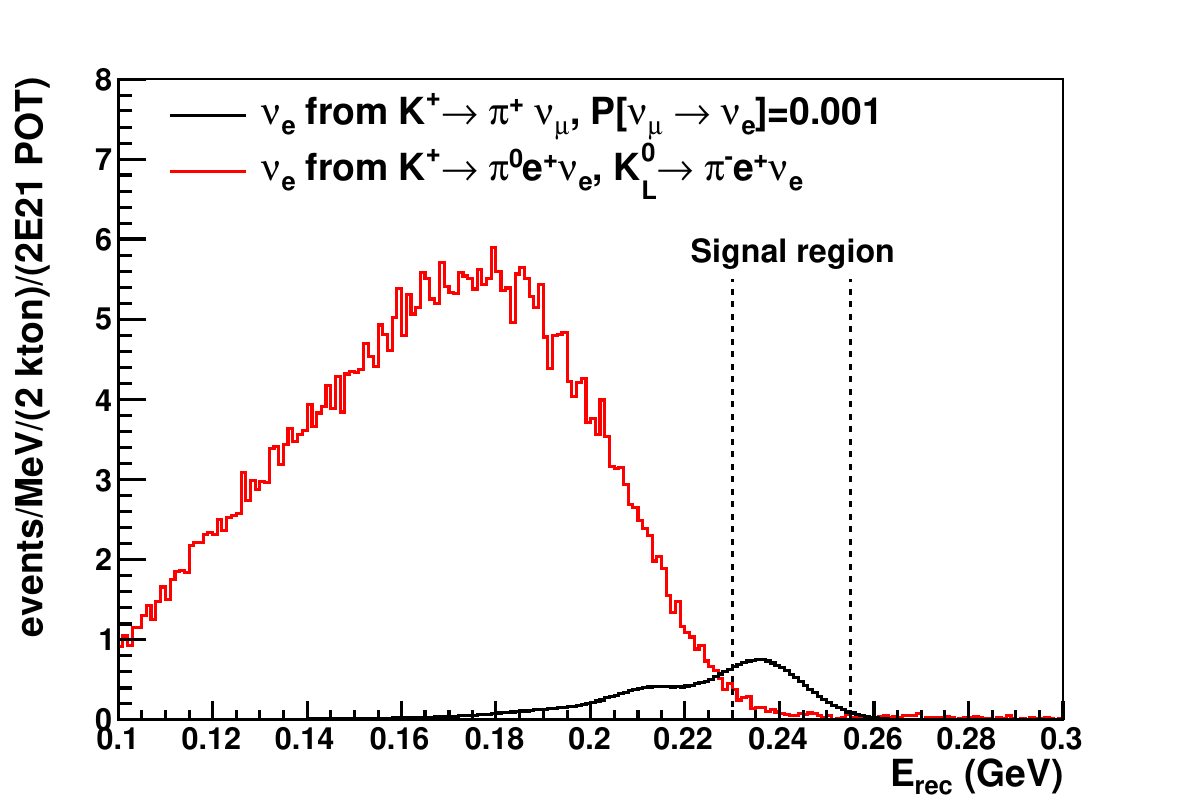}
\vspace{-.3cm}
\caption{\label{fig:rate_nuwro}The expected background rate alongside an injected electron neutrino appearance signal [$P(\nu_\mu \rightarrow \nu_e)=0.001$]. Note that the signal originates from oscillated monoenergetic muon neutrinos at 235.5~MeV. The reconstructed energy signal region is denoted by the vertical lines.}
\end{figure}

A ``signal region", around a reconstructed energy of 235.5~MeV, is defined in consideration of optimizing signal-to-background and having enough events left over to recognize a signal. The expected signal and background rates, after accounting for nuclear and detector resolution effects, are shown in Fig.~\ref{fig:rate_nuwro}. The signal region requirement, visible in the figure, leaves approximately 45\% of signal events. Although the chosen size and location of this region is largely based on the neutrino event generation nuclear models employed in this study as well as assumptions about the detector's calorimetric resolution, the actual signal region can be tuned once more information is available. Furthermore, the background prediction can be refined with \textit{in-situ} measurements below the signal region, most relevant for understanding the three body decay-at-rest spectrum ($K^{+} \rightarrow \pi^0 e^+ \nu_{e}$), and above the signal region, most relevant for understanding the charged and neutral kaon decay-in-flight induced background. The relationship between true neutrino energy and reconstructed energy, as well as relevant cross section measurements, can be obtained with these electron neutrinos along with the thousands of monoenergetic muon neutrinos expected.

\subsubsection*{Sensitivity}
A simple single bin, counting experiment is employed in an attempt to discern a potential oscillation signal from background inside the region of interest. Along with signal region fine tuning, a more sophisticated analysis beyond the counting only experiment can be employed in the future for oscillation sensitivity improvement.

A 2~kton detector and 2$\times 10^{21}$ POT exposure are considered for oscillation sensitivity studies. These values have been chosen in order to provide $\gtrsim$5$\sigma$ sensitivity to most of the LSND allowed region. As an example, approximately 16~$\nu_e$-like (signal+background) events above a background of 3~events are expected with an oscillation probability of $P(\nu_\mu \rightarrow \nu_e)=0.001$. The most relevant experimental parameters and assumptions can be seen in Table~\ref{assumptions}.

\begin{table}[t]
\label{base_opti}
  \begin{center}
    {\footnotesize
      \begin{tabular}{|c|c|} \hline 
        Proton target  & Copper  \\  \hline
        $\frac{\nu_\mu (235.5~\mathrm{MeV})}{\mathrm{proton}}$ yield at $T_{\mathrm{p}}=$8~GeV  & 0.038 \\  \hline
        Exposure  & 2$\times 10^{21}$ protons on target  \\  \hline
        Baseline  & 160~m  \\  \hline
        Neutrino target  & $^{40}$Ar (22 neutrons)  \\  \hline
        Neutrino target mass  & 2~kton  \\  \hline
        Detection efficiency  & 100\%  \\  \hline
        $\nu_e~\sigma_{\mathrm{CC}}$ at 235.5~MeV  & $1.9\times10^{-43}$~m$^{2}$/neutron \\
        \hline
        $\nu_\mu~\sigma_{\mathrm{CC}}$ at 235.5~MeV  & $1.3\times10^{-43}$~m$^{2}$/neutron \\
        \hline
        $\frac{\Delta E}{E}$ for e$^{-}$ reconstruction & 0.33/$\sqrt{E (\mathrm{MeV})}+0.012$  \\ \hline
        $\frac{\Delta T}{T}$ for proton reconstruction & 0.10  \\ \hline
        Background syst. uncertainty  & 25\%  \\  \hline
        Neutrino baseline spread  & $\pm$10~m  \\  \hline
      \end{tabular} 
      \caption{The relevant parameters and assumptions employed in this study.}\label{assumptions}
}
\end{center}
\end{table}

\begin{figure}
\includegraphics[width=0.6\textwidth]{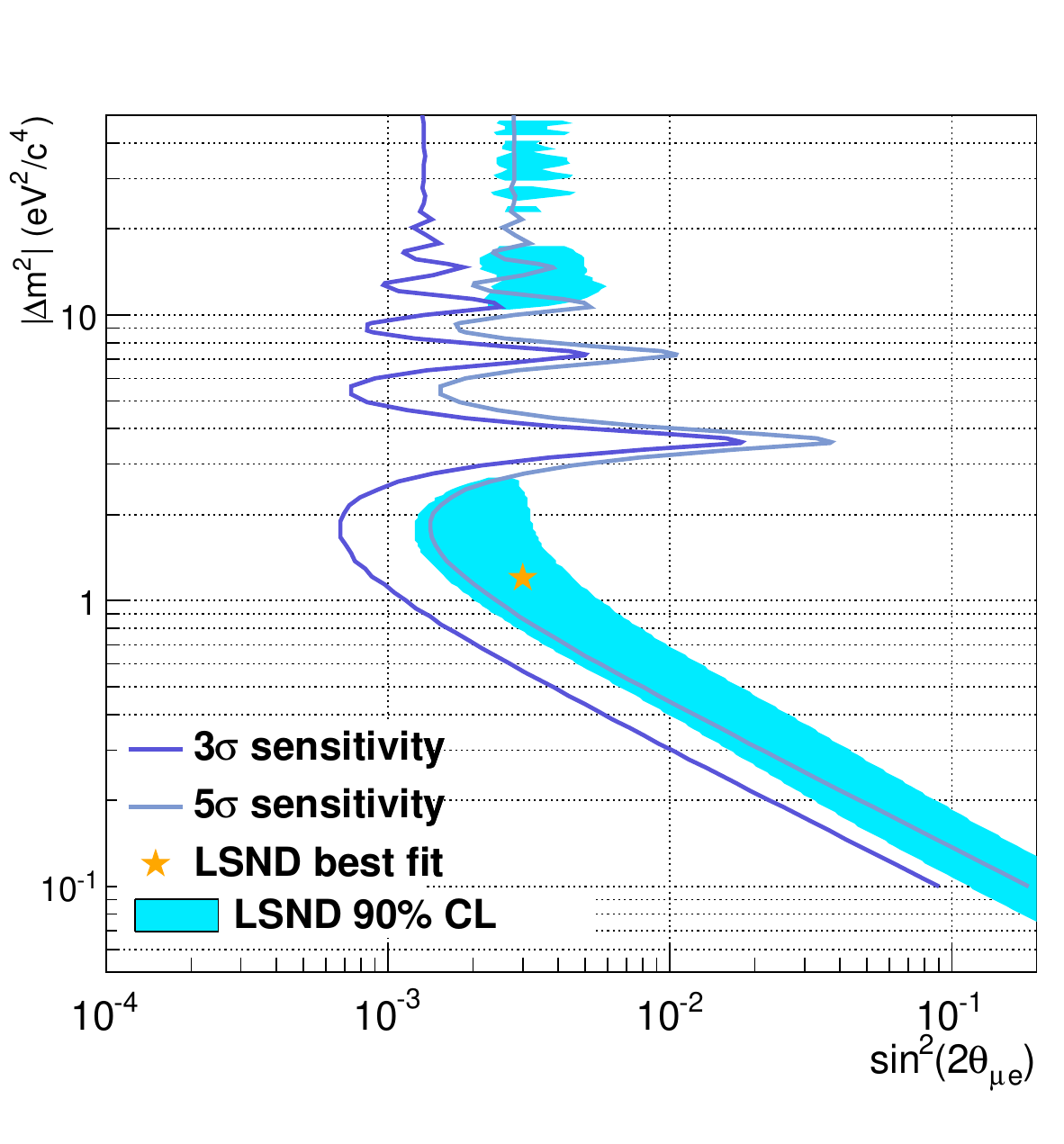}
\vspace{-.7cm}
\caption{\label{fig:sensitivity}The sensitivity of a kaon decay-at-rest based electron neutrino appearance experiment to the LSND allowed region.}
\end{figure}

The sensitivity of the experiment to the parameter space near $\Delta m^2\sim 1~\mathrm{eV}^2$, using a two neutrino oscillation probability, is shown in Fig.~\ref{fig:sensitivity}. The curves have been arrived at using fully frequent confidence intervals via the profile log-likelihood method~\cite{Rolke:2004mj,Lundberg:2009iu}. Sensitivity is defined based on the median upper limit that would be obtained by a set of experiments measuring just background, in the absence of a true signal~\cite{Feldman:1997qc}. A 2~kton detector and 2$\times 10^{21}$ POT exposure provides $>$5$\sigma$ sensitivity to most of the 90\% CL LSND allowed region. Although this mass and exposure can be considered optimistic, the idea of utilizing the monoenergetic muon neutrino from kaon decay-at-rest as a source for an electron neutrino appearance search is one that can be employed at a number of current and future proton facilities around the world.

\clearpage
\subsection{The MINOS+ Project\footnote{Proposed by the MINOS Collaboration \\
({\tt http://www-numi.fnal.gov/})}}
\label{subapp:MINOS+}

\begin{figure}[h!]
\includegraphics[width=0.495\linewidth]{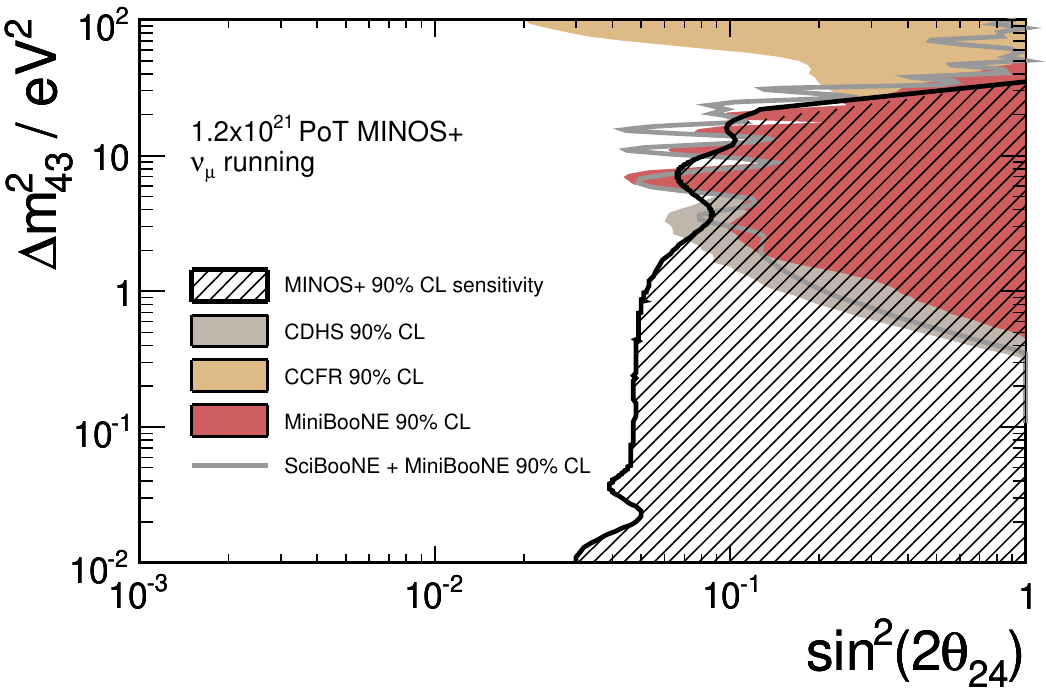} 
\includegraphics[width=0.495\linewidth]{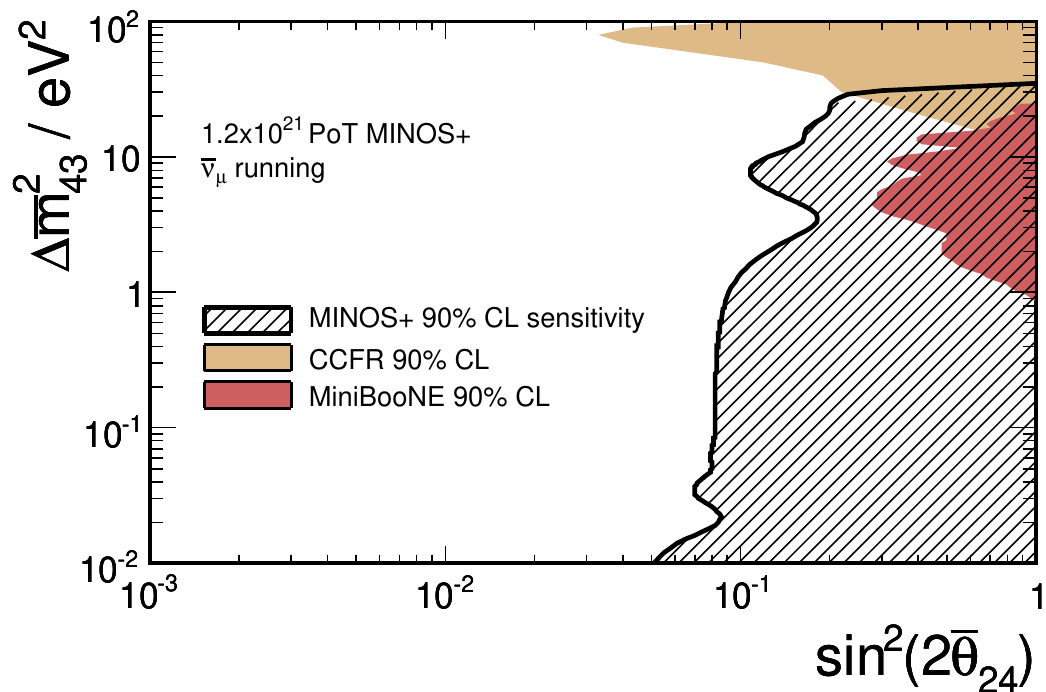}
\caption{Expected MINOS+ 90\% C.L. exclusion regions for the $\theta_{24}$ mixing angle 
compared to other disappearance measurements for neutrino(left) and antineutrino (right) 
running.}
\label{fig:minosplusneut} 
\end{figure}

The MINOS contributions to the body of knowledge on sterile neutrino searches will be 
continued and improved upon by the proposed MINOS+ project~\cite{Tzanankos:2011zz}. In 
MINOS+, both MINOS detectors will continue operation during the NO$\nu$A era (starting 
in 2013), using the NuMI beam upgraded to 700~kW of beam power. MINOS+ is expected to 
run for at least two~years in neutrino and one~year in antineutrino mode. The expected 
sensitivities to $\theta_{24}$ are shown in Fig.~\ref{fig:minosplusneut} for neutrino 
and antineutrino running. The 3+1 sterile neutrino oscillation model employed in producing 
these sensitivities includes the effects of oscillations at the Near Detector short 
baseline in the Far Detector predicted spectrum and waive the restrictions on the value 
of $\Delta m^2_{43}$ applied in the MINOS analysis described in Section~\ref{subsec:MINOS}. 
The MINOS+ sensitivities to neutrino disappearance show roughly a factor of 2 improvement 
over the current exclusion limits from MINOS. Compared to other experiments that study 
neutrino disappearance over short baselines, MINOS+ will place the most stringent limit on 
neutrino disappearance for $\Delta m_{43}^2\lsim 2 \rm{eV^2}$ and on antineutrino 
disappearance for $\Delta m_{43}^2\lsim 20 \rm{eV^2}$. 
 
\clearpage
\subsection{The BooNE Proposal\footnote{Proposed by the BooNE Collaboration}}

The MiniBooNE detector began to take data on September 1, 2002. The experiment was
designed to search for the appearance of excess electron (anti) neutrinos in a primarily muon (anti)
neutrino beam. While resources were not available to construct both a near and far detector at the
time, it was envisioned that a second detector would be constructed at a location appropriate to the
observed signal if MiniBooNE should see a signal. The second detector would be able to
differentiate between a true neutrino oscillation effect and an unforeseen new process or
background. A number of scientists now believe that a significant signal has been observed at
MiniBooNE, and a collaboration of scientists is forming in order to upgrade MiniBooNE to a twodetector
experiment: BooNE.

MiniBooNE has enjoyed a remarkable 9 years of smooth operation, during which an astounding
$6\times 10^{20}$ protons on target (POT) have been delivered in neutrino mode, and an even more
astounding $1\times 10^{21}$ POT have been delivered in antineutrino mode. The neutrino mode data has
yielded a low-energy excess of $129 \pm 20(stat) \pm 38(sys)$ events at reconstructed neutrino energies
below 475 MeV. That low-energy excess is not described well by a simple two-neutrino model, but
can be accommodated by an extended 3 active + 2 sterile neutrino model, fit to the world's relevant
neutrino data. While the statistical significance of the low-energy excess is ~ $6\sigma$, the overall
significance is limited to ~ $3\sigma$ by the systematic error in the estimation of the background, either in
the low energy range of 200-475 MeV or in the full range 200-1250 MeV.
That systematic error is related to the error in the detector acceptance or
efficiency for $\pi^0$ background events, and to a lessor extent, the flux of neutrinos, and the neutrinonucleus
cross sections. Similarly, an excess of is observed in antineutrino mode of $54.9 \pm 17.4(stat) \pm 16.3(sys)$ 
 events events, consistent with the neutrino-mode data. The antineutrino-mode excess is
limited in statistical power to ~ $3\sigma$ and appears to have a higher energy component of ~ 500-600 MeV.

We now believe we have explored all the possible avenues for explaining the excess events by
conventional processes and have exhausted the possible ways to reduce the systematic errors via
further analysis. We believe the construction of a near detector, BooNE\cite{BooNE1}, at ~ 200 meters from the Booster
Neutrino Beam proton target to be the most expedient way understand whether or not the excess
events observed by MiniBooNE are caused by an oscillation process or some other process that
scales more conventionally by $L^{-2}$. The primary motivation for building a near detector, rather than a
detector further away, is that the neutrino interaction rate will be over 7 times larger, and the
measurement will precisely determine the neutrino-related backgrounds within 6 months of running.
A far-detector would take much longer to accumulate sufficient statistics. The combination of the
present MiniBooNE neutrino-mode data, plus a 4-month ($1\times 10^{20}$ POT or ~ 700,000 neutrino
events) neutrino-mode run with a near detector, would result in a $5\sigma$ sensitivity to whether or not
the low energy excess is an oscillation effect. With MiniBooNE's anticipated $1.5\times 10^{21}$ POT in
antineutrino-mode, BooNE will provide a unique measurement of antineutrino appearance and
disappearance with an 8 month run ($2\times 10^{20}$ POT or ~ 140,000 events) required for comparable
statistics.

Furthermore, a two-detector BooNE experiment, in conjunction with the ultra-fine-grained
MicroBooNE liquid argon TPC, would be a tremendously powerful, oscillation-hunting
combination. While MicroBooNE does not anticipate any antineutrino-mode operation, the
operation of BooNE during the MicroBooNE neutrino-mode run would double the statistics of the
present MiniBooNE neutrino data to $1.2\times 10^{21}$ POT. That powerful trio of detectors would yield
precise measurements of both electron-neutrino appearance and muon neutrino disappearance, which are tightly
coupled in nearly all sterile-neutrino oscillation models

\clearpage
\subsection{Search for anomalies with muon spectrometers and large LArÐTPC imaging detectors at 
CERN\footnote{Proposed by the ICARUS + NESSiE Collaborations~\cite{Antonello:2012hf}.}}
\label{sec:future_exps_lar_nessie}

\begin{figure}[htbp]
\begin{center}
  \includegraphics[width=0.9\textwidth]{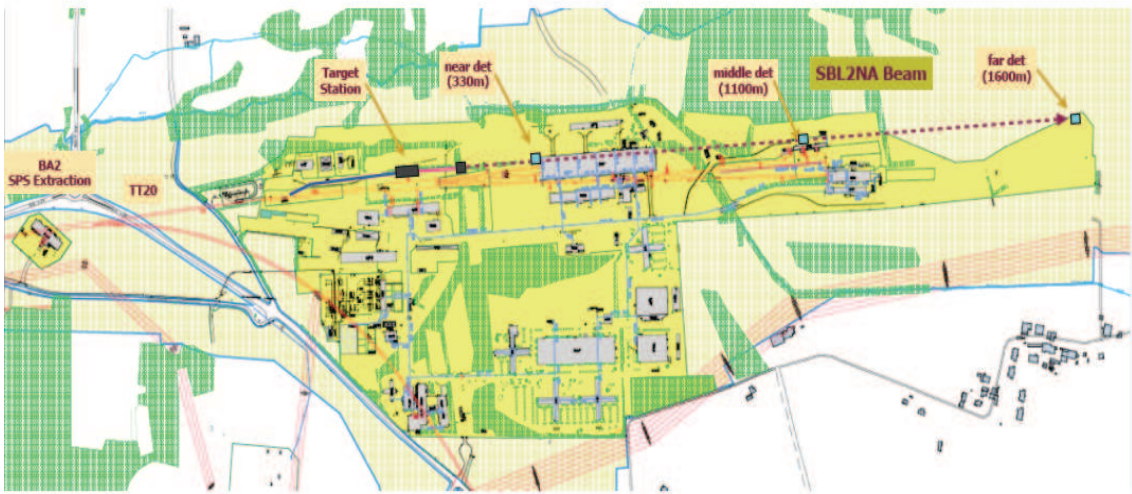}
    \caption{The new SPS North Area neutrino beam layout. Main parameters are: primary beam: 100 GeV; fast extracted from SPS; 
    target station next to TCC2, $\sim$11~m underground; decay pipe: 100~m, 3~m diameter; beam dump: 15~m of Fe with graphite core, 
    followed by muon stations; neutrino beam angle: pointing upwards; at $\sim$3 m in the far detector $\sim$5~mrad slope.}
    \label{larnessie_fig1}
\end{center}
\end{figure}

An experimental search for sterile neutrinos at new CERN-SPS neutrino beam has been recently proposed~\cite{Antonello:2012hf}. 
The experiment is based on two identical LAr-TPCÕs~\cite{Rubbia:2011a,Rubbia:2011b} followed by magnetized spectrometers~\cite{Bernardini:2011a}, 
observing the electron and muon neutrino events at the ÓFarÓ and ÒNearÓ positions 1600 and 300 m from the proton target, respectively 
(Figure~\ref{larnessie_fig1}). The project will exploit the ICARUS T600 detector, the largest LAr-TPC ever built with a size of about 600 t 
of imaging mass, now running in the LNGS underground laboratory exposed to the CNGS beam, moved to the CERN ÒFarÓ position. 
An additional 1/4 of the T600 detector (T150) will be constructed and located in the ``Near'' position. Two spectrometers will be placed 
downstream of the two LAr-TPC detectors to greatly complement the physics capabilities. Spectrometers will exploit a classical dipole 
magnetic field with iron slabs, and a new concept air-magnet, to perform charge identification and muon momentum measurements 
from low energy ($< 1$ GeV) in a wide energy range over a large transverse area ($>$ 50 m$^2$).

In the two positions, the shapes of the radial and energy spectra of the $\nu_e$ beam component are practically identical.
Comparing the two detectors, in absence 
of oscillations, all cross sections and experimental biases cancel out, and the two experimentally observed event distributions must be 
identical. Any difference of the event distributions at the locations of the two detectors might be attributed to the possible existence of 
$\nu$-oscillations, presumably due to additional neutrinos with a mixing angle $\sin^2 (2\theta$) and a larger mass difference 
$\Delta m^2$.

The superior quality of the LAr imaging TPC, now widely experimentally demonstrated, and in particular its unique {\em electron} - $\pi^0$ 
discrimination allows full rejection of backgrounds and offers a lossless $\nu_e$ detection capability. The determination of the muon charge 
with the spectrometers allows the full separation of $\nu_{\mu}$ from $\overline{\nu}_\mu$ and therefore controlling systematics from 
muon mis-identification largely at high momenta.

Two main anomalies will be explored with both neutrino and antineutrino focused beams. According to the first anomaly, emerged in 
radioactive sources and reactor neutrino experiments (see Sections~\ref{sec:gallium} and \ref{sec:reactor}, some of 
the $\nu_e$ ($\overline{\nu}_e$) and/or of the $\nu_\mu$ ($\overline{\nu}_\mu$) events might be converted into invisible (sterile) 
components, leading to observation of oscillatory, 
distance dependent disappearance rates. In a second anomaly (following LSND and MiniBooNE observations~\cite{Aguilar:2001ty,AguilarArevalo:2008rc,AguilarArevalo:2010wv,Zimmerman-PANIC2011}) 
some distance dependent $\nu_\mu \rightarrow \nu_e$ oscillations may be observed as $\nu_e$ excess, 
especially in the antineutrino channel. The disentangling of $\nu_{\mu}$ from $\overline{\nu}_\mu$ will allow to exploit the interplay 
of the different possible oscillation scenario, as well as the interplay between disappearance and appearance of different neutrino 
states and flavors. Moreover the NC/CC ratio will provide a sterile neutrino oscillation signal by itself and it will beautifully complement
the normalization and the systematics studies. This experiment will offer remarkable discovery potentialities, collecting a very large 
number of unbiased events both in the neutrino and antineutrino channels, largely adequate to definitely settle the origin of the 
$\nu$-related anomalies.

\subsubsection*{The new SPS neutrino facility}

\begin{figure}[htbp]
\begin{center}
  \includegraphics[width=0.45\textwidth]{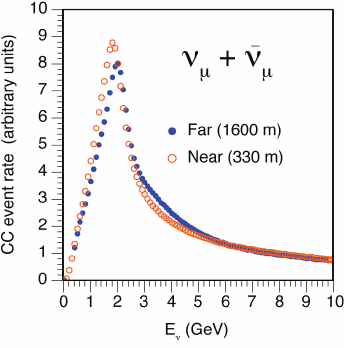}
  \includegraphics[width=0.45\textwidth]{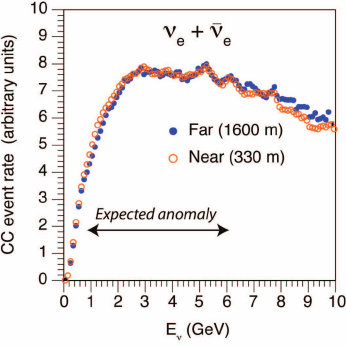}
    \caption{Muon (left) and electron (right) neutrino CC interaction spectra, at Near and Far positions.}
    \label{larnessie_fig2}
\end{center}
\end{figure}

To explore the interesting neutrino $\Delta m^2 \sim 1$~eV$^2$ region the ÒFarÓ distance has been chosen at 1.6 km with a central 
value of the onÐaxis neutrino beam energy spectrum around $E_{\nu} \sim$ 2~GeV (Figure~\ref{larnessie_fig2}). A proton beam 
intensity of $4.5\times 10^{19}$ pot/year at 100 GeV energy has been assumed as a conservative reference in order to produce 
high intensity $\nu$ beam and to minimize the beam related background expected at the Near Detector located at 300 m. 
A fast proton extraction from SPS is also required for the LAr-TPC operation at surface in order to time tag the beam related events 
among the cosmic ray background.

\subsubsection*{Expected sensitivities to neutrino oscillations}

A complete discussion of $\nu_\mu \rightarrow \nu_e$ oscillation search both in appearance and disappearance modes has been 
presented in the SPSC-P345 document~\cite{Rubbia:2011a,Rubbia:2011b}, that includes the genuine event selection and background rejection in the 
LAr- TPC. In particular, due to the excellent $\pi^0$ to electron separation, a $\pi^0$ rejection at $10^3$ level is obtained when 
requiring at least 90\% electron recognition efficiency. The effects of the high-energy event tail in the event selection has been carefully 
studied: the resulting background is negligible, of the same order of the residual NC background.

\begin{figure}[htbp]
\begin{center}
  \includegraphics[width=0.8\textwidth]{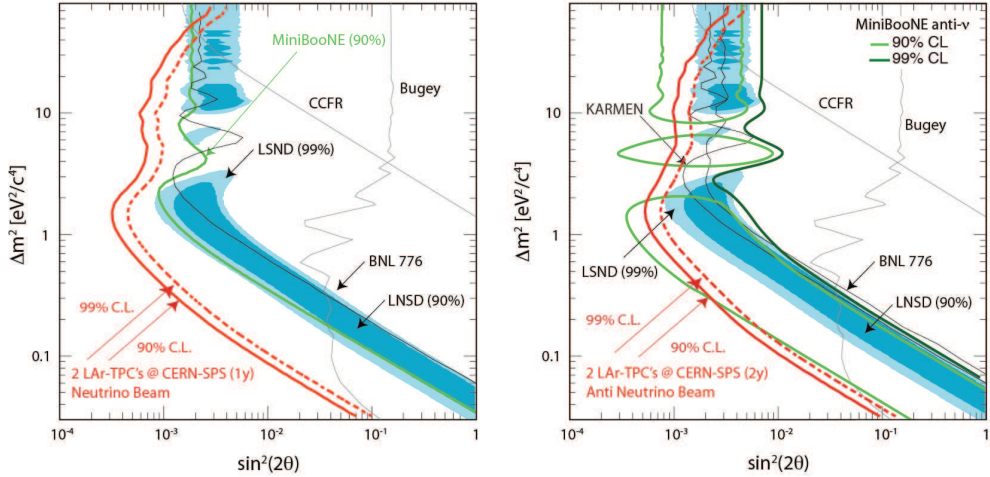}
    \caption{Expected sensitivity for the proposed experiment exposed at the CERN-SPS neutrino beam (left) and antineutrino (right) 
    for $4.5\times 10^{19}$ pot (1 year) and $9.0\times 10^{19}$ pot (2 years), respectively. The LSND allowed region is fully explored 
    in both cases.}
    \label{larnessie_fig3}
    \end{center}
\end{figure}

In addition to the $\nu_\mu \rightarrow \nu_e$ oscillation studies mentioned above, $\nu_{\mu}$ oscillation in disappearance mode 
was discussed at length in~\cite{Bernardini:2011a}, by using large mass spectrometers with high capabilities in charge identification and 
muon momentum measurement. It is important to note that all sterile models predict large $\nu_{\mu}$ disappearance effects together 
with $\nu_e$ appearance/disappearance. To fully constrain the oscillation searches, the $\nu_{\mu}$ disappearance studies have to 
be addressed. Much higher disappearance probabilities are expected than in appearance mode, where relative amplitudes as large 
as 10\% are possible. The spectrometers will be able to correctly identify about 40\% of all the CC events produced in, and escaped 
from, the LAr-TPCs, both in the near and far sites. That will greatly increase the fraction of CC events with charge identification and 
momentum measurement. Therefore complete measurement of the CC event spectra will be possible, along with the NC/CC event ratio 
(in synergy with the LAr-TPC), and the relative background systematics.

The large mass of the magnets will also allow an internal check of the NC/CC ratio in an extended energy range, and an independent 
measure of the CC oscillated events.

\begin{figure}[htbp]
\begin{center}
  \includegraphics[width=0.7\textwidth]{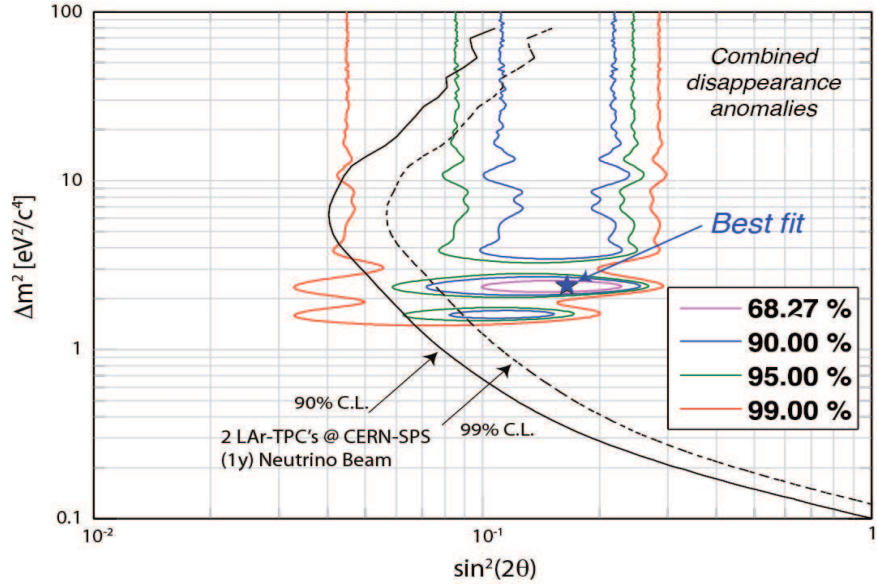}
     \caption{Oscillation sensitivity in $\sin^2 (2\theta$) vs $\Delta m^2$ distribution for 1 year data taking. 
     A 3\% systematic uncertainty on energy spectrum is included. Combined ÒanomaliesÓ from reactor neutrino, 
     Gallex and Sage experiments are also shown.}
    \label{larnessie_fig4}
    \end{center}
\end{figure}

A sensitivity of $\sin^2 (2\theta) > 3\times 10^{-4}$ (for $|\Delta m^2|  > 1.5$ eV$^2$) and $|\Delta m^2|  > 0.01$ 
eV$^2$ (for $\sin^2 (2\theta) = 1$) at 90\% C.L. is expected for the $\nu_\mu \rightarrow \nu_e$ transition with one year exposure 
($4.5\times 10^{19}$ pot) at the CERN-SPS $\nu_{\mu}$ beam (Figure~\ref{larnessie_fig3} left). The parameter space region allowed 
by the LSND experiment is fully covered, except for the highest $\Delta m^2$ region. The sensitivity has been computed according to 
the above described particle identification efficiency and assuming a 3\% systematic uncertainty in the prediction of ÒFarÓ to ÒNearÓ 
$\nu_e$ ratio. A further control of the overall systematics will be provided by the LAr and spectrometer combined measurement of 
CC spectra in the Near site and over the full energy range.

\begin{figure}[htbp]
\begin{center}
  \includegraphics[width=0.7\textwidth]{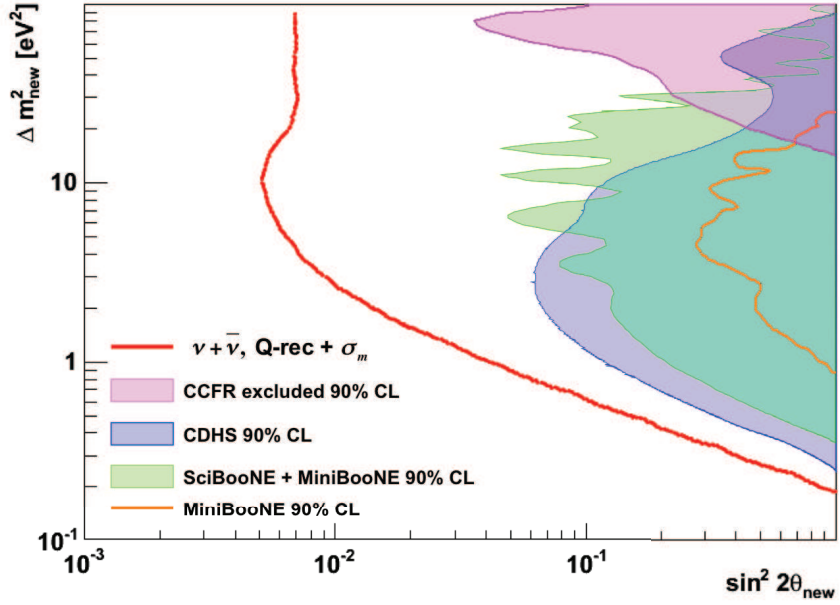}
     \caption{Sensitivity plot (at 90\% C.L.) considering 3 years of the CERN-SPS beam (2 years in antineutrino and 1 year in neutrino mode) 
     from CC events fully reconstructed in NESSiE$+$LAr. Red line: $\nu_{\mu}$ exclusion limit. The three filled areas correspond to the present 
     exclusion limits on the $\nu_{\mu}$ from CCFR, CDHS and SciBooNE+MiniBooNE experiments (at 90\% C.L.). 
     Orange line: recent exclusion limits on $\nu_{\mu}$ from MiniBooNE alone measurement.}
    \label{larnessie_fig5}
    \end{center}
\end{figure}

\begin{table}[htbp!]
  \begin{center}
  \begin{tabular}{lcccc}
  \hline
  & NEAR                            & NEAR             & FAR                        & FAR            \\
  &       (Negative foc.)           & (Positive foc.)  & (Negative foc.)            & (Positive foc.) \\
  \hline
 $\nu_e + \overline{\nu}_e$ (LAr) & 35 K & 54 K & 4.2 K & 6.4 K \\
 $\nu_{\mu} +  \overline{\nu}_{\mu}$ (LAr) & 2000 K & 5200 K & 270 K & 670 K \\ 
 Appearance Test Point & 590 & 1900 & 360 & 910 \\
                                            &        &         &      &       \\
 $\nu_{\mu}$ CC (NESSiE$+$LAr) & 230 K & 1200 K & 21 K & 110 K \\
  $\nu_{\mu}$ CC (NESSiE alone) & 1150 K & 3600 K & 94 K & 280 K \\
  $\overline{\nu}_{\mu}$ CC (NESSiE$+$LAr) & 370 K & 56 K & 33 K & 6.9 K \\
  $\overline{\nu}_{\mu}$ CC (NESSiE alone) & 1100 K & 300 K & 89 K & 22 K \\
 Disappearance Test Point & 1800 & 4700 & 1700 & 5000 \\
 \hline
  \end{tabular}
  \caption{The expected rates of interaction (LAr) and reconstructed (NESSiE) events 1 year of operation.Values for 
  $\Delta m^2$ around 2 eV$^2$ are reported as example.\label{larnessie_tab1}}
  \end{center}
\end{table}

In antineutrino focusing, twice as much exposure ($0.9\times 10^{20}$ pot) allows to cover both the LSND region and the new 
MiniBooNE results (Figure~\ref{larnessie_fig3} right)~\cite{Aguilar:2001ty,AguilarArevalo:2008rc,AguilarArevalo:2010wv,Zimmerman-PANIC2011}. Both favored MiniBooNE parameter sets, corresponding to 
two different energy regions in the MiniBooNE antineutrino analysis, fall well within the reach of this proposal.
In Figure~\ref{larnessie_fig4} the sensitivity for $\nu_e$ disappearance search in the $\sin^2 (2\theta)$, $\Delta m^2$ 
plane is shown for one year data taking. The oscillation parameter region related to the anomalies from the combination 
of the published reactor neutrino experiments, Gallex and Sage calibration sources experiments is completely explored.

\begin{table}[htbp!]
  \begin{center}
  \begin{tabular}{cll}
  \hline
  Osc. type & Neutrinos & Experiments \\
  \hline
  $\theta_{12}$ & $\nu_e$ (solar, reactors) & SNO, SK, Borexino, Kamland \\
  $\theta_{23}$ & $\nu_{\mu}$ (atmospheric,accelerators) & SK, Minos, T2K \\
  $\theta_{13}$ & $\nu_e$ (reactors) & Daya Bay, Reno, Double Chooz \\
   $\theta_{14}$ & $\nu_e$ (reactors, radioactive sources) & SBL Reactors, Gallex, Sage. {\bf This Proposal} \\
  $\theta_{24}$ & $\nu_{\mu}$ (accelerators) & CDHS, MiniBooNE. {\bf This Proposal} \\
  \hline
  \end{tabular}
  \caption{Measurements of the mixing angle as provided by different experiments.\label{larnessie_tab2}}
  \end{center}
\end{table}

The $\nu_{\mu}$ disappearance signal is well studied by the spectrometers, with a very large statistics and disentangling of $\nu_{\mu}$ and 
$\overline{\nu}_\mu$  interplay~\cite{Kopp:2011qd,Giunti:2011gz}. As an example, Figure~\ref{larnessie_fig5} shows the sensitivity plot (at 90\% C.L.) for two 
years negative-focusing plus one year positive-focusing. A large extension of the present limits for $\nu_{\mu}$ by CDHS and the 
recent SciBooNE+MiniBooNE will be achievable in $\sin^2 (2\theta),\; \Delta m^2$.

For 1 year of operation, either with negative or positive polarity beam, Table~\ref{larnessie_tab1} reports the expected interaction rates in the 
LAr-TPCs at the Near (effective 119 t) and Far locations (effective 476 t), and the expected rates of fully reconstructed events in the NESSiE 
spectrometers at the Near (effective 241 t) and Far locations (effective 661 t), with and without LAr contribution. Both $\nu_e$ and $\nu_{\mu}$ 
disappearance modes will be used to add conclusive  information on the sterile mixing angles as shown in the Table ~\ref{larnessie_tab2}.

\clearpage
\subsection{Liquid Argon Time Projection Chambers\footnote{Proposed by the MicroBooNE Collaboration \\ ({\tt http://www-microboone.fnal.gov/public/collaboration.html}).}}

Liquid Argon Time Projection Chambers (LArTPCs) use a recently developed technology for neutrino detection. These detectors, in addition to offering good neutrino efficiency, allow for excellent background rejection. Their good calorimetric reconstruction, their high resolution and their good particle ID, demonstrated both in simulations and measurments ($e.g.$ ref~\cite{Anderson:2011ce}), make these detectors a very attractive choice for high precision neutrino physics.

\subsubsection*{MicroBooNE} 

\begin{figure}[ht!]
\begin{center}
\includegraphics[angle=0,width=5.0cm]{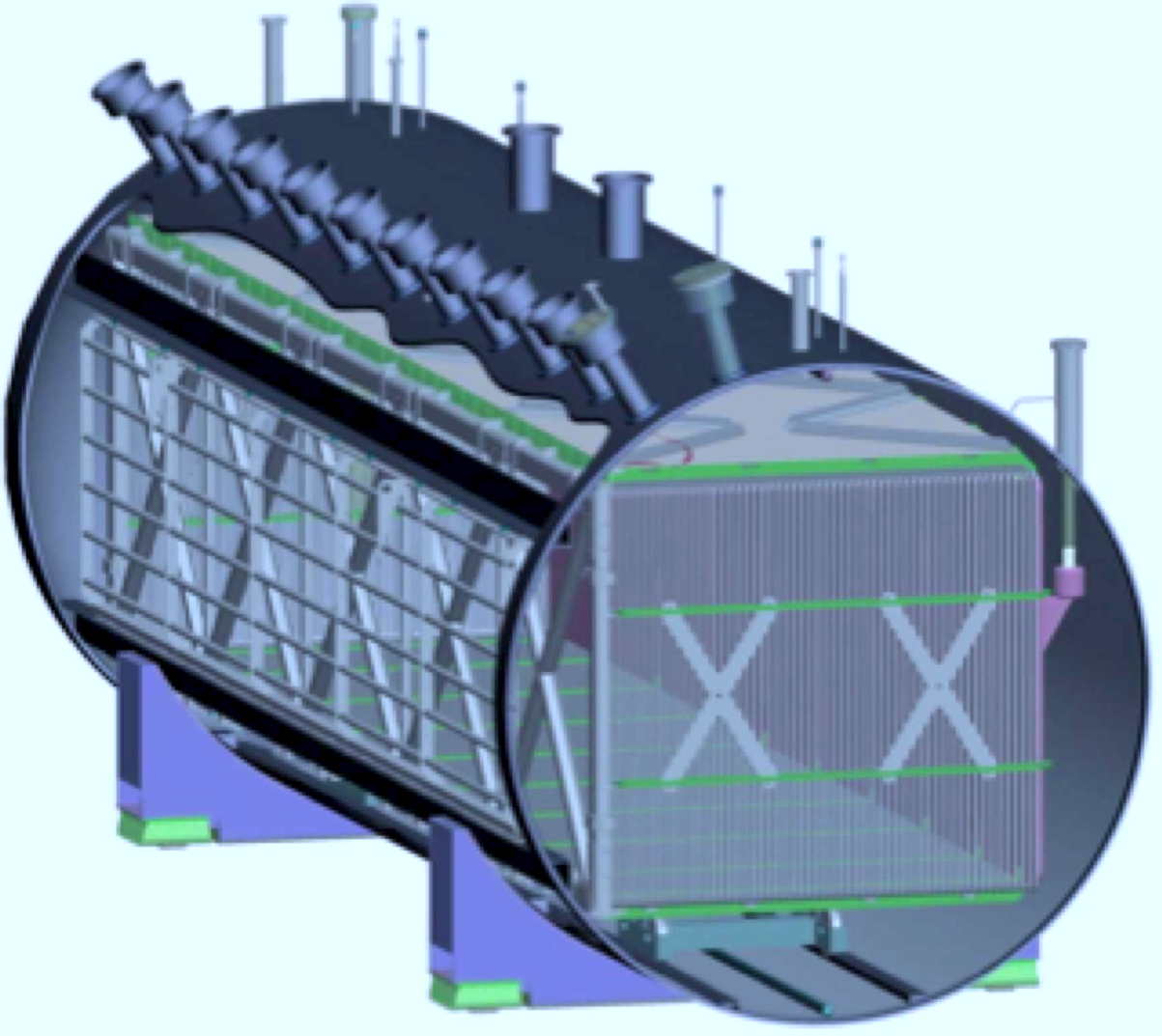}
\caption{The MicroBooNE detector. The 2.5m $\times$ 2.3m $\times$ 10.4m TPC is inserted into the cryostat. }
\label{detector}
\end{center}
\end{figure}

MicroBooNE is a new experiment currently under construction. The MicroBooNE detector is a 170 ton LArTPC (61.4 tons fiducial volume). The 2.5m $\times$ 2.3m $\times$ 10.4m TPC, shown in Figure \ref{detector}, will be located 470m downstream of the Booster Neutrino Beam (BNB) at Fermilab, and consequently be exposed to the same $\nu_{\mu}$-dominated beam as MiniBooNE. MicroBooNE will start taking data in 2013. More details on the MicroBooNE experiment can be found here \cite{uboone1,Jones:2011ci,Chen:2007ae}.

The main physics goal for the MicroBooNE experiment is to address the MiniBooNE low-energy excess \cite{AguilarArevalo:2010wv}. During it's neutrino-mode run, the MiniBooNE experiment observed a 3$\sigma$ excess of events in the 200-475 MeV range. The excess events are electron-like, which could be either electrons or photons. Water Cherenkov detectors do not have the capability to distinguish between electrons and photons, since both produce electromagnetic showers.

In order to address the excess, the MicroBooNE detector will be located close to MiniBooNE, so that the L/E conditions are nearly identical. The MicroBooNE detector will offer a better $\nu_{e}$ efficiency, being about two times that of MiniBooNE. In addition, using the ability of LArTPCs to differentiate between electrons and photons, by examining the dE/dx along the first few centimeters of the electromagnetic shower, MicroBooNE will be able to tell if the MiniBooNE excess comes from electron or photon events. MicroBooNE will have a $>$5$\sigma$ sensitivity (statistical) if the events are due to electrons and a 4$\sigma$ sensitivity (statistical) if the events are due to photons. Finally, MicroBooNE's low energy threshold (tens of MeV) will allow to study how the excess varies at lower energy.

\begin{figure}[ht!]
\begin{center}
\begin{tabular}{cc}
\hspace{0.5cm} \fbox{\bf{Neutrino mode}}& \hspace{1.5cm} \fbox{\bf{Antineutrino mode}}
\end{tabular}
\begin{tabular}{cc}
\includegraphics[angle=0,width=5.3cm]{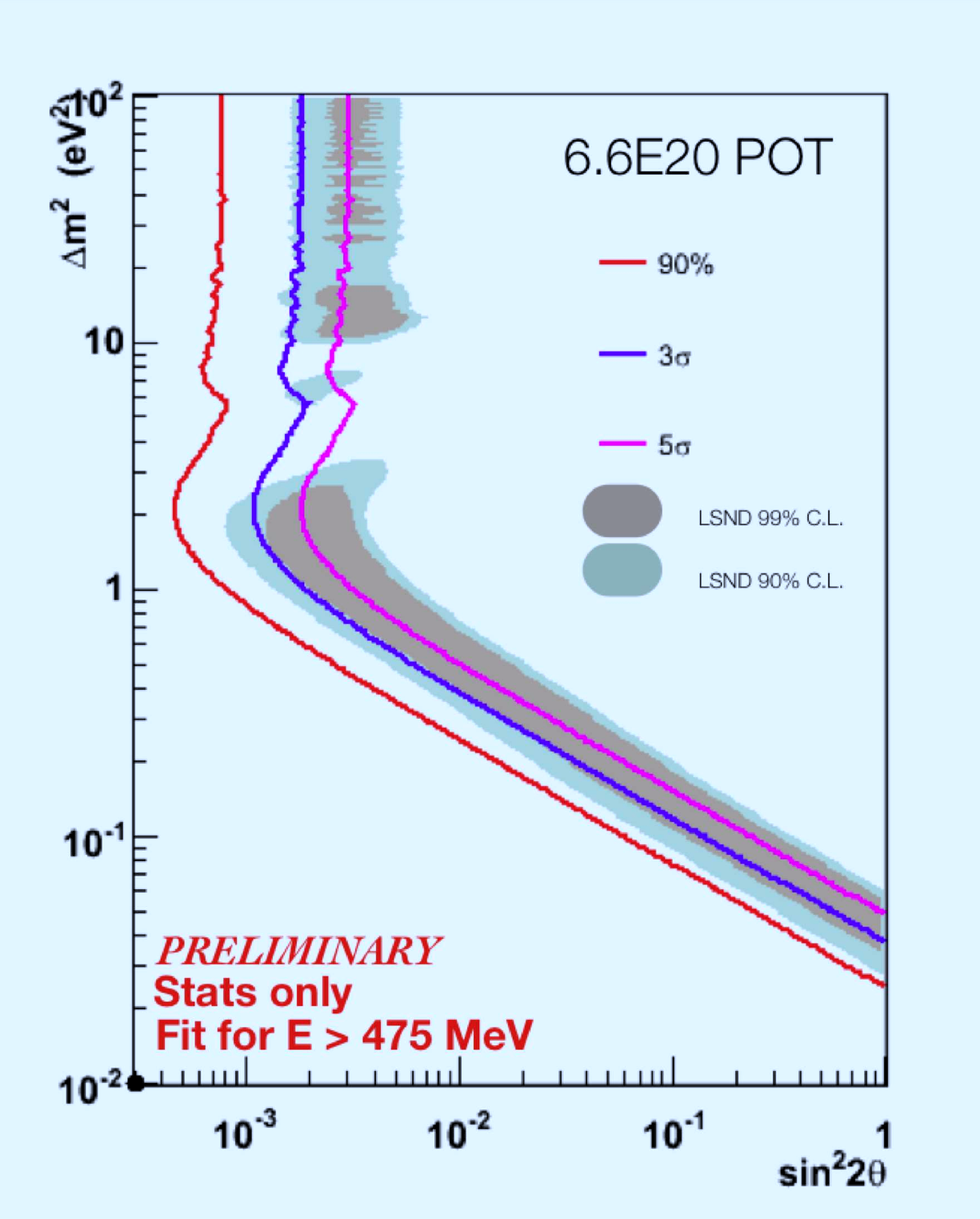} &
\includegraphics[angle=0,width=5.35cm]{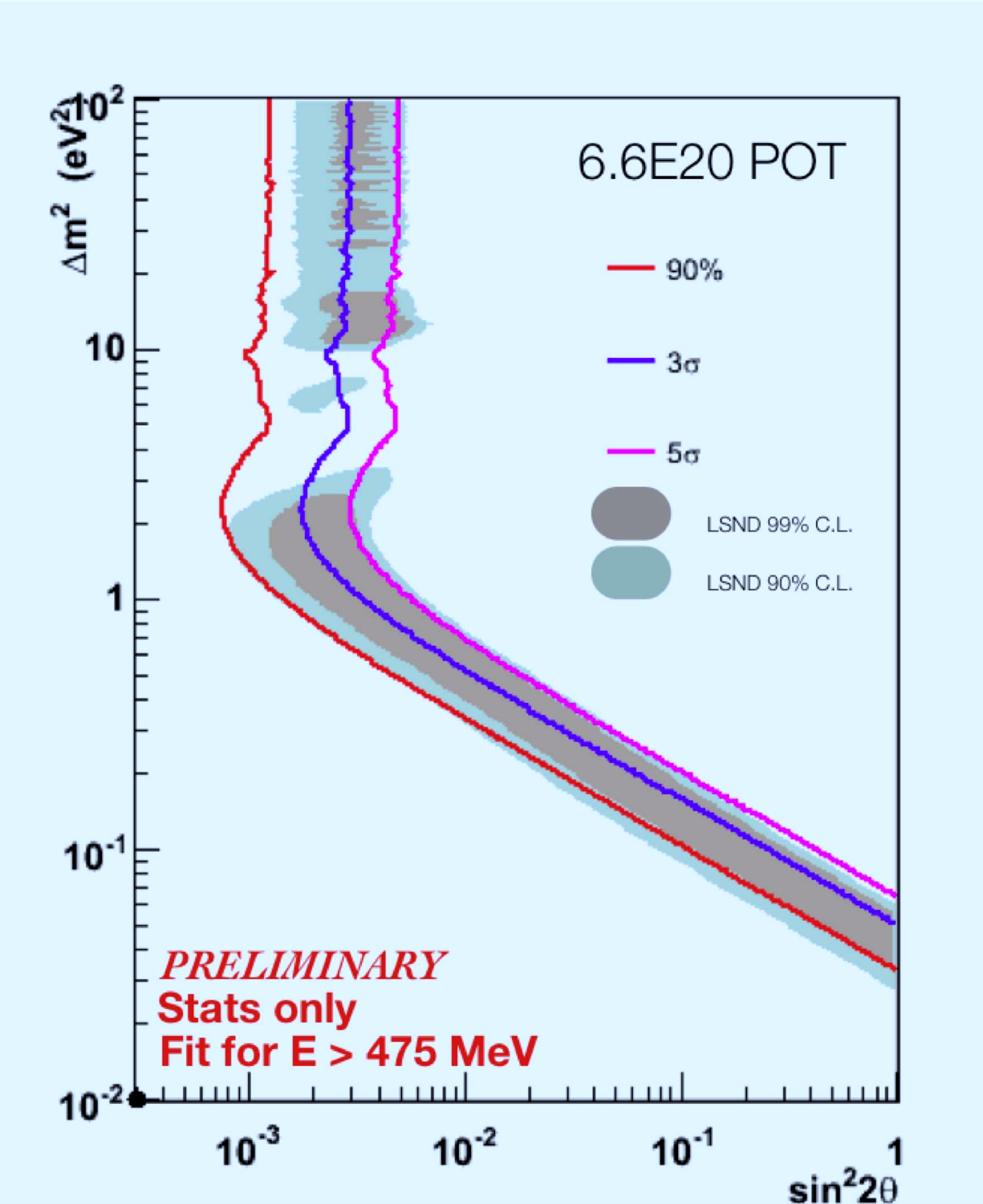}
\end{tabular}
\caption{Sensitivity curves (statistical errors only) for the MicroBooNE detector (61.4 tons fiducial volume, 470m from the BNB) to a 3+1 neutrino model in the ($\Delta m^{2},sin^{2}2\theta$) parameter space, for 6.6$\times 10^{20}$ POTs. The left plot is for neutrino mode and the right plot is for the antineutrino mode. The fits are done for energies above 475 MeV only.}
\label{uboone_sens}
\end{center}
\end{figure}

MicroBooNE is approved to run in neutrino mode for 3 years, accumulating 6.6$\times 10^{20}$ Protons On Target (POTs). With this amount of data, MicroBooNE will offer sensitivity to a sterile neutrino search considering a 3+1 neutrino model, which will be comparable to the MiniBooNE's neutrino-mode sensitivity. The left plot in Figure \ref{uboone_sens} shows the sensitivity to a 3+1 neutrino model (statistical errors only), for  6.6$\times 10^{20}$ POTs in neutrino mode, for MicroBooNE (61.4 tons fiducial volume).

It could, however, bo possible to run the MicroBooNE experiment in antineutrino mode as well, as a phase 2 for MicroBooNE. Motivation for antineutrino running could be justified by LSND's antineutrino result and MiniBooNE's (final) antineutrino results. The right plot in Figure \ref{uboone_sens} shows the sensitivity to a 3+1 neutrino model (statistical errors only), for 6.6$\times 10^{20}$ POTs in antineutrino mode.

MicroBooNE will have a definitive answer (5$\sigma$ level) to the MiniBooNE's low-energy excess, if the excess events are due to $\nu_{e}$. When looking a the simple 3+1 neutrino model to search for sterile neutrinos, MicroBooNE will offer good sensitivity in the region allowed by LSND. MicroBooNE, using a statistical error only analysis, will be able to exclude the allowed LSND region at almost 3$\sigma$ level in neutrino mode and at 90$\%$ C.L. in antineutrino mode. However, MicroBooNE will not have the sensitivity to exclude the LSND allowed region at the $5\sigma$ level.

\subsubsection*{Two LAr-detector experiment at FNAL}

An interesting and powerful way to probe the MiniBooNE/LSND anomalies would be to combined the MicroBooNE detector, described in the previous section, to another, larger, LArTPC (LarLAr) in a near/far configuration. A near/far configuration would  considerably reduce the systematic errors, while the size of the second detector would increase statistics significantly, which are expected to be the limiting factor for a MicroBooNE-only search.

The LBNE collaboration is currently designing a 1kt LArTPC as an engineering prototype \cite{lar1}. It has been pointed out that this detector could be instrumented and placed in the BNB at Fermilab to study short-baseline oscillations \cite{larlar1, larlar2}. 

Several configurations have been considered for this experiment. The MicroBooNE detector, used as the near detector, could be located either at 200m or 470m from the BNB. The far detector, LarLAr, could be placed either at 470m or 700m. Note that no further optimization has been done on the chosen detector locations, which leaves room for improvment. 

In the sensitivity studies presented here, the fiducial volumes assumed for MicroBooNE and LarLAr are 61.4t and 347.5t respectively. A flat $80\%$ $\nu_{e}$efficiency was assumed. All results shown below are for statistical errors only, which are assumed to be the dominant source of uncertainty.

\begin{figure}[ht!]
\begin{center}
\fbox{\bf{$\nu$ mode}}\\
\begin{tabular}{ccc}
 \fbox{\tiny{\bf{MicroBooNE 200m + LarLAr 470m}}}& \fbox{\tiny{\bf{MicroBooNE 200m + LarLAr 700m}}} & \fbox{\tiny{\bf{MicroBooNE 470m + LarLAr 700m}}}
\end{tabular}
\begin{tabular}{ccc}
\includegraphics[angle=0,width=4.75cm]{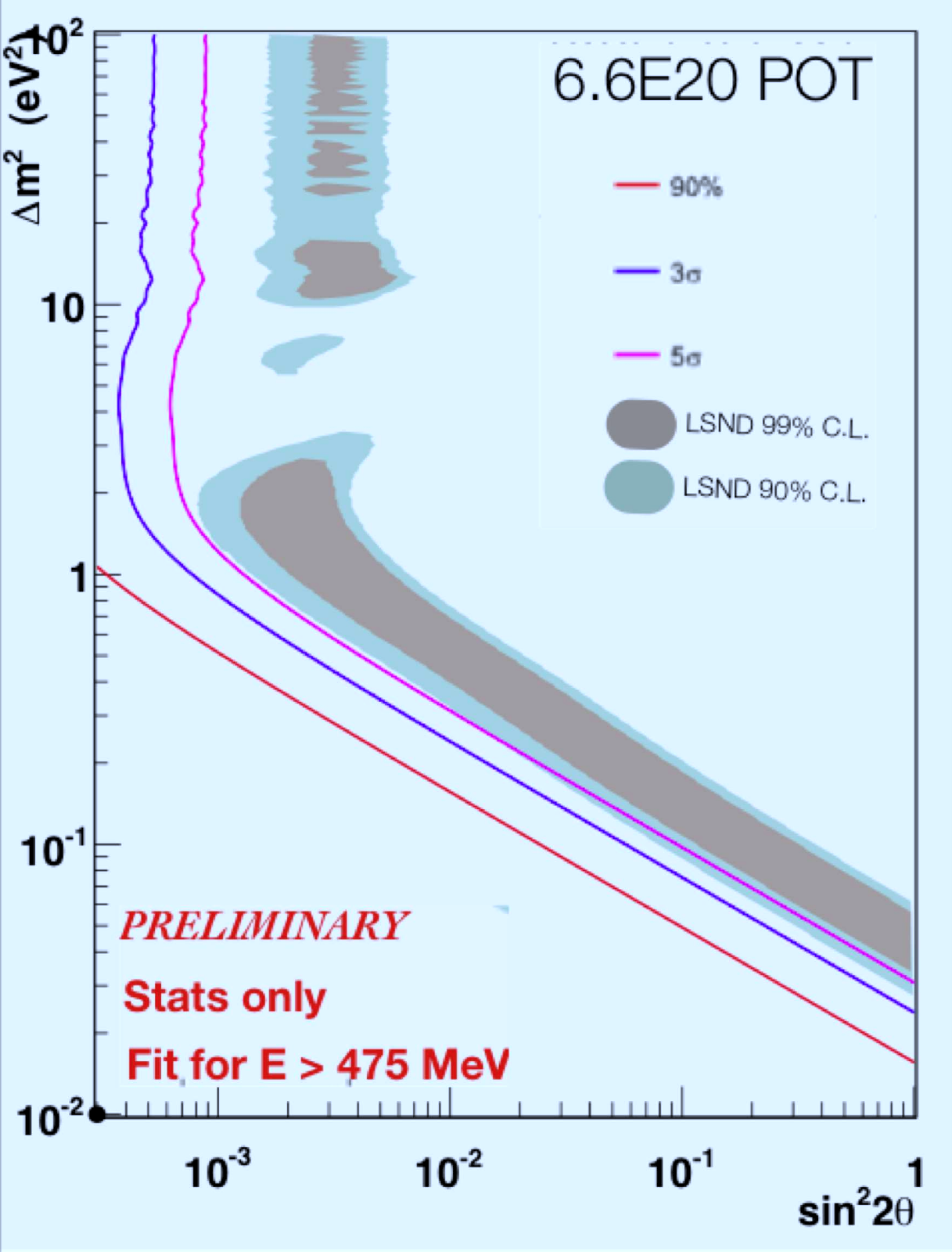} &
\includegraphics[angle=0,width=4.78cm]{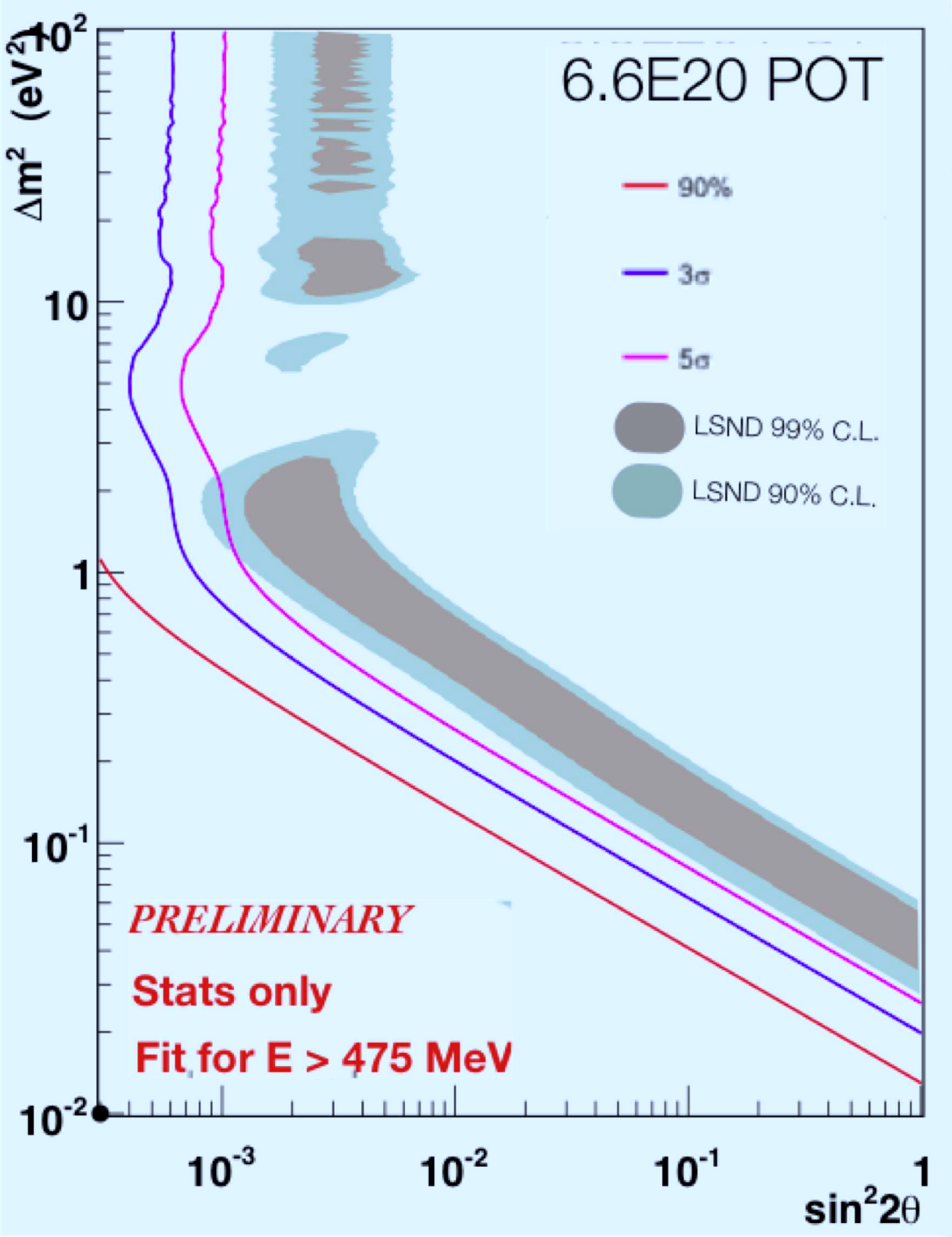} &
\includegraphics[angle=0,width=4.8cm]{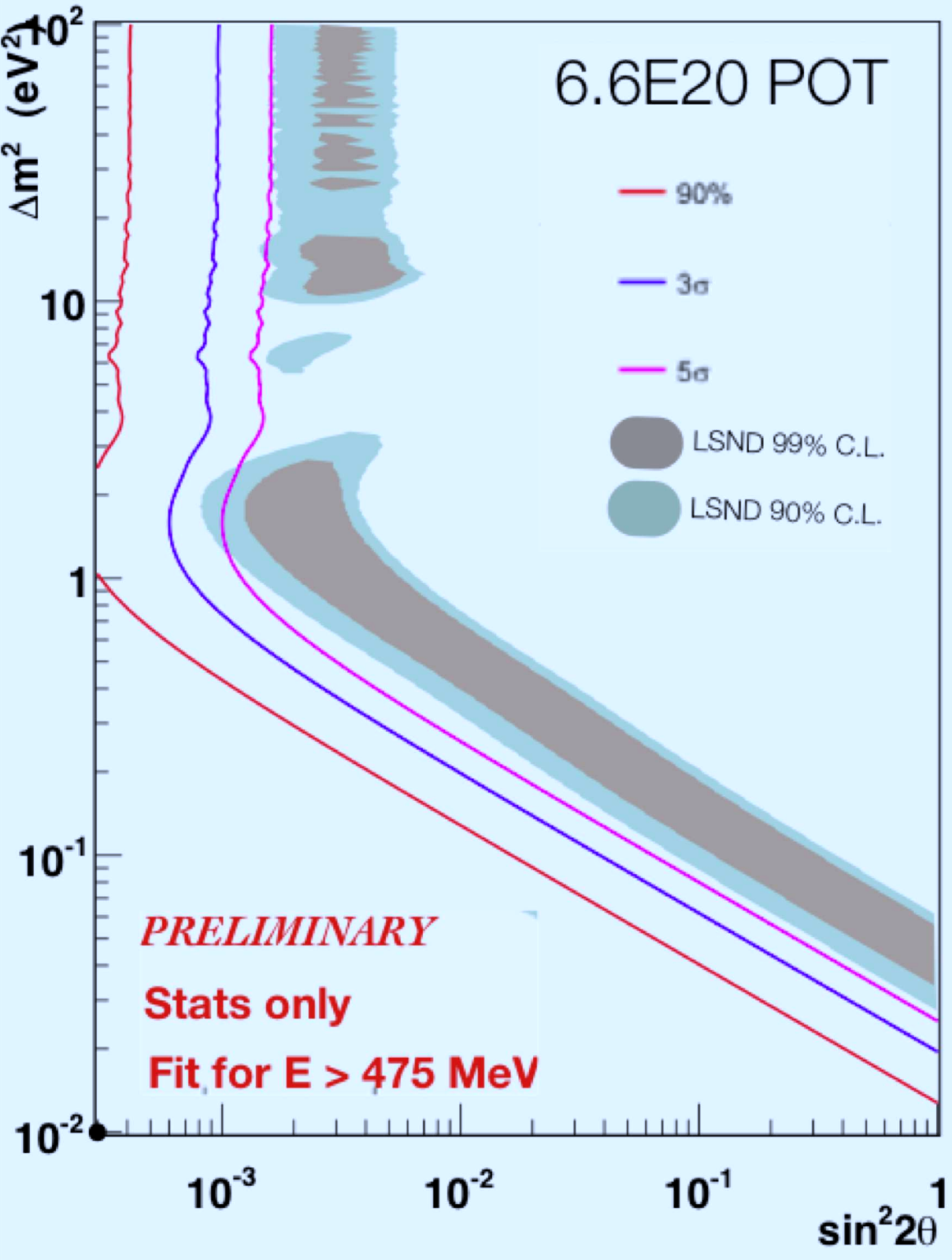}\\
\end{tabular} \\
\fbox{\bf{$\bar{\nu}$ mode}}\\
\begin{tabular}{ccc}
 \fbox{\tiny{\bf{MicroBooNE 200m + LarLAr 470m}}}& \fbox{\tiny{\bf{MicroBooNE 200m + LarLAr 700m}}} & \fbox{\tiny{\bf{MicroBooNE 470m + LarLAr 700m}}}
\end{tabular}
\begin{tabular}{ccc}
\includegraphics[angle=0,width=4.75cm]{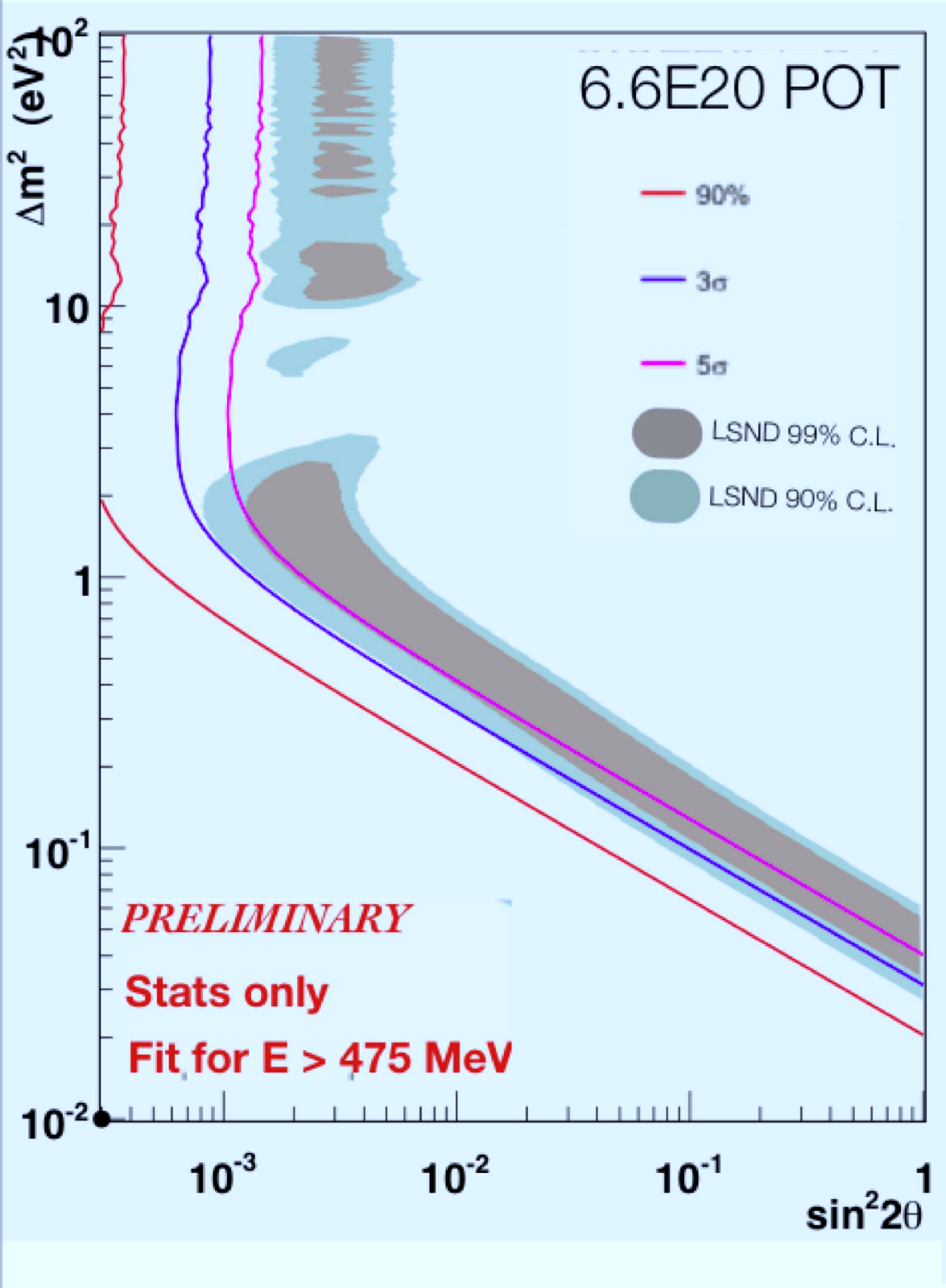} &
\includegraphics[angle=0,width=4.9cm]{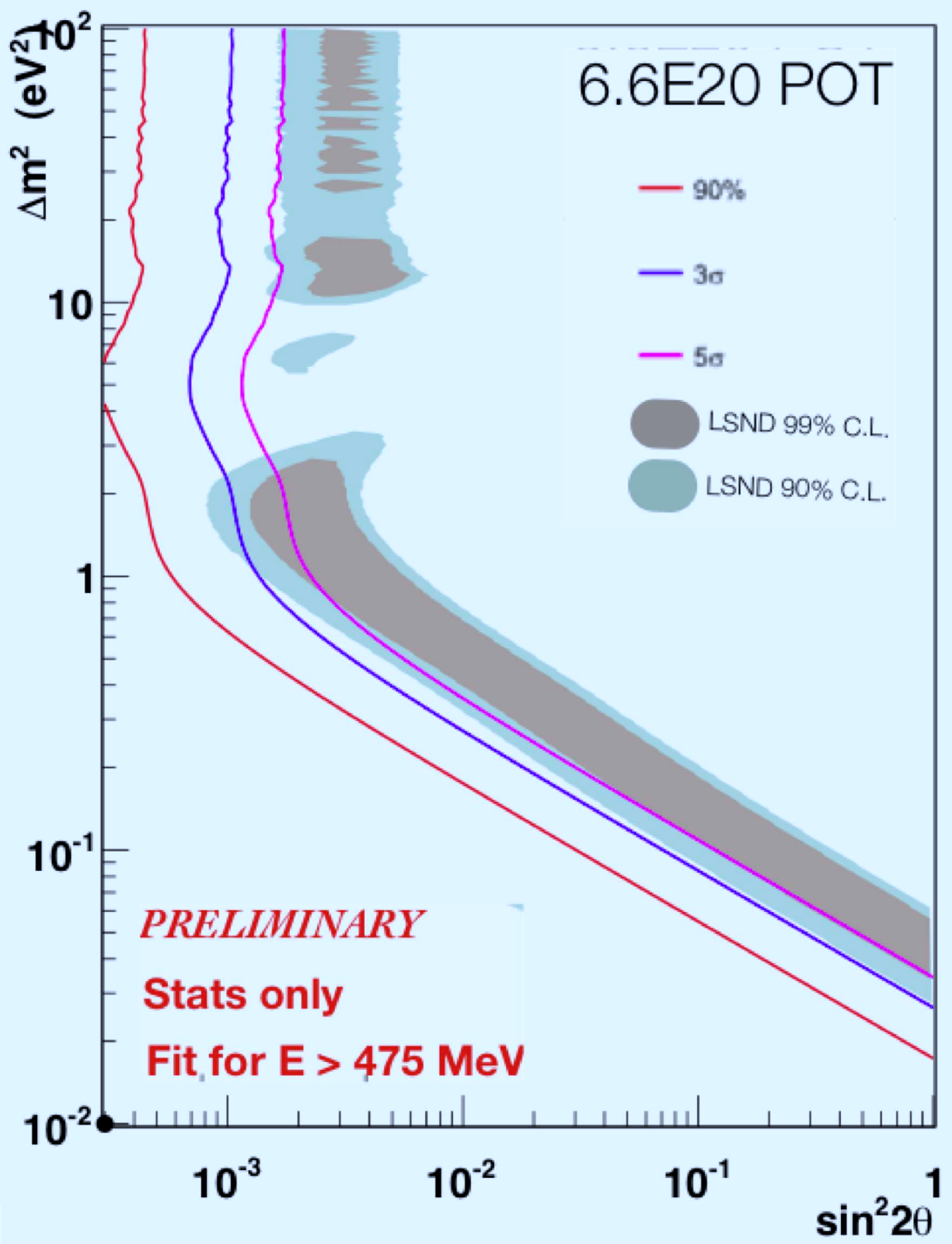} &
\includegraphics[angle=0,width=4.85cm]{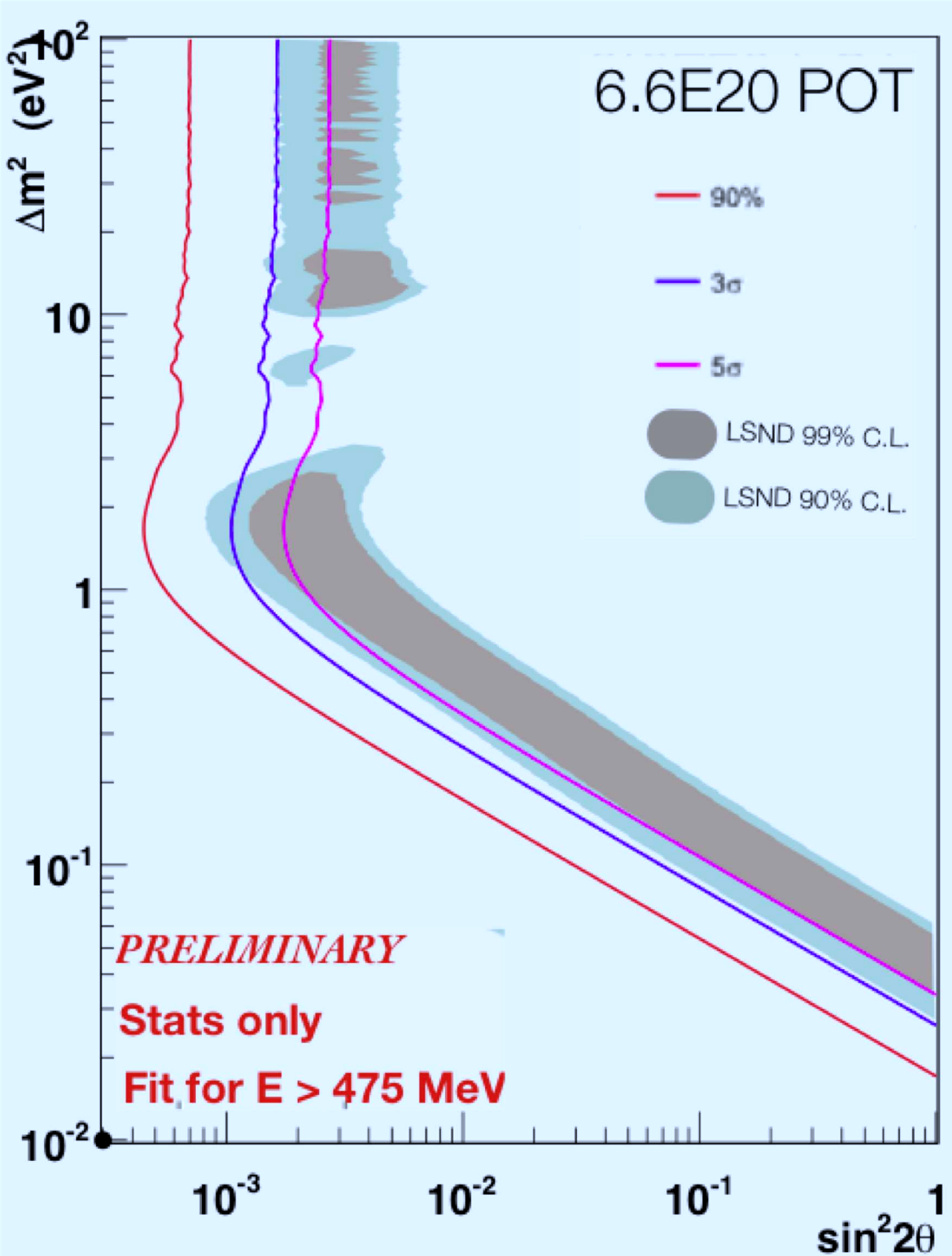}
\end{tabular}
\caption{Sensitivity curves for a combined configuration with the MicroBooNE detector (as the near detector) and the LarLAr one (as the far detector) to a 2-neutrino model in the $\Delta m^{2}/sin^{2}2\theta$ parameter space. The top plots show the sensitivities for neutrino mode and the bottom plots show the antineutrino mode. The left plots are for a configuration where the MicroBooNE detector is located at 200m and LarLAr at 470m. The middle plots are for a configuration with MicroBooNE at 200m and LarLAr at 700m. And finally the right plots are for a configuration with the MicroBooNE detector at 470m and LarLAr at 700m.}
\label{comb_sens}
\end{center}
\end{figure}

Figure \ref{comb_sens} shows sensitivity curves to a 3+1 neutrino model, for different configurations with both MicroBooNE and LarLAr detectors combined in neutrino and antineutrino modes, for a total of 6.6$\times 10^{20}$ POT in each mode.

It is clear from these studies that combining two LAr detectors is a very powerful way to probe short-baseline oscillations. If systematic uncertainities can be reasonably mitigated, this two LAr-detector experiment would offer definitive measurements (at the 5$\sigma$ level) of the MiniBooNE/LSND anomalies in both neutrino and antineutrino modes. Note that in the antineutrino case, more than 6.6$\times 10^{20}$ POTs would be required to reach the 5$\sigma$ level for the whole allowed parameter space.

\clearpage
\subsection{Very-Low Energy Neutrino Factory (VLENF)\footnote{Proposed by 
M.~Ellis, P.~Kyberd (Brunel University),
C.~M.~Ankenbrandt, S.~J.~Brice, A.~D.~Bross, L.~Coney, S.~Geer, J.~Kopp, N.~V.~Mokhov, J.~G.~Morfin, D.~Neuffer, M.~Popovic, T.~Roberts, S.~Striganov, G.~P.~Zeller (Fermi National Accelerator Laboratory),
A.~Blondel, A.~Bravar (University of Geneva),
R.~Bayes, F.~J.~P.~Soler (University of Glasgow),
A.~Dobbs, K.~R.~Long, J.~Pasternak, E.~Santos, M.~O.~Wascko (Imperial College, London),
S.~A.~Bogacz (Jefferson Laboratory),
J.~B.~Lagrange, Y.~Mori (Kyoto University)
A.~P.~T.~Palounek (Los Alamos National Laboratory),
A.~de~Gouvêa (Northwestern University),
Y.~Kuno, A.~Sato (Osaka University),
C.~D.~Tunnell (University of Oxford),
K.~T.~McDonald (Princton University),
S.~K.~Agarwalla    (Universitat de Val\`encia),
P.~Huber, J.~M.~Link (Virginia Tech), and
W.~Winter (Universit\"at W\"urzburg) 
}}
\label{VLENF}

The idea of using a muon storage ring to produce a high-energy ($\simeq$ 50 GeV) neutrino beam for experiments
was first discussed by Koshkarev \cite{Koshkarev:1974my}.  However, a detailed description of the concept for neutrino oscillation
experiments was first produced by Neuffer \cite{NeufferTelmark} in 1980.  The Very-Low Energy Neutrino Factory (VLENF) is essentially the same 
facility proposed in 1980 and would utilize a 2-3 GeV/c muon storage ring to study eV-scale oscillation physics and, in addition, could add 
significantly to our understanding of $\nu_e$ and $\nu_\mu$ cross sections.  In particular the facility can:

\begin{enumerate}
\item Address the large $\Delta$m$^2$ oscillation regime and add significantly to the study of sterile neutrinos.
\item Make precision $\nu_e$ and $\bar{\nu}_e$ cross-section measurements. 
\item Provide a technology ($\mu$ decay ring) test demonstration and $\mu$ beam diagnostics test bed.
\item Provide a precisely understood $\nu$ beam for detector studies
\end{enumerate}  

\begin{figure*}[b]
    \includegraphics[width=0.6\textwidth]{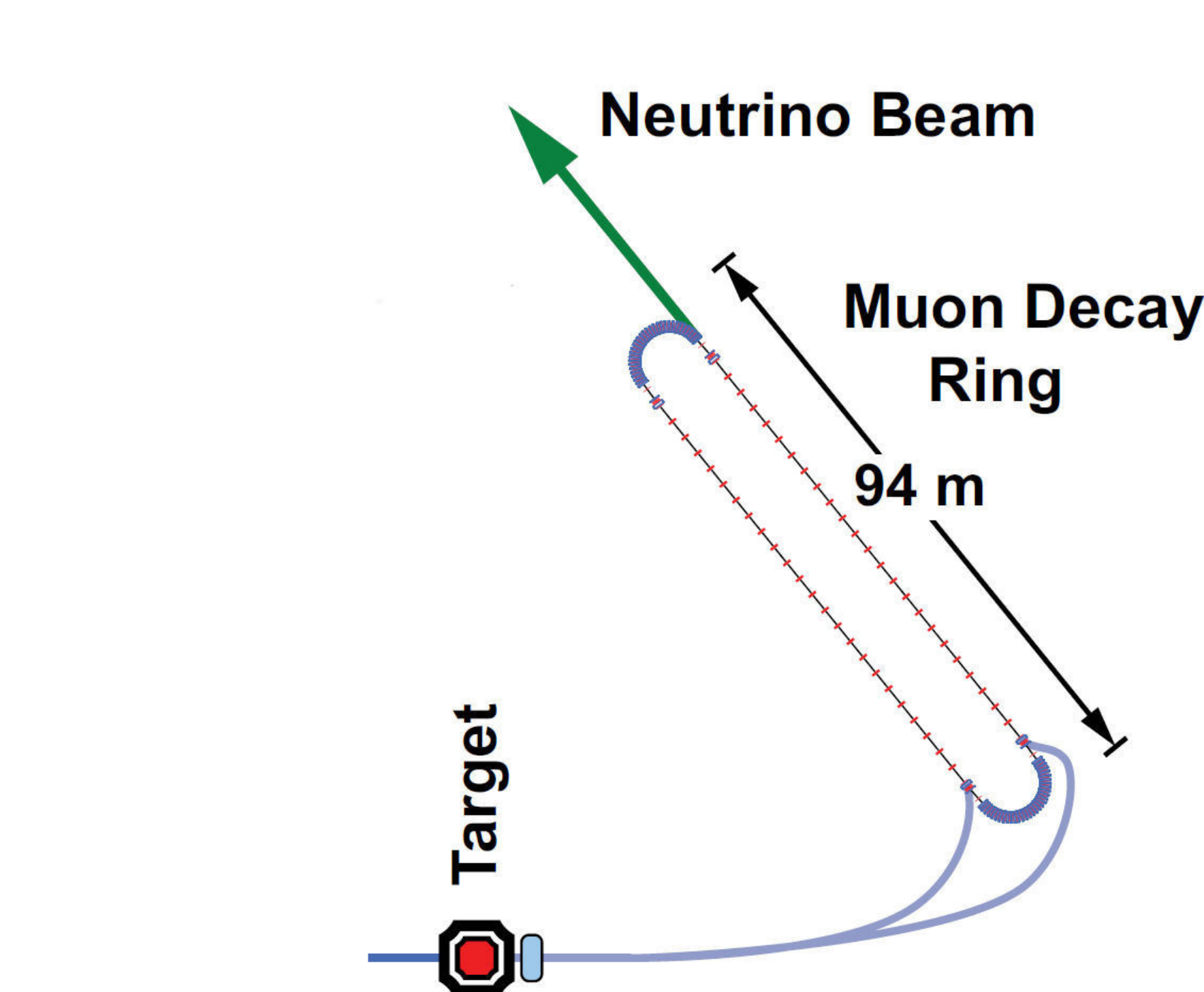}
  \caption{Schematic of Very-Low Energy Neutrino Factory}
  \label{fig:VLENF}
\end{figure*}

Pions are collected from a target, then transported to and injected into a storage ring where they decay to 
muons. The muons then subsequently decay into electrons and neutrinos.  We are starting with a storage ring design that
is optimized for 2 GeV/c muon momentum. In this case, the energy is optimized for the needs of both the oscillation and the cross section
physics.  See Fig.~\ref{fig:VLENF} for a schematic of the facility.

For positive muons, the decay,  $\mu^+ \rightarrow e^+ + \bar{\nu}_\mu + \nu_e$,
yields a neutrino beam of precisely known flavor content.  In addition, if the circulating muon flux in the ring is measured
accurately (with beam-current transformers, for example) then the neutrino beam flavor content and flux are precisely known.
Near and far detectors are placed along one of the straight sections of the racetrack decay ring.
The near detector can be placed at 20-50~meters from
the end of the straight and will measure neutrino-nucleon
cross sections that are potentially important for future long-baseline experiments. 
This would include the first precision measurements of $\nu_e$ and $\bar{\nu}_e$
cross sections. A far detector at 800-1000 m would study neutrino oscillation physics and would be capable
of performing searches in both appearance and disappearance channels (see Ref.~\cite{Winter:2012sk} for a recent disappearance proof of
principle).
The experiment will take advantage of the ``golden channel" of oscillation appearance $\nu_e \rightarrow \nu_\mu$, where the resulting 
final state has a muon of the wrong-sign from interactions of the $\nu_\mu$ in the beam.
In the case of $\mu^+$s stored in the ring, this would mean the observation of an event with a $\mu^-$.
This detector would need to be magnetized for the
wrong-sign muon appearance channel, as is the case for the baseline Neutrino Factory detector \cite{NF:2011aa}.
A number of possibilities for the far detector exist.  However, a magnetized iron detector similar to that used
in MINOS is likely to be the most straightforward approach for the far detector design.
For the purposes of the VLENF oscillation physics, a detector inspired by MINOS, but with 
thinner plates and much larger excitation current (larger B field) is assumed.

Both conventional  (FODO) and fixed field alternating 
gradient (FFAG) lattices are being investigated.
The racetrack FFAG \cite{MoriRFFAG} is a very promising possibility for this application due to its large momentum
acceptance ($\simeq$ 20\%).  A schematic of the current racetrack FFAG concept is shown in Fig.~\ref{fig:RFFAG}.
Both the FODO design implemented in our simulation (see below) and the FFAG referenced above would require only normal conducting
magnets which would simplify construction, commissioning, and operations. 

\begin{figure*}
    \includegraphics[width=1.0\textwidth]{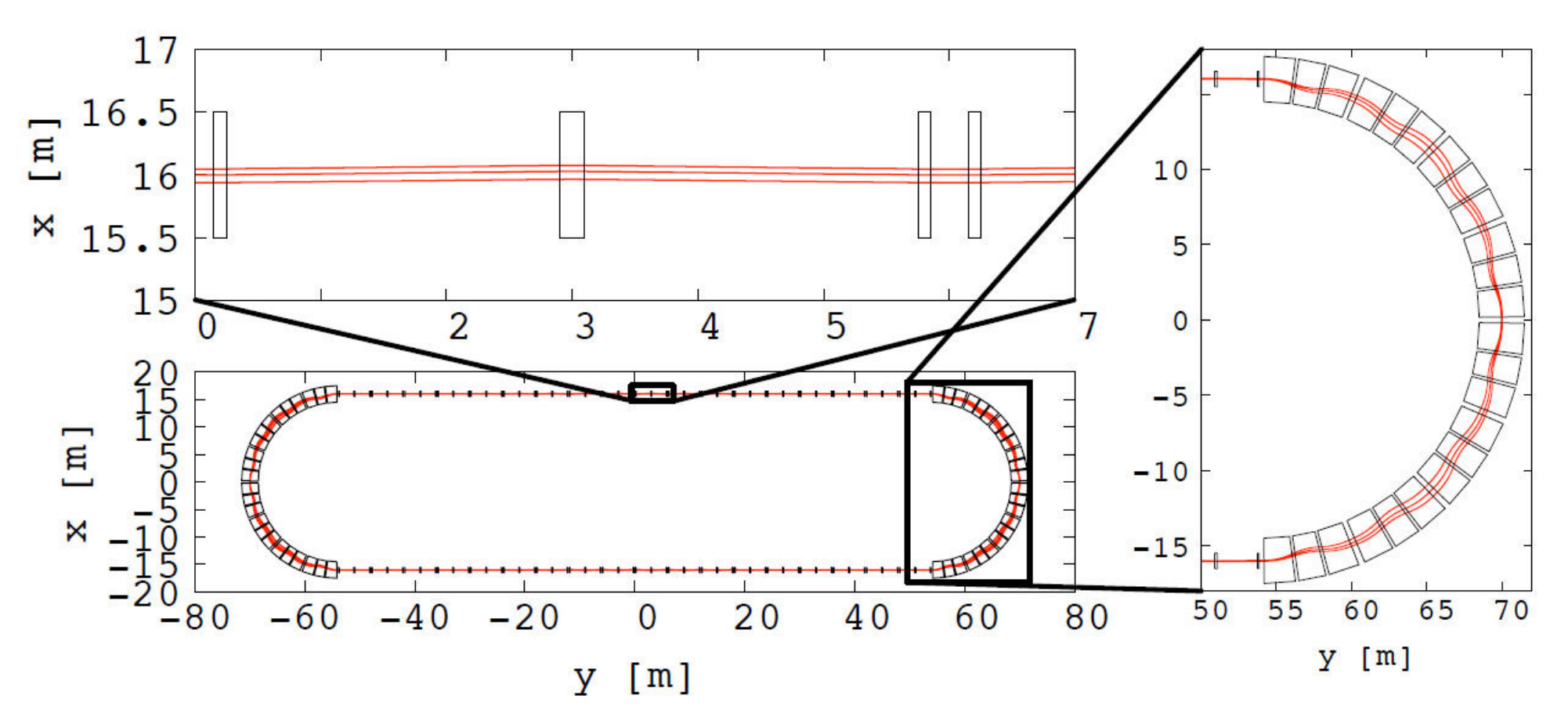}
  \caption{Top view of the racetrack FFAG lattice (bottom left scheme). The top left scheme shows a
     zoom on the straight section and the right scheme a zoom on the arc section.}
  \label{fig:RFFAG}
\end{figure*}

\subsubsection*{Physics Reach}

At present, the VLENF has been simulated using G4Beamline \cite{Kaplan:2007zza} only for a conventional (FODO) ring lattice 
with a central $\mu$ momentum of 3.0 GeV/c.  The design input to this simulation included:

\begin{enumerate}
\item{8 GeV protons on a 2 $\lambda_I$ Be target.}
\item{Decay ring tuned for muons with kinetic energy of 3.0 GeV.}
\item{Decay ring acceptance of $\frac{\delta p}{p}$ = 2\% rms}.
\end{enumerate}

The flux of the circulating muon beam (after 10 turns to allow all pions to decay) determined by this simulation was $1.1 \times 10^8$ muons 
per $10^{12}$ protons on target .  This yields a figure of merit of $1.1 \times 10^{-4}$ stored $\mu$ per proton on target.  However, the 
simulation does not yet include pion capture, transport and injection into the decay ring.  For this analysis, it was assumed all pions 
(within a 100 mrad forward cone from the target) make it into the ring.  Although a conceptual scheme for capture, transport and 
injection exists, a full design and its simulation will be part of future work.

In the sensitivity
calculations that follow, we have assumed the racetrack FFAG decay ring performance described in \cite{MoriRFFAG}:  a center momentum 
of 2 GeV and a momentum acceptance of 16\% (full width).  We also have assumed a total exposure of $10^{21}$ protons on target, which yields
$\sim 10^{18}$ muons.  Finally, a detector fiducial mass of 1kT is assumed.  Background
 levels are based on extrapolations from the MIND analysis \cite{NF:2011aa} (see detector considerations below).  The background rejection 
for the wrong-sign muons is assumed to be $10^{-4}$ in  neutral current (NC) events and $10^{-5}$ for $\mu$ charge misidentified
charged current (CC) events.  The 
backgrounds from $\nu_e \to \nu_e$ CC and NC are negligible. Sensitivities have been calculated for both $\nu$ and $\bar{\nu}$ running.    
Finally, a 35\% background normalization error has been assumed as well as a 2\% 
signal systematic error.

Tables of the raw event rates are given below. The first row corresponds to the ``golden channel" appearance signal. The other rows are 
potential backgrounds to the signal.  The backgrounds that drive the analyses are charged-current $\bar{\nu}_\mu \to \bar{\nu}_{\mu}$
 and $\nu_\mu \to \nu_\mu$, for neutrinos and antineutrinos, respectively.

\begin{table}
     \centering
     	  \begin{tabular}{c|r}
     	     	  \multicolumn{2}{c}{$\bar{\nu}$-mode with stored $\mu^-$}\\
     	     	  Channel name & Number Events\\
     	     	  \hline
     	     	  $\bar{\nu}_e \to \bar{\nu}_\mu$ CC & 23\\
     	     	  $\bar{\nu}_e \to \bar{\nu}_e$ CC & 24824\\
     	     	  $\nu_\mu \to \nu_\mu$ NC & 26657\\
     	     	  $\nu_\mu \to \nu_\mu$ CC & 72539\\
     	  \end{tabular}
	  \hspace{1.5cm}
     	  \begin{tabular}{c|r}
     	     	  \multicolumn{2}{c}{$\nu$-mode with stored $\mu^+$}\\
     	     	  Channel name & Number Events\\
     	     	  \hline
     	     	  $\nu_e \to \nu_\mu$ CC & 60\\
     	     	  $\nu_e \to \nu_e$ CC & 61448\\
     	     	  $\bar{\nu}_\mu \to \bar{\nu}_\mu$ NC & 12456\\
     	     	  $\bar{\nu}_\mu \to \bar{\nu}_\mu$ CC & 30596\\
     	  \end{tabular}

     \caption{VLENF raw event rates for $10^{21}$ POT.}
\end{table}

\begin{figure*}
     \includegraphics[width=0.45\textwidth]{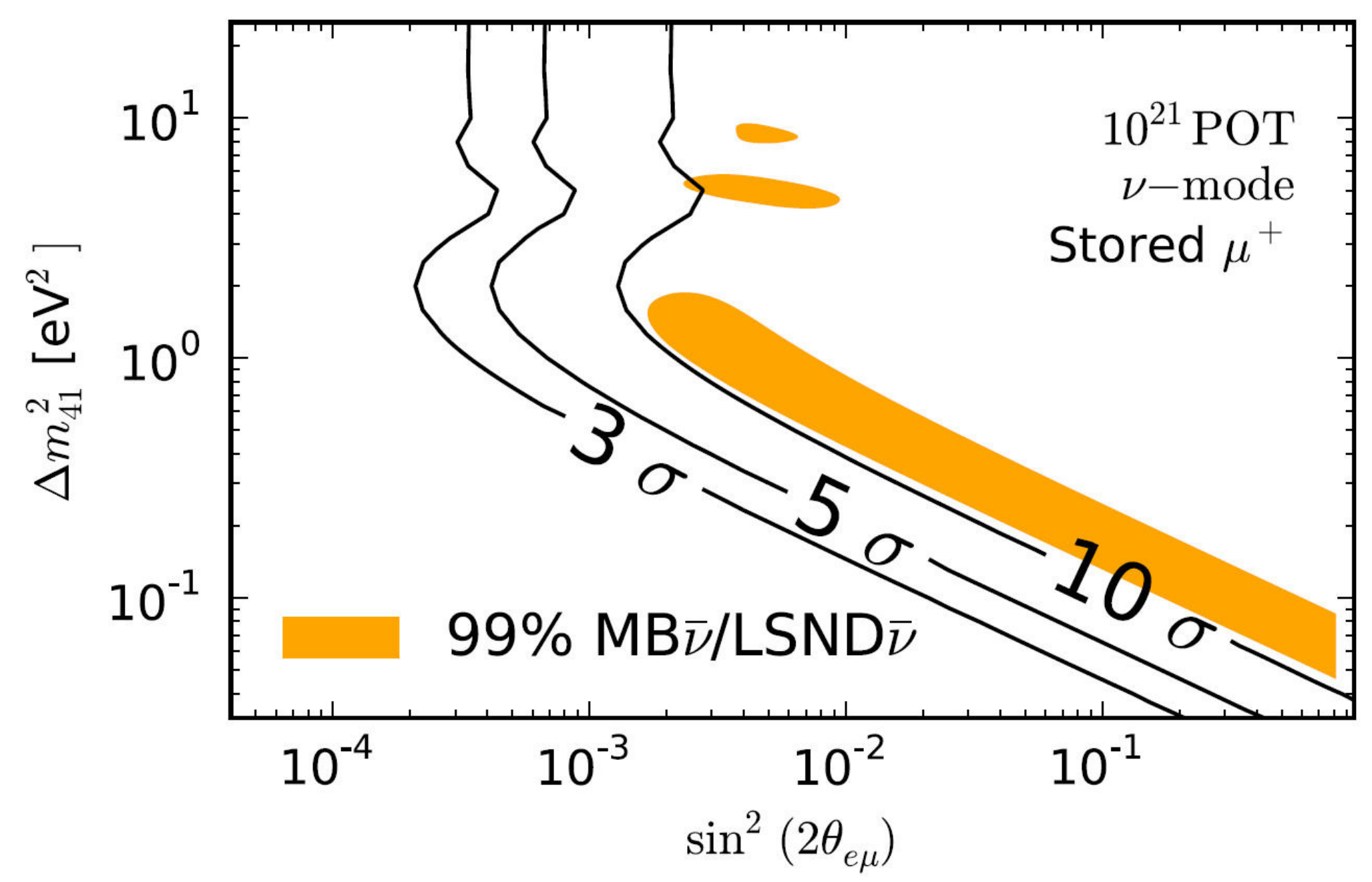} 
      \includegraphics[width=0.45\textwidth]{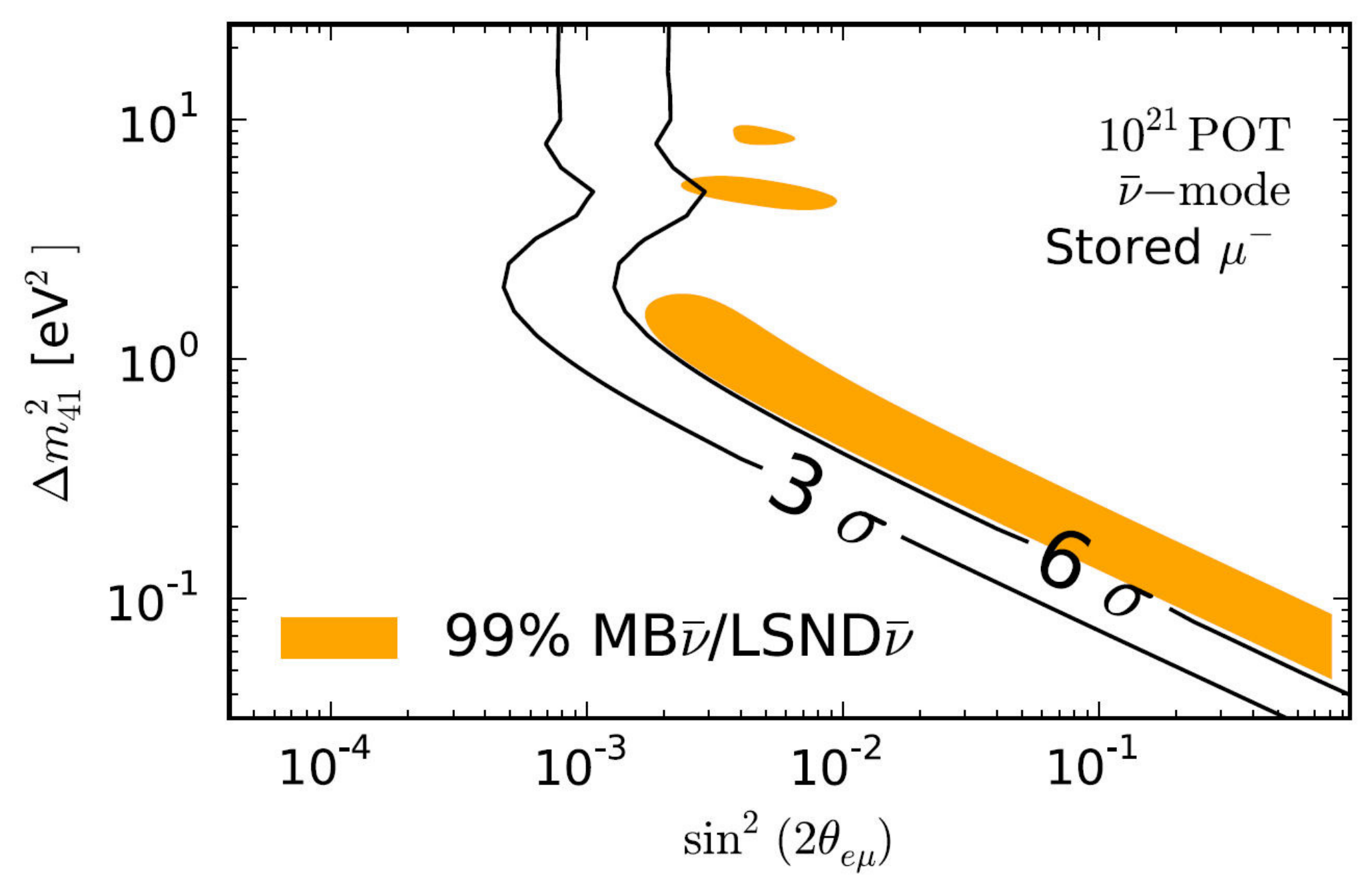}
  \caption{Appearance channel sensitivity plots. Left for $\nu$ running, right for $\bar{\nu}$ running.}
  \label{fig:Sens}
\end{figure*}

The oscillation sensitivities have been computed using the GLoBES software (version 3.1.10)  \cite{Huber:2004ka,Huber:2007ji}.  Since 
GLoBES, by default, only allows for a $3 \times 3$ mixing matrix, the SNU (version 1.1) add-on  \cite{Kopp:2006wp,Kopp:2007ne} is used 
to extend computations in GLoBES to $4\times4$ mixing matrices.  See \cite{Tunnell:2011ya} for details regarding the analysis.
Fig.~\ref{fig:Sens} shows the appearance sensitivity for this facility under the above assumptions.

\subsubsection*{Detector Considerations}

\begin{figure*}
    \includegraphics[width=0.7\textwidth]{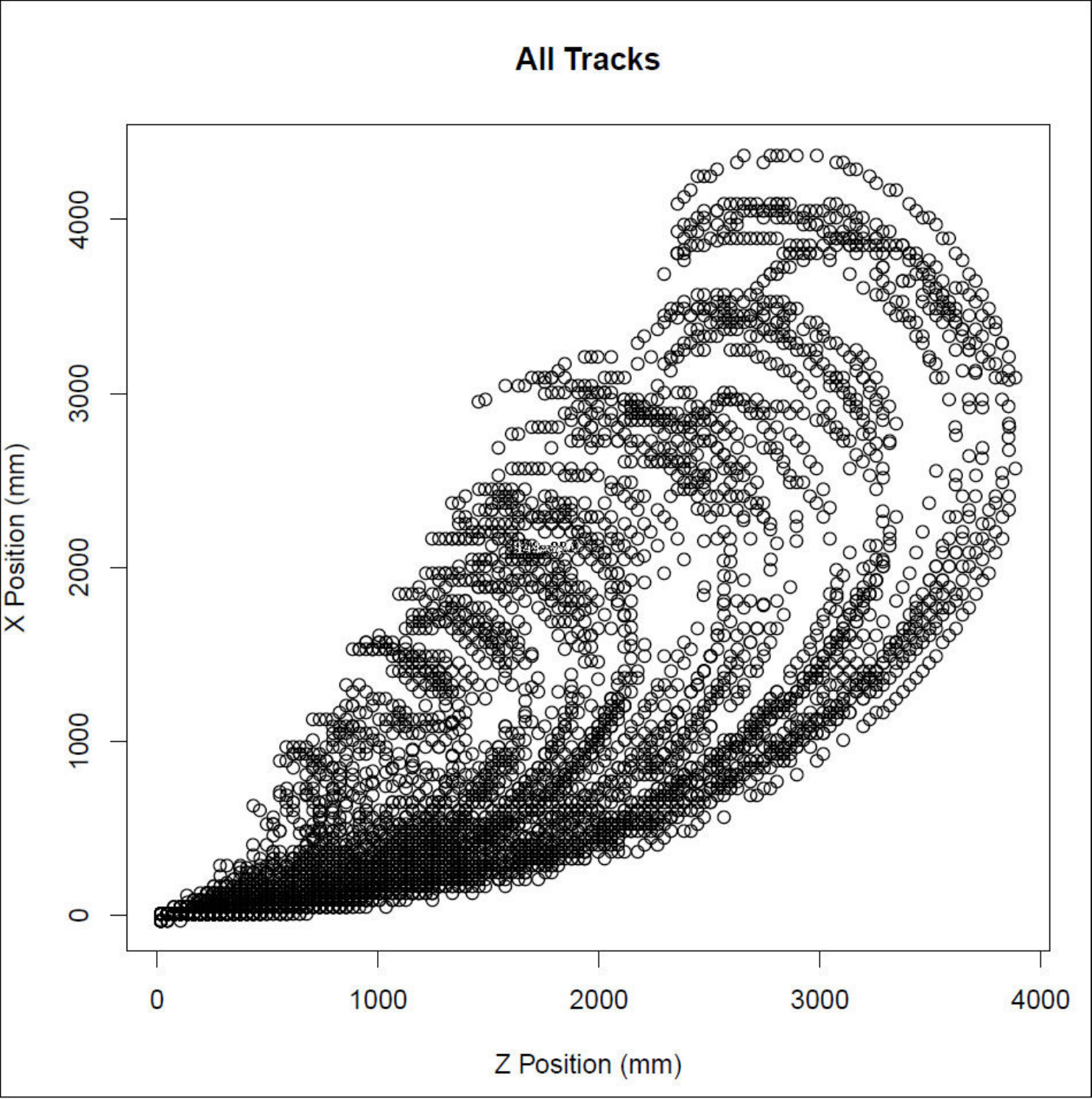}
  \caption{Plot of muon events (XY projection) for 10 muons in each momentum bin where the momentum range is from 0.1 GeV/c to 3.0 GeV/c.}
  \label{fig:SuperBIND}
\end{figure*}

The concept for the far detector for the VLENF that we are exploring is basically a modified MINOS near detector \cite{Nelson:2001gm}.  
However, in our case we have reduced the plate thickness to 1 cm (from the 1" used in MINOS), are using two layers of scintillator between
 Fe plates (X and Y measurement) and will increase the excitation current to roughly 270 kA-turns by using multiple turns of the superconducting 
transmission line \cite{Ambrosio:2001ej}.  A preliminary ANSYS analysis indicates that for a 6 m diameter plate, the magnetic field is $\geq$ 1.8T 
throughout the entire plate volume.   In addition, a very preliminary Geant4 simulation indicates that the $\mu$ charge mis-ID rate is at or below 
$10^{-4}$ for p$_\mu \geq$ 250 MeV/c.  See Fig.~\ref{fig:SuperBIND}.  This gives us confidence that with a full Geant4 simulation with an 
optimized tracking algorithm, we will reach the $10^{-5}$ specification given above.

\subsubsection*{Conclusions}

As can been see from Fig.~\ref{fig:Sens}, this facility has the potential to give unprecedented performance in the large $\Delta$m$^2$ 
oscillation regime providing the opportunity to explore physics beyond the $\nu$SM.  If new physics exists, the VLENF will allow for a 
detailed study of these processes.  In addition, the $\mu$ storage ring presents the only way to obtain large samples of $\nu_e$ events 
for cross-section measurements and offers the only experiment that can measure both $\nu_e$ and $\nu_\mu$ cross-sections with 
high precision.   The next generation of long-baseline oscillation experiments will face a significant challenge in order to get their 
systematic errors to the 1\% level.  Gaining a better understanding of these cross sections will be essential for them to meet this goal.

\clearpage
\subsection{Probing active-sterile oscillations with the atmospheric neutrino signal in large iron/liquid argon detectors\footnote{Contributed by Pomita Ghoshal and Raj Gandhi.}}
\label{Ghoshal}
The down-going atmospheric $\nu_{\mu}$ and ${\bar{\nu_{\mu}}}$ fluxes can be significantly altered due to the presence of eV$^2$-scale active-sterile oscillations.
We study the sensitivity of  a large Liquid Argon detector and  a large magnetized iron detector
like the proposed ICAL at INO to these oscillations. Using the allowed sterile parameter ranges in a 3+1 mixing framework,
we perform a fit assuming active-sterile oscillations 
in both the muon neutrino and antineutrino sectors,
and compute
oscillation exclusion limits  
 using atmospheric down-going muon neutrino and antineutrino events. We find that (for both $\nu_{\mu}$ and ${\bar{\nu_{\mu}}}$) a Liquid Argon detector, an ICAL-like detector or a combined analysis of both detectors with an exposure of 1 Mt yr provides significant sensitivity to regions of parameter space in the range  $0.5 < \Delta m^2 < 5$ eV$^2$ 
for $\sin^2 2\theta_{\mu\mu}\geq 
0.07$. Thus atmospheric neutrino experiments can provide complementary coverage in these regions, improving sensitivity limits 
in combination with bounds from other experiments on these parameters.

We perform our statistical analysis using two kinds of proposed detectors:

\begin{itemize}

\item A large Liquid Argon detector\cite{Rubbia:2004tz,Bueno:2007um,Cline:2006st}, which can detect 
charged particles with very good resolution over the energy range of MeV to multi GeV,  
with magnetization over a 100 kT
volume with a magnetic field of about 1 tesla \cite{Ereditato:2005yx}.
For the down-going events, with a baseline range of 15 to 130 Kms,  and energies upto 20 GeV (above which the flux is very small), the L/E range is exactly the one relevant to sterile-parameter induced oscillations.  
We  assume the following
energy resolutions over the  ranges relevant to our
calculations 
\cite{Bueno:2007um}: $\sigma_{E_e} = 0.01$, $\sigma_{E_{\mu}} = 0.01$, $\sigma_{E_{had}} = \sqrt{(0.15)^2/E_{had} + (0.03)^2}$,
$\sigma_{\theta_e} = 0.03~{\mathrm{radians}} = 1.72^o$, $\sigma_{\theta_{\mu}} = 0.04~{\mathrm{radians}} = 2.29^o$,   
$\sigma_{\theta_{had}} = 0.04~{\mathrm{radians}} = 2.29^o$.  
Here $E_{had}$ is the hadron energy in GeV, $\sigma_E$ are the energy resolutions and $\sigma_{\theta}$ are the angular resolutions 
of electrons, muons and hadrons as indicated. 
The energy and angular resolution of the detector in terms of the neutrino energy and zenith angle can be derived from the above.
In our computation, we take the average rapidity in the GeV energy region 
to be 0.45 for neutrinos and 0.3 for antineutrinos \cite{Gandhi:1995tf}. 
The energy threshold and ranges in which charge identification is feasible  are 
$E_{threshold}  = 800$ MeV for muons and
$E_{electron}  = 1-5$ GeV  for electrons.
Charged lepton detection and separation (e vs $\mu$) without charge identification 
is possible for $E_{lepton}>$ few MeV.

\item An iron calorimeter detector like ICAL \cite{ino}, which, like the above detector, offers the advantage of muon charge discrimination using magnetization with a field of 1.3 tesla, allowing a separate observation of atmospheric muon neutrino and antineutrino events.  
For this detector, standard resolutions of 10$^o$ in angle and 15$\%$ in energy are assumed. 

\end{itemize}

The muon event rates are a function of both $P_{{{\mu}}{{\mu}}}$ and  $P_{{{e}}{{\mu}}}$, but  $P_{{{e}}{{\mu}}}$ is highly suppressed due to the smallness of the parameter $\sin^2 2\theta_{e\mu}$. Thus  it is reasonable to anticipate that the downgoing muon event spectrum should reflect the behaviour of $P_{{{\mu}}{{\mu}}}$, and hence show signatures of oscillations due to the sterile parameters $\theta_{\mu\mu}$ and $\Delta m^2$ defined above.

A 1 Mt yr exposure is assumed for both types of detectors, and flux uncertainties and systematic errors
are incorporated by the pull method \cite{GonzalezGarcia:2004wg}. The values of uncertainties are chosen as in \cite{Gandhi:2007td}. 
We take a double binning in energy and zenith angle, with 20 energy bins in the range 1-20 GeV and 18 $\cos \theta_z$ bins in the range 0.1 to 1.0.   
The atmospheric fluxes are taken from the 3-dimensional calculation in \cite{Honda:2004yz}.
The earth matter profile defined in \cite{Dziewonski:1981xy} is used to take into account matter effects 
on the oscillation probabilities.    

One can extract the statistical sensitivity with which experimental-set ups like the ones described above may be able to exclude sterile-scale oscillations and constrain sterile parameters using the downgoing muon and anti-muon event spectra as the signal.
We perform this study in two stages: 

a) The best exclusion limits possible from this analysis are determined using simultaneously the downgoing muon neutrino and antineutrino event spectra for both kinds of detectors and doing a combined fit.

b) In order to test the MiniBooNE/LSND antineutrino results, the atmospheric downgoing muon antineutrino event spectra with sterile oscillations for both kinds of detectors are analysed to determine the bounds for the sterile parameters, and compared with the bounds from MiniBooNE.      

\subsubsection*{Exclusion limits with a combined $\nu_{\mu}$, ${\bar{\nu}_{\mu}}$ analysis}

\begin{figure}[t]
{\centerline{
\includegraphics[width=.45\textwidth]{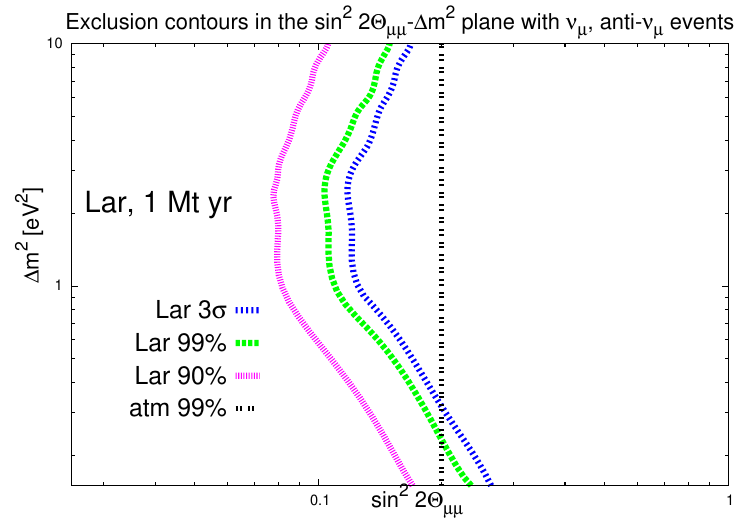}
\hspace{0.05\textwidth}
\includegraphics[width=.45\textwidth]{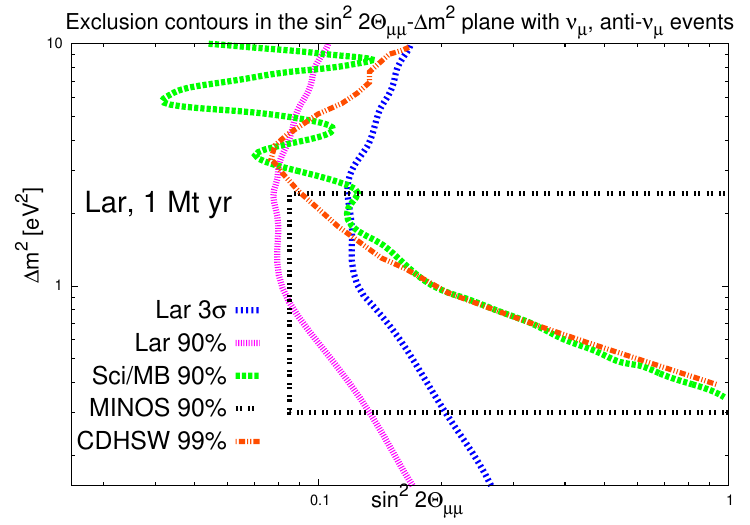}}
}
\hskip 0.7cm
{\bf (a)}
\hskip 7.3cm
{\bf (b)}
\caption[]{\footnotesize {Exclusion curves in the $\sin^2 2\theta_{\mu\mu}-\Delta m^2$ plane using Liquid Argon (1 Mt yr) downgoing muon and anti-muon events with sterile oscillations - (a) Comparison with 99$\%$ c.l. limit from atmospheric analysis \cite{Giunti:2011gz,Maltoni:2007zf}, 
(b) Comparison with 90$\%$ limits from SciBooNE/MiniBooNE {\cite{Mahn:2011ea}} and MINOS {\cite{Adamson:2011ku}}, 99$\%$ c.l. limit from CDHSW {\cite{Giunti:2011gz,Dydak:1983zq}}.}}
\label{fig1} 
\end{figure}

For this study, 
the sterile oscillation exclusion limits are computed by combining both the atmospheric downgoing muon and antimuon event spectra
with sterile-scale oscillations. 
This involves  taking i) the 'expected' spectrum $N_{th}$, 
in which sterile oscillations are included and the test values of the sterile parameters are varied, 
and ii) the 'observed' spectrum $N_{ex}{\mathrm{(no-osc)}}$, where 'no-osc' indicates no sterile-scale oscillations.
Since homogeneity is expected between the $\nu$ and ${\bar{\nu}}$ sectors in the 3+1 scenario,
we assume identical sterile-scale oscillations in both sectors.
The exclusion limits are presented in Figure 1 and 2 for a Liquid Argon detector and an ICAL detector respectively, with an exposure of 1 Mt yr for both. Figure 3 shows the results obtained with a combined analysis of ICAL (1 Mt yr) + Liquid Argon (1 Mt yr).
In each case, the left panel gives a comparison of the bounds obtained from our analysis with the 99$\%$ c.l. exclusion region from atmospheric neutrino data \cite{Giunti:2011gz,Maltoni:2007zf}, and the right panel compares our results with 
the 90$\%$ limits from SciBooNE/MiniBooNE \cite{Mahn:2011ea} and MINOS \cite{Adamson:2011ku} and the 99$\%$ limit from the CDHSW disappearance analysis \cite{Giunti:2011gz,Dydak:1983zq}.
Figure 4 compares the 90$\%$ c.l. bounds from our Liquid Argon and combined Liquid Argon/ICAL studies with the allowed regions at 90$\%$ c.l. from the most recent global analysis in \cite{Giunti:2011cp}.  

 With a Liquid Argon detector and an exposure of 1 Mt yr,
regions greater than $\sin^2 2\theta_{\mu\mu}\sim 0.08$ can be excluded at 90$\%$ c.l. with a combination of muon and anti-muon events over most of the allowed $\Delta m^2$ range.
An ICAL-like detector with a similar exposure gives a weaker 90$\%$ c.l. exclusion bound 
at $\sin^2 2\theta_{\mu\mu} \sim 0.12$ with this combination.  
A combined analysis of the two experiments gives a 3$\sigma$ exclusion bound for $\sin^2 2\theta_{\mu\mu}\geq 0.1$, 
and a 90$\%$ c.l. limit for $\sin^2 2\theta_{\mu\mu}\geq 0.07$,
which is seen to be an improvement over the earlier bounds obtained from atmospheric neutrinos, as well as those from CDHSW, SciBooNE/MiniBooNE and MINOS, over significant regions of the parameter space. 
Also, Figure 4 shows that the exclusion limits given by Liquid Argon or a Liquid Argon/ICAL combination can significantly affect the fits and allowed regions given by the current global analysis. 

\begin{figure}[t]
{\centerline{
\includegraphics[width=.45\textwidth]{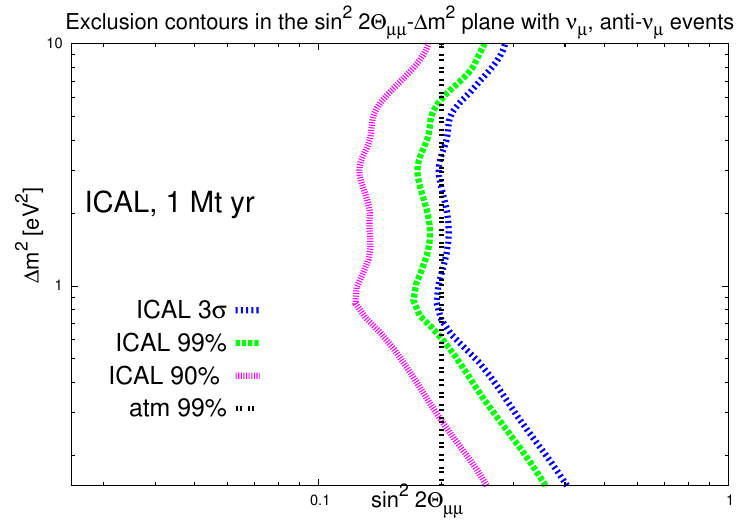}
\hspace{0.05\textwidth}
\includegraphics[width=.45\textwidth]{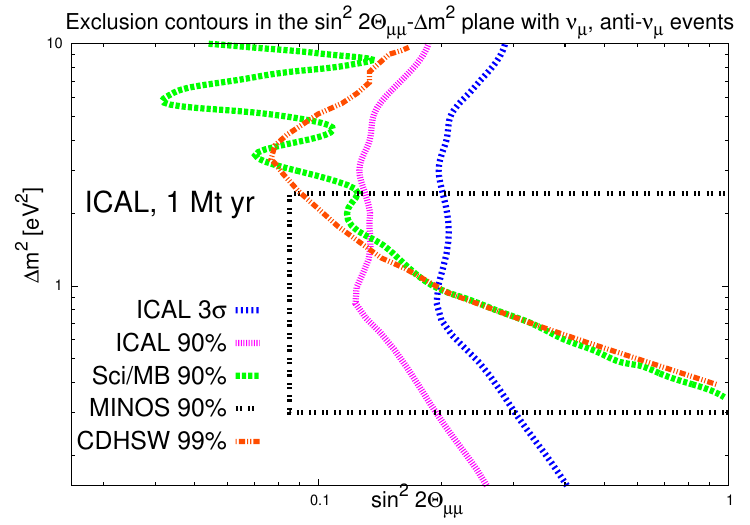}}
}
\hskip 0.7cm
{\bf (a)}
\hskip 7.3cm
{\bf (b)}
\caption[]{\footnotesize {Same as Figure 1 using ICAL (1 Mt yr) downgoing muon and antimuon events.}}
\label{fig2} 
\end{figure}

\begin{figure}[t]
{\centerline{
\includegraphics[width=.45\textwidth]{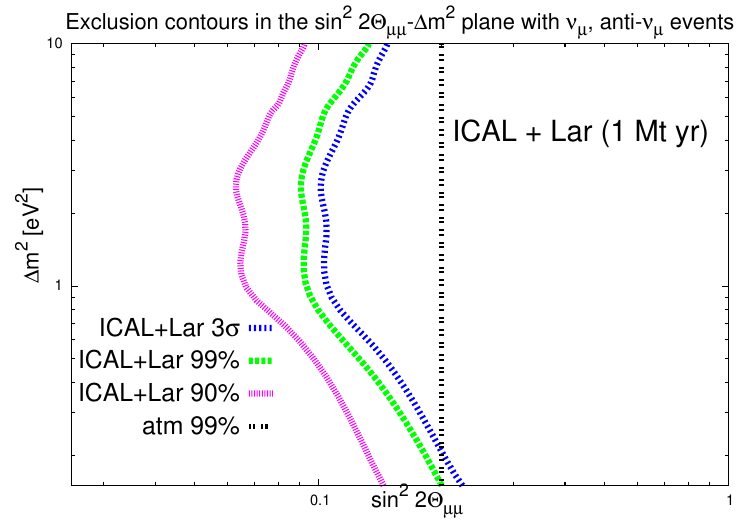}
\hspace{0.05\textwidth}
\includegraphics[width=.45\textwidth]{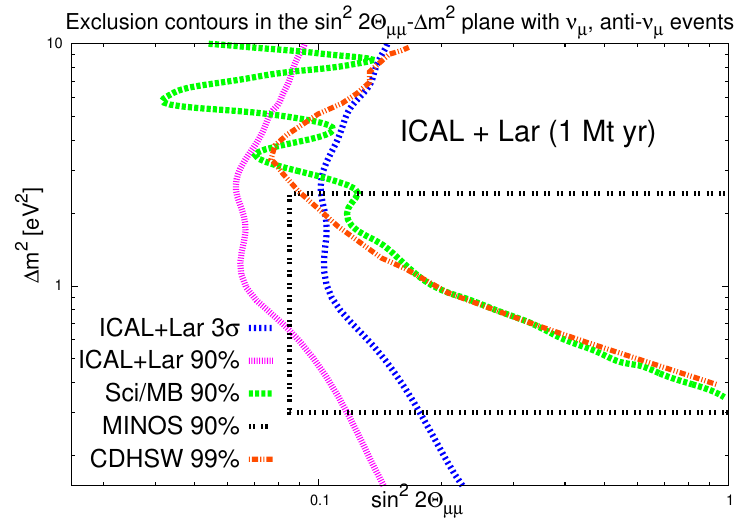}}
}
\hskip 0.7cm
{\bf (a)}
\hskip 7.3cm
{\bf (b)}
\caption[]{\footnotesize {Same as Figure 1 using ICAL (1 Mt yr) $+$ Liquid Argon (1 Mt yr) downgoing muon and antimuon events.}}
\label{fig3} 
\end{figure}

\begin{figure}[t]
{\centerline{
\includegraphics[width=.45\textwidth]{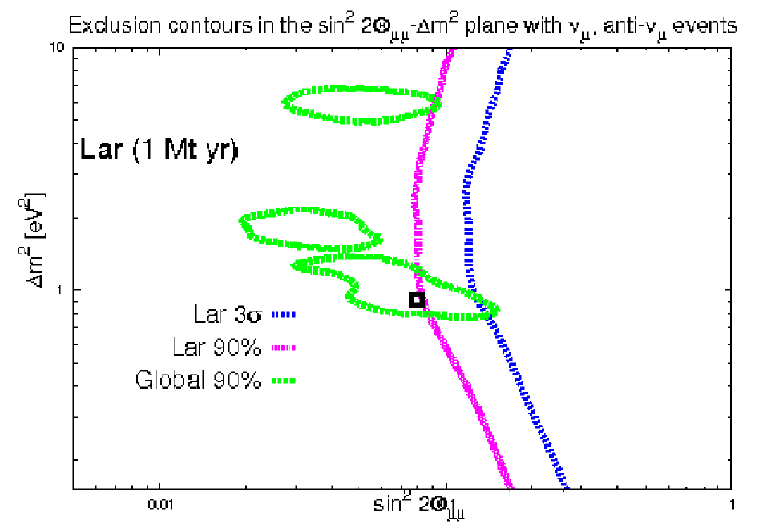}
\hspace{0.05\textwidth}
\includegraphics[width=.45\textwidth]{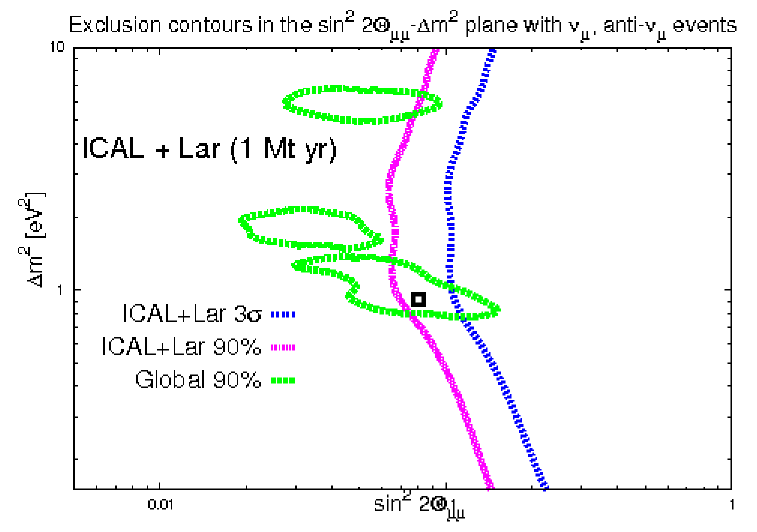}}
}
\hskip 0.7cm
{\bf (a)}
\hskip 7.3cm
{\bf (b)}
\caption[]{\footnotesize {Exclusion curves in the $\sin^2 2\theta_{\mu\mu}-\Delta m^2$ plane using (a) Liquid Argon (1 Mt yr) and (b) ICAL (1 Mt yr) + Liquid Argon (1 Mt yr) downgoing muon and anti-muon events.  Comparison with 90$\%$ c.l.
allowed regions from the GLO-LOW global analysis in \cite{Giunti:2011cp}. The small squares denote the best-fit point from the global analysis.}}
\label{fig4} 
\end{figure}

\begin{figure}[t]
{\centerline{
\includegraphics[width=.45\textwidth]{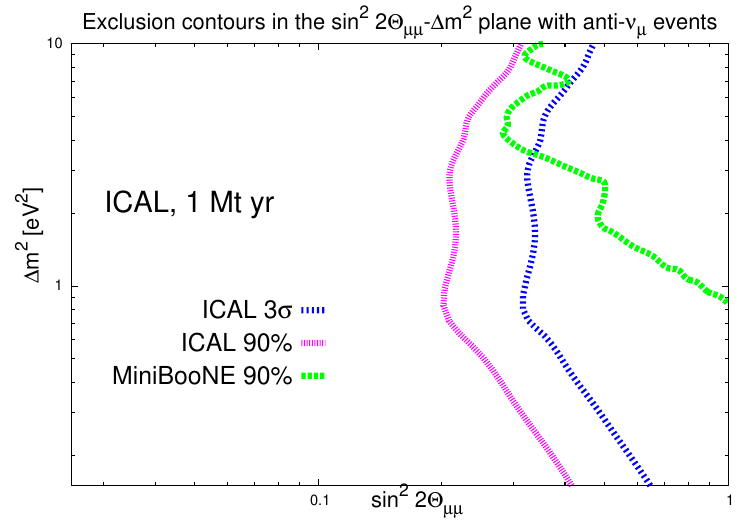}
\hspace{0.05\textwidth}
\includegraphics[width=.45\textwidth]{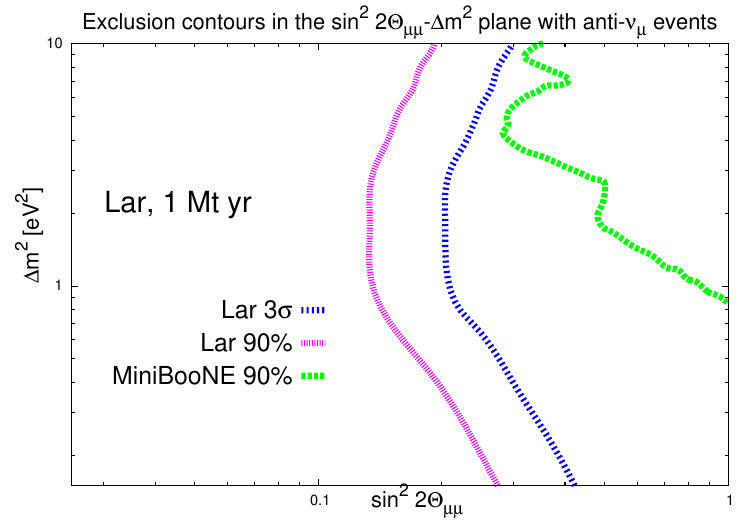}}
}
\hskip 0.7cm
{\bf (a)}
\hskip 7.3cm
{\bf (b)}
\caption[]{\footnotesize {Exclusion curves in the $\sin^2 2\theta_{\mu\mu}-\Delta m^2$ plane using (a) ICAL (1 Mt yr) downgoing anti-muon events and (b) Liquid Argon (1 Mt yr) downgoing anti-muon events with sterile oscillations. 
oscillations (upper row),
The limits at 90$\%$ c.l. and 3$\sigma$ are indicated.
The 90$\%$ exclusion limit from the MiniBooNE \cite{AguilarArevalo:2009yj} antineutrino analysis is also shown.}}
\label{fig5} 
\end{figure}

\subsubsection*{Testing LSND and Miniboone results with sterile oscillations in the ${\bar{\nu}}$ sector}
Here we compute the   
 sterile oscillation exclusion limits using the downgoing antimuon event spectrum
with sterile-scale oscillations, for comparison with the results from MiniBooNE/LSND antineutrino data. 
The bounds obtained from this analysis are presented in Figure 5.
The left and right panels correspond to the exclusion bounds for the parameters $\Delta m^2$ and 
$\sin^2 2\theta_{\mu\mu}$ with the downgoing ${\bar{\nu}}_{\mu}$ spectrum for
an ICAL detector and a Liquid Argon detector respectively, both with an exposure of 1 Mt yr.   
The 90$\%$ exclusion limit from MiniBooNE \cite{AguilarArevalo:2009yj} is superimposed on the figure in both cases. 
It can be seen that this set-up provides a 90$\%$ c.l. exclusion capacity with 
an ICAL-like detector with an exposure of 1 Mt yr for about $\sin^2 2\theta_{\mu\mu} > 0.2$ for a range $0.5 < \Delta m^2 < 5$ eV$^2$, and for a Liquid Argon detector with an exposure of 1 Mt yr for about $\sin^2 2\theta_{\mu\mu} > 0.15$ for a range $0.5 < \Delta m^2 < 5$ eV$^2$. These are stronger bounds than those from the MiniBooNE antineutrino analysis.

\subsubsection*{Summary}

We have studied the  possible sensitivity of atmospheric neutrino data in a large magnetized iron calorimeter detector
like the proposed ICAL at INO and a large Liquid Argon detector, to eV$^2$-scale active-sterile neutrino and antineutrino oscillations. 
With the present sterile parameter ranges 
in a 3+1 mixing framework, 
down-going atmospheric ${{\nu}_\mu}$ and ${\bar{\nu}_\mu}$ events can show signatures of eV$^2$-scale oscillations, due to their suitable energy and baseline range (neutrinos with multi-GeV energies and baselines ranging from about 10 to 100 Kms). The results of our analysis are as follows:

\begin{itemize}

\item {\bf{Active-sterile oscillation exclusion limits using both $\nu_{\mu}$ and ${\bar{\nu_{\mu}}}$ events:}}

a) With a Liquid Argon detector and an exposure of 1 Mt yr,
regions greater than $\sin^2 2\theta_{\mu\mu}\sim 0.08$ can be excluded at 90$\%$ c.l. with a combination of muon and anti-muon events, over most of the $\Delta m^2$ range. 
A 3$\sigma$ exclusion bound is possible for $\sin^2 2\theta_{\mu\mu}\sim 0.15$.

b) With an ICAL-like detector and an exposure of 1 Mt yr,
a weaker 90 $\%$ c.l. exclusion bound is obtained at $\sin^2 2\theta_{\mu\mu} \sim  
0.12$ with a combination of muon and anti-muon events.

c) With a combined analysis of ICAL (1 Mt yr) and Liquid Argon (1 Mt yr), a 90$\%$ c.l. exclusion limit is obtained for 
$\sin^2 2\theta_{\mu\mu} \geq 0.07$ and a 3$\sigma$ bound for $\sin^2 2\theta_{\mu\mu} \geq  
0.1$, which compares favorably with present limits from CDHSW, MINOS, MiniBooNE and atmospheric neutrinos (Figure 3),
and is able to exclude a significant part of the allowed region from the most recent global analysis (Figure 4). 

\item {\bf{Fit with active-sterile oscillations in the muon antineutrino sector for comparison with MiniBooNE:}}

Such oscillations can be excluded by this set-up at 90$\%$ c.l. with 

a) an ICAL-like detector with an exposure of 1 Mt yr for about $\sin^2 2\theta_{\mu\mu} > 0.2$ for a range $0.5 < \Delta m^2 < 5$ eV$^2$, 

b) A Liquid Argon detector with an exposure of 1 Mt yr for about $\sin^2 2\theta_{\mu\mu} > 0.15$ for a range $0.5 < \Delta m^2 < 5$ eV$^2$.  

\end{itemize}

The limits for both detectors from an exposure of 1 Mt yr may be accessible 
in a time-frame of about 10-15 years. 

In conclusion, a down-going event analysis using large future atmospheric  detectors may be helpful in providing 
significant complementary constraints on the sterile parameters, which can strengthen existing bounds 
when combined with other experimental signatures of sterile-scale oscillations.  Evidence (or the lack of it) from such detectors has the advantage of originating in a sector which is different from those currently providing clues pointing to the existence of sterile neutrinos (i.e short-baseline experiments). Additionally, it permits access to a wide-band of $L/E$, which is important if oscillatory behaviour is to be unambiguously tested.

\end{appendix}
\clearpage
%
\bibliography{whitepaper}
%
\end{document}